\documentclass[preprint,authoryear,12pt,portrait]{elsarticle}
\usepackage{graphicx}
\usepackage{amssymb}
\usepackage{lineno}

\journal{J. Theor. Biol.}

\usepackage{ifthen}
\usepackage{xifthen}
\usepackage{ifthenx}
\newcommand{\EQ}{eq}
\newcommand{\TBL}{tbl}
\newcommand{\FIG}{fig}

\usepackage{graphicx}
\usepackage[extendedchar]{grffile}

\usepackage{color}
\usepackage{xcolor}
\usepackage{times}
\usepackage{txfonts}
\usepackage{bm}

\usepackage[hidelinks]{hyperref}        

\newcommand{\Detail}[1]{}

\newcommand{\Skip}[1]{}                 
\newcommand{\script}[1]{{\mbox{\scriptsize #1}}}
\newcommand{\CITE}[1]{ \citep{#1}}
\newcommand{\VEC}[1]{\mathbf{#1}}
\newcommand{\Eqno}[1]{{(#1)}}
\newcommand{\REF}[1]{\Eqno{\ref{#1}}}
\newcommand{\Eq}[1]{Eq. \Eqno{#1}}
\newcommand{\Eqs}[1]{Eqs. {#1}}

\newcommand{\EqPunc}[1]{}
\newcommand{\EqPeriod}[1]{}
\newcommand{\Fig}[1]{Fig. {#1}}

\newcommand{\Figs}[1]{Figs. {#1}}

\newcommand{\Table}[1]{Table {#1}}
\newcommand{\Tables}[1]{Tables {#1}}

\newcommand{\SkipFigure}[1]{}
\newcommand{\FigureInText}[1]{}
\newcommand{\FigureLegends}[1]{#1}
\newcommand{\FigureInLegends}[1]{}
\newcommand{\TableInText}[1]{}
\newcommand{\TableLegends}[1]{#1}
\newcommand{\TableInLegends}[1]{#1}

\newcommand{\text}[1]{\textrm{#1}}
\renewcommand{\VEC}[1]{\text{\boldmath{$#1$}}} 
\newcommand{\VECS}[1]{\text{\boldmath{$#1$}}}

\newcommand{\BF}[1]{\textbf{#1}}

\newcommand{\RED}[1]{\textcolor{red}{#1}}
\newcommand{\REDa}[1]{\textcolor{blue}{#1}}    
\newcommand{\REDb}[1]{\textcolor{blue}{#1}}

\renewcommand{\RED}[1]{#1}

\ifdefined\text
\else
\newcommand{\text}[1]{\textrm{#1}}
\fi

\renewcommand{\FigureInText}[1]{}

\renewcommand{\FigureInLegends}[1]{#1}
\newcommand{\NoFigureInLegends}[1]{#1}
\FigureInLegends{
\renewcommand{\NoFigureInLegends}[1]{}
}

\renewcommand{\FigureLegends}[1]{#1}	

\newcommand{\NoFigureInText}[1]{#1}
\FigureInText{
\renewcommand{\NoFigureInText}[1]{}
}

\renewcommand{\TableInText}[1]{}

\renewcommand{\TableInLegends}[1]{#1}
\newcommand{\NoTableInLegends}[1]{#1}
\TableInLegends{
\renewcommand{\NoTableInLegends}[1]{}
}

\renewcommand{\TableLegends}[1]{#1}
\newcommand{\NoTableInText}[1]{#1}
\TableInText{
\renewcommand{\NoTableInText}[1]{}
}

\newcommand{\TextFig}[1]{#1}
\newcommand{\SupFig}[1]{}
\newcommand{\TextTable}[1]{#1}
\newcommand{\SupTable}[1]{}

\newcommand{\SupplementaryBibliography}[1]{#1}
\newcommand{\SupplementaryMaterial}[1]{#1}
\newcommand{\TextMaterial}[1]{#1}
\TextMaterial{
\renewcommand{\SupplementaryBibliography}[1]{}
}%  TextMaterial

\ifdefined\SUPPLEMENT
\else
\newcommand{\SUPPLEMENT}[1]{}
\fi
\ifdefined\TEXT
\else
\newcommand{\TEXT}[1]{#1}
\fi

\newcommand{\SkipFigsInText}[1]{}
\FigureInText{
\renewcommand{\SkipFigsInText}[1]{#1}
}
\newcommand{\SkipTablesInText}[1]{}
\TableInText{
\renewcommand{\SkipTablesInText}[1]{#1}
}
\begin{document}

\TextMaterial{

%\linenumbers

\begin{frontmatter}

%% Title, authors and addresses

%% use the tnoteref command within \title for footnotes;
%% use the tnotetext command for the associated footnote;
%% use the fnref command within \author or \address for footnotes;
%% use the fntext command for the associated footnote;
%% use the corref command within \author for corresponding author footnotes;
%% use the cortext command for the associated footnote;
%% use the ead command for the email address,
%% and the form \ead[url] for the home page:
%%
%% \title{Title\tnoteref{label1}}
%% \tnotetext[label1]{}
%% \author{Name\corref{cor1}\fnref{label2}}
%% \ead{email address}
%% \ead[url]{home page}
%% \fntext[label2]{}
%% \cortext[cor1]{}
%% \address{Address\fnref{label3}}
%% \fntext[label3]{}

%\title{{Selection maintaining protein stability at equilibrium}}
%\title{{Protein fitness originating from protein stability}}
%\title{{Relationships between protein folding free energy, fitness, and sequence distribution: 
%selection originating from protein stability}}
%\title{{Selection maintaining protein foldability/stability:
\title{Selection originating from protein stability/foldability:
\\
Relationships between protein folding free energy, sequence ensemble, and fitness
%I. A new method to estimate the effective temperature of natural selection and folding free energy of protein
%II. Folding free energy, sequence ensemble, and fitness
%II. Relationships between protein folding free energy,
%
\\
{\footnotesize
(J. Theoretical Biol., 433, 21-38, 2017 (DOI:10.1016/j.jtbi.2017.08.018) with some revisions)
}
}

%% use optional labels to link authors explicitly to addresses:
%% \author[label1,label2]{<author name>}
%% \address[label1]{<address>}
%% \address[label2]{<address>}

\author{Sanzo Miyazawa}
\ead{sanzo.miyazawa@gmail.com}
%\address{Gunma University, Graduate School of Engineering}
\address{6--5--607 Miyanodai, Sakura, Chiba 285--0857, Japan}

\begin{abstract}
%% Text of abstract
%\input{abst_JTB.tex}
% \input{abst_1+2.tex}

Assuming that mutation and fixation processes are
reversible Markov processes, 
we prove that the equilibrium ensemble of sequences obeys
a Boltzmann distribution with 
$\exp (4N_e m (1 - 1/(2N)))$,
where $m$ is 
Malthusian fitness and $N_e$ and $N$ are 
effective and actual population sizes.
On the other hand,
the probability distribution of sequences with maximum entropy that satisfies
a given amino acid composition at each site
and a given pairwise amino acid frequency at each site pair
is a
Boltzmann distribution with 
$\exp(-\psi_N)$, where the 
evolutionary statistical energy
$\psi_N$
is represented as the sum of one body 
($h$) (compositional)
and 
pairwise 
($J$) (covariational) 
interactions over all sites and site pairs.
A protein folding theory based on the random energy model (REM) indicates that
the equilibrium ensemble of natural protein sequences 
is well represented by a canonical ensemble
characterized by $\exp(-\Delta G_{ND}/k_B T_s)$
or by $\exp(- G_{N}/k_B T_s)$ if an amino acid composition is kept constant,
where $\Delta G_{ND} \equiv G_N - G_D$, 
$G_N$ and $G_D$ are the native and denatured free energies, and
$T_s$ is the effective temperature 
representing the strength of selection pressure.
Thus, $4N_e m (1 - 1/(2N))$, $-\Delta \psi_{ND} (\equiv -\psi_{N} + \psi_{D})$, and $-\Delta G_{ND}/k_B T_s$
must be equivalent to each other.
With $h$ and $J$ estimated by the DCA program, 
the changes ($\Delta \psi_N$) of $\psi_N$ due to single nucleotide nonsynonymous substitutions
are analyzed. 
The results indicate that the standard deviation of $\Delta G_N (= k_B T_s \Delta \psi_N)$ is approximately constant 
irrespective of protein families, and therefore can be used to estimate the relative value
of $T_s$. 
Glass transition temperature $T_g$ and
$\Delta G_{ND}$ are estimated from estimated $T_s$ and experimental melting temperature ($T_m$) 
for 14 protein domains.  The estimates of $\Delta G_{ND}$ agree with their experimental values for 5 proteins,
and those of $T_s$ and $T_g$ are all within a reasonable range.
In addition, approximating the probability density function (PDF) of $\Delta \psi_N$
by a log-normal distribution, PDFs of $\Delta \psi_N$ 
and $K_a/K_s$, which is the ratio of nonsynonymous to synonymous substitution rate per site,
in all and in fixed mutants are estimated. The equilibrium values of $\psi_N$,
at which the average of $\Delta \psi$ in fixed mutants is equal to zero,
well match $\psi_N$ averaged over homologous sequences,
confirming that 
the present methods for a fixation process of mutations 
and for the equilibrium ensemble of $\psi_N$ give a consistent result with each other.
The PDFs of $K_a/K_s$
at equilibrium confirm that 
$T_s$ negatively correlates with the amino acid substitution rate (the mean of $K_a/K_s$) of protein.
Interestingly,
stabilizing mutations are significantly fixed by positive selection, and
balance with destabilizing mutations fixed by random drift, although 
most of them are removed from population.
\RED{
Supporting the nearly neutral theory,
neutral selection is not significant
even in fixed mutants.
}%  RED

% End of abst_1+2.tex

%%% temporarily removed 
% \input{highlights.tex}

\vspace*{1em}
\noindent
Highlights
\begin{itemize}
\item A Boltzmann distribution with protein fitness is derived.
\item Relationships between folding free energy, inverse statistical potential and fitness. 
\item Selective temperature, glass transition temperature and folding free energy are estimated.
\item Relationship between selective temperature and substitution rate ($K_a/K_s$).
\item Protein stability/foldability is kept in a balance of positive selection and random drift.
\end{itemize}
% End of highlights.tex

\end{abstract}

%\Skip{
%%% temporarily removed 

\begin{keyword}
%% keywords here, in the form: keyword \sep keyword
% max 5
%stability at equilibrium
%\sep
%protein stability/foldability
%\sep
%folding free energy
%\sep
%sequence ensemble %distribution
folding free energy change
\sep
%Potts problem
inverse statistical potential
%\sep
%equilibrium distribution 
\sep
Boltzmann distribution
% with fitness 
\sep
selective temperature
%\sep
%neutral theory
\sep
positive selection
%random drift
%\sep
%protein evolution
%\sep

%% MSC codes here, in the form: \MSC code \sep code
%% or \MSC[2008] code \sep code (2000 is the default)

\end{keyword}

%} % Skip

\end{frontmatter}

% \linenumbers
%\linenumbers

%% main text
%\section{}
%\label{}
%\input{contents_JTB.tex}
% \input{contents_1+2.tex}

\newpage
\section{Introduction}

Natural proteins can fold their sequences into unique structures.
Protein's stability and foldability result from
natural selection and are not typical
characteristics of random polymers\CITE{BW:87,SG:93b,SG:93a,RS:94,PGT:97}.
Natural selection maintains
protein's stability and foldability
over evolutionary timescales.
On the basis of the random energy model (REM) for protein folding,
it was discussed\CITE{SG:93b,SG:93a,RS:94} that the equilibrium ensemble
of natural protein sequences in sequence space
is well represented by a canonical ensemble characterized by a Boltzmann factor
$\exp(- \Delta G_{ND}(\VEC{\sigma})/k_B T_s)$,
where $\Delta G_{ND}(\VEC{\sigma}) (\equiv G_N(\VEC{\sigma}) - G_D(\VEC{\sigma}))$ is
the folding free energy of sequence $\VEC{\sigma}$,
$G_N$ and $G_D$ are the free energies of the native and denatured states,
$k_B$ is the Boltzmann constant, and $T_s$ is the effective temperature
representing the strength of selection pressure 
and must satisfy $T_s < T_g < T_m$ for natural proteins to fold
into unique native structures; $T_g$ is glass transition temperature
and $T_m$ is melting temperature.
The REM also indicates that
the free energy of denatured conformations ($G_D$) is
a function of amino acid frequencies only and does not depend on amino acid order,
and therefore the Boltzmann factor will be taken as $\exp(- G_{N}(\VEC{\sigma})/k_B T_s)$,
if amino acid frequencies are kept constant.
It was shown by lattice Monte Carlo simulations\CITE{S:94} that
lattice protein sequences selected with this Boltzmann factor 
were not trapped by competing structures but 
could fold into unique native structures.
Selective temperatures were also estimated\CITE{DS:01} for actual proteins 
to yield good correlations of sequence entropy
between actual protein families and sequences designed 
with this type of Boltzmann factor.

On the other hand, the maximum entropy principle insists that
the probability distribution of sequences in sequence space,
which satisfies constraints on amino acid compositions at all sites and on amino acid pairwise frequencies
for all site pairs, is a Boltzmann distribution
with the Boltzmann factor $\exp(- \psi_{N}(\VEC{\sigma}))$,
where 
the total interaction $\psi_{N}(\VEC{\sigma})$ of 
a sequence $\VEC{\sigma}$
is represented as the sum of
one-body ($h$) (compositional) and pairwise ($J$) (covariational) interactions 
between residues in the sequence; 
$\psi_{N}(\VEC{\sigma})$ is called 
the evolutionary statistical energy by Hopf et al.\CITE{HIPSSSM:17}.
The inverse statistical potentials,
the one-body ($h$) and pairwise ($J$) interactions,
that satisfy those constraints for homologous sequences
have been estimated\CITE{MPLBMSZOHW:11,MCSHPZS:11,ELLWA:13,EHA:14}
as one of inverse Potts problems and successfully employed
to predict contacting residue pairs in protein structures
\CITE{MPLBMSZOHW:11,MCSHPZS:11,ELLWA:13,EHA:14,M:13,SMWHO:12,HCSRSM:12}.
Morcos et al.\CITE{MSCOW:14}
noticed that the $\psi_N$ in the Boltzmann factor
is the dimensionless energy corresponding to $G_{N}/k_B T_s$,
and estimated selective temperatures ($T_s$) for
several protein families by comparing the difference ($\Delta \psi_{ND}$) of
$\psi$ between the native and the molten globule states
with folding free energies ($\Delta G_{ND}$) estimated
with associative-memory, water-mediated, structure, and energy model (AWSEM)\CITE{DSZCWP:12}.

A purpose of the present study is to establish relationships between protein foldability/stability,
sequence distribution, and protein fitness. 
First, we prove that if mutation and fixation processes in protein evolution
are reversible Markov processes, the equilibrium ensemble of genes will obey a Boltzmann distribution
with the Boltzmann factor $\exp(4N_e m (1- 1/(2N)))$,
where $N_e$ and $N$ are effective and actual population sizes, and $m$ is the Malthusian fitness of a gene.
In other words,
correspondences between $-\Delta G_{ND}/k_B T_s$, $-\Delta \psi_{ND} (\equiv \psi_N - \psi_D)$
and $4N_e m (1- 1/(2N))$ are obtained
by equating these three Boltzmann distributions with each other; 
$\psi_D \simeq G_D / k_B T_s + \text{constant}$.

The second purpose is to 
analyze
the effects ($\Delta \psi_N$)
of single amino acid substitutions on the 
evolutionary statistical energy
of a protein,
and to estimate from the distribution of $\Delta \psi_N$
the effective temperature of natural selection ($T_s$) and then
glass transition temperature ($T_g$) and
folding free energy ($\Delta G_{ND}$) of protein.
We estimate the one-body ($h$) and pairwise ($J$) interactions with
the DCA program, which is available at ``http://dca.rice.edu/portal/dca/home'', and then
analyze the changes ($\Delta \psi_N$) of the 
evolutionary statistical energy
($\psi_N$) of 
a natural sequence
due to single amino acid substitutions caused by single nucleotide changes. 
The data of $\Delta \psi_N$ due to single nucleotide nonsynonymous substitutions 
for 14 protein domains
show that the standard deviation of $\Delta \psi_N$ over all the substitutions at all sites
hardly depends on the 
evolutionary statistical energy
($\psi_N$) of each homologous sequence and 
is nearly constant for each protein family, indicating that
the standard deviation of $\Delta G_N \simeq k_B T_s \Delta \psi_N$ is nearly constant irrespective of protein families.
From this finding, $T_s$ for each protein family  has been estimated 
in relative to $T_s$ for the PDZ family, which is determined by directly comparing
$\Delta\Delta \psi_{ND} (\equiv \Delta (\psi_N - \psi_D) \simeq \Delta \psi_N)$ with 
the experimental values of folding free energy changes, $\Delta\Delta G_{ND}$,
due to single amino acid substitutions.
Also 
$T_g$
and
$\Delta G_{ND}$ for each protein family are estimated on the basis of the REM from 
the estimated $T_s$ and an experimental melting temperature $T_m$.
The estimates of $T_s$ and $T_g$ are all within a reasonable range, and
those of $\Delta G_{ND}$ are well compared with experimental $\Delta G_{ND}$ for 5 protein families.
The present method for estimating $T_s$ is simpler than the method\CITE{MSCOW:14} using AWSEM,
and also is useful for the prediction of $\Delta G_{ND}$, because 
the experimental data of $\Delta G_{ND}$ are limited in comparison with $T_m$,
and also experimental conditions such as temperature and {pH} tend
to be different among them.
In addition, it has been revealed that $\Delta \psi_N$ averaged over all single nucleotide nonsynonymous substitutions
is a linear function of $\psi_N / L$ of each homologous sequence,
where $L$ is sequence length;
the average of $\Delta \psi_N$ decreases as $\psi_N / L$ increases.
This characteristic is required for homologous proteins to stay
at the equilibrium state of
the native conformational energy $G_N \simeq k_B T_s \psi_N$,
and indicates a weak dependency\CITE{SRS:12,M:16} of $\Delta\Delta G_{ND}$ 
on $\Delta G_{ND} / L$ of protein across protein families.

The third purpose is to study an amino acid substitution process
in protein evolution, which is characterized by the fitness,
$m = - \Delta \psi_{ND} / (4N_e(1-1/(2N)))$.
We employ a monoclonal approximation for mutation and fixation processes of genes, in which
protein evolution proceeds with single amino acid substitutions fixed at a time in a population.
In this approximation, 
$\psi_N$ of a protein gene attains
the equilibrium, $\psi_N = \psi_N^{\script{eq}}$, 
when
the average of $\Delta \psi_{N} (\simeq \Delta\Delta \psi_{ND})$ over singe nucleotide nonsynonymous mutations
fixed in a population is equal to zero.
Approximating the distribution of $\Delta \psi_{N}$ due to singe nucleotide nonsynonymous mutations
by a  log-normal distribution,
their distribution for fixed mutants
is numerically calculated and used to calculate the averages of various quantities
and also the probability density functions (PDF) of
$K_a/K_s$ in all arising mutants and also in fixed mutants only; 
$K_a/K_s$ is defined as
the ratio of nonsynonymous to synonymous substitution rate per site.
There is a good agreement 
between the time average ($\psi_N^{\script{eq}}$) and ensemble average ($\langle \psi_N \rangle_{\VEC{\sigma}}$),
which is equal to the sample average, $\overline{\psi_N}$, of $\psi_N$
over homologous sequences, supporting
the constancy of the standard deviation of $\Delta\psi_N$ assumed in the monoclonal approximation.

We also study protein evolution at equilibrium, $\psi_N = \psi_N^{\script{eq}}$.
The common understanding of protein evolution
has been that amino acid substitutions observed in homologous proteins are
neutral \CITE{K:68,K:69,KO:71,KO:74} or slightly deleterious
\CITE{O:73,O:92}, and random drift is a primary force to fix amino acid
substitutions in population.
The PDFs of $K_a/K_s$ in all arising mutations and in
their fixed mutations are examined to see how significant
each of positive, neutral, slightly negative,and negative selections is.
Interestingly, stabilizing mutations are significantly fixed in population by positive selection,
and balance with
destabilizing mutations that are also significantly fixed by random drift, although
most negative mutations are removed from population.
Contrary to the neutral theory\CITE{K:68,K:69,KO:71,KO:74}
and supporting the nearly neutral theory\CITE{O:73,O:92,O:02},
the proportion of neutral selection is not large even in fixed mutants.
It is also confirmed that the effective temperature ($T_s$) of selection
negatively correlates with the amino acid substitution rate ($K_a/K_s$) of protein at equilibrium.

% End of intro_1+2.tex

\section{Methods}

\ifdefined\SkipDETAIL
\else
\newcommand{\SkipDETAIL}[1]{}
\fi
\ifdefined\SUPPLEMENT
\else
\newcommand{\SUPPLEMENT}[1]{}
\fi
\ifdefined\TEXT
\else
\newcommand{\TEXT}[1]{#1}
\fi

\subsection{Knowledge of protein folding}
\TEXT{
\label{protein_folding_theory}
}%  TEXT

A protein folding theory\CITE{SG:93b,SG:93a,RS:94,PGT:97}, which is based on a random energy model (REM), 
indicates that the equilibrium ensemble of amino acid sequences, $\VECS{\sigma} \equiv (\sigma_1, \cdots, \sigma_L)$ 
where $\sigma_i$ is the type of amino acid at site $i$ and $L$ is sequence length,
can be well approximated by
a canonical ensemble with a Boltzmann factor 
consisting of the folding free energy, $\Delta G_{ND}(\VEC{\sigma}, T)$
and an effective temperature $T_s$ representing the strength of selection pressure.
\begin{eqnarray}
	P(\VEC{\sigma}) 
		&\propto&
		P^{\script{mut}}(\VEC{\sigma}) \exp (\frac{- \Delta G_{ND}(\VEC{\sigma}, T)}{k_B T_s}) 
		\label{\EQ: canonical_selection}
		\\
		&\propto&
			\exp (\frac{- G_N(\VEC{\sigma})}{k_B T_s})	
				\hspace*{2em} \textrm{ if } \VEC{f}(\VEC{\sigma}) = \textrm{constant}
		\label{\EQ: canonical_selection_for_constant_composition}
		\\
	\Delta G_{ND}(\VEC{\sigma}, T) &\equiv& G_N(\VEC{\sigma}) - G_D(\VEC{f}(\VEC{\sigma}), T)
\end{eqnarray}
where 
$p^{\script{mut}}(\VEC{\sigma})$ is 
the probability of a sequence ($\VEC{\sigma}$) randomly occurring in a mutational process
and depends only on the amino acid frequencies $\VEC{f}(\VEC{\sigma})$,
$k_B$ is the Boltzmann constant, 
$T$ is a growth temperature, 
and $G_N$ and $G_D$ are the free energies of 
the native conformation and denatured state, respectively.
Selective temperature $T_s$
quantifies how strong the folding constraints 
are in protein evolution,
and is specific to the protein structure and function.
The free energy $G_D$ of the denatured state
does not depend on the amino acid order
but the amino acid composition, $\VEC{f}(\VEC{\sigma})$, in a sequence\CITE{SG:93b,SG:93a,RS:94,PGT:97}.
It is reasonable to assume that mutations independently occur between sites, and
therefore the equilibrium frequency of a sequence in the mutational process is equal to the product of
the equilibrium frequencies over sites; 
$
	P^{\script{mut}}(\VEC{\sigma}) = \prod_i p^{\script{mut}}(\sigma_i)
$, where $p^{\script{mut}}(\sigma_i)$ is the equilibrium frequency of $\sigma_i$ at site $i$ in the mutational process. 

The distribution of conformational energies
in the denatured state (molten globule state), which
consists of conformations as compact as the native conformation,
is approximated in 
the random energy model (REM), particularly the independent 
interaction model (IIM) \CITE{PGT:97}, to be equal to
the energy distribution of randomized sequences, 
which is then approximated by a Gaussian distribution,
in the native conformation.
That is, the partition function $Z$ for the denatured state is written as follows with 
the energy density $n(E)$ of conformations that is approximated by a product   
of a Gaussian probability density and the total number of conformations 
whose logarithm is proportional to the chain length.
\begin{eqnarray}
	Z &=& \int \exp (\frac{- E}{k_B T}) \, n(E) dE
		\\
	n(E) &\approx& \exp ( \omega L ) \mathcal{N}(\bar{E}(\VEC{f}(\VECS{\sigma})), \delta E^2(\VEC{f}(\VECS{\sigma})) )
\end{eqnarray}
where $\omega$ is the conformational entropy per residue 
in the compact denatured state, 
and $\mathcal{N}(\bar{E}(\VEC{f}(\VECS{\sigma})), \delta E^2(\VEC{f}(\VECS{\sigma})) )$ is
the Gaussian probability density with mean $\bar{E}$ and variance $\delta E^2$, 
which depend only on the amino acid composition of the protein sequence.
The free energy of the denatured state is approximated as follows.
\begin{eqnarray}
	G_D(\VEC{f}(\VECS{\sigma}),T) 
		&\approx&
		\bar{E}(\VEC{f}(\VECS{\sigma}))
	- \frac{\delta E^2(\VEC{f}(\VECS{\sigma}))}{2 k_B T} 
	- k_B T \omega L
	\\
	&=& \bar{E}(\VEC{f}(\VECS{\sigma}))
	- \delta E^2(\VEC{f}(\VECS{\sigma}))
		\frac{\vartheta(T/T_g)}{k_B T}
	\\
 \vartheta(T/T_g) &\equiv& \left\{ \begin{array}{ll}
			\frac{1}{2}(1 + \frac{T^2}{T_g^2}) & \textrm{ for } T > T_g \\
			\frac{T}{T_g} 			& \textrm{ for } T \leq T_g \\
		\end{array} 
	    \right. 
	\label{\EQ: free_energy_of_denatured_state}
\end{eqnarray}
where $\bar{E}$ and $\delta E^2$ are estimated as the mean and variance of
interaction energies of randomized sequences in the native conformation.
$T_g$ is the glass transition temperature of the protein
\TEXT{
at which entropy becomes zero\CITE{SG:93b,SG:93a,RS:94,PGT:97}; 
$ - \partial G_D / \partial T |_{T=T_g} = 0$.
}%  TEXT
\SUPPLEMENT{
at which entropy becomes zero\CITE{SG:93b,SG:93a,RS:94,PGT:97}. 
\begin{eqnarray}
	- \frac{\partial G_D}{\partial T} |_{T=T_g} &=& 0
\end{eqnarray}
}%  SUPPLEMENT
The conformational entropy per residue $\omega$ in the compact denatured state
can be represented with 
\TEXT{
$T_g$; $\omega L = \delta E^2 / (2 (k_B T_g)^2) $.
}%  TEXT
\SUPPLEMENT{
$T_g$.
\begin{eqnarray}
	\omega L &=& \frac{ \delta E^2}{ 2 (k_B T_g)^2 }
\end{eqnarray}
}%  SUPPLEMENT
Thus, unless $T_g < T_m$, a protein will be trapped at local minima 
on a rugged free energy landscape before 
it can fold into a unique native structure.

\subsection{Probability distribution of homologous sequences with the same native fold in sequence space}
\TEXT{
\label{homologous_sequences}
}%  TEXT

The probability distribution $P(\VEC{\sigma})$ of homologous sequences
with the same native fold,
$\VECS{\sigma}= (\sigma_1, \cdots, \sigma_L)$
where $\sigma_i \in \{ \text{amino acids, deletion} \}$,
in sequence space 
with maximum entropy, 
which satisfies a given amino acid frequency at each site and
a given pairwise amino acid frequency at each site pair,
is a Boltzmann distribution\CITE{MPLBMSZOHW:11,MCSHPZS:11}. 
\begin{eqnarray}
P(\VECS{\sigma}) &\propto& \exp( - \psi_N(\VECS{\sigma}) )
        \label{\EQ: potts_model}
	\\
\psi_N(\VECS{\sigma}) &\equiv& - (\sum_i^L (h_i(\sigma_i) + \sum_{j>i} J_{ij}(\sigma_i, \sigma_j)) )
        \label{\EQ: total_interaction_in_potts_model}
\end{eqnarray}
where $h_i$ and $J_{ij}$ are one-body (compositional) and two-body (covariational) interactions 
and must satisfy the following constraints.
\RED{
\begin{eqnarray}
	\sum_{\VEC{\sigma}} P(\VEC{\sigma}) \, \delta_{\sigma_i a_k} &=& P_i(a_k) 
	\\
	\sum_{\VEC{\sigma}} P(\VEC{\sigma}) \, \delta_{\sigma_i a_k} \delta_{\sigma_j a_l} &=& P_{ij}(a_k, a_l) 
\end{eqnarray}
where $\delta_{\sigma_i a_k}$ is the Kronecker delta,
$P_i(a_k)$ is the frequency of amino acid $a_k$ at site $i$, and
}%  RED
$P_{ij}(a_k, a_l)$ is the frequency of amino acid pair, $a_k$ at $i$ and 
$a_l$ at $j$; $a_k \in \{ \text{amino acids, deletion} \}$.
The pairwise interaction matrix $J$ satisfies 
	$J_{ij}(a_k, a_l) =  J_{ji}(a_l, a_k)$  
and
	$J_{ii}(a_k, a_l) = 0$.
Interactions $h_i$ and $J_{ij}$ can be well estimated
from a multiple sequence alignment (MSA) in the mean field approximation\CITE{MPLBMSZOHW:11,MCSHPZS:11},
or by maximizing a pseudo-likelihood\CITE{ELLWA:13,EHA:14}. 
Because
$\psi_N(\VEC{\sigma})$ 
has been estimated under the constraints on amino acid compositions at all sites, 
only sequences with a given amino acid composition contribute significantly to the partition function,
and other sequences may be ignored.

Hence, from \Eqs{\REF{\EQ: canonical_selection_for_constant_composition} and \REF{\EQ: potts_model}},
\begin{eqnarray}
	\psi_N(\VEC{\sigma}) &\simeq& 
 		G_N(\VEC{\sigma}) / (k_B T_s) + \textrm{function of} \VEC{f}(\VEC{\sigma})
        	\label{\EQ: expression_of_Gn}
		\\
	\psi_D(\VEC{f}(\VEC{\sigma}), T) &\simeq&	
		G_D(\VEC{f}(\VEC{\sigma}), T)  / (k_B T_s) + \textrm{function of} \VEC{f}(\VEC{\sigma})
        	\label{\EQ: expression_of_Gd}
		\\
	\Delta \psi_{ND}(\VEC{\sigma}, T) &\simeq&	
		\Delta G_{ND}(\VEC{\sigma}, T) / (k_B T_s)
        	\label{\EQ: expression_of_dG}
		\\
	\Delta \psi_{ND}(\VEC{\sigma}, T) &\equiv& \psi_{N}(\VEC{\sigma}) - \psi_{D}(\VEC{f}(\VEC{\sigma}), T)
        	\label{\EQ: expression_of_dPsi}
		\\
	\psi_D(\VEC{f}(\VEC{\sigma}), T) &\approx&
			\bar{\psi}(\VEC{f}(\VEC{\sigma})) - \delta \psi^2(\VEC{f}(\VEC{\sigma}))
			\vartheta(T/T_g) T_s / T
        	\label{\EQ: expression_of_psi_D}
		\\
	\omega &=&  (T_s/ T_g)^2 \delta \psi^2 / (2 L)
        \label{\EQ: expression_of_entropy}
\end{eqnarray}
where 
the $\bar{\psi}$ and ${\delta \psi}^2$ are estimated 
as the mean and variance of $\psi_N$ over randomized sequences;
$\bar{E} \simeq k_B T_s \bar{\psi}$ and ${\delta E}^2 \simeq (k_B T_s)^2 {\delta \psi}^2$.

% End of methods_for_protein_folding.tex

% \input{methods_for_mutation_fixation_process.tex}

\subsection{The equilibrium distribution of sequences in a mutation-fixation process}

Here we assume that the mutational process is a reversible Markov process.
That is,
the mutation rate per gene, $M_{\VEC{\mu}\VEC{\nu}}$, 
from sequence $\VEC{\mu} \equiv (\mu_1, \cdots, \mu_L)$ to $\VEC{\nu}$ satisfies
the detailed balance condition
\begin{eqnarray}
	P^{\script{mut}}(\VEC{\mu}) M_{\VEC{\mu}\VEC{\nu}} &=& P^{\script{mut}}(\VEC{\nu}) M_{\VEC{\nu}\VEC{\mu}} 
	\label{\EQ: detailed_balance_for_mutation}
\end{eqnarray}
where $P^{\script{mut}}(\VEC{\nu})$ is the equilibrium frequency 
of sequence $\VEC{\nu}$ in a mutational process, $M_{\VEC{\mu}\VEC{\nu}}$.
The mutation rate per population is equal to $2 N M_{\VEC{\mu}\VEC{\nu}}$
for a diploid population, where $N$ is the population size.
The substitution rate 
$R_{\VEC{\mu}\VEC{\nu}}$ from $\VEC{\mu}$ to $\VEC{\nu}$
is equal to the product of
the mutation rate and the fixation probability 
with which a single mutant gene becomes to fully occupy the population\CITE{CK:70}.
\begin{eqnarray}
	R_{\VEC{\mu}\VEC{\nu}}
	&=& 
	2 N M_{\VEC{\mu}\VEC{\nu}} u(s(\VEC{\mu} \rightarrow \VEC{\nu}))
\end{eqnarray}
where $u(s(\VEC{\mu} \rightarrow \VEC{\nu}))$ is the fixation probability of mutants 
from $\VEC{\mu}$ to $\VEC{\nu}$ the selective advantage of which is equal to $s$. 

For genic selection (no dominance) or gametic selection in a Wright-Fisher population of diploid,
the fixation probability, $u$, of a single mutant gene, the selective advantage of which is equal to $s$ 
and the frequency of which in a population is equal to $q_m = 1/(2N)$, was estimated\CITE{CK:70} as
\begin{eqnarray}
	2N u(s) &=& 2N \frac{1 - e^{-4N_e s q_{m}}}{1 - e^{- 4N_e s}}
	\label{\EQ: fixation_probability}
	\\
	&=& \frac{u(s)}{u(0)}
	\hspace*{2em} \text{ with } \hspace*{2em} q_{\script{m}} = \frac{1}{2N} 	
	\label{\EQ: def_qm}
\end{eqnarray}
where $N_e$ is effective population size.
\RED{
\Eq{\ref{\EQ: fixation_probability}} will be also valid for haploid population if $2N_e$ and $2N$ are replaced by $N_e$ and $N$, respectively.
Also, for Moran population of haploid, $4N_e$ and $2N$ should be replaced by
$N_e$ and $N$, respectively.
Fixation probabilities for various selection models, which are compiled from p. 192 and p. 424--427 of Crow and Kimura (1970)
and from 
Moran (1958) and Ewens (1979),
are listed in \Table{\ref{stbl: fixation_prob}}.
}%  RED
The selective advantage of a mutant sequence $\VEC{\nu}$ 
to a wildtype $\VEC{\mu}$ 
is equal to
\begin{eqnarray}
	s(\VEC{\mu} \rightarrow \VEC{\nu}) &=& m(\VEC{\nu}) - m(\VEC{\mu}) 
\end{eqnarray}
where $m(\VEC{\nu})$ is the Malthusian fitness of a mutant sequence,
and $m(\VEC{\mu})$ is for the wildtype. 

This Markov process of substitutions in sequence
is reversible, and the equilibrium frequency of sequence $\VEC{\mu}$, 
$P^{\script{eq}}(\VEC{\mu})$, 
in the total process consisting of mutation and fixation processes
is represented by
\begin{eqnarray}
	P^{\script{eq}}(\VEC{\mu}) &=& 
	\frac{P^{\script{mut}}(\VEC{\mu}) \exp( 4N_e m(\VEC{\mu}) (1-q_m) ) }
		{ \sum_{\VEC{\nu}} P^{\script{mut}}(\VEC{\nu}) 
			\exp( 4N_e m(\VEC{\nu} ) (1-q_m) )}
	\label{\EQ: equilibrium_of_mutation_fixation_process}
\end{eqnarray}
because both the mutation and fixation processes satisfy the detailed balance conditions,
\Eq{\ref{\EQ: detailed_balance_for_mutation}} and the following equation, respectively.
\begin{eqnarray}
\lefteqn{
	\exp( 4N_e m(\VEC{\mu} ) (1-q_m) )
	\,
	u(s(\VEC{\mu} \rightarrow \VEC{\nu}))
}
	\nonumber
	\\
	&=& 
	\frac{\exp(-4N_e m(\VEC{\mu}  ) q_m) - \exp(-4N_e m(\VEC{\nu} ) q_m) }
	{\exp(-4N_e m(\VEC{\mu})) - \exp(-4N_e m(\VEC{\nu})) }
	\\
	&=&
	\exp( 4N_e m(\VEC{\nu}) (1-q_m) )
	\, u(s(\VEC{\nu} \rightarrow \VEC{\mu}))
\end{eqnarray}
As a result, the ensemble of homologous sequences in molecular evolution 
obeys a Boltzmann distribution.

\subsection{Relationships between $m(\VEC{\sigma})$, $\psi_N(\VEC{\sigma})$, and
$\Delta G_{ND}(\VEC{\sigma})$ of protein sequence}

From \Eqs{
\REF{\EQ: canonical_selection},
\REF{\EQ: potts_model},
and
\REF{\EQ: equilibrium_of_mutation_fixation_process}
},
we can get the following relationships among the Malthusian fitness $m$,
the folding free energy $\Delta G_{ND}$
and $\Delta \psi_{ND}$ of protein sequence.
\begin{eqnarray}
	P^{\script{eq}}(\VEC{\mu}) &=& \frac{ P^{\script{mut}}(\VEC{\mu}) \exp( 4N_e m( \VEC{\mu} ) ( 1 - q_m )) }
		{ \sum_\nu P^{\script{mut}}(\VEC{\nu}) 
			\exp( 4N_e m(\VEC{\nu}) ( 1 - q_m )) )}
		\label{\EQ: equilibrium_distr_of_seq_of_m}
		\\
	&=& \frac{ P^{\script{mut}}(\overline{\VEC{\mu}}) \exp ( - (\psi_{N}(\VEC{\mu}) - \psi_{D}(\overline{\VEC{f}(\VEC{\mu})}, T) )) }
		{ \sum_{\VEC{\nu}} P^{\script{mut}}(\overline{\VEC{\nu}}) 
			\exp ( - (\psi_{N}(\VEC{\nu}) - \psi_{D}(\overline{\VEC{f}(\VEC{\nu})}, T) ) ) }
		\label{\EQ: equilibrium_distr_of_seq}
		\label{\EQ: equilibrium_distr_of_seq_of_dPsi}
		\\
	&\simeq& 
		\frac{ P^{\script{mut}}(\VEC{\mu}) \exp( - \Delta G_{ND}(\VEC{\mu}, T ) / (k_B T_s) ) }
		{ \sum_\nu P^{\script{mut}}(\VEC{\nu}) 
			\exp( - \Delta G_{ND}(\VEC{\nu}, T) / (k_B T_s) )}
		\label{\EQ: equilibrium_distr_of_seq_of_dG}
\end{eqnarray}
where 
$\overline{\VEC{f}(\VEC{\sigma}) } \equiv \sum_{\VEC{\sigma}} \VEC{f}(\VEC{\sigma}) P(\VEC{\sigma})$ and
$\log P^{\script{mut}}(\overline{\VEC{\sigma}}) \equiv \sum_{\VEC{\sigma}} P(\VEC{\sigma}) \log (\prod_i P^{\script{mut}}(\sigma_i))$.
Then, the following relationships are derived for sequences for which $f(\VEC{\mu}) = \overline{f(\VEC{\mu})}$.
\begin{eqnarray}
	4N_e m(\VEC{\mu}) (1 - q_m)
	&=& - \Delta \psi_{ND}(\VEC{\mu},T) + \mathrm{constant}
	\label{\EQ: relationship_between_m_and_dPsi}
        \\
	&\simeq& 
	\frac{- \Delta G_{ND}(\VEC{\mu},T)}{k_B T_s} + \mathrm{constant}
	\label{\EQ: relationship_between_m_and_dG}
\end{eqnarray}
The selective advantage of $\VEC{\nu}$ to $\VEC{\mu}$ is represented as follows for 
$f(\VEC{\mu}) = f(\VEC{\nu}) = \overline{f(\VEC{\sigma})}$.
\begin{eqnarray}
\lefteqn{
	4N_e s(\VEC{\mu} \rightarrow \VEC{\nu}) ( 1 - q_m )
}
	\nonumber
	\\
	&=&
	(4N_e m(\VEC{\nu})
	- 4N_e m( \VEC{\mu} ) ) ( 1 - q_m )
	\\
	&=& - (\Delta \psi_{ND}(\VEC{\nu}, T) - \Delta \psi_{ND}(\VEC{\mu}, T) )
	= 
	- ( \psi_N(\VEC{\nu}) - \psi_N(\VEC{\mu}) )
	\label{\EQ: s_vs_dPsi}
	\label{\EQ: selective_advantage_and_dPsi_N}
	\\
	&\simeq& 
	- (\Delta G_{ND}(\VEC{\nu}, T) - \Delta G_{ND}(\VEC{\mu}, T) ) / (k_B T_s)	 
	\label{\EQ: s_vs_dG}
	= 
	- ( G_N(\VEC{\nu}) - G_N(\VEC{\mu}) ) / (k_B T_s)
\end{eqnarray}
It should be noted here that
only sequences for which $f(\VEC{\sigma}) = \overline{f(\VEC{\sigma})}$ 
contribute significantly to the partition functions in 
\Eq{\ref{\EQ: equilibrium_distr_of_seq_of_dPsi}}, and other sequences may be ignored.

\RED{
\Eq{\ref{\EQ: s_vs_dPsi}} indicates that evolutionary statistical energy $\psi$ 
should be proportional to effective population size $N_e$, 
and therefore it is ideal to estimate one-body ($h$) and two-body ($J$) interactions
from homologous sequences of species that 
do not significantly differ in effective population size. 
Also, \Eq{\ref{\EQ: s_vs_dG}} indicates that 
selective temperature $T_s$ is inversely proportional to the effective population size $N_e$;
$T_s \propto 1 / N_e$,
because free energy is a physical quantity and should not depend on effective population size.
}%  RED

% End of methods_for_mutation_fixation_process.tex

% \input{methods_for_ensemble_averages.tex}

\subsection{The ensemble average of folding free energy, $\Delta G_{ND}(\VEC{\sigma}, T)$, over sequences}
\TEXT{
\label{ensemble_average}
}%  TEXT

The ensemble average of $\Delta G_{ND}(\VEC{\sigma}, T)$ over sequences 
with \Eq{\ref{\EQ: canonical_selection}} is
\begin{eqnarray}
\lefteqn{
	\langle \Delta G_{ND}(\VEC{\sigma}, T) \rangle_{\VEC{\sigma}} 
} 
	\\
	&\equiv& 
	\, [ \, 
	\sum_{\VEC{\sigma}} \Delta G_{ND}(\VEC{\sigma}, T) 
		P^{\script{mut}}(\VEC{\sigma}) \exp( - \frac{\Delta G_{ND}(\VEC{\sigma}, T)}{k_B T_s} ) 
	\, ] \,
	/ 
	\, [ \, 
	\sum_{\VEC{\sigma}}
		P^{\script{mut}}(\VEC{\sigma}) \exp( - \frac{\Delta G_{ND}(\VEC{\sigma}, T)}{k_B T_s} ) 
	\, ] \,
	\\
	&\approx&
		\, [ \,
		\sum_{\VEC{\sigma}\, | \, \VEC{f}(\VEC{\sigma}) = \overline{\VEC{f}(\VEC{\sigma}_N)} }  
		G_{N}(\VEC{\sigma} ) 
		\exp( - \frac{G_{N}(\VEC{\sigma})}{k_B T_s} ) 
		\, ] \,  
		/ 
		\, [ \,
		\sum_{\VEC{\sigma}\, | \, \VEC{f}(\VEC{\sigma}) = \overline{\VEC{f}(\VEC{\sigma}_N)} }  
		\exp( - \frac{G_{N}(\VEC{\sigma})}{k_B T_s} ) 
		\, \, ] \,  
		- G_{D}( \overline{\VEC{f}(\VEC{\sigma}_N)} , T)
	\label{\EQ: native_sequence_approximation}
	\\
	&=& \langle G_N(\VEC{\sigma}) \rangle_{\VEC{\sigma}} - G_D( \overline{\VEC{f}(\VEC{\sigma}_N)} , T)
\end{eqnarray}
where $\VEC{\sigma}_N$ denotes 
a natural sequence,  
and $\overline{\VEC{f}(\VEC{\sigma_N})}$ denotes the average of 
amino acid frequencies $\VEC{f}(\VEC{\sigma_N})$ over homologous sequences.
In \Eq{\ref{\EQ: native_sequence_approximation}}, 
the sum over all sequences is approximated by the sum over sequences the amino acid composition
of which is the same as that over the 
natural sequences.

The ensemble averages of $G_N$
and
$\psi_N(\VEC{\sigma})$ 
are estimated in the Gaussian approximation\CITE{PGT:97}.
\begin{eqnarray}
	\langle G_N(\VEC{\sigma}) \rangle_{\VEC{\sigma}} 	
	&\approx& \frac{ \int E \exp (- E / (k_B T_s)) \, n(E) \, dE }  
		{\int \exp (- E / (k_B T_s)) \, n(E) \, dE }
	\\
	&=& \bar{E}( \overline{\VEC{f}(\VEC{\sigma_N})} ) - {\delta E}^2( \overline{\VEC{f}(\VEC{\sigma_N})} ) / (k_B T_s)
	\label{\EQ: ensemble_ave_of_G}
\end{eqnarray}
\begin{eqnarray}
\langle \psi_N(\VEC{\sigma}) \rangle_{\VEC{\sigma}} 	
	&\equiv& 
	\, [ \, 
	\sum_{\VEC{\sigma}} \psi_{ND}(\VEC{\sigma}) 
		\exp( - \psi_{N}(\VEC{\sigma}) )
	\, ] \,
	/ 
	\, [ \, 
	\sum_{\VEC{\sigma}}
		\exp( - \psi_{N}(\VEC{\sigma}) ) 
	\, ] \,
	\\
	&\approx& \bar{\psi}( \overline{\VEC{f}(\VEC{\sigma_N})} ) - {\delta \psi}^2( \overline{\VEC{f}(\VEC{\sigma_N})} )
	\label{\EQ: ensemble_ave_of_psi}
\end{eqnarray}
The ensemble averages of $\Delta G_{ND}(\VEC{\sigma}, T)$ and $\psi_N(\VEC{\sigma})$ over sequences
are observable as
the sample averages of $\Delta G_{ND}(\VEC{\sigma_N}, T)$ and $\psi_N(\VEC{\sigma_N})$ over 
homologous sequences fixed in protein evolution,
respectively. 
\begin{eqnarray}
\overline{ \Delta G_{ND}(\VEC{\sigma_N}, T) } / (k_B T_s)
	&=& 
	\langle \Delta G_{ND}(\VEC{\sigma}, T) \rangle_{\VEC{\sigma}} / (k_B T_s) 
	\\
\SUPPLEMENT{
	&\approx& \, [ \, {\delta E}^2( \overline{\VEC{f}(\VEC{\sigma_N})} ) 
	\, [ \, \vartheta(T/T_g) T_s / T - 1 \, ] / ( k_B T_s )^2
	\\
	&=&
}%  SUPPLEMENT
\TEXT{
	&\approx& 
}%  TEXT
	{\delta \psi}^2( \overline{\VEC{f}(\VEC{\sigma_N})} ) \, [ \,
	\vartheta(T/T_g) T_s / T - 1 \, ]
	\label{\EQ: ensemble_ave_of_ddG}
	\\
\SUPPLEMENT{
	&=& \overline{ \Delta G_{ND}(\VEC{\sigma_N}, T_g) } \, / \, ( k_B T^{\prime}_s )
	\\
	T^{\prime}_s &=& T_s (T_s/T - 1) / ( \vartheta(T/T_g) T_s / T - 1 )
	\\
}%  SUPPLEMENT
	\overline{ \psi_N(\VEC{\sigma_N}) } 
	&\equiv &
		\frac{\sum_{\VEC{\sigma}_N} w_{\VEC{\sigma}_N} \psi_N(\VEC{\sigma}_N) }{\sum_{\VEC{\sigma}_N} w_{\VEC{\sigma}_N} } 
	\label{\EQ: def_sample_ave_of_psi}
		\\
	&=&
	\langle \psi_N(\VEC{\sigma}) \rangle_{\VEC{\sigma}} 	
	\label{\EQ: sample_ave_of_psi}
\end{eqnarray}
where the overline denotes a sample average with a sample weight 
$w_{\VEC{\sigma}_N}$ for each homologous sequence,
which is used to reduce phylogenetic biases in the set of homologous sequences.
\SUPPLEMENT{
$\Delta G_{ND}(\VEC{\sigma_N}, T_g)$ corresponds to the energy gap\CITE{SG:93a} 
between the native and the glass states, and
$T^{\prime}_s$ will be the selective temperature
if $\Delta G_{ND}(\VEC{\sigma_N}, T_g)$ is used for selection instead of $\Delta G_{ND}(\VEC{\sigma_N}, T)$.
}%  SUPPLEMENT

The folding free energy becomes equal to zero at the melting temperature $T_m$; 
$\langle \Delta G_{ND}(\VEC{\sigma_N}, T_m) \rangle_{\VEC{\sigma}} = 0$.  Thus, the following relationship must be 
satisfied\CITE{SG:93b,SG:93a,RS:94,PGT:97}.
\begin{eqnarray}
	\vartheta(T_m/T_g) \frac{T_s}{T_m} &=& \frac{T_s}{2T_m}(1 + \frac{T_m^2}{T_g^2}) = 1
	\hspace*{1em} \textrm{ with } T_s \leq T_g \leq T_m
	\label{\EQ: relationship_among_characteristic_T}
\end{eqnarray}

% End of methods_for_ensemble_averages.tex

\TEXT{

\subsection{Probability distributions of selective advantage, fixation rate and $K_a/K_s$}

\SUPPLEMENT{
Now, we can consider 
}%  SUPPLEMENT
\TEXT{
Let us consider 
}%  TEXT
the probability distributions of
characteristic quantities that describe the evolution of genes. 
First of all, the probability density function (PDF) of 
selective advantage $s$, $p(s)$, of mutant genes can be calculated
from the PDF of 
the change of $\Delta \psi_{ND}$ due to a mutation from $\VEC{\mu}$ to $\VEC{\nu}$, 
$\Delta\Delta \psi_{ND} (\equiv \Delta \psi_{ND}(\VEC{\nu}, T) - \Delta \psi_{ND}(\VEC{\mu}, T) )$.
The PDF of $4N_e s$, $p(4N_e s) = p(s)/(4N_e) $, may be more useful than $p(s)$.
\begin{eqnarray}
p(4N_e s)
	&=& p(\Delta \Delta \psi_{ND}) \, | \frac{d \Delta \Delta \psi_{ND}}{d 4N_e s} |
	=
	p(\Delta\Delta \psi_{ND}) (1 - q_m)
	\label{\EQ: pdf_of_4Nes}
	\label{\EQ: pdf_of_s}
\end{eqnarray}
where $\Delta\Delta \psi_{ND}$ must be regarded as a function of $4N_es$, that is,
$
\Delta\Delta \psi_{ND} = - 4N_e s (1 - q_m)
$; see \Eq{\ref{\EQ: s_vs_dPsi}}.

The PDF of fixation probability $u$ can be represented by
\begin{eqnarray}
p(u)	&=& p(4N_e s) \frac{d 4N_e s}{d u} 
	= 
	p(4N_e s)
	 \frac{ ( e^{4N_e s} - 1)^2 e^{4N_e s (q_m - 1)} }
	{ q_m( e^{4N_e s} - 1) - (e^{4N_e s q_m} - 1) }
		\label{\EQ: pdf_of_fixation_prob}
\end{eqnarray}
where $4N_e s$ must be regarded as a function of $u$. 

The ratio of the substitution rate per nonsynonymous site ($K_a$) for nonsynonymous
substitutions with selective advantage s to the substitution rate per
synonymous site ($K_s$) for synonymous substitutions with s = 0 is
\begin{eqnarray}
	\frac{K_a}{K_s} &=& \frac{u(s)}{u(0)} = \frac{u(s)}{q_m}
		\label{\EQ: def_Ka_over_Ks}
\end{eqnarray}
assuming that synonymous substitutions are completely neutral 
and mutation rates at both types of sites are the same.
The PDF of $K_a/K_s$ is
\begin{eqnarray}
	p(K_a/K_s) &=& 
		p(u) \frac{d u}{d (K_a/K_s)} 
	= p(u) \, q_m  
		\label{\EQ: pdf_of_Ka_over_Ks}
\end{eqnarray}

\subsection{Probability distributions of $\Delta\Delta \psi_{ND}$, $4N_e s$, $u$, and $K_a/K_s$ in fixed mutant genes}

\SUPPLEMENT{
Now, let us consider fixed mutant genes. 
}%  SUPPLEMENT
The PDF of $\Delta\Delta \psi_{ND}$ in fixed mutants is
proportional to that multiplied by the fixation probability.
\begin{eqnarray}
p(\Delta\Delta \psi_{ND, \script{fixed}})
	&=& p(\Delta\Delta \psi_{ND}) 
	\frac{u(s(\Delta \Delta \psi_{ND}))}{\langle u(s(\Delta \Delta \psi_{ND})) \rangle}
	\label{\EQ: pdf_of_fixed_ddPsi}
	\\
  \langle u \rangle &\equiv& \int_{-\infty}^{\infty} u(s) p(\Delta\Delta \psi_{ND}) d\Delta\Delta \psi_{ND}
	\label{\EQ: ave_of_u}
\end{eqnarray}
Likewise, the PDF of selective advantage 
in fixed mutants is 
\begin{eqnarray}
p(4N_e s_{\script{fixed}}) &=&
	p(4N_e s)
	\frac{u(s)}{\langle u(s) \rangle}
\end{eqnarray}
and those of the $u$ and $K_a/K_s$ in fixed mutants are
\begin{eqnarray}
	p( u_{\script{fixed}} ) &=&
		p(u) \frac{u}{\langle u \rangle}
		\\
	p( (\frac{K_a}{K_s})_{\script{fixed}}) &=&
		p(\frac{K_a}{K_s}) \frac{u}{\langle u \rangle}
		=
		p(\frac{K_a}{K_s}) \frac{\frac{K_a}{K_s}}{\langle \frac{K_a}{K_s} \rangle}
\end{eqnarray}
The average of $K_a/K_s$ in fixed mutants is equal to the ratio of the second moment to the first moment of $K_a/K_s$
\TEXT{
in all arising mutants; 
$
\langle K_a/K_s \rangle_{\script{fixed}} = \langle (K_a / K_s)^2 \rangle / \langle K_a/K_s \rangle
$.
}%  TEXT
\SUPPLEMENT{
in all arising mutants.
\begin{eqnarray}
\langle \frac{K_a}{K_s} \rangle_{\script{fixed}} &=& \langle (\frac{K_a}{K_s})^2 \rangle / \langle \frac{K_a}{K_s} \rangle
\end{eqnarray}
}%  SUPPLEMENT

% End of methods_for_monoclonal_approximation.tex
}%  TEXT

\SUPPLEMENT{

\subsection{ Estimation of $\bar{\psi}(\VEC{f}(\VECS{\sigma}))$ and $\delta {\psi}^2(\VEC{f}(\VECS{\sigma}))$ }
\label{estimation_psi_distr}

The mean $\bar{\psi}(\VEC{f}(\VECS{\sigma}))$ and the variance $\delta {\psi}^2(\VEC{f}(\VECS{\sigma}))$ 
in the Gaussian approximation for the distribution of conformational energies at the denatured state are estimated 
as the mean and variance of $\psi_N$ of random sequences
in the native conformation\CITE{PGT:97}.
\begin{eqnarray}
	\bar{\psi}(\VEC{f}(\VECS{\sigma})) &=& 
	- \sum_i [ \hat{h}_i(::) +  \sum_{j>i} \hat{J}_{ij}(::,::) ] 
\end{eqnarray}
where $\hat{h}_i(::)$ and $\hat{J}_{ij}(::,::)$ are the means of one-body and two-body interactions in random sequences.
\begin{eqnarray}
	\hat{h}_i(::) &\equiv& \sum_k \hat{h}_i(a_k)f_{a_k}(\VECS{\sigma}) 
	\\
	\hat{J}_{ij}(::,::) &\equiv& \sum_k \sum_l \hat{J}_{ij}(a_k, a_l)f_{a_k}(\VECS{\sigma}) f_{a_l}(\VECS{\sigma})
\end{eqnarray}
where $f_{a_k}(\VECS{\sigma})$ is the composition of amino acid $a_k$ in the sequence $\VECS{\sigma}$.
\begin{eqnarray}
	f_{a_k}(\VECS{\sigma}) &=& \frac{1}{L} \sum_{i=1}^{L} \delta_{\sigma_i a_k}
\end{eqnarray}
where $\delta_{\sigma_i a_k}$ is the Kronecker delta.
The variance, $\delta\psi^2(\VEC{f}(\VEC{\sigma}))$, is
\begin{eqnarray}
	\delta {\psi}^2(\VEC{f}(\VECS{\sigma})) &=& 
	\sum_k \, f_{a_k}(\VECS{\sigma}) \, \sum_i \, [ \, \delta \hat{h}_i(a_k)^2
	+ \sum_{j \neq i} \, \{ \, 2 \delta \hat{h}_i(a_k) \delta \hat{J}_{ij}(a_k,::)
	\\
	&+& \sum_{m \neq \{i, j\}} \delta \hat{J}_{ij}(a_k, ::) \delta \hat{J}_{i m}(a_k,::) 
	+ \frac{1}{2} \sum_l \delta \hat{J}_{ij}(a_k, a_l)^2 f_{a_l}(\VECS{\sigma}) \, \} \, 
	\, ] 
\end{eqnarray}
where
\begin{eqnarray}
	\delta \hat{h}_i(a_k) &\equiv& \hat{h}_i(a_k) - \hat{h}_i(::)
	\\
	\delta \hat{J}_{ij}(a_k, ::) &\equiv& \hat{J}_{ij}(a_k, ::) - \hat{J}_{ij}(::, ::)
	\\
	\delta \hat{J}_{ij}(a_k, a_l) &\equiv& \hat{J}_{ij}(a_k, a_l) - \hat{J}_{ij}(::, ::)
\end{eqnarray}

% End of methods_to_estimate_Gaussian_parameters.tex
}%  SUPPLEMENT

\SUPPLEMENT{

\subsection{Estimation of one-body ($h$) and pairwise ($J$) interactions}

  The estimates of $h$ and $J$\CITE{MPLBMSZOHW:11,MCSHPZS:11} are noisy as a result of estimating
many interaction parameters from a relatively small number of sequences.
Therefore, only pairwise interactions within a certain distance are taken into account;
the estimate of $J$ is modified as follows,
according to Morcos et al.\CITE{MSCOW:14}.

\begin{eqnarray}
	\hat{J}^{\script{q}}_{ij}(a_k, a_l) &=& J^{\script{q}}_{ij}(a_k, a_l) H( r_{\script{cutoff}} - r_{ij} )
		\label{\EQ: estimation_of_J}
\end{eqnarray}
where 
$\hat{J}^{\script{q}}$ is the statistical estimate of $J$ in the mean field approximation in which
the amino acid $a_q$ is the reference state,
$H$ is the Heaviside step function, and $r_{ij}$ is the distance between the centers of amino acid side chains in protein structure,
and $r_{\script{cutoff}}$ is a distance threshold for residue pairwise interactions.
Maximum interaction ranges employed for pairwise interactions are
$r_{\script{cutoff}} \sim 8$ and $15.5$ \AA\ , 
which correspond to the first and second interaction shells between residues, respectively.
Here it should be noticed that the 
total interaction 
$\psi_N(\VECS{\sigma})$ 
defined by \Eq{\ref{\EQ: total_interaction_in_potts_model}} does not depend on any gauge 
unless the interaction range for pairwise interactions is limited, but 
a gauge conversion in which interconversions between $h$ and $J$ 
occur must not be done before calculating $\hat{J}$, because
it may change the estimate of $\psi_N(\VECS{\sigma})$
in the present scheme of \Eq{\ref{\EQ: estimation_of_J}} 
in which pairwise interactions are cut off at a certain distance.
Thus, a natural gauge must be used before calculating $\hat{J}$. 

For example, let us think about the Ising gauge\CITE{EHA:14}, in which $h^I$ and $J^I$ can be
calculated from $h^g$ and $J^g$ in any gauge through the following conversions.
\begin{eqnarray}
	J^{\script{I}}_{ij}(a_k, a_l) &=& J^{\script{g}}_{ij}(a_k, a_l) - J^{\script{g}}_{ij}(a_k, :)
				- J^{\script{g}}_{ij}(:, a_l) + J^{\script{g}}_{ij}(:, :) 
		\label{\EQ: to_Ising_gauge_1}
			\\
	h^{\script{I}}(a_k) &=& h^{\script{g}}_{i}(a_k) - h^{\script{g}}_{i}(:)
			+ \sum_{j \neq i} (J^{\script{g}}_{ij}(a_k, :) - J^{\script{g}}_{ij}(:, :) ) 
		\label{\EQ: to_Ising_gauge_2}
\end{eqnarray}
where
\begin{eqnarray}
	h_{i}(:) &\equiv& \frac{1}{q} \sum_{k=1}^q h_{i}(a_k)
		\\
	J_{ij}(:, :) &\equiv& \frac{1}{q^2} \sum_{k=1}^q \sum_{l=1}^q J_{ij}(a_k, a_l)
\end{eqnarray}
where $q$ is equal to the total number of amino acid types including deletion, that is, $q=21$.
Thus, the gauge conversion of $\hat{J}$ does not affect the total interaction $\psi_N(\VECS{\sigma})$
but the gauge conversion before calculating $\hat{J}$ may significantly change the total interaction.

In the DCA\CITE{MPLBMSZOHW:11,MCSHPZS:11}, 
the interaction terms are estimated in the mean field approximation as follows.
\begin{eqnarray}
	J^{\script{q}}_{ij}(a_k, a_l) &=& - (C^{-1})_{ij}(a_k, a_l)		\\	
	J^{\script{q}}_{ij}(a_q, a_l) &=& J^{\script{q}}_{ij}(a_k, a_q) = J^{\script{q}}_{ij}(a_q, a_q) = 0
\end{eqnarray}
where $i \neq j$ and $1 \leq k, l \leq q - 1$, and the covariance matrix $C$ is defined as
\begin{eqnarray}
	C_{ij}(a_k,a_l) &\equiv & P_{ij}(a_k, a_l) - P_i(a_k) P_j(a_l)	
\end{eqnarray}
Here, one ($a_q$) of the amino acid types including deletion is used as the reference state; 
$J^{\script{q}}$ denotes the $J$ in this gauge, which is called the $q$ gauge here.
According to Morcos et al.\CITE{MPLBMSZOHW:11},
the probability $P_i(a_k)$ of amino acid $a_k$ at site $i$ and 
the joint probabilities $P_{ij}(a_k, a_l)$ of amino acids, $a_k$ at site $i$ and $a_l$ at site $j$, 
are
evaluated by 
\begin{eqnarray}
	P_i(a_k) &=& (1 - p_c) f_i(a_k) + p_c \frac{1}{q} 
	\label{\EQ: pseudocount_for_Pi}
	\\
	P_{ij}(a_k, a_l) &=& (1 - p_c) f_{ij}(a_k, a_l) + p_c \frac{1}{q^2} 
		\hspace*{2em} \text{ for } \hspace*{1em} i \neq j
	\label{\EQ: pseudocount_for_Pij}
	\\
	P_{ii}(a_k, a_l) &=& P_i(a_k) \delta_{a_k a_l}  
\end{eqnarray}
where   
$0 \leq p_c \leq 1$ is the ratio of pseudocount, and
$f_i(a_k)$ is the frequency of amino acid $a_k$ at site $i$ and $f_{ij}(a_k, a_l)$
is the frequency of the site pair, $a_k$ at $i$ and $a_l$ at $j$, in an alignment; 
$f_{ii}(a_k, a_l)$ is defined as $f_{ii}(a_k, a_l) = f_i(a_k) \delta_{a_k a_l}$.

In the mean field approximation, one body interactions $h^q_i(a_k)$ 
in the $q$ gauge are estimated by $\hat{h}^q_i(a_k) = \log (P_i(a_k)/P_i(a_q)) - \sum_{j \neq i} \sum_{l \neq q} \hat{J}^q_{ij}(a_k, a_l) P_j(a_l)$.
Here, instead the one body interactions $h_i(a_k)$ are estimated in the isolated two-state model\CITE{MPLBMSZOHW:11}, 
that is,
\begin{eqnarray}
	P_i(a_k) &\propto& \exp \, [ \, h^{\script{q}}_{ij}(a_k) + J^{\script{q}}_{ij}(a_k, a_l) + h^{\script{q}}_{ji}(a_l) \, ]
		\\
	\hat{h}^{\script{q}}_i(a_k) &=& \frac{1}{L-1} \sum_{j \neq i} h^{\script{q}}_{ij}(a_k)
\end{eqnarray}
These $\hat{h}^q$ and $\hat{J}^q$ in the $q$ gauge are converted to a new gauge, which
is called the zero-sum gauge here, 
\begin{eqnarray}
	\hat{h}^{\script{s}}_{i}(a_k) &=& \hat{h}^{\script{q}}_{i}(a_k) - \hat{h}^{\script{q}}_{i}(:)
	\label{\EQ: simple_gauge_1}	\\
	\hat{J}^{\script{s}}_{ij}(a_k, a_l) &=& \hat{J}^{\script{q}}_{ij}(a_k, a_l) - \hat{J}^{\script{q}}_{ij}(:, :)
	\label{\EQ: simple_gauge_2}
\end{eqnarray}
In this gauge, the reference state is the average state over amino acids including deletion, instead of
a specific amino acid ($a_q$) in the $q$ gauge.

% End of methods_for_h_and_J_estimations.tex
}%  SUPPLEMENT

\SUPPLEMENT{

\subsection{Distribution of $\Delta\Delta \psi_{ND} \simeq \Delta \psi_N$ due to single nucleotide nonsynonymous substitutions }

The probability density function (PDF) of $\Delta\Delta \psi_{ND}$, $p(\Delta\Delta \psi_{ND})$, 
due to single nucleotide nonsynonymous substitutions is 
approximated by the PDF of $\Delta \psi_{N}$, $p(\Delta \psi_{N})$, because $\Delta \psi_{D} \simeq 0$ 
for single amino acid substitutions.
\begin{eqnarray}
	\Delta\Delta \psi_{ND} &\simeq& \Delta \psi_{N}
		\\
	p(\Delta\Delta \psi_{ND}) &\simeq& p(\Delta \psi_N)
\end{eqnarray}
for single nucleotide nonsynonymous substitutions.

For simplicity, a log-normal distribution, 
$\ln\mathcal{N}(x; \mu, \sigma)$, for which $x, \mu$ and $\sigma$ are defined as follows, 
is arbitrarily employed here to reproduce observed PDFs of $\Delta \psi_N$, 
particularly in the domain of $\Delta \psi_N < \overline{\Delta \psi_N}$,   
although other distributions such as
inverse $\Gamma$ distributions can equally reproduce the observed ones, too.
\begin{eqnarray}
	p(\Delta \psi_N) &\approx&
	\ln\mathcal{N}(x; \mu, \sigma) \equiv \frac{1}{x} \mathcal{N}(\ln x; \mu, \sigma)
	\label{\EQ: log-normal}
	\\
	x &\equiv& \max ( \Delta \psi_{N} - \Delta \psi_{N}^{\mbox{o}}, 0)
	\\
	\exp(\mu + \sigma^2/2) &=& \overline{\Delta \psi_{N}} - \Delta \psi_{N}^{\mbox{o}}
	\\
	\exp(2\mu + \sigma^2) ( \exp(\sigma^2) - 1 ) &=& \overline{ (\Delta \psi_{N} - \overline{\Delta \psi_{N}})^2 })
	\\
	\Delta \psi_{N}^{\mbox{o}} &\equiv& \min (\overline{\Delta \psi_{N}} 
		- n_{\mbox{shift}} \overline{ (\Delta \psi_{N} - \overline{\Delta \psi_{N}})^2 })^{1/2}, 0)
	\label{\EQ: statistics_for_log-normal}
\end{eqnarray}
where $\Delta \psi_{N}^{\mbox{o}}$ is the origin for the log-normal distribution 
and the shifting factor $n_{\mbox{shift}}$ is taken to be equal to $2$, unless specified.

% End of methods_for_pdf_of_ddPsi.tex
% \input{methods_for_monoclonal_approximation.tex}

\subsection{Probability distributions of selective advantage, fixation rate and $K_a/K_s$}

\SUPPLEMENT{
Now, we can consider 
}%  SUPPLEMENT
\TEXT{
Let us consider 
}%  TEXT
the probability distributions of
characteristic quantities that describe the evolution of genes. 
First of all, the probability density function (PDF) of 
selective advantage $s$, $p(s)$, of mutant genes can be calculated
from the PDF of 
the change of $\Delta \psi_{ND}$ due to a mutation from $\VEC{\mu}$ to $\VEC{\nu}$, 
$\Delta\Delta \psi_{ND} (\equiv \Delta \psi_{ND}(\VEC{\nu}, T) - \Delta \psi_{ND}(\VEC{\mu}, T) )$.
The PDF of $4N_e s$, $p(4N_e s) = p(s)/(4N_e) $, may be more useful than $p(s)$.
\begin{eqnarray}
p(4N_e s)
	&=& p(\Delta \Delta \psi_{ND}) \, | \frac{d \Delta \Delta \psi_{ND}}{d 4N_e s} |
	=
	p(\Delta\Delta \psi_{ND}) (1 - q_m)
	\label{\EQ: pdf_of_4Nes}
	\label{\EQ: pdf_of_s}
\end{eqnarray}
where $\Delta\Delta \psi_{ND}$ must be regarded as a function of $4N_es$, that is,
$
\Delta\Delta \psi_{ND} = - 4N_e s (1 - q_m)
$; see \Eq{\ref{\EQ: s_vs_dPsi}}.

The PDF of fixation probability $u$ can be represented by
\begin{eqnarray}
p(u)	&=& p(4N_e s) \frac{d 4N_e s}{d u} 
	= 
	p(4N_e s)
	 \frac{ ( e^{4N_e s} - 1)^2 e^{4N_e s (q_m - 1)} }
	{ q_m( e^{4N_e s} - 1) - (e^{4N_e s q_m} - 1) }
		\label{\EQ: pdf_of_fixation_prob}
\end{eqnarray}
where $4N_e s$ must be regarded as a function of $u$. 

The ratio of the substitution rate per nonsynonymous site ($K_a$) for nonsynonymous
substitutions with selective advantage s to the substitution rate per
synonymous site ($K_s$) for synonymous substitutions with s = 0 is
\begin{eqnarray}
	\frac{K_a}{K_s} &=& \frac{u(s)}{u(0)} = \frac{u(s)}{q_m}
		\label{\EQ: def_Ka_over_Ks}
\end{eqnarray}
assuming that synonymous substitutions are completely neutral 
and mutation rates at both types of sites are the same.
The PDF of $K_a/K_s$ is
\begin{eqnarray}
	p(K_a/K_s) &=& 
		p(u) \frac{d u}{d (K_a/K_s)} 
	= p(u) \, q_m  
		\label{\EQ: pdf_of_Ka_over_Ks}
\end{eqnarray}

\subsection{Probability distributions of $\Delta\Delta \psi_{ND}$, $4N_e s$, $u$, and $K_a/K_s$ in fixed mutant genes}

\SUPPLEMENT{
Now, let us consider fixed mutant genes. 
}%  SUPPLEMENT
The PDF of $\Delta\Delta \psi_{ND}$ in fixed mutants is
proportional to that multiplied by the fixation probability.
\begin{eqnarray}
p(\Delta\Delta \psi_{ND, \script{fixed}})
	&=& p(\Delta\Delta \psi_{ND}) 
	\frac{u(s(\Delta \Delta \psi_{ND}))}{\langle u(s(\Delta \Delta \psi_{ND})) \rangle}
	\label{\EQ: pdf_of_fixed_ddPsi}
	\\
  \langle u \rangle &\equiv& \int_{-\infty}^{\infty} u(s) p(\Delta\Delta \psi_{ND}) d\Delta\Delta \psi_{ND}
	\label{\EQ: ave_of_u}
\end{eqnarray}
Likewise, the PDF of selective advantage 
in fixed mutants is 
\begin{eqnarray}
p(4N_e s_{\script{fixed}}) &=&
	p(4N_e s)
	\frac{u(s)}{\langle u(s) \rangle}
\end{eqnarray}
and those of the $u$ and $K_a/K_s$ in fixed mutants are
\begin{eqnarray}
	p( u_{\script{fixed}} ) &=&
		p(u) \frac{u}{\langle u \rangle}
		\\
	p( (\frac{K_a}{K_s})_{\script{fixed}}) &=&
		p(\frac{K_a}{K_s}) \frac{u}{\langle u \rangle}
		=
		p(\frac{K_a}{K_s}) \frac{\frac{K_a}{K_s}}{\langle \frac{K_a}{K_s} \rangle}
\end{eqnarray}
The average of $K_a/K_s$ in fixed mutants is equal to the ratio of the second moment to the first moment of $K_a/K_s$
\TEXT{
in all arising mutants; 
$
\langle K_a/K_s \rangle_{\script{fixed}} = \langle (K_a / K_s)^2 \rangle / \langle K_a/K_s \rangle
$.
}%  TEXT
\SUPPLEMENT{
in all arising mutants.
\begin{eqnarray}
\langle \frac{K_a}{K_s} \rangle_{\script{fixed}} &=& \langle (\frac{K_a}{K_s})^2 \rangle / \langle \frac{K_a}{K_s} \rangle
\end{eqnarray}
}%  SUPPLEMENT

% End of methods_for_monoclonal_approximation.tex
}%  SUPPLEMENT

\SUPPLEMENT{
}%  SUPPLEMENT

% End of methods_1+2.tex

\section{Materials}

\subsection{Sequence data}

We study the single domains of 8 Pfam\CITE{FCEEMMPPQSSTB:16} families and both the single domains 
and multi-domains from 3 Pfam families.
In \Table{\ref{\TBL: Proteins_studied}},
their Pfam ID for a multiple sequence alignment, and
UniProt ID and PDB ID with the starting- and ending-residue positions of the domains are listed.
The full alignments for their families at the Pfam are used to estimate one-body interactions $h$ and 
pairwise interactions $J$ with the DCA program from ``http://dca.rice.edu/portal/dca/home''\CITE{MCSHPZS:11,MPLBMSZOHW:11}.
To estimate the sample ($\overline{\psi_N}$) and ensemble ($\langle \psi_N \rangle_{\VEC{\sigma}}$) averages of the 
evolutionary statistical energy,
$M$ unique sequences with no deletions are used.  
In order to reduce
phylogenetic biases in the set of homologous sequences,
we employ
a sample weight ($w_{\VEC{\sigma}_N}$) for each sequence, which is equal to the inverse of the number of sequences
that are less than 20\% different from a given sequence in a given set of homologous sequences. 
Only representatives of unique sequences with no deletions, which are at least 20\% different from each other, are used to calculate 
the changes of the 
evolutionary statistical energy
($\Delta \psi_N$)
due to single nucleotide nonsynonymous substitutions;
the number of the representatives is almost equal to the effective number of sequences ($M_{\script{eff}}$) in
\Table{\ref{\TBL: Proteins_studied}}.

% End of materials_1.tex
% End of materials_1+2.tex

\section{Results}

First, We describe how one-body and pairwise interactions, $h$ and $J$,
are estimated. 
Then, the changes of evolutionary statistical energy ($\Delta \psi_{N}$)
due to single nucleotide nonsynonymous changes on natural sequences
are analyzed with respect to dependences on the $\psi_{N}$ of the wildtype
sequences.  
The results indicate
that the standard deviation of $\Delta G_{N} \simeq k_B T_s \Delta \psi_{N}$ 
is almost constant over protein families. 
Hence, the 
selective
temperatures, $T_s$, of various protein families 
can be estimated in a relative scale
from the standard deviation of $\Delta \psi_{N}$. 
The $T_s$ of a reference protein is estimated by
comparing the expected values of $\Delta \Delta G_{ND}$ with their
experimental values.
Folding free energies $\Delta G_{ND}$ 
are estimated from estimated $T_s$ and 
experimental melting temperature $T_m$, and compared with their experimental 
values for 5 protein families.
Glass transition temperatures $T_g$ are also estimated from $T_s$ and $T_m$.

Secondly, based on the distribution of $\Delta \psi_N$,
protein evolution is studied.
Evolutionary statistical energy ($\psi_N$) attains the equilibrium 
when the average of $\Delta \psi_N$ over fixed mutations is equal to zero.
The PDF of $\Delta \psi_N$ is approximated by log-normal distributions.
The basic relationships are that 1) the standard deviation of $\Delta \psi_N$ is constant specific to 
a protein family, and 2) the mean of $\Delta \psi_N$ linearly depends on $\psi_N$.
The equilibrium value of $\psi_N$ is shown to agree with the mean of $\psi_N$
over homologous proteins in each protein family.
In the present approximation,
the standard deviation of $\Delta \psi_N$
and
selective
temperature $T_s$ 
at the equilibrium are simple functions of the equilibrium value of 
mean $\Delta \psi_N$, $\overline{\Delta \psi_N}^{\script{eq}}$.
Lastly, the probability distribution of $K_a/K_s$, which is the ratio of nonsynonymous to 
synonymous substitution rate per site, is analyzed 
as a function of $\overline{\Delta \psi_N}^{\script{eq}}$,
in order to examine how significant neutral selection is 
in the selection 
maintaining protein stability and foldability.
Also, it is confirmed that 
selective
temperature $T_s$ negatively correlates with the mean of $K_a/K_s$,
which represents the evolutionary rate of protein.

\subsection{Important parameters in the estimations of one-body and pairwise interactions, $h$ and $J$, and of the 
evolutionary statistical energy,
$\psi_N(\VEC{\sigma})$}

The one-body ($h$) and pairwise ($J$) interactions for amino acid order in a protein sequence 
are estimated here by the DCA method\CITE{MCSHPZS:11,MPLBMSZOHW:11}, 
although there are multiple methods for estimating them\CITE{ELLWA:13,EHA:14}.
In the case of the DCA method,
the ratio of pseudocount ($0 \leq p_c \leq 1$) defined in \Eqs{\REF{seq: pseudocount_for_Pi} and \REF{seq: pseudocount_for_Pij}} 
is a parameter and controls the values of the ensemble and sample averages of $\psi_N$ in sequence space,  
$\langle \psi_N(\VEC{\sigma}) \rangle_{\VEC{\sigma}}$
in \Eq{\ref{\EQ: ensemble_ave_of_psi}}
and
$\overline{\psi_N(\VEC{\sigma}_N)}$ 
in \Eq{\ref{\EQ: def_sample_ave_of_psi}};
a weight for observed counts is defined to be equal to $(1-p_c)$.
Sample average means the average over all homologous sequences with a weight for each sequence to reduce phylogenetic biases.
An appropriate value must be chosen for the ratio of pseudocount in a reasonable manner. 

Another problem is that the estimates of $h$ and $J$\CITE{MPLBMSZOHW:11,MCSHPZS:11} 
may be noisy as a result of estimating
many interaction parameters from a relatively small number of sequences.
Therefore,
only pairwise interactions within a certain distance are taken into account;
the estimate of $J$ is modified as follows,
according to Morcos et al.\CITE{MSCOW:14}.
\begin{eqnarray}
	\hat{J}^{\script{q}}_{ij}(a_k, a_l) &=& J^{\script{q}}_{ij}(a_k, a_l) H( r_{\script{cutoff}} - r_{ij} )
		\label{\EQ: estimation_of_J}
\end{eqnarray}
where 
$J^{\script{q}}$ is the statistical estimate of $J$ in the mean field approximation in which
the amino acid $a_q$ is the reference state,
$H$ is the Heaviside step function, and $r_{ij}$ is the distance between the centers of amino acid side chains 
at sites $i$ and $j$ in a protein structure,
and $r_{\script{cutoff}}$ is a distance threshold for residue pairwise interactions.
The one-body interactions $h_i(a_k)$ are estimated in the isolated two-state model\CITE{MPLBMSZOHW:11}
rather than the mean field approximation; see the Method section in the Supplement for details.
The zero-sum gauge is employed to represent $h$ and $J$; 
$\sum_{k} \hat{h}^{\script{s}}_i(a_k) = \sum_k\sum_l \hat{J}^{\script{s}}_{ij}(a_k, a_l) = 0$ in the zero-sum gauge.

Candidates for the cutoff distance may be about 8 \AA\  for the first interaction 
shell and 15--16 \AA\  for the second interaction shell between residues;
distance between the centers of side chain atoms is employed for residue distance. 
Here both the distances are tested for the cutoff distance.
Pseudocount in the Bayesian statistics is determined usually as 
a function of the number of samples (sequences), although 
the ratio of pseudocount $p_c=0.5$ was used for all proteins in the contact prediction\CITE{MPLBMSZOHW:11}.
Here, an appropriate value for the ratio of pseudocount for the certain cutoff distance, 
either about 8 \AA\  or 15--16 \AA,
is chosen for each protein family in such a way that
the sample average of the 
evolutionary statistical energies
must be equal to the ensemble average, 
$\overline{\psi_N} = \langle \psi_N \rangle_{\VEC{\sigma}}$;
see \Eqs{\REF{\EQ: ensemble_ave_of_psi} and \REF{\EQ: sample_ave_of_psi}}.
As shown in 
\Fig{\ref{sfig: 1gm1-a:16-96_full_non_del_dca0_205_0_20_simple-gauge_ensemble_vs_sample_ave_phi_vs_r}},
the value of $r_{\script{cutoff}}$, where $\overline{\psi_N} = \langle \psi_N \rangle_{\VEC{\sigma}}$
is satisfied, monotonously changes as a function of the ratio of pseudocount $p_c$.
The values of $p_c$, 
where $\overline{\psi_N} = \langle \psi_N \rangle_{\VEC{\sigma}}$ 
is satisfied near the specified values of $r_{\script{cutoff}}$, 8 \AA\  and 15.5 \AA,  
are employed for $r_{\script{cutoff}} \simeq 8$ \AA\ and $15.5$ \AA, respectively.
In the present multiple sequence alignment for the PDZ domain, 
with the ratios of pseudocount $p_c = 0.205$ and $p_c = 0.33$,
the sample and ensemble averages agree with each other at 
the cutoff distances $r_{\script{cutoff}} \sim 8$ \AA\  and $r_{\script{cutoff}} \sim 15.5$ \AA, respectively;
see 
\Fig{\ref{sfig: 1gm1-a:16-96_full_non_del_dca0_205_0_20_simple-gauge_ensemble_vs_sample_ave_phi_vs_r}}.
In \Fig{\ref{sfig: 1gm1-a:16-96_full_non_del_dca0_205_0_20_simple-gauge_ddG-dPhi_vs_r}},
the reflective correlation and regression coefficients
between the experimental $\Delta\Delta G_{ND}$\CITE{GGCJVTVB:07} and $\Delta \psi_N$ due to single 
amino acid substitutions
are plotted against the cutoff distance for pairwise interactions in the PDZ domain.
The reflective correlation coefficient has the maximum at the $r_{\script{cutoff}} \sim 8$ \AA\  for
$p_c = 0.205$ and at $r_{\script{cutoff}} \sim 15.5$ \AA\  for $p_c = 0.33$ , indicating that
these cutoff distances are appropriate for these ratios of pseudocount.
The ratio of pseudocount and a cutoff distance employed are listed for each protein family 
in \Tables{\ref{\TBL: ddPsi_with_8A} and \ref{stbl: ddPsi_with_16A}} for $r_{\script{cutoff}} \sim 8$ and $15.5$ \AA, 
respectively. 
The ratios of pseudocount employed here are all smaller than $0.5$, which was reported to be 
appropriate for contact prediction;
by using strong regularization,
contact prediction is improved but the generative power of the inferred model is degraded\CITE{BLCC:16}.
In the text, only results with $r_{\script{cutoff}} \sim 8$ \AA\ are shown. 
In a supplement, results with $r_{\script{cutoff}} \sim 15.5$ \AA\  are provided and
discussed in comparison with the results of $r_{\script{cutoff}} \sim 8$ \AA.

\subsection{Changes of the 
evolutionary statistical energy,
$\Delta \psi_{N}$, by single nucleotide nonsynonymous substitutions}

The changes of the 
evolutionary statistical energy, $\Delta \psi_N$ and $\Delta \psi_D$,
due to a single amino acid substitution 
from $\sigma^{N}_i$ to $\sigma_i$
at site $i$ in 
a natural sequence 
$\VEC{\sigma}_N$ are defined as
\begin{eqnarray}
	\Delta \psi_N(\sigma^N_{j \neq i}, \sigma^N_i\rightarrow \sigma_i) &\equiv& \psi_N(\sigma^N_{j \neq i}, \sigma_i) - \psi_N(\VEC{\sigma}_N)
		\\
	\Delta \psi_D(\sigma^N_{j \neq i}, \sigma^N_i\rightarrow \sigma_i, T) &\equiv& \psi_D(\sigma^N_{j \neq i}, \sigma_i, T) - \psi_D(\VEC{\sigma}_N, T) 
		\\
	\Delta \Delta \psi_{ND} (\sigma^N_{j \neq i}, \sigma^N_i \rightarrow \sigma_i) 
		&\equiv& \Delta \psi_N(\sigma^N_{j \neq i}, \sigma^N_i\rightarrow \sigma_i) 
			- \Delta \psi_D(\sigma^N_{j \neq i}, \sigma^N_i\rightarrow \sigma_i)
		\\
		&\simeq& \Delta \psi_N(\sigma^N_{j \neq i}, \sigma^N_i\rightarrow \sigma_i)
		\hspace*{1em} \mbox{ because } \VEC{f}(\VEC{\sigma}_N) \approx  \VEC{f}(\sigma^N_{j \neq i}, \sigma_i) 
\end{eqnarray}
Here, single amino acid substitutions caused by single nucleotide nonsynonymous mutations
are taken into account, unless specified. 
Let us use a single overline to denote the average of the changes of interaction 
over all types of single nucleotide nonsynonymous mutations at all sites in a specific native sequence, 
and a double overline to denote their averages
over all homologous sequences in a protein family.

We calculated the $\psi_N$ of 
the wildtype
and $\Delta \psi_N$ due to all types of single nucleotide nonsynonymous substitutions
for all homologous sequences, and their means and variances.
We have examined the dependence of $\overline{\Delta \Delta \psi_{ND}} \simeq \overline{\Delta \psi_{N}}$ on 
the $\psi_{N}$ of each homologous sequence in each protein family. 
\Fig{\ref{fig: 1gm1-a:16-96_dca0_205_simple-gauge_ddPhi_at_opt}} 
for the PDZ family
and
\Figs{\ref{sfig: 1r69-a:6-58.full_non_del.dca0_18.0_20.simple-gauge.ddPhi_at_opt} to 
\ref{sfig: 5azu-a:4-128.full_non_del.dca0_23.0_20.simple-gauge.BD.ddPhi_at_opt}}
for all proteins show that $\overline{ \Delta \psi_N }$ is negatively proportional to    
the $\psi_{N}/L$ of the wildtype, that is,
\begin{eqnarray}
\lefteqn{
	\overline{ \Delta \Delta \psi_{ND}(\sigma^N_{j \neq i}, \sigma^N_i\rightarrow \sigma_i) } 
	\simeq \overline{ \Delta \psi_N(\sigma^N_{j \neq i}, \sigma^N_i\rightarrow \sigma_i) } 
}
	\nonumber
	\\
        &\approx&  
	\alpha_{\psi_N} \frac{\psi_N(\VEC{\sigma}_N) - \overline{\psi_N(\VEC{\sigma}_N)} }{L} + 
	\overline{ \overline{        \Delta \psi_{N}(\sigma^N_{j \neq i}, \sigma^N_i\rightarrow \sigma_i) } } 
	\hspace*{2em} \text{ with } \alpha_{\psi_N} < 0
	\label{\EQ: regression_of_dPsi_on_Psi}
\end{eqnarray}
where $L$ is sequence length.
This relationship is found in all of the protein families examined here; 
the correlation and regression coefficients 
for $r_{\script{cutoff}} \sim 8$ and $15.5$ \AA\  are listed 
in \Tables{\ref{\TBL: ddPsi_with_8A} and \ref{stbl: ddPsi_with_16A}}, respectively.
Most of the correlation coefficients are larger than 0.95, and all are greater than 0.9. 
It is reasonable that the change of the 
evolutionary statistical energy
($\Delta \psi_{N}$) 
depends on interaction per residue ($\psi_{N}/L$) 
rather than the 
evolutionary statistical energy
($\psi_{N}$), because
interactions change only for one residue substituted in the sequence.
Note that
the average interactions including a single residue 
will be equal to $2 \psi_N / L $ if all interactions are two-body.
The important fact is that
the linear dependence of $\Delta \psi_N$ on $ \psi_N / L$
shown in 
\Fig{\ref{fig: 1gm1-a:16-96_dca0_205_simple-gauge_ddPhi_at_opt}} 
and \Tables{\ref{\TBL: ddPsi_with_8A} and \ref{stbl: ddPsi_with_16A}} 
is equivalent to
the linear dependence of free energy changes caused by single amino acid substitutions
on the native conformational energy of the wildtype protein, because
the selective temperatures $T_S$ of homologous sequences in a protein family 
are approximated to be equal.

Is the same type of dependence on $\psi_{N}/L$ found for the standard deviation 
of $\Delta \psi_N$ over single nucleotide nonsynonymous substitutions at all sites?
\Fig{\ref{fig: 1gm1-a:16-96_dca0_205_simple-gauge_ddPhi_at_opt}}, 
\Figs{\ref{sfig: 1r69-a:6-58.full_non_del.dca0_18.0_20.simple-gauge.ddPhi_at_opt} to 
\ref{sfig: 5azu-a:4-128.full_non_del.dca0_23.0_20.simple-gauge.BD.ddPhi_at_opt}}
and \Tables{\ref{\TBL: ddPsi_with_8A} and \ref{stbl: ddPsi_with_16A}} show that 
the correlation between the standard deviation of $\Delta \psi_N$ and
$\psi_N$ of the wildtype is very weak except for Nitroreductase, SBP\_bac\_3 
and LysR\_substrate families.
Even for these protein families, 
the standard deviations of $\text{Sd}(\Delta\psi_N)$ are less than 10\% of the mean, $\overline{\text{Sd}(\Delta\psi_N)}$;
see \Tables{\ref{\TBL: ddPsi_with_8A} and \ref{stbl: ddPsi_with_16A}}. 
Thus, it is indicated that in general 
the variance/standard deviation of $\Delta \psi_N$ due to single amino acid substitutions
is almost constant irrespectively of the $\psi_N$ across homologous sequences.
The standard deviations of $\text{Sd}(\Delta\psi_N)$ is relatively large for the HTH\_3, because
in \Fig{\ref{sfig: 1r69-a:6-58.full_non_del.dca0_18.0_20.simple-gauge.ddPhi_at_opt}} 
there is a minor sequence group that has a distinguishable value of $\text{Sd}(\Delta\psi_N)$
from the major sequence group.

\subsection{Effective temperature $T_s$ of selection estimated from 
the changes of interaction, $\Delta \psi_{N}$, by single nucleotide nonsynonymous substitutions}

In the previous section, it has been shown that
the standard deviation of $\Delta \psi_N$ hardly depends on $\psi_N$ of the wildtype and
is nearly constant across homologous sequences in every protein family 
that has its own characteristic temperature ($T_s$) for selection pressure, 
indicating that $\text{Sd}(\Delta \psi_N)$ must be approximated by a function of only $k_B T_s$.
On the other hand, the free energy of the native structure, $\Delta G_{N}$, 
must not explicitly depend on $k_B T_s$, although
it may be approximated by a function of $G_N$.
In other words, the following relationships are derived.
\begin{eqnarray}
	\text{Sd}( \Delta \psi_{N}(\sigma^N_{j \neq i}, \sigma^N_i\rightarrow \sigma_i) )
	&\approx& \text{independent of } \psi_N \text{ and } 
		\nonumber
		\\
	&\ & \text{constant across homologous sequences in every protein family}	
		\nonumber
	\\
	&=& \text{function of } k_B T_s
	\label{\EQ: var_of_dPsi}
	\\
\mbox{Sd}(\Delta G_{N}(\sigma^N_{j \neq i}, \sigma^N_i\rightarrow \sigma_i))
	&=& \text{function that must not explicitly depend on } k_B T_s \text{ but } G_N
\end{eqnarray}
From the equations above, we obtain the important relation that 
the standard deviation 
of $\Delta G_N (= k_B T_s \Delta \psi_N)$ 
does not depend on $G_N$ and is nearly constant irrespective of protein families. 
\begin{eqnarray}
\mbox{Sd}(\Delta G_{N}(\sigma^N_{j \neq i}, \sigma^N_i\rightarrow \sigma_i)) 
		&\simeq&
		k_B T_s \, \mbox{Sd}( \Delta \psi_{N}(\sigma^N_{j \neq i}, \sigma^N_i \rightarrow \sigma_i) )
		\nonumber
		\\
		&\approx& \mbox{ constant }
		\label{\EQ: var_of_ddG}
		\label{\EQ: sd_of_ddG}
\end{eqnarray}
This relationship is consistent with the observation that
the standard deviation of 
$\Delta\Delta G_{ND} (\simeq \Delta G_{N})$
is nearly constant irrespectively of protein families\CITE{TSSST:07}.
This relationship allows us to estimate
a selective temperature ($T_s$) for a protein family 
in a scale relative to 
that of a reference protein from the ratio of the standard deviation of $\Delta \psi_N$.
The PDZ family is employed here as a reference protein,
and its $T_s$ is estimated by a direct comparison of $\Delta \psi_{N}$ and 
experimental $\Delta\Delta G_{ND}$;
the amino acid pair types and site locations of single amino acid substitutions are the most various, and 
also the correlation between the experimental $\Delta\Delta G_{ND}$ and $\Delta \psi_N$ is the best 
for the PDZ family in the present set of protein families, 
SH3\_1\CITE{GRSB:98}, ACBP\CITE{KONSKKP:99}, PDZ\CITE{GCAVBT:05,GGCJVTVB:07}, and Copper-bind\CITE{WW:05}; 
see \Tables{\ref{\TBL: Ts_with_8A} and \ref{stbl: Ts_with_16A}}. 
\begin{eqnarray}
	k_B\hat{T}_s &=& k_B \hat{T}_{s\script{, PDZ}} \, 
		[ \, 
			{ 
		\overline{ \mbox{Sd}( \Delta \psi_{\script{PDZ}}(\sigma^N_{j \neq i}, \sigma^N_i\rightarrow \sigma_i) ) }
			}
			\, / \, 
			{
		\overline{ \mbox{Sd}( \Delta \psi_{N}(\sigma^N_{j \neq i}, \sigma^N_i\rightarrow \sigma_i) ) }
			}
			\, ]
		\label{\EQ: Ts}
\end{eqnarray}
where the overline denotes the average over all homologous sequences.
Here, the averages of 
standard deviations
over all homologous sequences are employed, because
$T_s$ for all homologous sequences are approximated to be equal.
It will be confirmed in the later section, 
``the equilibrium value of $\psi_N$ in protein evolution'', 
that the assumption of the constant value specific to each protein family 
for $\text{Sd}(\Delta\psi_N)$ 
is appropriate.

\subsection{A direct Comparison of the changes of interaction, $\Delta \psi_{N} (\simeq \Delta\Delta \psi_{ND})$, 
with the experimental $\Delta\Delta G_{ND}$ due to single amino acid substitutions}

In order to determine the $T_s$ for a reference protein, 
the experimental values\CITE{GGCJVTVB:07} of $\Delta\Delta G_{ND}$ due to single amino acid substitutions in the PDZ domain  are plotted
against the changes of interaction, $\Delta \psi_{N}$, for the same types of substitutions
in \Figs{\ref{fig: 1gm1-a:16-96_dca0_205_simple-gauge_ddG-dPhi_at_opt} and \ref{sfig: 1gm1-a:16-96_dca0_33_simple-gauge_ddG-dPhi_at_opt}}.
The slope of the least-squares regression line through the origin, 
which is an estimate of $k_B T_s$, is equal to $k_B\hat{T}_s = 0.279$ kcal/mol,
and the reflective correlation coefficient is equal to $0.93$.
This estimate of $k_B T_s$
for the PDZ 
yield $\overline{ \mbox{Sd}(\Delta\Delta G_{ND}) } \simeq k_B \hat{T}_s \overline{\text{Sd}(\Delta \psi_N)} = 1.30$ kcal/mol,
which corresponds to $76\%$ of $1.7$ kcal/mol\CITE{SRS:12} estimated from 
ProTherm database or $80\%$ of $1.63$ kcal/mol\CITE{TSSST:07} computationally predicted for single nucleotide mutations 
by using the FoldX.
Using $\overline{ \mbox{Sd}(\Delta\Delta G_{ND}) } =1.30$ estimated 
from the $T_s$ for PDZ, the absolute values of $T_s$ for other proteins
are calculated by \Eq{\ref{\EQ: Ts}} and listed in \Tables{\ref{\TBL: Ts_with_8A}}; 
see \Table{\ref{stbl: Ts_with_16A}} 
for $r_{\script{cutoff}} \sim 15.5$ \AA. 
The $T_s$ estimated with $r_{\script{cutoff}} \sim 8$ and $15.5$\AA\ are compared with each other
in \Fig{\ref{sfig: Ts_relative_to_Tpdz_8_vs_16A}}.
Morcos et al.\CITE{MSCOW:14} estimated $T_s$ by comparing 
$\Delta \psi_{ND}$ with
$\Delta G_{ND}$ 
estimated by the associative-memory,
water-mediated, structure, and energy model (AWSEM).
They estimated $\psi_N$ with $r_{\script{cutoff}} = 16$ \AA\ and
probably $p_c = 0.5$.
In \Fig{\ref{sfig: Ts_relative_to_Tpdz_Wolynes_vs_8_and_16A}},
the present estimates of $T_s$ are compared with those by Morcos et al.\CITE{MSCOW:14}. 
The Morcos's estimates of $T_s$ with some exceptions tend to be located between
the present estimates with $r_{\script{cutoff}} \sim 8$ \AA\ and $15.5$\AA\,
which correspond to upper and lower limits for $T_s$ as discussed in the Discussion and the supplement.

\subsection{Relationship among $\overline{\overline{\Delta \psi_N}}$ of protein families;
weak dependency of $\Delta\Delta G_{ND}$ on $\Delta G_{ND}/L$}

The weak dependence of $\Delta\Delta G_{ND}$ on $\Delta G_{ND}$ 
was found\CITE{SRS:12,M:16} from the analysis of
stability changes due to single amino acid substitutions
in proteins, which are collected in the ProTherm database\CITE{KBGPKUS:06}. 
To understand this weak dependence,
let us consider the average of $\overline{\Delta \psi_{N}}$ over homologous sequences in each protein family.
The following regression line with $\alpha_{\overline{\psi_N}} = -1.74$ is shown in \Fig{\ref{fig: dpsi2_over_L_vs_dPsi}}. 
\begin{eqnarray}
	\overline{\overline{ \Delta \psi_N(\sigma^N_{j \neq i}, \sigma^N_i\rightarrow \sigma_i) } } 
		&\approx&  \alpha_{\overline{\psi_N}} \, \frac{ \overline{\psi_N(\VEC{\sigma}_N) } - \bar{\psi}(\overline{\VEC{f}(\VEC{\sigma}_N)}) }{L} + \beta_{\overline{\psi_N}} 
		\\
		&=& 	\alpha_{\overline{\psi_N}} \, \frac{ - {\delta \psi}^2( \overline{\VEC{f}(\VEC{\sigma}_N) } )  }{L} + \beta_{\overline{\psi_N}}
		\label{\EQ: regression_of_dPsiN_on_PsiN}
		\\
	\alpha_{\overline{\psi_N}} < 0
	\hspace*{2em} &,& \hspace*{2em}
	\beta_{\overline{\psi_N}} \approx 0
\end{eqnarray}
Here, $\overline{\psi_N(\VEC{\sigma}_N)}$ is reduced by $\bar{\psi}$
because the origin of the $\psi_N$ scale is not unique.
The correlation between $\overline{\overline{ \Delta \psi_N}}$ and ${\delta \psi}^2 / L$ is significant; the correlation coefficient is larger than 0.99.
\RED{
The 
}%  RED
intercept $\beta_{\overline{\psi_N}}$ should be equal to 0,
because if $T_s \rightarrow \infty$ then ${\delta \psi}^2 \rightarrow 0$ and $\Delta \psi_N \rightarrow 0$.
Actually, \Fig{\ref{fig: dpsi2_over_L_vs_dPsi}} shows that $\beta_{\overline{\psi_N}}$ is nearly equal to 0.

Finally, the regression of $\Delta\Delta G_{ND}$ on $\Delta G_{ND}$ would be derived if
$T_g$, $T_s$ and $T$ were constant.
\begin{eqnarray}
	 \overline{\overline{ \Delta\Delta G_{ND}(\sigma^N_{j \neq i}, \sigma^N_i\rightarrow \sigma_i ) } }
		&\approx& - \, \alpha_{\overline{\psi_N}}  k_B T_s \frac{ {\delta \psi}^2(\overline{\VEC{f}(\VEC{\sigma}_N)}) }{L} + 
		k_B T_s \, \beta_{\overline{\psi_N}}
		\\
		&=& 
		\alpha_{\Delta G_{ND}} k_B T_s \frac{ {\delta \psi}^2( \overline{\VEC{f}(\VEC{\sigma}_N))} }{L}(\vartheta(T/T_g) \frac{T_s}{T} - 1) + 
		\beta_{\Delta G_{ND}}
		\\
		&=& \alpha_{\Delta G_{ND}} \frac{\langle \Delta G_{ND}(\VEC{\sigma}_N, T) \rangle} {L} + 
		\beta_{\Delta G_{ND}}
		\label{\EQ: regression_of_ddG_on_dG}
\end{eqnarray}
In general, $T_s$ and $T_g$ are different among protein families, so that the correlation
between $\overline{\overline{\Delta\Delta G_{ND}}}$ and $\langle \Delta G_{ND} \rangle / L$ 
cannot be strong.
In \Fig{\ref{fig: dG_over_L_vs_ddG}}, $\overline{\overline{\Delta\Delta G_{ND}}}$ for the present proteins are plotted
against $\langle \Delta G_{ND} \rangle / L$.
However, it should be noted that the correlation is not expected for 
$\overline{\overline{\Delta\Delta G_{ND}}}$ and $\langle \Delta G_{ND} \rangle$ but 
for $\overline{\overline{\Delta\Delta G_{ND}}}$ and $\langle \Delta G_{ND} \rangle / L$ .

\subsection{Estimation of $T_g$, $\omega$, and $\langle \Delta G_{ND}(\VEC{\sigma}) \rangle_{\VEC{\sigma}}$ from $T_s$ and $T_m$}

To estimate glass transition temperature $T_g$,
the conformational entropy per residue $\omega$ in the compact denatured state, 
and the ensemble average of folding free energy in sequence space $\langle \Delta G_{ND} \rangle_{\VEC{\sigma}}$, 
melting temperature $T_m$ 
must be known for each protein; see 
\Eqs{\REF{\EQ: relationship_among_characteristic_T}, 
\REF{\EQ: expression_of_entropy}, 
and \REF{\EQ: ensemble_ave_of_ddG}}
for $T_g$, $\omega$ and $\langle \Delta G_{ND} \rangle_{\VEC{\sigma}}$, respectively.
The experimental value of $T_m$ 
\CITE{GDBCJMS:09,SZMSS:06,DSVSSAMRT:05,PLO:06,WPLLSLCOD:02,SRSSO:08,AUFCBG:04,GRGBLG:10,KMCBKVSL:98,OKKSW:15,TES:12,RMGGS:95}
employed for each protein is listed
in \Tables{\ref{\TBL: Ts_with_8A} and \ref{stbl: Ts_with_16A}}.
For comparison, temperature $T$ is set up to be equal to the experimental temperature for $\Delta G_{ND}$ or
to $298 ^\circ$K if unknown.

An estimate of glass transition temperature, $\hat{T}_g$,
has been calculated with $\hat{T}_s$ and $T_m$ 
by \Eq{\ref{\EQ: relationship_among_characteristic_T}}, 
and is listed in \Tables{\ref{\TBL: Ts_with_8A} and \ref{stbl: Ts_with_16A}} for each protein.
In \Fig{\ref{fig: Tm_over_Tg_vs_Ts_over_Tg}}, $\hat{T}_s/ \hat{T}_g$ is plotted against $T_m/ \hat{T}_g$ 
for each protein family.
Unless $T_g < T_m$, a protein will be trapped at local minima
on a rugged free energy landscape before it folds into a unique native structure.
Protein foldability increases as $T_m / T_g$ increases.
A condition, $\Delta G_{ND} = 0$ at $T = T_m$, for the first order transition
requires that \Eq{\ref{\EQ: relationship_among_characteristic_T}}, 
which is indicated by a dotted curve in \Fig{\ref{fig: Tm_over_Tg_vs_Ts_over_Tg}},
must be satisfied. As a result, $T_s/ T_g$ must be lowered to increase $T_m / T_g$;
in other words, proteins must be selected at lower $T_s$.
The present estimates of $T_s$ and $T_g$ would be 
within a reasonable range\CITE{OWLS:95,PGT:00,MSCOW:14}
of values required for protein foldability.

In \Tables{\ref{\TBL: Ts_with_8A} and \ref{stbl: Ts_with_16A}},
the ensemble average of $\Delta G_{ND}(\VEC{\sigma})$ over sequences
calculated by \Eq{\ref{\EQ: ensemble_ave_of_ddG}},
and the conformational entropy per residue $\omega$ in the compact denatured state
by \Eq{\ref{\EQ: expression_of_entropy}} are also listed for each protein.
\Fig{\ref{fig: dG_exp_vs_8_and_16A}} shows the comparison of their ensemble averages,
$\langle \Delta G_{ND}(\VEC{\sigma}) \rangle_{\VEC{\sigma}}$, 
and the experimental values of $\Delta G_{ND}(\VEC{\sigma}_N)$ 
\CITE{RSTPMF:99,GRSB:98,KONSKKP:99,GCAVBT:05,GGCJVTVB:07,WW:05}
listed in \Table{\ref{stbl: experimental_data}}.
The correlation in the case of $r_\script{cutoff} \sim 8$ \AA\ 
is quite good, indicating that
the constancy approximation (\Eq{\ref{\EQ: var_of_ddG}}) for the variance of $\Delta G_{N}$ 
is appropriate. 
The conformational entropy per residue in the compact denatured state, 
$\hat{\omega}$ in \Eq{\ref{\EQ: expression_of_entropy}},
estimated from the condition for the first order transition 
falls into the 
range 
of $0.60 $--$ 1.13 k_B$ for $r_{\script{cutoff}} \sim 8$ \AA,  
which agrees well with
the range estimated by Morcos et al. (2014).	

% End of results_1.tex
% \input{results_2.tex}

\subsection{The equilibrium value of evolutionary statistical energy $\psi_N$
in the mutation--fixation process of amino acid substitutions}

Let us consider the fixation process of amino acid substitutions in a monoclonal approximation,
in which protein evolution is assumed to proceed with single amino acid substitutions 
fixed at a time in a population.
In this approximation, 
$\Delta \psi_{ND}$ and $\psi_N$ are at equilibrium and 
the ensemble of protein sequences attains to the equilibrium state, 
when the average of $\Delta\Delta \psi_{ND} \simeq \Delta \psi_N$ over singe nucleotide nonsynonymous mutations
fixed in a population is equal to zero; an amino acid composition is assumed to be constant in protein evolution.
\begin{eqnarray}
	\langle \Delta\Delta \psi_{ND} \rangle_{\script{fixed}} \simeq \langle \Delta \psi_N \rangle_{\script{fixed}} = 0 
		&\Longleftrightarrow&  \Delta \psi_{ND} \text{ and } \psi_N \text{ are at equilibrium. }
	\label{\EQ: equilibrium_condition}
\end{eqnarray}
The average of $\Delta \psi_N$ over fixed mutations, $\langle \Delta \psi_N \rangle_{\script{fixed}}$,
is calculated numerically with the probability density function (PDF) of 
$\Delta\Delta \psi_{ND} (\simeq \Delta \psi_N$) 
for single nucleotide nonsynonymous mutations; 
see \Eqs{\REF{\EQ: pdf_of_fixed_ddPsi} and \REF{\EQ: ave_of_u}}.
$N = 10^6$ is employed.

The PDF of $\Delta \Delta G_{ND}$ were approximated with a normal distribution\CITE{SRS:12} or a bi-normal distribution\CITE{TSSST:07}.
\Figs{\ref{fig: 1gm1-a:16-96_full_non_del_dca0_205_0_20_simple-gauge_dPhiN_distr}, 
\ref{sfig: 1gm1-a:16-96_full_non_del_dca0_205_0_20_simple-gauge_dPhiN_distr},
and \ref{sfig: 1n2x-a:8-292_full_non_del_dca0_13_0_20_simple-gauge_dPhiN_distr} },
however, 
show that a single normal distribution with the observed mean and standard deviation 
cannot well reproduce the observed distribution of $\Delta \psi_{N}$ due to 
single nucleotide nonsynonymous mutations.
For simplicity, a log-normal distribution, 
$\ln\mathcal{N}(x; \mu, \sigma)$, for which $x, \mu$ and $\sigma$ defined as follows, 
is arbitrarily used here to better reproduce observed distributions of $\Delta \psi_N$,
particularly in the domain of $\Delta \psi_N < \overline{\Delta \psi_N}$,
although other distributions such as
inverse $\Gamma$ distributions can equally well reproduce the observed ones, too.

\begin{eqnarray}
	p(\Delta \psi_N) 
	&\approx&
	\ln\mathcal{N}(x; \mu, \sigma) \equiv \frac{1}{x} \mathcal{N}(\ln x; \mu, \sigma)
	\label{\EQ: log-normal}
	\\
	x &\equiv& \max ( \Delta \psi_{N} - \Delta \psi_{N}^{\mbox{o}}, 0)
	\\
	\exp(\mu + \sigma^2/2) &=& \overline{\Delta \psi_{N}} - \Delta \psi_{N}^{\mbox{o}}
	\\
	\exp(2\mu + \sigma^2) ( \exp(\sigma^2) - 1 ) &=& \overline{ (\Delta \psi_{N} - \overline{\Delta \psi_{N}})^2 })
	\\
	\Delta \psi_{N}^{\mbox{o}} &\equiv& \min (\overline{\Delta \psi_{N}} 
		- n_{\mbox{shift}} \overline{ (\Delta \psi_{N} - \overline{\Delta \psi_{N}})^2 })^{1/2}, 0)
	\label{\EQ: statistics_for_log-normal}
\end{eqnarray}
where $\Delta \psi_{N}^{\mbox{o}}$ is the origin for the log-normal distribution 
and the shifting factor $n_{\mbox{shift}}$ is taken to be equal to $2$, unless specified.
It is shown in
\Figs{\ref{fig: 1gm1-a:16-96_full_non_del_dca0_205_0_20_simple-gauge_dPhiN_distr}, 
\ref{sfig: 1gm1-a:16-96_full_non_del_dca0_205_0_20_simple-gauge_dPhiN_distr}, 
and \ref{sfig: 1n2x-a:8-292_full_non_del_dca0_13_0_20_simple-gauge_dPhiN_distr} }
that log-normal distributions can better reproduce the observed 
distribution of $\Delta \psi_N$ due to 
single nucleotide nonsynonymous mutations except in the tails.
Disagreements between the log-normal and observed distributions 
in the domain of $\Delta \psi_N > \overline{\Delta \psi_N}$ do not much affect
the PDF of $\Delta \psi_N$ in fixed mutants,
because fixation probabilities for $\Delta \psi_N (> \overline{\Delta \psi_N})$
are too low.

The average of $\Delta \psi_N$ over fixed mutants is uniquely determined by the distribution of 
$\Delta \Delta \psi_N (\simeq \Delta \psi_{N})$,
which is approximated here by a log-normal distribution 
estimated from the mean and variance of $\Delta \psi_N$;
it depends also on $q_m$, which is assumed to be constant, 
through fixation probability, because $2N_e s \simeq - \Delta \psi_N / (1 - q_m)$.
In other words, $\langle \Delta \psi_N \rangle_{\script{fixed}}$ is
uniquely determined by the mean and variance of $\Delta \psi_N$.  
Therefore, under the equilibrium condition $\langle \Delta \psi_N \rangle_{\script{fixed}} = 0$,
only one of the mean and variance can be freely specified, and the other is uniquely determined.
We employ $\overline{\Delta \psi_N}$ or $\psi_N$ as a parameter,
because $\overline{\Delta \psi_N}$ depends on $\psi_N$, and only one of them can be specified.
We define $\overline{\Delta \psi}_N^{\script{eq}}$ as $\overline{\Delta \psi}_N$ 
at which $\langle \Delta \psi_N \rangle_{\script{fixed}} = 0$. 

Suppose that
the regression equation, \Eq{\ref{\EQ: regression_of_dPsi_on_Psi}}, of $\Delta \psi_N$ on $\psi_N$ 
is exact, and the standard deviation of $\Delta \psi_N$ is constant irrespective of $\psi_N$;
the slope ($\alpha_{\psi_N}$), $\overline{\overline{\Delta \psi_N}}$, 
$\overline{\text{Sd}(\Delta \psi_N)}$,
and $\overline{\psi_N}$ that are estimated with
$r_{\script{cutoff}} \sim 8$ \AA\ for the PDZ and listed in \Table{\ref{\TBL: ddPsi_with_8A}}
are employed here.
In \Fig{\ref{fig: ave_ddPhi_vs_dPhiN_of_fixed_mutants_logG-2sd_PDZ_8A}},
the average of $\Delta \psi_N$ over single nucleotide nonsynonymous substitutions fixed in a population, 
$\langle \Delta \psi_N \rangle_{\script{fixed}}$, 
is plotted against $\psi_N / L$ of a wildtype for the PDZ protein family.
This figure shows that $\langle \Delta \psi_N \rangle_{\script{fixed}}$ changes its value from positive to negative
as $\psi_N$ increases, that is, the value of $\psi_N$ at which 
$\langle \Delta \psi_N \rangle_{\script{fixed}} = 0$, $\psi_N^{\script{eq}}$, is
the stable equilibrium value for $\psi_N$.
In order for protein to have such a stable equilibrium value for
folding free energy ($\Delta G_{ND} = k_B T_s \Delta \psi_{ND}$),
the regression coefficient of $\overline{\Delta \psi_N}$ on $\psi_N$ 
must be more negative than that of the standard deviation, $\text{Sd}(\Delta \psi_N)$,
because otherwise stabilizing mutations increase as $\psi_N$ decreases.
This condition is, of course, satisfied for all protein families studied here, because 
the mean of $\Delta \psi_N$ over all substitutions at all sites is negatively proportional to $\psi_N$ of a wildtype,
but its standard deviation is nearly constant irrespective of $\psi_N$ across homologous sequences;
see \Tables{\ref{\TBL: ddPsi_with_8A} and \ref{stbl: ddPsi_with_16A}}.

The equilibrium value of $\psi_N$ for each protein domain 
is calculated
with the estimated values of 
$\alpha_{\psi_N}$, 
$\overline{\psi_N}$, 
$\overline{\overline{\Delta \psi_{N}}}$, 
and $\overline{\script{Sd}(\Delta \psi_N)}$
listed in \Tables{\ref{\TBL: ddPsi_with_8A} and \ref{stbl: ddPsi_with_16A}};
it should be noticed here that $\overline{\script{Sd}(\Delta \psi_N)}$ is assumed to be constant.
In 
\Figs{\ref{fig: PhiNe_obs_vs_exp_logG_8A} and \ref{sfig: PhiNe_obs_vs_exp_logG_16A}},
the equilibrium values of $\psi_N / L$ estimated with $n_{\script{shift}} = 1.5, 2$, and $2.5$
in the monoclonal approximation are plotted against the average of $\psi_N / L$ over homologous sequences 
for each protein family.
The agreement between the time average ($\psi_N^{\script{eq}}$) and ensemble average ($\langle \psi_N \rangle_{\VEC{\sigma}} (= \overline{\psi_N}$) )
is better for $r_{\script{cutoff}} \sim 8$ \AA\ than for $r_{\script{cutoff}} \sim 15.5$ \AA\ 
and is not bad in the case of $r_{\script{cutoff}} \sim 8$ \AA, 
indicating that
the present methods for the fixation process of amino acid substitutions
and for the equilibrium ensemble of $\psi_N$ give a consistent result with each other,
and also that it is a good approximation to assume
the standard deviation of $\Delta \psi_N$ not to depend on $\psi_N$ in each protein family.

\subsection{Relationships between $\overline{\Delta \psi_N} (= \overline{\Delta \psi}_N^{\script{eq}})$ 
and the standard deviation of $\Delta \psi_N$, $\hat{T}_s$, and $\Delta\Delta \hat{G}_{ND}$ at equilibrium } 

In the present model, 
the equilibrium values, 
$\psi_{N}^{\script{eq}}$ and the corresponding $\overline{\Delta \psi}_{N}^{\script{eq}}$, are 
functions of the mean and standard deviation of $\Delta \psi_{N}$ only,
because the distribution of $\Delta\Delta \psi_{ND} (\simeq  \Delta \psi_{N})$ is approximately estimated with
the mean and standard deviation of $\Delta \psi_{N}$.
On the other hand, 
$\psi_{N}^{\script{eq}}$ and
$\overline{\Delta \psi}_{N}^{\script{eq}}$ should be equal to $\overline{\psi_{N}} = \langle \psi_{N} \rangle$ 
and $\overline{\overline{\Delta \psi_{N}}}$,
respectively; the time average and ensemble average should be consistent.
Actually $\psi_N^{\script{eq}}$ almost agrees with $\overline{\psi_{N}}$ 
as shown in \Fig{\ref{fig: PhiNe_obs_vs_exp_logG_8A}}.
Therefore the standard deviation of $\Delta \psi_{N}$ is uniquely determined 
from its mean as long as $\psi_N$ and $\overline{\Delta \psi_{N}}$ 
are at equilibrium;
conversely the equilibrium value of $\overline{\Delta \psi_{N}}$ is determined by $\text{Sd}(\Delta\psi_N)$. 
In \Fig{\ref{fig: ddPhi_mean_vs_sd_at_equil}}, the standard deviation of $\Delta \psi_{N}$ is plotted
against $\overline{\Delta \psi_{N}}(= \overline{\Delta \psi}_{N}^{\script{eq}})$.
Likewise the estimate of effective temperature of selection, 
$\hat{T}_s (= (\hat{T}_s \overline{Sd}(\Delta \psi_N))_{PDZ}/Sd(\Delta \psi_N))$, and that of
folding free energy change, 
$\Delta\Delta \hat{G}_{ND} (= k_B (\hat{T}_s \overline{Sd}(\Delta \psi_N))_{PDZ}/Sd(\Delta \psi_N) \cdot \overline{\Delta \psi_N} )$,
are plotted as a function of 
$\overline{\Delta \psi_{N}}(= \overline{\Delta \psi}_{N}^{\script{eq}})$ in 
\Fig{\ref{fig: ddPhi_mean_vs_Ts_at_equil}}. 
\SUPPLEMENT{
It should be noted that in \Figs{\ref{\FIG: ddPhi_mean_vs_Ts_at_equil}}
the curves for $r_{\script{cutoff}} \sim 8$ and $15.5$\AA\ almost overlap with each other,
because the estimates of $(\hat{T}_s \overline{Sd}(\Delta \psi_N))_{PDZ}$ for the PDZ
with $r_{\script{cutoff}} \sim 8$ and $15.5$\AA\  are almost equal to each other.
}%  SUPPLEMENT
These figures show that the averages,
$\overline{\overline{\Delta \psi_N}}$ and 
$\overline{\text{Sd}(\psi_N)}$,
over homologous sequences scatter along the expected curves.

\subsection{Protein evolution at equilibrium, $\langle \Delta \psi_N \rangle_{\script{fixed}} = 0$} 

The common understanding of protein evolution
has been that amino acid substitutions observed in homologous proteins are
neutral \CITE{K:68,K:69,KO:71,KO:74} or slightly deleterious
\CITE{O:73,O:92}, and random drift is a primary force to fix amino acid
substitutions in population. In order to see how significant neutral/slightly deleterious 
substitutions are in protein evolution, 
the PDFs of $K_a/K_s$
in all single nucleotide nonsynonymous mutations and in their fixed mutations 
are calculated;
$K_a / K_s$ is the ratio of nonsynonymous to synonymous substitution rate per site \CITE{MY:80} and 
defined here as $K_a/K_s \equiv u(s) / u(0)$, where $u(s)$ is a fixation probability for selective advantage $s$; 
see \Eq{\ref{\EQ: def_Ka_over_Ks}}.

First let us see the distributions of $\Delta \psi_N$ at equilibrium, 
$\langle \Delta \psi_N \rangle_{\script{fixed}} = 0$.
\Fig{ \ref{fig: pdf_of_ddPhi_fixed_at_equil_for_mean_ddPhi} }
shows the PDFs of $\Delta \psi_N$ in all single nucleotide nonsynonymous mutations and in their fixed mutations
as a function of $\overline{\Delta \psi_N} (= \overline{\Delta \psi}_N^{\script{eq}})$, respectively.
Because $4N_es(1-q_m) = - \Delta \Delta \psi_{ND} \simeq  - \Delta \psi_{N}$, the PDFs of $\Delta \psi_N$ 
can be regarded as the PDFs of $-4N_es(1-q_m)$. 
At equilibrium, the distribution of $\Delta \psi_N$ in all single nucleotide nonsynonymous mutants 
becomes wider as the mean of $\Delta \psi_N$ increases, however,
that in fixed mutants remains to be narrow with a peak near zero.

The PDFs of $K_a/K_s$
in all single nucleotide nonsynonymous mutations and in their fixed mutations 
are shown in
\Fig{\ref{fig: pdf_of_ka_over_ks_fixed_at_equil_for_mean_ddPhi}}.
The blue line on the landscape of the PDF shows the averages of $K_a/K_s$.
The averages of $K_a/K_s$ in all single nucleotide nonsynonymous mutations and in their fixed mutations 
are also shown in \Fig{\ref{fig: ave_ka_over_ks_fixed_at_equil_for_mean_ddPhi}}.
The average of $K_a/K_s$ in all the arising mutants is less than 1 
and decreases as $\overline{\Delta \psi_N} \simeq \overline{\Delta \psi}_N^{\script{eq}}$ increases, indicating
that negative mutants significantly occur and increase as $\overline{\Delta \psi_N}$ increases.
On the other hand, $\langle K_a/K_s \rangle_{\script{fixed}}$ in fixed mutants  
is larger than 1 and increases as $\Delta \psi_N^{\script{eq}}$ increases, 
indicating that positive mutants fix significantly in population and 
increase as equilibrium folding free energy change increases, that is, equilibrium protein stability decreases.
To see each contribution of positive, neutral, slightly negative and negative selections,
the value of $K_a/K_s$ is divided arbitrarily into four categories, $K_a/K_s > 1.05$, $1.05 > K_a/K_s > 0.95$, $0.95 > K_a/K_s > 0.5$, and $0.5 > K_a/K_s$ 
for their selection categories, respectively.
The probabilities of each selection category in all single nucleotide nonsynonymous mutations and 
in their fixed mutations are shown in 
\Fig{\ref{fig: prob_of_each_selection_category_at_equil_for_mean_ddPhi}}.
The almost 50\% of fixed mutations are
stabilizing mutations fixed by positive selection ($1.05 < K_a/K_s$),
and another 50\% are destabilizing mutations
fixed by random drift. 
They are balanced with each other, and the stability of protein is maintained.
\RED{
Contrary to the neutral theory\CITE{K:68,K:69,KO:71,KO:74},
}%  RED
the proportion of neutral selection is not large even in fixed mutations,
and slightly negative mutations are significantly fixed.
Neutral mutations fixed with $0.95< K_a/K_s < 1.05$
are only less than 10\%, and slightly negative mutations fixed with $0.5< K_a/K_s < 0.95$ and
negative mutations fixed with $K_a/K_s < 0.5$ are both from 10 to 30 \%.
\RED{
The nearly neutral theory\CITE{O:73,O:92,O:02} insists that most fixed mutations satisfy $| N_e s | \leq 2$.
This condition corresponds to $0.003\leq K_a/K_s(=u(s)/u(0))\leq 8$; see \Eqs{\REF{\EQ: fixation_probability} and \REF{\EQ: def_Ka_over_Ks}}.
The PDF of $K_a/K_s$ shown in \Fig{\ref{\FIG: prob_of_each_selection_category_fixed_at_equil_for_mean_ddPhi}}
indicates that this condition is satisfied, supporting the nearly neutral theory.
}%  RED

\subsection{Relationship between $T_s$ and $K_a/K_s$}

The effective temperature ($T_s$) of protein for selection,  
which is defined in \Eq{\ref{\EQ: canonical_selection}}, 
represents
the strength of selection 
originating from
protein stability and foldability.
Thus, it must be related with the evolutionary rate (amino acid substitution rate) of protein.
As the effective temperature of selection ($T_s$) decreases, 
the mean change of 
evolutionary statistical energy
($\overline{\Delta \psi}_N^{\script{eq}}$) due to single amino acid substitutions increases;
see \Fig{\ref{fig: ddPhi_mean_vs_Ts_at_equil}}.
Therefore, destabilizing mutations increase, and an amino acid substitution rate is expected to decrease.
\Fig{\ref{fig: ave_ka_over_ks_at_equil_for_mean_ddPhi}} shows that the average of $K_a/K_s$ decreases as 
$\overline{\Delta \psi}_N^{\script{eq}}$ increases.
The direct relationship between substitution rate and 
$T_s (= (T_s \overline{Sd}(\Delta\psi_N))_{PDZ} /Sd(\Delta\psi_N) )$ 
is shown in \Fig{\ref{fig: ave_ka_over_ks_at_equil_for_Ts}};
the average of $K_a/K_s$ decreases as $T_s$ increases.
In the selection maintaining protein foldability/stability, 
the effective temperature of selection is directly reflected in the average amino acid substitution rate.

% End of results_2.tex

% End of results_1+2.tex

% \input{discussion_1+2.tex}

\section{Discussion}

A main purpose of the present study is
to formulate protein fitness originating 
from
protein foldability and stability. 
From a phenomenological viewpoint,
Drummond and Wilke (2008)
took notice of toxicity of
misfolded proteins as well as diversion of protein synthesis resources, and formulated
a Malthusian fitness of a genome to be negatively proportional
to the total amount of misfolded proteins, which must be produced to obtain the
necessary amount of folded proteins\CITE{SRS:12}.
They also formulated a Malthusian fitness based on protein dispensability to be
negatively proportional to the ratio of unfolded proteins.
These formulas of protein fitness can be well approximated
by a generic form, $m = - \kappa \exp (\Delta G_{ND}/(k_B T))$, where
$T$ is growth temperature, and $\kappa (\geq 0)$ is a parameter that depends on
protein disability and cellular abundance of protein\CITE{M:16}.

In the comparison of this 
generic formula
of protein fitness with the present one,
it may be interpreted that $4N_e(1-q_m) \kappa / T \sim 1 / T_s$,
if $| \Delta G_{ND} / (k_B T) | \ll 1$,
however, the growth temperature $T$ and folding free energy do not always satisfy this condition.
These two types of selection should be considered to be the different types of selection,
although both are related with protein stability ($\Delta G_{ND}$).
Selective advantage of mutant is not upper-bounded
in the present scheme of a Malthusian fitness but
in the case of $m = - \kappa \exp (\Delta G_{ND}/(k_B T))$.
As a result, PDFs of $K_a/K_s$ in all arising mutations and in fixed mutations
have very different shapes between these two formulas of fitness\CITE{M:16}.
Selection modeled here is one that 
yields the distribution of homologous sequences in protein evolution.
In other word, the present formula for protein fitness 
models
natural selection 
maintaining protein's stability, foldability, and function
over the evolutionary time scale,
which is much longer than the time scale 
for the selection originating from toxicity of misfolded proteins.

The present formulas for protein fitness,
\Eqs{\REF{\EQ: relationship_between_m_and_dG} and \REF{\EQ: relationship_between_m_and_dPsi} },  have been
derived on the basis of
a protein folding theory, particularly the random energy model, and
the maximum entropy principle for
the distribution of homologous sequences with the same fold in sequence space, respectively.
The former indicates that the equilibrium ensemble of sequences can be well approximated
by a canonical ensemble with the Boltzmann factor $\exp(-\Delta G_{ND}/k_B T_s)$,
and the latter insists
that the probability distribution of homologous sequences,
which satisfies a given amino acid composition at each site and a given pairwise amino acid
frequency at each site pair,
can be represented as a Boltzmann distribution with $\exp(- \psi_{N})$,
in which the 
evolutionary statistical energy
($\psi_N$)
is represented as the sum of one-body (compositional) and pairwise (covariational) interactions between sites.
On the other hand, assuming mutation and fixation processes to be reversible Markov processes
leads us to a formulation that
the equilibrium ensemble of sequences also obeys a Boltzmann distribution with $\exp(4N_e m (1-q_m))$.
As a result, we obtain the correspondences between folding free energy ($-\Delta G_{ND}/k_B T_s$), and
$-\Delta \psi_{ND}$ and protein fitness ($4N_e m (1- q_m)$):
the equality between the latter two variables (\Eq{\ref{\EQ: selective_advantage_and_dPsi_N}}), 
which indicates that $\Delta\psi_N$ is proportional to fitness ($s$),
and the approximate equality between the former two variables (\Eq{\ref{\EQ: s_vs_dG}})
since a canonical ensemble with $\Delta G_{ND}/(k_B T_s)$ is 
an approximate for the sequence ensemble under natural selection.
A discrepancy between evolutionary statistical energies $J_{ij}$ and actual interaction energies  
was pointed out for non-contacting residue pairs in Monte Carlo simulations of lattice proteins\CITE{JGSCM:16}.
Also, 
the ratio of $- J_{ij}(a_k,a_l)$ to the corresponding actual contact energy was shown to differ among contact site pairs. 
On the other hand,
Hopf et al.\CITE{HIPSSSM:17} 
successfully predicted mutation effects with evolutionary statistical energy and
showed that
the change of evolutionary statistical energy ($\Delta\psi_N$) due to amino acid substitutions
can capture experimental fitness landscapes and identify deleterious human variants.

In the analysis of
the interaction changes ($\Delta \psi_N$) due to single nucleotide nonsynonymous substitutions,
we have employed the cutoff distances for pairwise interactions, $r_{\script{cutoff}} \sim 8$ and $15.5$ \AA, which
correspond to the first and second interaction shells between residues, respectively.
Both the cutoff distances yield similar values 
for $T_s / T_{s,PDZ} = \overline{Sd}(\Delta \psi_{N,PDZ}) / \overline{Sd}(\Delta \psi_N)$; 
see \Fig{\ref{sfig: Ts_relative_to_Tpdz_8_vs_16A}}.
Thus, the differences in the estimation of $T_s$ between these two cutoff distances
principally originate in the estimation of $T_s$ for the reference protein, PDZ.
The absolute value of $k_B \hat{T}_{s,PDZ}$ for the PDZ has been estimated
to be equal to the slope
of the reflective regression line of $\Delta\Delta G_{ND}$ on $\Delta \psi_{N}$.
Therefore, 
as long as the correlation between $\Delta\Delta G_{ND}$ and $\Delta \psi_{N}$ is good enough as shown in 
\Figs{\ref{fig: 1gm1-a:16-96_dca0_205_simple-gauge_ddG-dPhi_at_opt} and \ref{sfig: 1gm1-a:16-96_dca0_33_simple-gauge_ddG-dPhi_at_opt}},
$k_B (\hat{T}_{s} \overline{\text{Sd}(\psi_N)})_{PDZ}$ takes a similar value irrespective of $r_{\script{cutoff}}$,
and the estimate $\hat{T}_{s,PDZ}$ differs depending on $\overline{\text{Sd}(\psi_{N,PDZ})}$.
Thus, $\Delta \psi_N$ must correlate with experimental $\Delta\Delta G_{ND}$, but 
on the basis of the correlation coefficient one cannot determine which estimation of $\Delta \psi_N$ is better.
Larger the standard deviation of $\Delta \psi_N$ is, the smaller the estimate of $T_s$ 
from a direct comparison between $\Delta\Delta G_{ND}$ and $\Delta \psi_N$ is.
Including the longer range of pairwise interactions tend to increase the variance of $\Delta \psi_N$.
The range of interactions must be limited to a realistic value, either the first interaction shell or
the second interaction shell. 
Thus, the estimates of $T_s$ with $r_{\script{cutoff}} \sim 8$ \AA\ and $15.5$\AA\ would be upper 
and lower limits, respectively.
Unfortunately $T_s$ is not directly observable.
Comparison of the estimates of folding free energies with their experimental values may be appropriate to
judge which value is more appropriate for the cutoff distance, although 
the number of experimental data is limited.  
Actual values of $T_s$ may be closer to the estimates with
$r_{\script{cutoff}} \sim 8$\AA, because
contact predictions based on the estimate of pairwise interactions $J$ succeed for close contacts 
within the first interaction shell.
Also, the estimation of $\Delta G_{ND}$
and the correlation between $\psi_N^{\script{eq}}$ and $\overline{\psi_N}$
are slightly better with 
$r_{\script{cutoff}} \sim 8$\AA\ than $15.5$\AA\ ;
see \Figs{\ref{fig: dG_exp_vs_8_and_16A}, \ref{fig: PhiNe_obs_vs_exp_logG_8A},
\ref{sfig: dG_exp_vs_8_and_16A}, and \ref{sfig: PhiNe_obs_vs_exp_logG_16A}}.

On the basis of
the random energy model(REM)\CITE{SG:93a,SG:93b,PGT:97},
glass transition temperatures ($T_g$)
and folding free energies ($\Delta G_{ND}$) for 14 protein domains
are estimated 
under the condition of 
$\overline{\psi_N} = \langle \psi_N \rangle_{\VEC{\sigma}}$.
The first order transition for protein folding is assumed 
to estimate the folding free energies by \Eq{\ref{\EQ: ensemble_ave_of_ddG}}.
Selective temperature,
$T_s$, is estimated in the empirical approximation that the standard deviation of $\Delta \psi_N$
is constant across homologous sequences with different $\psi_N$, 
so that their estimates may be more coarse-grained, however,
this method is easier and faster than the method\CITE{MSCOW:14} using the AWSEM\CITE{DSZCWP:12}. 
Experimental data for $\Delta G_{ND}$ are very limited, and also experimental conditions
such as temperature and {pH} tend to be different among them.
A prediction method for folding free energy would be useful in such a situation,
although the present method requires the knowledge of melting temperature ($T_m$) besides sequence data,
however, experimental data of $T_m$
are more available than for $\Delta G_{ND}$.

For proteins to have a stable equilibrium value of $\psi_N$ in protein evolution,
the regression coefficient of
 mean interaction change ($\overline{\Delta \psi_N}$) on $\psi_N$
must be more negative
than that of their standard deviation ($\text{Sd}(\Delta \psi_N)$), otherwise 
stabilizing mutations increase as $\psi_N$ decreases.
Actually \Tables{\ref{\TBL: ddPsi_with_8A} and \ref{stbl: ddPsi_with_16A}}
show that their mean over all the substitutions at all sites is
negatively proportional to $\psi_N$ of a wildtype,
but their standard deviation is nearly constant irrespective of $\psi_N$ across homologous sequences.
The equilibrium value $\psi_N^{\script{eq}}$,
where the average of $\Delta \psi_N$ over fixed mutants is equal to zero,
is calculated with
the approximation of the distribution of $\Delta \psi_N$
by a log-normal distribution and
the empirical rules of \Eqs{\REF{\EQ: regression_of_dPsi_on_Psi}
and \REF{\EQ: var_of_dPsi}}.
In the monoclonal approximation,
it has been confirmed that
the time average ($\psi_N^{\script{eq}}$) and
ensemble average ($\overline{\psi_N} = \langle \psi_N \rangle_{\VEC{\sigma}}$)
of 
evolutionary statistical energy
($\psi_N$) almost agree with each other.
Therefore, this result also supports these approximations and empirical rules,
particularly \Eq{\ref{\EQ: var_of_dPsi}}, that is,
the constancy of the standard deviation of
$\Delta \psi_N$ 
across homologous sequences.
In the log-normal distribution approximation,
$\overline{\Delta \psi}_N^{\script{eq}}$,
$\text{Sd}(\Delta \psi_N)^{\script{eq}}$,
$\hat{T}_s$, and $\Delta\Delta \hat{G}_{ND}$
can be determined as a function of any one of them.
Here they have been shown
as a function of $\overline{\Delta \psi}_N^{\script{eq}}$.

We have also studied the evolution of protein at equilibrium, at which
the ensemble of homologous sequences obeys a Boltzmann distribution with 
$\exp (-\psi_{N}) (\simeq \exp (-\Delta \psi_{ND})) $, and
the ensemble averages of 
evolutionary statistical energy
($\psi_{N} \simeq G_{N}/(k_B T_s)$) 
and its change due to a mutation
($\Delta \psi_N \simeq \Delta\Delta \psi_{ND} \simeq \Delta\Delta G_{ND} / (k_B T_s)$)
agree with their steady values;
$\langle \psi_N \rangle_{\VEC{\sigma}} = \overline{\psi_N} = \psi_N^{\script{eq}}$ and
$\langle \overline{\Delta \psi_N} \rangle{\VEC{\sigma}}  = \overline{\overline{\Delta \psi_N}}  = \overline{\Delta \psi}_N^{\script{eq}}$.
The PDFs of $\Delta \psi_N$ and $K_a/K_s$
in all the mutants and in their fixed mutants have been estimated.
It is confirmed that the effective temperature ($T_s$)
of selection negatively correlates with the amino acid substitution rate ($K_a/K_s$) of protein.

New alleles can become fixed owing to random drift 
or to positive selection of substantially advantageous
mutations\CITE{K:83,G:91,O:02}.
The present study indicates
that the stability of protein is maintained in such a way that 
stabilizing mutations are significantly fixed by positive
selection, and balance with destabilizing mutations 
fixed by random drift.
As shown in \Fig{\REF{\FIG: prob_of_each_selection_category_fixed_at_equil_for_mean_ddPhi}}, 
the almost 50\% of fixed mutations are 
stabilizing mutations fixed by positive selection ($1.05 < K_a/K_s$),
and another 50\% are destabilizing mutations 
fixed by random drift. 
An interesting fact is that
contrary to the neutral theory\CITE{K:68,K:69,KO:71,KO:74},
the proportion of neutral selection is not large even in fixed mutants.
In the selection to maintain protein stability/foldability,
neutral mutations fixed with $0.95< K_a/K_s < 1.05$ 
are only less than 10\%, and slightly negative mutations fixed with $0.5< K_a/K_s < 0.95$ and 
negative mutations fixed with $K_a/K_s < 0.5$ are both from 10 to 30 \%.
As a result, at equilibrium
the average of $K_a/K_s$ in all the mutants is less than 1,
but that in their fixed mutants is larger than 1.
\RED{
The PDF of $K_a/K_s$ shown in \Fig{\ref{\FIG: prob_of_each_selection_category_fixed_at_equil_for_mean_ddPhi}}
supports the nearly neutral theory\CITE{O:73,O:92,O:02}, 
which insists that most fixed mutations satisfy $| N_e s | \leq 2$ corresponding to $0.003\leq K_a/K_s(=u(s)/u(0)) \leq 8$.
}%  RED
It should be noted that these conclusions
based on the PDFs of $\Delta \psi_N$ and $K_s/K_s$
require
only an equilibrium condition of $\overline{\Delta \psi_N}$ = $\overline{\Delta \psi}_N^{\script{eq}}$,
but does not require
the approximation of constancy for the variance of $\Delta \psi_{N}$ across homologous sequences, 
which is used only to estimate $T_s$ and $\psi_N^{\script{eq}}$
and other relations based on $T_s$.

\RED{
In the present study, we have analyzed the mutation-fixation process in equilibrium.
The equilibrium state
will vary if an environmental condition varies.  
The evolutionary statistical energy $\psi_N$ and the inverse of selective temperature $1/T_s$ 
are linearly proportional to the effective population size $N_e$,
as indicated by \Eq{\ref{\EQ: s_vs_dPsi}}.
Thus, the equilibrium values, $\psi_N^{\script{eq}}$, $\overline{\Delta \psi_N}^{\script{eq}}$ and  $\text{Sd}(\Delta \psi_N)^{\script{eq}}$,
are all linearly proportional to the effective population size $N_e$.
On the other hand, $\text{Sd}(\Delta \psi_N)^{\script{eq}}$
is not linearly proportional to $\overline{\Delta \psi_N}^{\script{eq}}$ 
but downward-concave, as shown in \Fig{\ref{\FIG: ddPhi_mean_vs_sd_at_equil}}.
As a result, as $N_e$ decreases,  
$k_B T_s \, \overline{\Delta \psi_N}^{\script{eq}} \simeq
k_B T_s \, \overline{\Delta \Delta \psi_{ND}}^{\script{eq}} (\simeq \overline{\Delta \Delta G_{ND}}^{\script{eq}})$ 
decreases.
In other words, the equilibrium value of the mean folding free energy change
becomes less positive and therefore that of folding free energy 
($\overline{\Delta G_{ND}}^{\script{eq}} \simeq k_B T_s \, \overline{\Delta \psi_{ND}}^{\script{eq}}$) 
is expected to be
less negative (less stable) for a smaller number of effective population size $N_e$; 
see \Eq{\ref{\EQ: regression_of_ddG_on_dG}}.
}%  RED

% End of discussion_1+2.tex

% \input{acknowledge_1+2.tex}

\section*{Acknowledgement}

I would like to thank a reviewer for his excellent comments
and suggestions that have helped me improve the paper considerably.
% End of acknowledge_1+2.tex

% End of contents_1+2.tex

{
\vspace*{2em}
%\newline
\noindent
\section*{Supplementary document}
  \subsection*{File 1 --- Supplementary methods, tables, and figures}
        A PDF file in which the details of the methods are described and additional tables and figures are provided;
	methods, tables, and figures provided in the text are also included as part of their full descriptions for reader's convenience.

\vspace*{2em}
%\newline
\noindent
\textbf{Funding}
%\newline
\noindent
This research did not receive any specific grant from funding agencies in the public, commercial, or
not-for-profit sectors.
}

\vspace*{2em}
%\newline
\noindent
\section*{Publication}
%\textbf{Publication}

\noindent
The original version was published in J. Theoretical Biol., 433, 21-38, 2017 (DOI:10.1016/j.jtbi.2017.08.018);
in this version, Table 3 (Table S.3) and Table S.5 are revised by employing thermochemical calorie (1~cal~=~4.184~J)
rather than steam table calorie (1~cal~=~4.1868~J); the revised values in these tables are printed in blue.

\newpage

%\noindent
%\textbf{References}

\bibliographystyle{elsarticle-harv}
\bibliography{jnames_with_dots,MolEvol,Protein,Bioinfo,SM}

%% Authors are advised to submit their bibtex database files. They are
%% requested to list a bibtex style file in the manuscript if they do
%% not want to use elsarticle-harv.bst.

%% References without bibTeX database:

% \begin{thebibliography}{00}

%% \bibitem must have one of the following forms:
%%   \bibitem[Jones et al.(1990)]{key}...
%%   \bibitem[Jones et al.(1990)Jones, Baker, and Williams]{key}...
%%   \bibitem[Jones et al., 1990]{key}...
%%   \bibitem[\protect\citeauthoryear{Jones, Baker, and Williams}{Jones
%%       et al.}{1990}]{key}...
%%   \bibitem[\protect\citeauthoryear{Jones et al.}{1990}]{key}...
%%   \bibitem[\protect\astroncite{Jones et al.}{1990}]{key}...
%%   \bibitem[\protect\citename{Jones et al., }1990]{key}...
%%   \harvarditem[Jones et al.]{Jones, Baker, and Williams}{1990}{key}...
%%

% \bibitem[ ()]{}

% \end{thebibliography}

\NoFigureInText{
\small

\renewcommand{\TextTable}[1]{#1}
\renewcommand{\SupTable}[1]{}
\setcounter{table}{0}

\TextTable{

\ifdefined\TableZ
\else
\TableInText{
\newcommand{\TableZ}[1]{}
}%  TableInText
\NoTableInText{
\newcommand{\TableZ}[1]{#1}
}%  NoTableInText
\fi

\ifdefined\TableA
\else
\TableInText{
\newcommand{\TableA}[1]{}
}%  TableInText
\NoTableInText{
\newcommand{\TableA}[1]{#1}
}%  NoTableInText
\fi

\ifdefined\TableB
\else
\TableInText{
\newcommand{\TableB}[1]{}
}%  TableInText
\NoTableInText{
\newcommand{\TableB}[1]{#1}
}%  NoTableInText
\fi

\renewcommand{\SUPPLEMENT}[1]{}
\ifdefined\CLEARPAGE
\TableInText{
\renewcommand{\CLEARPAGE}{\TableInLegends{}}
}%  TableInText
\NoTableInText{
\renewcommand{\CLEARPAGE}{\TableInLegends{\clearpage \newpage}}
}%  NoTableInText
\else
\TableInText{
\newcommand{\CLEARPAGE}{\TableInLegends{}}
}%  TableInText
\NoTableInText{
\newcommand{\CLEARPAGE}{\TableInLegends{\clearpage \newpage}}
}%  NoTableInText
\fi
}%  TextTable

\SupTable{

\ifdefined\TableZ
\renewcommand{\TableZ}[1]{#1}
\else
\newcommand{\TableZ}[1]{#1}
\fi

\ifdefined\TableA
\renewcommand{\TableA}[1]{#1}
\else
\newcommand{\TableA}[1]{#1}
\fi

\ifdefined\TableB
\renewcommand{\TableB}[1]{#1}
\else
\newcommand{\TableB}[1]{#1}
\fi

\renewcommand{\SUPPLEMENT}[1]{#1}
\ifdefined\CLEARPAGE
\renewcommand{\CLEARPAGE}{\TableInLegends{\clearpage \newpage}}
\else
\newcommand{\CLEARPAGE}{\TableInLegends{\clearpage \newpage}}
\fi
}%  SupTable

\TableZ{

\CLEARPAGE
\setlength{\textwidth}{17.5cm}
\setlength{\oddsidemargin}{0cm}
\setlength{\evensidemargin}{0cm}

\begin{table}[!ht]
\caption{
\SUPPLEMENT{
\label{stbl: Proteins_studied}
}%  SUPPLEMENT
\TextTable{
\label{tbl: Proteins_studied}
}%  TextTable
\BF{
Protein families, and structures studied.
}%  BF
}%  caption
\vspace*{2em}
\TableInLegends{

\small
\footnotesize

\begin{tabular}{llrrrrrl}
\hline
	Pfam family	& UniProt ID		& $N$ $^a$		& $N_{\script{eff}}$ $^{bc}$		& $M$ $^d$	& $M_{\script{eff}}$ $^{ce}$		& $L$ $^f$	& PDB ID			\\		
										\hline
HTH\_3		& RPC1\_BP434/7-59	& 15315(15917)	& 11691.21 	& 6286	& 4893.73	& 53	& 1R69-A:6-58		\\
Nitroreductase	& Q97IT9\_CLOAB/4-76	& 6008(6084)	& 4912.96	& 1057	& 854.71	& 73	& 3E10-A/B:4-76 $^g$	\\
SBP\_bac\_3 $^h$	& GLNH\_ECOLI/27-244	& 9874(9972)	& 7374.96	& 140	& 99.70	& 218	& 1WDN-A:5-222		\\
SBP\_bac\_3	& GLNH\_ECOLI/111-204	& 9712(9898)	& 7442.85	& 829	& 689.64	& 94	& 1WDN-A:89-182		\\
OmpA		& PAL\_ECOLI/73-167	& 6035(6070)	& 4920.44	& 2207	& 1761.24	& 95	& 1OAP-A:52-146		\\
DnaB		& DNAB\_ECOLI/31-128	& 1929(1957)	& 1284.94	& 1187	& 697.30	& 98	& 1JWE-A:30-127		\\
LysR\_substrate $^h$	& BENM\_ACIAD/90-280	& 25138(25226)	& 20707.06 	& 85(1)	& 67.00	& 191	& 2F6G-A/B:90-280 $^g$	\\
LysR\_substrate	& BENM\_ACIAD/163-265	& 25032(25164)	& 21144.74 	& 121(1) & 99.27	& 103	& 2F6G-A/B:163-265 $^g$	\\
Methyltransf\_5 $^h$	& RSMH\_THEMA/8-292	& 1942(1953)	& 1286.67	& 578(2) & 357.97	& 285	& 1N2X-A:8-292		\\
Methyltransf\_5	& RSMH\_THEMA/137-216	& 1877(1911)	& 1033.35	& 975(2) & 465.53	& 80	& 1N2X-A:137-216	\\
SH3\_1		& SRC\_HUMAN:90-137	& 9716(16621)	& 3842.47	& 1191	& 458.31	& 48	& 1FMK-A:87-134		\\
ACBP		& ACBP\_BOVIN/3-82	& 2130(2526)	& 1039.06	& 161	& 70.72	& 80	& 2ABD-A:2-81		\\
PDZ		& PTN13\_MOUSE/1358-1438	& 13814(23726)	& 4748.76	& 1255	& 339.99	& 81	& 1GM1-A:16-96		\\
Copper-bind	& AZUR\_PSEAE:24-148	& 1136(1169)	& 841.56	& 67(1)	& 45.23	& 125	& 5AZU-B/C:4-128 $^g$		\\
\hline
\end{tabular}

\vspace*{1em}
\noindent
$^a$ The number of unique sequences and the total number of sequences in parentheses; 
	the full alignments in the Pfam\CITE{FCEEMMPPQSSTB:16} are used.

\noindent
$^b$ The effective number of sequences.

\noindent
$^c$ A sample weight ($w_{\VEC{\sigma}_N}$) for a given sequence is
equal to the inverse of the number of sequences 
that are less than 20\% different from the given sequence.

\noindent
$^d$ The number of unique sequences that include no deletion unless specified.
	The number in parentheses indicates the maximum number of deletions allowed. 

\noindent
$^e$ The effective number of unique sequences that include no deletion
or at most the specified number of deletions.

\noindent
$^f$ The number of residues.

\noindent
$^g$ Contacts are calculated in the homodimeric state for these protein.

\noindent
$^h$ These proteins consist of two domains, and other ones are single domains.

% End of TABLES/Proteins.tex
}%  TableInLegends
\end{table}

}%  TableZ

\TableA{

\CLEARPAGE
\setlength{\textwidth}{17.5cm}
\setlength{\oddsidemargin}{-0.5cm}
\setlength{\evensidemargin}{-0.5cm}

\begin{table}[!ht]
\caption{
\SUPPLEMENT{
\label{stbl: ddPsi_with_8A}
}%  SUPPLEMENT
\TextTable{
\label{tbl: ddPsi_with_8A}
}%  TextTable
\BF{
Parameter values
for $r_{\script{cutoff}} \sim 8$ \AA\
}%  BF
employed for each protein family, and
the averages of the 
evolutionary statistical energies
($\overline{\psi_N}$)
over all homologous sequences
and of the means and 
the standard deviations of interaction changes ($\overline{\overline{\Delta \psi_N}}$ and $\overline{\text{Sd}(\Delta \psi_N)}$) 
due to single nucleotide nonsynonymous mutations 
at all sites over all homologous sequences in each protein family.
}%  caption
\vspace*{2em}
\TableInLegends{

\footnotesize
\scriptsize

\begin{tabular}{lrrrrrrrrrrrrr}
\hline
\\
Pfam family & $L$ & $p_c$ & $n_c$ $^a$ & $r_{\script{cutoff}}$ & 
		$\bar{\psi}/L$ $^b$	& ${\delta \psi}^2/L$ $^b$	& $\overline{\psi_N}/L$ $^b$ & 
		$\overline{\overline{\Delta \psi_N}}$ $^c$	&
		$\overline{\mbox{Sd}(\Delta \psi_N) } \, \pm$ $^c$ &
		$r_{\psi_N}$	& $\alpha_{\psi_N}$	& $r_{\psi_N}$	& $\alpha_{\psi_N} $		\\
		&	&	&	& (\AA\ )	&	&	&	&	& $\mbox{Sd}(\mbox{Sd}(\Delta \psi_N) )$	&
		\multicolumn{2}{c}{for $\overline{\Delta \psi_N}$ $^d$}	& \multicolumn{2}{c}{for $\mbox{Sd}(\Delta \psi_N)$ $^e$}
\vspace*{2mm}
		\\
\hline	
HTH\_3	& $53$	& $0.18$	& $7.43$	& $8.22$	& $-0.1997$	& $2.7926$	& $-2.9861$	& $4.2572$	& $5.3503 \pm 0.5627$	& $-0.961$	& $-1.5105$	& $-0.598$	& $-0.9888$ \\
Nitroreductase	& $73$	& $0.23$	& $6.38$	& $8.25$	& $-0.1184$	& $2.1597$	& $-2.2788$	& $3.3115$	& $3.6278 \pm 0.2804$	& $-0.939$	& $-1.3371$	& $-0.426$	& $-0.3721$ \\
SBP\_bac\_3	& $218$	& $0.25$	& $9.23$	& $8.10$	& $-0.1000$	& $2.1624$	& $-2.2618$	& $3.2955$	& $3.4496 \pm0.2742$	& $-0.980$	& $-1.5286$	& $-0.841$	& $-0.7876$ \\
SBP\_bac\_3	& $94$	& $0.37$	& $8.00$	& $7.90$	& $-0.1634$	& $1.2495$	& $-1.4054$	& $1.9291$	& $2.3436 \pm 0.1901$	& $-0.959$	& $-1.3938$	& $-0.634$	& $-0.4815$ \\
OmpA	& $95$	& $0.169$	& $8.00$	& $8.20$	& $-0.2457$	& $3.9093$	& $-4.1542$	& $6.5757$	& $7.6916 \pm 0.3078$	& $-0.957$	& $-1.5694$	& $-0.410$	& $-0.3804$ \\
DnaB	& $98$	& $0.235$	& $9.65$	& $8.17$	& $-0.2284$	& $3.9976$	& $-4.2291$	& $6.3502$	& $6.1244 \pm 0.3245$	& $-0.965$	& $-1.4509$	& $-0.495$	& $-0.4198$ \\
LysR\_substrate	& $191$	& $0.235$	& $8.59$	& $7.98$	& $-0.2241$	& $1.4888$	& $-1.7173$	& $2.2784$	& $2.6519 \pm 0.1445$	& $-0.964$	& $-1.3347$	& $-0.541$	& $-0.5664$ \\
LysR\_substrate	& $103$	& $0.265$	& $8.84$	& $8.25$	& $-0.2244$	& $1.4144$	& $-1.6379$	& $2.2110$	& $2.7371 \pm 0.2055$	& $-0.982$	& $-1.4159$	& $-0.727$	& $-0.5307$ \\
Methyltransf\_5	& $285$	& $0.13$	& $7.99$	& $7.78$	& $-0.1462$	& $7.2435$	& $-7.3887$	& $12.4689$	& $10.9352 \pm 0.3030$	& $-0.981$	& $-1.9140$	& $-0.122$	& $-0.0783$ \\
Methyltransf\_5	& $80$	& $0.18$	& $6.78$	& $7.85$	& $-0.1763$	& $5.5162$	& $-5.6896$	& $8.9849$	& $7.6133 \pm 0.4382$	& $-0.944$	& $-1.4824$	& $0.125$	& $0.1141$ \\
SH3\_1	& $48$	& $0.14$	& $6.42$	& $8.01$	& $-0.1348$	& $3.9109$	& $-4.0434$	& $5.5792$	& $6.1426 \pm 0.2935$	& $-0.919$	& $-1.4061$	& $-0.196$	& $-0.1718$ \\
ACBP	& $80$	& $0.22$	& $9.17$	& $8.24$	& $-0.0525$	& $4.6411$	& $-4.7084$	& $7.7612$	& $7.1383 \pm 0.2970$	& $-0.972$	& $-1.5884$	& $-0.335$	& $-0.2235$ \\
PDZ	& $81$	& $0.205$	& $9.06$	& $8.16$	& $-0.2398$	& $3.1140$	& $-3.3572$	& $4.7589$	& $4.6605 \pm 0.2255$	& $-0.954$	& $-1.5282$	& $-0.369$	& $-0.3042$ \\
Copper-bind	& $125$	& $0.23$	& $9.50$	& $8.27$	& $-0.0940$	& $4.2450$	& $-4.3272$	& $7.2650$	& $6.9283 \pm 0.2316$	& $-0.980$	& $-1.8915$	& $-0.282$	& $-0.2352$ \\
% End of ./TABLES/r=8A/Table-8A_1.proto.tex
\hline

\end{tabular}

\vspace*{1em}
\noindent
$^a$ The average number of contact residues per site within the cutoff distance;
the center of side chain is used to represent a residue.

\noindent
$^b$ $M$ unique sequences with no deletions are used with a sample weight ($w_{\VEC{\sigma}_N}$) for each sequence;
$w_{\VEC{\sigma}_N}$ is equal to the inverse of the number of sequences
that are less than 20\% different from a given sequence.
The $M$ and the effective number $M_{\script{eff}}$ of the sequences are listed for each protein family 
in \Table{\ref{\TBL: Proteins_studied}}.

\noindent
$^c$ The averages of $\overline{\Delta \psi_N}$ and $\mbox{Sd}(\Delta \psi_N)$, 
which are the mean and the standard deviation of $\Delta \psi_N$ for a sequence, and
the standard deviation of $\mbox{Sd}(\Delta \psi_N)$ over homologous sequences.
Representatives of unique sequences with no deletions, which are at least 20\% different from each other, are used; 
     the number of the representatives used is almost equal to $M_{\script{eff}}$.  

\noindent
$^d$ The correlation and regression coefficients of $\overline{\Delta \psi_N}$ on $\psi_N/L$; see \Eq{\ref{\EQ: regression_of_dPsi_on_Psi}}.

\noindent
$^e$ The correlation and regression coefficients of $\mbox{Sd}(\Delta \psi_N)$ on $\psi_N/L$.

% End of TABLES/r=8A/Table-8A_1.tex
}%  TableInLegends
\end{table}

}%  TableA

\TableB{

\CLEARPAGE
\setlength{\textwidth}{17.5cm}
\setlength{\oddsidemargin}{0cm}
\setlength{\evensidemargin}{0cm}

\begin{table}[!ht]
\caption{
\SUPPLEMENT{
\label{stbl: Ts_with_8A}
}%  SUPPLEMENT
\TextTable{
\label{tbl: Ts_with_8A}
}%  TextTable
\BF{
Thermodynamic quantities estimated with $r_{\script{cutoff}} \sim 8$ \AA. 
}%  BF
}%  caption
\vspace*{2em}
\TableInLegends{

\small
\begin{tabular}{lrrrrrrrr}
\hline
	&
	&
		& 	& Experimental	&	& 	 &	&
	\\
Pfam family
	& $r$ $^a$
 	& $k_B \hat{T}_s$ $^a$ & $\hat{T}_s$ & $T_m$ & $\hat{T}_g$ & $\hat{\omega}$ $^b$ &$T$ $^c$ & $\langle \Delta G_{ND} \rangle$ $^d$
	\\ 
	&	&(kcal/mol)	& ($^\circ$K) & ($^\circ$K) & ($^\circ$K) & ($k_B$) & ($^\circ$K) & (kcal/mol)
\vspace*{2mm}
	\\ 
\hline
HTH\_3	& -- 	& -- 	& $122.\REDa{5}$	& $343.7$	& $160.\REDa{0}$	& $0.81\REDa{78}$	& $298$	& $-2.95$ \\
Nitroreductase	& -- 	& --	& $180.\REDa{6}$	& $337$	& $20\REDa{3.9}$	& $0.847\REDa{3}$	& $298$	& $-2.81$ \\
SBP\_bac\_3	& --	& --	& $190.\REDa{0}$	& $336.1$	& $21\REDa{0.9}$	& $0.87\REDa{68}$	& $298$	& $-8.0\REDa{4}$ \\
SBP\_bac\_3	& --	& --	& $279.\REDa{6}$	& $336.1$	& $283.\REDa{6}$	& $0.607\REDa{1}$	& $298$	& $-0.85$ \\
OmpA	& -- 	& --	& $85.2$	& $320$	& $125.4$	& $0.902\REDa{2}$	& $298$	& $-3.13$ \\
DnaB	& --	& --	& $107.\REDa{0}$	& $312.8$	& $142.1$	& $1.13\REDa{35}$	& $298$	& $-2.56$ \\
LysR\_substrate	& --	& -- 	& $247.\REDa{1}$	& $338$	& $256.\REDa{5}$	& $0.690\REDa{6}$	& $298$	& $-3.63$ \\
LysR\_substrate	& --	& --	& $239.\REDa{4}$	& $338$	& $250.\REDa{3}$	& $0.647\REDa{0}$	& $298$	& $-2.00$ \\
Methyltransf\_5	& --	& --	& $\REDa{59.9}$	& $375$	& $110.5$	& $1.065\REDa{0}$	& $298$	& $-41.3\REDa{7}$ \\
Methyltransf\_5	& --	& --	& $86.1$	& $375$	& $135.\REDa{0}$	& $1.12\REDa{08}$	& $298$	& $-11.48$ \\
SH3\_1	& $0.865$	& $0.1583$	& $106.7$	& $344$	& $147.4$	& $1.02\REDa{47}$	& $295$	& $-3.76$ \\
ACBP	& $0.825$	& $0.1169$	& $91.\REDa{8}$	& $324.4$	& $131.7$	& $1.12\REDa{75}$	& $278$	& $-6.72$ \\
PDZ	& $0.931$	& $0.2794$	& $140.\REDa{6}$	& $312.88$	& $168.\REDa{4}$	& $1.08\REDa{49}$	& $298$	& $-1.81$ \\
Copper-bind	& $0.828$	& $0.1781$	& $94.6$	& $359.3$	& $139.9$	& $0.970\REDa{3}$	& $298$	& $-12.0\REDa{8}$ \\
% End of ./TABLES/r=8A/Table-8A_2.proto_revised.tex
\hline
\end{tabular}

\vspace*{1em}
\noindent
$^a$ Reflective correlation ($r$) and regression ($k_B \hat{T}_s$) coefficients 
	for least-squares regression lines of experimental $\Delta\Delta G_{ND}$ on $\Delta \psi_N$ through the origin.

\noindent
$^b$ Conformational entropy per residue, in $k_B$ units, in the denatured molten-globule state; see \Eq{\ref{\EQ: expression_of_entropy}}.

\noindent
$^c$ Temperatures are set up for comparison to be equal to the experimental temperatures for $\Delta G_{ND}$ or to $298 ^\circ$K if unavailable; 
see \Table{\ref{stbl: experimental_data}} for the experimental data.

\noindent
$^d$ Folding free energy in kcal/mol units; see \Eq{\ref{\EQ: ensemble_ave_of_ddG}}.
% End of TABLES/r=8A/Table-8A_2.revised.tex
}%  TableInLegends
\end{table}

}%  TableB

\SUPPLEMENT{

\CLEARPAGE

\begin{table}[!ht]
\caption{
\label{stbl: experimental_data}
\label{tbl: experimental_data}
\BF{
Experimental data used.
}%  BF
}%  caption
\vspace*{2em}
\TableInLegends{

\small
\footnotesize

\begin{tabular}{lrrrll}
\hline
			& \multicolumn{3}{c}{experimental values} &	&	\\
	Pfam family	& $T_m$	& $T$ & $\Delta G_{ND}$	
	& ref. for $T_m$ & ref. for $\Delta G_{ND}$ and $\Delta\Delta G_{ND}$
		\\
	& ($^\circ$K) & ($^\circ$K) & (kcal/mol)
\vspace*{2mm}
		\\
\hline
HTH\_3	& $343.7$ & $298$ & $-5.33 \pm 0.36$	& \CITE{GDBCJMS:09}	& \CITE{RSTPMF:99}	\\ 
Nitroreductase	& $337.0$ & - & -	& \CITE{SZMSS:06}	&	\\ 
SBP\_bac\_3	& $336.1$ & - & -	& \CITE{DSVSSAMRT:05}	&	\\ 
OmpA	& $320.0$ & - & -		& \CITE{PLO:06}	&	\\ 
DnaB	& $312.8$ & - & -		& \CITE{WPLLSLCOD:02}	&	\\ 
LysR\_substrate	& $338.0$  & - & -	& \CITE{SRSSO:08}	&	\\ 
Methyltransf\_5	& $375.0$ & - & -	& \CITE{AUFCBG:04}	&	\\ 
		&	& 	&	& \CITE{GRGBLG:10}	&	\\
SH3\_1	& $344.0$ & $295$ & $-3.70$	& \CITE{KMCBKVSL:98}	& \CITE{GRSB:98}	\\ 
ACBP	& $324.4$ & $278$ & $-8.08 \pm 0.08$	& \CITE{OKKSW:15}	& \CITE{KONSKKP:99}	\\ 
PDZ	& $312.9$ & $298$ & $-2.9$ \REDb{$\pm$} \REDb{$0.2$}	& \CITE{TES:12}	& \CITE{GCAVBT:05,GGCJVTVB:07}	\\ 
Copper-bind	& $359.3$ & $298$ & $-12.90 \pm 0.36$	& \CITE{RMGGS:95}	& \CITE{WW:05}	\\ 
% End of ./TABLES/r=8A/Table-Exp_1.proto.tex
\hline
\end{tabular}
% End of TABLES/r=8A/Table-Exp_1.tex
}%  TableInLegends
\end{table}

}%  SUPPLEMENT

\SUPPLEMENT{

\CLEARPAGE
\setlength{\textwidth}{17.5cm}
\setlength{\oddsidemargin}{-0.5cm}
\setlength{\evensidemargin}{-0.5cm}

\begin{table}[!ht]
\caption{
\label{stbl: ddPsi_with_16A}
\label{tbl: ddPsi_with_16A}
\BF{
Parameter values
for $r_{\script{cutoff}} \sim 15.5$ \AA\
}%  BF
employed for each protein family, and
the averages of the 
evolutionary statistical energies
($\overline{\psi_N}$)
over all homologous sequences
and of the means and 
the standard deviations of interaction changes ($\overline{\overline{\Delta \psi_N}}$ and $\overline{\text{Sd}(\Delta \psi_N)}$) 
due to single nucleotide nonsynonymous mutations 
at all sites over all homologous sequences in each protein family.
}%  caption
\vspace*{2em}
\TableInLegends{

\footnotesize
\scriptsize

\begin{tabular}{lrrrrrrrrrrrrr}
\hline
\\
Pfam family & $L$ & $p_c$ & $n_c$ $^a$ & $r_{\script{cutoff}}$ & 
		$\bar{\psi}/L$ $^b$	& ${\delta \psi}^2/L$ $^b$	& $\overline{\psi_N}/L$ $^b$ & 
		$\overline{\overline{\Delta \psi_N}}$ $^c$	&
		$\overline{\mbox{Sd}(\Delta \psi_N) } \, \pm$ $^c$ &
		$r_{\psi_N}$	& $\alpha_{\psi_N}$	& $r_{\psi_N}$	& $\alpha_{\psi_N} $		\\
		&	&	&	& (\AA\ ) &	&	&	&	& $\mbox{Sd}(\mbox{Sd}(\Delta \psi_N) )$	&
		\multicolumn{2}{c}{for $\overline{\Delta \psi_N}$ $^d$}	& \multicolumn{2}{c}{for $\mbox{Sd}(\Delta \psi_N)$ $^e$}
\vspace*{2mm}
		\\
\hline	
HTH\_3	& $53$	& $0.245$	& $32.90$	& $15.67$	& $-0.2548$	& $4.0057$	& $-4.2642$	& $6.8512$	& $6.9544 \pm 0.5309$	& $-0.955$	& $-1.5717$	& $-0.519$	& $-0.5727$ \\
Nitroreductase	& $73$	& $0.315$	& $28.71$	& $15.75$	& $-0.1476$	& $3.7093$	& $-3.8565$	& $6.3226$	& $5.6267 \pm 0.5440$	& $-0.953$	& $-1.5765$	& $-0.694$	& $-0.6640$ \\
SBP\_bac\_3	& $218$	& $0.35$	& $55.48$	& $15.90$	& $-0.0669$	& $3.4004$	& $-3.4674$	& $5.7978$	& $4.8666 \pm 0.4517$	& $-0.971$	& $-1.6708$	& $-0.821$	& $-0.8874$ \\
SBP\_bac\_3	& $94$	& $0.455$	& $42.81$	& $15.45$	& $-0.1628$	& $2.3208$	& $-2.4831$	& $4.0963$	& $3.7760 \pm 0.3970$	& $-0.968$	& $-1.6628$	& $-0.770$	& $-0.6408$ \\
OmpA	& $95$	& $0.235$	& $35.58$	& $15.69$	& $-0.2552$	& $5.8175$	& $-6.0757$	& $10.4102$	& $11.8829 \pm 0.4108$	& $-0.948$	& $-1.6212$	& $-0.354$	& $-0.3599$ \\
DnaB	& $98$	& $0.35$	& $46.65$	& $15.57$	& $-0.2351$	& $6.1890$	& $-6.4167$	& $10.7294$	& $8.0204 \pm 0.3493$	& $-0.894$	& $-1.5176$	& $-0.311$	& $-0.3037$ \\
LysR\_substrate	& $191$	& $0.335$	& $52.30$	& $15.58$	& $-0.2826$	& $2.5962$	& $-2.8789$	& $4.4194$	& $4.1701 \pm 0.1782$	& $-0.963$	& $-1.6196$	& $-0.613$	& $-0.4726$ \\
LysR\_substrate	& $103$	& $0.37$	& $44.33$	& $15.60$	& $-0.2816$	& $2.4438$	& $-2.7239$	& $4.1276$	& $4.2029 \pm 0.3674$	& $-0.984$	& $-1.5436$	& $-0.769$	& $-0.5462$ \\
Methyltransf\_5	& $285$	& $0.175$	& $53.52$	& $15.53$	& $-0.1687$	& $12.8982$	& $-13.0658$	& $23.6376$	& $18.7982 \pm 0.4701$	& $-0.952$	& $-1.9804$	& $-0.171$	& $-0.1630$ \\
Methyltransf\_5	& $80$	& $0.24$	& $37.02$	& $15.11$	& $-0.1632$	& $9.9944$	& $-10.1576$	& $17.5749$	& $13.9124 \pm 0.4756$	& $-0.862$	& $-1.6406$	& $-0.290$	& $-0.2822$ \\
SH3\_1	& $48$	& $0.165$	& $28.46$	& $15.76$	& $-0.1350$	& $7.6161$	& $-7.7523$	& $11.9725$	& $13.3845 \pm 0.4719$	& $-0.896$	& $-1.5944$	& $-0.255$	& $-0.2420$ \\
ACBP	& $80$	& $0.28$	& $36.27$	& $15.34$	& $-0.0235$	& $7.4707$	& $-7.4947$	& $13.1892$	& $9.7188 \pm 0.4242$	& $-0.911$	& $-1.7087$	& $0.085$	& $0.0861$ \\
PDZ	& $81$	& $0.33$	& $40.82$	& $15.77$	& $-0.3022$	& $5.2295$	& $-5.5313$	& $8.6909$	& $7.9383 \pm 0.2930$	& $-0.966$	& $-1.7215$	& $-0.316$	& $-0.2328$ \\
Copper-bind	& $125$	& $0.295$	& $45.22$	& $15.32$	& $-0.0999$	& $8.5521$	& $-8.6592$	& $15.5941$	& $9.6566 \pm 0.3556$	& $-0.951$	& $-1.7441$	& $-0.175$	& $-0.1981$ \\
% End of ./TABLES/r=16A/Table-16A_1.proto.tex
\hline

\end{tabular}

\vspace*{1em}
\noindent
$^a$ The average number of contact residues per site within the cutoff distance;
the center of side chain is used to represent a residue.

\noindent
$^b$ $M$ unique sequences without deletions are used with a sample weight ($w_{\VEC{\sigma}_N}$) for each sequence;
$w_{\VEC{\sigma}_N}$ is equal to the inverse of the number of sequences
that are less than 20\% different from a given sequence.
The $M$ and the effective number $M_{\script{eff}}$ of the sequences are listed for each protein family 
in \Table{\ref{\TBL: Proteins_studied}}.

\noindent
$^c$ The averages of $\overline{\Delta \psi_N}$ and $\mbox{Sd}(\Delta \psi_N)$,
which are the mean and the standard deviation of $\Delta \psi_N$ for a sequence, and
the standard deviation of $\mbox{Sd}(\Delta \psi_N)$ over homologous sequences.
Representatives of unique sequences without deletions, which are at least 20\% different from each other, are used;
     the number of the representatives used is almost equal to $M_{\script{eff}}$.

\noindent
$^d$ The correlation and regression coefficients of $\overline{\Delta \psi_N}$ on $\psi_N/L$;see \Eq{\ref{\EQ: regression_of_dPsi_on_Psi}}.

\noindent
$^e$ The correlation and regression coefficients of $\mbox{Sd}(\Delta \psi_N)$ on $\psi_N/L$.

% End of TABLES/r=16A/Table-16A_1.tex
}%  TableInLegends
\end{table}

\CLEARPAGE
\setlength{\textwidth}{17.5cm}
\setlength{\oddsidemargin}{0cm}
\setlength{\evensidemargin}{0cm}

\begin{table}[!ht]
\caption{
\label{stbl: Ts_with_16A}
\label{tbl: Ts_with_16A}
\BF{
Thermodynamic quantities estimated with $r_{\script{cutoff}} \sim 15.5$ \AA. 
}%  BF
}%  caption
\vspace*{2em}
\TableInLegends{

\small
\begin{tabular}{lrrrrrrrr}
\hline
	&
	&
		& 	& Experimental	&	& 	 &	&
	\\
Pfam family
	& $r$ $^a$
 	& $k_B \hat{T}_s$ $^a$ & $\hat{T}_s$ & $T_m$ & $\hat{T}_g$ & $\hat{\omega}$ $^b$ &$T$ $^c$ & $\langle \Delta G_{ND} \rangle$ $^d$
	\\ 
	&	&(kcal/mol)	& ($^\circ$K) & ($^\circ$K) & ($^\circ$K) & ($k_B$) & ($^\circ$K) & (kcal/mol)
\vspace*{2mm}
	\\ 
\hline
HTH\_3	& -- 	& -- 	& $93.\REDa{0}$	& $343.7$	& $136.0$	& $0.937\REDa{3}$	& $298$	& $-3.70$ \\
Nitroreductase	& --	& -- 	& $115.0$	& $337$	& $152.\REDa{8}$	& $1.0\REDa{495}$	& $298$	& $-4.56$ \\
SBP\_bac\_3	& --	& --	& $13\REDa{2.9}$	& $336.1$	& $166.9$	& $1.07\REDa{88}$	& $298$	& $-12.8\REDa{6}$ \\
SBP\_bac\_3	& --	& --	& $171.\REDa{3}$	& $336.1$	& $196.6$	& $0.881\REDa{4}$	& $298$	& $-3.85$ \\
OmpA	& -- 	& --	& $54.\REDa{4}$	& $320$	& $97.6$	& $0.90\REDa{54}$	& $298$	& $-3.3\REDa{9}$ \\
DnaB	& --	& -- 	& $80.\REDa{6}$	& $312.8$	& $120.\REDa{3}$	& $1.390\REDa{0}$	& $298$	& $-3.38$ \\
LysR\_substrate	& --	& -- 	& $155.\REDa{1}$	& $338$	& $184.\REDa{4}$	& $0.918\REDa{0}$	& $298$	& $-9.2\REDa{3}$ \\
LysR\_substrate	& --	& --	& $15\REDa{3.9}$	& $338$	& $183.\REDa{5}$	& $0.859\REDa{4}$	& $298$	& $-4.68$ \\
Methyltransf\_5	& --	& --	& $34.4$	& $375$	& $82.\REDa{2}$	& $1.129\REDa{2}$	& $298$	& $-46.26$ \\
Methyltransf\_5	& --	& --	& $46.5$	& $375$	& $96.4$	& $1.16\REDa{23}$	& $298$	& $-13.04$ \\
SH3\_1	& $0.836$	& $0.0821$	& $48.\REDa{3}$	& $344$	& $94.6$	& $0.99\REDa{48}$	& $295$	& $-4.24$ \\
ACBP	& $0.823$	& $0.0689$	& $66.6$	& $324.4$	& $109.7$	& $1.37\REDa{55}$	& $278$	& $-8.79$ \\
PDZ	& $0.944$	& $0.1619$	& $81.5$	& $312.88$	& $121.1$	& $1.18\REDa{46}$	& $298$	& $-2.39$ \\
Copper-bind	& $0.888$	& $0.1015$	& $67.0$	& $359.3$	& $115.2$	& $1.44\REDa{57}$	& $298$	& $-19.2\REDa{9}$ \\
% End of ./TABLES/r=16A/Table-16A_2.proto_revised.tex
\hline
\end{tabular}

\vspace*{1em}
\noindent
$^a$ Reflective correlation ($r$) and regression ($k_B \hat{T}_s$) coefficients 
	for least-squares regression lines of experimental $\Delta\Delta G_{ND}$ on $\Delta \psi_N$ through the origin.

\noindent
$^b$ Conformational entropy per residue, in $k_B$ units, in the denatured molten-globule state; see \Eq{\ref{\EQ: expression_of_entropy}}.

\noindent
$^c$ Temperatures are set up for comparison to be equal to the experimental temperatures for $\Delta G_{ND}$ or to $298 ^\circ$K if unavailable; 
see \Table{\ref{stbl: experimental_data}} for the experimental data.

\noindent
$^d$ Folding free energy in kcal/mol units; see \Eq{\ref{\EQ: ensemble_ave_of_ddG}}.
% End of TABLES/r=16A/Table-16A_2.revised.tex
}%  TableInLegends
\end{table}

}%  SUPPLEMENT

% End of tables_JTB_1.tex
% \input{tables_JTB_2.tex}

\SUPPLEMENT{

\CLEARPAGE

\begin{table}[!ht]
\caption{
\label{stbl: fixation_prob}
\label{tbl: fixation_prob}
\BF{
Fixation probabilities of a single mutant in various models.
}%  BF
}%  caption
\vspace*{2em}
\TableInLegends{

\begin{tabular}{lllllll}
A) & \multicolumn{6}{l}{For Wright-Fisher population; compiled from p. 192 and pp. 424--427 of Crow and Kimura (1970).}	
		\\
		\\
  &	Fitness/Selection $^\script{a}$  & $h$ $^\script{a}$	& $M_{\delta x}$ $^\script{b}$ & $V_{\delta x}$ $^\script{c}$ & $u$ $^\script{de}$	& $q_m$ $^\script{f}$
		\\
		\hline
 & No dominance	  & $1/2$	& $sx(1-x)$	& $x(1-x)/(2N_e)$	& $(1- e^{-4N_esq_m})/ (1- e^{-4N_es})$ & $1/(2N)$
		\\
 & Dominance favored & $1$	& $2sx(1-x)^2$	& $x(1-x)/(2N_e)$	& \hspace*{5em}$^\script{e}$	  & $1/(2N)$
		\\
 & Recessive favored & $0$	& $2sx^2(1-x)$	& $x(1-x)/(2N_e)$	& \hspace*{5em}$^\script{e}$	  & $1/(2N)$
		\\
 & Gametic selection &	& $sx(1-x)$	& $x(1-x)/(2N_e)$ & $(1- e^{-4N_esq_m})/(1- e^{-4N_es})$	& $1/(2N)$
		\\
 & Haploid  	& 	& $sx(1-x)$	& $x(1-x)/N_e$	& $(1- e^{-2N_esq_m})/ (1- e^{-2N_es})$	& $1/N$
		\\
		\hline
		\\
B) & \multicolumn{6}{l}{For Moran population\CITE{M:58,E:79}}	
		\\
		\\
  & Fitness/Selection $^\script{a}$  & 	& $M_{\delta x}$ 	& $V_{\delta x}$ $^\script{c}$	& $u$ $^\script{de}$	& $q_m$ $^\script{f}$
		\\
		\hline
 & Haploid  	& 	& $s x(1-x)/N_e$ & $2x(1-x)/N_e^2$	& $(1- e^{-N_esq_m})/ (1- e^{-N_es})$	& $1/N$
		\\
		\hline
\end{tabular}

\vspace*{1em}
\noindent
$^\script{a}$ For zygotic selection, $2s$ and $2sh$ are the selective advantages of mutant homogeneous and heterogeneous zygotes, respectively.
For others, $s$ is the selective advantage of mutant gene.

\noindent
$^\script{b}$ Mean in the rate of the change of gene frequency per generation; $M_{\delta x} = 2sx(1-x)(h + (1-2h)x)$ for zygotic selection.

\noindent
$^\script{c}$ Variance in the rate of the change of gene frequency per generation.

\noindent
$^\script{d}$ Fixation probability. 

\noindent 
$^\script{e}$ $u(q_m) = F(q_m)/F(1)$
where $F(q_m)=\int_0^{q_m} G(x)dx$ and $G(x)=\exp (-\int 2M_{\delta x}/V_{\delta x} dx)$. 

\noindent
$^\script{f}$ Frequency of a single mutant gene.
% End of TABLES/fixation_prob.tex
}%  TableInLegends
\end{table}

}%  SUPPLEMENT
% End of tables_JTB_2.tex
% End of tables_1+2.tex

\clearpage

\renewcommand{\TextFig}[1]{#1}
\renewcommand{\SupFig}[1]{}

\setcounter{figure}{0}

\TextFig{

\ifdefined\FigA
\else
\NoFigureInText{
\newcommand{\FigA}[1]{#1}
}%  NoFigureInText
\FigureInText{
\newcommand{\FigA}[1]{}
}%  FigureInText
\fi

\ifdefined\FigB
\else
\NoFigureInText{
\newcommand{\FigB}[1]{#1}
}%  NoFigureInText
\FigureInText{
\newcommand{\FigB}[1]{}
}%  FigureInText
\fi

\ifdefined\FigC
\else
\NoFigureInText{
\newcommand{\FigC}[1]{#1}
}%  NoFigureInText
\FigureInText{
\newcommand{\FigC}[1]{}
}%  FigureInText
\fi

\ifdefined\FigD
\else
\NoFigureInText{
\newcommand{\FigD}[1]{#1}
}%  NoFigureInText
\FigureInText{
\newcommand{\FigD}[1]{}
}%  FigureInText
\fi

\ifdefined\FigE
\else
\NoFigureInText{
\newcommand{\FigE}[1]{#1}
}%  NoFigureInText
\FigureInText{
\newcommand{\FigE}[1]{}
}%  FigureInText
\fi

\ifdefined\FigF
\else
\NoFigureInText{
\newcommand{\FigF}[1]{#1}
}%  NoFigureInText
\FigureInText{
\newcommand{\FigF}[1]{}
}%  FigureInText
\fi

\ifdefined\FigG
\else
\NoFigureInText{
\newcommand{\FigG}[1]{#1}
}%  NoFigureInText
\FigureInText{
\newcommand{\FigG}[1]{}
}%  FigureInText
\fi

\renewcommand{\SUPPLEMENT}[1]{}

\ifdefined\CLEARPAGE

\NoFigureInText{
\renewcommand{\CLEARPAGE}{\FigureLegends{\clearpage\newpage}}
}%  NoFigureInText
\FigureInText{
\renewcommand{\CLEARPAGE}{}
}%  FigureInText

\else

\NoFigureInText{
\newcommand{\CLEARPAGE}{\FigureLegends{\clearpage\newpage}}
}%  NoFigureInText
\FigureInText{
\newcommand{\CLEARPAGE}{}
}%  FigureInText

\fi

}%  TextFig

\SupFig{

\ifdefined\FigA
\renewcommand{\FigA}[1]{#1}
\else
\newcommand{\FigA}[1]{#1}
\fi

\ifdefined\FigB
\renewcommand{\FigB}[1]{#1}
\else
\newcommand{\FigB}[1]{#1}
\fi

\ifdefined\FigC
\renewcommand{\FigC}[1]{#1}
\else
\newcommand{\FigC}[1]{#1}
\fi

\ifdefined\FigD
\renewcommand{\FigD}[1]{#1}
\else
\newcommand{\FigD}[1]{#1}
\fi

\ifdefined\FigE
\renewcommand{\FigE}[1]{#1}
\else
\newcommand{\FigE}[1]{#1}
\fi

\ifdefined\FigF
\renewcommand{\FigF}[1]{#1}
\else
\newcommand{\FigF}[1]{#1}
\fi

\ifdefined\FigG
\renewcommand{\FigG}[1]{#1}
\else
\newcommand{\FigG}[1]{#1}
\fi

\renewcommand{\SUPPLEMENT}[1]{#1}
\ifdefined\CLEARPAGE
\renewcommand{\CLEARPAGE}{\FigureLegends{\clearpage\newpage}}
\else
\newcommand{\CLEARPAGE}{\FigureLegends{\clearpage\newpage}}
\fi
}%  SupFig

\renewcommand{\SkipFigure}[1]{}

\SUPPLEMENT{

\CLEARPAGE
 
\begin{figure*}[h!]
\FigureInLegends{
\noindent
(a) $p_c=0.205$

\includegraphics*[width=82mm,angle=0]{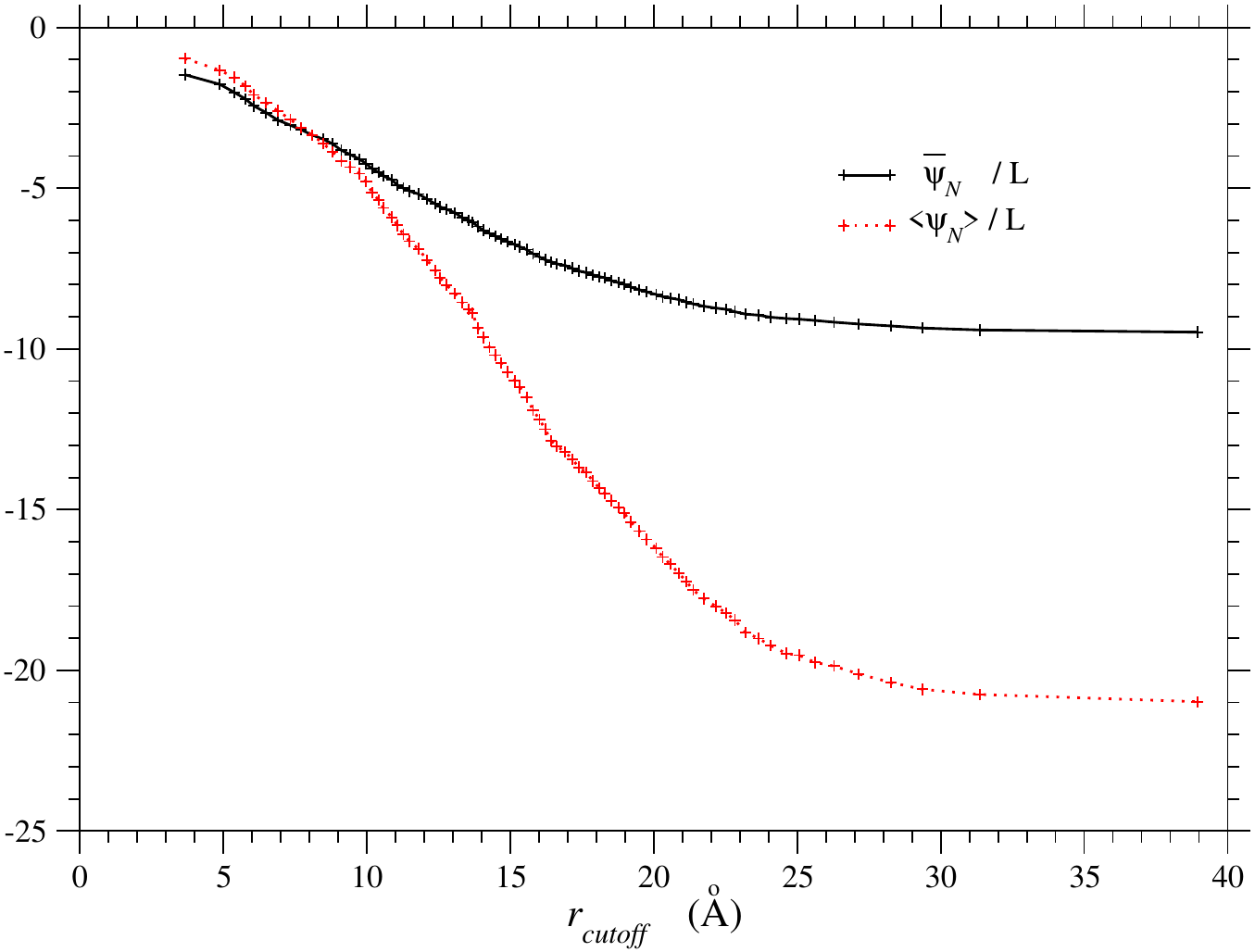}

{
\noindent
(b) $p_c=0.33$

\includegraphics*[width=82mm,angle=0]{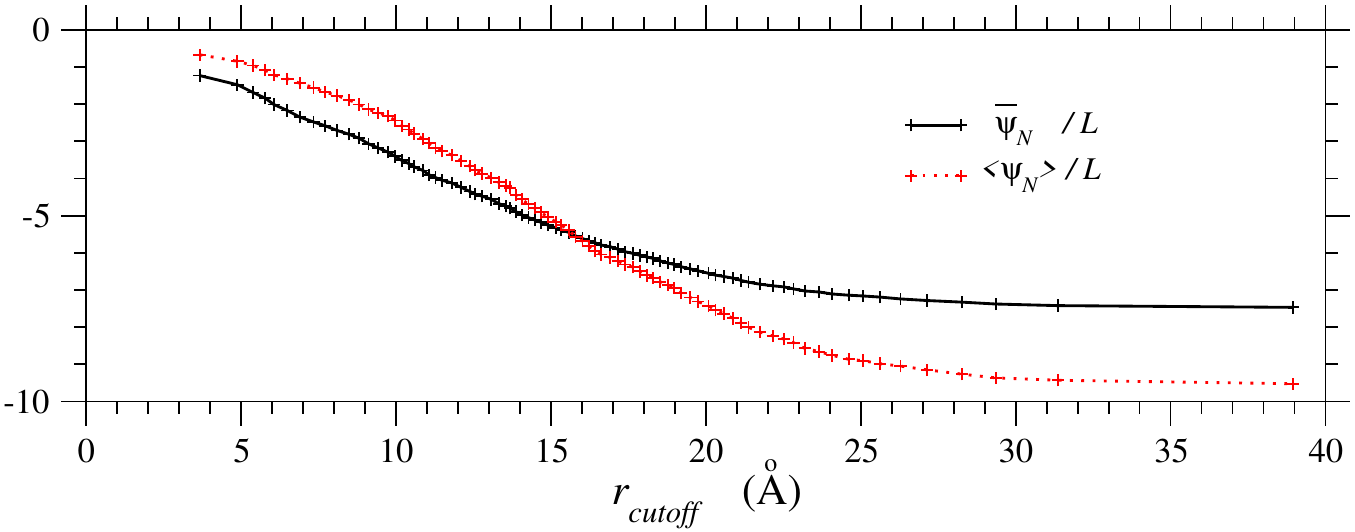}

\noindent
(c) $p_c=0.5$

\includegraphics*[width=82mm,angle=0]{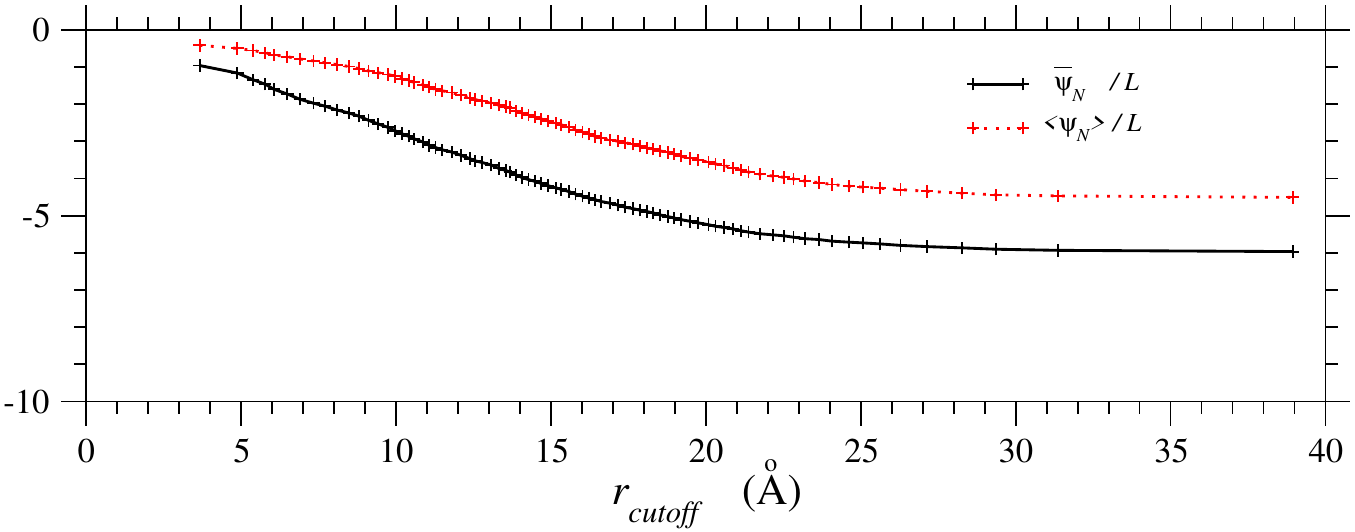}
}
}%  FigureInLegends
\vspace*{1em}
\caption{
\FigureLegends{
\label{sfig: 1gm1-a:16-96_full_non_del_dca0_205_0_20_simple-gauge_ensemble_vs_sample_ave_phi_vs_r}
\label{fig: 1gm1-a:16-96_full_non_del_dca0_205_0_20_simple-gauge_ensemble_vs_sample_ave_phi_vs_r}
\label{sfig: 1gm1-a:16-96_full_non_del_dca0_33_0_20_simple-gauge_ensemble_vs_sample_ave_phi_vs_r}
\label{fig: 1gm1-a:16-96_full_non_del_dca0_33_0_20_simple-gauge_ensemble_vs_sample_ave_phi_vs_r}
\label{sfig: 1gm1-a:16-96_full_non_del_dca0_5_0_20_simple-gauge_ensemble_vs_sample_ave_phi_vs_r}
\label{fig: 1gm1-a:16-96_full_non_del_dca0_5_0_20_simple-gauge_ensemble_vs_sample_ave_phi_vs_r}
\BF{
Dependences of
the sample ($\overline{\psi_N} / L$) and ensemble ($\langle \psi_N \rangle_{\VEC{\sigma}} / L$) averages of 
evolutionary statistical energy
per residue
on the cutoff distance for pairwise interactions 
in the PDZ domain.
}
The ratios of pseudocount $p_c = 0.205$ and $0.33$
are employed here
for the cutoff distance $r_{\script{cutoff}} \sim 8$ and $15.5$ \AA, respectively.  
The black solid and red dotted lines indicate the sample and ensemble averages, respectively.
}%  FigureLegends
}
\end{figure*}

\CLEARPAGE
 
\begin{figure*}[h!]
\FigureInLegends{
\noindent
\hspace*{1em} (a) $p_c=0.205$	\hspace*{14em} (b) $p_c=0.33$

\centerline{
\includegraphics*[width=82mm,angle=0]{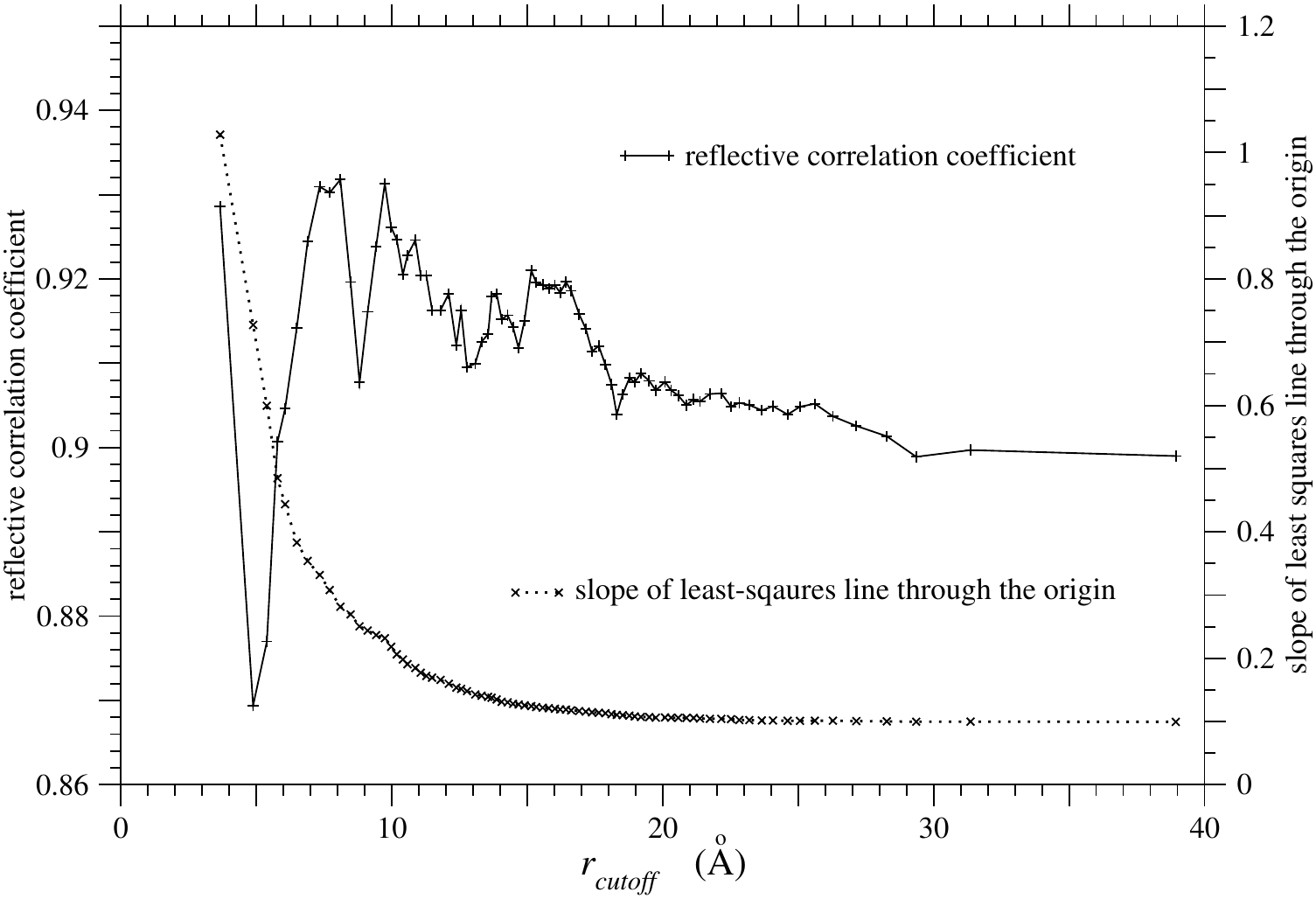}
\includegraphics*[width=82mm,angle=0]{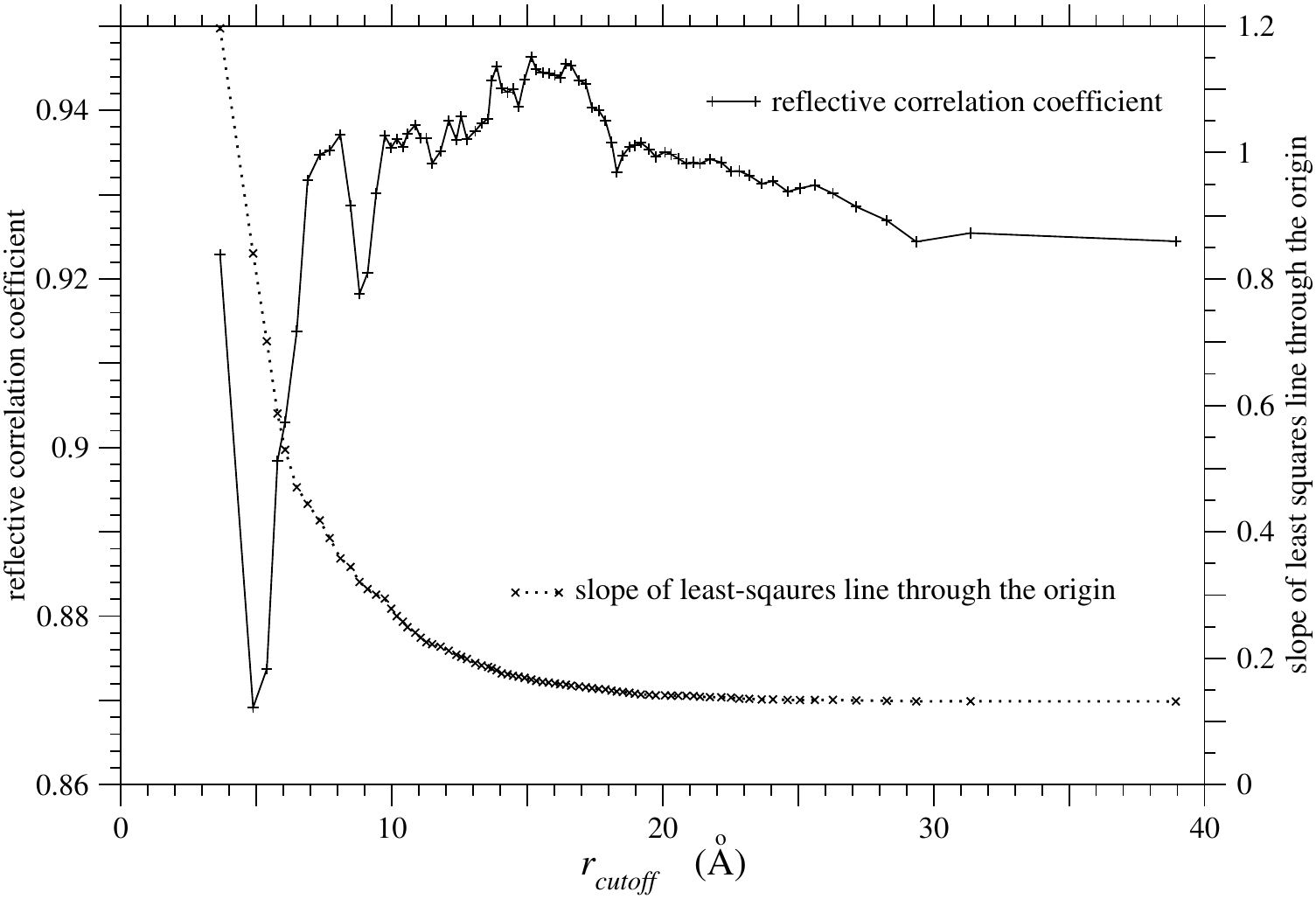}
}
}%  FigureInLegends
\vspace*{1em}
\caption{
\FigureLegends{
\label{sfig: 1gm1-a:16-96_full_non_del_dca0_205_0_20_simple-gauge_ddG-dPhi_vs_r}
\label{fig: 1gm1-a:16-96_full_non_del_dca0_205_0_20_simple-gauge_ddG-dPhi_vs_r}
\label{sfig: 1gm1-a:16-96_full_non_del_dca0_33_0_20_simple-gauge_ddG-dPhi_vs_r}
\label{fig: 1gm1-a:16-96_full_non_del_dca0_33_0_20_simple-gauge_ddG-dPhi_vs_r}
\BF{
Dependences of
the reflective correlation and regression coefficients
between the experimental $\Delta\Delta G_{ND}$\CITE{GGCJVTVB:07} 
and $\Delta \psi_N$ due to single amino acid substitutions
on the cutoff distance for pairwise interactions in the PDZ domain.
}
The left and right figures are for the ratios of pseudocount, $p_c=0.205$ and $0.33$, respectively.
The solid and dotted lines show the reflective correlation and 
regression coefficients
for the least-squares regression line through the origin, respectively.
The sample ($\overline{\psi_N} / L$) and ensemble ($\langle \psi_N \rangle_{\VEC{\sigma}} / L$) averages of 
evolutionary statistical energy
agree with each other at the cutoff distance $r_{\script{cutoff}} \sim 8$ \AA\  for $p_c = 0.205$ 
and $r_{\script{cutoff}} \sim 15.5$ \AA\  for $p_c = 0.33$,
where the reflective correlation coefficients attain to the maximum.
}%  FigureLegends
}
\end{figure*}

}%  SUPPLEMENT

\SUPPLEMENT{

\CLEARPAGE
\begin{figure*}[h!]
\FigureInLegends{
\noindent
\hspace*{1em} (a) $r_{\script{cutoff}} \sim 8$ \AA\  \hspace*{14em} (b) $r_{\script{cutoff}} \sim 15.5$ \AA\ 

\centerline{
\includegraphics*[width=82mm,angle=0]{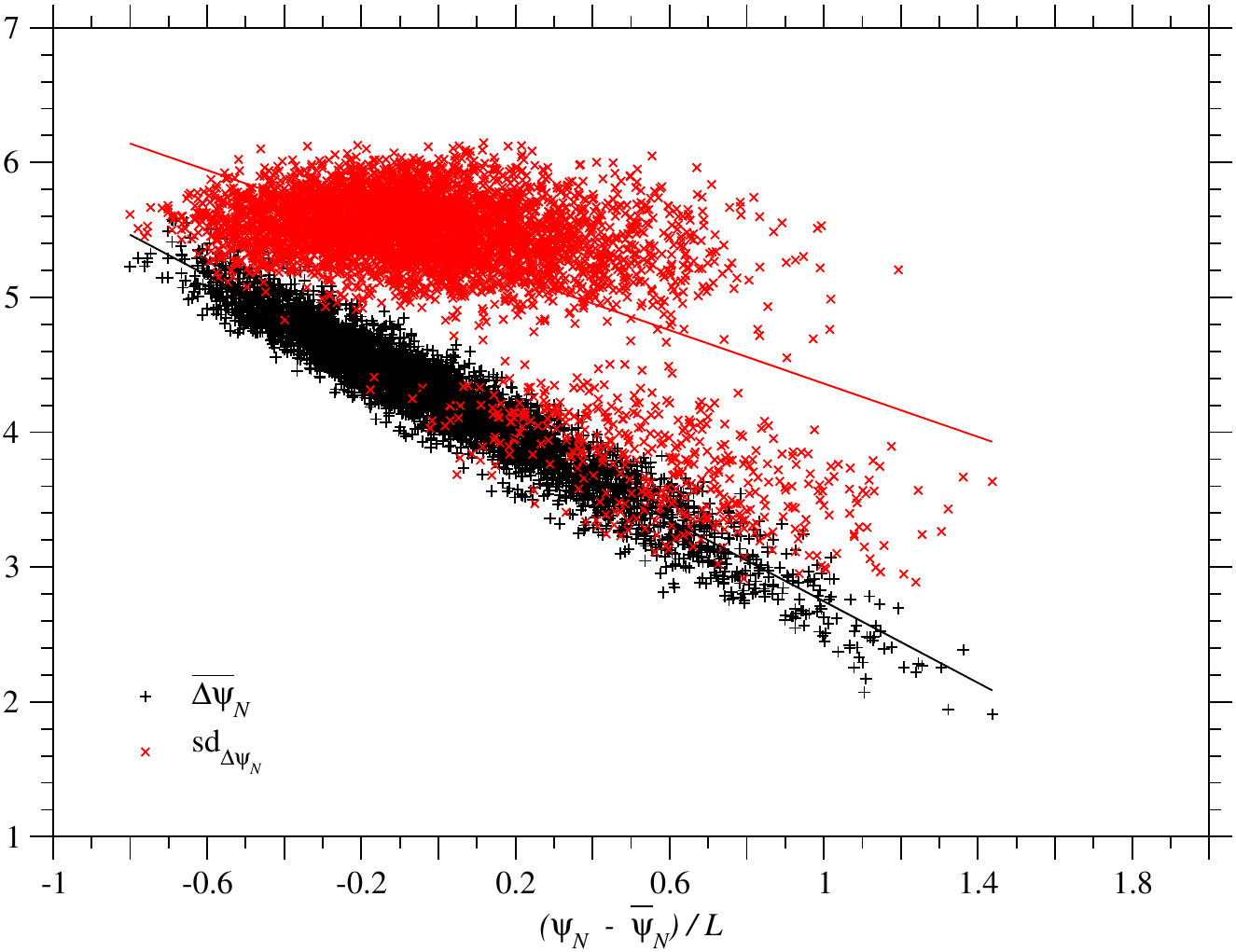}
\includegraphics*[width=82mm,angle=0]{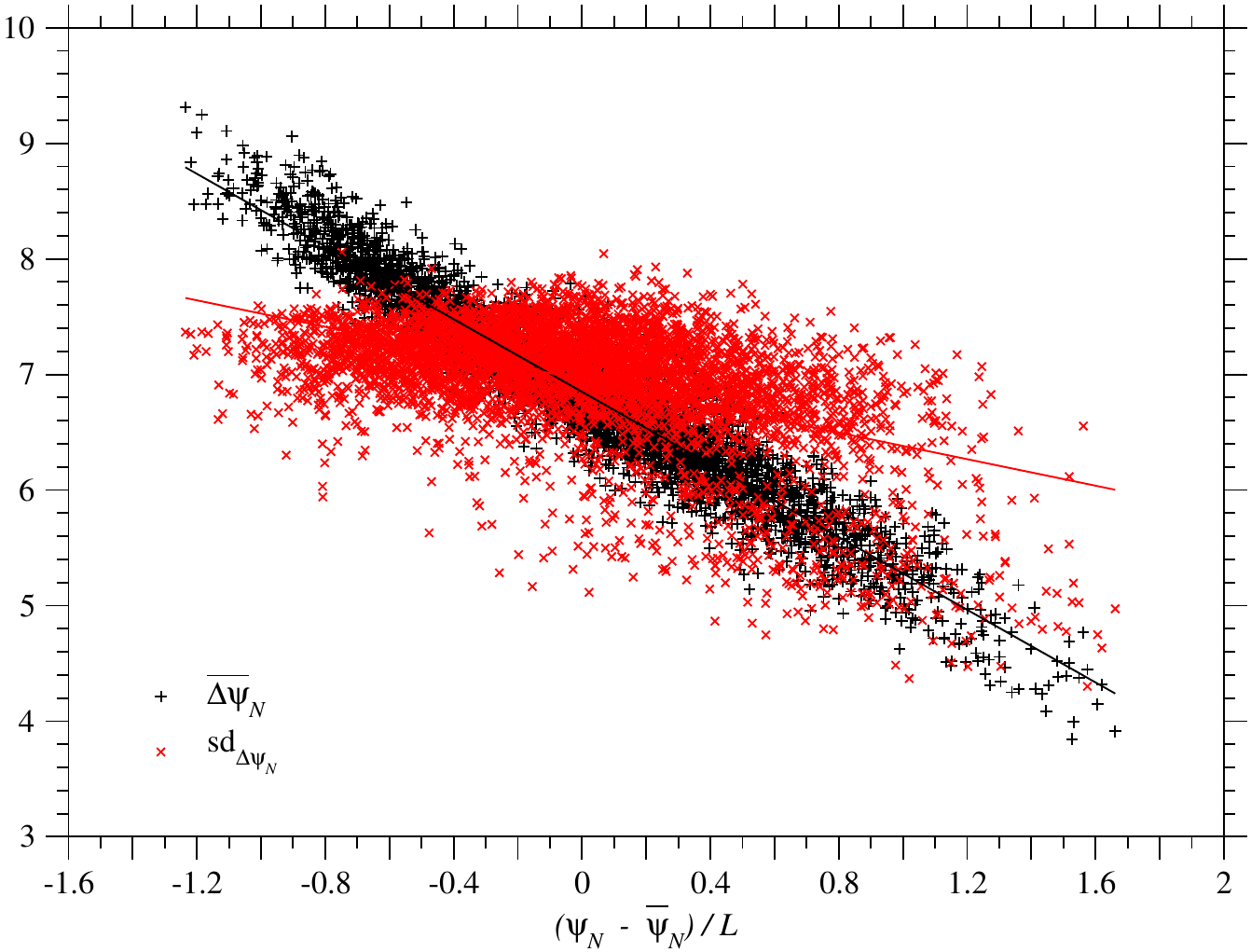}
}
}%  FigureInLegends
\vspace*{1em}
\caption{
\FigureLegends{
\label{sfig: 1r69-a:6-58.full_non_del.dca0_18.0_20.simple-gauge.ddPhi_at_opt}
\label{fig: 1r69-a:6-58.full_non_del.dca0_18.0_20.simple-gauge.ddPhi_at_opt}
\BF{
Correlation between $\Delta \psi_N$ due to single nucleotide nonsynonymous substitutions and
$\psi_N$ of homologous sequences in the HTH\_3 family of the domain, 1R69-A:6-58.
}
\protect
The left and right figures correspond to the cutoff distance $r_{\script{cutoff}} \sim 8$ and $15.5$ \AA,
respectively.
Each of the black plus or red cross marks corresponds to the mean or the standard deviation 
of $\Delta \psi_N$ due to
all types of single nucleotide nonsynonymous substitutions
over all sites in each of the homologous sequences.
Representatives of unique sequences,
which are at least 20\% different from each other, are employed;
the number of the representatives is almost equal to $M_{\script{eff}}$ in \Table{\ref{\TBL: Proteins_studied}}.
The solid lines show the regression lines for the mean and the standard deviation of $\Delta \psi_N$.
% End of figures_mean_and_sd_legends.tex
}%  FigureLegends
}
\end{figure*}

\CLEARPAGE
\begin{figure*}[h!]
\FigureInLegends{
\noindent
\hspace*{1em} (a) $r_{\script{cutoff}} \sim 8$ \AA\  \hspace*{14em} (b) $r_{\script{cutoff}} \sim 15.5$ \AA\ 

\centerline{
\includegraphics*[width=82mm,angle=0]{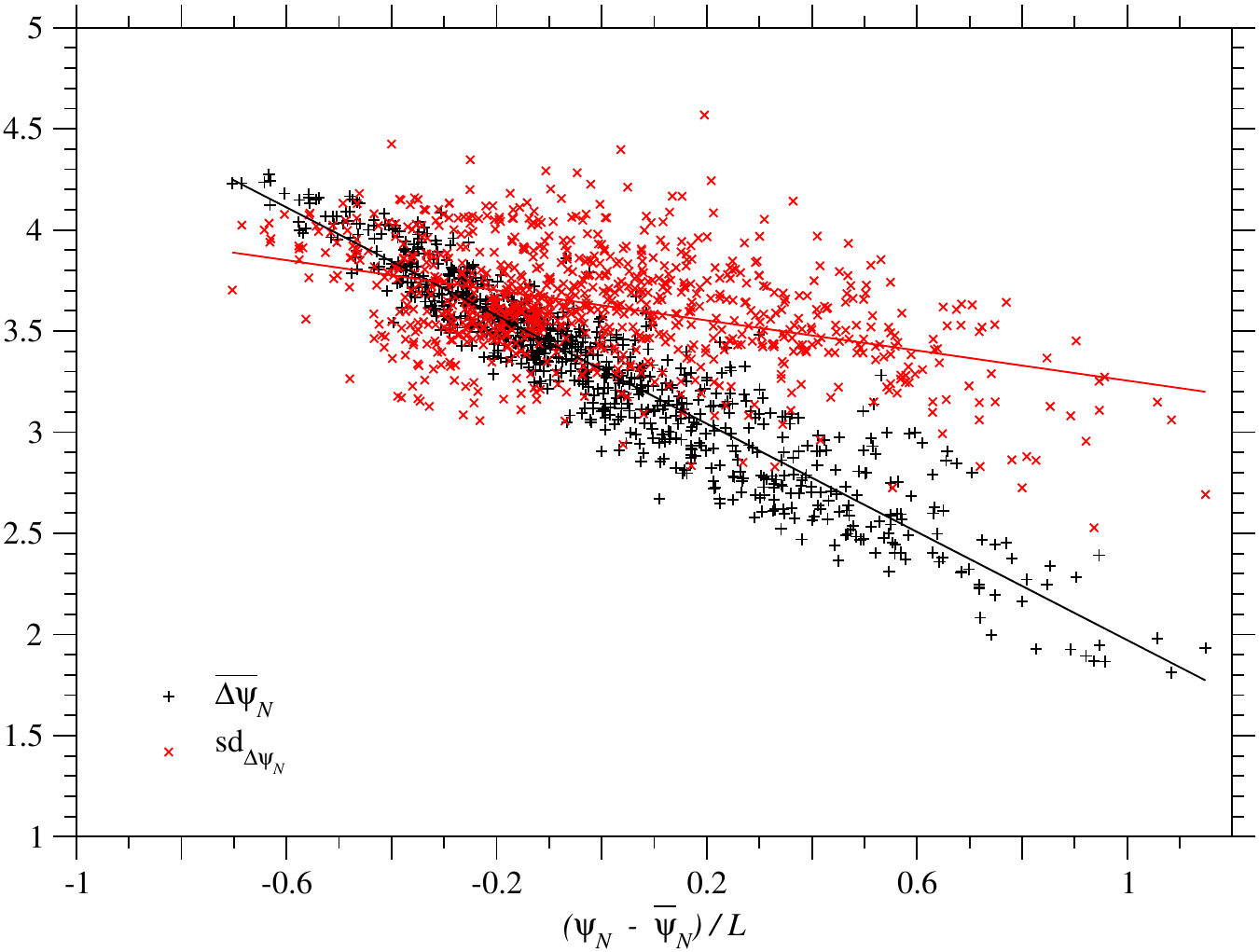}
\includegraphics*[width=82mm,angle=0]{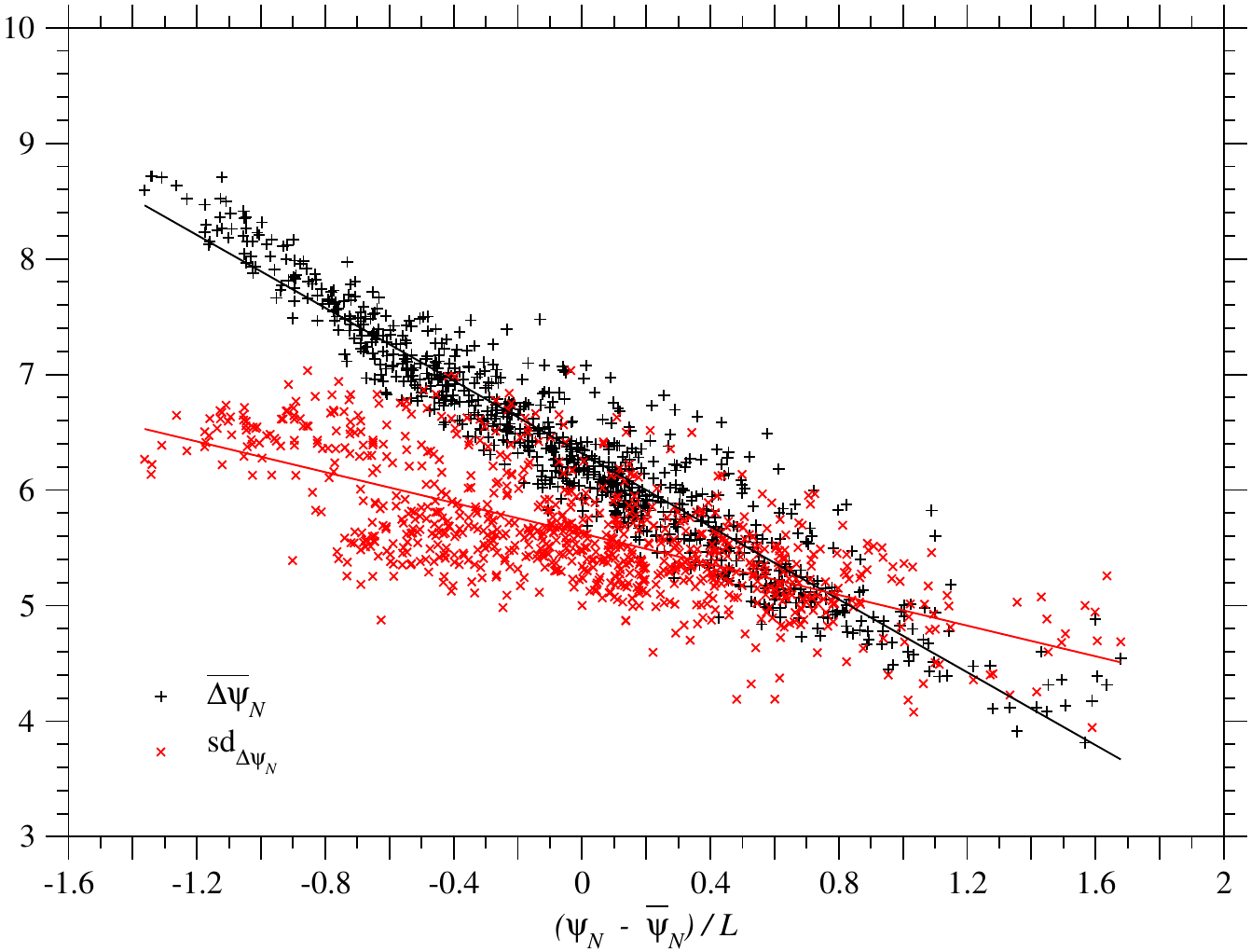}
}
}%  FigureInLegends
\vspace*{1em}
\caption{
\FigureLegends{
\label{sfig: 3e10:4-76.full_non_del.dca0_23.0_20.simple-gauge.AB.ddPhi_at_opt}
\label{fig: 3e10:4-76.full_non_del.dca0_23.0_20.simple-gauge.AB.ddPhi_at_opt}
\BF{
Correlation between $\Delta \psi_N$ due to single nucleotide nonsynonymous substitutions and
$\psi_N$ of homologous sequences in the Nitroreductase family of the domain, 3E10-A/B:4-76.
}
\protect
The left and right figures correspond to the cutoff distance $r_{\script{cutoff}} \sim 8$ and $15.5$ \AA,
respectively.
Each of the black plus or red cross marks corresponds to the mean or the standard deviation 
of $\Delta \psi_N$ due to
all types of single nucleotide nonsynonymous substitutions
over all sites in each of the homologous sequences.
Representatives of unique sequences,
which are at least 20\% different from each other, are employed;
the number of the representatives is almost equal to $M_{\script{eff}}$ in \Table{\ref{\TBL: Proteins_studied}}.
The solid lines show the regression lines for the mean and the standard deviation of $\Delta \psi_N$.
% End of figures_mean_and_sd_legends.tex
}%  FigureLegends
}
\end{figure*}

\CLEARPAGE
\begin{figure*}[h!]
\FigureInLegends{
\noindent
\hspace*{1em} (a) $r_{\script{cutoff}} \sim 8$ \AA\  \hspace*{14em} (b) $r_{\script{cutoff}} \sim 15.5$ \AA\ 

\centerline{
\includegraphics*[width=82mm,angle=0]{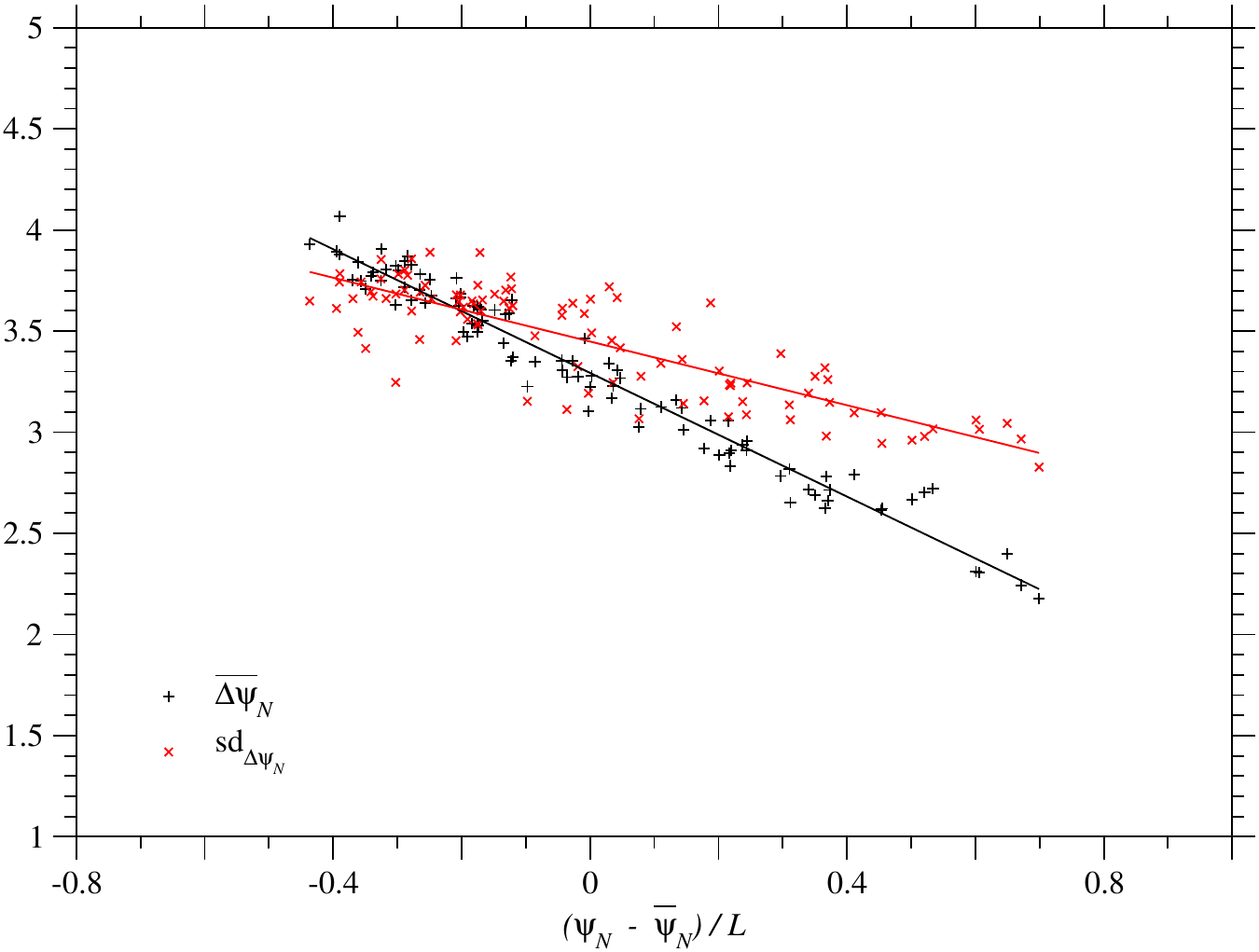}
\includegraphics*[width=82mm,angle=0]{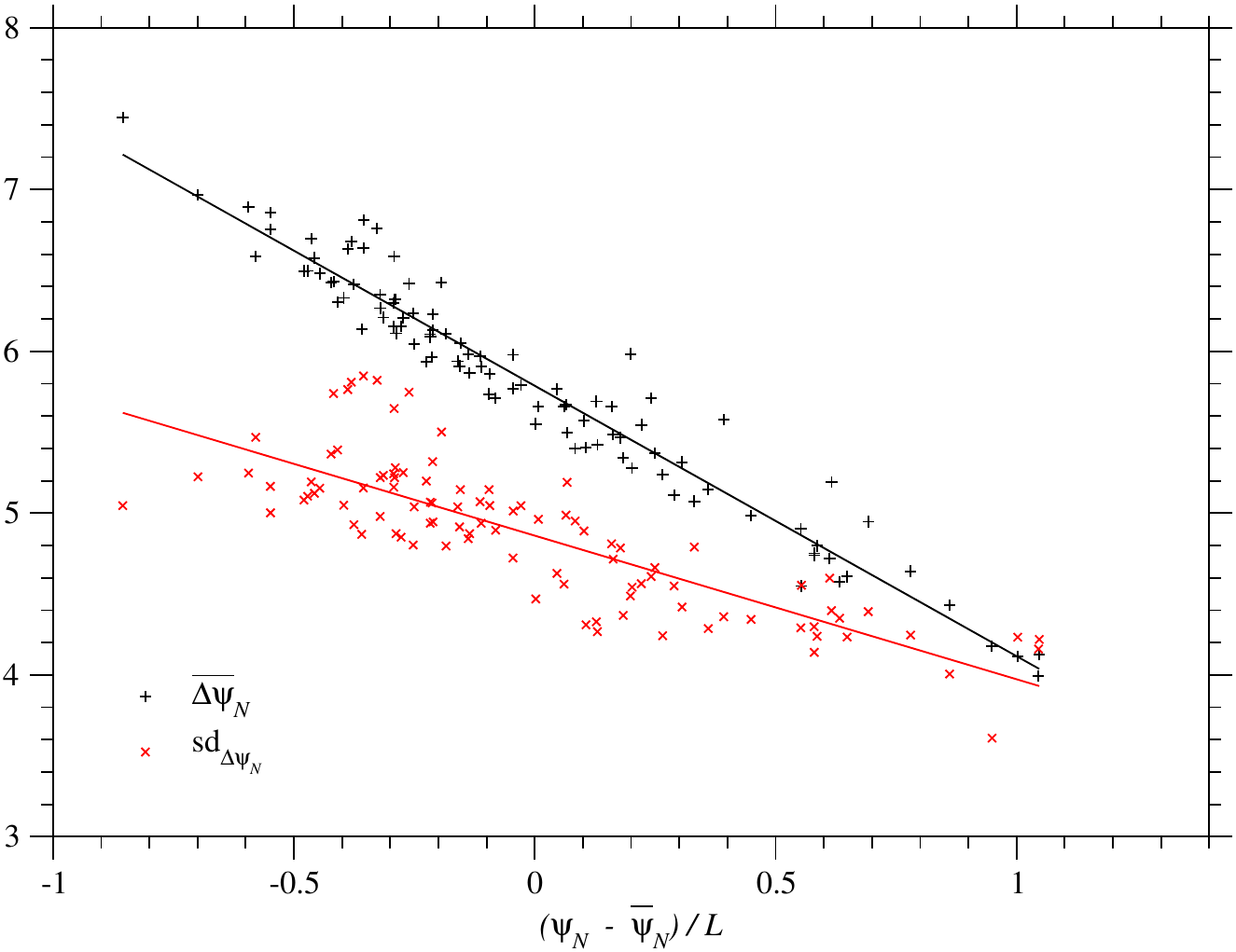}
}
\centerline{
\includegraphics*[width=82mm,angle=0]{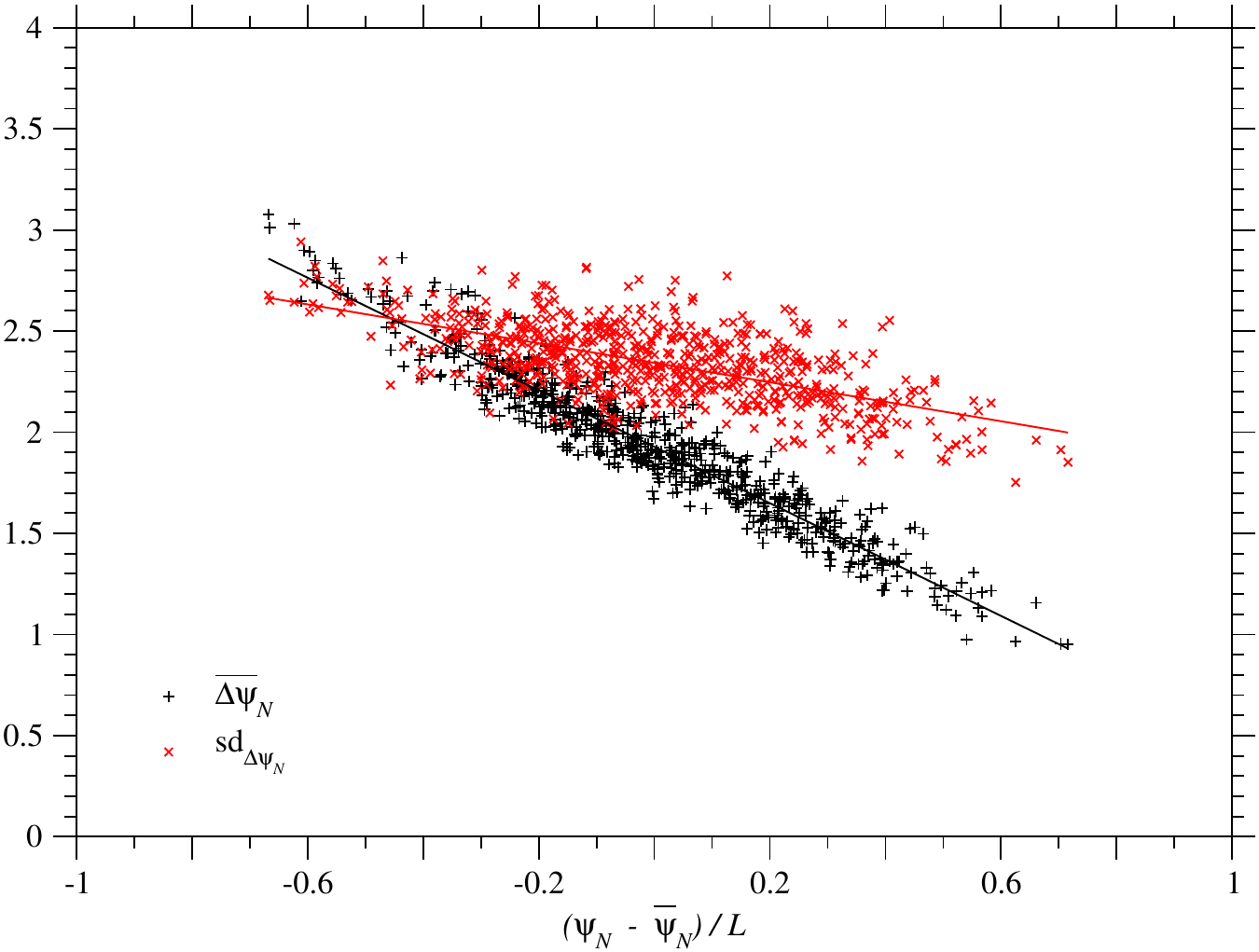}
\includegraphics*[width=82mm,angle=0]{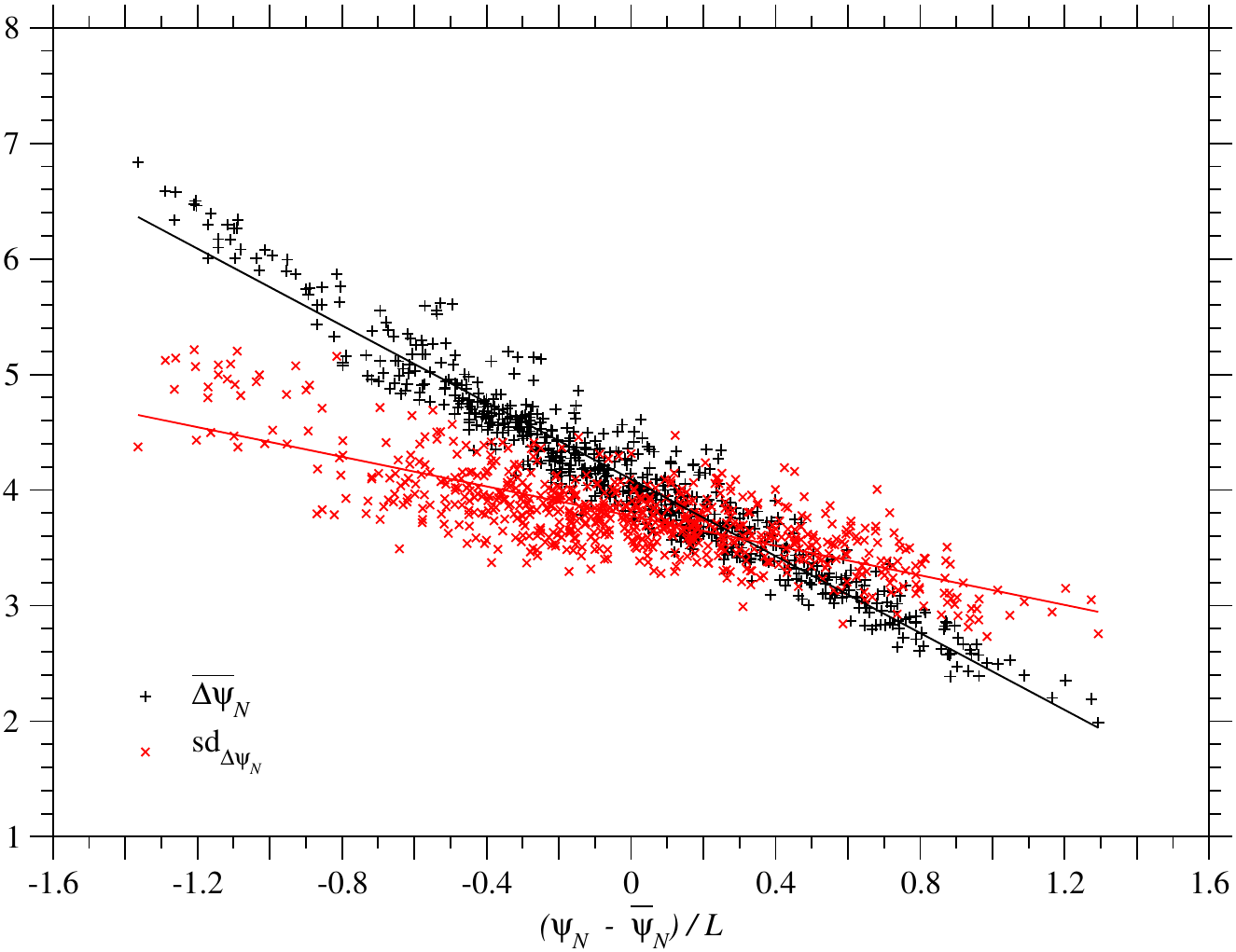}
}
}%  FigureInLegends
\vspace*{1em}
\caption{
\FigureLegends{
\label{sfig: 1wdn-a:5-222.full_non_del.dca0_25.0_20.simple-gauge.ddPhi_at_opt}
\label{fig: 1wdn-a:5-222.full_non_del.dca0_25.0_20.simple-gauge.ddPhi_at_opt}
\BF{
Correlation between $\Delta \psi_N$ due to single nucleotide nonsynonymous substitutions and
$\psi_N$ of homologous sequences in the SBP\_bac\_3 family of the domains, 1WDN-A:5-222 (upper) and 1WDN-A:89-182 (lower).
}
\protect
The left and right figures correspond to the cutoff distance $r_{\script{cutoff}} \sim 8$ and $15.5$ \AA,
respectively.
Each of the black plus or red cross marks corresponds to the mean or the standard deviation 
of $\Delta \psi_N$ due to
all types of single nucleotide nonsynonymous substitutions
over all sites in each of the homologous sequences.
Representatives of unique sequences,
which are at least 20\% different from each other, are employed;
the number of the representatives is almost equal to $M_{\script{eff}}$ in \Table{\ref{\TBL: Proteins_studied}}.
The solid lines show the regression lines for the mean and the standard deviation of $\Delta \psi_N$.
% End of figures_mean_and_sd_legends.tex
}%  FigureLegends
}
\end{figure*}

\CLEARPAGE
\begin{figure*}[h!]
\FigureInLegends{
\noindent
\hspace*{1em} (a) $r_{\script{cutoff}} \sim 8$ \AA\  \hspace*{14em} (b) $r_{\script{cutoff}} \sim 15.5$ \AA\ 

\centerline{
\includegraphics*[width=82mm,angle=0]{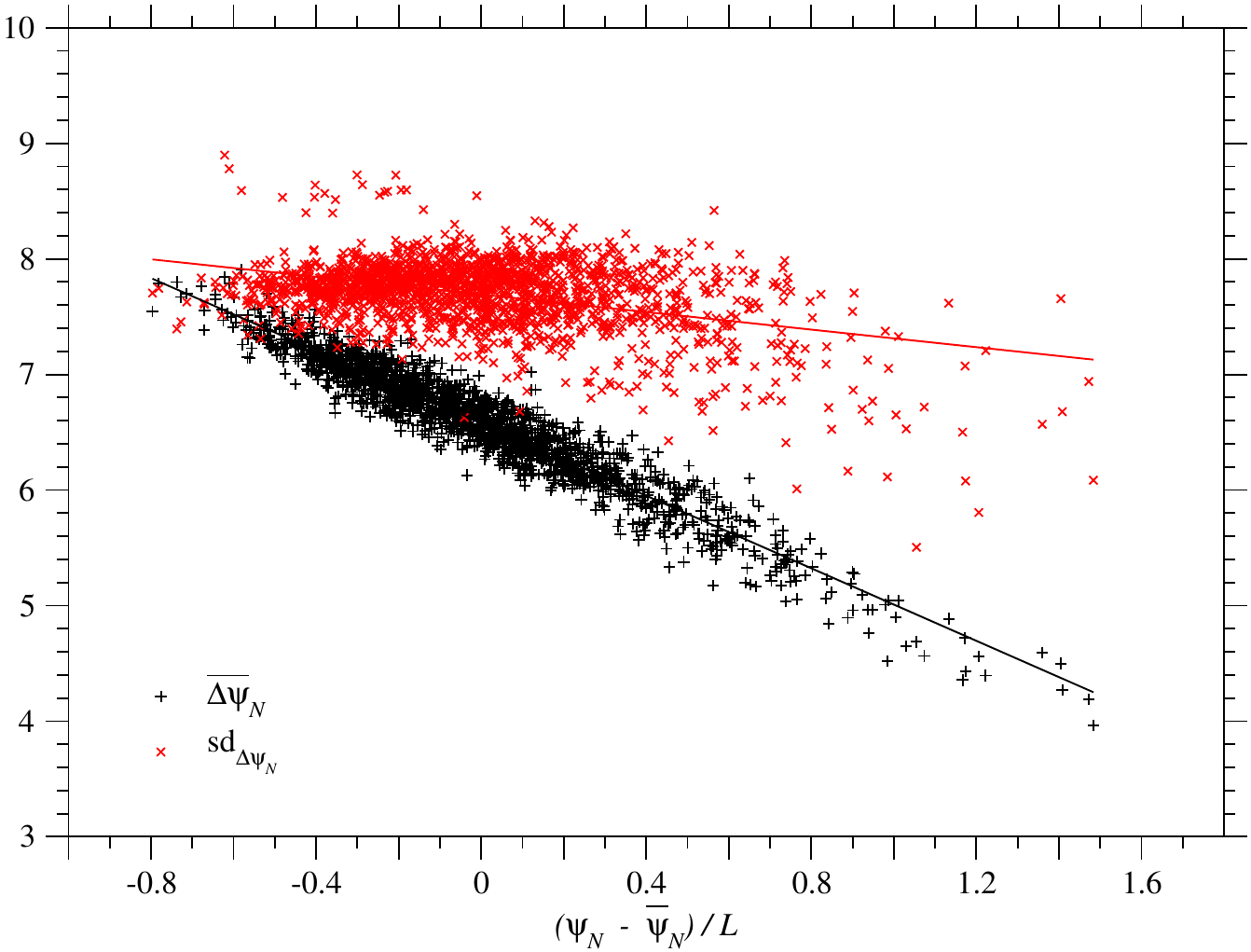}
\includegraphics*[width=82mm,angle=0]{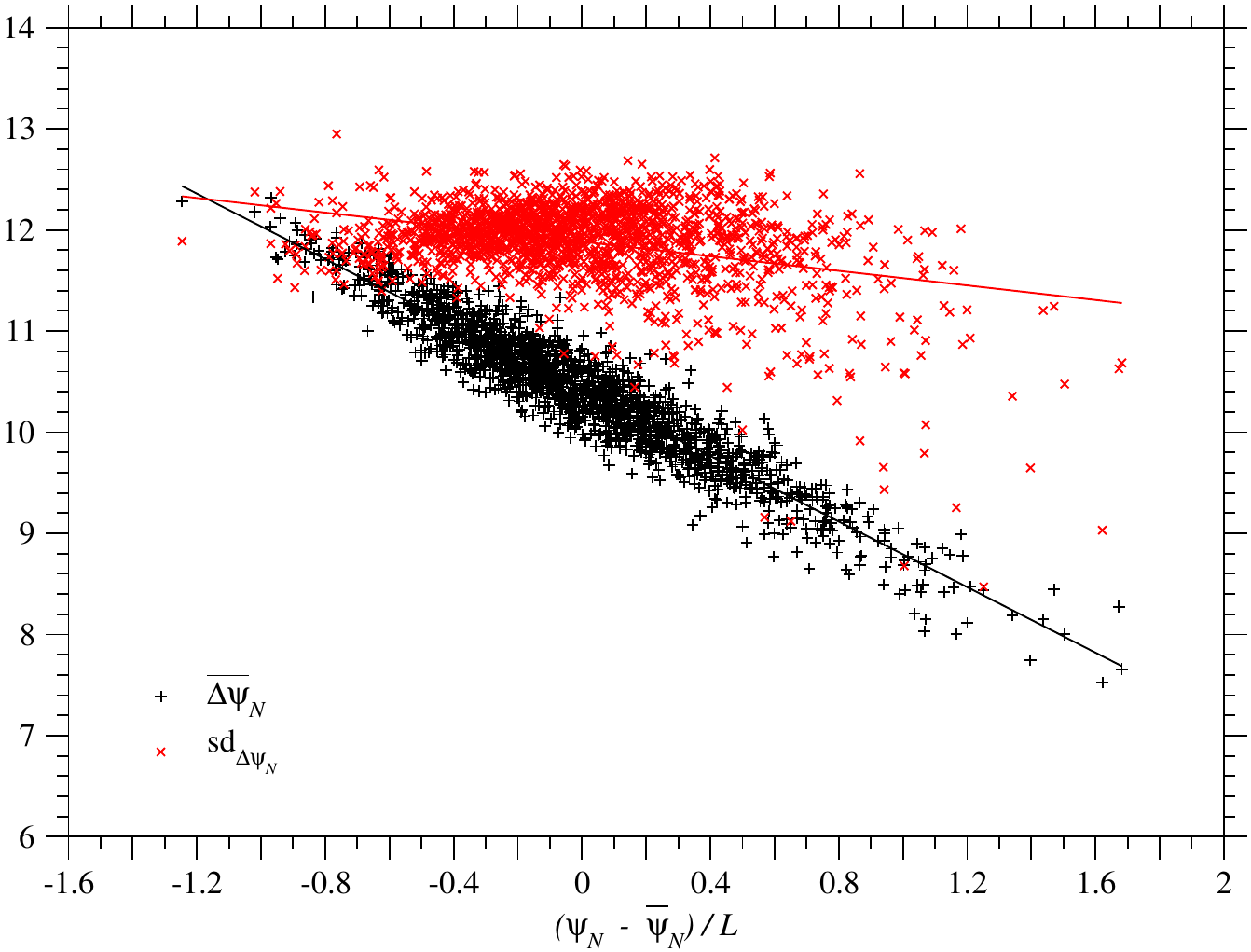}
}
}%  FigureInLegends
\vspace*{1em}
\caption{
\FigureLegends{
\label{sfig: 1oap-a:52-146.full_non_del.dca0_169.0_20.simple-gauge.ddPhi_at_opt}
\label{fig: 1oap-a:52-146.full_non_del.dca0_169.0_20.simple-gauge.ddPhi_at_opt}
\BF{
Correlation between $\Delta \psi_N$ due to single nucleotide nonsynonymous substitutions and
$\psi_N$ of homologous sequences in the OmpA family of the domains, 1OAP-A:52-146.
}
\protect
The left and right figures correspond to the cutoff distance $r_{\script{cutoff}} \sim 8$ and $15.5$ \AA,
respectively.
Each of the black plus or red cross marks corresponds to the mean or the standard deviation 
of $\Delta \psi_N$ due to
all types of single nucleotide nonsynonymous substitutions
over all sites in each of the homologous sequences.
Representatives of unique sequences,
which are at least 20\% different from each other, are employed;
the number of the representatives is almost equal to $M_{\script{eff}}$ in \Table{\ref{\TBL: Proteins_studied}}.
The solid lines show the regression lines for the mean and the standard deviation of $\Delta \psi_N$.
% End of figures_mean_and_sd_legends.tex
}%  FigureLegends
}
\end{figure*}

\CLEARPAGE
\begin{figure*}[h!]
\FigureInLegends{
\noindent
\hspace*{1em} (a) $r_{\script{cutoff}} \sim 8$ \AA\  \hspace*{14em} (b) $r_{\script{cutoff}} \sim 15.5$ \AA\ 

\centerline{
\includegraphics*[width=82mm,angle=0]{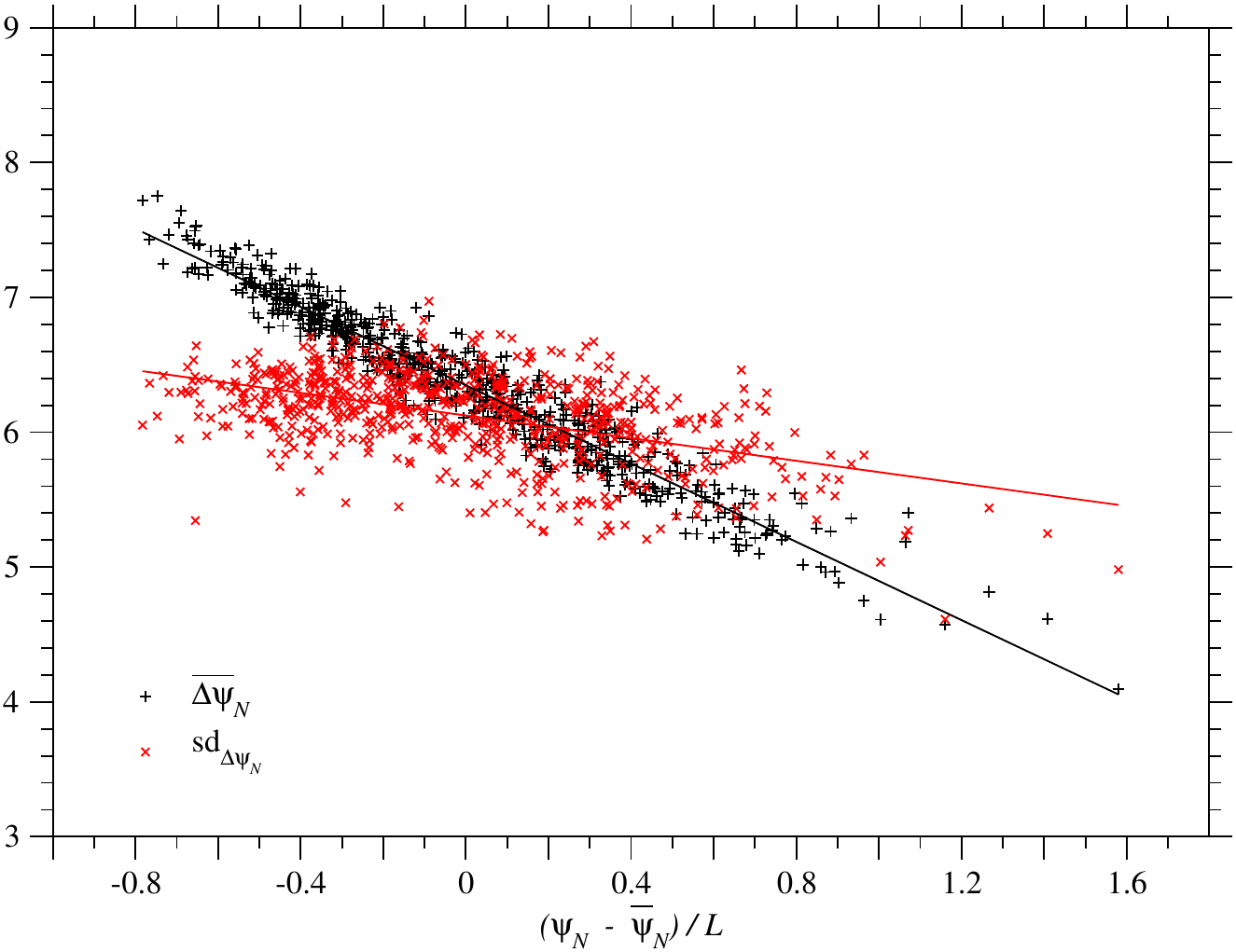}
\includegraphics*[width=82mm,angle=0]{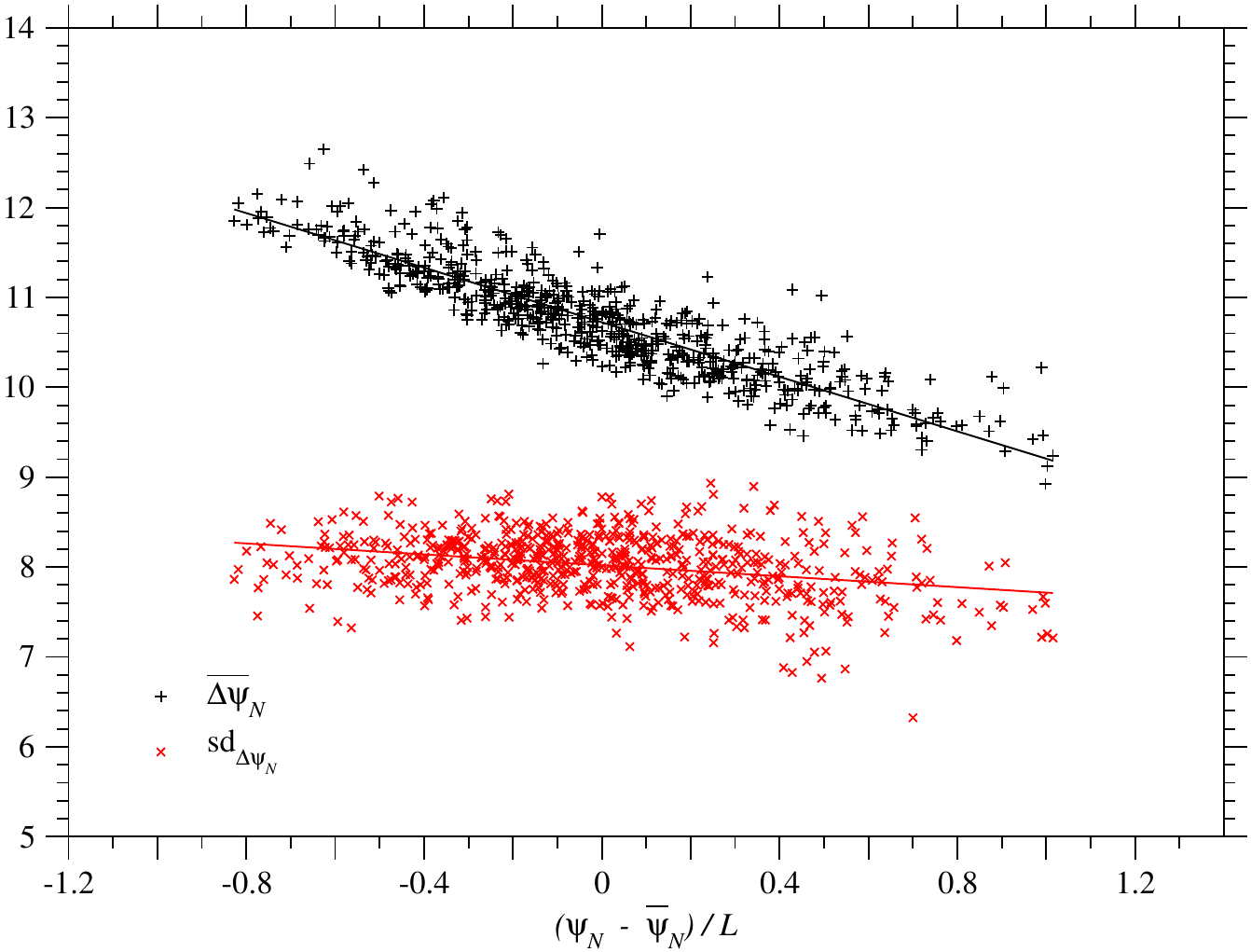}
}
}%  FigureInLegends
\vspace*{1em}
\caption{
\FigureLegends{
\label{sfig: 1jwe-a:30-127.full_non_del.dca0_235.0_20.simple-gauge.ddPhi_at_opt}
\label{fig: 1jwe-a:30-127.full_non_del.dca0_235.0_20.simple-gauge.ddPhi_at_opt}
\BF{
Correlation between $\Delta \psi_N$ due to single nucleotide nonsynonymous substitutions and
$\psi_N$ of homologous sequences in the DnaB family of the domains, 1JWE-A:30-127.
}
\protect
The left and right figures correspond to the cutoff distance $r_{\script{cutoff}} \sim 8$ and $15.5$ \AA,
respectively.
Each of the black plus or red cross marks corresponds to the mean or the standard deviation 
of $\Delta \psi_N$ due to
all types of single nucleotide nonsynonymous substitutions
over all sites in each of the homologous sequences.
Representatives of unique sequences,
which are at least 20\% different from each other, are employed;
the number of the representatives is almost equal to $M_{\script{eff}}$ in \Table{\ref{\TBL: Proteins_studied}}.
The solid lines show the regression lines for the mean and the standard deviation of $\Delta \psi_N$.
% End of figures_mean_and_sd_legends.tex
}%  FigureLegends
}
\end{figure*}

\CLEARPAGE
\begin{figure*}[h!]
\FigureInLegends{
\noindent
\hspace*{1em} (a) $r_{\script{cutoff}} \sim 8$ \AA\  \hspace*{14em} (b) $r_{\script{cutoff}} \sim 15.5$ \AA\ 

\centerline{
\includegraphics*[width=82mm,angle=0]{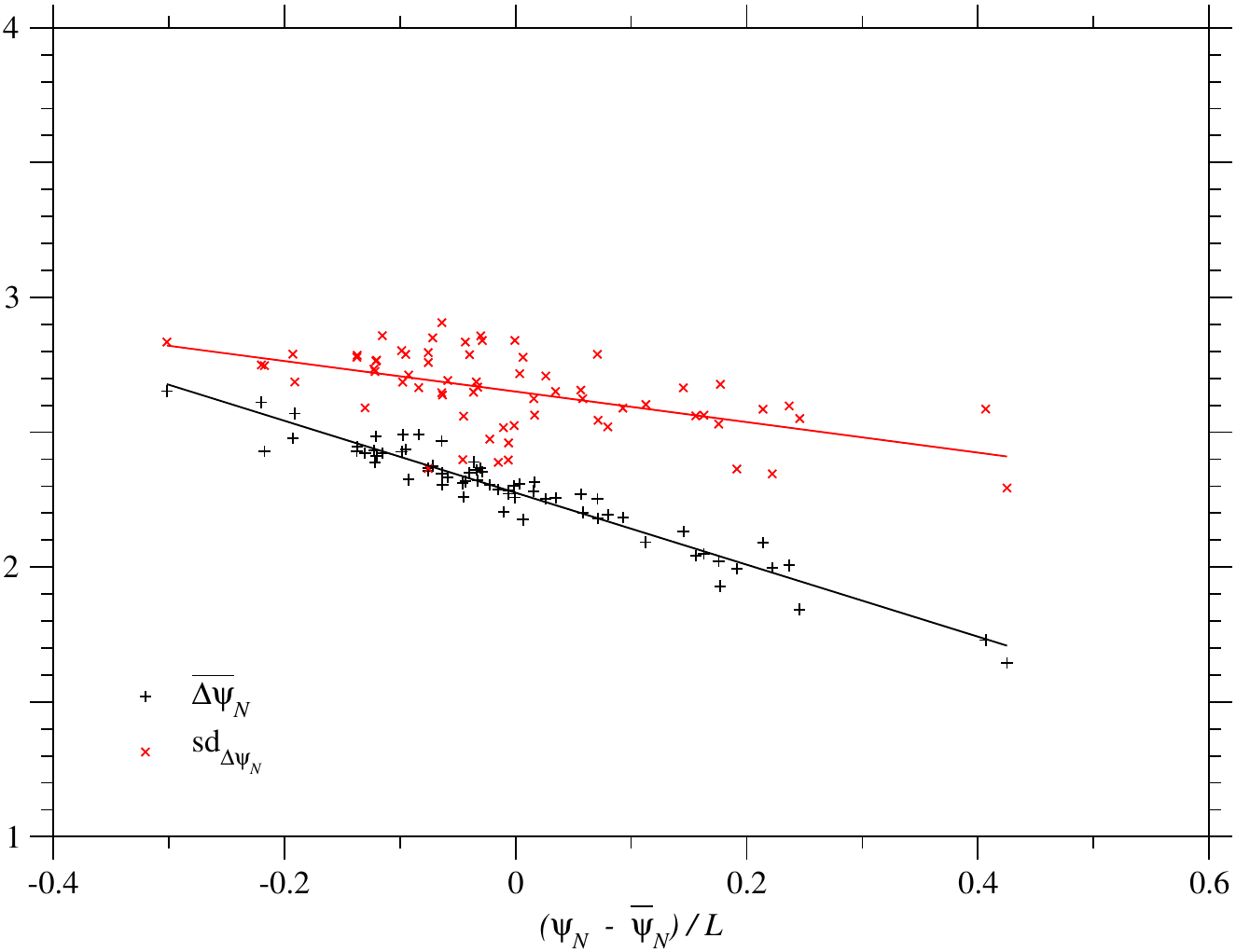}
\includegraphics*[width=82mm,angle=0]{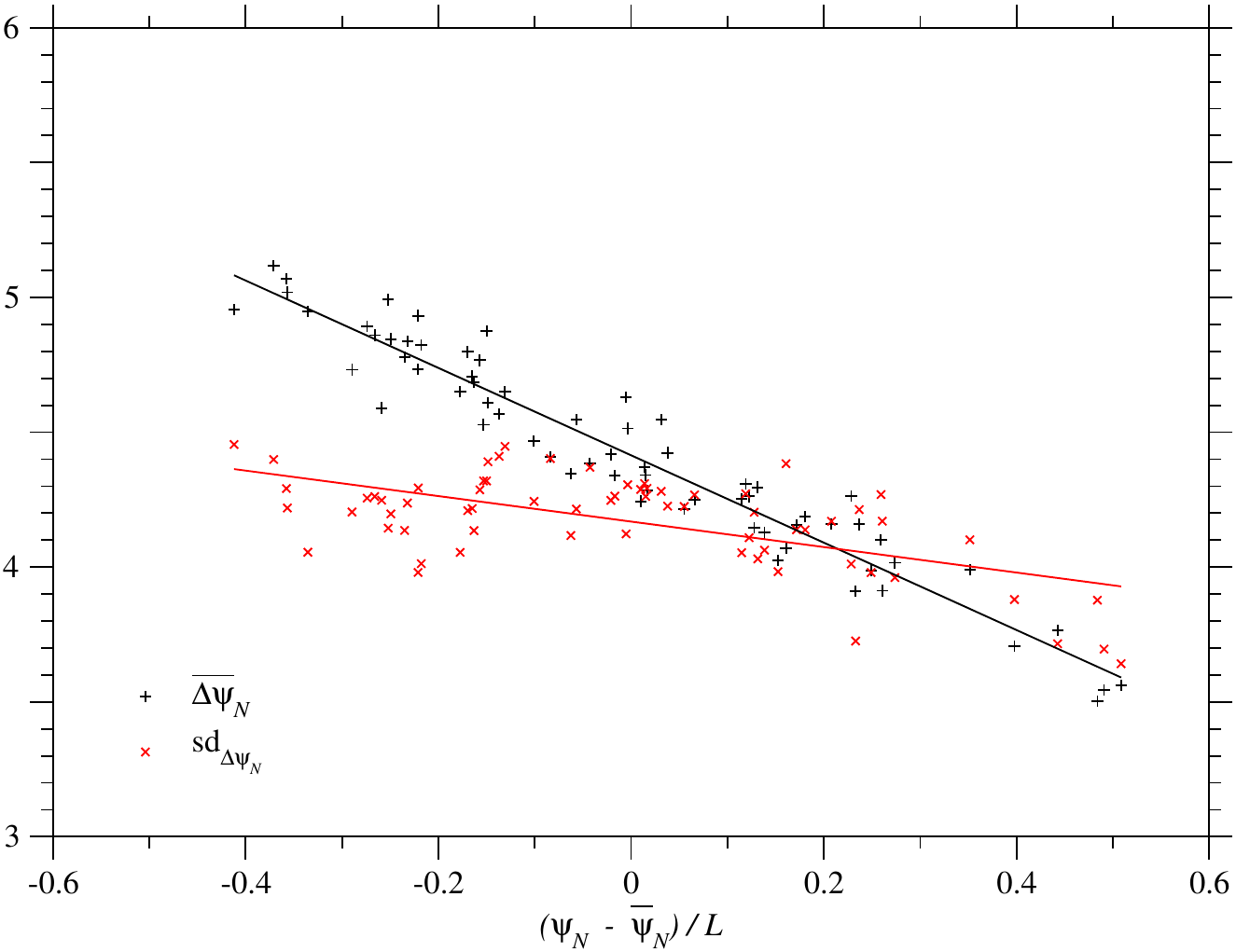}
}
\centerline{
\includegraphics*[width=82mm,angle=0]{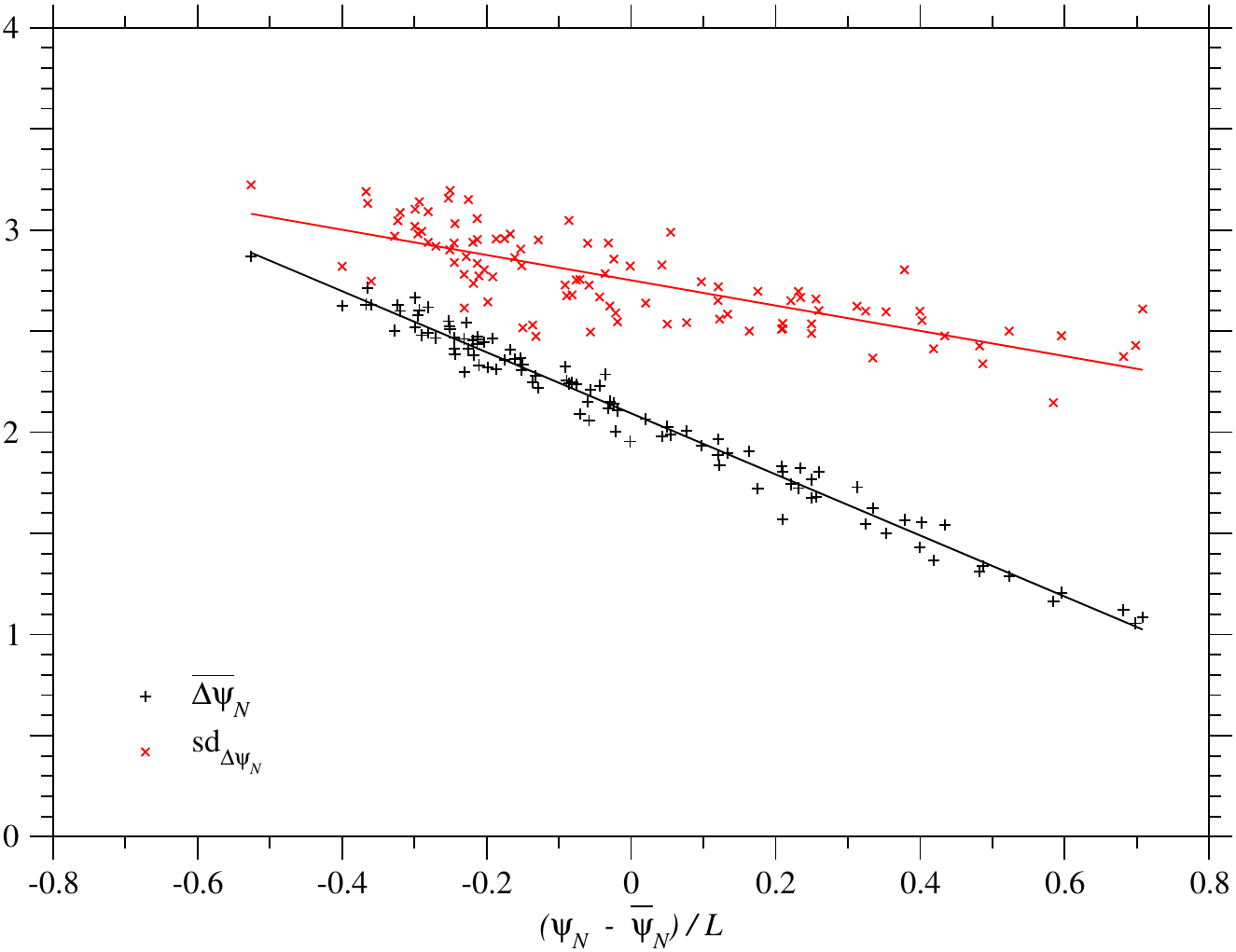}
\includegraphics*[width=82mm,angle=0]{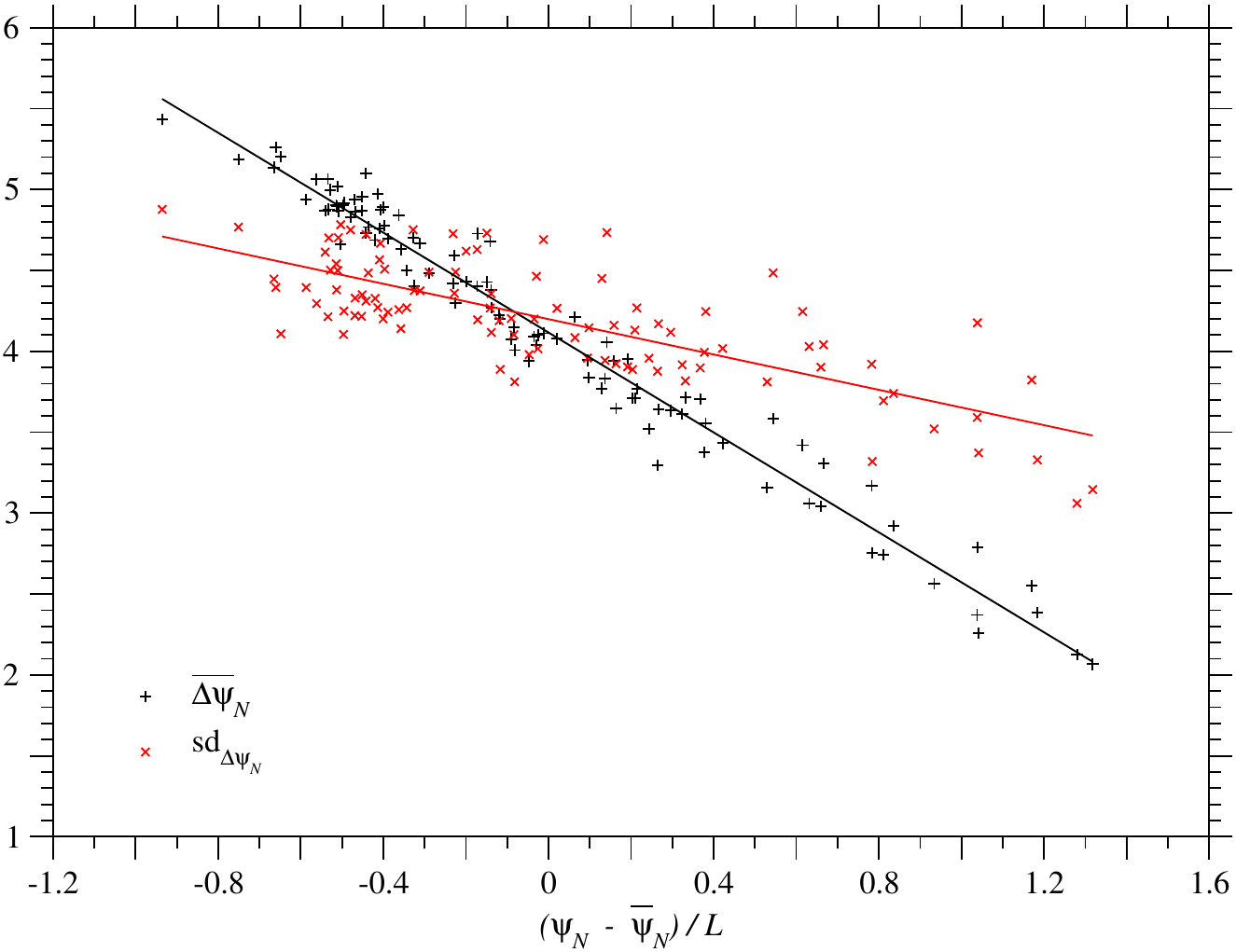}
}
}%  FigureInLegends
\vspace*{1em}
\caption{
\FigureLegends{
\label{sfig: 2f6g:90-280.full_non_del.dca0_235.0_20.simple-gauge.AB.ddPhi_at_opt}
\label{fig: 2f6g:90-280.full_non_del.dca0_235.0_20.simple-gauge.AB.ddPhi_at_opt}
\BF{
Correlation between $\Delta \psi_N$ due to single nucleotide nonsynonymous substitutions and
$\psi_N$ of homologous sequences in the LysR\_substrate family of the domains, 2F6G-A:90-280 (above) and 2F6G-A:163-265 (below).
}
\protect
The left and right figures correspond to the cutoff distance $r_{\script{cutoff}} \sim 8$ and $15.5$ \AA,
respectively.
Each of the black plus or red cross marks corresponds to the mean or the standard deviation 
of $\Delta \psi_N$ due to
all types of single nucleotide nonsynonymous substitutions
over all sites in each of the homologous sequences.
Representatives of unique sequences,
which are at least 20\% different from each other, are employed;
the number of the representatives is almost equal to $M_{\script{eff}}$ in \Table{\ref{\TBL: Proteins_studied}}.
The solid lines show the regression lines for the mean and the standard deviation of $\Delta \psi_N$.
% End of figures_mean_and_sd_legends.tex
}%  FigureLegends
}
\end{figure*}

\CLEARPAGE
\begin{figure*}[h!]
\FigureInLegends{
\noindent
\hspace*{1em} (a) $r_{\script{cutoff}} \sim 8$ \AA\  \hspace*{14em} (b) $r_{\script{cutoff}} \sim 15.5$ \AA\ 

\centerline{
\includegraphics*[width=82mm,angle=0]{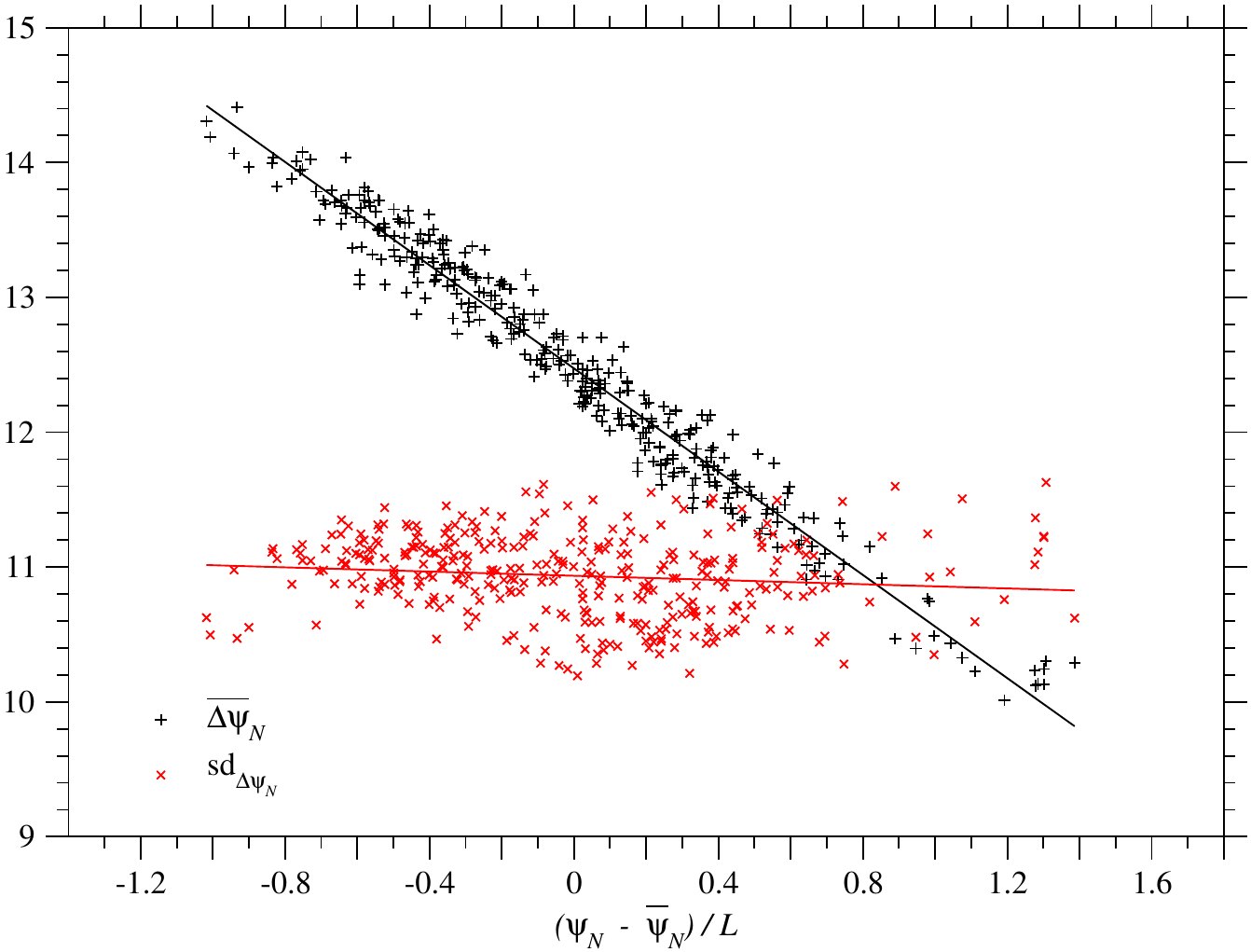}
\includegraphics*[width=82mm,angle=0]{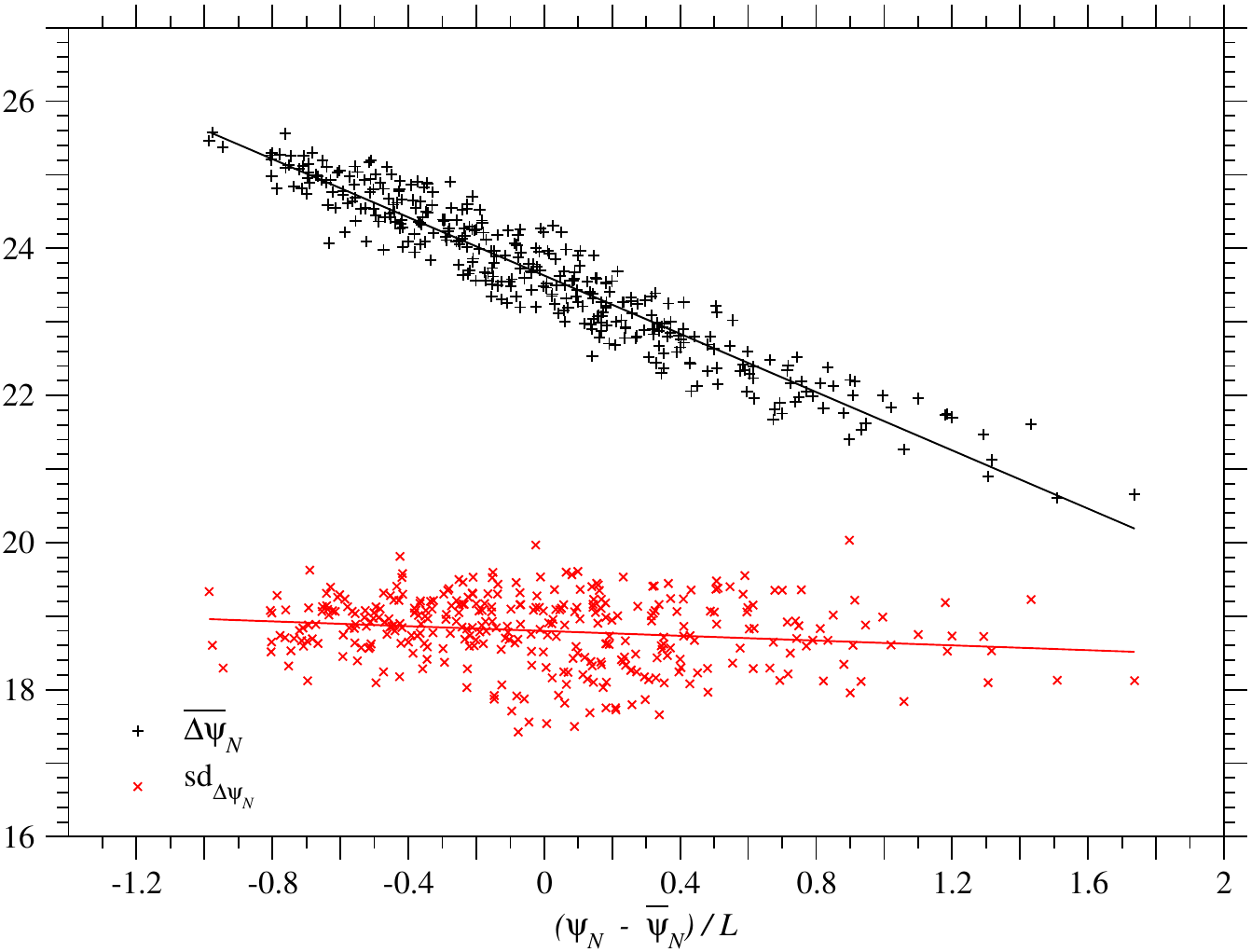}
}
\centerline{
\includegraphics*[width=82mm,angle=0]{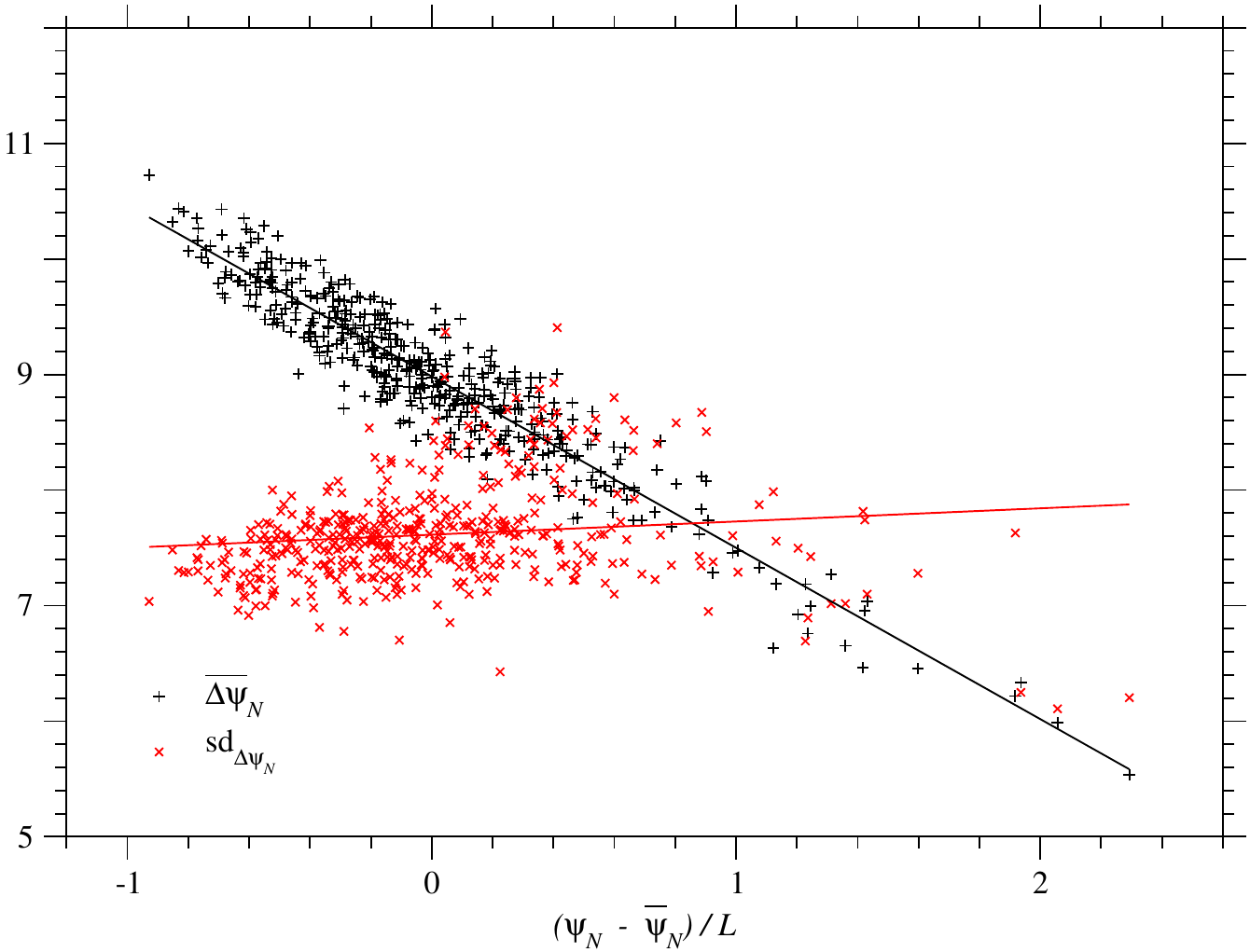}
\includegraphics*[width=82mm,angle=0]{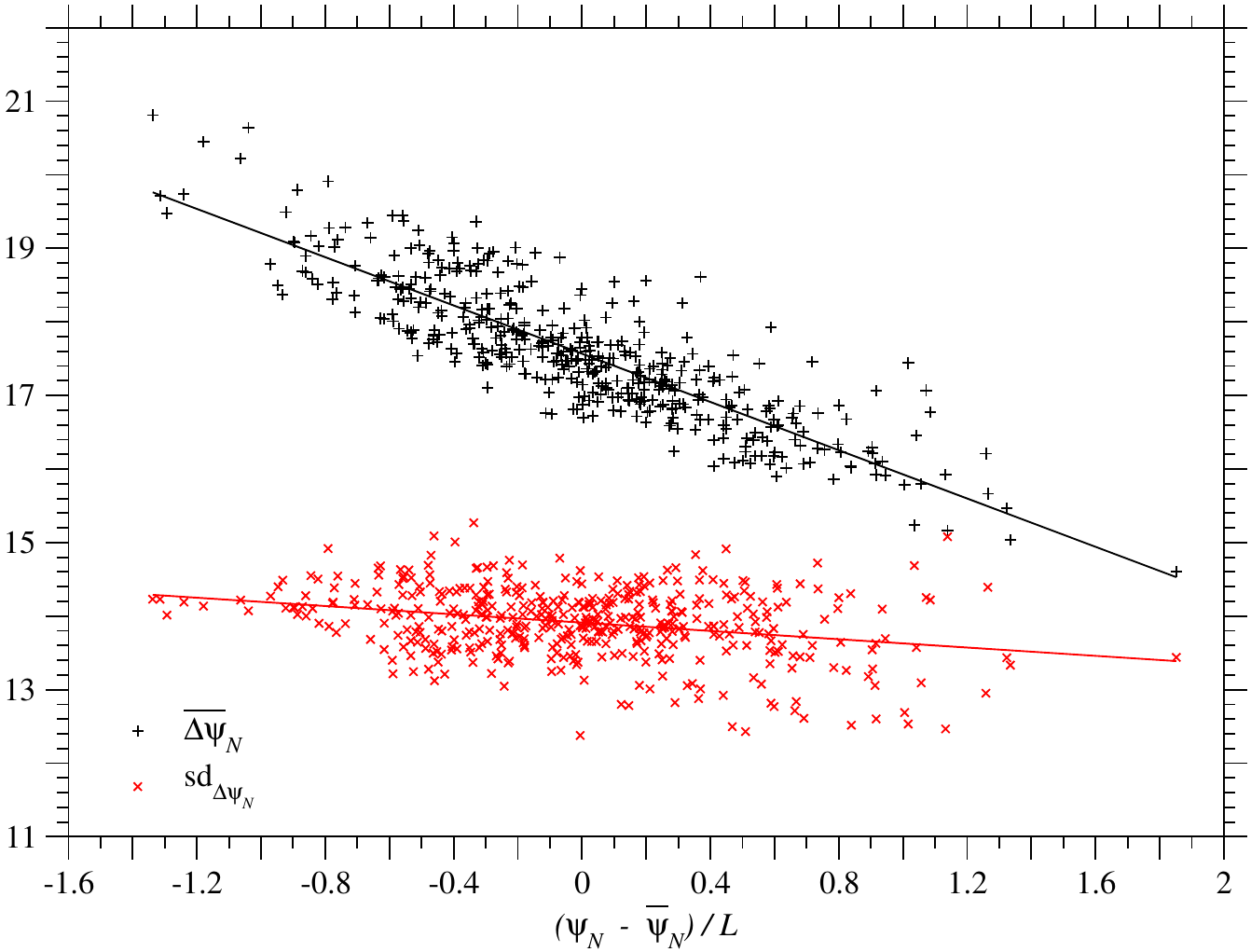}
}
}%  FigureInLegends
\vspace*{1em}
\caption{
\FigureLegends{
\label{sfig: 1n2x-a:8-292.full_non_del.dca0_13.0_20.simple-gauge.ddPhi_at_opt}
\label{fig: 1n2x-a:8-292.full_non_del.dca0_13.0_20.simple-gauge.ddPhi_at_opt}
\BF{
Correlation between $\Delta \psi_N$ due to single nucleotide nonsynonymous substitutions and
$\psi_N$ of homologous sequences in the Methyltransf\_5 family of the domains, 1N2X-A:8-292 (above) and 1N2X-A:137-216 (below).
}
\protect
The left and right figures correspond to the cutoff distance $r_{\script{cutoff}} \sim 8$ and $15.5$ \AA,
respectively.
Each of the black plus or red cross marks corresponds to the mean or the standard deviation 
of $\Delta \psi_N$ due to
all types of single nucleotide nonsynonymous substitutions
over all sites in each of the homologous sequences.
Representatives of unique sequences,
which are at least 20\% different from each other, are employed;
the number of the representatives is almost equal to $M_{\script{eff}}$ in \Table{\ref{\TBL: Proteins_studied}}.
The solid lines show the regression lines for the mean and the standard deviation of $\Delta \psi_N$.
% End of figures_mean_and_sd_legends.tex
}%  FigureLegends
}
\end{figure*}

\CLEARPAGE
\begin{figure*}[h!]
\FigureInLegends{
\noindent
\hspace*{1em} (a) $r_{\script{cutoff}} \sim 8$ \AA\  \hspace*{14em} (b) $r_{\script{cutoff}} \sim 15.5$ \AA\ 

\centerline{
\includegraphics*[width=82mm,angle=0]{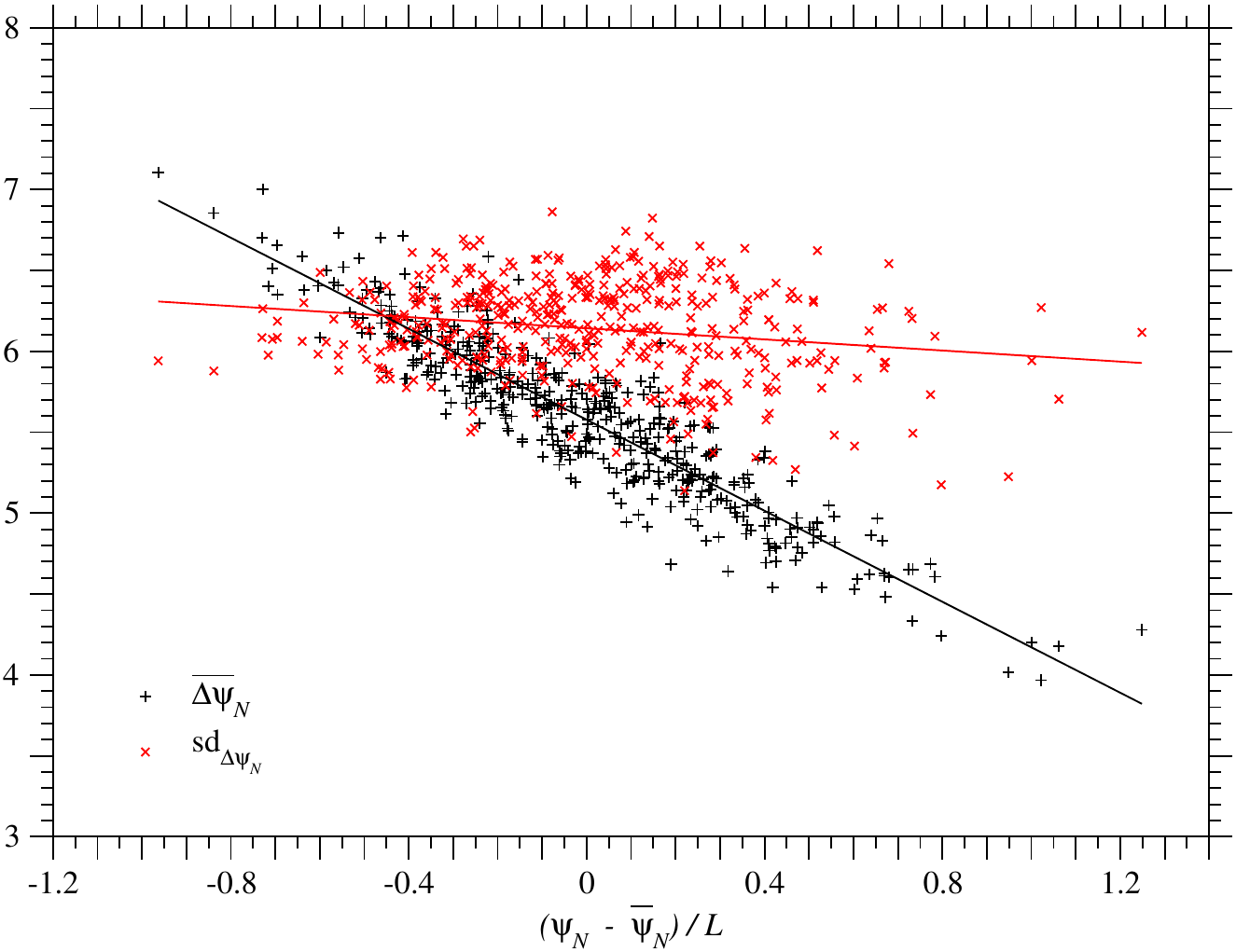}
\includegraphics*[width=82mm,angle=0]{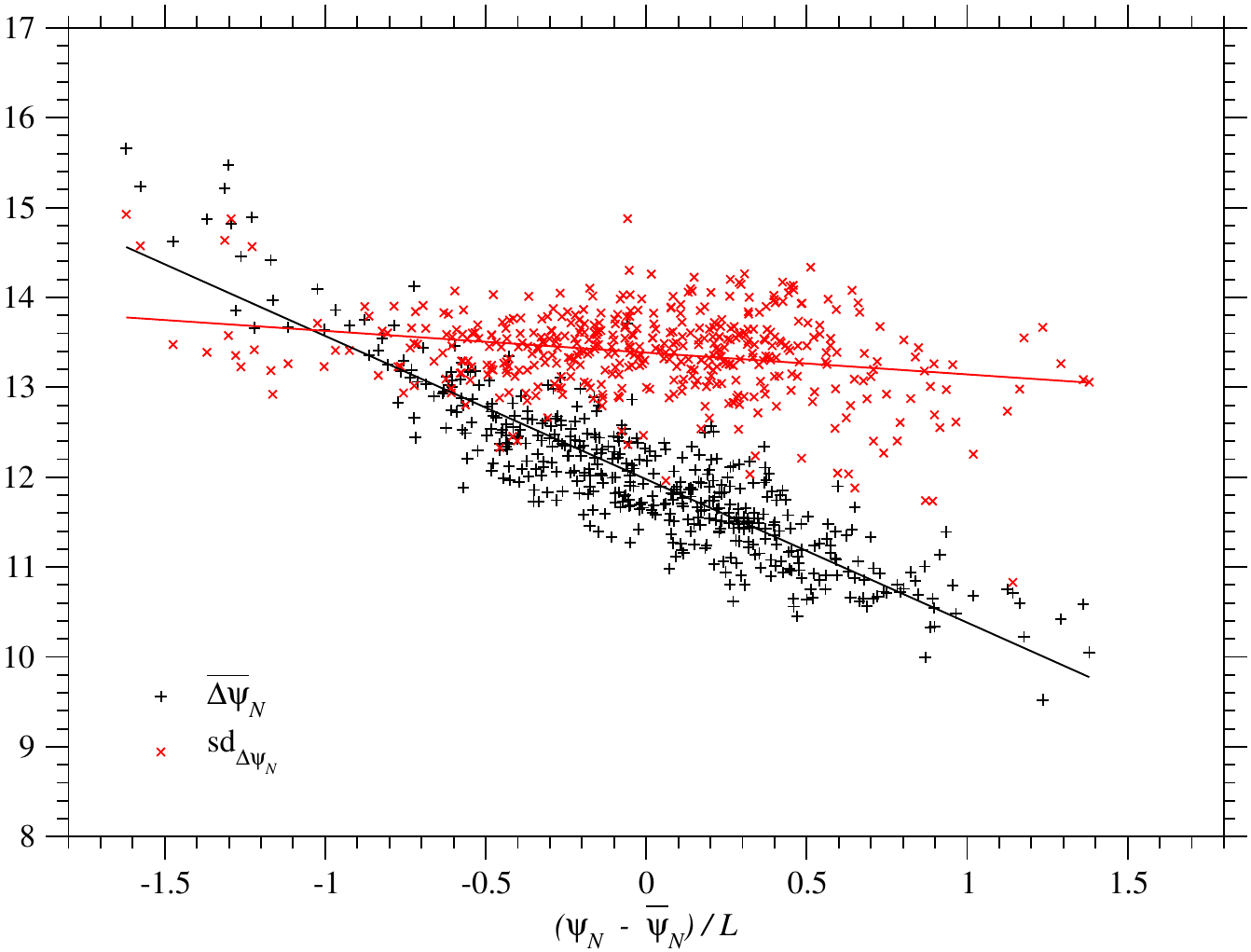}
}
}%  FigureInLegends
\vspace*{1em}
\caption{
\FigureLegends{
\label{sfig: 1fmk-a:87-134.full_non_del.dca0_14.0_20.simple-gauge.ddPhi_at_opt}
\label{fig: 1fmk-a:87-134.full_non_del.dca0_14.0_20.simple-gauge.ddPhi_at_opt}
\BF{
Correlation between $\Delta \psi_N$ due to single nucleotide nonsynonymous substitutions and
$\psi_N$ of homologous sequences in the SH3\_1 family of the domain, 1FMK-A:87-134.
}
\protect
The left and right figures correspond to the cutoff distance $r_{\script{cutoff}} \sim 8$ and $15.5$ \AA,
respectively.
Each of the black plus or red cross marks corresponds to the mean or the standard deviation 
of $\Delta \psi_N$ due to
all types of single nucleotide nonsynonymous substitutions
over all sites in each of the homologous sequences.
Representatives of unique sequences,
which are at least 20\% different from each other, are employed;
the number of the representatives is almost equal to $M_{\script{eff}}$ in \Table{\ref{\TBL: Proteins_studied}}.
The solid lines show the regression lines for the mean and the standard deviation of $\Delta \psi_N$.
% End of figures_mean_and_sd_legends.tex
}%  FigureLegends
}
\end{figure*}

\CLEARPAGE
\begin{figure*}[h!]
\FigureInLegends{
\noindent
\hspace*{1em} (a) $r_{\script{cutoff}} \sim 8$ \AA\  \hspace*{14em} (b) $r_{\script{cutoff}} \sim 15.5$ \AA\ 

\centerline{
\includegraphics*[width=82mm,angle=0]{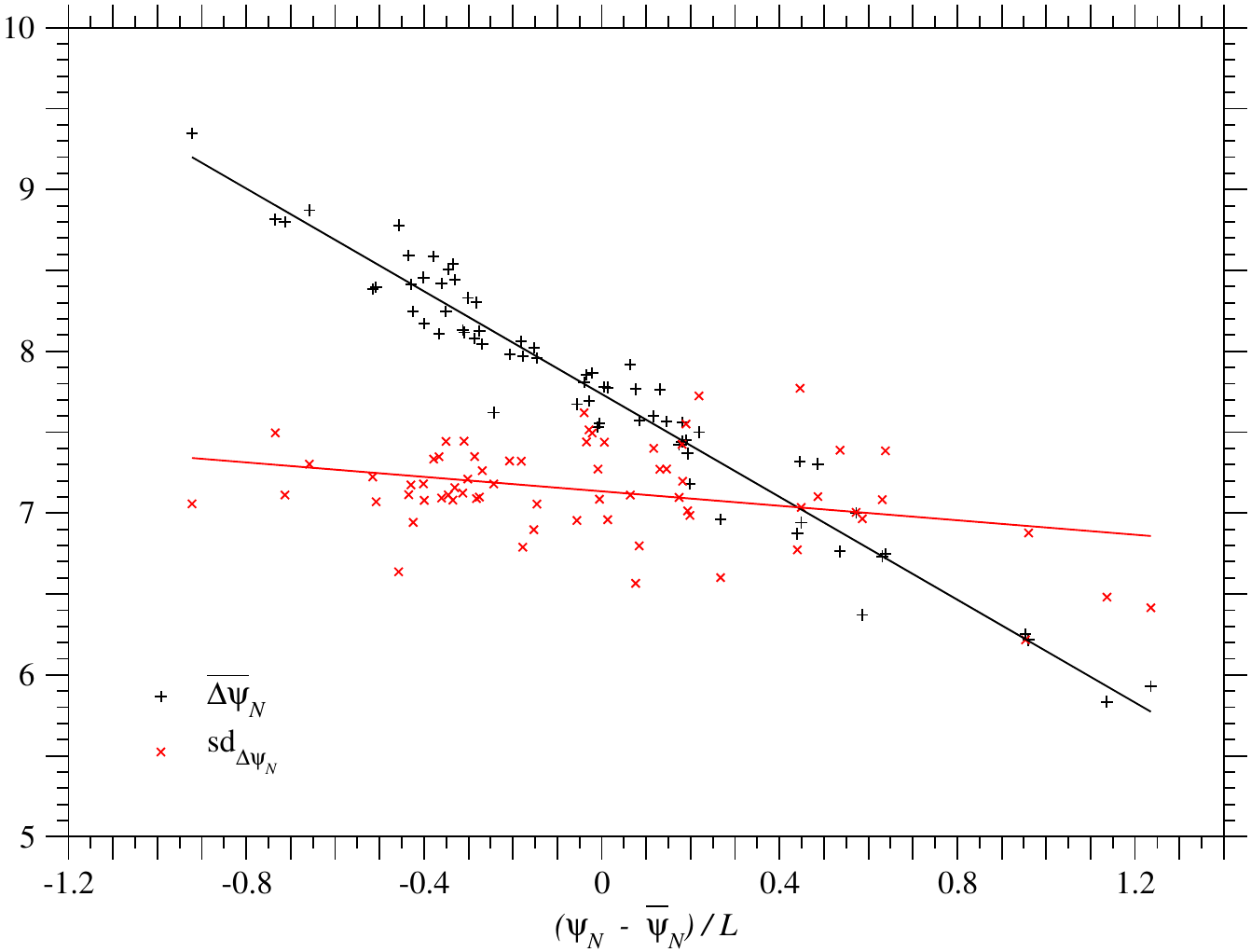}
\includegraphics*[width=82mm,angle=0]{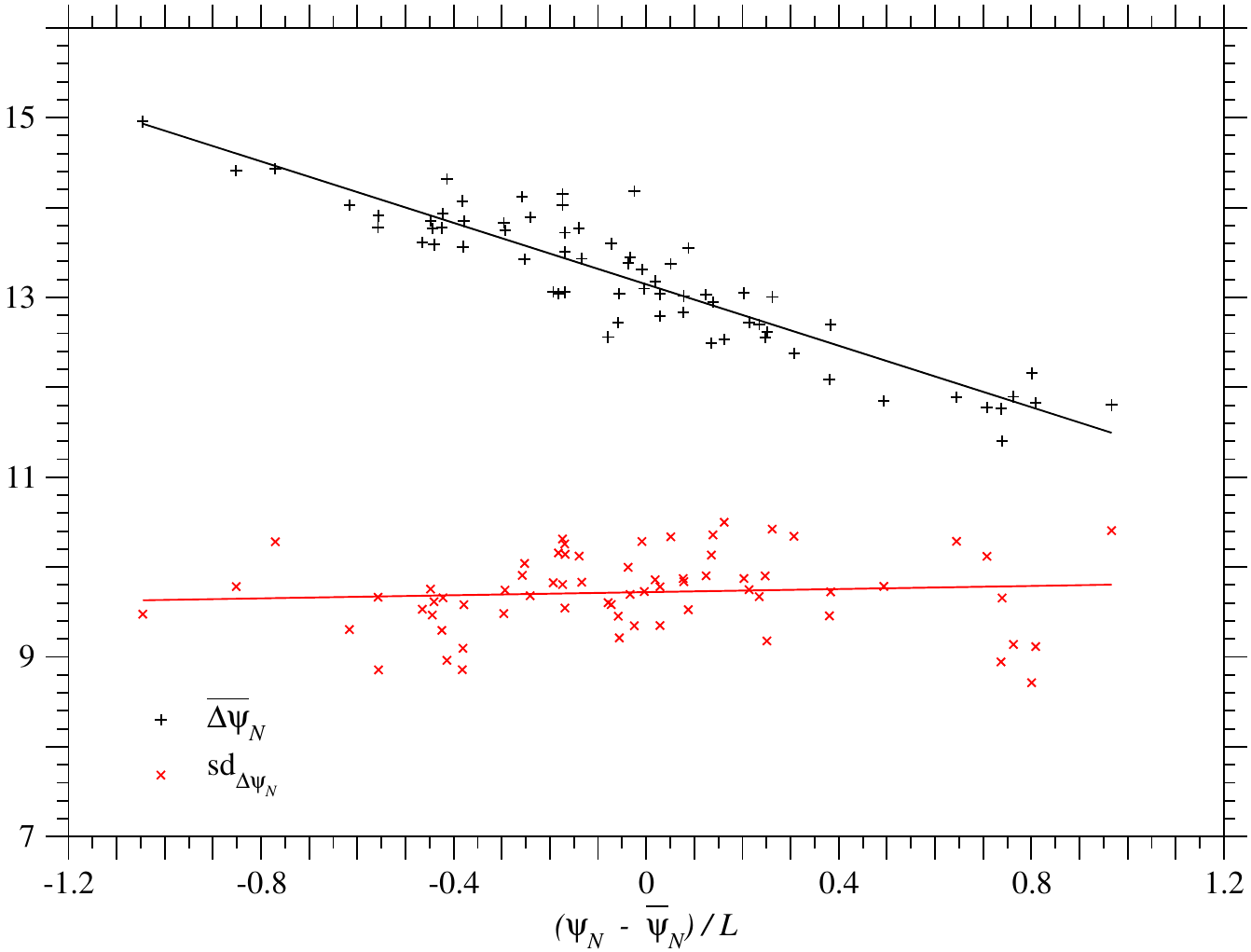}
}
}%  FigureInLegends
\vspace*{1em}
\caption{
\FigureLegends{
\label{sfig: 2abd-a:2-81.full_non_del.dca0_22.0_20.simple-gauge.ddPhi_at_opt}
\label{fig: 2abd-a:2-81.full_non_del.dca0_22.0_20.simple-gauge.ddPhi_at_opt}
\BF{
Correlation between $\Delta \psi_N$ due to single nucleotide nonsynonymous substitutions and
$\psi_N$ of homologous sequences in the ACBP family of the domain, 2ABD-A:2-81.
}
\protect
The left and right figures correspond to the cutoff distance $r_{\script{cutoff}} \sim 8$ and $15.5$ \AA,
respectively.
Each of the black plus or red cross marks corresponds to the mean or the standard deviation 
of $\Delta \psi_N$ due to
all types of single nucleotide nonsynonymous substitutions
over all sites in each of the homologous sequences.
Representatives of unique sequences,
which are at least 20\% different from each other, are employed;
the number of the representatives is almost equal to $M_{\script{eff}}$ in \Table{\ref{\TBL: Proteins_studied}}.
The solid lines show the regression lines for the mean and the standard deviation of $\Delta \psi_N$.
% End of figures_mean_and_sd_legends.tex
}%  FigureLegends
}
\end{figure*}

}%  SUPPLEMENT

\FigA{

\CLEARPAGE
 
\begin{figure*}[h!]
\FigureInLegends{
\noindent
\SUPPLEMENT{
\hspace*{1em} (a) $r_{\script{cutoff}} \sim 8$ \AA\  \hspace*{14em} (b) $r_{\script{cutoff}} \sim 15.5$ \AA\ 

}%  SUPPLEMENT
\TEXT{
\centerline{
{\small{$r_{\script{cutoff}} \sim 8$ \AA\ }}
}
}%  TEXT
\SUPPLEMENT{
\centerline{
\includegraphics*[width=82mm,angle=0]{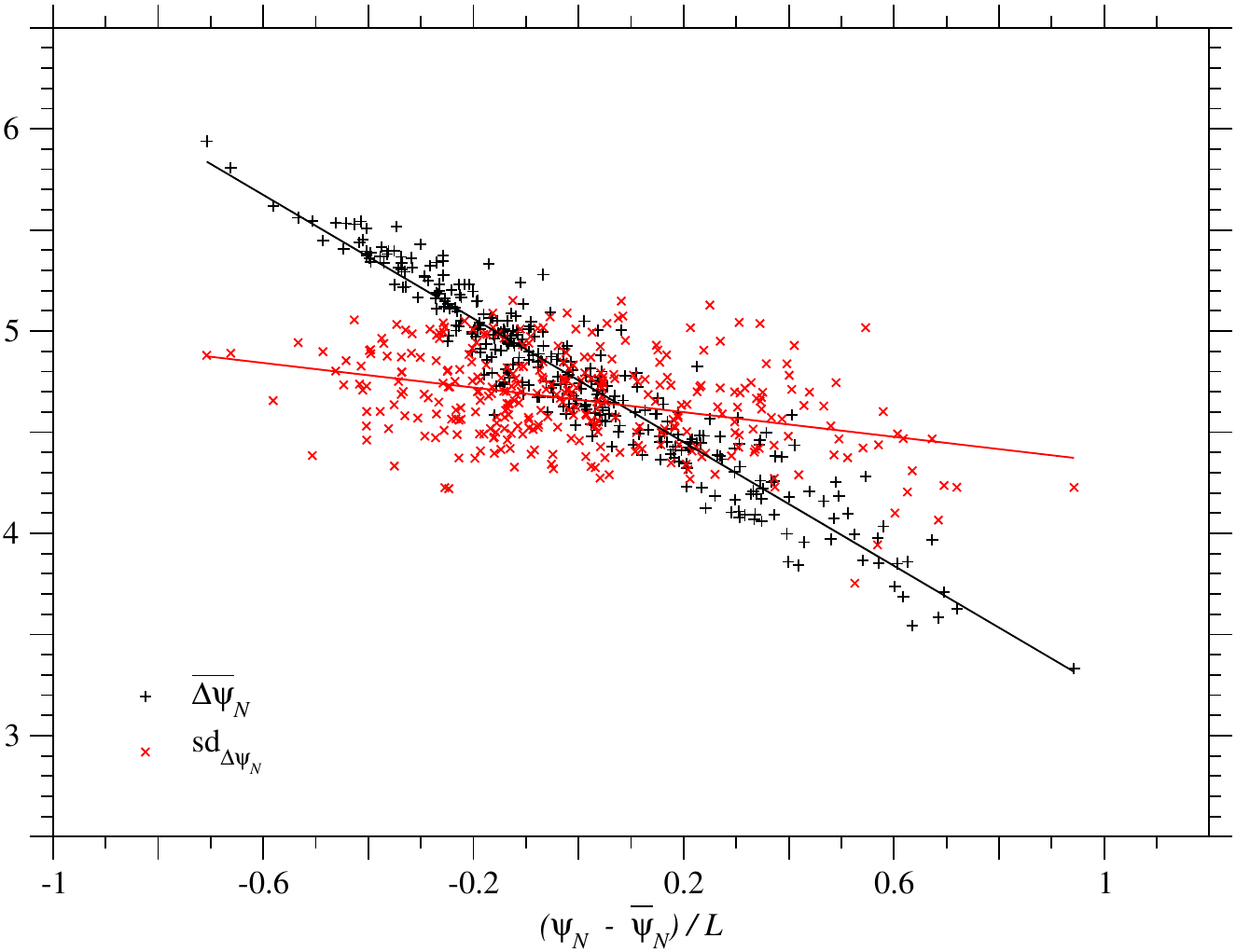}
\includegraphics*[width=82mm,angle=0]{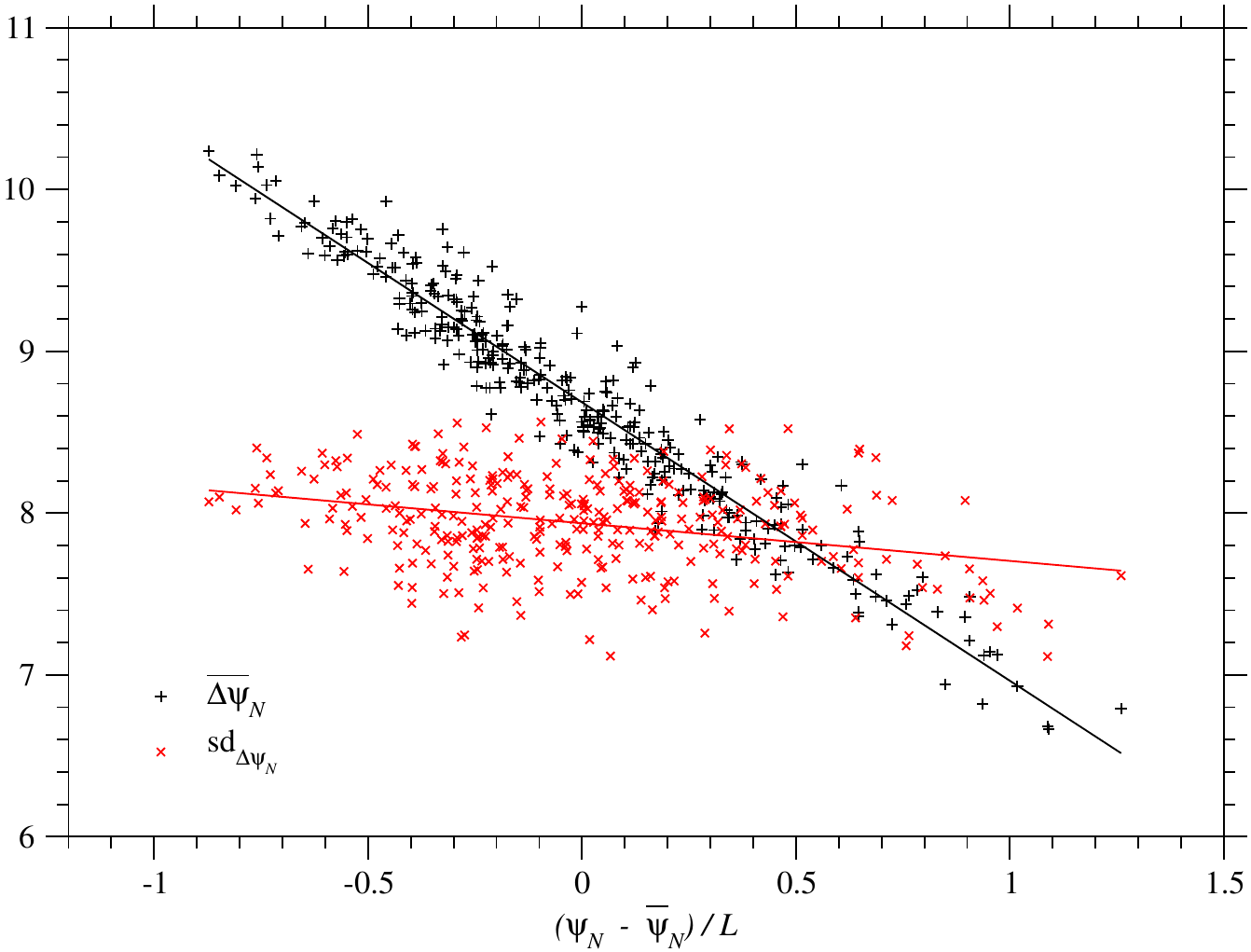}
}
}%  SUPPLEMENT
\TEXT{
\centerline{
\includegraphics*[width=82mm,angle=0]{FIGS/PDZ/1gm1-a_16-96_full_non_del_dca0_205_0_20_simple-gauge_ddPhi_at_opt}
}
}%  TEXT
}%  FigureInLegends
\vspace*{1em}
\caption{
\FigureLegends{
\SUPPLEMENT{
\label{sfig: 1gm1-a:16-96_dca0_205_simple-gauge_ddPhi_at_opt}
\label{sfig: 1gm1-a:16-96_dca0_33_simple-gauge_ddPhi_at_opt}
}%  SUPPLEMENT
\TEXT{
\label{fig: 1gm1-a:16-96_dca0_205_simple-gauge_ddPhi_at_opt}
}%  TEXT
\BF{
Correlation between $\Delta \psi_N$ due to single nucleotide nonsynonymous substitutions and
$\psi_N$ of homologous sequences in the PDZ domain family.
}
\SUPPLEMENT{
The left and right figures correspond to the cutoff distance $r_{\script{cutoff}} \sim 8$ and $15.5$ \AA,
respectively.
}%  SUPPLEMENT
\TEXT{
This figure corresponds to the cutoff distance $r_{\script{cutoff}} \sim 8$ \AA;
see \Fig{\ref{sfig: 1gm1-a:16-96_dca0_33_simple-gauge_ddPhi_at_opt}} for $r_{\script{cutoff}} \sim 15.5$ \AA.
}%  TEXT
Each of the black plus or red cross marks corresponds to the mean or the standard deviation 
of $\Delta \psi_N$ due to
all types of single nucleotide nonsynonymous substitutions
over all sites in each of the homologous sequences of the PDZ domain family.
Only 335 representatives of unique sequences with no deletions, which are at least 20\% different from each other, are employed;
the number of the representatives is almost equal to $M_{\script{eff}}$ in \Table{\ref{\TBL: Proteins_studied}}.
The solid lines show the regression lines for the mean and the standard deviation of $\Delta \psi_N$.
}%  FigureLegends
}
\end{figure*}
}%  FigA

\SUPPLEMENT{

\CLEARPAGE
\begin{figure*}[h!]
\FigureInLegends{
\noindent
\hspace*{1em} (a) $r_{\script{cutoff}} \sim 8$ \AA\  \hspace*{14em} (b) $r_{\script{cutoff}} \sim 15.5$ \AA\ 

\centerline{
\includegraphics*[width=82mm,angle=0]{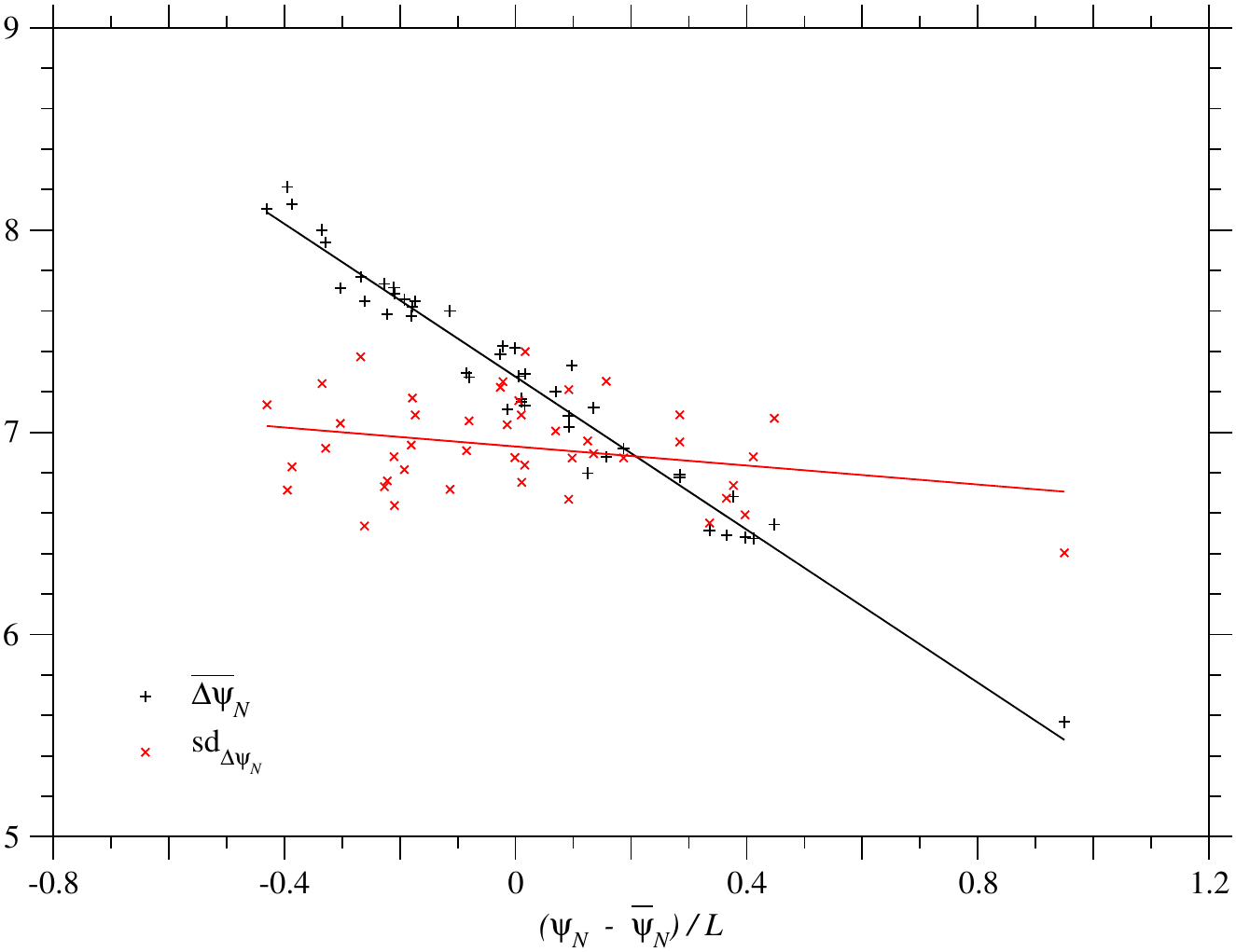}
\includegraphics*[width=82mm,angle=0]{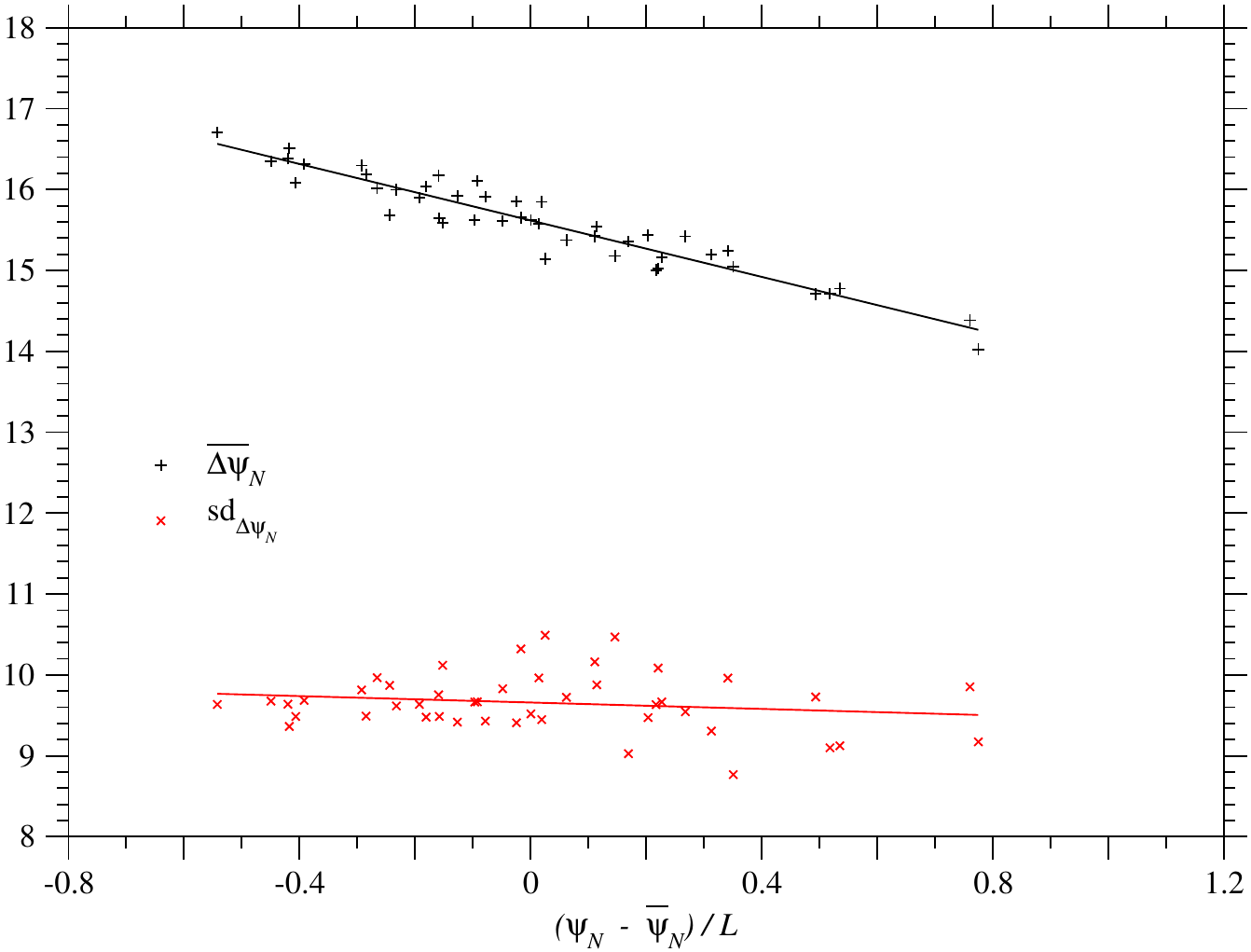}
}
}%  FigureInLegends
\vspace*{1em}
\caption{
\FigureLegends{
\label{sfig: 5azu-a:4-128.full_non_del.dca0_23.0_20.simple-gauge.BD.ddPhi_at_opt}
\label{fig: 5azu-a:4-128.full_non_del.dca0_23.0_20.simple-gauge.BD.ddPhi_at_opt}
\BF{
Correlation between $\Delta \psi_N$ due to single nucleotide nonsynonymous substitutions and
$\psi_N$ of homologous sequences in the Copper-bind family of the domain, 5AZU-B/D:4-128.
}
\protect
The left and right figures correspond to the cutoff distance $r_{\script{cutoff}} \sim 8$ and $15.5$ \AA,
respectively.
Each of the black plus or red cross marks corresponds to the mean or the standard deviation 
of $\Delta \psi_N$ due to
all types of single nucleotide nonsynonymous substitutions
over all sites in each of the homologous sequences.
Representatives of unique sequences,
which are at least 20\% different from each other, are employed;
the number of the representatives is almost equal to $M_{\script{eff}}$ in \Table{\ref{\TBL: Proteins_studied}}.
The solid lines show the regression lines for the mean and the standard deviation of $\Delta \psi_N$.
% End of figures_mean_and_sd_legends.tex
}%  FigureLegends
}
\end{figure*}

}%  SUPPLEMENT

% End of figures_mean_and_sd.tex

\FigB{
\CLEARPAGE

\begin{figure*}[h!]
\FigureInLegends{
\noindent
\SUPPLEMENT{
(a) $r_{\script{cutoff}} \sim 8$ \AA\  \hspace*{13em} (b) $r_{\script{cutoff}} \sim 15.5$ \AA\ 

}%  SUPPLEMENT
\TEXT{
\centerline{
{\small{$r_{\script{cutoff}} \sim 8$ \AA\ }}
}
}%  TEXT
\SUPPLEMENT{
\centerline{
\includegraphics*[height=75mm,angle=0]{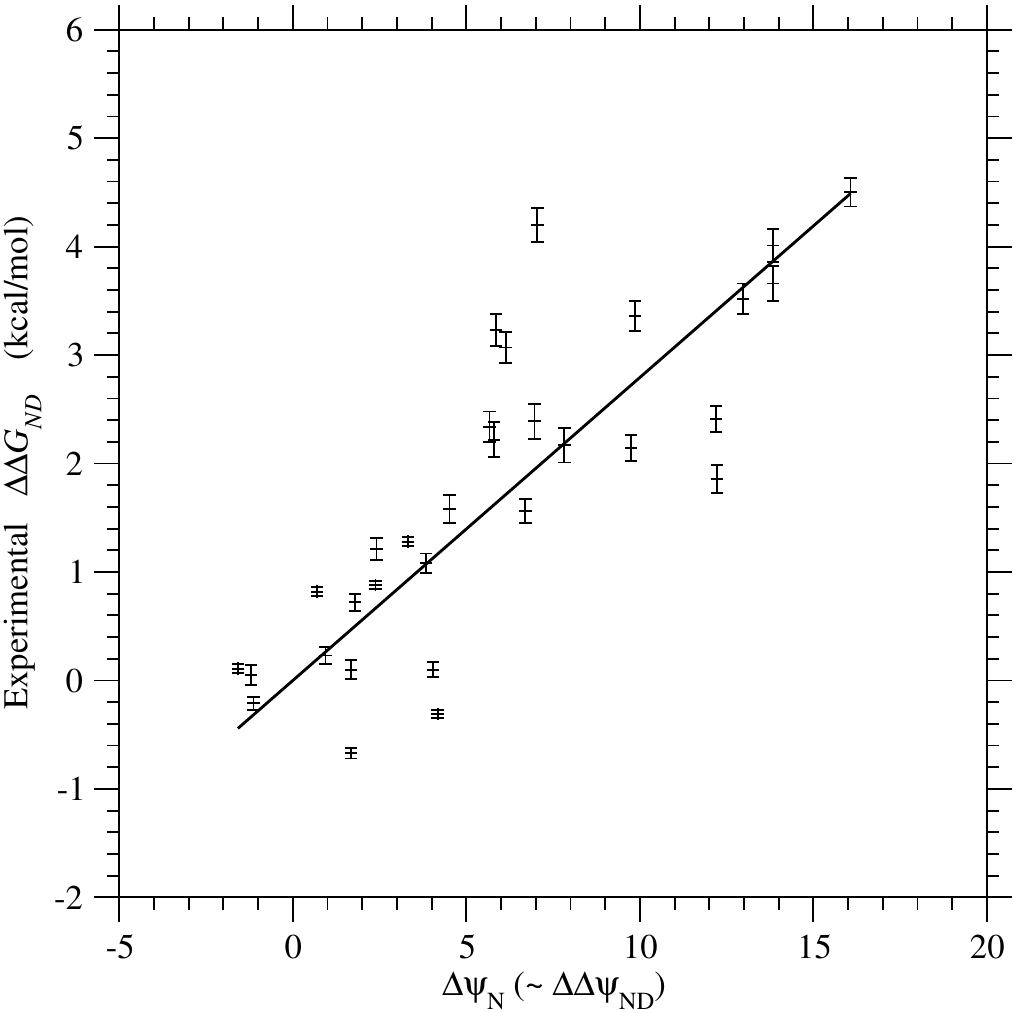}
\includegraphics*[height=75mm,angle=0]{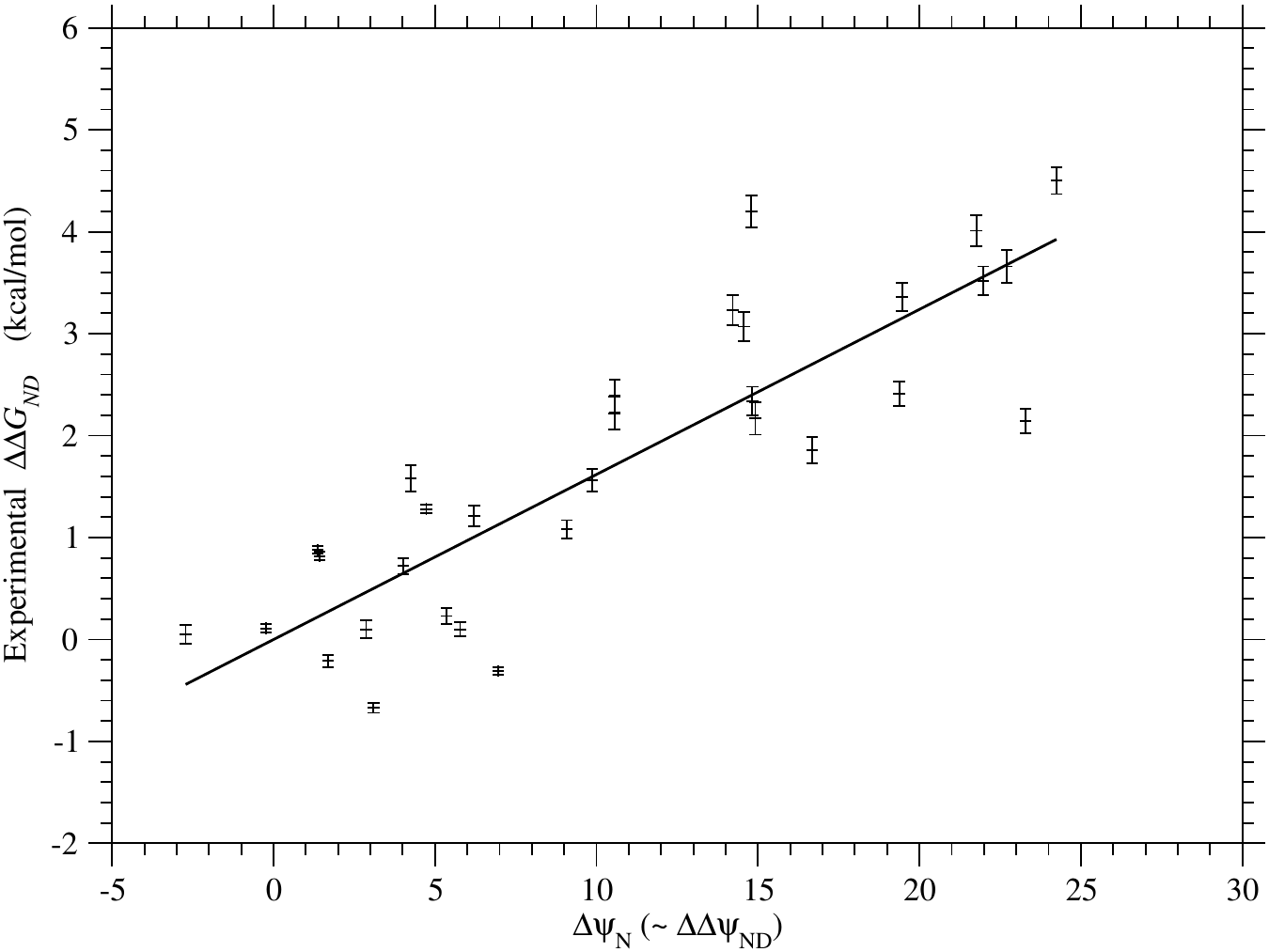}
}
}%  SUPPLEMENT
\TEXT{
\centerline{
\includegraphics*[height=75mm,angle=0]{FIGS/PDZ/1gm1-a_16-96_full_non_del_dca0_205_0_20_simple-gauge_ddG-dPhi_at_opt}
}
}%  TEXT
}%  FigureInLegends
\vspace*{1em}
\caption{
\FigureLegends{
\SUPPLEMENT{
\label{sfig: 1gm1-a:16-96_dca0_205_simple-gauge_ddG-dPhi_at_opt}
\label{sfig: 1gm1-a:16-96_dca0_33_simple-gauge_ddG-dPhi_at_opt}
}%  SUPPLEMENT
\TEXT{
\label{fig: 1gm1-a:16-96_dca0_205_simple-gauge_ddG-dPhi_at_opt}
}%  TEXT
\BF{
Regression of the experimental values\CITE{GGCJVTVB:07} of folding free energy changes ($\Delta\Delta G_{ND}$) 
due to single amino acid substitutions on
$\Delta \psi_N (\simeq \Delta\Delta \psi_{ND})$ for the same types of substitutions in the PDZ domain. 
}
\SUPPLEMENT{
The left and right figures correspond to the cutoff distance $r_{\script{cutoff}} \sim 8$ and $15.5$ \AA, respectively.
The solid lines show the least-squares regression lines through the origin with the slopes, $0.279$ kcal/mol for $r_{\script{cutoff}} \sim 8$ \AA\ 
and $0.162$ kcal/mol for $r_{\script{cutoff}} \sim 15.5$ \AA, which are the estimates of $k_B T_s$.
The reflective correlation coefficients for them are equal to $0.93$ and $0.94$, respectively.
}%  SUPPLEMENT
\TEXT{
This figure corresponds to the cutoff distance $r_{\script{cutoff}} \sim 8$ \AA;
see \Fig{\ref{sfig: 1gm1-a:16-96_dca0_33_simple-gauge_ddG-dPhi_at_opt}} for $r_{\script{cutoff}} \sim 15.5$ \AA.
The solid line shows the least-squares regression line through the origin with the slope, $0.279$ kcal/mol, which is the estimates of $k_B T_s$.
The reflective correlation coefficient is equal to $0.93$.
}%  TEXT
The free energies are in kcal/mol units.
}%  FigureLegends
}
\end{figure*}
}%  FigB

\SUPPLEMENT{
\FigC{

\CLEARPAGE
 
\begin{figure*}[h!]
\FigureInLegends{
\centerline{
\includegraphics*[width=82mm,angle=0]{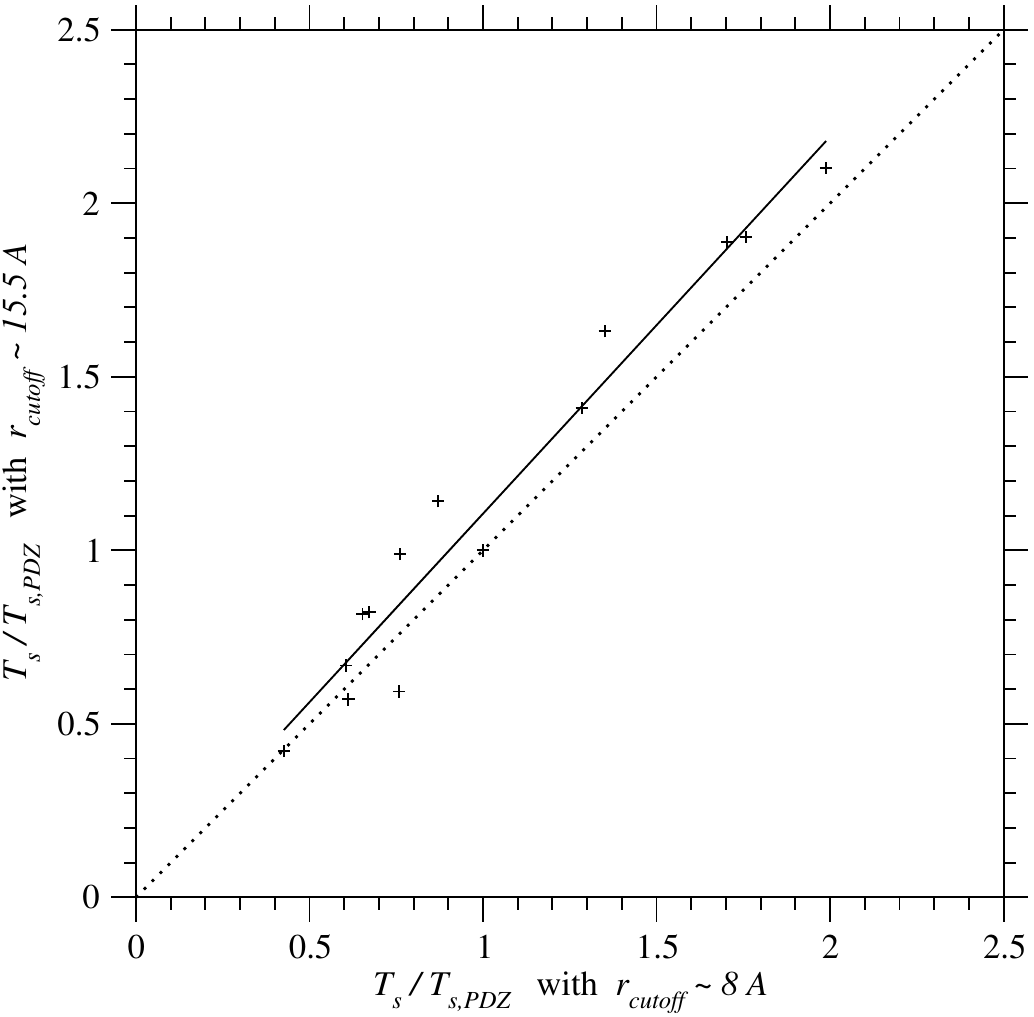}
\includegraphics*[width=82mm,angle=0]{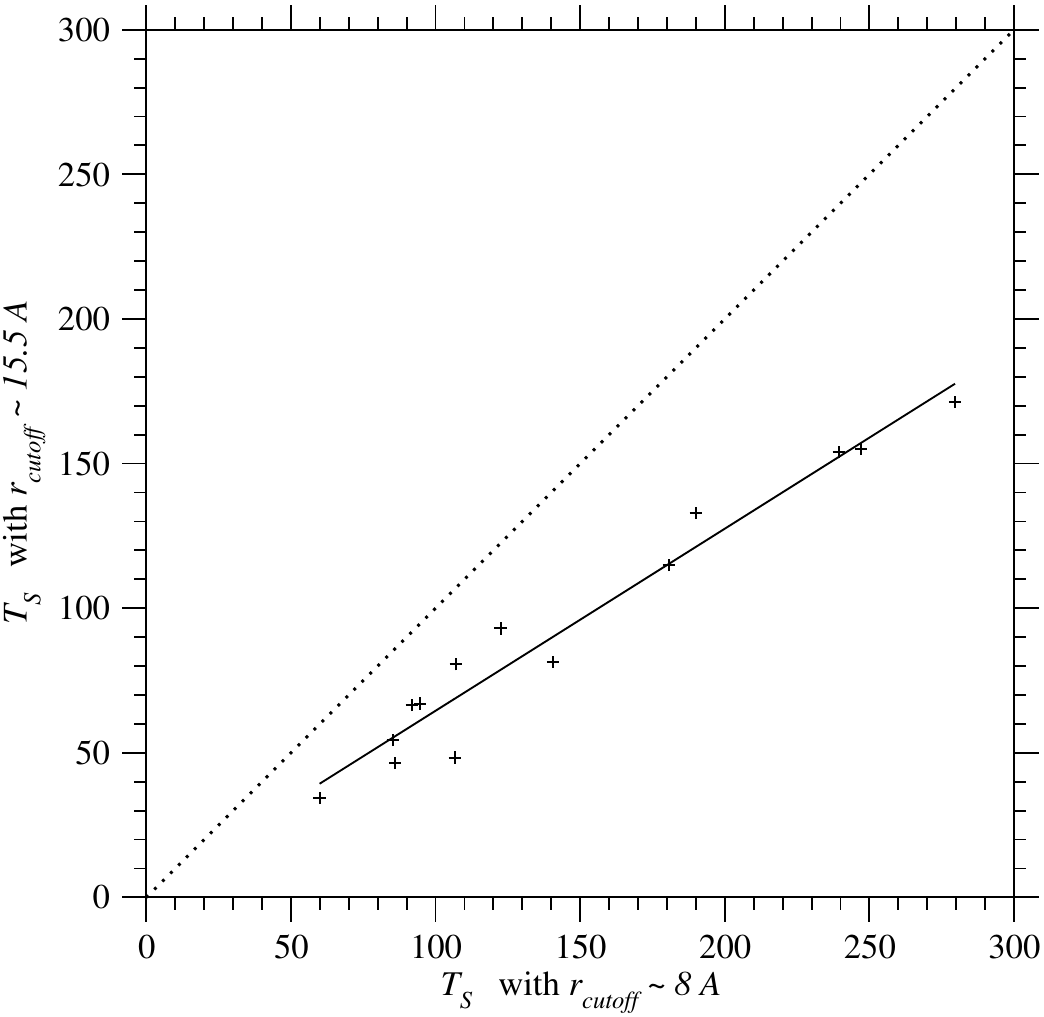}
}
}%  FigureInLegends
\vspace*{1em}
\caption{
\SUPPLEMENT{
\label{sfig: Ts_relative_to_Tpdz_8_vs_16A}
\label{sfig: Ts_8_vs_16A}
\label{fig: Ts_relative_to_Tpdz_8_vs_16A}
\label{fig: Ts_8_vs_16A}
}%  SUPPLEMENT
\TEXT{
\label{fig: Ts_relative_to_Tpdz_8_vs_16A}
\label{fig: Ts_8_vs_16A}
}%  TEXT
\FigureLegends{
\BF{Comparison of selective temperatures ($T_s$)
estimated with different cutoff distances by the present method. 
}
The abscissa and ordinate correspond to the cases of $r_{\script{cutoff}} \sim 8$ and $15.5$ \AA,
respectively.
The $T_s$ is in $^\circ$K units.
The solid lines show the regression lines,
$(T_s/T_{s,PDZ})_{15.5A} = 1.09 (T_s/T_{s,PDZ})_{8A}  + 0.02$ and
$(T_s)_{15.5A} = 0.630 (T_s)_{8A}  + 1.57$.
The correlation coefficients are equal to 0.98 for both.
}%  FigureLegends
}
\end{figure*}

}%  FigC
}%  SUPPLEMENT

\SUPPLEMENT{

\CLEARPAGE
 
\begin{figure*}[h!]
\FigureInLegends{
\centerline{
\includegraphics*[width=82mm,angle=0]{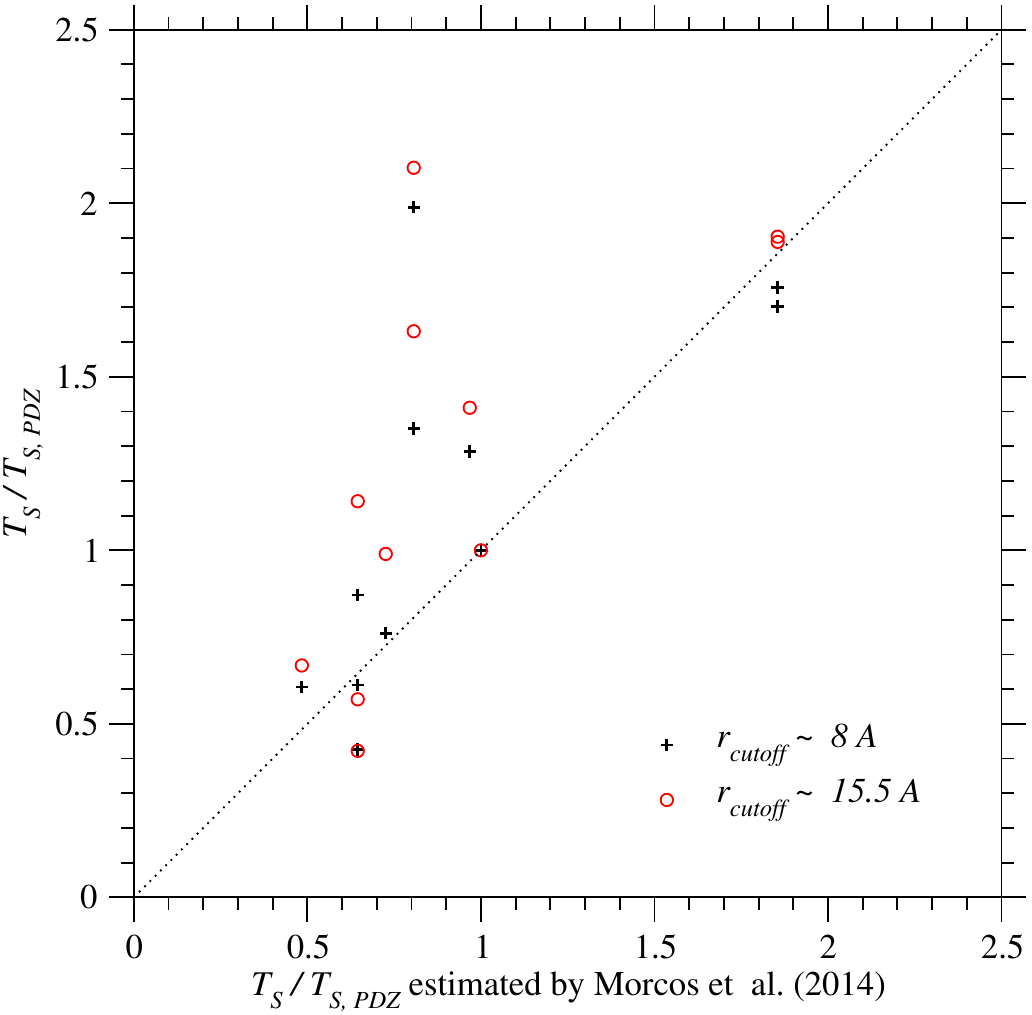}
\includegraphics*[width=82mm,angle=0]{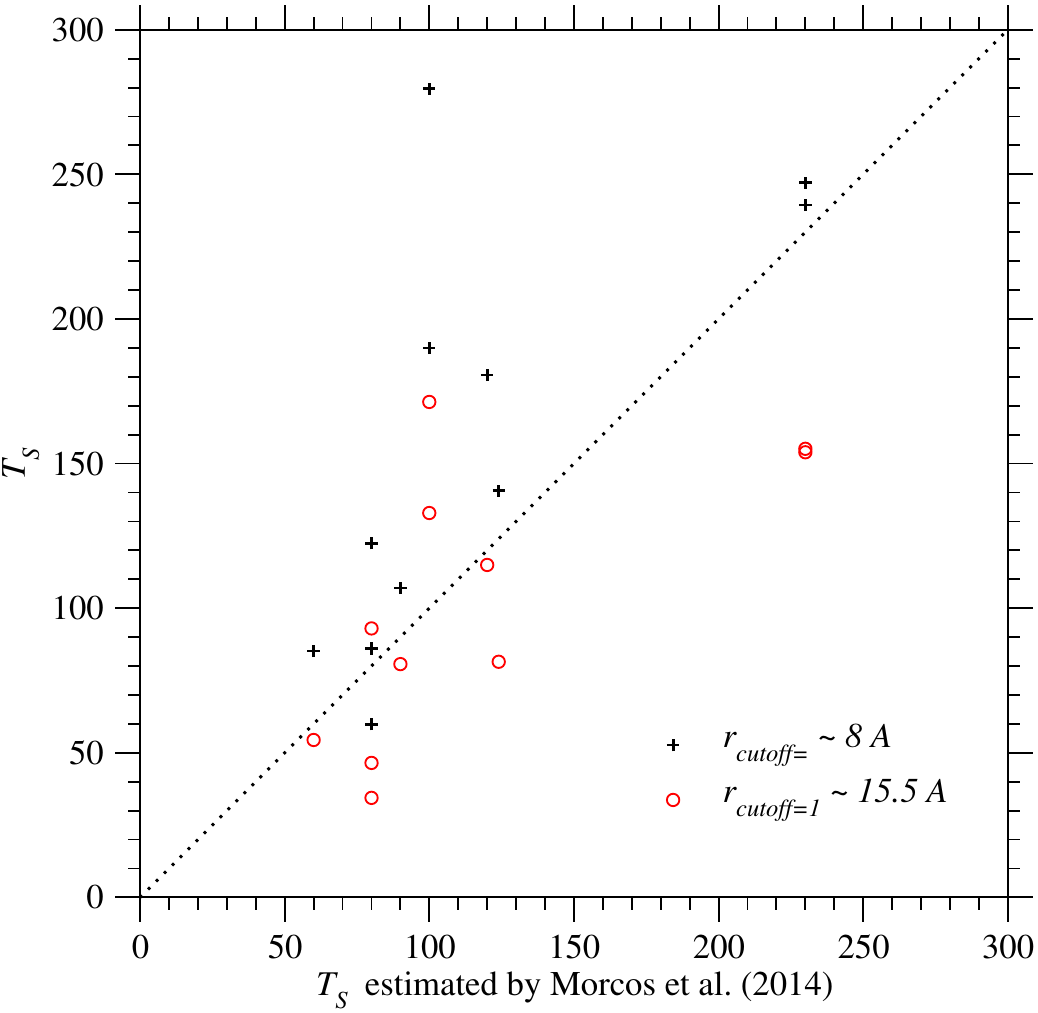}
}
}%  FigureInLegends
\vspace*{1em}
\caption{
\label{sfig: Ts_relative_to_Tpdz_Wolynes_vs_8_and_16A}
\label{fig: Ts_relative_to_Tpdz_Wolynes_vs_8_and_16A}
\label{sfig: Ts_Wolynes_vs_8_and_16A}
\label{fig: Ts_Wolynes_vs_8_and_16A}
\FigureLegends{
\BF{Selective temperatures ($T_s$) 
estimated by the present method are plotted against those estimated by Morcos et al.\CITE{MSCOW:14};
their estimated values of $T_s$ tend to fall between the upper ($r_{\script{cutoff}} \sim 8$) 
and lower ($r_{\script{cutoff}} \sim 15.5$ \AA) estimates of $T_s$. 
Plus and open circle marks correspond to
the cases of $r_{\script{cutoff}} \sim 8$ and $15.5$ \AA,
respectively.
}
}%  FigureLegends
}
\end{figure*}

}%  SUPPLEMENT

\SUPPLEMENT{

\CLEARPAGE
 
\begin{figure*}[h!]
\FigureInLegends{
\centerline{
\includegraphics*[width=82mm,angle=0]{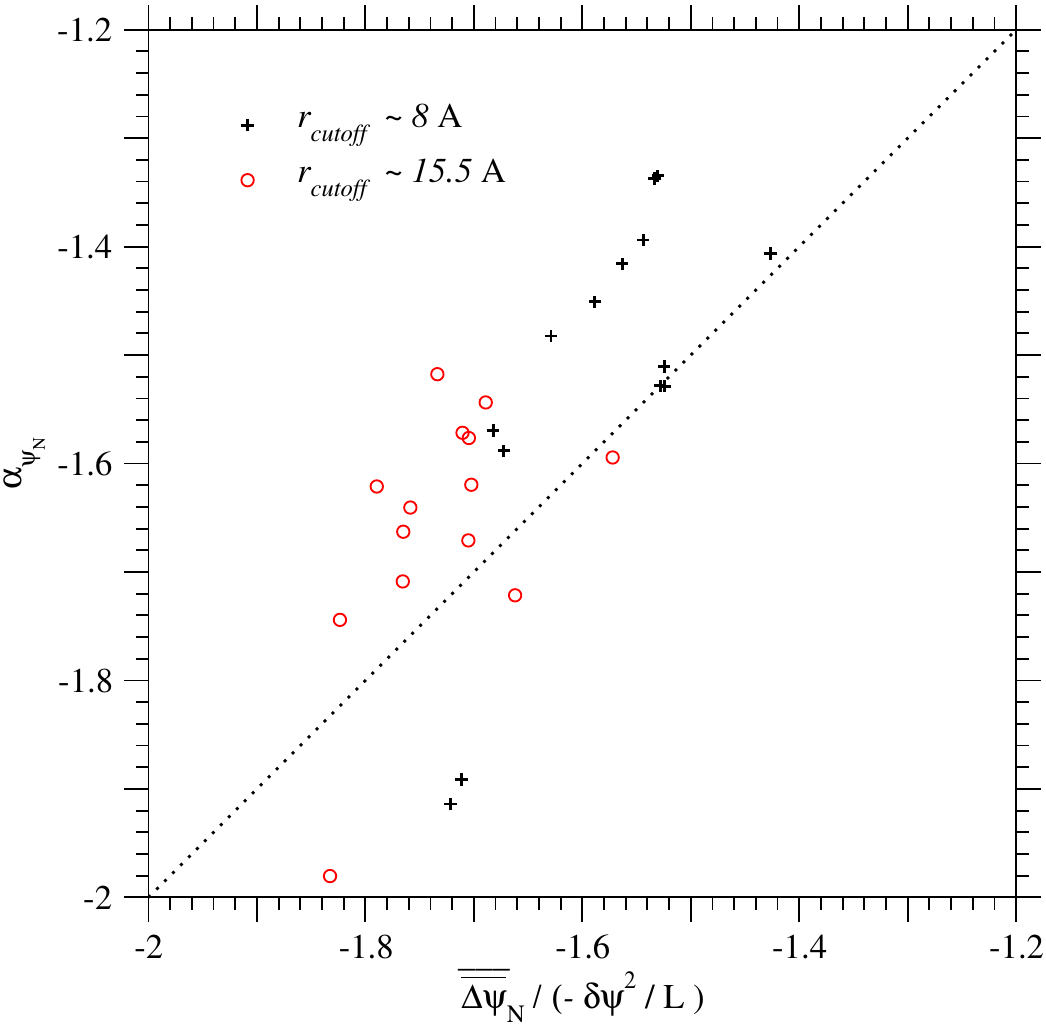}
}
}%  FigureInLegends
\vspace*{1em}
\caption{
\label{sfig: mean_over_dpsi2_vs_slope}
\label{fig: mean_over_dpsi2_vs_slope}
\FigureLegends{
\BF{
Comparison of $\alpha_{\psi_N}$,
which is the regression coefficient of $\overline{\Delta \psi_N}$ on $\psi_N / L$,
with $\overline{\overline{\Delta \psi_N}} / ( -{\delta \psi}^2 / L) $
for each protein family.
}
Plus and open circle marks correspond to
the cases of $r_{\script{cutoff}} \sim 8$ and $15.5$ \AA,
respectively.
}%  FigureLegends
}
\end{figure*}

}%  SUPPLEMENT

\FigD{

\CLEARPAGE
 
\begin{figure*}[h!]
\FigureInLegends{
\SUPPLEMENT{
\centerline{
\includegraphics*[width=82mm,angle=0]{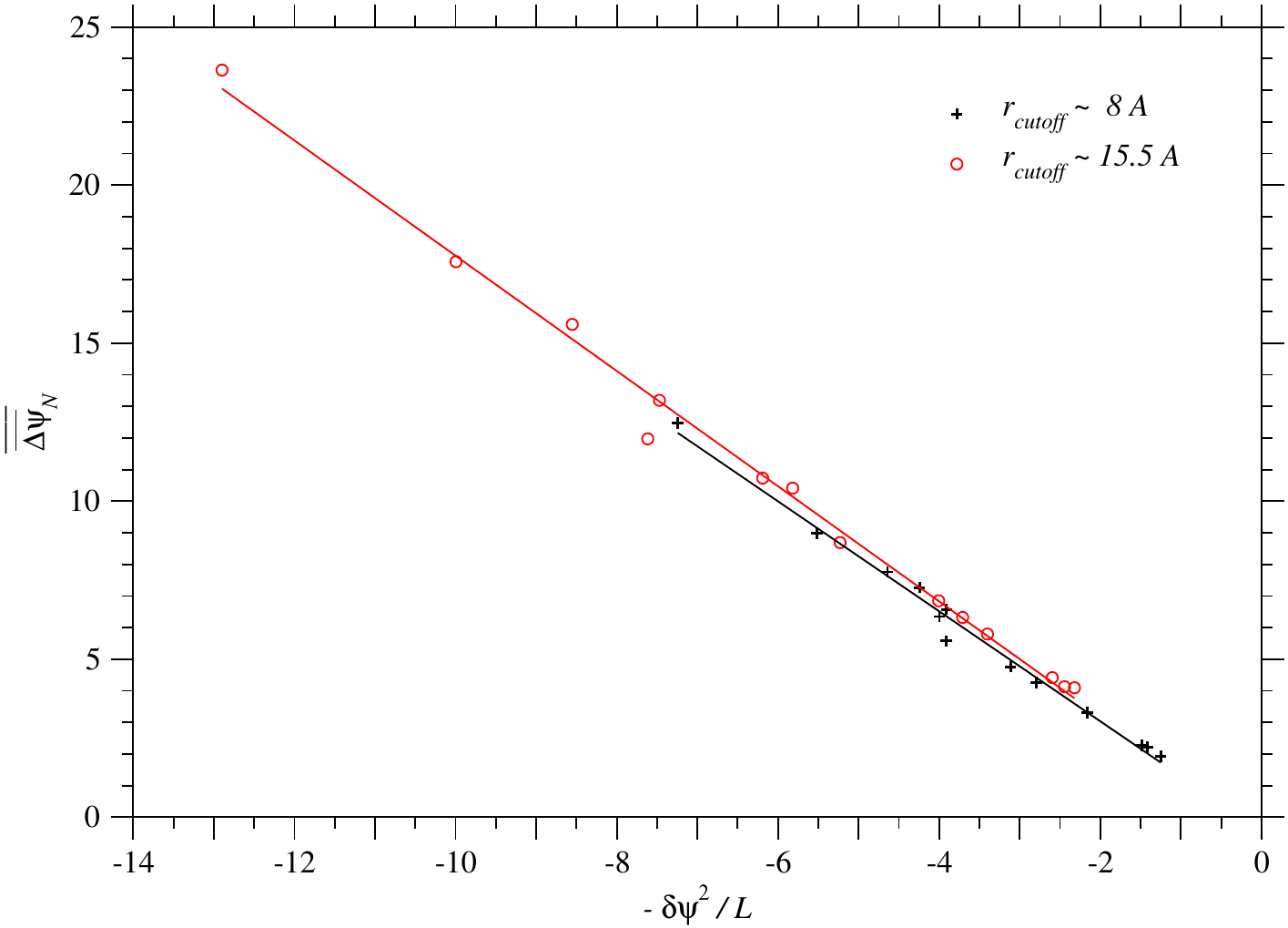}
}
}%  SUPPLEMENT
\TEXT{
\centerline{
\includegraphics*[width=82mm,angle=0]{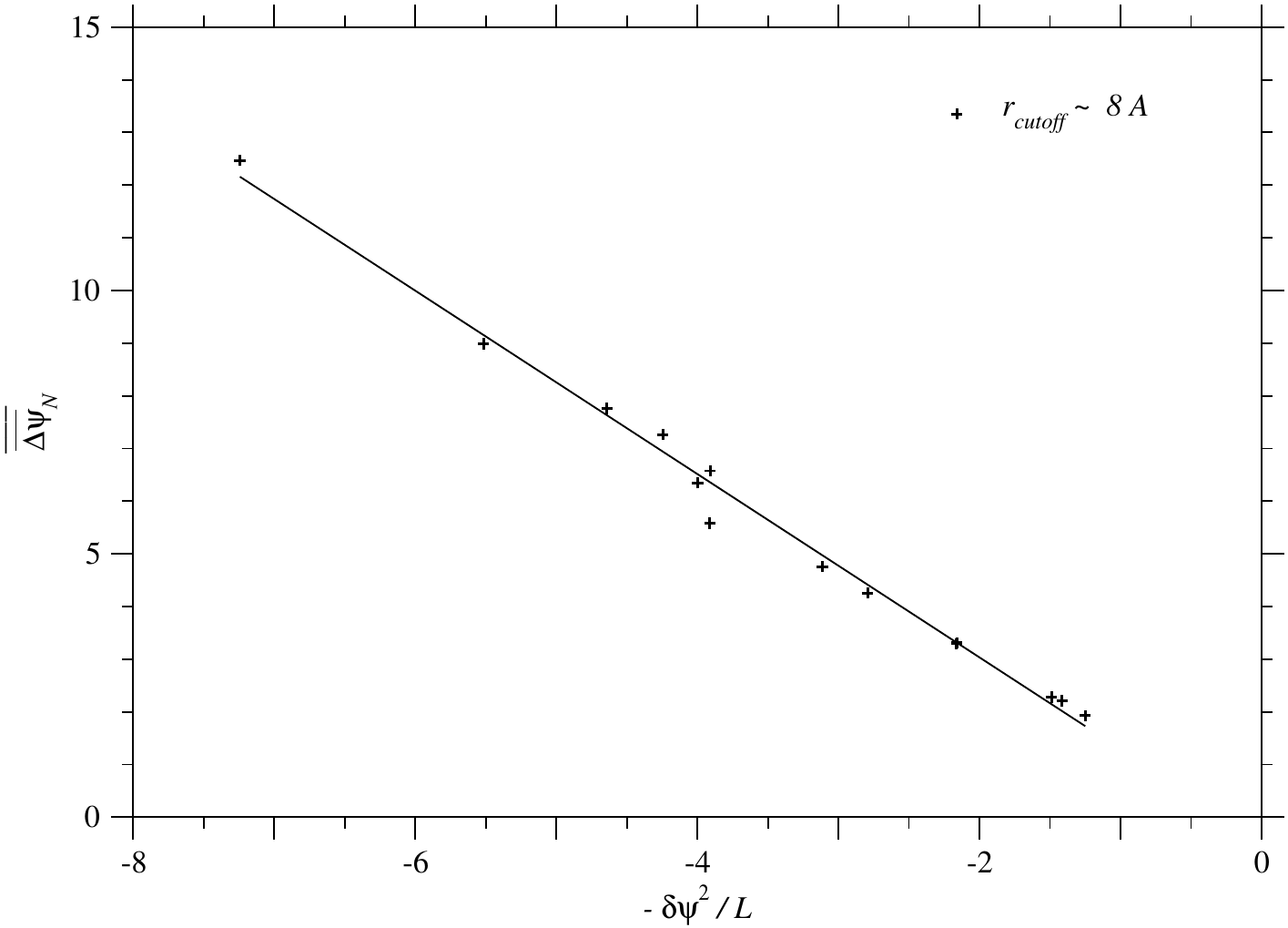}
}
}%  TEXT
}%  FigureInLegends
\vspace*{1em}
\caption{
\SUPPLEMENT{
\label{sfig: dpsi2_over_L_vs_ddPsi}
\label{sfig: dpsi2_over_L_vs_dPsi}
}%  SUPPLEMENT
\TEXT{
\label{fig: dpsi2_over_L_vs_ddPsi}
\label{fig: dpsi2_over_L_vs_dPsi}
}%  TEXT
\FigureLegends{
\BF{
Dependence of the average of
$\overline{\Delta \psi_N}$ 
due to single nucleotide nonsynonymous substitutions 
over homologous sequences
on $-{\delta \psi}^2 / L$ across protein families.
}
\SUPPLEMENT{
Plus and open circle marks indicate the values for
each protein family 
in the cases of $r_{\script{cutoff}} \sim 8$ and $15.5$ \AA,
respectively.
}%  SUPPLEMENT
\TEXT{
Plus marks indicate the value for
each protein family 
in the case of $r_{\script{cutoff}} \sim 8$ \AA.
The correlation coefficient is
equal to 0.995,
and the regression line is
$\overline{\overline{ \Delta \psi_N(\sigma^N_{j \neq i}, \sigma^N_i\rightarrow \sigma_i) } }
= - 1.74 (- {\delta \psi}^2 / L) - 0.445$.
See \Fig{\ref{sfig: dpsi2_over_L_vs_ddPsi}} for $r_{\script{cutoff}} \sim 15.5$ \AA.
}%  TEXT
\SUPPLEMENT{
In the case of the cutoff distance 8 \AA, 
the correlation coefficient is
equal to 0.995,
and the regression line is
$\overline{\overline{ \Delta \psi_N(\sigma^N_{j \neq i}, \sigma^N_i\rightarrow \sigma_i) } }
= - 1.74 (- {\delta \psi}^2 / L) - 0.445$.
In the case of $r_{\script{cutoff}} \sim 15.5$ \AA, 
the correlation coefficient is
equal to 0.996,
and the regression line is
$\overline{\overline{ \Delta \psi_N(\sigma^N_{j \neq i}, \sigma^N_i\rightarrow \sigma_i) } }
= - 1.82 (- {\delta \psi}^2 / L) - 0.466$.
}%  SUPPLEMENT
}%  FigureLegends
}
\end{figure*}

}%  FigD

\FigE{

\CLEARPAGE
 
\begin{figure*}[h!]
\FigureInLegends{
\SUPPLEMENT{
\centerline{
\includegraphics*[width=82mm,angle=0]{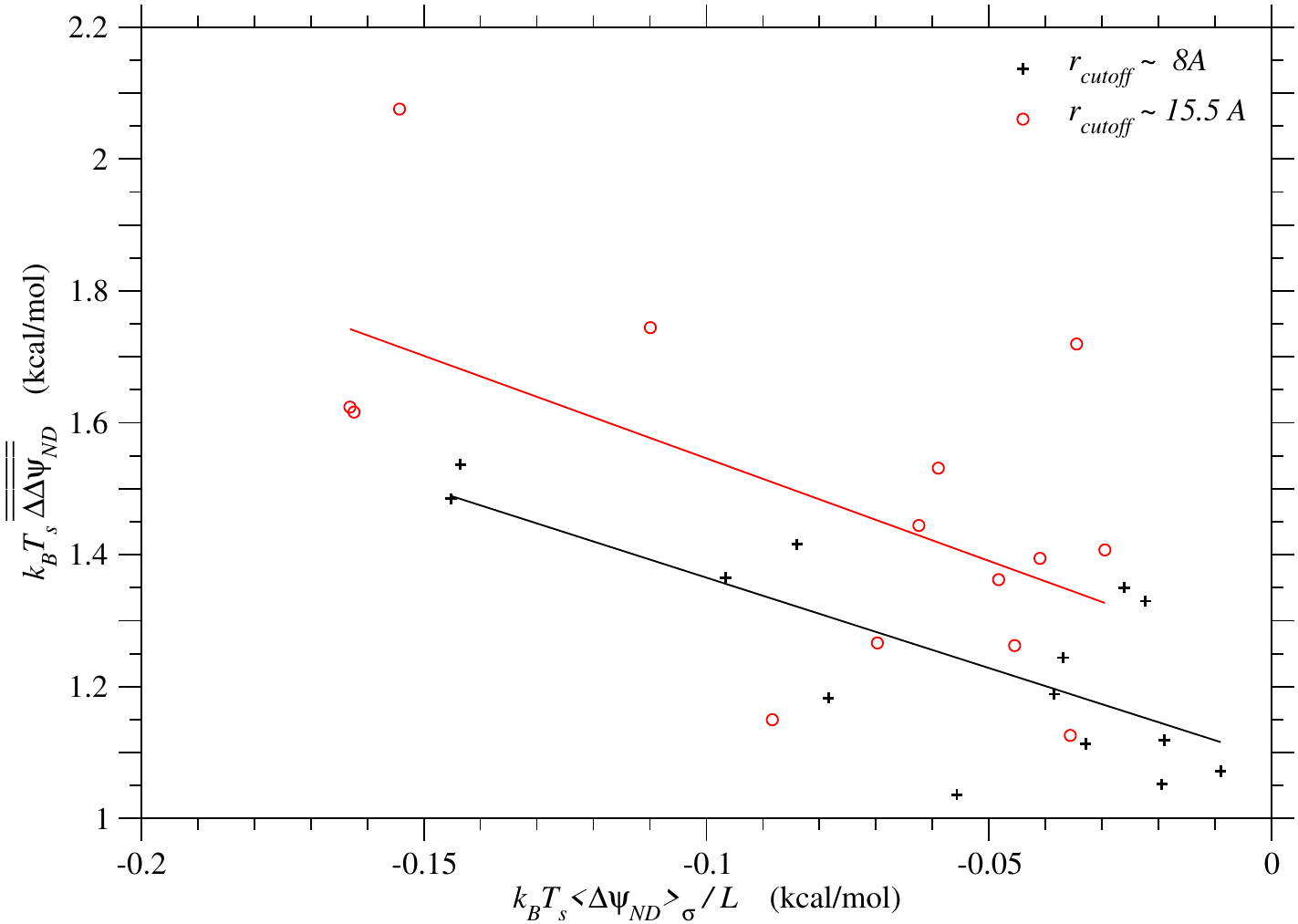}
}
}%  SUPPLEMENT
\TEXT{
\centerline{
\includegraphics*[width=82mm,angle=0]{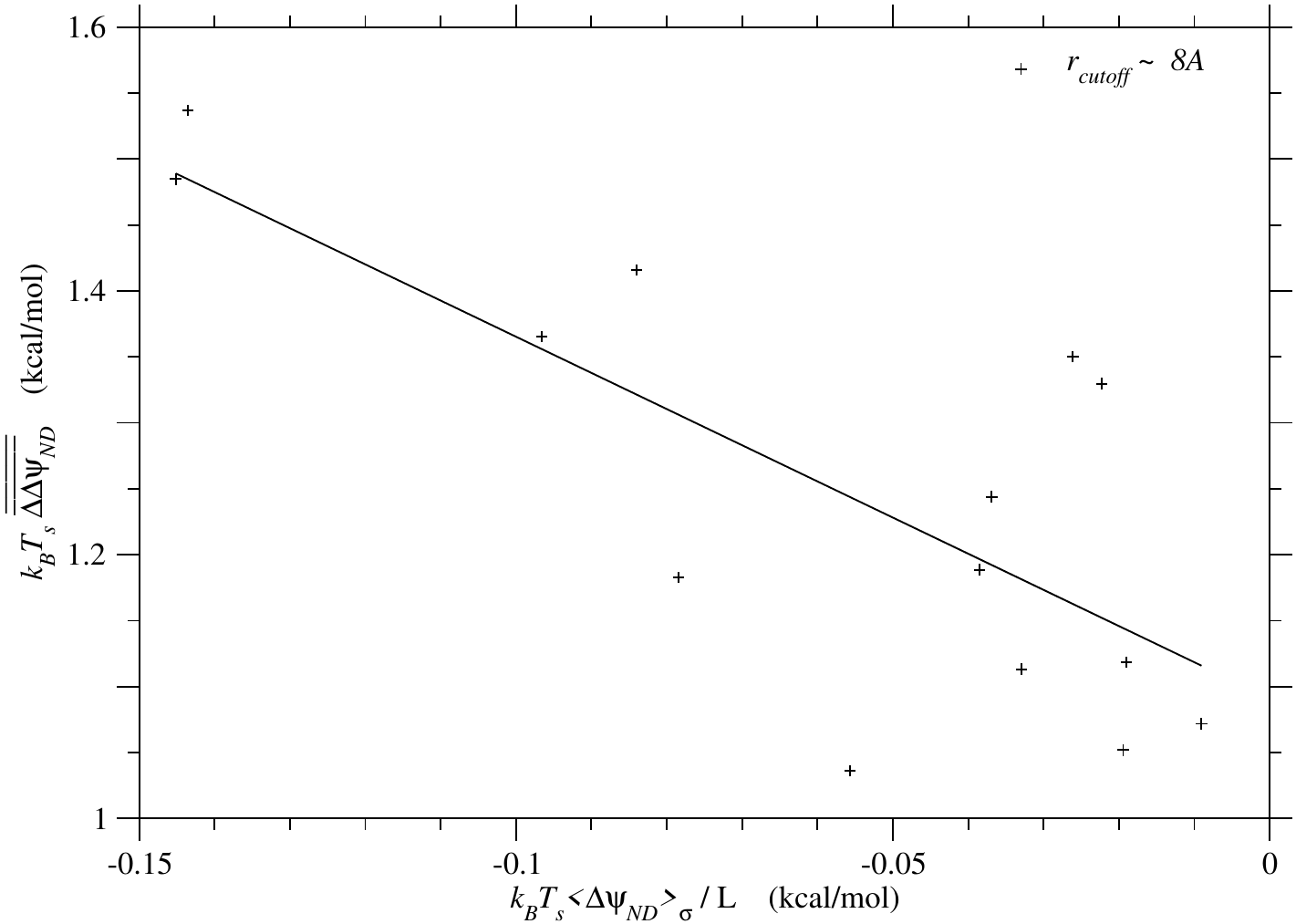}
}
}%  TEXT
}%  FigureInLegends
\vspace*{1em}
\caption{
\SUPPLEMENT{
\label{sfig: dG_over_L_vs_ddG}
}%  SUPPLEMENT
\TEXT{
\label{fig: dG_over_L_vs_ddG}
}%  TEXT
\FigureLegends{
\BF{
The sample average of folding free energy change, 
$\overline{\overline{\Delta\Delta G_{ND}}} \simeq k_B T_s \overline{\overline{\Delta\Delta \psi_{ND}}}$,
is plotted against the ensemble average of folding free energy per residue, 
$\langle \Delta G_{ND} \rangle_{\VEC{\sigma}} / L \simeq k_B T_s \langle \Delta \psi_{ND} \rangle_{\VEC{\sigma}} / L$,
for each protein family.
}
\SUPPLEMENT{
In the case of the cutoff distance 8 \AA, 
the correlation coefficient is $r = -0.75$, and the regression line is 
$\overline{\overline{ \Delta \Delta G_{ND}(\sigma^N_{j \neq i}, \sigma^N_i\rightarrow \sigma_i) } }
= - 2.74 \langle \Delta G_{ND} \rangle_{\VEC{\sigma}} / L + 1.09$.
In the case of $r_{\script{cutoff}} \sim 15.5$ \AA, 
the correlation coefficient is $r = -0.59$, and the regression line is 
$\overline{\overline{ \Delta \Delta G_{ND}(\sigma^N_{j \neq i}, \sigma^N_i\rightarrow \sigma_i) } }
= - 3.11 \langle \Delta G_{ND} \rangle_{\VEC{\sigma}} / L + 1.24$.
}%  SUPPLEMENT
\TEXT{
The correlation coefficient is $r = -0.75$, and the regression line is 
$\overline{\overline{ \Delta \Delta G_{ND}(\sigma^N_{j \neq i}, \sigma^N_i\rightarrow \sigma_i) } }
= - 2.74 \langle \Delta G_{ND} \rangle_{\VEC{\sigma}} / L + 1.09$.
See \Fig{\ref{sfig: dG_over_L_vs_ddG}} for $r_{\script{cutoff}} \sim 15.5$ \AA.
}%  TEXT
The free energies are in kcal/mol units.
}%  FigureLegends
}
\end{figure*}

}%  FigE

\FigF{

\CLEARPAGE
 
\begin{figure*}[h!]
\FigureInLegends{
\SUPPLEMENT{
\centerline{
\includegraphics*[width=82mm,angle=0]{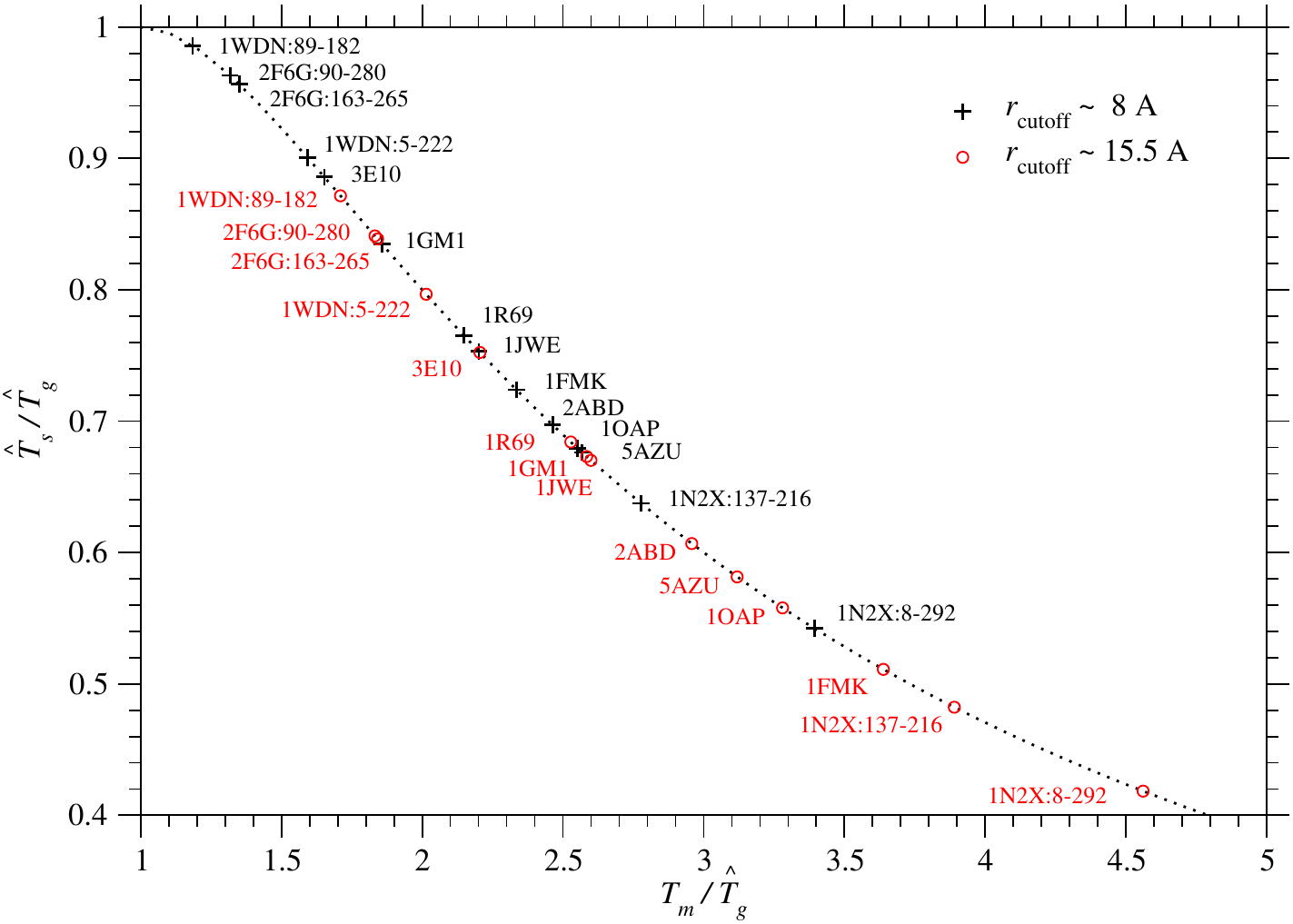}
}
}%  SUPPLEMENT
\TEXT{
\centerline{
\includegraphics*[width=82mm,angle=0]{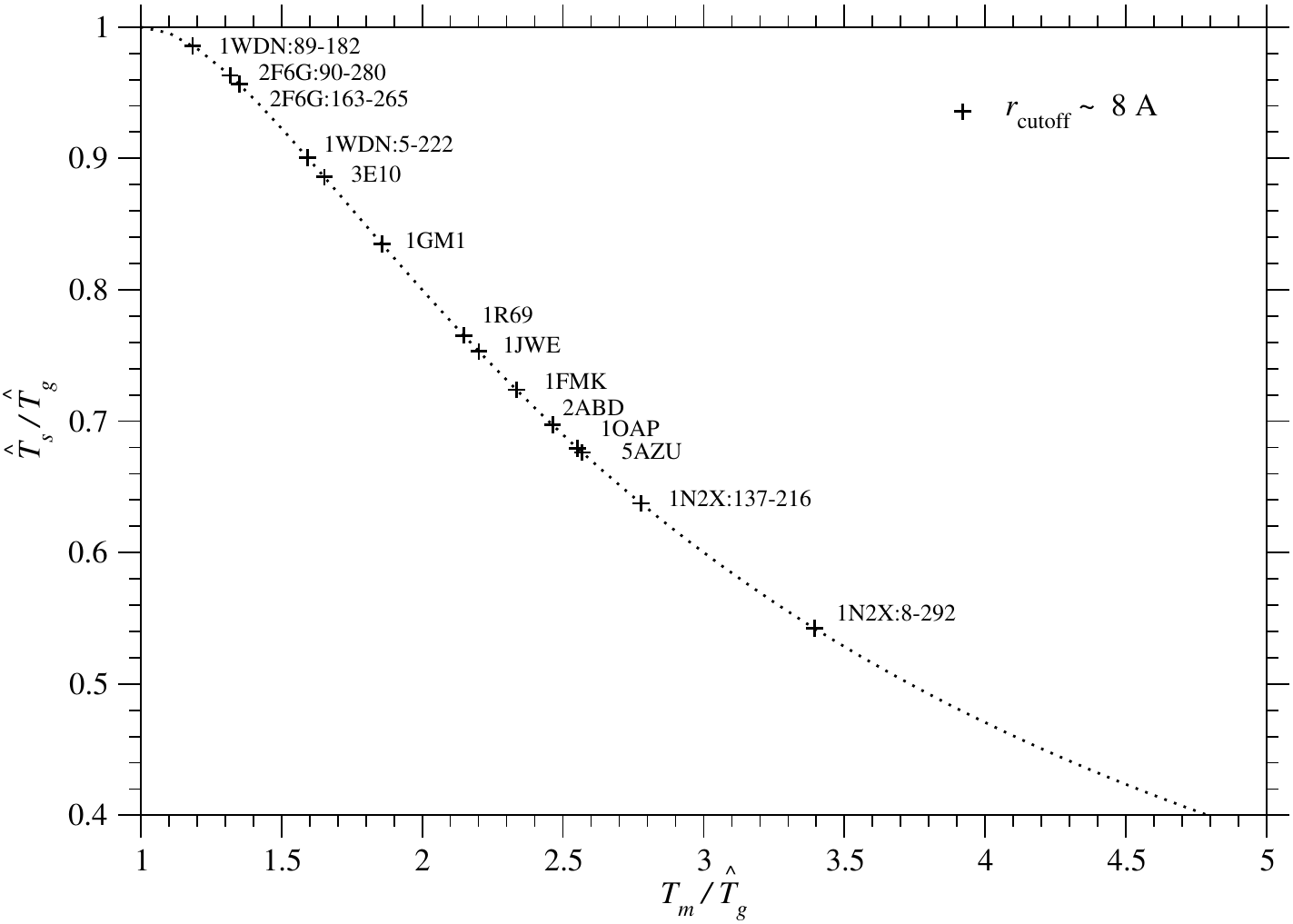}
}
}%  TEXT
}%  FigureInLegends
\vspace*{1em}
\caption{
\SUPPLEMENT{
\label{sfig: Tm_over_Tg_vs_Ts_over_Tg}
}%  SUPPLEMENT
\TEXT{
\label{fig: Tm_over_Tg_vs_Ts_over_Tg}
}%  TEXT
\FigureLegends{
\BF{$\hat{T}_s/\hat{T}_g$ is plotted against $T_m / \hat{T}_g$ for each protein domain.
}
A dotted curve corresponds to 
\Eq{\ref{\EQ: relationship_among_characteristic_T}},
$\hat{T}_s/\hat{T}_g = 2 (T_m / \hat{T}_g) / ((T_m/\hat{T}_g)^2 + 1)$.
\SUPPLEMENT{
Plus and open circle marks indicate the values estimated with $r_{\script{cutoff}} \sim 8$ and $15.5$ \AA,
respectively.
}%  SUPPLEMENT
\TEXT{
Plus marks indicate the values estimated with $r_{\script{cutoff}} \sim 8$ \AA.
See \Fig{\ref{sfig: Tm_over_Tg_vs_Ts_over_Tg}} for $r_{\script{cutoff}} \sim 15.5$ \AA.
}%  TEXT
The effective temperature $T_s$ for selection and glass transition temperature $T_g$ must satisfy $T_s < T_g < T_m$ 
for proteins to be able to fold into unique native structures.
}%  FigureLegends
}
\end{figure*}

}%  FigF

\FigG{

\CLEARPAGE
 
\begin{figure*}[h!]
\FigureInLegends{
\SUPPLEMENT{
\centerline{
\includegraphics*[width=82mm,angle=0]{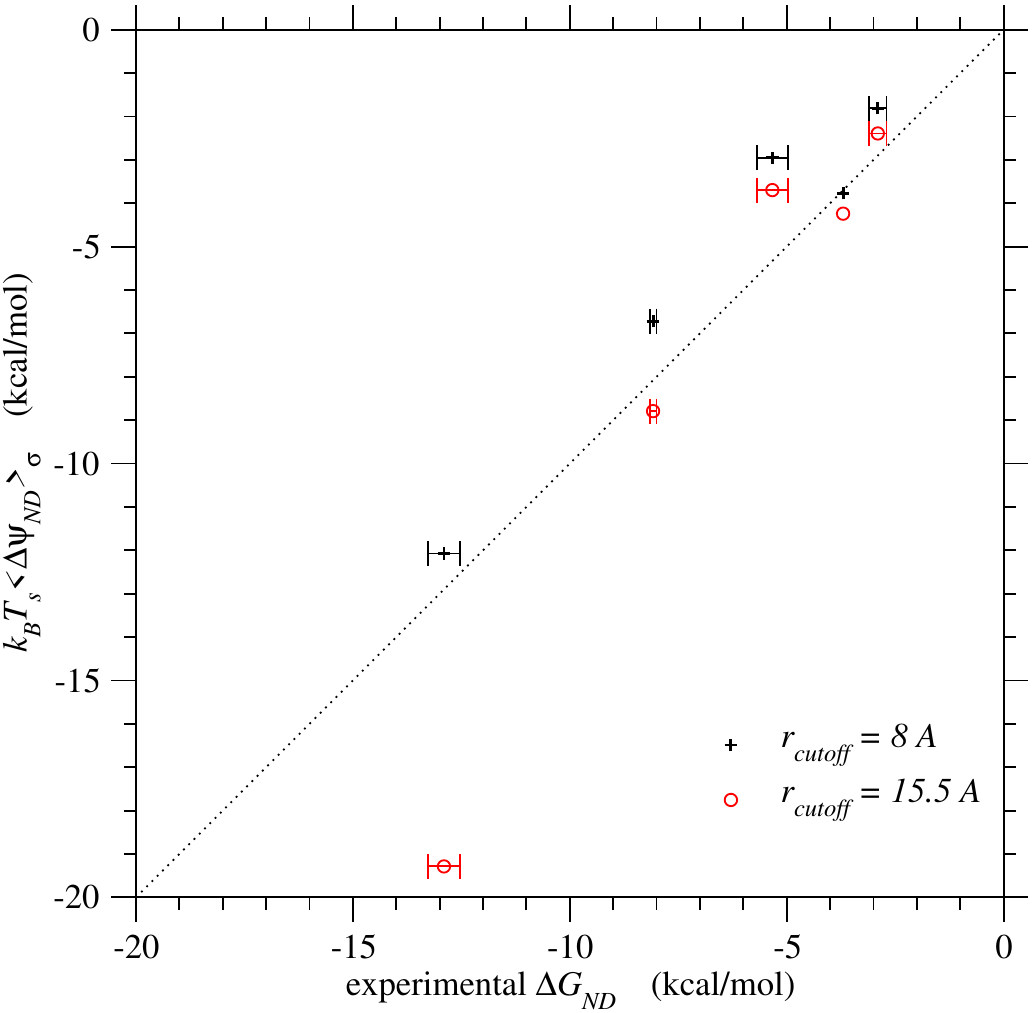}
}
}%  SUPPLEMENT
\TEXT{
\centerline{
\includegraphics*[width=82mm,angle=0]{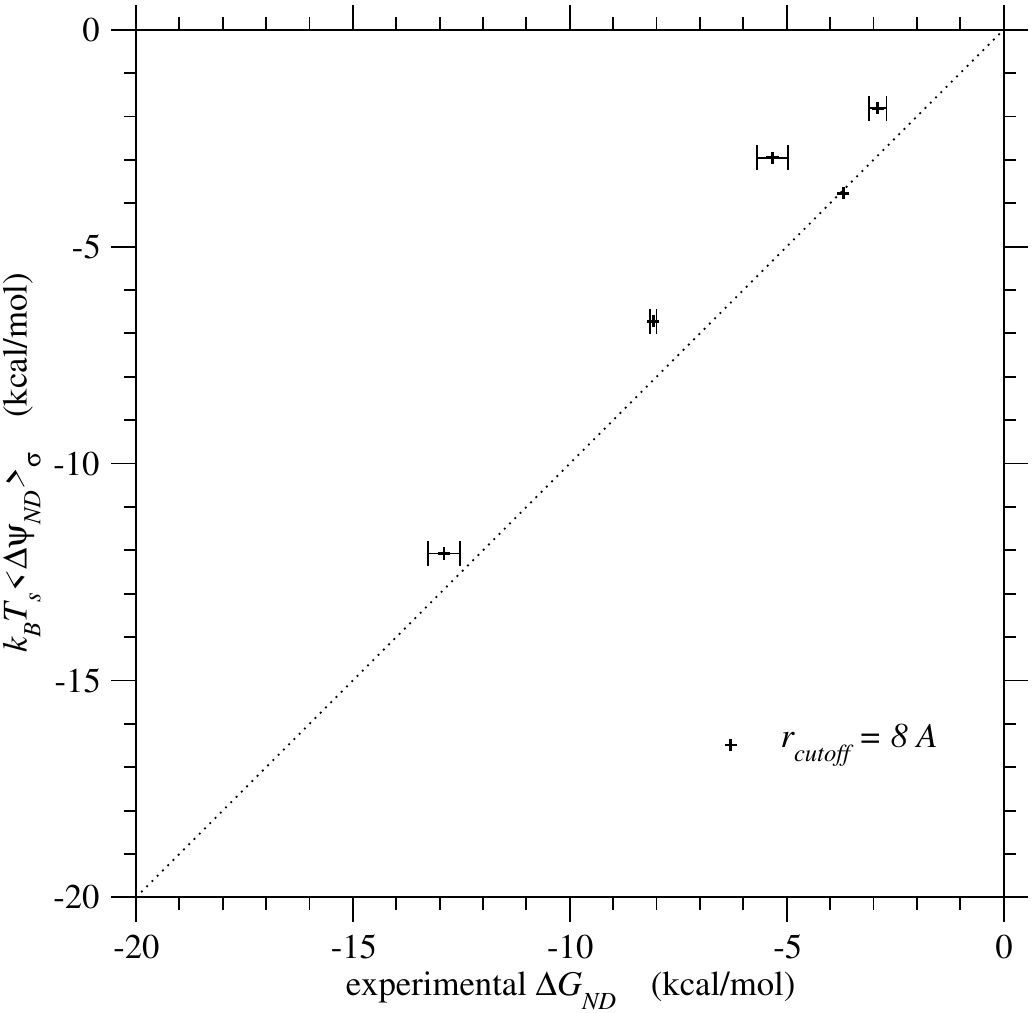}
}
}%  TEXT
}%  FigureInLegends
\vspace*{1em}
\caption{
\SUPPLEMENT{
\label{sfig: dG_exp_vs_8_and_16A}
}%  SUPPLEMENT
\TEXT{
\label{fig: dG_exp_vs_8_and_16A}
}%  TEXT
\FigureLegends{
\BF{Folding free energies, 
$\langle \Delta G_{ND} \rangle_{\VEC{\sigma}} \simeq k_B T_s \langle \Delta \psi_{ND} \rangle_{\VEC{\sigma}}$,
predicted by the present method are plotted against their experimental values, $\Delta G_{ND}(\VEC{\sigma_N})$.
}
\SUPPLEMENT{
Plus and open circle marks 
indicate the values estimated with $r_{\script{cutoff}} \sim 8$ and $15.5$ \AA,
respectively.
}%  SUPPLEMENT
\TEXT{
Plus marks 
indicate the values estimated with $r_{\script{cutoff}} \sim 8$ \AA.
See \Fig{\ref{sfig: dG_exp_vs_8_and_16A}} for $r_{\script{cutoff}} \sim 15.5$ \AA.
}%  TEXT
The free energies are in kcal/mol units.
}%  FigureLegends
}
\end{figure*}

}%  FigG

% End of figures_JTB_1.tex
% \input{figures_JTB_2.tex}

\TextFig{

\ifdefined\FigH
\else
\NoFigureInText{
\newcommand{\FigH}[1]{#1}
}%  NoFigureInText
\FigureInText{
\newcommand{\FigH}[1]{}
}%  FigureInText
\fi

\ifdefined\FigI
\else
\NoFigureInText{
\newcommand{\FigI}[1]{#1}
}%  NoFigureInText
\FigureInText{
\newcommand{\FigI}[1]{}
}%  FigureInText
\fi

\ifdefined\FigJ
\else
\NoFigureInText{
\newcommand{\FigJ}[1]{#1}
}%  NoFigureInText
\FigureInText{
\newcommand{\FigJ}[1]{}
}%  FigureInText
\fi

\ifdefined\FigK
\else
\NoFigureInText{
\newcommand{\FigK}[1]{#1}
}%  NoFigureInText
\FigureInText{
\newcommand{\FigK}[1]{}
}%  FigureInText
\fi

\ifdefined\FigL
\else
\NoFigureInText{
\newcommand{\FigL}[1]{#1}
}%  NoFigureInText
\FigureInText{
\newcommand{\FigL}[1]{}
}%  FigureInText
\fi

\ifdefined\FigM
\else
\NoFigureInText{
\newcommand{\FigM}[1]{#1}
}%  NoFigureInText
\FigureInText{
\newcommand{\FigM}[1]{}
}%  FigureInText
\fi

\ifdefined\FigN
\else
\NoFigureInText{
\newcommand{\FigN}[1]{#1}
}%  NoFigureInText
\FigureInText{
\newcommand{\FigN}[1]{}
}%  FigureInText
\fi

\ifdefined\FigO
\else
\NoFigureInText{
\newcommand{\FigO}[1]{#1}
}%  NoFigureInText
\FigureInText{
\newcommand{\FigO}[1]{}
}%  FigureInText
\fi

\ifdefined\FigP
\else
\NoFigureInText{
\newcommand{\FigP}[1]{#1}
}%  NoFigureInText
\FigureInText{
\newcommand{\FigP}[1]{}
}%  FigureInText
\fi

\renewcommand{\SUPPLEMENT}[1]{}

\ifdefined\CLEARPAGE

\NoFigureInText{
\renewcommand{\CLEARPAGE}{\FigureLegends{\clearpage\newpage}}
}%  NoFigureInText
\FigureInText{
\renewcommand{\CLEARPAGE}{}
}%  FigureInText

\else

\NoFigureInText{
\newcommand{\CLEARPAGE}{\FigureLegends{\clearpage\newpage}}
}%  NoFigureInText
\FigureInText{
\newcommand{\CLEARPAGE}{}
}%  FigureInText

\fi

}%  TextFig

\SupFig{

\ifdefined\FigH
\renewcommand{\FigH}[1]{#1}
\else
\newcommand{\FigH}[1]{#1}
\fi

\ifdefined\FigI
\renewcommand{\FigI}[1]{#1}
\else
\newcommand{\FigI}[1]{#1}
\fi

\ifdefined\FigJ
\renewcommand{\FigJ}[1]{#1}
\else
\newcommand{\FigJ}[1]{#1}
\fi

\ifdefined\FigK
\renewcommand{\FigK}[1]{#1}
\else
\newcommand{\FigK}[1]{#1}
\fi

\ifdefined\FigL
\renewcommand{\FigL}[1]{#1}
\else
\newcommand{\FigL}[1]{#1}
\fi

\ifdefined\FigM
\renewcommand{\FigM}[1]{#1}
\else
\newcommand{\FigM}[1]{#1}
\fi

\ifdefined\FigN
\renewcommand{\FigN}[1]{#1}
\else
\newcommand{\FigN}[1]{#1}
\fi

\ifdefined\FigO
\renewcommand{\FigO}[1]{#1}
\else
\newcommand{\FigO}[1]{#1}
\fi

\ifdefined\FigP
\renewcommand{\FigP}[1]{#1}
\else
\newcommand{\FigP}[1]{#1}
\fi

\renewcommand{\SUPPLEMENT}[1]{#1}
\ifdefined\CLEARPAGE
\renewcommand{\CLEARPAGE}{\FigureLegends{\clearpage\newpage}}
\else
\newcommand{\CLEARPAGE}{\FigureLegends{\clearpage\newpage}}
\fi
}%  SupFig

\renewcommand{\SkipFigure}[1]{}

\FigH{

\CLEARPAGE
 
\begin{figure*}[h!]
\FigureInLegends{
\noindent
\SUPPLEMENT{
\hspace*{1em} (a) $r_{\script{cutoff}} \sim 8$\AA\ 
\hspace*{14em} (b) $r_{\script{cutoff}} \sim 15.5$\AA\ 

}%  SUPPLEMENT
\TEXT{
\centerline{
{\small{$r_{\script{cutoff}} \sim 8$\AA\ }}
}
}%  TEXT
\SUPPLEMENT{
\centerline{
\includegraphics*[width=82mm,angle=0]{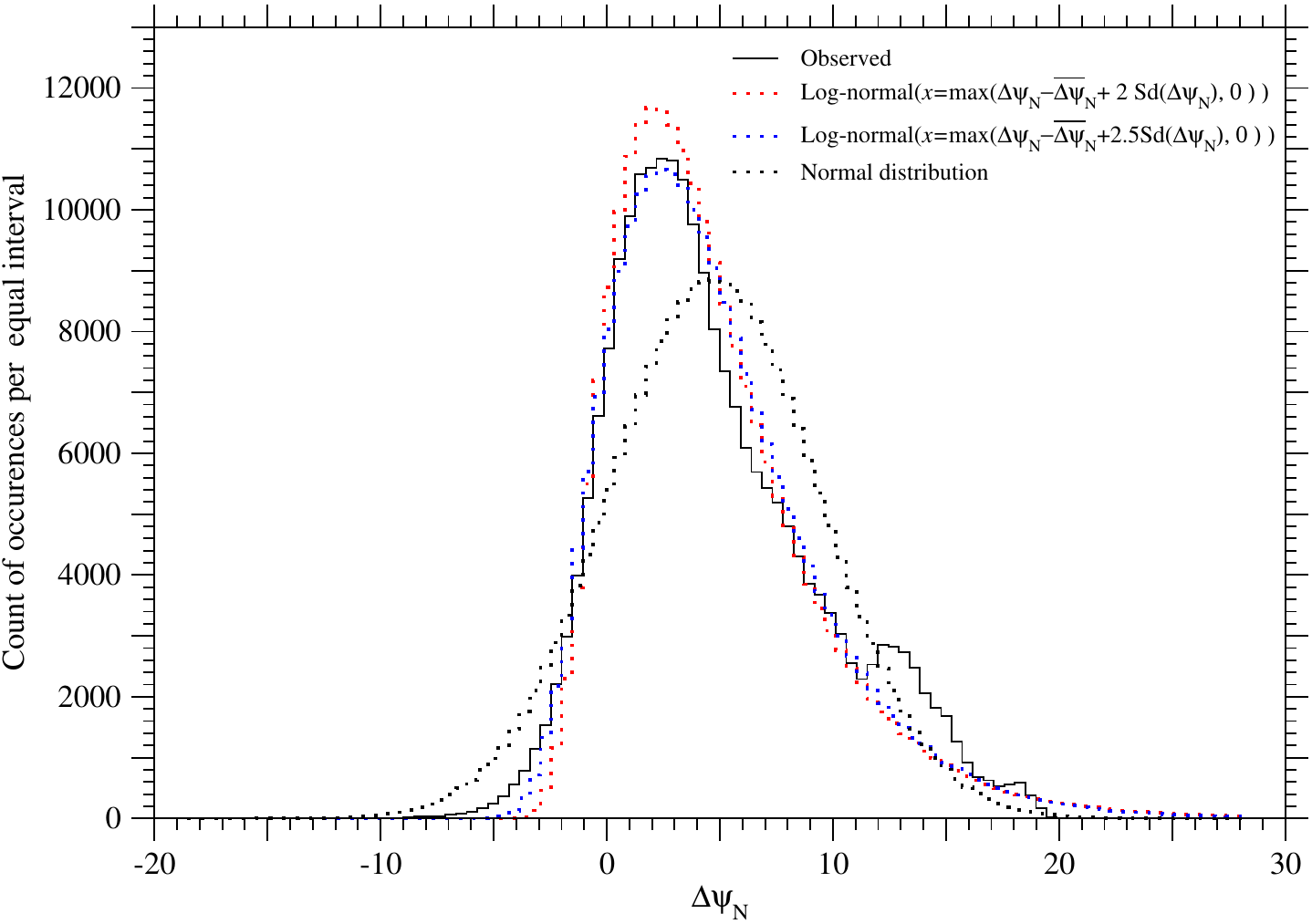}
\includegraphics*[width=82mm,angle=0]{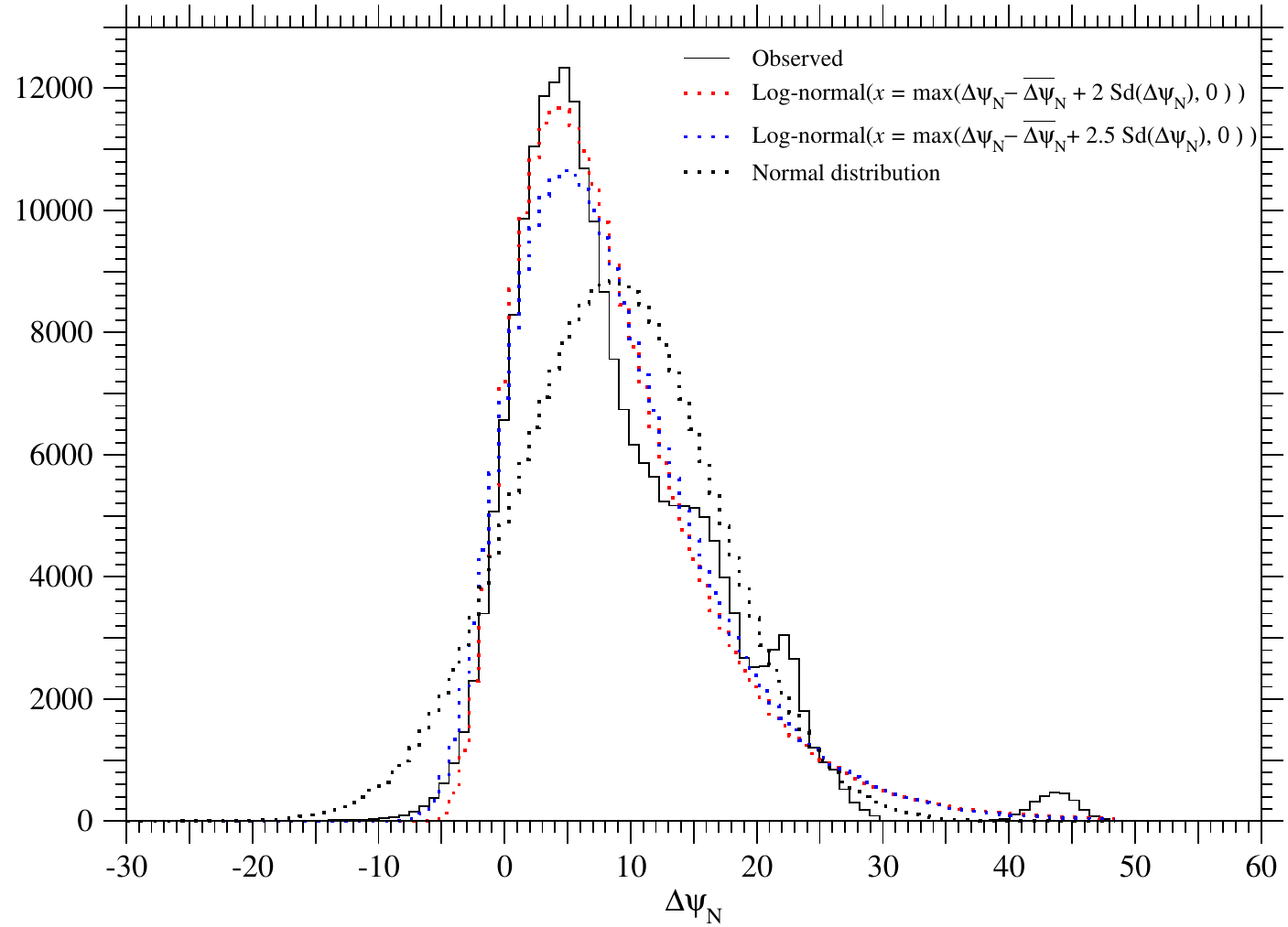}
}
}%  SUPPLEMENT
\TEXT{
\centerline{
\includegraphics*[width=82mm,angle=0]{FIGS/PDZ/1gm1-a_16-96_full_non_del_dca0_205_0_20_simple-gauge_dPhiN_distr}
}
}%  TEXT
}%  FigureInLegends
\vspace*{1em}
\caption{
\SUPPLEMENT{
\label{sfig: 1gm1-a:16-96_full_non_del_dca0_205_0_20_simple-gauge_dPhiN_distr}
\label{sfig: 1gm1-a:16-96_full_non_del_dca0_33_0_20_simple-gauge_dPhiN_distr}
}%  SUPPLEMENT
\TEXT{
\label{fig: 1gm1-a:16-96_full_non_del_dca0_205_0_20_simple-gauge_dPhiN_distr}
}%  TEXT
\FigureLegends{
\BF{
The observed frequency distribution and the fitted distributions of $\Delta \psi_N$ in the PDZ protein family.
}
A black solid line indicates the observed frequency distribution of $\Delta \psi_N$ per equal interval in homologous sequences of the PDZ protein family,
and red dotted and blue dotted lines indicate the total frequencies of log-normal distributions with $n_{\script{shift}} = 2$ or $2.5$
and parameters estimated with the mean and variance of the observed distribution for each protein;
see \Eqs{\REF{\EQ: log-normal} to \REF{\EQ: statistics_for_log-normal}}.
A black dotted line indicates the total frequencies of normal distributions the mean and variance of which are equal to those of 
the observed distribution for each protein.
Only representatives of unique sequences with no deletions, which are at least 20\% different from each other, are employed;
the total count is equal to 
222,466 over 335 homologous sequences,
which is almost equal to $M_{\script{eff}}$ in \Table{\ref{\TBL: Proteins_studied}}.
\TEXT{
See \Fig{\ref{sfig: 1gm1-a:16-96_full_non_del_dca0_33_0_20_simple-gauge_dPhiN_distr}} for $r_{\script{cutoff}} \sim 15.5$ \AA.
}%  TEXT
}%  FigureLegends
}
\end{figure*}

}%  FigH

\SUPPLEMENT{

\CLEARPAGE
 
\begin{figure*}[h!]
\FigureInLegends{
\noindent
\hspace*{1em}
(a) $r_{\script{cutoff}} \sim 8$\AA\ 
\hspace*{14em}
(b) $r_{\script{cutoff}} \sim 15.5$\AA\ 

\centerline{
\includegraphics*[width=82mm,angle=0]{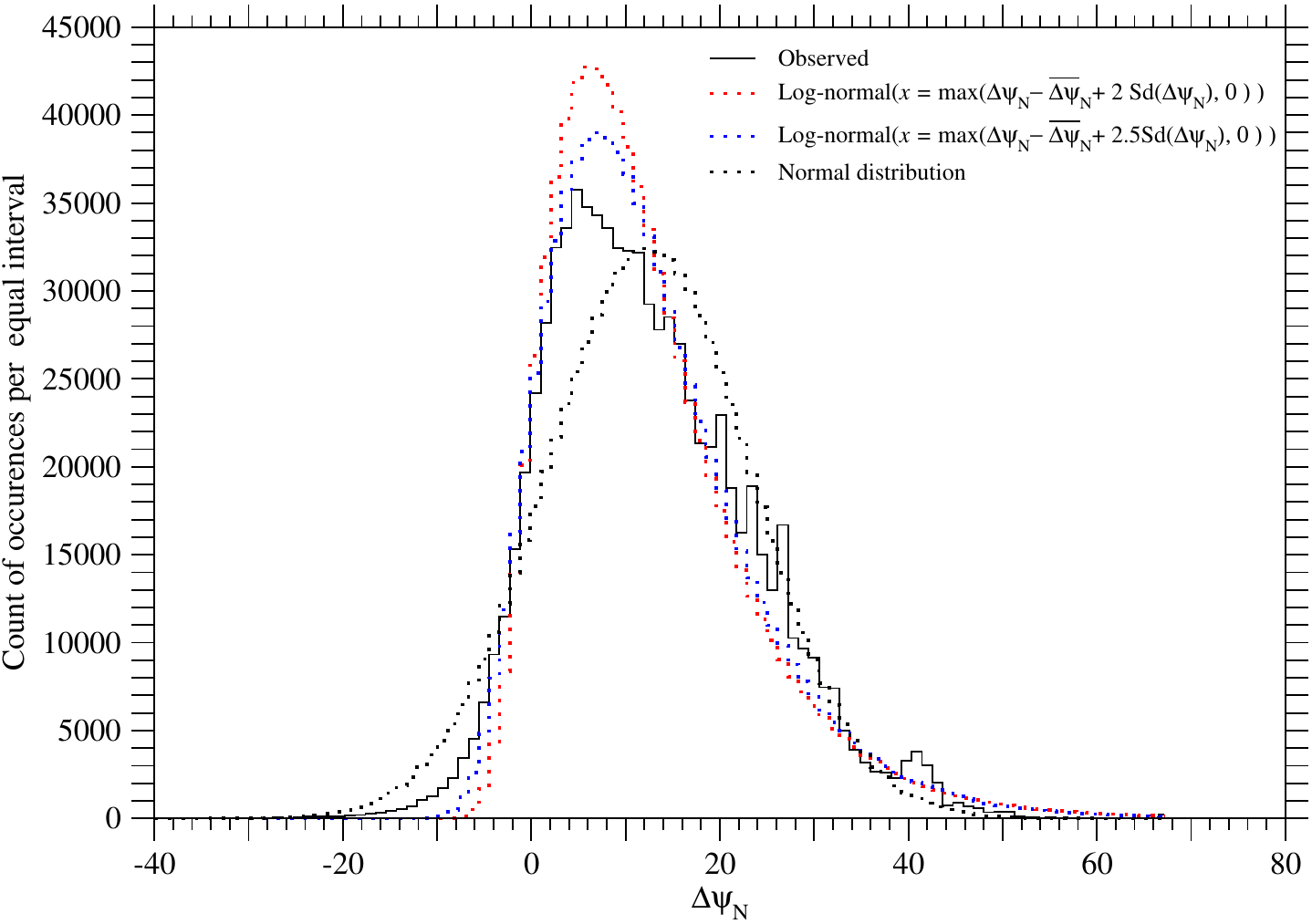}
\includegraphics*[width=82mm,angle=0]{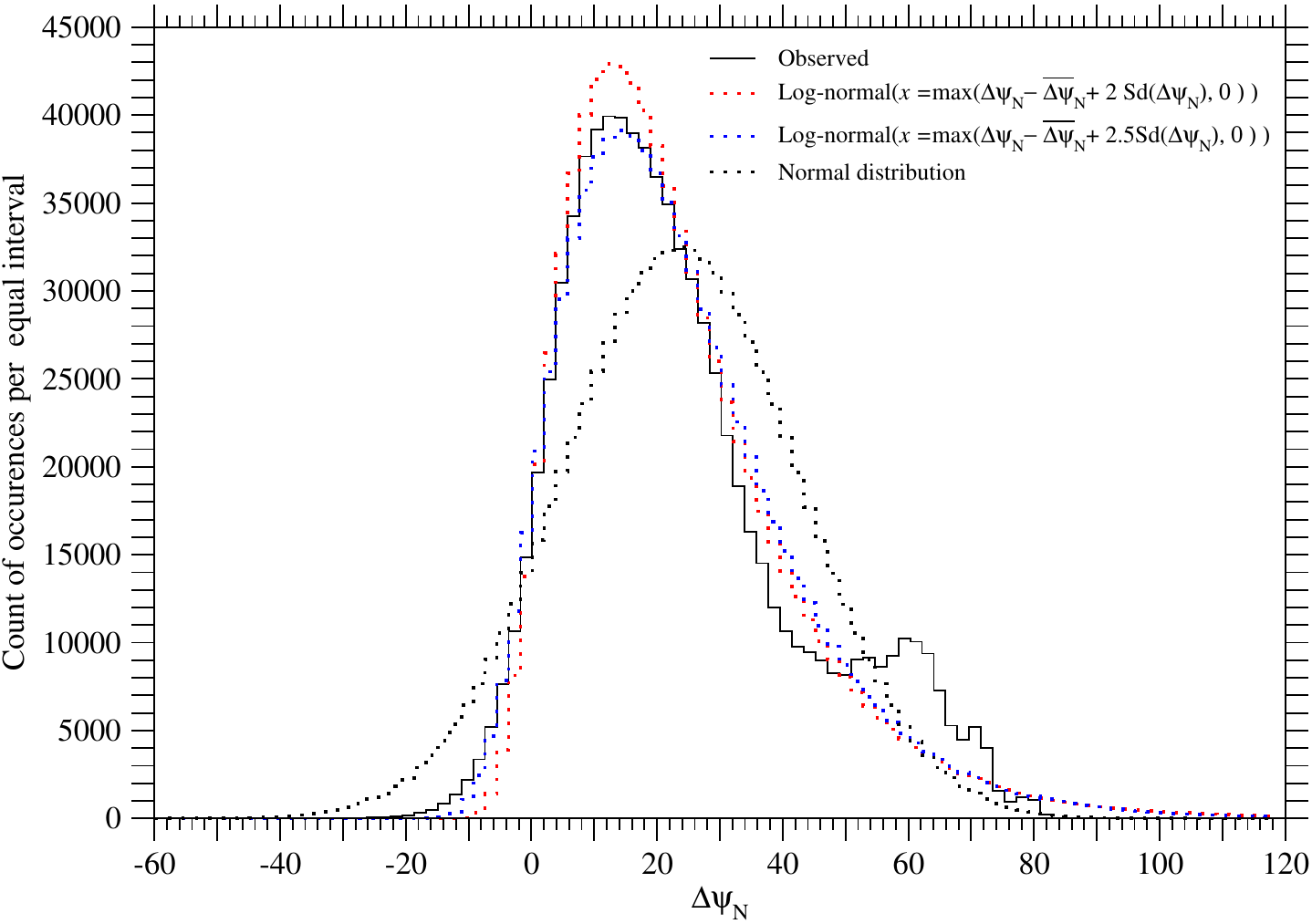}
}
}%  FigureInLegends
\vspace*{1em}
\caption{
\label{sfig: 1n2x-a:8-292_full_non_del_dca0_13_0_20_simple-gauge_dPhiN_distr}
\label{fig: 1n2x-a:8-292_full_non_del_dca0_13_0_20_simple-gauge_dPhiN_distr}
\label{sfig: 1n2x-a:8-292_full_non_del_dca0_175_0_20_simple-gauge_dPhiN_distr}
\label{fig: 1n2x-a:8-292_full_non_del_dca0_175_0_20_simple-gauge_dPhiN_distr}
\FigureLegends{
\BF{
The observed frequency distribution and the fitted distribution of $\Delta \psi_N$ in the Methyltransf\_5 family of the domain, 1N2X-A:8-292.
}
A black solid line indicates the observed frequency distribution of $\Delta \psi_N$ per equal interval 
in homologous sequences of the Methyltransf\_5 protein family,
and red dotted and blue dotted lines indicate the total frequencies of log-normal distributions 
with $n_{\script{shift}} = 2$ or $2.5$
and parameters estimated with the mean and variance of the observed distribution for each protein;
see \Eqs{\REF{\EQ: log-normal} to \REF{\EQ: statistics_for_log-normal}}.
A black dotted line indicates the total frequencies of normal distributions the mean and variance of which are equal to those of 
the observed distribution for each protein.
Only representatives of unique sequences, which are at least 20\% different from each other, are employed;
the total count is equal to 
814549 over 354 homologous sequences,
which is almost equal to $M_{\script{eff}}$ in \Table{\ref{\TBL: Proteins_studied}}.
}%  FigureLegends
}
\end{figure*}

}%  SUPPLEMENT

\FigI{

\CLEARPAGE
 
\begin{figure*}[h!]
\FigureInLegends{
\centerline{
\includegraphics*[width=82mm,angle=0]{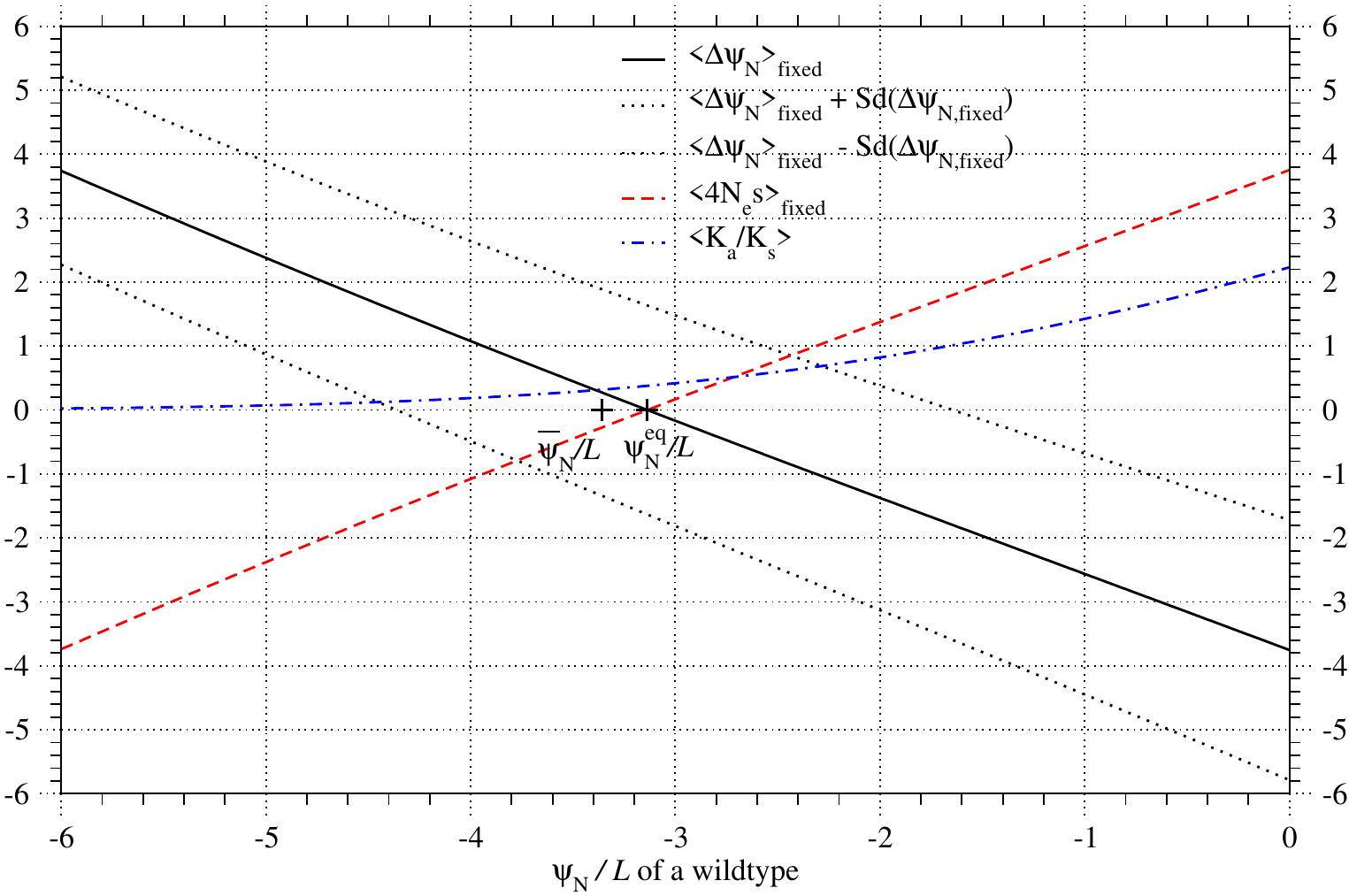}
}
}%  FigureInLegends
\vspace*{1em}
\caption{
\SUPPLEMENT{
\label{sfig: ave_ddPhi_vs_dPhiN_of_fixed_mutants_logG-2sd_PDZ_8A}
}%  SUPPLEMENT
\TEXT{
\label{fig: ave_ddPhi_vs_dPhiN_of_fixed_mutants_logG-2sd_PDZ_8A}
}%  TEXT
\FigureLegends{
\BF{
The average 
of $\Delta \psi_{N} (\simeq \Delta\Delta \psi_{ND})$ 
over fixed single nucleotide nonsynonymous mutations versus $\psi_{N} / L$ of 
a wildtype for the PDZ protein family.
}
The averages of $\Delta \psi_{N} (\simeq \Delta\Delta \psi_{ND})$ 
and $4N_e s$ over the fixed mutants,
and the average of $K_a/K_s (\equiv u(s)/u(0))$ over all the mutants are
plotted against $\psi_{N} / L$ of a wildtype by solid, broken, and dash-dot lines,
respectively; $q_m = 1 / (2 \times 10^6)$ is assumed.
Dotted lines show the values of $\langle \Delta \psi_{N} \rangle_{\script{fixed}} \pm \textrm{sd}$,
where the $\textrm{sd}$ is the standard deviation of $\Delta \psi_{N}$ over fixed mutants.
Fixation probability has been calculated with $\Delta\Delta \psi_{ND} \simeq \Delta \psi_{N}$;
see \Eqs{\REF{\EQ: fixation_probability} and \REF{\EQ: selective_advantage_and_dPsi_N}}.
Here the empirical relationships of \Eqs{\REF{\EQ: regression_of_dPsi_on_Psi} and \REF{\EQ: var_of_dPsi}}
are assumed; that is,
the mean of $\Delta \psi_N$ linearly decreases as $\psi_N$ increases,
but the standard deviation of $\Delta \psi_N$ is constant irrespective of $\psi_N$.
The slope ($\alpha_{\psi_N}$) and intercept 
($- \alpha_{\psi_N}\overline{\psi_N}/L + \overline{\overline{\Delta \psi_N}}$)
and the average of $\text{Sd}(\Delta \psi_N)$ over homologous sequences  
that are estimated with $r_{\script{cutoff}} \sim 8$\AA\ for the PDZ
and listed in \Table{\ref{\TBL: ddPsi_with_8A}} are employed here. 
The distribution of $\Delta \psi_N$ due to single nucleotide nonsynonymous mutations is approximated by
a log-normal distribution with $n_{\script{shift}} = 2.0$;
see \Eqs{\REF{\EQ: log-normal} to \REF{\EQ: statistics_for_log-normal}}.
The $\psi_{N}^{\script{eq}}$, where $\langle \Delta\Delta \psi_{ND} \rangle_{\script{fixed}} \simeq \langle \Delta \psi_{N} \rangle_{\script{fixed}} = 0$, 
is the stable equilibrium value
of $\psi_{N}$ in the protein evolution of the PDZ protein family.
The $\psi_{N}^{\script{eq}}$ is close to the average of $\psi_N$ over homologous sequences ($\overline{\psi_N}$),
indicating that the present approximations for $\psi_{N}^{\script{eq}}$ and
for $\overline{\psi_N} = \langle \psi_N \rangle_{\VEC{\sigma}}$ are consistent to each other.
}%  FigureLegends
}
\end{figure*}

}%  FigI

\SUPPLEMENT{

\CLEARPAGE
 
\begin{figure*}[h!]
\FigureInLegends{
\centerline{
\includegraphics*[width=82mm,angle=0]{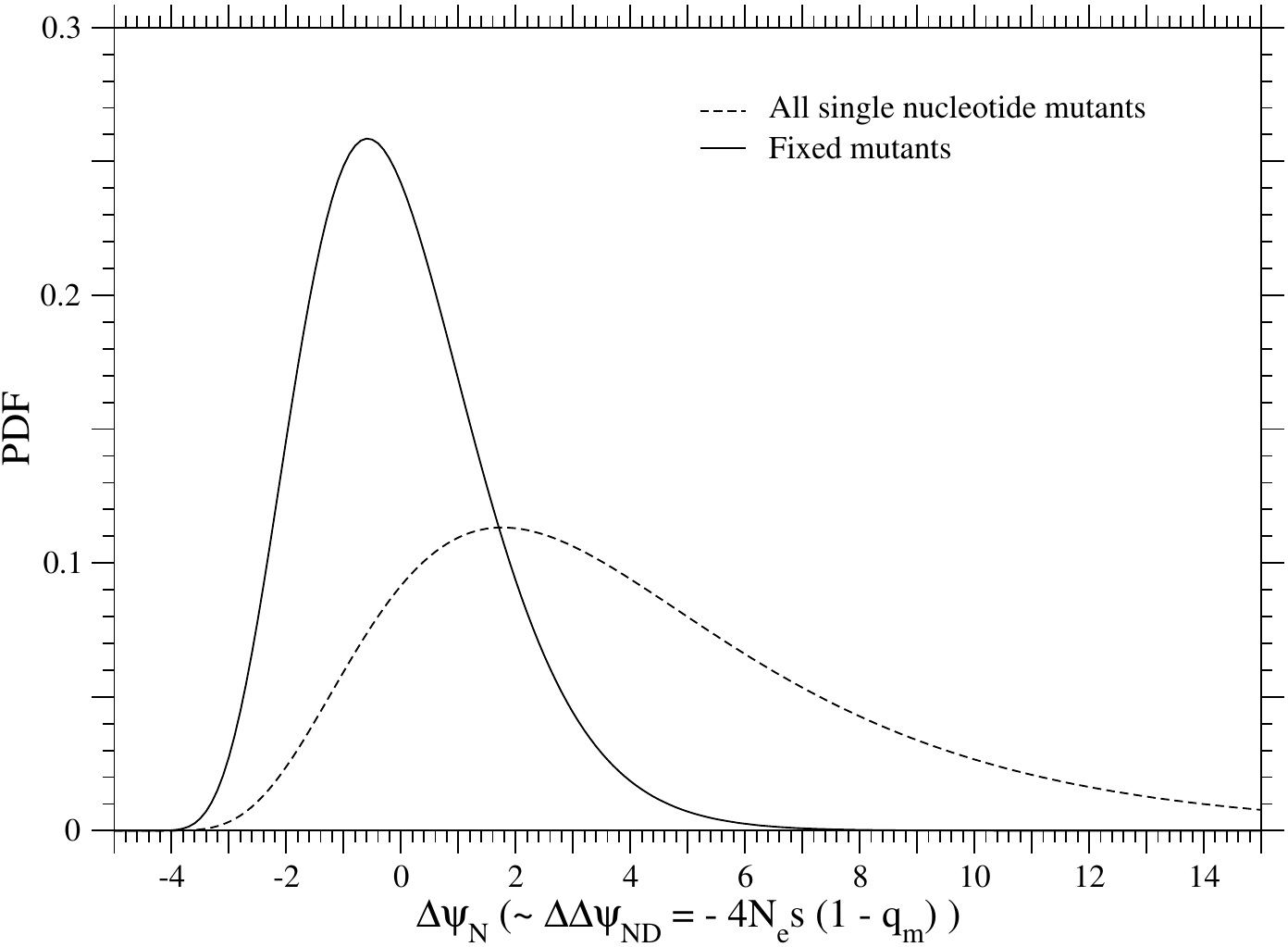}
\includegraphics*[width=82mm,angle=0]{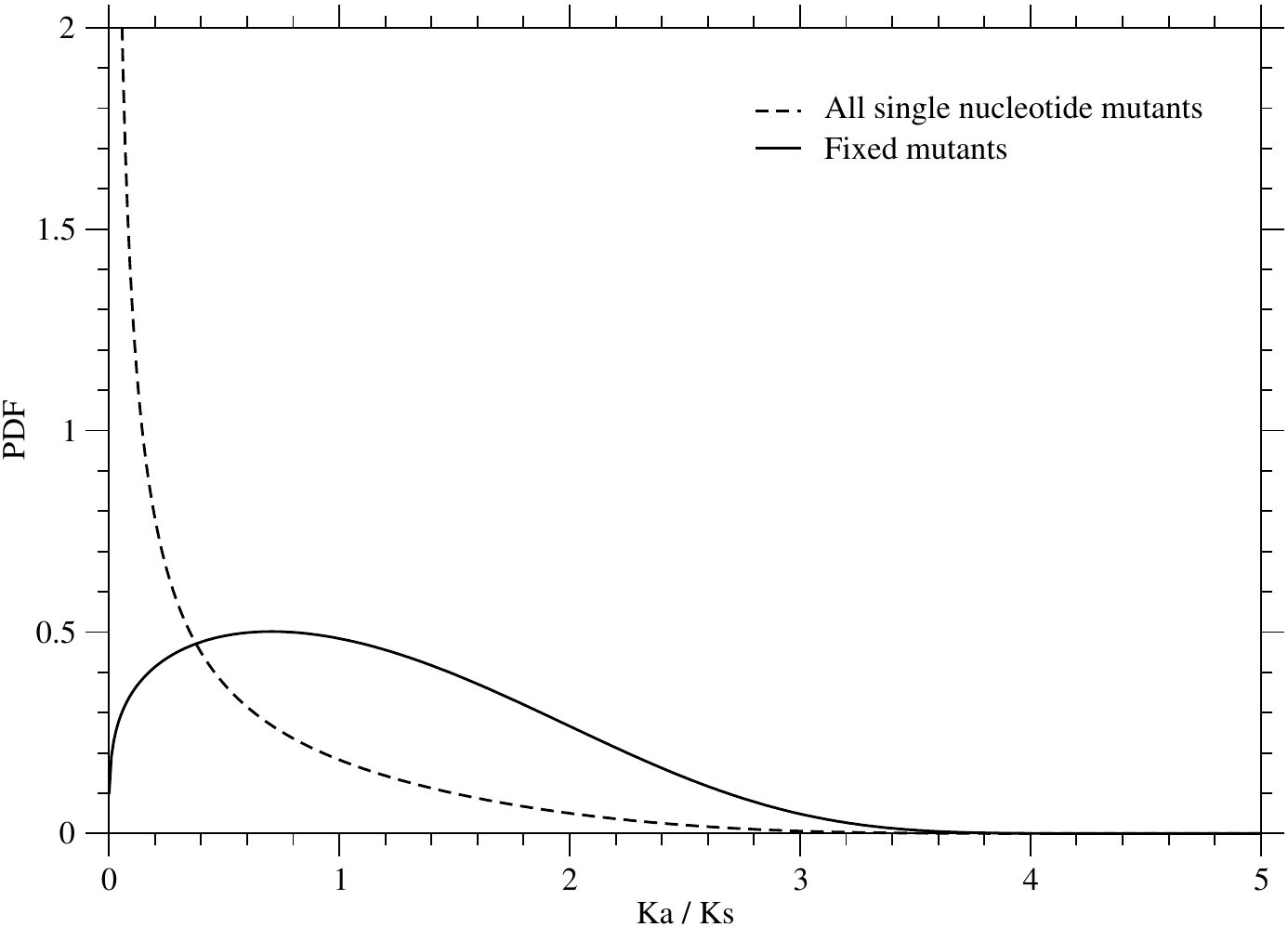}
}
}%  FigureInLegends
\vspace*{1em}
\caption{
\label{sfig: pdf_of_ddPhi_at_PhiNe_logG-2sd_PDZ_8A}
\label{fig: pdf_of_ddPhi_at_PhiNe_logG-2sd_PDZ_8A}
\label{sfig: pdf_of_ka_over_ks_at_PhiNe_logG-2sd_PDZ_8A}
\label{fig: pdf_of_ka_over_ks_at_PhiNe_logG-2sd_PDZ_8A}
\FigureLegends{
\BF{
PDFs of $\Delta \psi_N (\simeq \Delta\Delta \psi_{ND} = - 4N_e s (1 - q_m) )$ 
and of $K_a/K_s$ 
for all single nucleotide nonsynonymous mutants and for their fixed mutants 
at equilibrium 
($\langle \Delta \psi_N \rangle_{\script{fixed}} = 0$)
for the PDZ protein family.
}
$K_a/K_s$ is defined as the ratio of nonsynonymous to synonymous substitution rate per site, $u(s)/u(0)$;
see \Eq{\ref{seq: def_Ka_over_Ks}}.
Fixation probability has been calculated with $\Delta\Delta \psi_{ND} \simeq \Delta \psi_{N}$;
see \Eqs{\REF{\EQ: fixation_probability} and \REF{\EQ: selective_advantage_and_dPsi_N}}.
The equilibrium value $\psi_N^{\script{eq}}$, 
where $\langle \Delta\Delta \psi_{ND} \rangle_{\script{fixed}} \simeq \langle \Delta \psi_{N} \rangle_{\script{fixed}} = 0$, 
is calculated
by using 
the linear dependency 
of $\overline{\Delta \psi_N}$ on $\psi_N$ (\Eq{\ref{\EQ: regression_of_dPsi_on_Psi}})
and 
estimated values
with $r_{\script{cutoff}} \sim 8$\AA\  for the PDZ 
in \Tables{\ref{\TBL: ddPsi_with_8A}}.
The standard deviation of $\Delta \psi_N$ is approximated to be constant
and equal to $\overline{\mbox{Sd}(\Delta \psi_N)}$; see \Eq{\ref{\EQ: var_of_dPsi}}.
The distribution of $\Delta \psi_N$ due to single nucleotide nonsynonymous mutations is approximated by
a log-normal distribution with $n_{\script{shift}} = 2.0$;
see \Eqs{\REF{\EQ: log-normal} to \REF{\EQ: statistics_for_log-normal}}.
}%  FigureLegends
}
\end{figure*}

}%  SUPPLEMENT

\FigJ{

\CLEARPAGE
 
\begin{figure*}[h!]
\FigureInLegends{
\noindent
\SUPPLEMENT{
\hspace*{1em}
(a) $r_{\script{cutoff}} \sim 8$\AA\ 
\hspace*{14em}
(b) $r_{\script{cutoff}} \sim 15.5$\AA\ 

}%  SUPPLEMENT
\TEXT{
\centerline{
{\small{$r_{\script{cutoff}} \sim 8$\AA\ }}
}
}%  TEXT
\SUPPLEMENT{
\centerline{
\includegraphics*[width=82mm,angle=0]{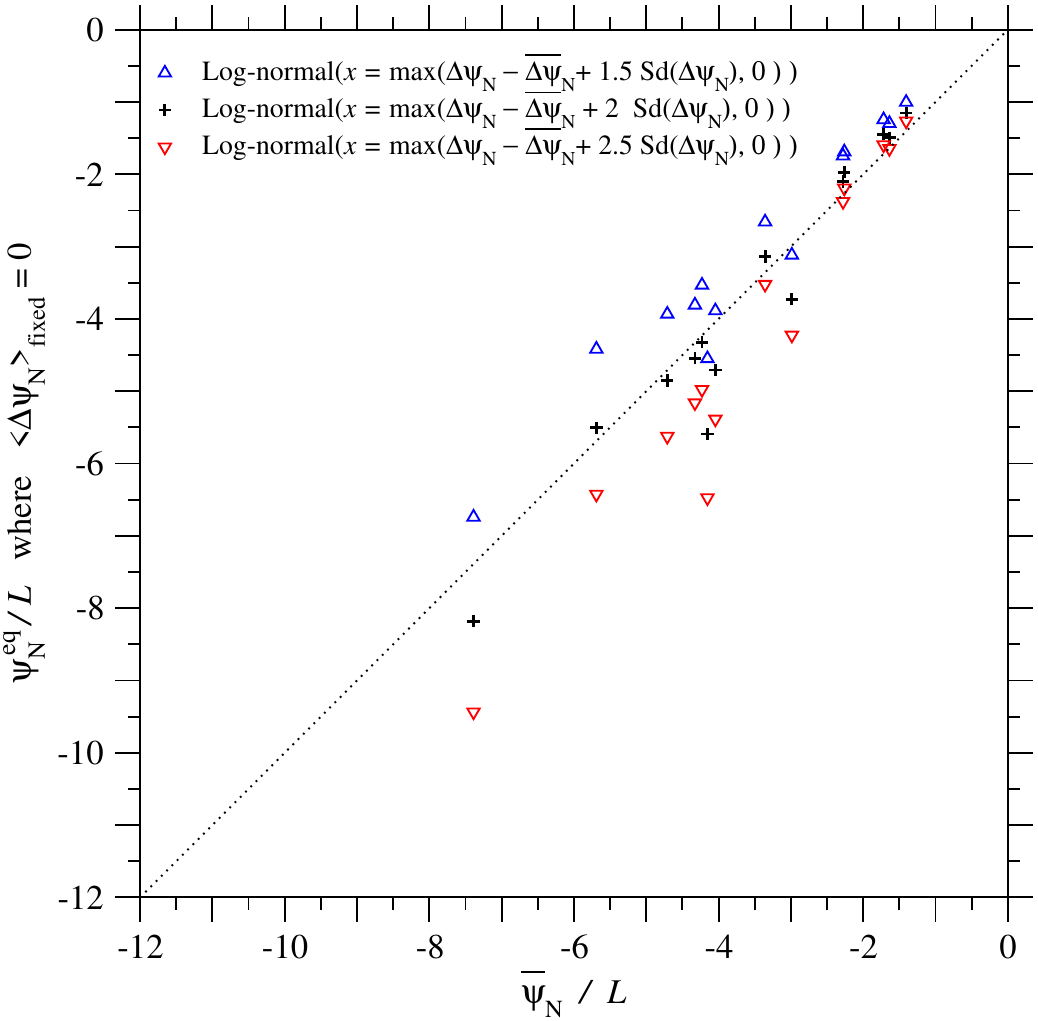}
\includegraphics*[width=82mm,angle=0]{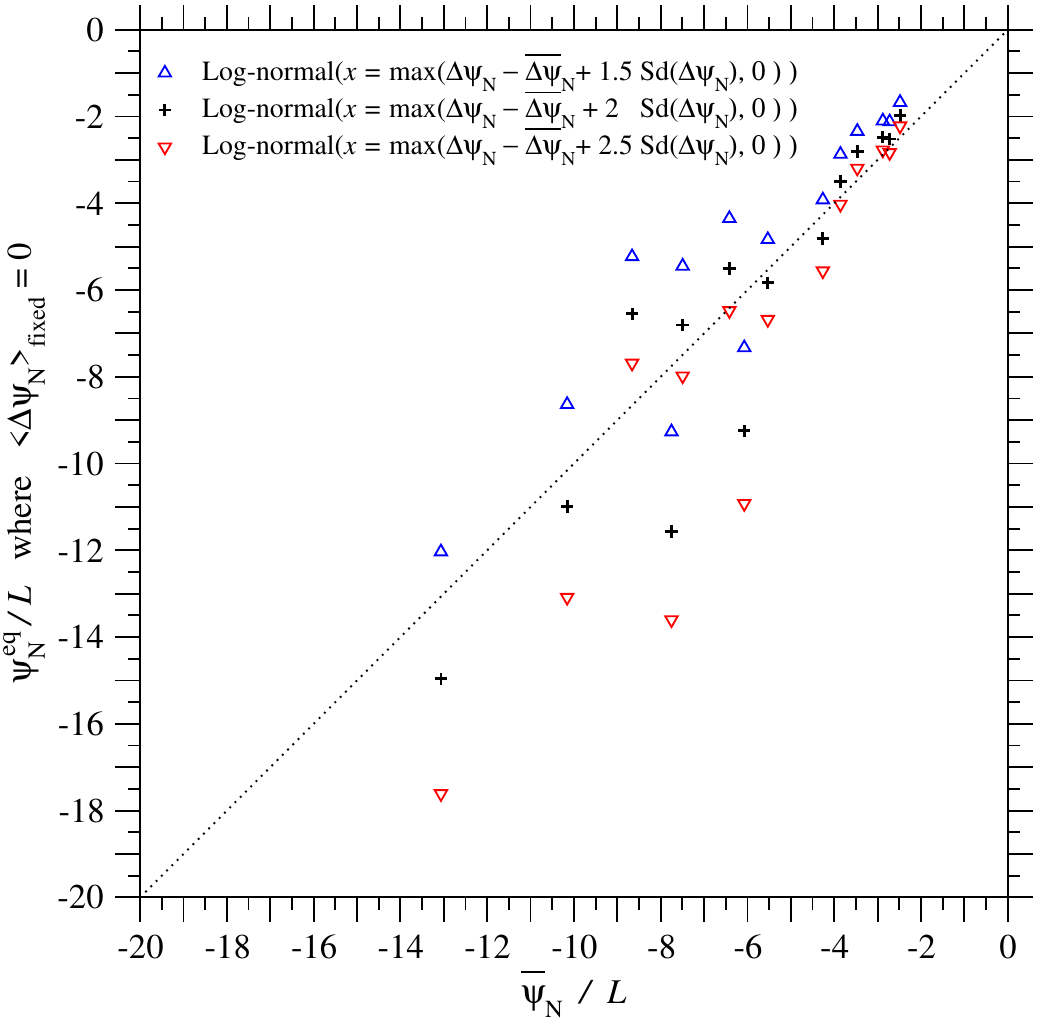}
}
}%  SUPPLEMENT
\TEXT{
\centerline{
\includegraphics*[width=82mm,angle=0]{FIGS/Protein_Evolution/PhiNe_obs_vs_exp_logG_8A}
}
}%  TEXT
}%  FigureInLegends
\vspace*{1em}
\caption{
\SUPPLEMENT{
\label{sfig: PhiNe_obs_vs_exp_logG_8A}
\label{sfig: PhiNe_obs_vs_exp_logG_16A}
}%  SUPPLEMENT
\TEXT{
\label{fig: PhiNe_obs_vs_exp_logG_8A}
}%  TEXT
\FigureLegends{
\BF{
The equilibrium value of $\psi_N / L$, where $\langle \Delta \psi_N \rangle_{\script{fixed}} = 0$,
is plotted against the average of $\psi_N / L$ over homologous sequences for each protein family.
}
\SUPPLEMENT{
The cutoff distances, 
(a) $r_{\script{cutoff}} = 8$\AA\ and (b) $r_{\script{cutoff}} = 15.5$\AA, are employed to estimate $\psi_N$ of
each protein family.
}%  SUPPLEMENT
\TEXT{
The cutoff distance
$r_{\script{cutoff}} = 8$\AA\ is employed to estimate $\psi_N$ of each protein family;
see \Fig{\ref{sfig: PhiNe_obs_vs_exp_logG_16A}} for $r_{\script{cutoff}} = 15.5$\AA.
}%  TEXT
The equilibrium values $\psi_N^{\script{eq}}$, 
where $\langle \Delta \psi_N \rangle_{\script{fixed}} = 0$,
are calculated
by using 
the linear dependency 
of $\overline{\Delta \psi_N}$ on $\psi_N$ (\Eq{\ref{\EQ: regression_of_dPsi_on_Psi}})
and
estimated values 
\SUPPLEMENT{
with $r_{\script{cutoff}} \sim 8$ or $15.5$\AA\  
in \Tables{\ref{\TBL: ddPsi_with_8A} or \ref{\TBL: ddPsi_with_16A}}.
}%  SUPPLEMENT
\TEXT{
with $r_{\script{cutoff}} \sim 8$
in \Tables{\ref{\TBL: ddPsi_with_8A}}.
}%  TEXT
The standard deviation of $\Delta \psi_N$ is approximated to be constant
and equal to $\overline{\mbox{Sd}(\Delta \psi_N)}$; see \Eq{\ref{\EQ: var_of_dPsi}}.
Plus, upper triangle, and lower triangle marks indicate the cases of
log-normal distributions with $n_{\script{shift}} = 1.5, 2.0$, and $2.5$ 
employed to approximate
the distribution of $\Delta \psi_N$, respectively; 
see \Eqs{\REF{\EQ: log-normal} to \REF{\EQ: statistics_for_log-normal}}. 
}%  FigureLegends
}
\end{figure*}

}%  FigJ

\FigK{

\CLEARPAGE
 
\begin{figure*}[h!]
\FigureInLegends{
\SUPPLEMENT{
\centerline{
\includegraphics*[width=82mm,angle=0]{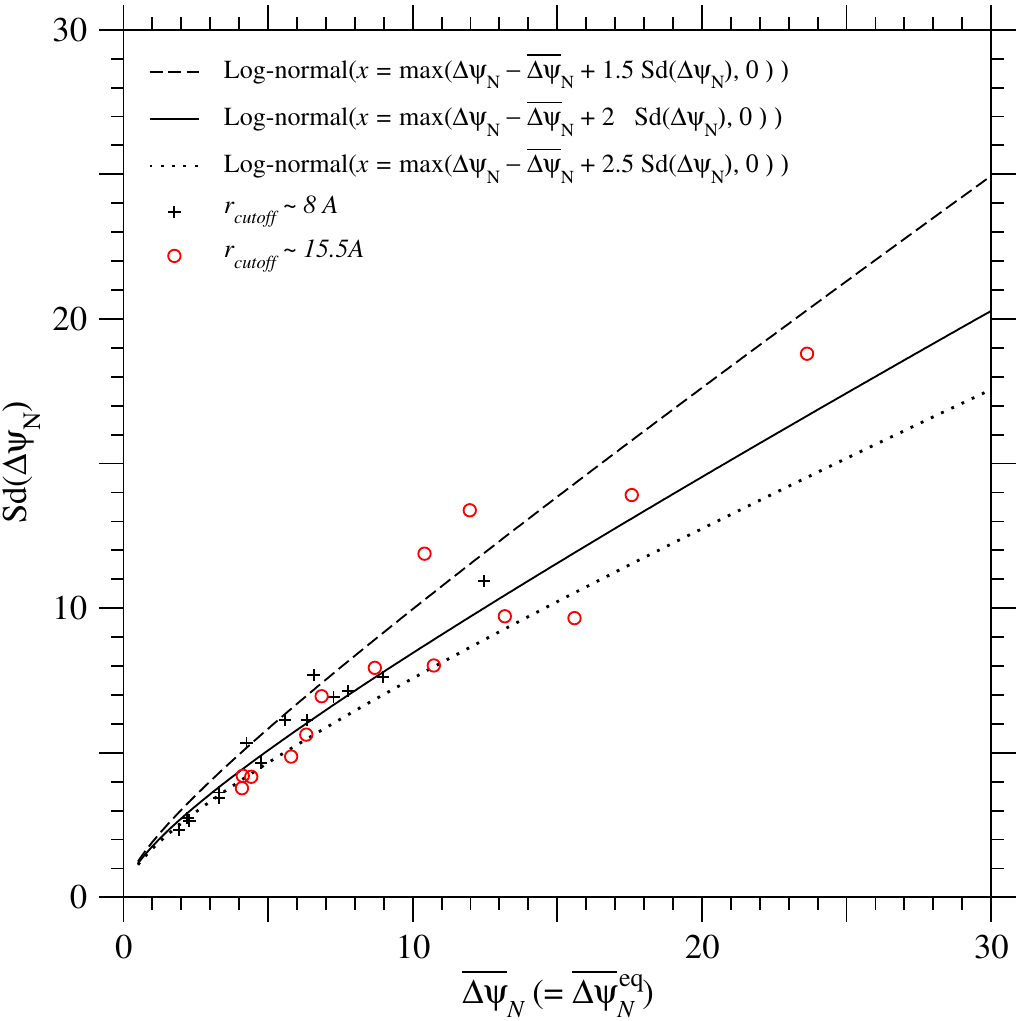}
}
}%  SUPPLEMENT
\TEXT{
\centerline{
\includegraphics*[width=82mm,angle=0]{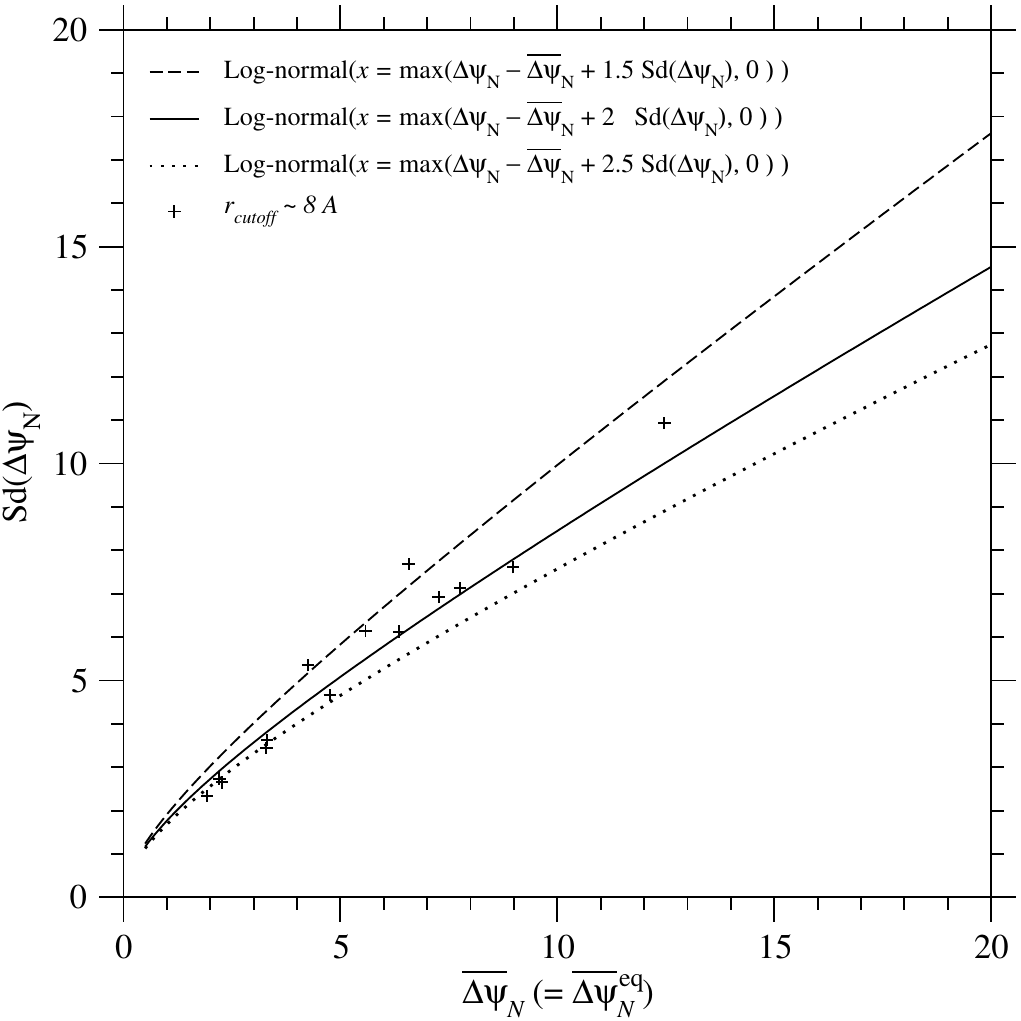}
}
}%  TEXT
}%  FigureInLegends
\vspace*{1em}
\caption{
\SUPPLEMENT{
\label{sfig: ddPhi_mean_vs_sd_at_equil}
}%  SUPPLEMENT
\TEXT{
\label{fig: ddPhi_mean_vs_sd_at_equil}
}%  TEXT
\FigureLegends{
\BF{
Relationship between the mean and the standard deviation of $\Delta \psi_N$ due to single nucleotide nonsynonymous mutations 
at equilibrium,
$\langle \Delta \psi_{N} \rangle_{\script{fixed}} = 0$
}
The standard deviation of $\Delta \psi_N$ 
that satisfies $\langle \Delta \psi_{N} \rangle_{\script{fixed}} = 0$
is plotted against
its mean, $\overline{\Delta \psi_N}$.
Broken, solid, and dotted lines indicate the cases of
log-normal distributions with $n_{\script{shift}} = 1.5, 2.0$ and $2.5$ 
employed to approximate the distribution of $\Delta \psi_N$, respectively; 
see \Eqs{\REF{\EQ: log-normal} to \REF{\EQ: statistics_for_log-normal}}. 
\SUPPLEMENT{
Plus and open circle marks indicate the averages, 
$\overline{\overline{\Delta \psi_N}}$ and $\overline{\text{Sd}(\Delta \psi_N)}$,
over homologous sequences in each protein family
for $r_{\script{cutoff}} \sim 8$ and $15.5$\AA , respectively; 
see \Tables{\ref{\TBL: ddPsi_with_8A} and \ref{\TBL: ddPsi_with_16A}}.
}%  SUPPLEMENT
\TEXT{
Plus marks indicate the averages, 
$\overline{\overline{\Delta \psi_N}}$ and $\overline{\text{Sd}(\Delta \psi_N)}$,
over homologous sequences in each protein family
for $r_{\script{cutoff}} \sim 8$ \AA, 
which are listed in \Tables{\ref{\TBL: ddPsi_with_8A}}.
See \Fig{\ref{sfig: ddPhi_mean_vs_sd_at_equil}} for $r_{\script{cutoff}} \sim 15.5$ \AA.
}%  TEXT
}%  FigureLegends
}
\end{figure*}

}%  FigK

\FigL{

\CLEARPAGE
 
\begin{figure*}[h!]
\FigureInLegends{
\SUPPLEMENT{
\centerline{
\includegraphics*[width=82mm,angle=0]{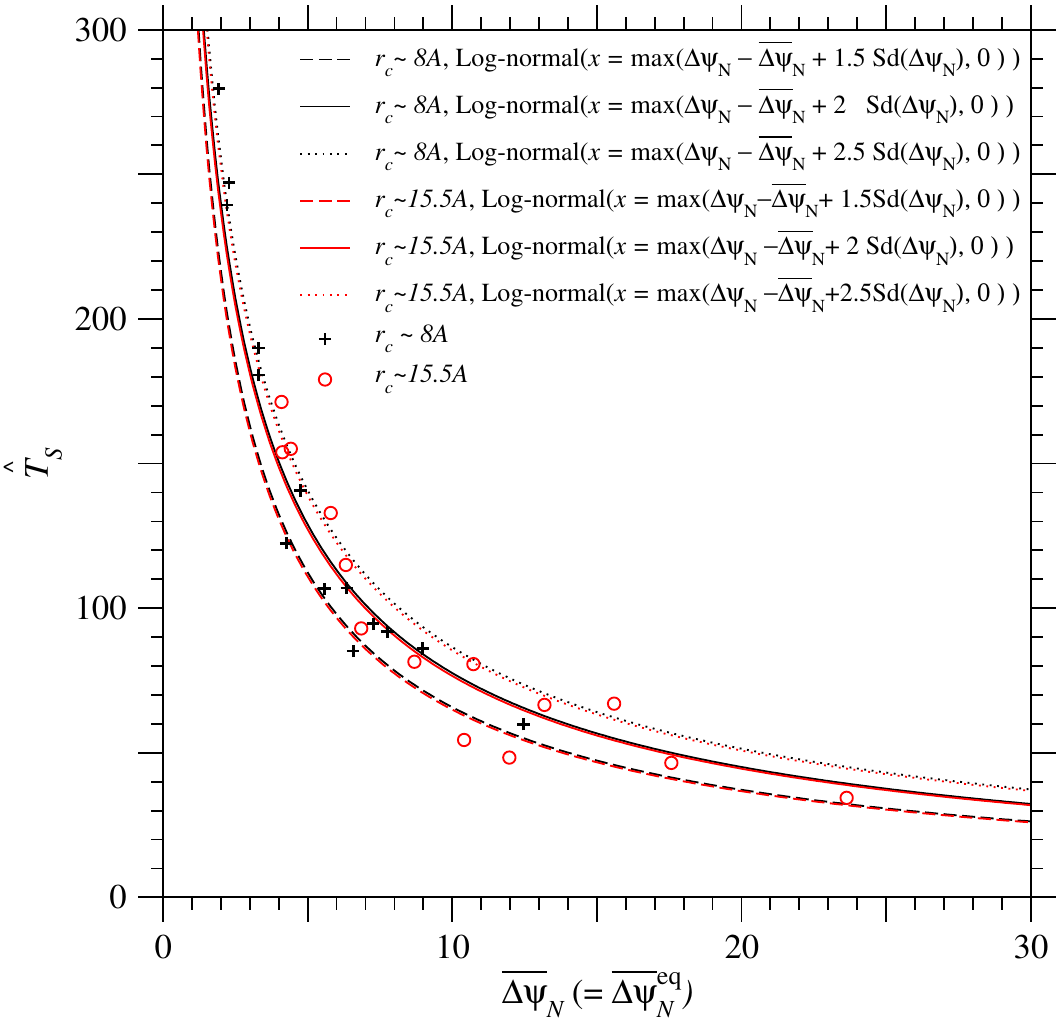}
\includegraphics*[width=82mm,angle=0]{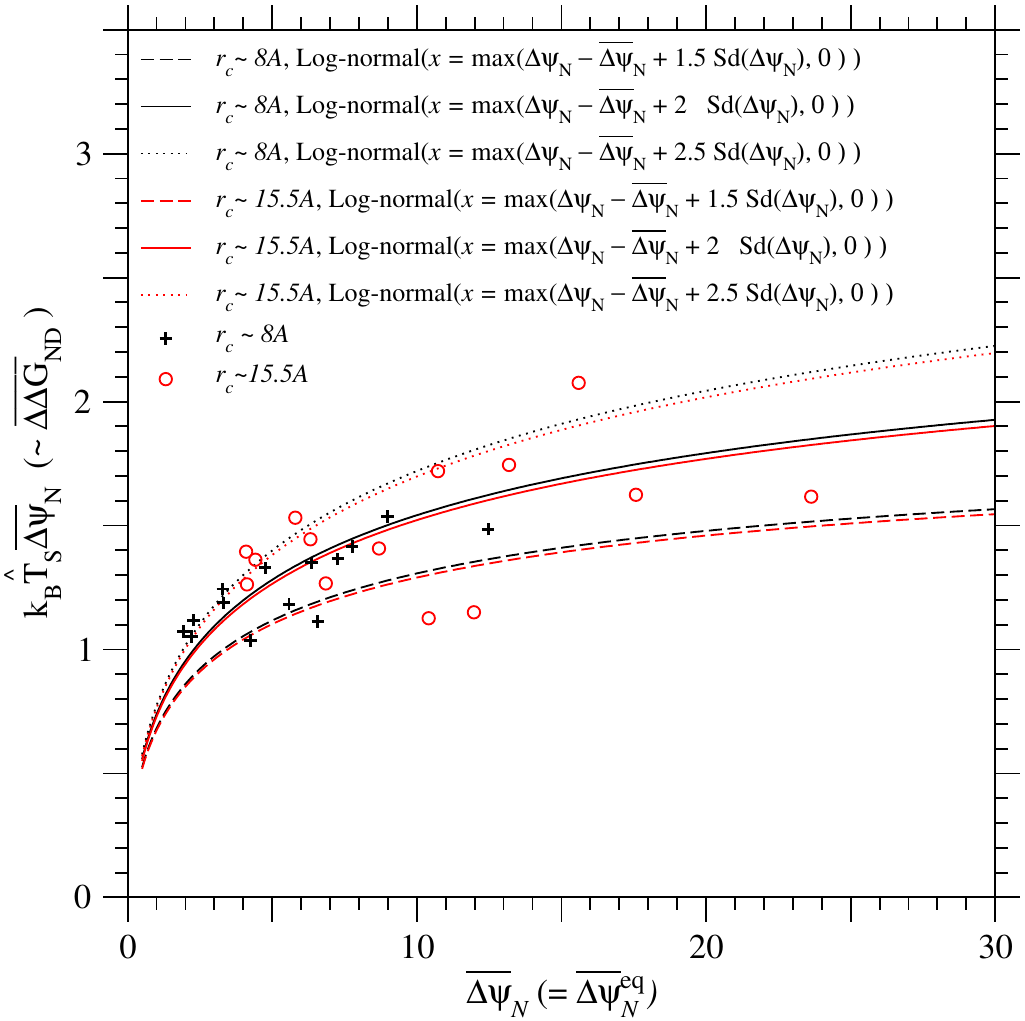}
}
}%  SUPPLEMENT
\TEXT{
\centerline{
\includegraphics*[width=82mm,angle=0]{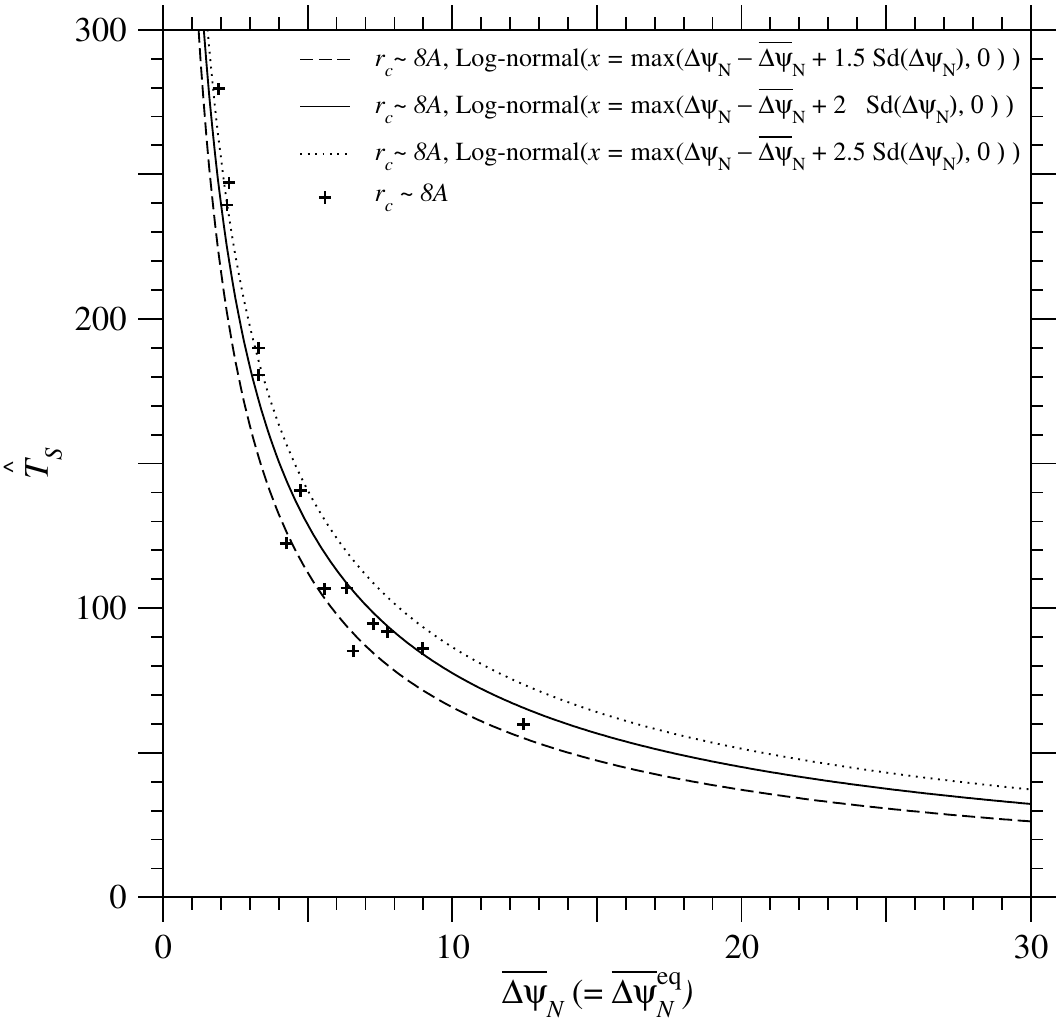}
\includegraphics*[width=82mm,angle=0]{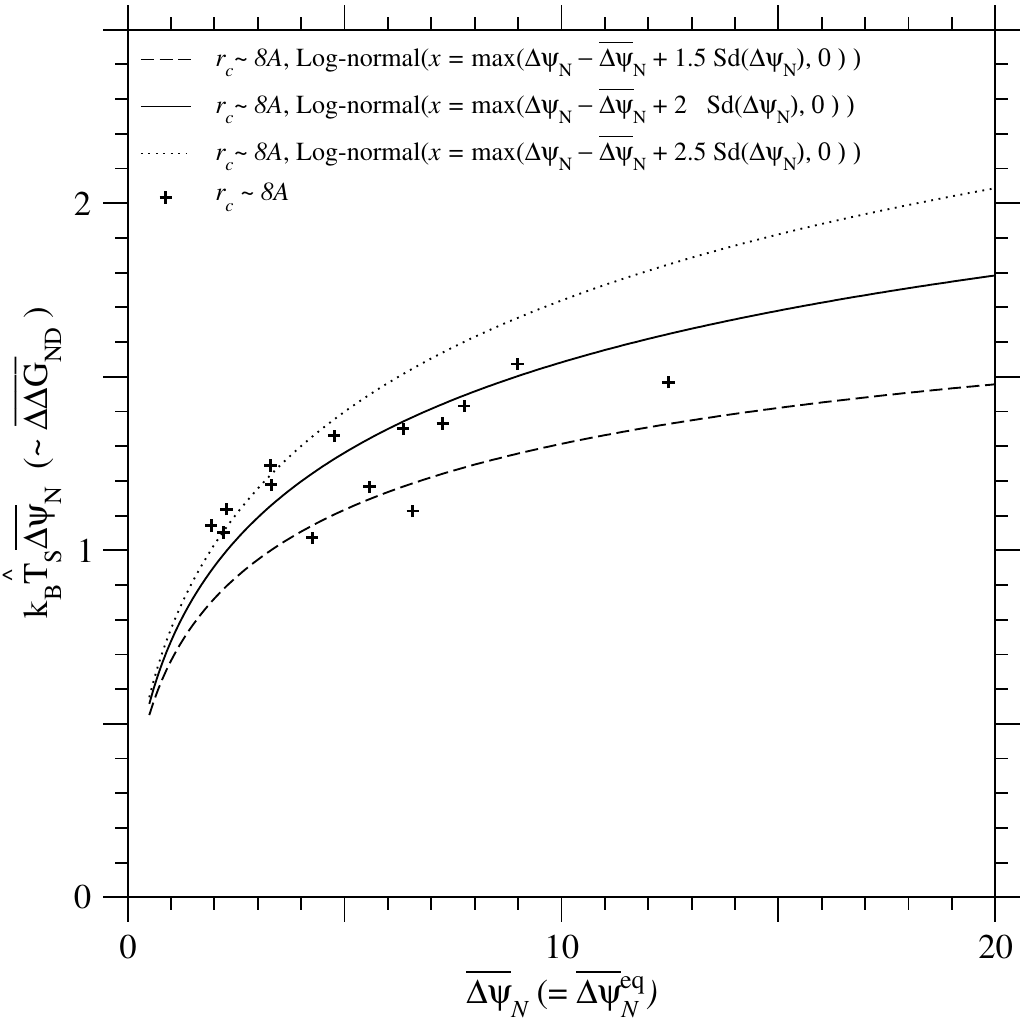}
}
}%  TEXT
}%  FigureInLegends
\vspace*{1em}
\caption{
\SUPPLEMENT{
\label{sfig: ddPhi_mean_vs_Ts_at_equil}
\label{sfig: ddPhi_mean_vs_ddG_at_equil}
}%  SUPPLEMENT
\TEXT{
\label{fig: ddPhi_mean_vs_Ts_at_equil}
\label{fig: ddPhi_mean_vs_ddG_at_equil}
}%  TEXT
\FigureLegends{
\BF{
Relationships between $\hat{T}_s$ and $\overline{\Delta \psi_N}$ and between 
$k_B \hat{T}_s \overline{\Delta \psi_N} (\simeq \overline{\Delta\Delta G_{ND}} )$ 
and $\overline{\Delta \psi_N}$
at equilibrium, 
$\langle \Delta \psi_{N} \rangle_{\script{fixed}} = 0$.
}
The estimate $\hat{T}_s (= (\hat{T}_s \overline{Sd}(\Delta \psi_N))_{PDZ}/Sd(\Delta \psi_N))$ of effective temperature for selection 
and the estimate of mean folding free energy change,
\SUPPLEMENT{
$k_B \hat{T}_s \overline{\Delta \psi_N} (= k_B (\hat{T}_s \overline{Sd}(\Delta \psi_N))_{PDZ}/Sd(\Delta \psi_N) \cdot \overline{\Delta \psi_N} 
\simeq \overline{\Delta\Delta G_{ND}} )$,
}%  SUPPLEMENT
\TEXT{
$k_B \hat{T}_s \overline{\Delta \psi_N} (\simeq \overline{\Delta\Delta G_{ND}} )$,
}%  TEXT
are plotted against $\overline{\Delta \psi_N}$
under the condition of 
$\langle \Delta \psi_{N} \rangle_{\script{fixed}} = 0$.
The $T_s$ is estimated in relative to the $T_s$ of the PDZ family   
in the approximation
that the standard deviation of $\Delta G_{N}$ due to single nucleotide
nonsynonymous mutations is constant irrespective of protein families;
see \Eq{\ref{\EQ: var_of_ddG}}.
Broken, solid, and dotted lines indicate the cases of
log-normal distributions with $n_{\script{shift}} = 1.5, 2.0$ and $2.5$ 
employed to approximate the distribution of $\Delta \psi_N$, respectively; 
see \Eqs{\REF{\EQ: log-normal} to \REF{\EQ: statistics_for_log-normal}}. 
\SUPPLEMENT{
Plus and open circle marks indicate
those estimates against 
the average of $\overline{\Delta \psi_N}$ over homologous sequences
for each protein family
with $r_{\script{cutoff}} \sim 8$ and $15.5$\AA , respectively; 
see \Tables{\ref{\TBL: ddPsi_with_8A} and \ref{\TBL: ddPsi_with_16A}}.
The curves for $r_{\script{cutoff}} \sim 8$ and $15.5$\AA\ almost overlap with each other,
because the estimates of $(\hat{T}_s \overline{Sd}(\Delta \psi_N))_{PDZ}$ for the PDZ 
with $r_{\script{cutoff}} \sim 8$ and $15.5$\AA\  are almost equal to each other.
}%  SUPPLEMENT
\TEXT{
Plus marks indicate
those estimates against 
the average of $\overline{\Delta \psi_N}$ over homologous sequences
for each protein family
with $r_{\script{cutoff}} \sim 8$\AA, 
which are listed in \Tables{\ref{\TBL: ddPsi_with_8A} and \ref{\TBL: Ts_with_8A}}.
See \Fig{\ref{sfig: ddPhi_mean_vs_Ts_at_equil}} for $r_{\script{cutoff}} \sim 15.5$\AA.
}%  TEXT
}%  FigureLegends
}
\end{figure*}

}%  FigL

\FigM{

\CLEARPAGE
 
\begin{figure*}[h!]
\FigureInLegends{
\centerline{
\includegraphics*[width=82mm,angle=0]{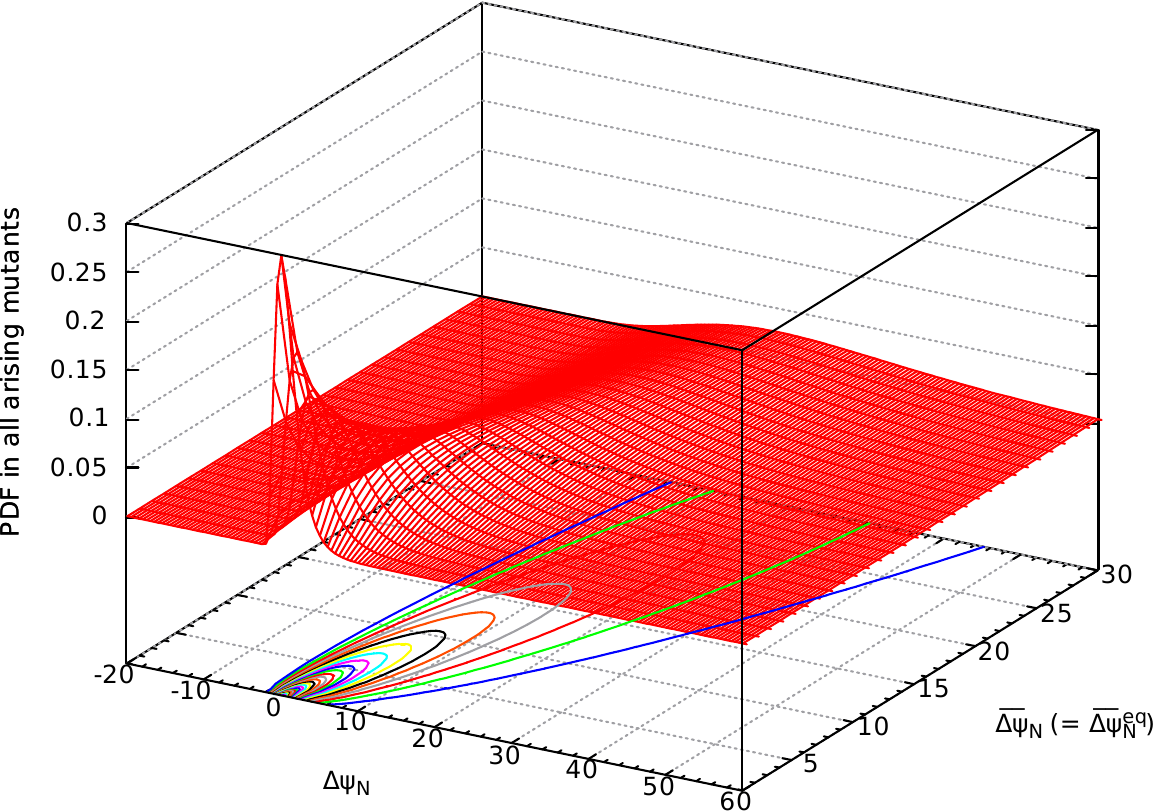}
\includegraphics*[width=82mm,angle=0]{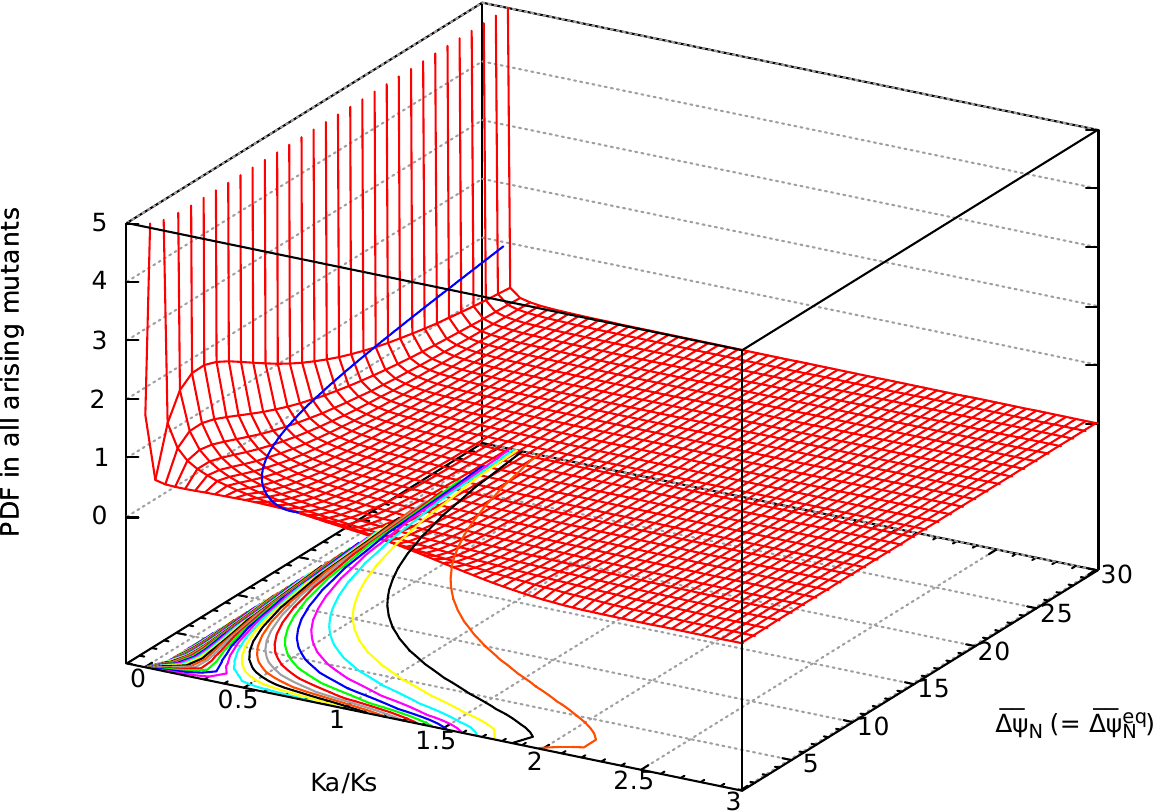}
}
\vspace*{1em}
\centerline{
\includegraphics*[width=82mm,angle=0]{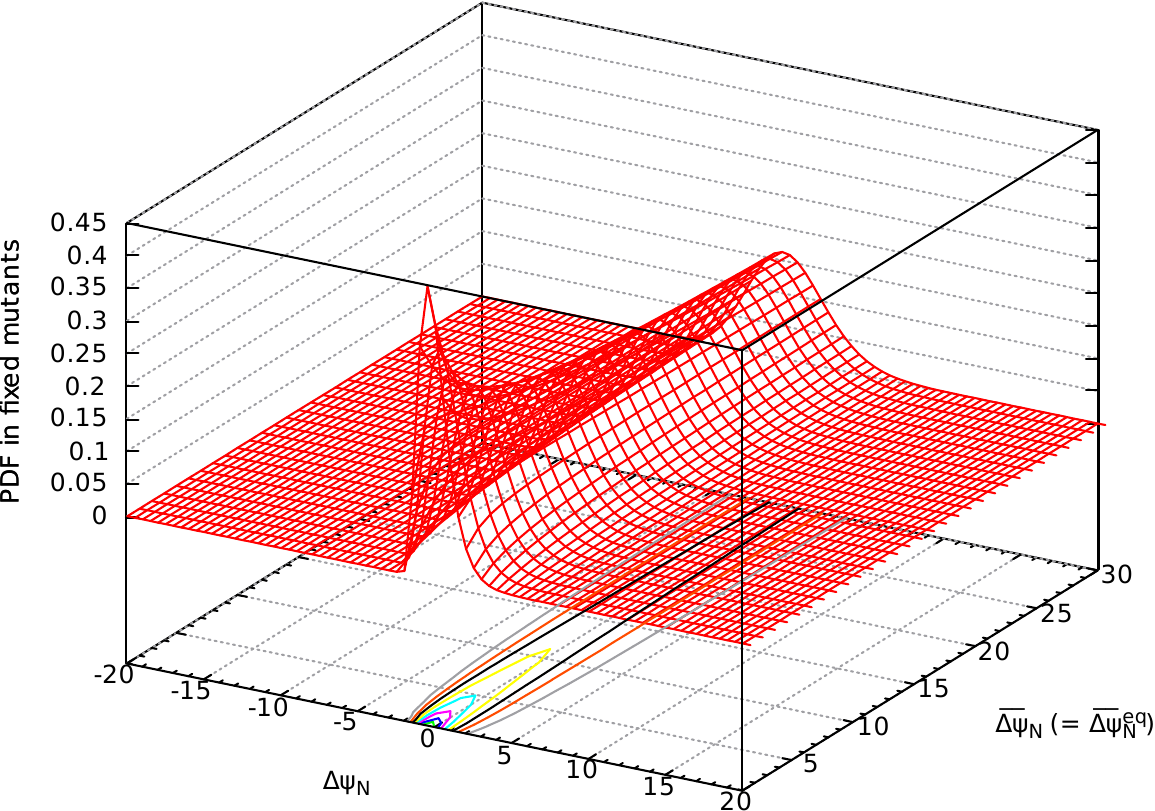}
\includegraphics*[width=82mm,angle=0]{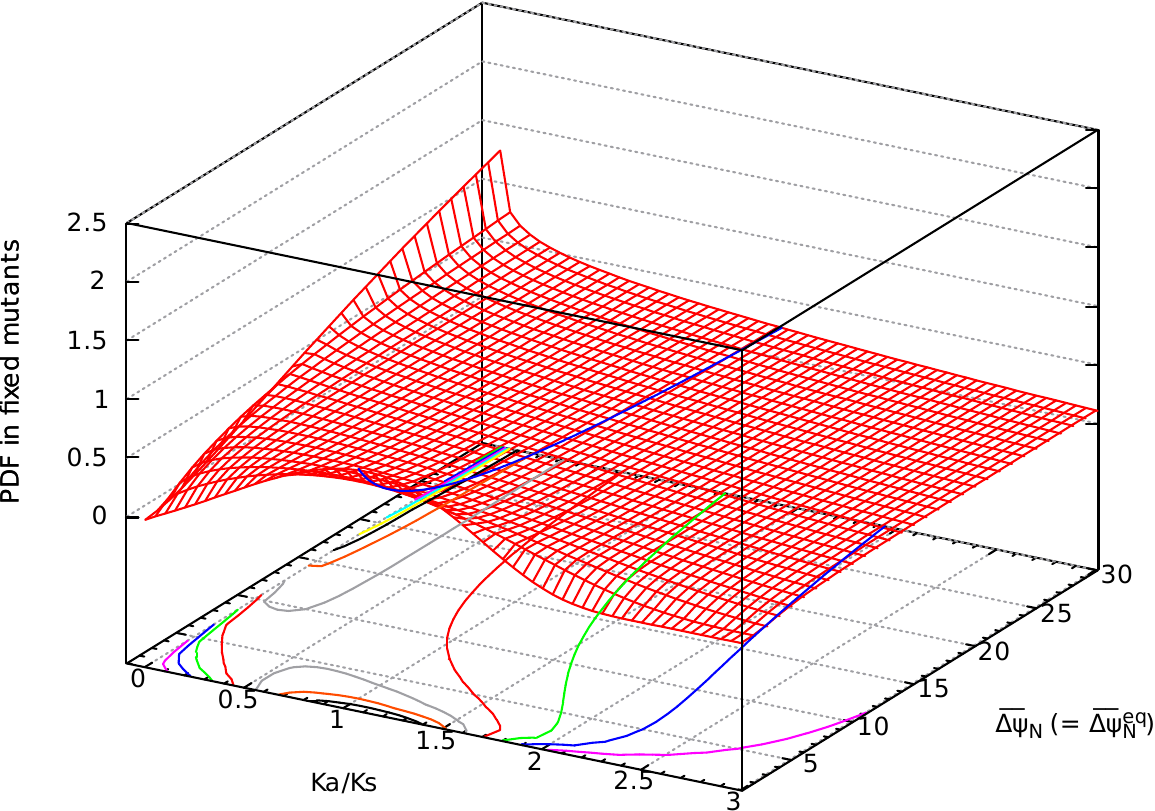}
}
}%  FigureInLegends
\vspace*{1em}
\caption{
\SUPPLEMENT{
\label{sfig: pdf_of_ddPhi_at_equil_for_mean_ddPhi}
\label{sfig: pdf_of_ddPhi_fixed_at_equil_for_mean_ddPhi}
\label{sfig: pdf_of_ka_over_ks_at_equil_for_mean_ddPhi}
\label{sfig: pdf_of_ka_over_ks_fixed_at_equil_for_mean_ddPhi}
}%  SUPPLEMENT
\TEXT{
\label{fig: pdf_of_ddPhi_at_equil_for_mean_ddPhi}
\label{fig: pdf_of_ddPhi_fixed_at_equil_for_mean_ddPhi}
\label{fig: pdf_of_ka_over_ks_at_equil_for_mean_ddPhi}
\label{fig: pdf_of_ka_over_ks_fixed_at_equil_for_mean_ddPhi}
}%  TEXT
\FigureLegends{
\BF{
PDFs of $\Delta \psi_N$ (left) and $K_a/K_s$ (right)
in all singe nucleotide nonsynonymous mutants (upper) and in their fixed mutants (lower)
as a function of 
$\overline{\Delta \psi_N}$
at equilibrium, $\langle \Delta \psi_{N} \rangle_{\script{fixed}} = 0$.
}
Fixation probability has been calculated with $\Delta\Delta \psi_{ND} \simeq \Delta \psi_{N}$;
see \Eqs{\REF{\EQ: fixation_probability} and \REF{\EQ: selective_advantage_and_dPsi_N}}.
The distribution of $\Delta \psi_N$ due to single nucleotide nonsynonymous mutations is approximated by
a log-normal distribution with $n_{\script{shift}} = 2.0$;
see \Eqs{\REF{\EQ: log-normal} to \REF{\EQ: statistics_for_log-normal}}.
The standard deviation of $\Delta \psi_N$ is determined to
satisfy $\langle \Delta \psi_{N} \rangle_{\script{fixed}} = 0$ at
$\overline{\Delta \psi_N} = \overline{\Delta \psi}_N^{\script{eq}}$. 
}%  FigureLegends
}
\end{figure*}

}%  FigM

\FigN{

\CLEARPAGE
 
\begin{figure*}[h!]
\FigureInLegends{
\centerline{
\includegraphics*[width=82mm,angle=0]{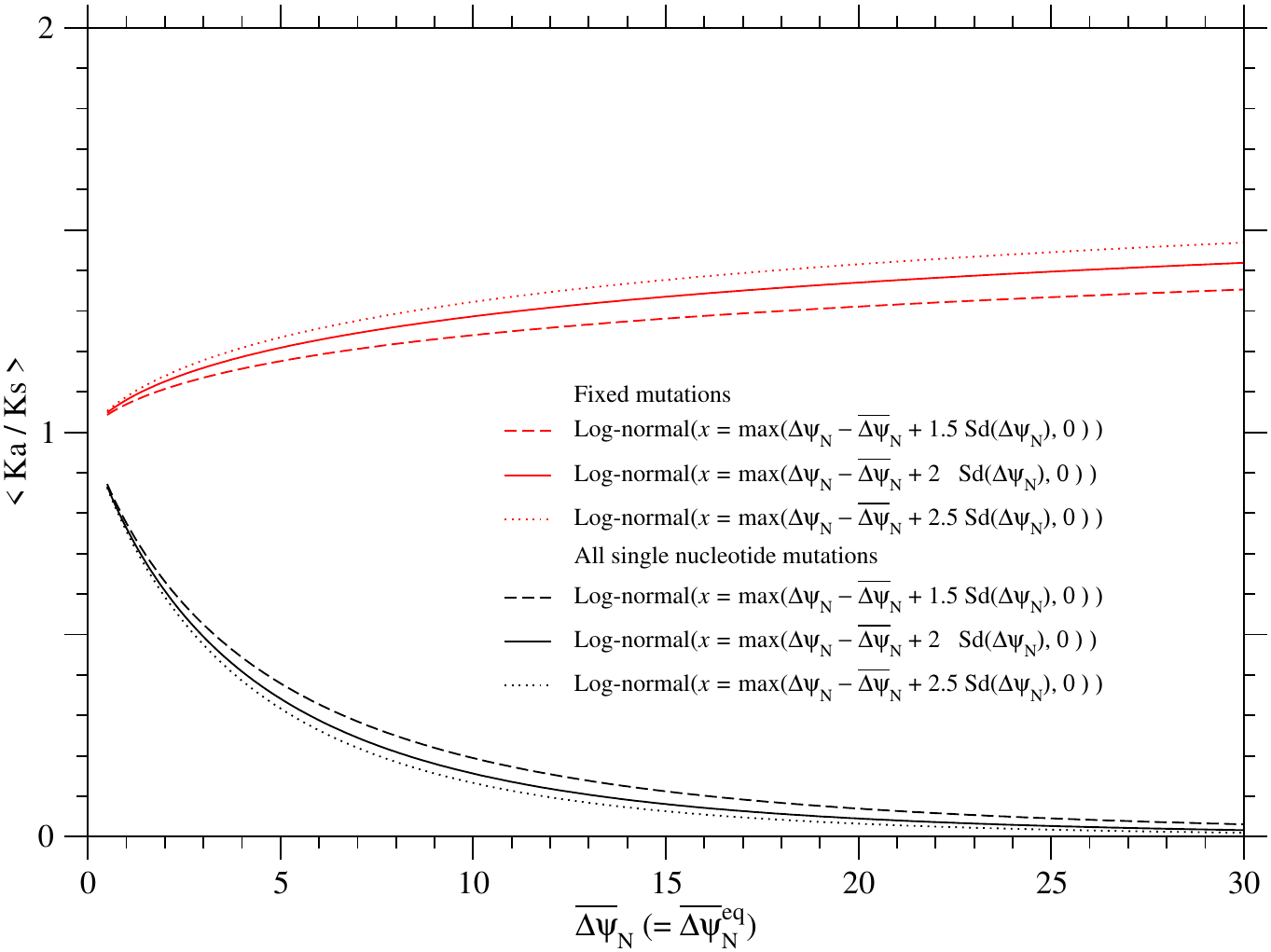}
}
}%  FigureInLegends
\vspace*{1em}
\caption{
\SUPPLEMENT{
\label{sfig: ave_ka_over_ks_at_equil_for_mean_ddPhi}
\label{sfig: ave_ka_over_ks_fixed_at_equil_for_mean_ddPhi}
}%  SUPPLEMENT
\TEXT{
\label{fig: ave_ka_over_ks_at_equil_for_mean_ddPhi}
\label{fig: ave_ka_over_ks_fixed_at_equil_for_mean_ddPhi}
}%  TEXT
\FigureLegends{
\BF{
The averages of $K_a/Ks$ over all single nucleotide nonsynonymous mutations and over
their fixed mutations as a function of $\overline{\Delta \psi_N}$
at equilibrium, $\langle \Delta \psi_{N} \rangle_{\script{fixed}} = 0$.
}
Black and red lines indicate $\langle K_a/K_s \rangle$ and $\langle K_a/K_s \rangle_{\script{fixed}}$,
respectively.
Fixation probability has been calculated with $\Delta\Delta \psi_{ND} \simeq \Delta \psi_{N}$;
see \Eqs{\REF{\EQ: fixation_probability} and \REF{\EQ: selective_advantage_and_dPsi_N}}.
Broken, solid, and dotted lines indicate the cases of
log-normal distributions with $n_{\script{shift}} = 1.5, 2.0$ and $2.5$ 
employed to approximate the distribution of $\Delta \psi_N$, respectively; 
see \Eqs{\REF{\EQ: log-normal} to \REF{\EQ: statistics_for_log-normal}}. 
The standard deviation of $\Delta \psi_N$ is determined to
satisfy 
$\langle \Delta \psi_{N} \rangle_{\script{fixed}} = 0$ at 
$\overline{\Delta \psi_N} = \overline{\Delta \psi}_N^{\script{eq}}$. 
}%  FigureLegends
}
\end{figure*}

}%  FigN

\FigO{

\CLEARPAGE
 
\begin{figure*}[h!]
\FigureInLegends{
\centerline{
\includegraphics*[width=82mm,angle=0]{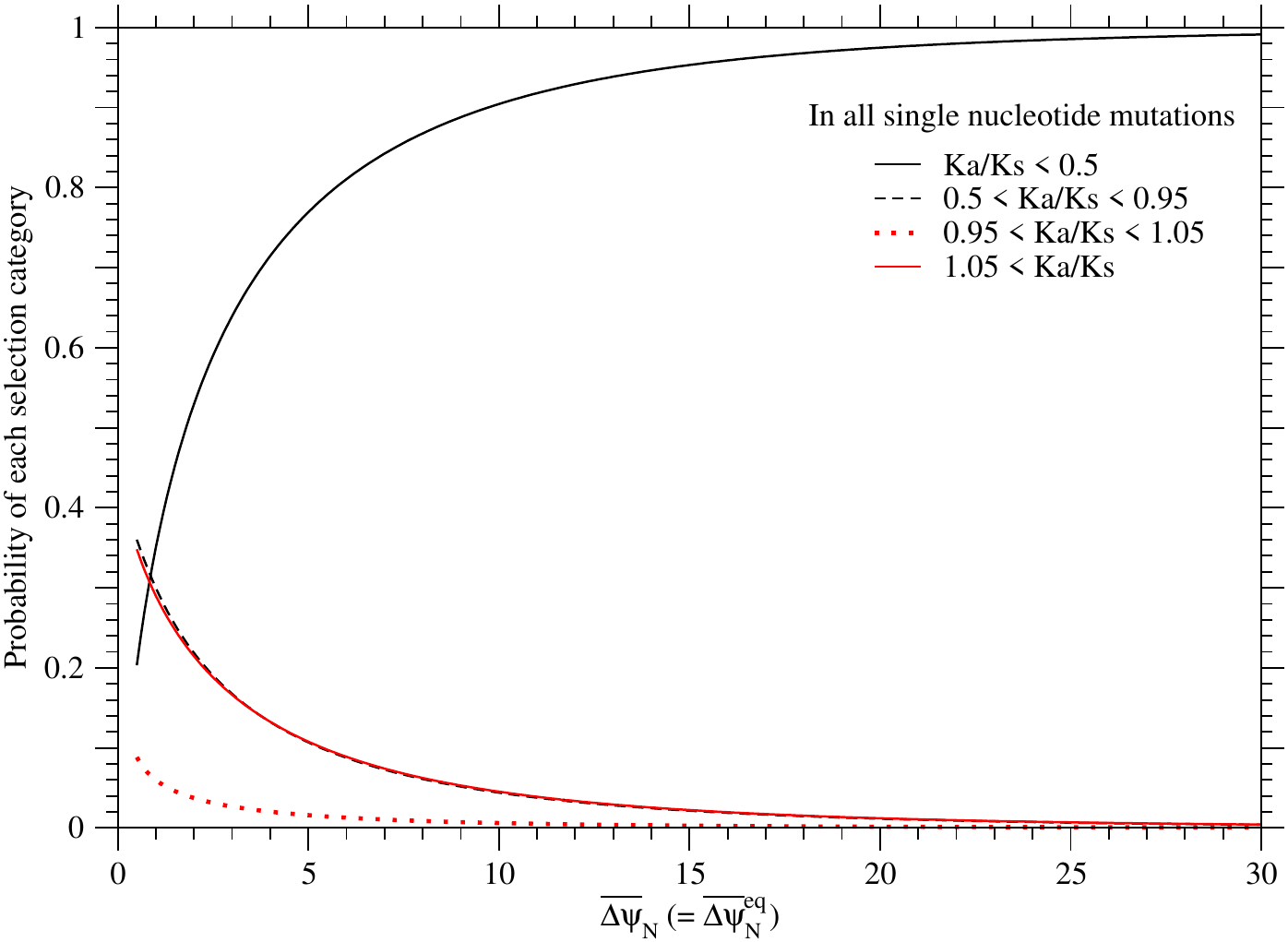}
\includegraphics*[width=82mm,angle=0]{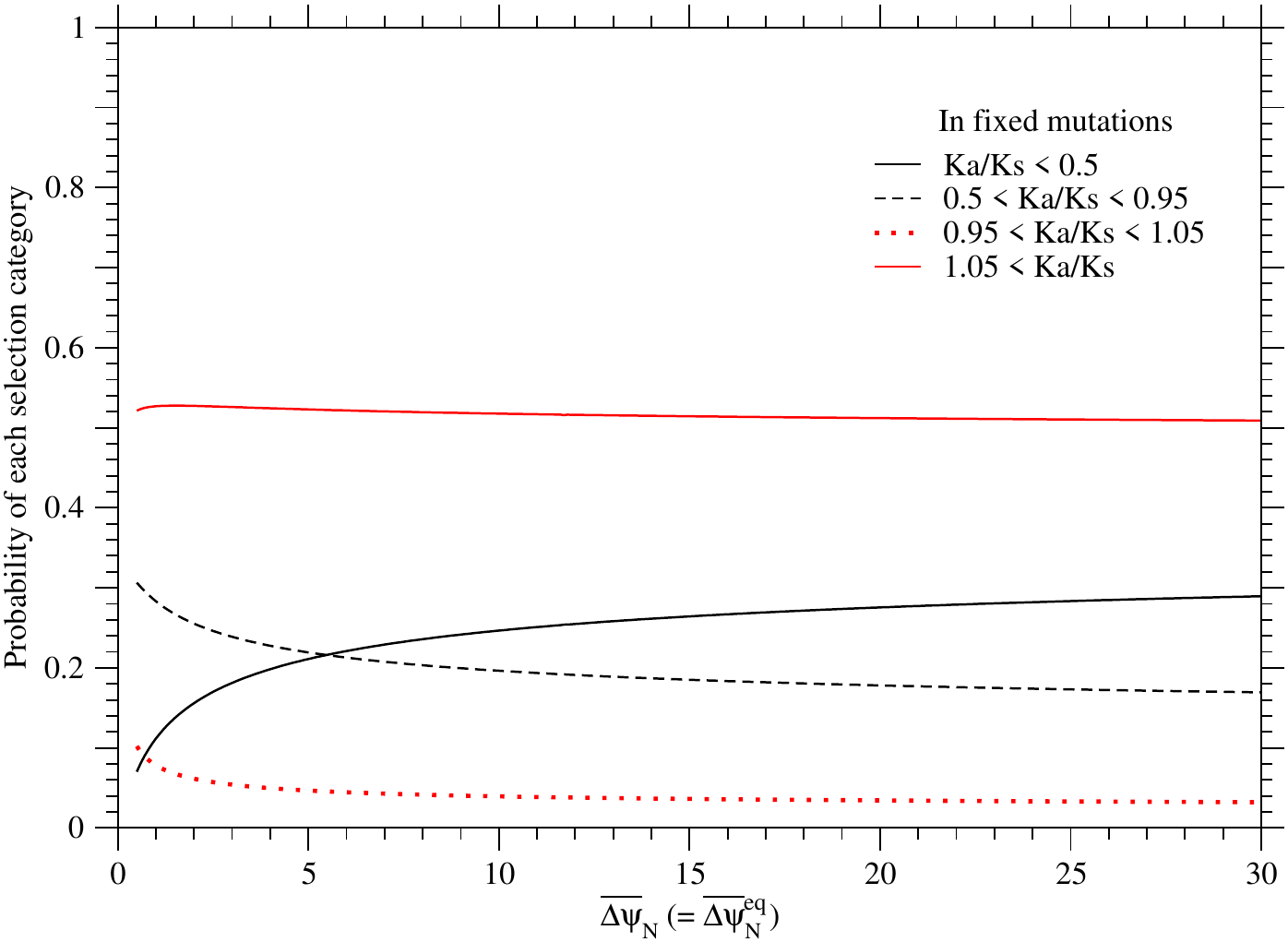}
}
}%  FigureInLegends
\vspace*{1em}
\caption{
\SUPPLEMENT{
\label{sfig: prob_of_each_selection_category_at_equil_for_mean_ddPhi}
\label{sfig: prob_of_each_selection_category_fixed_at_equil_for_mean_ddPhi}
}%  SUPPLEMENT
\TEXT{
\label{fig: prob_of_each_selection_category_at_equil_for_mean_ddPhi}
\label{fig: prob_of_each_selection_category_fixed_at_equil_for_mean_ddPhi}
}%  TEXT
\FigureLegends{
\BF{
The probabilities of each selection category in all single nucleotide nonsynonymous mutations and in
their fixed mutations
as a function of $\overline{\Delta \psi_N}$ 
at equilibrium, $\langle \Delta \psi_{N} \rangle_{\script{fixed}} = 0$.
}
The left and right figures are for single nucleotide nonsynonymous mutations and
for their fixed mutations, respectively.
Red solid, red dotted, black broken, and black solid lines indicate 
positive, neutral, slightly negative and negative selection categories, respectively;
the values of $K_a/K_s$ are divided arbitrarily into four categories,
$K_a/K_s > 1.05$, $1.05 > K_a/K_s > 0.95$, $0.95 > K_a/K_s > 0.5$, and $0.5 > K_a/K_s$,
which correspond to their selection categories, respectively.
Fixation probability has been calculated with $\Delta\Delta \psi_{ND} \simeq \Delta \psi_{N}$;
see \Eqs{\REF{\EQ: fixation_probability} and \REF{\EQ: selective_advantage_and_dPsi_N}}.
The distribution of $\Delta \psi_N$ due to single nucleotide nonsynonymous mutations is approximated by
a log-normal distribution with $n_{\script{shift}} = 2.0$;
see \Eqs{\REF{\EQ: log-normal} to \REF{\EQ: statistics_for_log-normal}}.
The standard deviation of $\Delta \psi_N$ is determined to
satisfy
$\langle \Delta \psi_{N} \rangle_{\script{fixed}} = 0$ at 
$\overline{\Delta \psi_N} (= \overline{\Delta \psi}_N^{\script{eq}} )$.
}%  FigureLegends
}
\end{figure*}

}%  FigO

\FigP{

\CLEARPAGE
 
\begin{figure*}[h!]
\FigureInLegends{
\centerline{
\includegraphics*[width=82mm,angle=0]{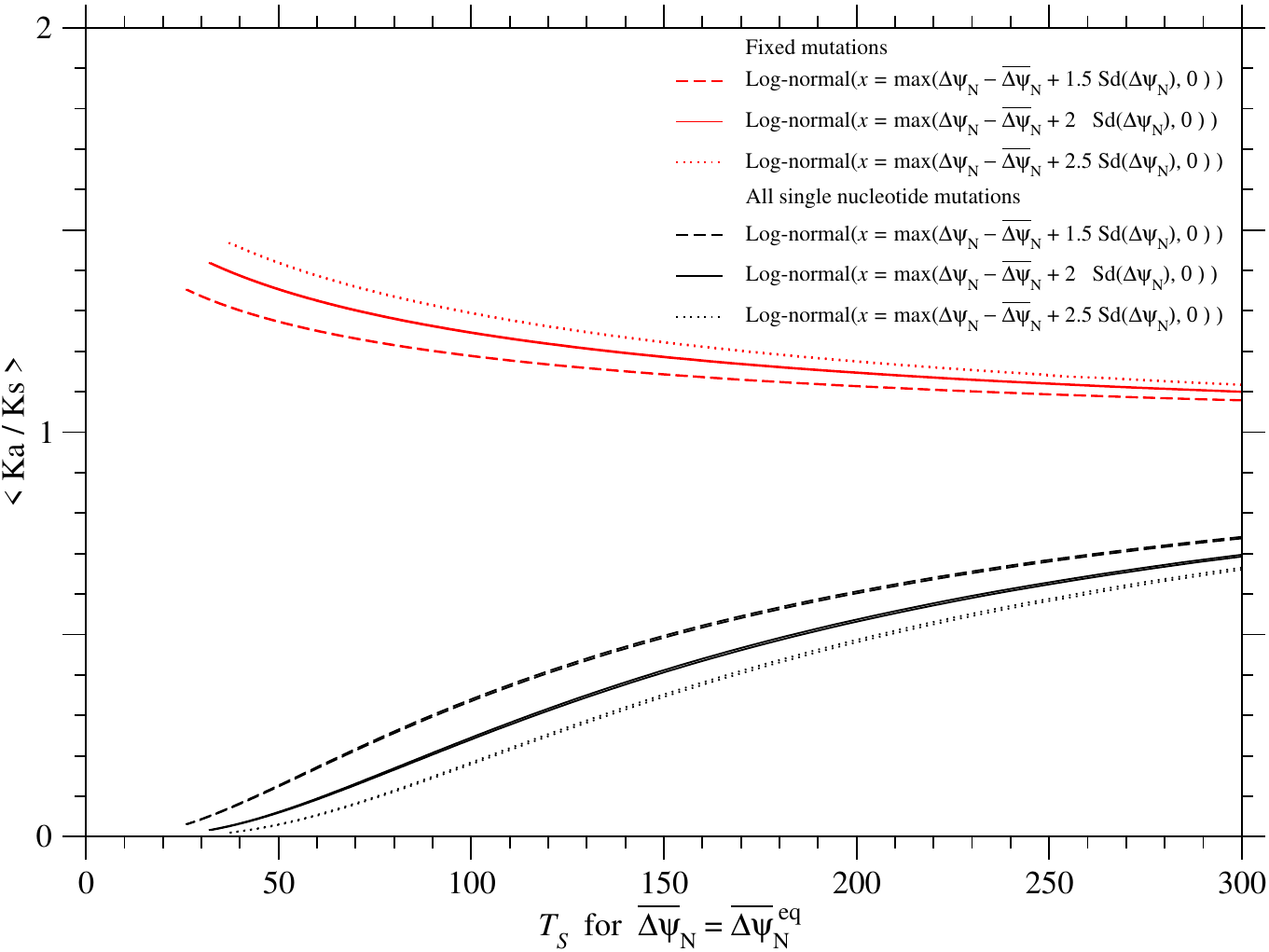}
}
}%  FigureInLegends
\vspace*{1em}
\caption{
\SUPPLEMENT{
\label{sfig: ave_ka_over_ks_at_equil_for_Ts}
\label{sfig: ave_ka_over_ks_fixed_at_equil_for_Ts}
}%  SUPPLEMENT
\TEXT{
\label{fig: ave_ka_over_ks_at_equil_for_Ts}
\label{fig: ave_ka_over_ks_fixed_at_equil_for_Ts}
}%  TEXT
\FigureLegends{
\BF{
The averages of $K_a/Ks$ over all single nucleotide nonsynonymous mutations and over
their fixed mutations as a function of the effective temperature of selection, 
$T_s (= (T_s \overline{Sd}(\Delta\psi_N))_{PDZ} /Sd(\Delta\psi_N) )$, 
at equilibrium, 
$\langle \Delta \psi_{N} \rangle_{\script{fixed}} = 0$.
}
Black and red lines indicate $\langle K_a/K_s \rangle$ and $\langle K_a/K_s \rangle_{\script{fixed}}$,
respectively.
Fixation probability has been calculated with $\Delta\Delta \psi_{ND} \simeq \Delta \psi_{N}$;
see \Eqs{\REF{\EQ: fixation_probability} and \REF{\EQ: selective_advantage_and_dPsi_N}}.
The distribution of $\Delta \psi_N$ due to single nucleotide nonsynonymous mutations is approximated by
a log-normal distribution with $n_{\script{shift}} = 2.0$;
see \Eqs{\REF{\EQ: log-normal} to \REF{\EQ: statistics_for_log-normal}}.
The standard deviation of $\Delta \psi_N$ is determined to
satisfy 
$\langle \Delta \psi_{N} \rangle_{\script{fixed}} = 0$ at 
$\overline{\Delta \psi_N} (= \overline{\Delta \psi}_N^{\script{eq}} )$.
The $T_s$ is estimated in the scale relative to the $T_s$ of the PDZ family   
in the approximation
that the standard deviation of $\Delta G_{N}$ due to single nucleotide
nonsynonymous mutations is constant irrespective of protein families;
see \Eq{\ref{\EQ: var_of_ddG}}.
Broken, solid, and dotted lines indicate the cases of
log-normal distributions with $n_{\script{shift}} = 1.5, 2.0$ and $2.5$ 
employed to approximate the distribution of $\Delta \psi_N$, respectively; 
see \Eqs{\REF{\EQ: log-normal} to \REF{\EQ: statistics_for_log-normal}}. 
The curves for $r_{\script{cutoff}} \sim 8$ and $15.5$\AA\ almost overlap with each other,
because the estimates of $(\hat{T}_s \overline{Sd}(\Delta \psi_N))_{PDZ}$ for the PDZ
with $r_{\script{cutoff}} \sim 8$ and $15.5$\AA\  are almost equal to each other.
}%  FigureLegends
}
\end{figure*}

}%  FigP

% End of figures_JTB_2.tex
% End of figures_1+2.tex

\clearpage

% End of tables_figures_1+2.tex
}%  NoFigureInText

}%  TextMaterial

%%%%%%%%%%%
%\section*{SUPPLEMENTARY MATERIAL}

% \input{supplement_1+2.tex}

\SupplementaryMaterial{

\ifdefined\EQ
\renewcommand{\EQ}{seq}
\else
\newcommand{\EQ}{seq}
\fi

\ifdefined\TBL
\renewcommand{\TBL}{stbl}
\else
\newcommand{\TBL}{stbl}
\fi

\ifdefined\FIG
\renewcommand{\FIG}{sfig}
\else
\newcommand{\FIG}{sfig}
\fi

\renewcommand{\SUPPLEMENT}[1]{#1}
\renewcommand{\TEXT}[1]{}

\clearpage
\setlength{\textwidth}{17.5cm}
\setlength{\oddsidemargin}{-0.5cm}
\setlength{\evensidemargin}{-0.5cm}

\newpage
\setcounter{page}{1}
\renewcommand{\thepage}{S-\arabic{page}}

\setcounter{section}{0}

\setcounter{equation}{0}
\renewcommand{\theequation}{S.\arabic{equation}}

}%  SupplementaryMaterial

\SupplementaryMaterial{
\begin{center}
\textbf{Supplementary material}	\\
for 
\\
Selection originating from protein stability/foldability:
\\
Relationships between protein folding free energy, sequence ensemble, and fitness
\end{center}

\vspace*{1em}
\begin{center}
Sanzo Miyazawa	\\
sanzo.miyazawa@gmail.com
\end{center}
\begin{center}
2021-09-24
\end{center}

\vspace*{2em}
This supplementary document includes methods, tables and
figures provided in the text
as part of their full descriptions for reader's convenience.

\vspace*{2em}

\section{Methods and Materials}

\ifdefined\SkipDETAIL
\else
\newcommand{\SkipDETAIL}[1]{}
\fi
\ifdefined\SUPPLEMENT
\else
\newcommand{\SUPPLEMENT}[1]{}
\fi
\ifdefined\TEXT
\else
\newcommand{\TEXT}[1]{#1}
\fi

\subsection{Knowledge of protein folding}
\TEXT{
\label{protein_folding_theory}
}%  TEXT

A protein folding theory\CITE{SG:93b,SG:93a,RS:94,PGT:97}, which is based on a random energy model (REM), 
indicates that the equilibrium ensemble of amino acid sequences, $\VECS{\sigma} \equiv (\sigma_1, \cdots, \sigma_L)$ 
where $\sigma_i$ is the type of amino acid at site $i$ and $L$ is sequence length,
can be well approximated by
a canonical ensemble with a Boltzmann factor 
consisting of the folding free energy, $\Delta G_{ND}(\VEC{\sigma}, T)$
and an effective temperature $T_s$ representing the strength of selection pressure.
\begin{eqnarray}
	P(\VEC{\sigma}) 
		&\propto&
		P^{\script{mut}}(\VEC{\sigma}) \exp (\frac{- \Delta G_{ND}(\VEC{\sigma}, T)}{k_B T_s}) 
		\label{\EQ: canonical_selection}
		\\
		&\propto&
			\exp (\frac{- G_N(\VEC{\sigma})}{k_B T_s})	
				\hspace*{2em} \textrm{ if } \VEC{f}(\VEC{\sigma}) = \textrm{constant}
		\label{\EQ: canonical_selection_for_constant_composition}
		\\
	\Delta G_{ND}(\VEC{\sigma}, T) &\equiv& G_N(\VEC{\sigma}) - G_D(\VEC{f}(\VEC{\sigma}), T)
\end{eqnarray}
where 
$p^{\script{mut}}(\VEC{\sigma})$ is 
the probability of a sequence ($\VEC{\sigma}$) randomly occurring in a mutational process
and depends only on the amino acid frequencies $\VEC{f}(\VEC{\sigma})$,
$k_B$ is the Boltzmann constant, 
$T$ is a growth temperature, 
and $G_N$ and $G_D$ are the free energies of 
the native conformation and denatured state, respectively.
Selective temperature $T_s$
quantifies how strong the folding constraints 
are in protein evolution,
and is specific to the protein structure and function.
The free energy $G_D$ of the denatured state
does not depend on the amino acid order
but the amino acid composition, $\VEC{f}(\VEC{\sigma})$, in a sequence\CITE{SG:93b,SG:93a,RS:94,PGT:97}.
It is reasonable to assume that mutations independently occur between sites, and
therefore the equilibrium frequency of a sequence in the mutational process is equal to the product of
the equilibrium frequencies over sites; 
$
	P^{\script{mut}}(\VEC{\sigma}) = \prod_i p^{\script{mut}}(\sigma_i)
$, where $p^{\script{mut}}(\sigma_i)$ is the equilibrium frequency of $\sigma_i$ at site $i$ in the mutational process. 

The distribution of conformational energies
in the denatured state (molten globule state), which
consists of conformations as compact as the native conformation,
is approximated in 
the random energy model (REM), particularly the independent 
interaction model (IIM) \CITE{PGT:97}, to be equal to
the energy distribution of randomized sequences, 
which is then approximated by a Gaussian distribution,
in the native conformation.
That is, the partition function $Z$ for the denatured state is written as follows with 
the energy density $n(E)$ of conformations that is approximated by a product   
of a Gaussian probability density and the total number of conformations 
whose logarithm is proportional to the chain length.
\begin{eqnarray}
	Z &=& \int \exp (\frac{- E}{k_B T}) \, n(E) dE
		\\
	n(E) &\approx& \exp ( \omega L ) \mathcal{N}(\bar{E}(\VEC{f}(\VECS{\sigma})), \delta E^2(\VEC{f}(\VECS{\sigma})) )
\end{eqnarray}
where $\omega$ is the conformational entropy per residue 
in the compact denatured state, 
and $\mathcal{N}(\bar{E}(\VEC{f}(\VECS{\sigma})), \delta E^2(\VEC{f}(\VECS{\sigma})) )$ is
the Gaussian probability density with mean $\bar{E}$ and variance $\delta E^2$, 
which depend only on the amino acid composition of the protein sequence.
The free energy of the denatured state is approximated as follows.
\begin{eqnarray}
	G_D(\VEC{f}(\VECS{\sigma}),T) 
		&\approx&
		\bar{E}(\VEC{f}(\VECS{\sigma}))
	- \frac{\delta E^2(\VEC{f}(\VECS{\sigma}))}{2 k_B T} 
	- k_B T \omega L
	\\
	&=& \bar{E}(\VEC{f}(\VECS{\sigma}))
	- \delta E^2(\VEC{f}(\VECS{\sigma}))
		\frac{\vartheta(T/T_g)}{k_B T}
	\\
 \vartheta(T/T_g) &\equiv& \left\{ \begin{array}{ll}
			\frac{1}{2}(1 + \frac{T^2}{T_g^2}) & \textrm{ for } T > T_g \\
			\frac{T}{T_g} 			& \textrm{ for } T \leq T_g \\
		\end{array} 
	    \right. 
	\label{\EQ: free_energy_of_denatured_state}
\end{eqnarray}
where $\bar{E}$ and $\delta E^2$ are estimated as the mean and variance of
interaction energies of randomized sequences in the native conformation.
$T_g$ is the glass transition temperature of the protein
\TEXT{
at which entropy becomes zero\CITE{SG:93b,SG:93a,RS:94,PGT:97}; 
$ - \partial G_D / \partial T |_{T=T_g} = 0$.
}%  TEXT
\SUPPLEMENT{
at which entropy becomes zero\CITE{SG:93b,SG:93a,RS:94,PGT:97}. 
\begin{eqnarray}
	- \frac{\partial G_D}{\partial T} |_{T=T_g} &=& 0
\end{eqnarray}
}%  SUPPLEMENT
The conformational entropy per residue $\omega$ in the compact denatured state
can be represented with 
\TEXT{
$T_g$; $\omega L = \delta E^2 / (2 (k_B T_g)^2) $.
}%  TEXT
\SUPPLEMENT{
$T_g$.
\begin{eqnarray}
	\omega L &=& \frac{ \delta E^2}{ 2 (k_B T_g)^2 }
\end{eqnarray}
}%  SUPPLEMENT
Thus, unless $T_g < T_m$, a protein will be trapped at local minima 
on a rugged free energy landscape before 
it can fold into a unique native structure.

\subsection{Probability distribution of homologous sequences with the same native fold in sequence space}
\TEXT{
\label{homologous_sequences}
}%  TEXT

The probability distribution $P(\VEC{\sigma})$ of homologous sequences
with the same native fold,
$\VECS{\sigma}= (\sigma_1, \cdots, \sigma_L)$
where $\sigma_i \in \{ \text{amino acids, deletion} \}$,
in sequence space 
with maximum entropy, 
which satisfies a given amino acid frequency at each site and
a given pairwise amino acid frequency at each site pair,
is a Boltzmann distribution\CITE{MPLBMSZOHW:11,MCSHPZS:11}. 
\begin{eqnarray}
P(\VECS{\sigma}) &\propto& \exp( - \psi_N(\VECS{\sigma}) )
        \label{\EQ: potts_model}
	\\
\psi_N(\VECS{\sigma}) &\equiv& - (\sum_i^L (h_i(\sigma_i) + \sum_{j>i} J_{ij}(\sigma_i, \sigma_j)) )
        \label{\EQ: total_interaction_in_potts_model}
\end{eqnarray}
where $h_i$ and $J_{ij}$ are one-body (compositional) and two-body (covariational) interactions 
and must satisfy the following constraints.
\RED{
\begin{eqnarray}
	\sum_{\VEC{\sigma}} P(\VEC{\sigma}) \, \delta_{\sigma_i a_k} &=& P_i(a_k) 
	\\
	\sum_{\VEC{\sigma}} P(\VEC{\sigma}) \, \delta_{\sigma_i a_k} \delta_{\sigma_j a_l} &=& P_{ij}(a_k, a_l) 
\end{eqnarray}
where $\delta_{\sigma_i a_k}$ is the Kronecker delta,
$P_i(a_k)$ is the frequency of amino acid $a_k$ at site $i$, and
}%  RED
$P_{ij}(a_k, a_l)$ is the frequency of amino acid pair, $a_k$ at $i$ and 
$a_l$ at $j$; $a_k \in \{ \text{amino acids, deletion} \}$.
The pairwise interaction matrix $J$ satisfies 
	$J_{ij}(a_k, a_l) =  J_{ji}(a_l, a_k)$  
and
	$J_{ii}(a_k, a_l) = 0$.
Interactions $h_i$ and $J_{ij}$ can be well estimated
from a multiple sequence alignment (MSA) in the mean field approximation\CITE{MPLBMSZOHW:11,MCSHPZS:11},
or by maximizing a pseudo-likelihood\CITE{ELLWA:13,EHA:14}. 
Because
$\psi_N(\VEC{\sigma})$ 
has been estimated under the constraints on amino acid compositions at all sites, 
only sequences with a given amino acid composition contribute significantly to the partition function,
and other sequences may be ignored.

Hence, from \Eqs{\REF{\EQ: canonical_selection_for_constant_composition} and \REF{\EQ: potts_model}},
\begin{eqnarray}
	\psi_N(\VEC{\sigma}) &\simeq& 
 		G_N(\VEC{\sigma}) / (k_B T_s) + \textrm{function of} \VEC{f}(\VEC{\sigma})
        	\label{\EQ: expression_of_Gn}
		\\
	\psi_D(\VEC{f}(\VEC{\sigma}), T) &\simeq&	
		G_D(\VEC{f}(\VEC{\sigma}), T)  / (k_B T_s) + \textrm{function of} \VEC{f}(\VEC{\sigma})
        	\label{\EQ: expression_of_Gd}
		\\
	\Delta \psi_{ND}(\VEC{\sigma}, T) &\simeq&	
		\Delta G_{ND}(\VEC{\sigma}, T) / (k_B T_s)
        	\label{\EQ: expression_of_dG}
		\\
	\Delta \psi_{ND}(\VEC{\sigma}, T) &\equiv& \psi_{N}(\VEC{\sigma}) - \psi_{D}(\VEC{f}(\VEC{\sigma}), T)
        	\label{\EQ: expression_of_dPsi}
		\\
	\psi_D(\VEC{f}(\VEC{\sigma}), T) &\approx&
			\bar{\psi}(\VEC{f}(\VEC{\sigma})) - \delta \psi^2(\VEC{f}(\VEC{\sigma}))
			\vartheta(T/T_g) T_s / T
        	\label{\EQ: expression_of_psi_D}
		\\
	\omega &=&  (T_s/ T_g)^2 \delta \psi^2 / (2 L)
        \label{\EQ: expression_of_entropy}
\end{eqnarray}
where 
the $\bar{\psi}$ and ${\delta \psi}^2$ are estimated 
as the mean and variance of $\psi_N$ over randomized sequences;
$\bar{E} \simeq k_B T_s \bar{\psi}$ and ${\delta E}^2 \simeq (k_B T_s)^2 {\delta \psi}^2$.

% End of methods_for_protein_folding.tex

% \input{methods_for_mutation_fixation_process.tex}

\subsection{The equilibrium distribution of sequences in a mutation-fixation process}

Here we assume that the mutational process is a reversible Markov process.
That is,
the mutation rate per gene, $M_{\VEC{\mu}\VEC{\nu}}$, 
from sequence $\VEC{\mu} \equiv (\mu_1, \cdots, \mu_L)$ to $\VEC{\nu}$ satisfies
the detailed balance condition
\begin{eqnarray}
	P^{\script{mut}}(\VEC{\mu}) M_{\VEC{\mu}\VEC{\nu}} &=& P^{\script{mut}}(\VEC{\nu}) M_{\VEC{\nu}\VEC{\mu}} 
	\label{\EQ: detailed_balance_for_mutation}
\end{eqnarray}
where $P^{\script{mut}}(\VEC{\nu})$ is the equilibrium frequency 
of sequence $\VEC{\nu}$ in a mutational process, $M_{\VEC{\mu}\VEC{\nu}}$.
The mutation rate per population is equal to $2 N M_{\VEC{\mu}\VEC{\nu}}$
for a diploid population, where $N$ is the population size.
The substitution rate 
$R_{\VEC{\mu}\VEC{\nu}}$ from $\VEC{\mu}$ to $\VEC{\nu}$
is equal to the product of
the mutation rate and the fixation probability 
with which a single mutant gene becomes to fully occupy the population\CITE{CK:70}.
\begin{eqnarray}
	R_{\VEC{\mu}\VEC{\nu}}
	&=& 
	2 N M_{\VEC{\mu}\VEC{\nu}} u(s(\VEC{\mu} \rightarrow \VEC{\nu}))
\end{eqnarray}
where $u(s(\VEC{\mu} \rightarrow \VEC{\nu}))$ is the fixation probability of mutants 
from $\VEC{\mu}$ to $\VEC{\nu}$ the selective advantage of which is equal to $s$. 

For genic selection (no dominance) or gametic selection in a Wright-Fisher population of diploid,
the fixation probability, $u$, of a single mutant gene, the selective advantage of which is equal to $s$ 
and the frequency of which in a population is equal to $q_m = 1/(2N)$, was estimated\CITE{CK:70} as
\begin{eqnarray}
	2N u(s) &=& 2N \frac{1 - e^{-4N_e s q_{m}}}{1 - e^{- 4N_e s}}
	\label{\EQ: fixation_probability}
	\\
	&=& \frac{u(s)}{u(0)}
	\hspace*{2em} \text{ with } \hspace*{2em} q_{\script{m}} = \frac{1}{2N} 	
	\label{\EQ: def_qm}
\end{eqnarray}
where $N_e$ is effective population size.
\RED{
\Eq{\ref{\EQ: fixation_probability}} will be also valid for haploid population if $2N_e$ and $2N$ are replaced by $N_e$ and $N$, respectively.
Also, for Moran population of haploid, $4N_e$ and $2N$ should be replaced by
$N_e$ and $N$, respectively.
Fixation probabilities for various selection models, which are compiled from p. 192 and p. 424--427 of Crow and Kimura (1970)
and from 
Moran (1958) and Ewens (1979),
are listed in \Table{\ref{stbl: fixation_prob}}.
}%  RED
The selective advantage of a mutant sequence $\VEC{\nu}$ 
to a wildtype $\VEC{\mu}$ 
is equal to
\begin{eqnarray}
	s(\VEC{\mu} \rightarrow \VEC{\nu}) &=& m(\VEC{\nu}) - m(\VEC{\mu}) 
\end{eqnarray}
where $m(\VEC{\nu})$ is the Malthusian fitness of a mutant sequence,
and $m(\VEC{\mu})$ is for the wildtype. 

This Markov process of substitutions in sequence
is reversible, and the equilibrium frequency of sequence $\VEC{\mu}$, 
$P^{\script{eq}}(\VEC{\mu})$, 
in the total process consisting of mutation and fixation processes
is represented by
\begin{eqnarray}
	P^{\script{eq}}(\VEC{\mu}) &=& 
	\frac{P^{\script{mut}}(\VEC{\mu}) \exp( 4N_e m(\VEC{\mu}) (1-q_m) ) }
		{ \sum_{\VEC{\nu}} P^{\script{mut}}(\VEC{\nu}) 
			\exp( 4N_e m(\VEC{\nu} ) (1-q_m) )}
	\label{\EQ: equilibrium_of_mutation_fixation_process}
\end{eqnarray}
because both the mutation and fixation processes satisfy the detailed balance conditions,
\Eq{\ref{\EQ: detailed_balance_for_mutation}} and the following equation, respectively.
\begin{eqnarray}
\lefteqn{
	\exp( 4N_e m(\VEC{\mu} ) (1-q_m) )
	\,
	u(s(\VEC{\mu} \rightarrow \VEC{\nu}))
}
	\nonumber
	\\
	&=& 
	\frac{\exp(-4N_e m(\VEC{\mu}  ) q_m) - \exp(-4N_e m(\VEC{\nu} ) q_m) }
	{\exp(-4N_e m(\VEC{\mu})) - \exp(-4N_e m(\VEC{\nu})) }
	\\
	&=&
	\exp( 4N_e m(\VEC{\nu}) (1-q_m) )
	\, u(s(\VEC{\nu} \rightarrow \VEC{\mu}))
\end{eqnarray}
As a result, the ensemble of homologous sequences in molecular evolution 
obeys a Boltzmann distribution.

\subsection{Relationships between $m(\VEC{\sigma})$, $\psi_N(\VEC{\sigma})$, and
$\Delta G_{ND}(\VEC{\sigma})$ of protein sequence}

From \Eqs{
\REF{\EQ: canonical_selection},
\REF{\EQ: potts_model},
and
\REF{\EQ: equilibrium_of_mutation_fixation_process}
},
we can get the following relationships among the Malthusian fitness $m$,
the folding free energy $\Delta G_{ND}$
and $\Delta \psi_{ND}$ of protein sequence.
\begin{eqnarray}
	P^{\script{eq}}(\VEC{\mu}) &=& \frac{ P^{\script{mut}}(\VEC{\mu}) \exp( 4N_e m( \VEC{\mu} ) ( 1 - q_m )) }
		{ \sum_\nu P^{\script{mut}}(\VEC{\nu}) 
			\exp( 4N_e m(\VEC{\nu}) ( 1 - q_m )) )}
		\label{\EQ: equilibrium_distr_of_seq_of_m}
		\\
	&=& \frac{ P^{\script{mut}}(\overline{\VEC{\mu}}) \exp ( - (\psi_{N}(\VEC{\mu}) - \psi_{D}(\overline{\VEC{f}(\VEC{\mu})}, T) )) }
		{ \sum_{\VEC{\nu}} P^{\script{mut}}(\overline{\VEC{\nu}}) 
			\exp ( - (\psi_{N}(\VEC{\nu}) - \psi_{D}(\overline{\VEC{f}(\VEC{\nu})}, T) ) ) }
		\label{\EQ: equilibrium_distr_of_seq}
		\label{\EQ: equilibrium_distr_of_seq_of_dPsi}
		\\
	&\simeq& 
		\frac{ P^{\script{mut}}(\VEC{\mu}) \exp( - \Delta G_{ND}(\VEC{\mu}, T ) / (k_B T_s) ) }
		{ \sum_\nu P^{\script{mut}}(\VEC{\nu}) 
			\exp( - \Delta G_{ND}(\VEC{\nu}, T) / (k_B T_s) )}
		\label{\EQ: equilibrium_distr_of_seq_of_dG}
\end{eqnarray}
where 
$\overline{\VEC{f}(\VEC{\sigma}) } \equiv \sum_{\VEC{\sigma}} \VEC{f}(\VEC{\sigma}) P(\VEC{\sigma})$ and
$\log P^{\script{mut}}(\overline{\VEC{\sigma}}) \equiv \sum_{\VEC{\sigma}} P(\VEC{\sigma}) \log (\prod_i P^{\script{mut}}(\sigma_i))$.
Then, the following relationships are derived for sequences for which $f(\VEC{\mu}) = \overline{f(\VEC{\mu})}$.
\begin{eqnarray}
	4N_e m(\VEC{\mu}) (1 - q_m)
	&=& - \Delta \psi_{ND}(\VEC{\mu},T) + \mathrm{constant}
	\label{\EQ: relationship_between_m_and_dPsi}
        \\
	&\simeq& 
	\frac{- \Delta G_{ND}(\VEC{\mu},T)}{k_B T_s} + \mathrm{constant}
	\label{\EQ: relationship_between_m_and_dG}
\end{eqnarray}
The selective advantage of $\VEC{\nu}$ to $\VEC{\mu}$ is represented as follows for 
$f(\VEC{\mu}) = f(\VEC{\nu}) = \overline{f(\VEC{\sigma})}$.
\begin{eqnarray}
\lefteqn{
	4N_e s(\VEC{\mu} \rightarrow \VEC{\nu}) ( 1 - q_m )
}
	\nonumber
	\\
	&=&
	(4N_e m(\VEC{\nu})
	- 4N_e m( \VEC{\mu} ) ) ( 1 - q_m )
	\\
	&=& - (\Delta \psi_{ND}(\VEC{\nu}, T) - \Delta \psi_{ND}(\VEC{\mu}, T) )
	= 
	- ( \psi_N(\VEC{\nu}) - \psi_N(\VEC{\mu}) )
	\label{\EQ: s_vs_dPsi}
	\label{\EQ: selective_advantage_and_dPsi_N}
	\\
	&\simeq& 
	- (\Delta G_{ND}(\VEC{\nu}, T) - \Delta G_{ND}(\VEC{\mu}, T) ) / (k_B T_s)	 
	\label{\EQ: s_vs_dG}
	= 
	- ( G_N(\VEC{\nu}) - G_N(\VEC{\mu}) ) / (k_B T_s)
\end{eqnarray}
It should be noted here that
only sequences for which $f(\VEC{\sigma}) = \overline{f(\VEC{\sigma})}$ 
contribute significantly to the partition functions in 
\Eq{\ref{\EQ: equilibrium_distr_of_seq_of_dPsi}}, and other sequences may be ignored.

\RED{
\Eq{\ref{\EQ: s_vs_dPsi}} indicates that evolutionary statistical energy $\psi$ 
should be proportional to effective population size $N_e$, 
and therefore it is ideal to estimate one-body ($h$) and two-body ($J$) interactions
from homologous sequences of species that 
do not significantly differ in effective population size. 
Also, \Eq{\ref{\EQ: s_vs_dG}} indicates that 
selective temperature $T_s$ is inversely proportional to the effective population size $N_e$;
$T_s \propto 1 / N_e$,
because free energy is a physical quantity and should not depend on effective population size.
}%  RED

% End of methods_for_mutation_fixation_process.tex

% \input{methods_for_ensemble_averages.tex}

\subsection{The ensemble average of folding free energy, $\Delta G_{ND}(\VEC{\sigma}, T)$, over sequences}
\TEXT{
\label{ensemble_average}
}%  TEXT

The ensemble average of $\Delta G_{ND}(\VEC{\sigma}, T)$ over sequences 
with \Eq{\ref{\EQ: canonical_selection}} is
\begin{eqnarray}
\lefteqn{
	\langle \Delta G_{ND}(\VEC{\sigma}, T) \rangle_{\VEC{\sigma}} 
} 
	\\
	&\equiv& 
	\, [ \, 
	\sum_{\VEC{\sigma}} \Delta G_{ND}(\VEC{\sigma}, T) 
		P^{\script{mut}}(\VEC{\sigma}) \exp( - \frac{\Delta G_{ND}(\VEC{\sigma}, T)}{k_B T_s} ) 
	\, ] \,
	/ 
	\, [ \, 
	\sum_{\VEC{\sigma}}
		P^{\script{mut}}(\VEC{\sigma}) \exp( - \frac{\Delta G_{ND}(\VEC{\sigma}, T)}{k_B T_s} ) 
	\, ] \,
	\\
	&\approx&
		\, [ \,
		\sum_{\VEC{\sigma}\, | \, \VEC{f}(\VEC{\sigma}) = \overline{\VEC{f}(\VEC{\sigma}_N)} }  
		G_{N}(\VEC{\sigma} ) 
		\exp( - \frac{G_{N}(\VEC{\sigma})}{k_B T_s} ) 
		\, ] \,  
		/ 
		\, [ \,
		\sum_{\VEC{\sigma}\, | \, \VEC{f}(\VEC{\sigma}) = \overline{\VEC{f}(\VEC{\sigma}_N)} }  
		\exp( - \frac{G_{N}(\VEC{\sigma})}{k_B T_s} ) 
		\, \, ] \,  
		- G_{D}( \overline{\VEC{f}(\VEC{\sigma}_N)} , T)
	\label{\EQ: native_sequence_approximation}
	\\
	&=& \langle G_N(\VEC{\sigma}) \rangle_{\VEC{\sigma}} - G_D( \overline{\VEC{f}(\VEC{\sigma}_N)} , T)
\end{eqnarray}
where $\VEC{\sigma}_N$ denotes 
a natural sequence,  
and $\overline{\VEC{f}(\VEC{\sigma_N})}$ denotes the average of 
amino acid frequencies $\VEC{f}(\VEC{\sigma_N})$ over homologous sequences.
In \Eq{\ref{\EQ: native_sequence_approximation}}, 
the sum over all sequences is approximated by the sum over sequences the amino acid composition
of which is the same as that over the 
natural sequences.

The ensemble averages of $G_N$
and
$\psi_N(\VEC{\sigma})$ 
are estimated in the Gaussian approximation\CITE{PGT:97}.
\begin{eqnarray}
	\langle G_N(\VEC{\sigma}) \rangle_{\VEC{\sigma}} 	
	&\approx& \frac{ \int E \exp (- E / (k_B T_s)) \, n(E) \, dE }  
		{\int \exp (- E / (k_B T_s)) \, n(E) \, dE }
	\\
	&=& \bar{E}( \overline{\VEC{f}(\VEC{\sigma_N})} ) - {\delta E}^2( \overline{\VEC{f}(\VEC{\sigma_N})} ) / (k_B T_s)
	\label{\EQ: ensemble_ave_of_G}
\end{eqnarray}
\begin{eqnarray}
\langle \psi_N(\VEC{\sigma}) \rangle_{\VEC{\sigma}} 	
	&\equiv& 
	\, [ \, 
	\sum_{\VEC{\sigma}} \psi_{ND}(\VEC{\sigma}) 
		\exp( - \psi_{N}(\VEC{\sigma}) )
	\, ] \,
	/ 
	\, [ \, 
	\sum_{\VEC{\sigma}}
		\exp( - \psi_{N}(\VEC{\sigma}) ) 
	\, ] \,
	\\
	&\approx& \bar{\psi}( \overline{\VEC{f}(\VEC{\sigma_N})} ) - {\delta \psi}^2( \overline{\VEC{f}(\VEC{\sigma_N})} )
	\label{\EQ: ensemble_ave_of_psi}
\end{eqnarray}
The ensemble averages of $\Delta G_{ND}(\VEC{\sigma}, T)$ and $\psi_N(\VEC{\sigma})$ over sequences
are observable as
the sample averages of $\Delta G_{ND}(\VEC{\sigma_N}, T)$ and $\psi_N(\VEC{\sigma_N})$ over 
homologous sequences fixed in protein evolution,
respectively. 
\begin{eqnarray}
\overline{ \Delta G_{ND}(\VEC{\sigma_N}, T) } / (k_B T_s)
	&=& 
	\langle \Delta G_{ND}(\VEC{\sigma}, T) \rangle_{\VEC{\sigma}} / (k_B T_s) 
	\\
\SUPPLEMENT{
	&\approx& \, [ \, {\delta E}^2( \overline{\VEC{f}(\VEC{\sigma_N})} ) 
	\, [ \, \vartheta(T/T_g) T_s / T - 1 \, ] / ( k_B T_s )^2
	\\
	&=&
}%  SUPPLEMENT
\TEXT{
	&\approx& 
}%  TEXT
	{\delta \psi}^2( \overline{\VEC{f}(\VEC{\sigma_N})} ) \, [ \,
	\vartheta(T/T_g) T_s / T - 1 \, ]
	\label{\EQ: ensemble_ave_of_ddG}
	\\
\SUPPLEMENT{
	&=& \overline{ \Delta G_{ND}(\VEC{\sigma_N}, T_g) } \, / \, ( k_B T^{\prime}_s )
	\\
	T^{\prime}_s &=& T_s (T_s/T - 1) / ( \vartheta(T/T_g) T_s / T - 1 )
	\\
}%  SUPPLEMENT
	\overline{ \psi_N(\VEC{\sigma_N}) } 
	&\equiv &
		\frac{\sum_{\VEC{\sigma}_N} w_{\VEC{\sigma}_N} \psi_N(\VEC{\sigma}_N) }{\sum_{\VEC{\sigma}_N} w_{\VEC{\sigma}_N} } 
	\label{\EQ: def_sample_ave_of_psi}
		\\
	&=&
	\langle \psi_N(\VEC{\sigma}) \rangle_{\VEC{\sigma}} 	
	\label{\EQ: sample_ave_of_psi}
\end{eqnarray}
where the overline denotes a sample average with a sample weight 
$w_{\VEC{\sigma}_N}$ for each homologous sequence,
which is used to reduce phylogenetic biases in the set of homologous sequences.
\SUPPLEMENT{
$\Delta G_{ND}(\VEC{\sigma_N}, T_g)$ corresponds to the energy gap\CITE{SG:93a} 
between the native and the glass states, and
$T^{\prime}_s$ will be the selective temperature
if $\Delta G_{ND}(\VEC{\sigma_N}, T_g)$ is used for selection instead of $\Delta G_{ND}(\VEC{\sigma_N}, T)$.
}%  SUPPLEMENT

The folding free energy becomes equal to zero at the melting temperature $T_m$; 
$\langle \Delta G_{ND}(\VEC{\sigma_N}, T_m) \rangle_{\VEC{\sigma}} = 0$.  Thus, the following relationship must be 
satisfied\CITE{SG:93b,SG:93a,RS:94,PGT:97}.
\begin{eqnarray}
	\vartheta(T_m/T_g) \frac{T_s}{T_m} &=& \frac{T_s}{2T_m}(1 + \frac{T_m^2}{T_g^2}) = 1
	\hspace*{1em} \textrm{ with } T_s \leq T_g \leq T_m
	\label{\EQ: relationship_among_characteristic_T}
\end{eqnarray}

% End of methods_for_ensemble_averages.tex

\TEXT{

\subsection{Probability distributions of selective advantage, fixation rate and $K_a/K_s$}

\SUPPLEMENT{
Now, we can consider 
}%  SUPPLEMENT
\TEXT{
Let us consider 
}%  TEXT
the probability distributions of
characteristic quantities that describe the evolution of genes. 
First of all, the probability density function (PDF) of 
selective advantage $s$, $p(s)$, of mutant genes can be calculated
from the PDF of 
the change of $\Delta \psi_{ND}$ due to a mutation from $\VEC{\mu}$ to $\VEC{\nu}$, 
$\Delta\Delta \psi_{ND} (\equiv \Delta \psi_{ND}(\VEC{\nu}, T) - \Delta \psi_{ND}(\VEC{\mu}, T) )$.
The PDF of $4N_e s$, $p(4N_e s) = p(s)/(4N_e) $, may be more useful than $p(s)$.
\begin{eqnarray}
p(4N_e s)
	&=& p(\Delta \Delta \psi_{ND}) \, | \frac{d \Delta \Delta \psi_{ND}}{d 4N_e s} |
	=
	p(\Delta\Delta \psi_{ND}) (1 - q_m)
	\label{\EQ: pdf_of_4Nes}
	\label{\EQ: pdf_of_s}
\end{eqnarray}
where $\Delta\Delta \psi_{ND}$ must be regarded as a function of $4N_es$, that is,
$
\Delta\Delta \psi_{ND} = - 4N_e s (1 - q_m)
$; see \Eq{\ref{\EQ: s_vs_dPsi}}.

The PDF of fixation probability $u$ can be represented by
\begin{eqnarray}
p(u)	&=& p(4N_e s) \frac{d 4N_e s}{d u} 
	= 
	p(4N_e s)
	 \frac{ ( e^{4N_e s} - 1)^2 e^{4N_e s (q_m - 1)} }
	{ q_m( e^{4N_e s} - 1) - (e^{4N_e s q_m} - 1) }
		\label{\EQ: pdf_of_fixation_prob}
\end{eqnarray}
where $4N_e s$ must be regarded as a function of $u$. 

The ratio of the substitution rate per nonsynonymous site ($K_a$) for nonsynonymous
substitutions with selective advantage s to the substitution rate per
synonymous site ($K_s$) for synonymous substitutions with s = 0 is
\begin{eqnarray}
	\frac{K_a}{K_s} &=& \frac{u(s)}{u(0)} = \frac{u(s)}{q_m}
		\label{\EQ: def_Ka_over_Ks}
\end{eqnarray}
assuming that synonymous substitutions are completely neutral 
and mutation rates at both types of sites are the same.
The PDF of $K_a/K_s$ is
\begin{eqnarray}
	p(K_a/K_s) &=& 
		p(u) \frac{d u}{d (K_a/K_s)} 
	= p(u) \, q_m  
		\label{\EQ: pdf_of_Ka_over_Ks}
\end{eqnarray}

\subsection{Probability distributions of $\Delta\Delta \psi_{ND}$, $4N_e s$, $u$, and $K_a/K_s$ in fixed mutant genes}

\SUPPLEMENT{
Now, let us consider fixed mutant genes. 
}%  SUPPLEMENT
The PDF of $\Delta\Delta \psi_{ND}$ in fixed mutants is
proportional to that multiplied by the fixation probability.
\begin{eqnarray}
p(\Delta\Delta \psi_{ND, \script{fixed}})
	&=& p(\Delta\Delta \psi_{ND}) 
	\frac{u(s(\Delta \Delta \psi_{ND}))}{\langle u(s(\Delta \Delta \psi_{ND})) \rangle}
	\label{\EQ: pdf_of_fixed_ddPsi}
	\\
  \langle u \rangle &\equiv& \int_{-\infty}^{\infty} u(s) p(\Delta\Delta \psi_{ND}) d\Delta\Delta \psi_{ND}
	\label{\EQ: ave_of_u}
\end{eqnarray}
Likewise, the PDF of selective advantage 
in fixed mutants is 
\begin{eqnarray}
p(4N_e s_{\script{fixed}}) &=&
	p(4N_e s)
	\frac{u(s)}{\langle u(s) \rangle}
\end{eqnarray}
and those of the $u$ and $K_a/K_s$ in fixed mutants are
\begin{eqnarray}
	p( u_{\script{fixed}} ) &=&
		p(u) \frac{u}{\langle u \rangle}
		\\
	p( (\frac{K_a}{K_s})_{\script{fixed}}) &=&
		p(\frac{K_a}{K_s}) \frac{u}{\langle u \rangle}
		=
		p(\frac{K_a}{K_s}) \frac{\frac{K_a}{K_s}}{\langle \frac{K_a}{K_s} \rangle}
\end{eqnarray}
The average of $K_a/K_s$ in fixed mutants is equal to the ratio of the second moment to the first moment of $K_a/K_s$
\TEXT{
in all arising mutants; 
$
\langle K_a/K_s \rangle_{\script{fixed}} = \langle (K_a / K_s)^2 \rangle / \langle K_a/K_s \rangle
$.
}%  TEXT
\SUPPLEMENT{
in all arising mutants.
\begin{eqnarray}
\langle \frac{K_a}{K_s} \rangle_{\script{fixed}} &=& \langle (\frac{K_a}{K_s})^2 \rangle / \langle \frac{K_a}{K_s} \rangle
\end{eqnarray}
}%  SUPPLEMENT

% End of methods_for_monoclonal_approximation.tex
}%  TEXT

\SUPPLEMENT{

\subsection{ Estimation of $\bar{\psi}(\VEC{f}(\VECS{\sigma}))$ and $\delta {\psi}^2(\VEC{f}(\VECS{\sigma}))$ }
\label{estimation_psi_distr}

The mean $\bar{\psi}(\VEC{f}(\VECS{\sigma}))$ and the variance $\delta {\psi}^2(\VEC{f}(\VECS{\sigma}))$ 
in the Gaussian approximation for the distribution of conformational energies at the denatured state are estimated 
as the mean and variance of $\psi_N$ of random sequences
in the native conformation\CITE{PGT:97}.
\begin{eqnarray}
	\bar{\psi}(\VEC{f}(\VECS{\sigma})) &=& 
	- \sum_i [ \hat{h}_i(::) +  \sum_{j>i} \hat{J}_{ij}(::,::) ] 
\end{eqnarray}
where $\hat{h}_i(::)$ and $\hat{J}_{ij}(::,::)$ are the means of one-body and two-body interactions in random sequences.
\begin{eqnarray}
	\hat{h}_i(::) &\equiv& \sum_k \hat{h}_i(a_k)f_{a_k}(\VECS{\sigma}) 
	\\
	\hat{J}_{ij}(::,::) &\equiv& \sum_k \sum_l \hat{J}_{ij}(a_k, a_l)f_{a_k}(\VECS{\sigma}) f_{a_l}(\VECS{\sigma})
\end{eqnarray}
where $f_{a_k}(\VECS{\sigma})$ is the composition of amino acid $a_k$ in the sequence $\VECS{\sigma}$.
\begin{eqnarray}
	f_{a_k}(\VECS{\sigma}) &=& \frac{1}{L} \sum_{i=1}^{L} \delta_{\sigma_i a_k}
\end{eqnarray}
where $\delta_{\sigma_i a_k}$ is the Kronecker delta.
The variance, $\delta\psi^2(\VEC{f}(\VEC{\sigma}))$, is
\begin{eqnarray}
	\delta {\psi}^2(\VEC{f}(\VECS{\sigma})) &=& 
	\sum_k \, f_{a_k}(\VECS{\sigma}) \, \sum_i \, [ \, \delta \hat{h}_i(a_k)^2
	+ \sum_{j \neq i} \, \{ \, 2 \delta \hat{h}_i(a_k) \delta \hat{J}_{ij}(a_k,::)
	\\
	&+& \sum_{m \neq \{i, j\}} \delta \hat{J}_{ij}(a_k, ::) \delta \hat{J}_{i m}(a_k,::) 
	+ \frac{1}{2} \sum_l \delta \hat{J}_{ij}(a_k, a_l)^2 f_{a_l}(\VECS{\sigma}) \, \} \, 
	\, ] 
\end{eqnarray}
where
\begin{eqnarray}
	\delta \hat{h}_i(a_k) &\equiv& \hat{h}_i(a_k) - \hat{h}_i(::)
	\\
	\delta \hat{J}_{ij}(a_k, ::) &\equiv& \hat{J}_{ij}(a_k, ::) - \hat{J}_{ij}(::, ::)
	\\
	\delta \hat{J}_{ij}(a_k, a_l) &\equiv& \hat{J}_{ij}(a_k, a_l) - \hat{J}_{ij}(::, ::)
\end{eqnarray}

% End of methods_to_estimate_Gaussian_parameters.tex
}%  SUPPLEMENT

\SUPPLEMENT{

\subsection{Estimation of one-body ($h$) and pairwise ($J$) interactions}

  The estimates of $h$ and $J$\CITE{MPLBMSZOHW:11,MCSHPZS:11} are noisy as a result of estimating
many interaction parameters from a relatively small number of sequences.
Therefore, only pairwise interactions within a certain distance are taken into account;
the estimate of $J$ is modified as follows,
according to Morcos et al.\CITE{MSCOW:14}.

\begin{eqnarray}
	\hat{J}^{\script{q}}_{ij}(a_k, a_l) &=& J^{\script{q}}_{ij}(a_k, a_l) H( r_{\script{cutoff}} - r_{ij} )
		\label{\EQ: estimation_of_J}
\end{eqnarray}
where 
$\hat{J}^{\script{q}}$ is the statistical estimate of $J$ in the mean field approximation in which
the amino acid $a_q$ is the reference state,
$H$ is the Heaviside step function, and $r_{ij}$ is the distance between the centers of amino acid side chains in protein structure,
and $r_{\script{cutoff}}$ is a distance threshold for residue pairwise interactions.
Maximum interaction ranges employed for pairwise interactions are
$r_{\script{cutoff}} \sim 8$ and $15.5$ \AA\ , 
which correspond to the first and second interaction shells between residues, respectively.
Here it should be noticed that the 
total interaction 
$\psi_N(\VECS{\sigma})$ 
defined by \Eq{\ref{\EQ: total_interaction_in_potts_model}} does not depend on any gauge 
unless the interaction range for pairwise interactions is limited, but 
a gauge conversion in which interconversions between $h$ and $J$ 
occur must not be done before calculating $\hat{J}$, because
it may change the estimate of $\psi_N(\VECS{\sigma})$
in the present scheme of \Eq{\ref{\EQ: estimation_of_J}} 
in which pairwise interactions are cut off at a certain distance.
Thus, a natural gauge must be used before calculating $\hat{J}$. 

For example, let us think about the Ising gauge\CITE{EHA:14}, in which $h^I$ and $J^I$ can be
calculated from $h^g$ and $J^g$ in any gauge through the following conversions.
\begin{eqnarray}
	J^{\script{I}}_{ij}(a_k, a_l) &=& J^{\script{g}}_{ij}(a_k, a_l) - J^{\script{g}}_{ij}(a_k, :)
				- J^{\script{g}}_{ij}(:, a_l) + J^{\script{g}}_{ij}(:, :) 
		\label{\EQ: to_Ising_gauge_1}
			\\
	h^{\script{I}}(a_k) &=& h^{\script{g}}_{i}(a_k) - h^{\script{g}}_{i}(:)
			+ \sum_{j \neq i} (J^{\script{g}}_{ij}(a_k, :) - J^{\script{g}}_{ij}(:, :) ) 
		\label{\EQ: to_Ising_gauge_2}
\end{eqnarray}
where
\begin{eqnarray}
	h_{i}(:) &\equiv& \frac{1}{q} \sum_{k=1}^q h_{i}(a_k)
		\\
	J_{ij}(:, :) &\equiv& \frac{1}{q^2} \sum_{k=1}^q \sum_{l=1}^q J_{ij}(a_k, a_l)
\end{eqnarray}
where $q$ is equal to the total number of amino acid types including deletion, that is, $q=21$.
Thus, the gauge conversion of $\hat{J}$ does not affect the total interaction $\psi_N(\VECS{\sigma})$
but the gauge conversion before calculating $\hat{J}$ may significantly change the total interaction.

In the DCA\CITE{MPLBMSZOHW:11,MCSHPZS:11}, 
the interaction terms are estimated in the mean field approximation as follows.
\begin{eqnarray}
	J^{\script{q}}_{ij}(a_k, a_l) &=& - (C^{-1})_{ij}(a_k, a_l)		\\	
	J^{\script{q}}_{ij}(a_q, a_l) &=& J^{\script{q}}_{ij}(a_k, a_q) = J^{\script{q}}_{ij}(a_q, a_q) = 0
\end{eqnarray}
where $i \neq j$ and $1 \leq k, l \leq q - 1$, and the covariance matrix $C$ is defined as
\begin{eqnarray}
	C_{ij}(a_k,a_l) &\equiv & P_{ij}(a_k, a_l) - P_i(a_k) P_j(a_l)	
\end{eqnarray}
Here, one ($a_q$) of the amino acid types including deletion is used as the reference state; 
$J^{\script{q}}$ denotes the $J$ in this gauge, which is called the $q$ gauge here.
According to Morcos et al.\CITE{MPLBMSZOHW:11},
the probability $P_i(a_k)$ of amino acid $a_k$ at site $i$ and 
the joint probabilities $P_{ij}(a_k, a_l)$ of amino acids, $a_k$ at site $i$ and $a_l$ at site $j$, 
are
evaluated by 
\begin{eqnarray}
	P_i(a_k) &=& (1 - p_c) f_i(a_k) + p_c \frac{1}{q} 
	\label{\EQ: pseudocount_for_Pi}
	\\
	P_{ij}(a_k, a_l) &=& (1 - p_c) f_{ij}(a_k, a_l) + p_c \frac{1}{q^2} 
		\hspace*{2em} \text{ for } \hspace*{1em} i \neq j
	\label{\EQ: pseudocount_for_Pij}
	\\
	P_{ii}(a_k, a_l) &=& P_i(a_k) \delta_{a_k a_l}  
\end{eqnarray}
where   
$0 \leq p_c \leq 1$ is the ratio of pseudocount, and
$f_i(a_k)$ is the frequency of amino acid $a_k$ at site $i$ and $f_{ij}(a_k, a_l)$
is the frequency of the site pair, $a_k$ at $i$ and $a_l$ at $j$, in an alignment; 
$f_{ii}(a_k, a_l)$ is defined as $f_{ii}(a_k, a_l) = f_i(a_k) \delta_{a_k a_l}$.

In the mean field approximation, one body interactions $h^q_i(a_k)$ 
in the $q$ gauge are estimated by $\hat{h}^q_i(a_k) = \log (P_i(a_k)/P_i(a_q)) - \sum_{j \neq i} \sum_{l \neq q} \hat{J}^q_{ij}(a_k, a_l) P_j(a_l)$.
Here, instead the one body interactions $h_i(a_k)$ are estimated in the isolated two-state model\CITE{MPLBMSZOHW:11}, 
that is,
\begin{eqnarray}
	P_i(a_k) &\propto& \exp \, [ \, h^{\script{q}}_{ij}(a_k) + J^{\script{q}}_{ij}(a_k, a_l) + h^{\script{q}}_{ji}(a_l) \, ]
		\\
	\hat{h}^{\script{q}}_i(a_k) &=& \frac{1}{L-1} \sum_{j \neq i} h^{\script{q}}_{ij}(a_k)
\end{eqnarray}
These $\hat{h}^q$ and $\hat{J}^q$ in the $q$ gauge are converted to a new gauge, which
is called the zero-sum gauge here, 
\begin{eqnarray}
	\hat{h}^{\script{s}}_{i}(a_k) &=& \hat{h}^{\script{q}}_{i}(a_k) - \hat{h}^{\script{q}}_{i}(:)
	\label{\EQ: simple_gauge_1}	\\
	\hat{J}^{\script{s}}_{ij}(a_k, a_l) &=& \hat{J}^{\script{q}}_{ij}(a_k, a_l) - \hat{J}^{\script{q}}_{ij}(:, :)
	\label{\EQ: simple_gauge_2}
\end{eqnarray}
In this gauge, the reference state is the average state over amino acids including deletion, instead of
a specific amino acid ($a_q$) in the $q$ gauge.

% End of methods_for_h_and_J_estimations.tex
}%  SUPPLEMENT

\SUPPLEMENT{

\subsection{Distribution of $\Delta\Delta \psi_{ND} \simeq \Delta \psi_N$ due to single nucleotide nonsynonymous substitutions }

The probability density function (PDF) of $\Delta\Delta \psi_{ND}$, $p(\Delta\Delta \psi_{ND})$, 
due to single nucleotide nonsynonymous substitutions is 
approximated by the PDF of $\Delta \psi_{N}$, $p(\Delta \psi_{N})$, because $\Delta \psi_{D} \simeq 0$ 
for single amino acid substitutions.
\begin{eqnarray}
	\Delta\Delta \psi_{ND} &\simeq& \Delta \psi_{N}
		\\
	p(\Delta\Delta \psi_{ND}) &\simeq& p(\Delta \psi_N)
\end{eqnarray}
for single nucleotide nonsynonymous substitutions.

For simplicity, a log-normal distribution, 
$\ln\mathcal{N}(x; \mu, \sigma)$, for which $x, \mu$ and $\sigma$ are defined as follows, 
is arbitrarily employed here to reproduce observed PDFs of $\Delta \psi_N$, 
particularly in the domain of $\Delta \psi_N < \overline{\Delta \psi_N}$,   
although other distributions such as
inverse $\Gamma$ distributions can equally reproduce the observed ones, too.
\begin{eqnarray}
	p(\Delta \psi_N) &\approx&
	\ln\mathcal{N}(x; \mu, \sigma) \equiv \frac{1}{x} \mathcal{N}(\ln x; \mu, \sigma)
	\label{\EQ: log-normal}
	\\
	x &\equiv& \max ( \Delta \psi_{N} - \Delta \psi_{N}^{\mbox{o}}, 0)
	\\
	\exp(\mu + \sigma^2/2) &=& \overline{\Delta \psi_{N}} - \Delta \psi_{N}^{\mbox{o}}
	\\
	\exp(2\mu + \sigma^2) ( \exp(\sigma^2) - 1 ) &=& \overline{ (\Delta \psi_{N} - \overline{\Delta \psi_{N}})^2 })
	\\
	\Delta \psi_{N}^{\mbox{o}} &\equiv& \min (\overline{\Delta \psi_{N}} 
		- n_{\mbox{shift}} \overline{ (\Delta \psi_{N} - \overline{\Delta \psi_{N}})^2 })^{1/2}, 0)
	\label{\EQ: statistics_for_log-normal}
\end{eqnarray}
where $\Delta \psi_{N}^{\mbox{o}}$ is the origin for the log-normal distribution 
and the shifting factor $n_{\mbox{shift}}$ is taken to be equal to $2$, unless specified.

% End of methods_for_pdf_of_ddPsi.tex
% \input{methods_for_monoclonal_approximation.tex}

\subsection{Probability distributions of selective advantage, fixation rate and $K_a/K_s$}

\SUPPLEMENT{
Now, we can consider 
}%  SUPPLEMENT
\TEXT{
Let us consider 
}%  TEXT
the probability distributions of
characteristic quantities that describe the evolution of genes. 
First of all, the probability density function (PDF) of 
selective advantage $s$, $p(s)$, of mutant genes can be calculated
from the PDF of 
the change of $\Delta \psi_{ND}$ due to a mutation from $\VEC{\mu}$ to $\VEC{\nu}$, 
$\Delta\Delta \psi_{ND} (\equiv \Delta \psi_{ND}(\VEC{\nu}, T) - \Delta \psi_{ND}(\VEC{\mu}, T) )$.
The PDF of $4N_e s$, $p(4N_e s) = p(s)/(4N_e) $, may be more useful than $p(s)$.
\begin{eqnarray}
p(4N_e s)
	&=& p(\Delta \Delta \psi_{ND}) \, | \frac{d \Delta \Delta \psi_{ND}}{d 4N_e s} |
	=
	p(\Delta\Delta \psi_{ND}) (1 - q_m)
	\label{\EQ: pdf_of_4Nes}
	\label{\EQ: pdf_of_s}
\end{eqnarray}
where $\Delta\Delta \psi_{ND}$ must be regarded as a function of $4N_es$, that is,
$
\Delta\Delta \psi_{ND} = - 4N_e s (1 - q_m)
$; see \Eq{\ref{\EQ: s_vs_dPsi}}.

The PDF of fixation probability $u$ can be represented by
\begin{eqnarray}
p(u)	&=& p(4N_e s) \frac{d 4N_e s}{d u} 
	= 
	p(4N_e s)
	 \frac{ ( e^{4N_e s} - 1)^2 e^{4N_e s (q_m - 1)} }
	{ q_m( e^{4N_e s} - 1) - (e^{4N_e s q_m} - 1) }
		\label{\EQ: pdf_of_fixation_prob}
\end{eqnarray}
where $4N_e s$ must be regarded as a function of $u$. 

The ratio of the substitution rate per nonsynonymous site ($K_a$) for nonsynonymous
substitutions with selective advantage s to the substitution rate per
synonymous site ($K_s$) for synonymous substitutions with s = 0 is
\begin{eqnarray}
	\frac{K_a}{K_s} &=& \frac{u(s)}{u(0)} = \frac{u(s)}{q_m}
		\label{\EQ: def_Ka_over_Ks}
\end{eqnarray}
assuming that synonymous substitutions are completely neutral 
and mutation rates at both types of sites are the same.
The PDF of $K_a/K_s$ is
\begin{eqnarray}
	p(K_a/K_s) &=& 
		p(u) \frac{d u}{d (K_a/K_s)} 
	= p(u) \, q_m  
		\label{\EQ: pdf_of_Ka_over_Ks}
\end{eqnarray}

\subsection{Probability distributions of $\Delta\Delta \psi_{ND}$, $4N_e s$, $u$, and $K_a/K_s$ in fixed mutant genes}

\SUPPLEMENT{
Now, let us consider fixed mutant genes. 
}%  SUPPLEMENT
The PDF of $\Delta\Delta \psi_{ND}$ in fixed mutants is
proportional to that multiplied by the fixation probability.
\begin{eqnarray}
p(\Delta\Delta \psi_{ND, \script{fixed}})
	&=& p(\Delta\Delta \psi_{ND}) 
	\frac{u(s(\Delta \Delta \psi_{ND}))}{\langle u(s(\Delta \Delta \psi_{ND})) \rangle}
	\label{\EQ: pdf_of_fixed_ddPsi}
	\\
  \langle u \rangle &\equiv& \int_{-\infty}^{\infty} u(s) p(\Delta\Delta \psi_{ND}) d\Delta\Delta \psi_{ND}
	\label{\EQ: ave_of_u}
\end{eqnarray}
Likewise, the PDF of selective advantage 
in fixed mutants is 
\begin{eqnarray}
p(4N_e s_{\script{fixed}}) &=&
	p(4N_e s)
	\frac{u(s)}{\langle u(s) \rangle}
\end{eqnarray}
and those of the $u$ and $K_a/K_s$ in fixed mutants are
\begin{eqnarray}
	p( u_{\script{fixed}} ) &=&
		p(u) \frac{u}{\langle u \rangle}
		\\
	p( (\frac{K_a}{K_s})_{\script{fixed}}) &=&
		p(\frac{K_a}{K_s}) \frac{u}{\langle u \rangle}
		=
		p(\frac{K_a}{K_s}) \frac{\frac{K_a}{K_s}}{\langle \frac{K_a}{K_s} \rangle}
\end{eqnarray}
The average of $K_a/K_s$ in fixed mutants is equal to the ratio of the second moment to the first moment of $K_a/K_s$
\TEXT{
in all arising mutants; 
$
\langle K_a/K_s \rangle_{\script{fixed}} = \langle (K_a / K_s)^2 \rangle / \langle K_a/K_s \rangle
$.
}%  TEXT
\SUPPLEMENT{
in all arising mutants.
\begin{eqnarray}
\langle \frac{K_a}{K_s} \rangle_{\script{fixed}} &=& \langle (\frac{K_a}{K_s})^2 \rangle / \langle \frac{K_a}{K_s} \rangle
\end{eqnarray}
}%  SUPPLEMENT

% End of methods_for_monoclonal_approximation.tex
}%  SUPPLEMENT

\SUPPLEMENT{
}%  SUPPLEMENT

% End of methods_1+2.tex

% \input{materials_1+2.tex}

% \input{materials_1.tex}

\subsection{Sequence data}

We study the single domains of 8 Pfam\CITE{FCEEMMPPQSSTB:16} families and both the single domains 
and multi-domains from 3 Pfam families.
In \Table{\ref{\TBL: Proteins_studied}},
their Pfam ID for a multiple sequence alignment, and
UniProt ID and PDB ID with the starting- and ending-residue positions of the domains are listed.
The full alignments for their families at the Pfam are used to estimate one-body interactions $h$ and 
pairwise interactions $J$ with the DCA program from ``http://dca.rice.edu/portal/dca/home''\CITE{MCSHPZS:11,MPLBMSZOHW:11}.
To estimate the sample ($\overline{\psi_N}$) and ensemble ($\langle \psi_N \rangle_{\VEC{\sigma}}$) averages of the 
evolutionary statistical energy,
$M$ unique sequences with no deletions are used.  
In order to reduce
phylogenetic biases in the set of homologous sequences,
we employ
a sample weight ($w_{\VEC{\sigma}_N}$) for each sequence, which is equal to the inverse of the number of sequences
that are less than 20\% different from a given sequence in a given set of homologous sequences. 
Only representatives of unique sequences with no deletions, which are at least 20\% different from each other, are used to calculate 
the changes of the 
evolutionary statistical energy
($\Delta \psi_N$)
due to single nucleotide nonsynonymous substitutions;
the number of the representatives is almost equal to the effective number of sequences ($M_{\script{eff}}$) in
\Table{\ref{\TBL: Proteins_studied}}.

% End of materials_1.tex
% End of materials_1+2.tex

% \input{empirical_rules.tex}

\SUPPLEMENT{
\subsection{Estimation of effective temperature $T_s$ for selection}
}%  SUPPLEMENT

\SUPPLEMENT{
We have examined 
}%  SUPPLEMENT
\TEXT{
In the preceding manuscript\CITE{M:16B},
We examined 
}%  TEXT
the changes of $\psi_N$ due to single nucleotide nonsynonymous substitutions over
all sites in the homologous sequences of 14 protein families, and 
\SUPPLEMENT{
have found 
}%  SUPPLEMENT
\TEXT{
found 
}%  TEXT
the following regression equation.
\begin{eqnarray}
	\overline{ \Delta \psi_N }	
	&\approx&
	\alpha_{\psi_N} \frac{\psi_N - \overline{\psi_N} }{L} + 
	\overline{ \overline{        \Delta \psi_{N} }}	
	\hspace*{2em} \text{ with } \alpha_{\psi_N} < 0
	\label{\EQ: regression_of_dPsi_on_Psi}
\end{eqnarray}
with correlation coefficients, $r_{\psi_N} > 0.9$, where $L$ is sequence length,
$\overline{\psi_N}$ denotes the average of $\psi_N$ over all homologous sequences,
and $\overline{\Delta \psi_N}$ and $\overline{\overline{\Delta \psi_N}}$ denote
the average of $\Delta \psi_N$ over all single nucleotide synonymous substitutions at all sites
in a protein sequence and its total average over all homologous sequences in a protein family, respectively.
In addition, the following relationship for the standard deviation of $\Delta \psi_N$
\SUPPLEMENT{
has been found. 
}%  SUPPLEMENT
\TEXT{
was found.
}%  TEXT
\begin{eqnarray}
	\mbox{Sd}( \Delta \psi_{N} ) 	
	&\approx& \text{independent of } \psi_N \text{ and } 
		\nonumber
		\\
	&\ & \text{constant across homologous sequences in every protein family}	
	\\
	&=& \text{function of } k_B T_s
		\label{\EQ: var_of_dPsi}
\end{eqnarray}
Because
\begin{eqnarray}
	\mbox{Sd}(\Delta G_{N} )	
	&=& \text{function that must not explicitly depend on } k_B T_s \text{ but } G_N
\end{eqnarray}
the following important relationship, 
which can be used to estimate the relative value of $T_s$,
\SUPPLEMENT{
is derived.
}%  SUPPLEMENT
\TEXT{
was derived.
}%  TEXT
\begin{eqnarray}
	\mbox{Sd}(\Delta G_{N} )	
		&\simeq&
		k_B T_s  \mbox{Sd}( \Delta \psi_{N} )	
		\nonumber
		\\
		&\approx& \mbox{ constant }
		\label{\EQ: var_of_ddG}
		\label{\EQ: sd_of_ddG}
\end{eqnarray}
where $\mbox{Sd}(\Delta G_{N} )$ and $\mbox{Sd}( \Delta \psi_{N} )$ are the standard deviations
of $\Delta G_{N}$ and $\Delta \psi_N$ over all single nucleotide nonsynonymous substitutions at all sites, respectively.
These relationships, 
\Eqs{\ref{\EQ: regression_of_dPsi_on_Psi} and \ref{\EQ: var_of_ddG}},
are shown in 
\Figs{
\ref{sfig: 1r69-a:6-58.full_non_del.dca0_18.0_20.simple-gauge.ddPhi_at_opt}
to
\ref{sfig: 5azu-a:4-128.full_non_del.dca0_23.0_20.simple-gauge.BD.ddPhi_at_opt} },
and the regression coefficients ($\alpha_{\psi_N}$) and 
correlation coefficients ($r_{\psi_N}$)
are listed in \Tables{\ref{stbl: ddPsi_with_8A} and \ref{\TBL: ddPsi_with_16A}}.

The PDZ family is employed here as a reference protein for $T_s$,
and its $T_s$ is estimated by a direct comparison of $\Delta \psi_{N}$ and 
experimental $\Delta\Delta G_{ND}$;
the amino acid pair types and site locations of single amino acid substitutions are the most various, and 
also the correlation between the experimental $\Delta\Delta G_{ND}$ and $\Delta \psi_N$ is the best 
for the PDZ family in the present set of protein families, 
SH3\_1\CITE{GRSB:98}, ACBP\CITE{KONSKKP:99}, PDZ\CITE{GCAVBT:05,GGCJVTVB:07}, and Copper-bind\CITE{WW:05}; 
see \Tables{\ref{\TBL: Ts_with_8A} and \ref{stbl: Ts_with_16A}}. 
\begin{eqnarray}
	k_B\hat{T}_s &=& k_B \hat{T}_{s\script{, PDZ}} \, 
		[ \, 
			{ 
		\overline{ \mbox{Sd}( \Delta \psi_{\script{PDZ}} ) }
			}
			\, / \, 
			{
		\overline{ \mbox{Sd}( \Delta \psi_{N} ) }
			}
			\, ]
		\label{\EQ: Ts}
\end{eqnarray}
where the overline denotes the average over all homologous sequences.
Here, the averages of 
standard deviations
over all homologous sequences are employed, because
$T_s$ for all homologous sequences are approximated to be equal.
With estimated $T_s$ and experimental melting temperature $T_m$, 
glass transition temperature $T_g$ and
folding free energy $\Delta G_N$ 
\SUPPLEMENT{
have been estimated 
}%  SUPPLEMENT
\TEXT{
were estimated 
}%  TEXT
for each protein family
on the basis of the REM. 
The estimates of $T_s$ and $T_g$ are all within a reasonable range,
and the estimated values of $\Delta G_N$ agree well with
their experimental values for 5 protein families, justifying the estimates of $T_s$.

% End of empirical_rules.tex

% \input{results_8_vs_15.5A.tex}

\subsection{Comparison of results between $r_{\script{cutoff}} \sim 8$  and $15.5$ \AA\ }

In order to determine $T_s$ for a reference protein,
the experimental values\CITE{GGCJVTVB:07} of $\Delta\Delta G_{ND}$ due to single amino acid substitutions 
in the PDZ domain  are plotted
against the changes of interaction, $\Delta \psi_{N}$ for the same types of substitutions
in \Fig{\ref{sfig: 1gm1-a:16-96_dca0_205_simple-gauge_ddG-dPhi_at_opt}}
for $r_{\script{cutoff}} \sim 8$ and $15.5$ \AA\ .
The slopes of the least-squares regression lines through the origin, which are estimates of $k_B T_s$, are
equal to $k_B\hat{T}_s = 0.279$ kcal/mol
for  $r_{\script{cutoff}} \sim 8$ \AA\ and  $k_B\hat{T}_s = 0.162$ kcal/mol for $r_{\script{cutoff}} \sim 15.5$ \AA\ ,
and the reflective correlation coefficients are equal to $0.93$ and $0.94$, respectively.
These estimates of $k_B T_s$
for the PDZ
yield $\overline{ \mbox{Sd}(\Delta\Delta G_{ND}) } \simeq k_B \hat{T}_s \overline{\text{Sd}(\Delta \psi_N)} = 1.30$ kcal/mol
for $r_{\script{cutoff}} \sim 8$ \AA\ , and $1.29$ kcal/mol for $r_{\script{cutoff}} \sim 15.5$\AA\ .
The reason why $k_B \hat{T}_s \overline{\text{Sd}(\Delta \psi_N)}$ for the PDZ takes similar values
for both $r_{\script{cutoff}} \sim 8$ and $15.5$\AA\  is that
the correlation between the experimental $\Delta\Delta G_{ND}$ and $\Delta \psi_N$ is very good,
and the slopes of the regression lines
are very close to those of the reflective regression lines through the origin,
$0.25$ for $r_{\script{cutoff}} \sim 8$ \AA\ , and $0.16$ for $15.5$\AA\ ;
$k_B T_{s,PDZ} = $
$\langle \Delta\Delta G_{ND} \Delta \psi_N \rangle /\langle (\Delta \psi_N)^2 \rangle$
$\simeq \langle (\Delta\Delta G_{ND} - \overline{\Delta\Delta G_{ND}} )(\Delta \psi_N - \overline{\Delta \psi_N}) \rangle /\langle (\Delta \psi_N - \overline{\Delta \psi_N} )^2 \rangle$,
and so
$k_B (\hat{T}_{s} \overline{Sd}(\Delta \psi_N))_{\script{PDZ}} \simeq \text{constant}$ for $r_{\script{cutoff}} \sim 8$ and $15.5$ \AA\ .
In other words, as long as the correlation between the experimental $\Delta\Delta G_{ND}$ and $\Delta \psi_N$
is good, $k_B \hat{T}_{s} \overline{\text{Sd}}(\Delta \psi_N) \simeq \text{constant}$ irrespective of the cutoff value $r_{\script{cutoff}}$,
although the estimate of $T_s$ differs depending on $\text{Sd}(\Delta\psi_N)$.
This indicates that the correlation between experimental $\Delta\Delta G_{ND}$ and $\Delta \psi_N$ cannot be
a good measure for the correctness of estimated $\Delta \psi_N$, although it must be good enough.
Other comparisons are needed to judge which estimation of $T_s$ is better.

The estimate of $\overline{ \mbox{Sd}(\Delta\Delta G_{ND}) } = 1.30$ or $1.29$ kcal/mol
corresponds to $76\%$ of $1.7$ kcal/mol\CITE{SRS:12} estimated from
ProTherm database or $79$--$80\%$ of $1.63$ kcal/mol\CITE{TSSST:07} computationally predicted for single nucleotide mutations
by using the FoldX.
Using $\overline{ \mbox{Sd}(\Delta\Delta G_{ND}) } =1.30$ or $1.29$ kcal/mol estimated
from $T_s$ for PDZ, the absolute values of $T_s$ for other proteins
are calculated by \Eq{\ref{\EQ: Ts}} and listed in \Tables{\ref{\TBL: Ts_with_8A} and \ref{stbl: Ts_with_16A}}.
\Fig{\ref{fig: Ts_relative_to_Tpdz_8_vs_16A}} shows that
both $r_{\script{cutoff}} \sim 8$ and $15.5$\AA\
yield similar values for
$\hat{T}_s$ in a scale relative to the $\hat{T}_s$ of the PDZ,
because $T_s/T_{s,PDZ} = \overline{\text{Sd}}(\Delta\psi_{N,PDZ})/\overline{\text{Sd}}(\Delta\psi_{N})$.
In other words, the differences of the absolute values of $\hat{T}_s$
between $r_{\script{cutoff}} \sim 8$ and $15.5$\AA\
as shown in \Fig{\ref{fig: Ts_8_vs_16A}}
primarily originate in the difference of $\hat{T}_{s,PDZ}$ for the PDZ.

Larger the standard deviation of $\Delta \psi_N$ is, the smaller the estimate of $T_s$ is.
Including the longer range of pairwise interactions tend to increase the variance of $\Delta \psi_N$.
The range of interactions should be limited to a realistic value, either the first interaction shell or
the second interaction shell.
Thus, the estimates of $T_s$ with $r_{\script{cutoff}} \sim 8$ \AA\ and $15.5$\AA\ are upper
and lower limits, respectively.
Morcos et al.\CITE{MSCOW:14} estimated $T_s$ by comparing
$\Delta \psi_{ND}$ with
$\Delta G_{ND}$
estimated by the associative-memory,
water-mediated, structure, and energy model (AWSEM).
They estimated $\psi_N$ with $r_{\script{cutoff}} = 16$ \AA\ and
probably $p_c = 0.5$.
In \Fig{\ref{sfig: Ts_relative_to_Tpdz_Wolynes_vs_8_and_16A}},
the present estimates of $T_s$ are compared with those by Morcos et al.\CITE{MSCOW:14}.
The Morcos's estimates of $T_s$ with some exceptions tend to be located between
the present estimates with $r_{\script{cutoff}} \sim 8$ \AA\ and $15.5$\AA\ ,
which correspond to the upper and lower limits for $T_s$.

In \Figs{
\ref{sfig: dpsi2_over_L_vs_dPsi},   
\ref{sfig: dG_over_L_vs_ddG},
\ref{sfig: Tm_over_Tg_vs_Ts_over_Tg},
and
\ref{sfig: dG_exp_vs_8_and_16A}
}, and
\Figs{
\ref{sfig: 1gm1-a:16-96_full_non_del_dca0_33_0_20_simple-gauge_dPhiN_distr},
\ref{sfig: 1n2x-a:8-292_full_non_del_dca0_13_0_20_simple-gauge_dPhiN_distr},
\ref{sfig: PhiNe_obs_vs_exp_logG_16A},
\ref{sfig: ddPhi_mean_vs_sd_at_equil},
\ref{sfig: ddPhi_mean_vs_Ts_at_equil},
and
\ref{sfig: ave_ka_over_ks_at_equil_for_Ts}
},
various results are compared between $r_{\script{cutoff}} \sim 8$ \AA\ and $15.5$\AA\ .

% End of results_8_vs_15.5A.tex

\SupplementaryBibliography{
\newpage
\noindent
\textbf{References}
 
\bibliography{jnames_with_dots,MolEvol,Protein,Bioinfo,SM}
}

% End of support_1+2.tex

}%  SupplementaryMaterial

\SupplementaryMaterial{
\renewcommand{\TableInLegends}[1]{#1}
\renewcommand{\TableLegends}[1]{#1}
\renewcommand{\FigureInLegends}[1]{#1}
\renewcommand{\FigureLegends}[1]{#1}
}%  SupplementaryMaterial

\SupplementaryMaterial{

\renewcommand{\TextTable}[1]{}
\renewcommand{\SupTable}[1]{#1}
\setcounter{table}{0}

\renewcommand{\thetable}{S.\arabic{table}}

\TextTable{

\ifdefined\TableZ
\else
\TableInText{
\newcommand{\TableZ}[1]{}
}%  TableInText
\NoTableInText{
\newcommand{\TableZ}[1]{#1}
}%  NoTableInText
\fi

\ifdefined\TableA
\else
\TableInText{
\newcommand{\TableA}[1]{}
}%  TableInText
\NoTableInText{
\newcommand{\TableA}[1]{#1}
}%  NoTableInText
\fi

\ifdefined\TableB
\else
\TableInText{
\newcommand{\TableB}[1]{}
}%  TableInText
\NoTableInText{
\newcommand{\TableB}[1]{#1}
}%  NoTableInText
\fi

\renewcommand{\SUPPLEMENT}[1]{}
\ifdefined\CLEARPAGE
\TableInText{
\renewcommand{\CLEARPAGE}{\TableInLegends{}}
}%  TableInText
\NoTableInText{
\renewcommand{\CLEARPAGE}{\TableInLegends{\clearpage \newpage}}
}%  NoTableInText
\else
\TableInText{
\newcommand{\CLEARPAGE}{\TableInLegends{}}
}%  TableInText
\NoTableInText{
\newcommand{\CLEARPAGE}{\TableInLegends{\clearpage \newpage}}
}%  NoTableInText
\fi
}%  TextTable

\SupTable{

\ifdefined\TableZ
\renewcommand{\TableZ}[1]{#1}
\else
\newcommand{\TableZ}[1]{#1}
\fi

\ifdefined\TableA
\renewcommand{\TableA}[1]{#1}
\else
\newcommand{\TableA}[1]{#1}
\fi

\ifdefined\TableB
\renewcommand{\TableB}[1]{#1}
\else
\newcommand{\TableB}[1]{#1}
\fi

\renewcommand{\SUPPLEMENT}[1]{#1}
\ifdefined\CLEARPAGE
\renewcommand{\CLEARPAGE}{\TableInLegends{\clearpage \newpage}}
\else
\newcommand{\CLEARPAGE}{\TableInLegends{\clearpage \newpage}}
\fi
}%  SupTable

\TableZ{

\CLEARPAGE
\setlength{\textwidth}{17.5cm}
\setlength{\oddsidemargin}{0cm}
\setlength{\evensidemargin}{0cm}

\begin{table}[!ht]
\caption{
\SUPPLEMENT{
\label{stbl: Proteins_studied}
}%  SUPPLEMENT
\TextTable{
\label{tbl: Proteins_studied}
}%  TextTable
\BF{
Protein families, and structures studied.
}%  BF
}%  caption
\vspace*{2em}
\TableInLegends{

\small
\footnotesize

\begin{tabular}{llrrrrrl}
\hline
	Pfam family	& UniProt ID		& $N$ $^a$		& $N_{\script{eff}}$ $^{bc}$		& $M$ $^d$	& $M_{\script{eff}}$ $^{ce}$		& $L$ $^f$	& PDB ID			\\		
										\hline
HTH\_3		& RPC1\_BP434/7-59	& 15315(15917)	& 11691.21 	& 6286	& 4893.73	& 53	& 1R69-A:6-58		\\
Nitroreductase	& Q97IT9\_CLOAB/4-76	& 6008(6084)	& 4912.96	& 1057	& 854.71	& 73	& 3E10-A/B:4-76 $^g$	\\
SBP\_bac\_3 $^h$	& GLNH\_ECOLI/27-244	& 9874(9972)	& 7374.96	& 140	& 99.70	& 218	& 1WDN-A:5-222		\\
SBP\_bac\_3	& GLNH\_ECOLI/111-204	& 9712(9898)	& 7442.85	& 829	& 689.64	& 94	& 1WDN-A:89-182		\\
OmpA		& PAL\_ECOLI/73-167	& 6035(6070)	& 4920.44	& 2207	& 1761.24	& 95	& 1OAP-A:52-146		\\
DnaB		& DNAB\_ECOLI/31-128	& 1929(1957)	& 1284.94	& 1187	& 697.30	& 98	& 1JWE-A:30-127		\\
LysR\_substrate $^h$	& BENM\_ACIAD/90-280	& 25138(25226)	& 20707.06 	& 85(1)	& 67.00	& 191	& 2F6G-A/B:90-280 $^g$	\\
LysR\_substrate	& BENM\_ACIAD/163-265	& 25032(25164)	& 21144.74 	& 121(1) & 99.27	& 103	& 2F6G-A/B:163-265 $^g$	\\
Methyltransf\_5 $^h$	& RSMH\_THEMA/8-292	& 1942(1953)	& 1286.67	& 578(2) & 357.97	& 285	& 1N2X-A:8-292		\\
Methyltransf\_5	& RSMH\_THEMA/137-216	& 1877(1911)	& 1033.35	& 975(2) & 465.53	& 80	& 1N2X-A:137-216	\\
SH3\_1		& SRC\_HUMAN:90-137	& 9716(16621)	& 3842.47	& 1191	& 458.31	& 48	& 1FMK-A:87-134		\\
ACBP		& ACBP\_BOVIN/3-82	& 2130(2526)	& 1039.06	& 161	& 70.72	& 80	& 2ABD-A:2-81		\\
PDZ		& PTN13\_MOUSE/1358-1438	& 13814(23726)	& 4748.76	& 1255	& 339.99	& 81	& 1GM1-A:16-96		\\
Copper-bind	& AZUR\_PSEAE:24-148	& 1136(1169)	& 841.56	& 67(1)	& 45.23	& 125	& 5AZU-B/C:4-128 $^g$		\\
\hline
\end{tabular}

\vspace*{1em}
\noindent
$^a$ The number of unique sequences and the total number of sequences in parentheses; 
	the full alignments in the Pfam\CITE{FCEEMMPPQSSTB:16} are used.

\noindent
$^b$ The effective number of sequences.

\noindent
$^c$ A sample weight ($w_{\VEC{\sigma}_N}$) for a given sequence is
equal to the inverse of the number of sequences 
that are less than 20\% different from the given sequence.

\noindent
$^d$ The number of unique sequences that include no deletion unless specified.
	The number in parentheses indicates the maximum number of deletions allowed. 

\noindent
$^e$ The effective number of unique sequences that include no deletion
or at most the specified number of deletions.

\noindent
$^f$ The number of residues.

\noindent
$^g$ Contacts are calculated in the homodimeric state for these protein.

\noindent
$^h$ These proteins consist of two domains, and other ones are single domains.

% End of TABLES/Proteins.tex
}%  TableInLegends
\end{table}

}%  TableZ

\TableA{

\CLEARPAGE
\setlength{\textwidth}{17.5cm}
\setlength{\oddsidemargin}{-0.5cm}
\setlength{\evensidemargin}{-0.5cm}

\begin{table}[!ht]
\caption{
\SUPPLEMENT{
\label{stbl: ddPsi_with_8A}
}%  SUPPLEMENT
\TextTable{
\label{tbl: ddPsi_with_8A}
}%  TextTable
\BF{
Parameter values
for $r_{\script{cutoff}} \sim 8$ \AA\
}%  BF
employed for each protein family, and
the averages of the 
evolutionary statistical energies
($\overline{\psi_N}$)
over all homologous sequences
and of the means and 
the standard deviations of interaction changes ($\overline{\overline{\Delta \psi_N}}$ and $\overline{\text{Sd}(\Delta \psi_N)}$) 
due to single nucleotide nonsynonymous mutations 
at all sites over all homologous sequences in each protein family.
}%  caption
\vspace*{2em}
\TableInLegends{

\footnotesize
\scriptsize

\begin{tabular}{lrrrrrrrrrrrrr}
\hline
\\
Pfam family & $L$ & $p_c$ & $n_c$ $^a$ & $r_{\script{cutoff}}$ & 
		$\bar{\psi}/L$ $^b$	& ${\delta \psi}^2/L$ $^b$	& $\overline{\psi_N}/L$ $^b$ & 
		$\overline{\overline{\Delta \psi_N}}$ $^c$	&
		$\overline{\mbox{Sd}(\Delta \psi_N) } \, \pm$ $^c$ &
		$r_{\psi_N}$	& $\alpha_{\psi_N}$	& $r_{\psi_N}$	& $\alpha_{\psi_N} $		\\
		&	&	&	& (\AA\ )	&	&	&	&	& $\mbox{Sd}(\mbox{Sd}(\Delta \psi_N) )$	&
		\multicolumn{2}{c}{for $\overline{\Delta \psi_N}$ $^d$}	& \multicolumn{2}{c}{for $\mbox{Sd}(\Delta \psi_N)$ $^e$}
\vspace*{2mm}
		\\
\hline	
HTH\_3	& $53$	& $0.18$	& $7.43$	& $8.22$	& $-0.1997$	& $2.7926$	& $-2.9861$	& $4.2572$	& $5.3503 \pm 0.5627$	& $-0.961$	& $-1.5105$	& $-0.598$	& $-0.9888$ \\
Nitroreductase	& $73$	& $0.23$	& $6.38$	& $8.25$	& $-0.1184$	& $2.1597$	& $-2.2788$	& $3.3115$	& $3.6278 \pm 0.2804$	& $-0.939$	& $-1.3371$	& $-0.426$	& $-0.3721$ \\
SBP\_bac\_3	& $218$	& $0.25$	& $9.23$	& $8.10$	& $-0.1000$	& $2.1624$	& $-2.2618$	& $3.2955$	& $3.4496 \pm0.2742$	& $-0.980$	& $-1.5286$	& $-0.841$	& $-0.7876$ \\
SBP\_bac\_3	& $94$	& $0.37$	& $8.00$	& $7.90$	& $-0.1634$	& $1.2495$	& $-1.4054$	& $1.9291$	& $2.3436 \pm 0.1901$	& $-0.959$	& $-1.3938$	& $-0.634$	& $-0.4815$ \\
OmpA	& $95$	& $0.169$	& $8.00$	& $8.20$	& $-0.2457$	& $3.9093$	& $-4.1542$	& $6.5757$	& $7.6916 \pm 0.3078$	& $-0.957$	& $-1.5694$	& $-0.410$	& $-0.3804$ \\
DnaB	& $98$	& $0.235$	& $9.65$	& $8.17$	& $-0.2284$	& $3.9976$	& $-4.2291$	& $6.3502$	& $6.1244 \pm 0.3245$	& $-0.965$	& $-1.4509$	& $-0.495$	& $-0.4198$ \\
LysR\_substrate	& $191$	& $0.235$	& $8.59$	& $7.98$	& $-0.2241$	& $1.4888$	& $-1.7173$	& $2.2784$	& $2.6519 \pm 0.1445$	& $-0.964$	& $-1.3347$	& $-0.541$	& $-0.5664$ \\
LysR\_substrate	& $103$	& $0.265$	& $8.84$	& $8.25$	& $-0.2244$	& $1.4144$	& $-1.6379$	& $2.2110$	& $2.7371 \pm 0.2055$	& $-0.982$	& $-1.4159$	& $-0.727$	& $-0.5307$ \\
Methyltransf\_5	& $285$	& $0.13$	& $7.99$	& $7.78$	& $-0.1462$	& $7.2435$	& $-7.3887$	& $12.4689$	& $10.9352 \pm 0.3030$	& $-0.981$	& $-1.9140$	& $-0.122$	& $-0.0783$ \\
Methyltransf\_5	& $80$	& $0.18$	& $6.78$	& $7.85$	& $-0.1763$	& $5.5162$	& $-5.6896$	& $8.9849$	& $7.6133 \pm 0.4382$	& $-0.944$	& $-1.4824$	& $0.125$	& $0.1141$ \\
SH3\_1	& $48$	& $0.14$	& $6.42$	& $8.01$	& $-0.1348$	& $3.9109$	& $-4.0434$	& $5.5792$	& $6.1426 \pm 0.2935$	& $-0.919$	& $-1.4061$	& $-0.196$	& $-0.1718$ \\
ACBP	& $80$	& $0.22$	& $9.17$	& $8.24$	& $-0.0525$	& $4.6411$	& $-4.7084$	& $7.7612$	& $7.1383 \pm 0.2970$	& $-0.972$	& $-1.5884$	& $-0.335$	& $-0.2235$ \\
PDZ	& $81$	& $0.205$	& $9.06$	& $8.16$	& $-0.2398$	& $3.1140$	& $-3.3572$	& $4.7589$	& $4.6605 \pm 0.2255$	& $-0.954$	& $-1.5282$	& $-0.369$	& $-0.3042$ \\
Copper-bind	& $125$	& $0.23$	& $9.50$	& $8.27$	& $-0.0940$	& $4.2450$	& $-4.3272$	& $7.2650$	& $6.9283 \pm 0.2316$	& $-0.980$	& $-1.8915$	& $-0.282$	& $-0.2352$ \\
% End of ./TABLES/r=8A/Table-8A_1.proto.tex
\hline

\end{tabular}

\vspace*{1em}
\noindent
$^a$ The average number of contact residues per site within the cutoff distance;
the center of side chain is used to represent a residue.

\noindent
$^b$ $M$ unique sequences with no deletions are used with a sample weight ($w_{\VEC{\sigma}_N}$) for each sequence;
$w_{\VEC{\sigma}_N}$ is equal to the inverse of the number of sequences
that are less than 20\% different from a given sequence.
The $M$ and the effective number $M_{\script{eff}}$ of the sequences are listed for each protein family 
in \Table{\ref{\TBL: Proteins_studied}}.

\noindent
$^c$ The averages of $\overline{\Delta \psi_N}$ and $\mbox{Sd}(\Delta \psi_N)$, 
which are the mean and the standard deviation of $\Delta \psi_N$ for a sequence, and
the standard deviation of $\mbox{Sd}(\Delta \psi_N)$ over homologous sequences.
Representatives of unique sequences with no deletions, which are at least 20\% different from each other, are used; 
     the number of the representatives used is almost equal to $M_{\script{eff}}$.  

\noindent
$^d$ The correlation and regression coefficients of $\overline{\Delta \psi_N}$ on $\psi_N/L$; see \Eq{\ref{\EQ: regression_of_dPsi_on_Psi}}.

\noindent
$^e$ The correlation and regression coefficients of $\mbox{Sd}(\Delta \psi_N)$ on $\psi_N/L$.

% End of TABLES/r=8A/Table-8A_1.tex
}%  TableInLegends
\end{table}

}%  TableA

\TableB{

\CLEARPAGE
\setlength{\textwidth}{17.5cm}
\setlength{\oddsidemargin}{0cm}
\setlength{\evensidemargin}{0cm}

\begin{table}[!ht]
\caption{
\SUPPLEMENT{
\label{stbl: Ts_with_8A}
}%  SUPPLEMENT
\TextTable{
\label{tbl: Ts_with_8A}
}%  TextTable
\BF{
Thermodynamic quantities estimated with $r_{\script{cutoff}} \sim 8$ \AA. 
}%  BF
}%  caption
\vspace*{2em}
\TableInLegends{

\small
\begin{tabular}{lrrrrrrrr}
\hline
	&
	&
		& 	& Experimental	&	& 	 &	&
	\\
Pfam family
	& $r$ $^a$
 	& $k_B \hat{T}_s$ $^a$ & $\hat{T}_s$ & $T_m$ & $\hat{T}_g$ & $\hat{\omega}$ $^b$ &$T$ $^c$ & $\langle \Delta G_{ND} \rangle$ $^d$
	\\ 
	&	&(kcal/mol)	& ($^\circ$K) & ($^\circ$K) & ($^\circ$K) & ($k_B$) & ($^\circ$K) & (kcal/mol)
\vspace*{2mm}
	\\ 
\hline
HTH\_3	& -- 	& -- 	& $122.\REDa{5}$	& $343.7$	& $160.\REDa{0}$	& $0.81\REDa{78}$	& $298$	& $-2.95$ \\
Nitroreductase	& -- 	& --	& $180.\REDa{6}$	& $337$	& $20\REDa{3.9}$	& $0.847\REDa{3}$	& $298$	& $-2.81$ \\
SBP\_bac\_3	& --	& --	& $190.\REDa{0}$	& $336.1$	& $21\REDa{0.9}$	& $0.87\REDa{68}$	& $298$	& $-8.0\REDa{4}$ \\
SBP\_bac\_3	& --	& --	& $279.\REDa{6}$	& $336.1$	& $283.\REDa{6}$	& $0.607\REDa{1}$	& $298$	& $-0.85$ \\
OmpA	& -- 	& --	& $85.2$	& $320$	& $125.4$	& $0.902\REDa{2}$	& $298$	& $-3.13$ \\
DnaB	& --	& --	& $107.\REDa{0}$	& $312.8$	& $142.1$	& $1.13\REDa{35}$	& $298$	& $-2.56$ \\
LysR\_substrate	& --	& -- 	& $247.\REDa{1}$	& $338$	& $256.\REDa{5}$	& $0.690\REDa{6}$	& $298$	& $-3.63$ \\
LysR\_substrate	& --	& --	& $239.\REDa{4}$	& $338$	& $250.\REDa{3}$	& $0.647\REDa{0}$	& $298$	& $-2.00$ \\
Methyltransf\_5	& --	& --	& $\REDa{59.9}$	& $375$	& $110.5$	& $1.065\REDa{0}$	& $298$	& $-41.3\REDa{7}$ \\
Methyltransf\_5	& --	& --	& $86.1$	& $375$	& $135.\REDa{0}$	& $1.12\REDa{08}$	& $298$	& $-11.48$ \\
SH3\_1	& $0.865$	& $0.1583$	& $106.7$	& $344$	& $147.4$	& $1.02\REDa{47}$	& $295$	& $-3.76$ \\
ACBP	& $0.825$	& $0.1169$	& $91.\REDa{8}$	& $324.4$	& $131.7$	& $1.12\REDa{75}$	& $278$	& $-6.72$ \\
PDZ	& $0.931$	& $0.2794$	& $140.\REDa{6}$	& $312.88$	& $168.\REDa{4}$	& $1.08\REDa{49}$	& $298$	& $-1.81$ \\
Copper-bind	& $0.828$	& $0.1781$	& $94.6$	& $359.3$	& $139.9$	& $0.970\REDa{3}$	& $298$	& $-12.0\REDa{8}$ \\
% End of ./TABLES/r=8A/Table-8A_2.proto_revised.tex
\hline
\end{tabular}

\vspace*{1em}
\noindent
$^a$ Reflective correlation ($r$) and regression ($k_B \hat{T}_s$) coefficients 
	for least-squares regression lines of experimental $\Delta\Delta G_{ND}$ on $\Delta \psi_N$ through the origin.

\noindent
$^b$ Conformational entropy per residue, in $k_B$ units, in the denatured molten-globule state; see \Eq{\ref{\EQ: expression_of_entropy}}.

\noindent
$^c$ Temperatures are set up for comparison to be equal to the experimental temperatures for $\Delta G_{ND}$ or to $298 ^\circ$K if unavailable; 
see \Table{\ref{stbl: experimental_data}} for the experimental data.

\noindent
$^d$ Folding free energy in kcal/mol units; see \Eq{\ref{\EQ: ensemble_ave_of_ddG}}.
% End of TABLES/r=8A/Table-8A_2.revised.tex
}%  TableInLegends
\end{table}

}%  TableB

\SUPPLEMENT{

\CLEARPAGE

\begin{table}[!ht]
\caption{
\label{stbl: experimental_data}
\label{tbl: experimental_data}
\BF{
Experimental data used.
}%  BF
}%  caption
\vspace*{2em}
\TableInLegends{

\small
\footnotesize

\begin{tabular}{lrrrll}
\hline
			& \multicolumn{3}{c}{experimental values} &	&	\\
	Pfam family	& $T_m$	& $T$ & $\Delta G_{ND}$	
	& ref. for $T_m$ & ref. for $\Delta G_{ND}$ and $\Delta\Delta G_{ND}$
		\\
	& ($^\circ$K) & ($^\circ$K) & (kcal/mol)
\vspace*{2mm}
		\\
\hline
HTH\_3	& $343.7$ & $298$ & $-5.33 \pm 0.36$	& \CITE{GDBCJMS:09}	& \CITE{RSTPMF:99}	\\ 
Nitroreductase	& $337.0$ & - & -	& \CITE{SZMSS:06}	&	\\ 
SBP\_bac\_3	& $336.1$ & - & -	& \CITE{DSVSSAMRT:05}	&	\\ 
OmpA	& $320.0$ & - & -		& \CITE{PLO:06}	&	\\ 
DnaB	& $312.8$ & - & -		& \CITE{WPLLSLCOD:02}	&	\\ 
LysR\_substrate	& $338.0$  & - & -	& \CITE{SRSSO:08}	&	\\ 
Methyltransf\_5	& $375.0$ & - & -	& \CITE{AUFCBG:04}	&	\\ 
		&	& 	&	& \CITE{GRGBLG:10}	&	\\
SH3\_1	& $344.0$ & $295$ & $-3.70$	& \CITE{KMCBKVSL:98}	& \CITE{GRSB:98}	\\ 
ACBP	& $324.4$ & $278$ & $-8.08 \pm 0.08$	& \CITE{OKKSW:15}	& \CITE{KONSKKP:99}	\\ 
PDZ	& $312.9$ & $298$ & $-2.9$ \REDb{$\pm$} \REDb{$0.2$}	& \CITE{TES:12}	& \CITE{GCAVBT:05,GGCJVTVB:07}	\\ 
Copper-bind	& $359.3$ & $298$ & $-12.90 \pm 0.36$	& \CITE{RMGGS:95}	& \CITE{WW:05}	\\ 
% End of ./TABLES/r=8A/Table-Exp_1.proto.tex
\hline
\end{tabular}
% End of TABLES/r=8A/Table-Exp_1.tex
}%  TableInLegends
\end{table}

}%  SUPPLEMENT

\SUPPLEMENT{

\CLEARPAGE
\setlength{\textwidth}{17.5cm}
\setlength{\oddsidemargin}{-0.5cm}
\setlength{\evensidemargin}{-0.5cm}

\begin{table}[!ht]
\caption{
\label{stbl: ddPsi_with_16A}
\label{tbl: ddPsi_with_16A}
\BF{
Parameter values
for $r_{\script{cutoff}} \sim 15.5$ \AA\
}%  BF
employed for each protein family, and
the averages of the 
evolutionary statistical energies
($\overline{\psi_N}$)
over all homologous sequences
and of the means and 
the standard deviations of interaction changes ($\overline{\overline{\Delta \psi_N}}$ and $\overline{\text{Sd}(\Delta \psi_N)}$) 
due to single nucleotide nonsynonymous mutations 
at all sites over all homologous sequences in each protein family.
}%  caption
\vspace*{2em}
\TableInLegends{

\footnotesize
\scriptsize

\begin{tabular}{lrrrrrrrrrrrrr}
\hline
\\
Pfam family & $L$ & $p_c$ & $n_c$ $^a$ & $r_{\script{cutoff}}$ & 
		$\bar{\psi}/L$ $^b$	& ${\delta \psi}^2/L$ $^b$	& $\overline{\psi_N}/L$ $^b$ & 
		$\overline{\overline{\Delta \psi_N}}$ $^c$	&
		$\overline{\mbox{Sd}(\Delta \psi_N) } \, \pm$ $^c$ &
		$r_{\psi_N}$	& $\alpha_{\psi_N}$	& $r_{\psi_N}$	& $\alpha_{\psi_N} $		\\
		&	&	&	& (\AA\ ) &	&	&	&	& $\mbox{Sd}(\mbox{Sd}(\Delta \psi_N) )$	&
		\multicolumn{2}{c}{for $\overline{\Delta \psi_N}$ $^d$}	& \multicolumn{2}{c}{for $\mbox{Sd}(\Delta \psi_N)$ $^e$}
\vspace*{2mm}
		\\
\hline	
HTH\_3	& $53$	& $0.245$	& $32.90$	& $15.67$	& $-0.2548$	& $4.0057$	& $-4.2642$	& $6.8512$	& $6.9544 \pm 0.5309$	& $-0.955$	& $-1.5717$	& $-0.519$	& $-0.5727$ \\
Nitroreductase	& $73$	& $0.315$	& $28.71$	& $15.75$	& $-0.1476$	& $3.7093$	& $-3.8565$	& $6.3226$	& $5.6267 \pm 0.5440$	& $-0.953$	& $-1.5765$	& $-0.694$	& $-0.6640$ \\
SBP\_bac\_3	& $218$	& $0.35$	& $55.48$	& $15.90$	& $-0.0669$	& $3.4004$	& $-3.4674$	& $5.7978$	& $4.8666 \pm 0.4517$	& $-0.971$	& $-1.6708$	& $-0.821$	& $-0.8874$ \\
SBP\_bac\_3	& $94$	& $0.455$	& $42.81$	& $15.45$	& $-0.1628$	& $2.3208$	& $-2.4831$	& $4.0963$	& $3.7760 \pm 0.3970$	& $-0.968$	& $-1.6628$	& $-0.770$	& $-0.6408$ \\
OmpA	& $95$	& $0.235$	& $35.58$	& $15.69$	& $-0.2552$	& $5.8175$	& $-6.0757$	& $10.4102$	& $11.8829 \pm 0.4108$	& $-0.948$	& $-1.6212$	& $-0.354$	& $-0.3599$ \\
DnaB	& $98$	& $0.35$	& $46.65$	& $15.57$	& $-0.2351$	& $6.1890$	& $-6.4167$	& $10.7294$	& $8.0204 \pm 0.3493$	& $-0.894$	& $-1.5176$	& $-0.311$	& $-0.3037$ \\
LysR\_substrate	& $191$	& $0.335$	& $52.30$	& $15.58$	& $-0.2826$	& $2.5962$	& $-2.8789$	& $4.4194$	& $4.1701 \pm 0.1782$	& $-0.963$	& $-1.6196$	& $-0.613$	& $-0.4726$ \\
LysR\_substrate	& $103$	& $0.37$	& $44.33$	& $15.60$	& $-0.2816$	& $2.4438$	& $-2.7239$	& $4.1276$	& $4.2029 \pm 0.3674$	& $-0.984$	& $-1.5436$	& $-0.769$	& $-0.5462$ \\
Methyltransf\_5	& $285$	& $0.175$	& $53.52$	& $15.53$	& $-0.1687$	& $12.8982$	& $-13.0658$	& $23.6376$	& $18.7982 \pm 0.4701$	& $-0.952$	& $-1.9804$	& $-0.171$	& $-0.1630$ \\
Methyltransf\_5	& $80$	& $0.24$	& $37.02$	& $15.11$	& $-0.1632$	& $9.9944$	& $-10.1576$	& $17.5749$	& $13.9124 \pm 0.4756$	& $-0.862$	& $-1.6406$	& $-0.290$	& $-0.2822$ \\
SH3\_1	& $48$	& $0.165$	& $28.46$	& $15.76$	& $-0.1350$	& $7.6161$	& $-7.7523$	& $11.9725$	& $13.3845 \pm 0.4719$	& $-0.896$	& $-1.5944$	& $-0.255$	& $-0.2420$ \\
ACBP	& $80$	& $0.28$	& $36.27$	& $15.34$	& $-0.0235$	& $7.4707$	& $-7.4947$	& $13.1892$	& $9.7188 \pm 0.4242$	& $-0.911$	& $-1.7087$	& $0.085$	& $0.0861$ \\
PDZ	& $81$	& $0.33$	& $40.82$	& $15.77$	& $-0.3022$	& $5.2295$	& $-5.5313$	& $8.6909$	& $7.9383 \pm 0.2930$	& $-0.966$	& $-1.7215$	& $-0.316$	& $-0.2328$ \\
Copper-bind	& $125$	& $0.295$	& $45.22$	& $15.32$	& $-0.0999$	& $8.5521$	& $-8.6592$	& $15.5941$	& $9.6566 \pm 0.3556$	& $-0.951$	& $-1.7441$	& $-0.175$	& $-0.1981$ \\
% End of ./TABLES/r=16A/Table-16A_1.proto.tex
\hline

\end{tabular}

\vspace*{1em}
\noindent
$^a$ The average number of contact residues per site within the cutoff distance;
the center of side chain is used to represent a residue.

\noindent
$^b$ $M$ unique sequences without deletions are used with a sample weight ($w_{\VEC{\sigma}_N}$) for each sequence;
$w_{\VEC{\sigma}_N}$ is equal to the inverse of the number of sequences
that are less than 20\% different from a given sequence.
The $M$ and the effective number $M_{\script{eff}}$ of the sequences are listed for each protein family 
in \Table{\ref{\TBL: Proteins_studied}}.

\noindent
$^c$ The averages of $\overline{\Delta \psi_N}$ and $\mbox{Sd}(\Delta \psi_N)$,
which are the mean and the standard deviation of $\Delta \psi_N$ for a sequence, and
the standard deviation of $\mbox{Sd}(\Delta \psi_N)$ over homologous sequences.
Representatives of unique sequences without deletions, which are at least 20\% different from each other, are used;
     the number of the representatives used is almost equal to $M_{\script{eff}}$.

\noindent
$^d$ The correlation and regression coefficients of $\overline{\Delta \psi_N}$ on $\psi_N/L$;see \Eq{\ref{\EQ: regression_of_dPsi_on_Psi}}.

\noindent
$^e$ The correlation and regression coefficients of $\mbox{Sd}(\Delta \psi_N)$ on $\psi_N/L$.

% End of TABLES/r=16A/Table-16A_1.tex
}%  TableInLegends
\end{table}

\CLEARPAGE
\setlength{\textwidth}{17.5cm}
\setlength{\oddsidemargin}{0cm}
\setlength{\evensidemargin}{0cm}

\begin{table}[!ht]
\caption{
\label{stbl: Ts_with_16A}
\label{tbl: Ts_with_16A}
\BF{
Thermodynamic quantities estimated with $r_{\script{cutoff}} \sim 15.5$ \AA. 
}%  BF
}%  caption
\vspace*{2em}
\TableInLegends{

\small
\begin{tabular}{lrrrrrrrr}
\hline
	&
	&
		& 	& Experimental	&	& 	 &	&
	\\
Pfam family
	& $r$ $^a$
 	& $k_B \hat{T}_s$ $^a$ & $\hat{T}_s$ & $T_m$ & $\hat{T}_g$ & $\hat{\omega}$ $^b$ &$T$ $^c$ & $\langle \Delta G_{ND} \rangle$ $^d$
	\\ 
	&	&(kcal/mol)	& ($^\circ$K) & ($^\circ$K) & ($^\circ$K) & ($k_B$) & ($^\circ$K) & (kcal/mol)
\vspace*{2mm}
	\\ 
\hline
HTH\_3	& -- 	& -- 	& $93.\REDa{0}$	& $343.7$	& $136.0$	& $0.937\REDa{3}$	& $298$	& $-3.70$ \\
Nitroreductase	& --	& -- 	& $115.0$	& $337$	& $152.\REDa{8}$	& $1.0\REDa{495}$	& $298$	& $-4.56$ \\
SBP\_bac\_3	& --	& --	& $13\REDa{2.9}$	& $336.1$	& $166.9$	& $1.07\REDa{88}$	& $298$	& $-12.8\REDa{6}$ \\
SBP\_bac\_3	& --	& --	& $171.\REDa{3}$	& $336.1$	& $196.6$	& $0.881\REDa{4}$	& $298$	& $-3.85$ \\
OmpA	& -- 	& --	& $54.\REDa{4}$	& $320$	& $97.6$	& $0.90\REDa{54}$	& $298$	& $-3.3\REDa{9}$ \\
DnaB	& --	& -- 	& $80.\REDa{6}$	& $312.8$	& $120.\REDa{3}$	& $1.390\REDa{0}$	& $298$	& $-3.38$ \\
LysR\_substrate	& --	& -- 	& $155.\REDa{1}$	& $338$	& $184.\REDa{4}$	& $0.918\REDa{0}$	& $298$	& $-9.2\REDa{3}$ \\
LysR\_substrate	& --	& --	& $15\REDa{3.9}$	& $338$	& $183.\REDa{5}$	& $0.859\REDa{4}$	& $298$	& $-4.68$ \\
Methyltransf\_5	& --	& --	& $34.4$	& $375$	& $82.\REDa{2}$	& $1.129\REDa{2}$	& $298$	& $-46.26$ \\
Methyltransf\_5	& --	& --	& $46.5$	& $375$	& $96.4$	& $1.16\REDa{23}$	& $298$	& $-13.04$ \\
SH3\_1	& $0.836$	& $0.0821$	& $48.\REDa{3}$	& $344$	& $94.6$	& $0.99\REDa{48}$	& $295$	& $-4.24$ \\
ACBP	& $0.823$	& $0.0689$	& $66.6$	& $324.4$	& $109.7$	& $1.37\REDa{55}$	& $278$	& $-8.79$ \\
PDZ	& $0.944$	& $0.1619$	& $81.5$	& $312.88$	& $121.1$	& $1.18\REDa{46}$	& $298$	& $-2.39$ \\
Copper-bind	& $0.888$	& $0.1015$	& $67.0$	& $359.3$	& $115.2$	& $1.44\REDa{57}$	& $298$	& $-19.2\REDa{9}$ \\
% End of ./TABLES/r=16A/Table-16A_2.proto_revised.tex
\hline
\end{tabular}

\vspace*{1em}
\noindent
$^a$ Reflective correlation ($r$) and regression ($k_B \hat{T}_s$) coefficients 
	for least-squares regression lines of experimental $\Delta\Delta G_{ND}$ on $\Delta \psi_N$ through the origin.

\noindent
$^b$ Conformational entropy per residue, in $k_B$ units, in the denatured molten-globule state; see \Eq{\ref{\EQ: expression_of_entropy}}.

\noindent
$^c$ Temperatures are set up for comparison to be equal to the experimental temperatures for $\Delta G_{ND}$ or to $298 ^\circ$K if unavailable; 
see \Table{\ref{stbl: experimental_data}} for the experimental data.

\noindent
$^d$ Folding free energy in kcal/mol units; see \Eq{\ref{\EQ: ensemble_ave_of_ddG}}.
% End of TABLES/r=16A/Table-16A_2.revised.tex
}%  TableInLegends
\end{table}

}%  SUPPLEMENT

% End of tables_JTB_1.tex
% \input{tables_JTB_2.tex}

\SUPPLEMENT{

\CLEARPAGE

\begin{table}[!ht]
\caption{
\label{stbl: fixation_prob}
\label{tbl: fixation_prob}
\BF{
Fixation probabilities of a single mutant in various models.
}%  BF
}%  caption
\vspace*{2em}
\TableInLegends{

\begin{tabular}{lllllll}
A) & \multicolumn{6}{l}{For Wright-Fisher population; compiled from p. 192 and pp. 424--427 of Crow and Kimura (1970).}	
		\\
		\\
  &	Fitness/Selection $^\script{a}$  & $h$ $^\script{a}$	& $M_{\delta x}$ $^\script{b}$ & $V_{\delta x}$ $^\script{c}$ & $u$ $^\script{de}$	& $q_m$ $^\script{f}$
		\\
		\hline
 & No dominance	  & $1/2$	& $sx(1-x)$	& $x(1-x)/(2N_e)$	& $(1- e^{-4N_esq_m})/ (1- e^{-4N_es})$ & $1/(2N)$
		\\
 & Dominance favored & $1$	& $2sx(1-x)^2$	& $x(1-x)/(2N_e)$	& \hspace*{5em}$^\script{e}$	  & $1/(2N)$
		\\
 & Recessive favored & $0$	& $2sx^2(1-x)$	& $x(1-x)/(2N_e)$	& \hspace*{5em}$^\script{e}$	  & $1/(2N)$
		\\
 & Gametic selection &	& $sx(1-x)$	& $x(1-x)/(2N_e)$ & $(1- e^{-4N_esq_m})/(1- e^{-4N_es})$	& $1/(2N)$
		\\
 & Haploid  	& 	& $sx(1-x)$	& $x(1-x)/N_e$	& $(1- e^{-2N_esq_m})/ (1- e^{-2N_es})$	& $1/N$
		\\
		\hline
		\\
B) & \multicolumn{6}{l}{For Moran population\CITE{M:58,E:79}}	
		\\
		\\
  & Fitness/Selection $^\script{a}$  & 	& $M_{\delta x}$ 	& $V_{\delta x}$ $^\script{c}$	& $u$ $^\script{de}$	& $q_m$ $^\script{f}$
		\\
		\hline
 & Haploid  	& 	& $s x(1-x)/N_e$ & $2x(1-x)/N_e^2$	& $(1- e^{-N_esq_m})/ (1- e^{-N_es})$	& $1/N$
		\\
		\hline
\end{tabular}

\vspace*{1em}
\noindent
$^\script{a}$ For zygotic selection, $2s$ and $2sh$ are the selective advantages of mutant homogeneous and heterogeneous zygotes, respectively.
For others, $s$ is the selective advantage of mutant gene.

\noindent
$^\script{b}$ Mean in the rate of the change of gene frequency per generation; $M_{\delta x} = 2sx(1-x)(h + (1-2h)x)$ for zygotic selection.

\noindent
$^\script{c}$ Variance in the rate of the change of gene frequency per generation.

\noindent
$^\script{d}$ Fixation probability. 

\noindent 
$^\script{e}$ $u(q_m) = F(q_m)/F(1)$
where $F(q_m)=\int_0^{q_m} G(x)dx$ and $G(x)=\exp (-\int 2M_{\delta x}/V_{\delta x} dx)$. 

\noindent
$^\script{f}$ Frequency of a single mutant gene.
% End of TABLES/fixation_prob.tex
}%  TableInLegends
\end{table}

}%  SUPPLEMENT
% End of tables_JTB_2.tex
% End of stables_1+2.tex

\renewcommand{\TextFig}[1]{}
\renewcommand{\SupFig}[1]{#1}
\setcounter{figure}{0}

\renewcommand{\thefigure}{S.\arabic{figure}}

\TextFig{

\ifdefined\FigA
\else
\NoFigureInText{
\newcommand{\FigA}[1]{#1}
}%  NoFigureInText
\FigureInText{
\newcommand{\FigA}[1]{}
}%  FigureInText
\fi

\ifdefined\FigB
\else
\NoFigureInText{
\newcommand{\FigB}[1]{#1}
}%  NoFigureInText
\FigureInText{
\newcommand{\FigB}[1]{}
}%  FigureInText
\fi

\ifdefined\FigC
\else
\NoFigureInText{
\newcommand{\FigC}[1]{#1}
}%  NoFigureInText
\FigureInText{
\newcommand{\FigC}[1]{}
}%  FigureInText
\fi

\ifdefined\FigD
\else
\NoFigureInText{
\newcommand{\FigD}[1]{#1}
}%  NoFigureInText
\FigureInText{
\newcommand{\FigD}[1]{}
}%  FigureInText
\fi

\ifdefined\FigE
\else
\NoFigureInText{
\newcommand{\FigE}[1]{#1}
}%  NoFigureInText
\FigureInText{
\newcommand{\FigE}[1]{}
}%  FigureInText
\fi

\ifdefined\FigF
\else
\NoFigureInText{
\newcommand{\FigF}[1]{#1}
}%  NoFigureInText
\FigureInText{
\newcommand{\FigF}[1]{}
}%  FigureInText
\fi

\ifdefined\FigG
\else
\NoFigureInText{
\newcommand{\FigG}[1]{#1}
}%  NoFigureInText
\FigureInText{
\newcommand{\FigG}[1]{}
}%  FigureInText
\fi

\renewcommand{\SUPPLEMENT}[1]{}

\ifdefined\CLEARPAGE

\NoFigureInText{
\renewcommand{\CLEARPAGE}{\FigureLegends{\clearpage\newpage}}
}%  NoFigureInText
\FigureInText{
\renewcommand{\CLEARPAGE}{}
}%  FigureInText

\else

\NoFigureInText{
\newcommand{\CLEARPAGE}{\FigureLegends{\clearpage\newpage}}
}%  NoFigureInText
\FigureInText{
\newcommand{\CLEARPAGE}{}
}%  FigureInText

\fi

}%  TextFig

\SupFig{

\ifdefined\FigA
\renewcommand{\FigA}[1]{#1}
\else
\newcommand{\FigA}[1]{#1}
\fi

\ifdefined\FigB
\renewcommand{\FigB}[1]{#1}
\else
\newcommand{\FigB}[1]{#1}
\fi

\ifdefined\FigC
\renewcommand{\FigC}[1]{#1}
\else
\newcommand{\FigC}[1]{#1}
\fi

\ifdefined\FigD
\renewcommand{\FigD}[1]{#1}
\else
\newcommand{\FigD}[1]{#1}
\fi

\ifdefined\FigE
\renewcommand{\FigE}[1]{#1}
\else
\newcommand{\FigE}[1]{#1}
\fi

\ifdefined\FigF
\renewcommand{\FigF}[1]{#1}
\else
\newcommand{\FigF}[1]{#1}
\fi

\ifdefined\FigG
\renewcommand{\FigG}[1]{#1}
\else
\newcommand{\FigG}[1]{#1}
\fi

\renewcommand{\SUPPLEMENT}[1]{#1}
\ifdefined\CLEARPAGE
\renewcommand{\CLEARPAGE}{\FigureLegends{\clearpage\newpage}}
\else
\newcommand{\CLEARPAGE}{\FigureLegends{\clearpage\newpage}}
\fi
}%  SupFig

\renewcommand{\SkipFigure}[1]{}

\SUPPLEMENT{

\CLEARPAGE
 
\begin{figure*}[h!]
\FigureInLegends{
\noindent
(a) $p_c=0.205$

\includegraphics*[width=82mm,angle=0]{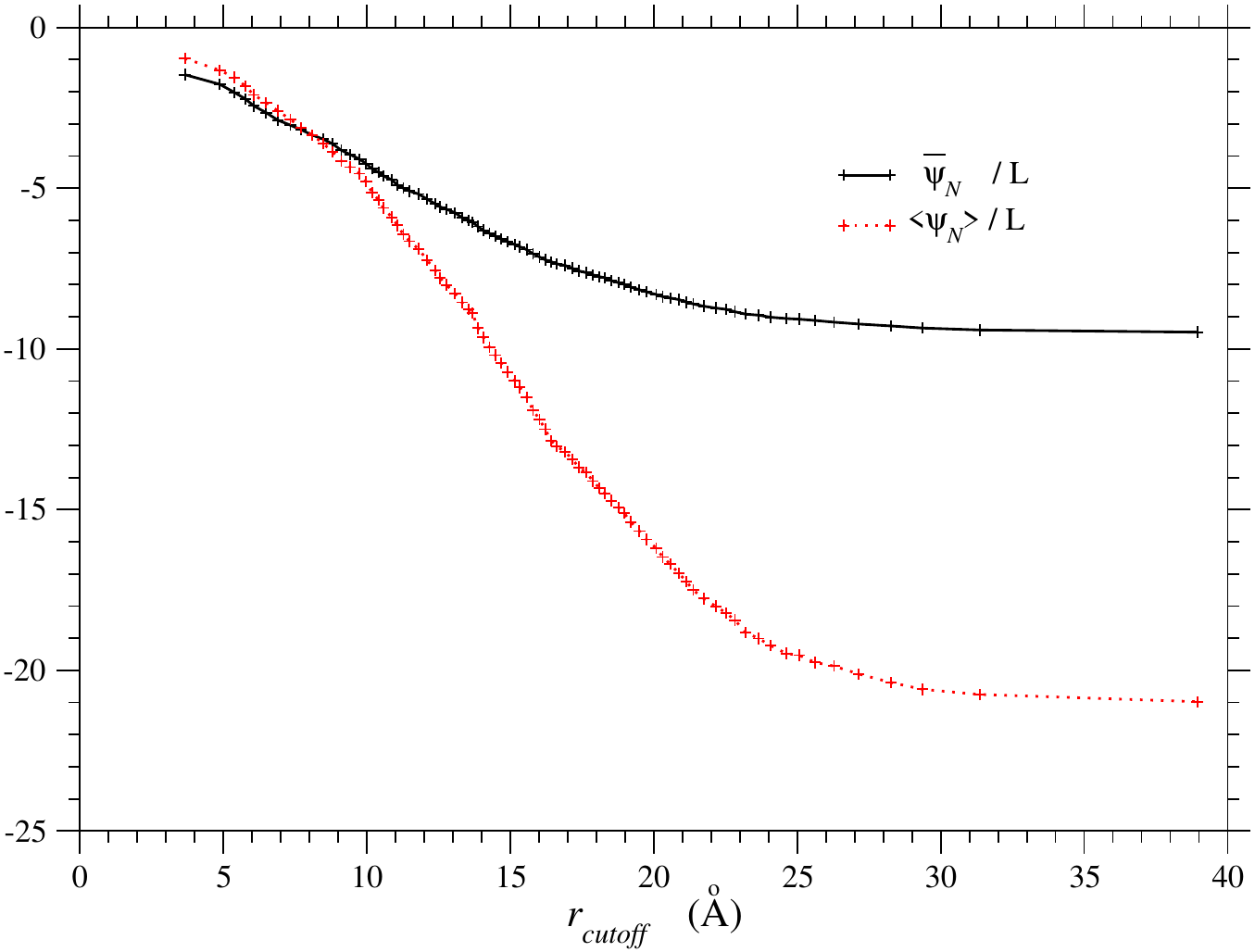}

{
\noindent
(b) $p_c=0.33$

\includegraphics*[width=82mm,angle=0]{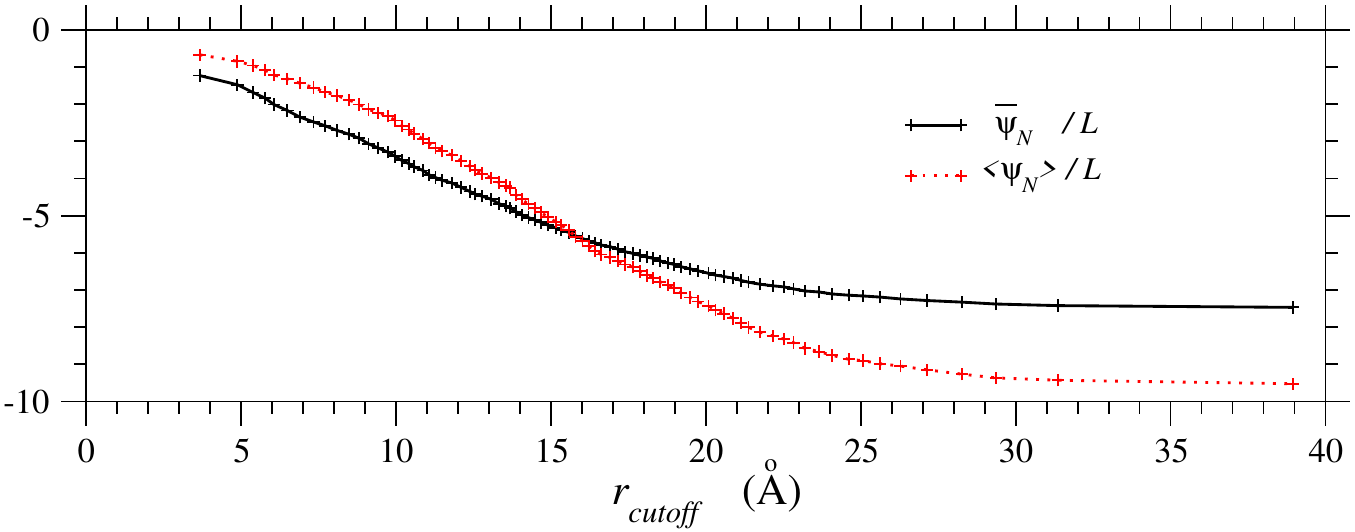}

\noindent
(c) $p_c=0.5$

\includegraphics*[width=82mm,angle=0]{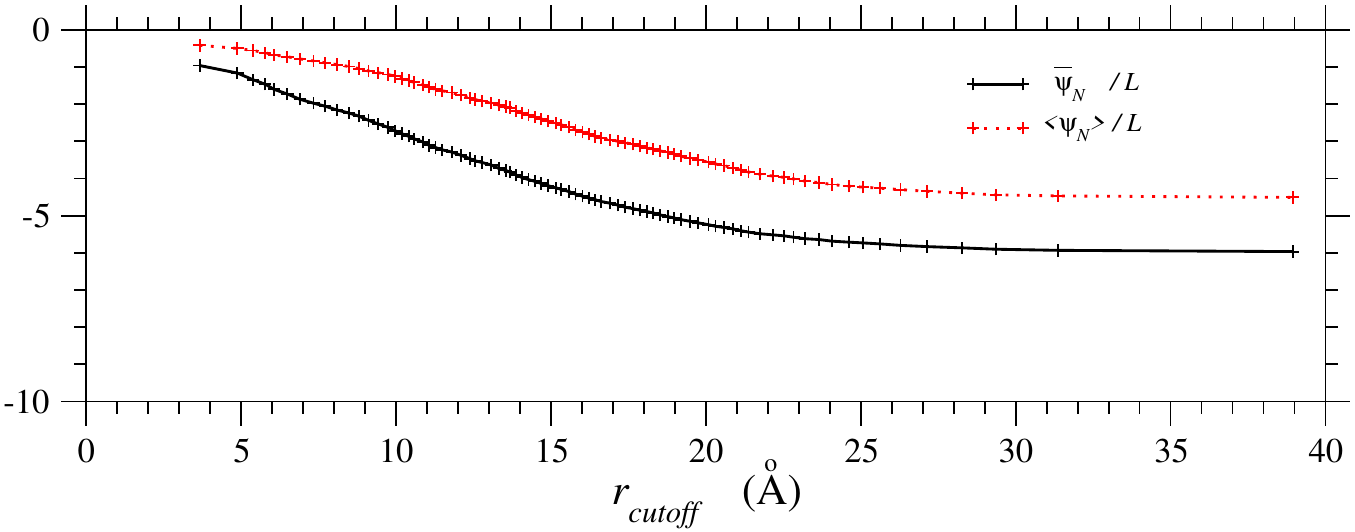}
}
}%  FigureInLegends
\vspace*{1em}
\caption{
\FigureLegends{
\label{sfig: 1gm1-a:16-96_full_non_del_dca0_205_0_20_simple-gauge_ensemble_vs_sample_ave_phi_vs_r}
\label{fig: 1gm1-a:16-96_full_non_del_dca0_205_0_20_simple-gauge_ensemble_vs_sample_ave_phi_vs_r}
\label{sfig: 1gm1-a:16-96_full_non_del_dca0_33_0_20_simple-gauge_ensemble_vs_sample_ave_phi_vs_r}
\label{fig: 1gm1-a:16-96_full_non_del_dca0_33_0_20_simple-gauge_ensemble_vs_sample_ave_phi_vs_r}
\label{sfig: 1gm1-a:16-96_full_non_del_dca0_5_0_20_simple-gauge_ensemble_vs_sample_ave_phi_vs_r}
\label{fig: 1gm1-a:16-96_full_non_del_dca0_5_0_20_simple-gauge_ensemble_vs_sample_ave_phi_vs_r}
\BF{
Dependences of
the sample ($\overline{\psi_N} / L$) and ensemble ($\langle \psi_N \rangle_{\VEC{\sigma}} / L$) averages of 
evolutionary statistical energy
per residue
on the cutoff distance for pairwise interactions 
in the PDZ domain.
}
The ratios of pseudocount $p_c = 0.205$ and $0.33$
are employed here
for the cutoff distance $r_{\script{cutoff}} \sim 8$ and $15.5$ \AA, respectively.  
The black solid and red dotted lines indicate the sample and ensemble averages, respectively.
}%  FigureLegends
}
\end{figure*}

\CLEARPAGE
 
\begin{figure*}[h!]
\FigureInLegends{
\noindent
\hspace*{1em} (a) $p_c=0.205$	\hspace*{14em} (b) $p_c=0.33$

\centerline{
\includegraphics*[width=82mm,angle=0]{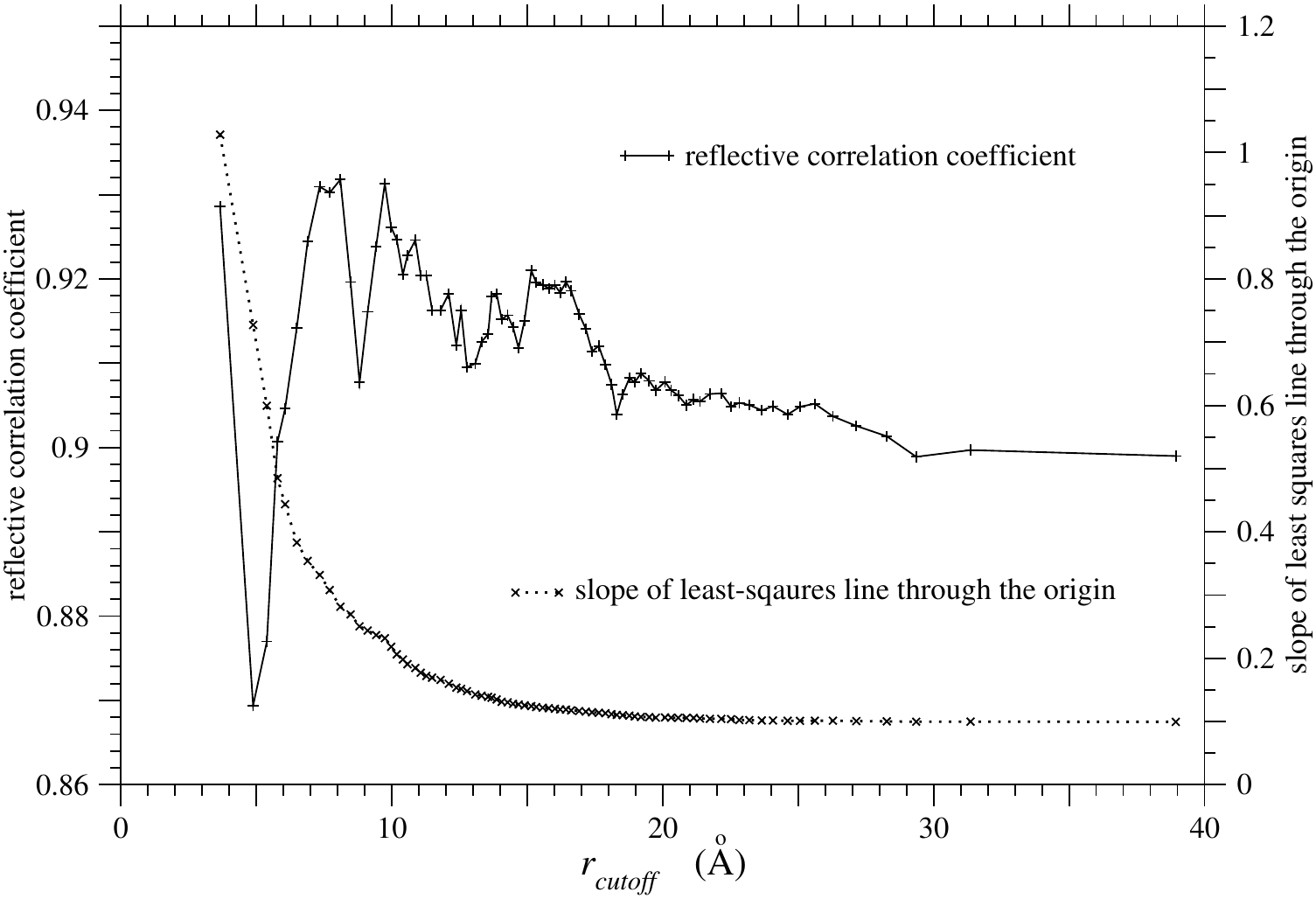}
\includegraphics*[width=82mm,angle=0]{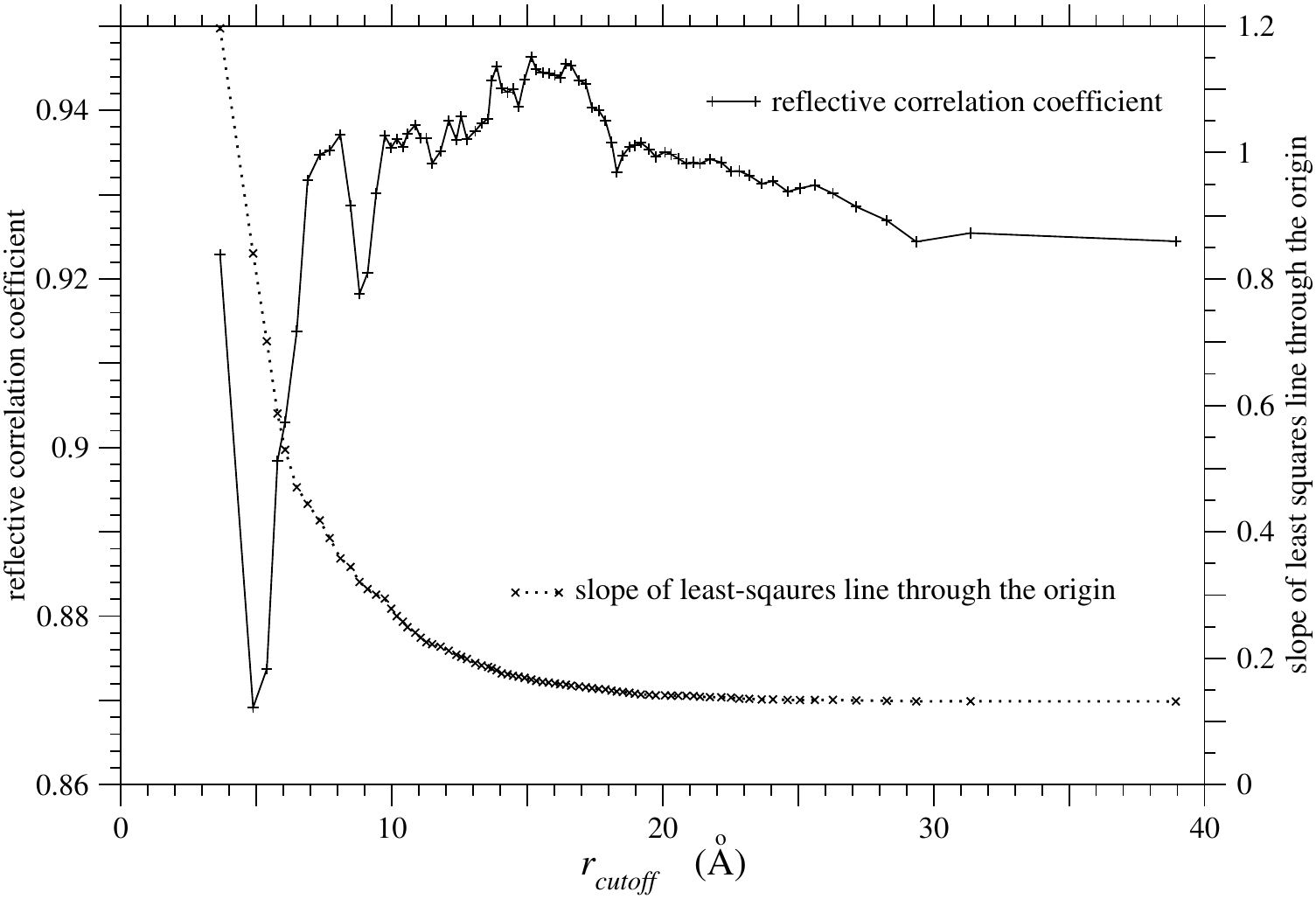}
}
}%  FigureInLegends
\vspace*{1em}
\caption{
\FigureLegends{
\label{sfig: 1gm1-a:16-96_full_non_del_dca0_205_0_20_simple-gauge_ddG-dPhi_vs_r}
\label{fig: 1gm1-a:16-96_full_non_del_dca0_205_0_20_simple-gauge_ddG-dPhi_vs_r}
\label{sfig: 1gm1-a:16-96_full_non_del_dca0_33_0_20_simple-gauge_ddG-dPhi_vs_r}
\label{fig: 1gm1-a:16-96_full_non_del_dca0_33_0_20_simple-gauge_ddG-dPhi_vs_r}
\BF{
Dependences of
the reflective correlation and regression coefficients
between the experimental $\Delta\Delta G_{ND}$\CITE{GGCJVTVB:07} 
and $\Delta \psi_N$ due to single amino acid substitutions
on the cutoff distance for pairwise interactions in the PDZ domain.
}
The left and right figures are for the ratios of pseudocount, $p_c=0.205$ and $0.33$, respectively.
The solid and dotted lines show the reflective correlation and 
regression coefficients
for the least-squares regression line through the origin, respectively.
The sample ($\overline{\psi_N} / L$) and ensemble ($\langle \psi_N \rangle_{\VEC{\sigma}} / L$) averages of 
evolutionary statistical energy
agree with each other at the cutoff distance $r_{\script{cutoff}} \sim 8$ \AA\  for $p_c = 0.205$ 
and $r_{\script{cutoff}} \sim 15.5$ \AA\  for $p_c = 0.33$,
where the reflective correlation coefficients attain to the maximum.
}%  FigureLegends
}
\end{figure*}

}%  SUPPLEMENT

\SUPPLEMENT{

\CLEARPAGE
\begin{figure*}[h!]
\FigureInLegends{
\noindent
\hspace*{1em} (a) $r_{\script{cutoff}} \sim 8$ \AA\  \hspace*{14em} (b) $r_{\script{cutoff}} \sim 15.5$ \AA\ 

\centerline{
\includegraphics*[width=82mm,angle=0]{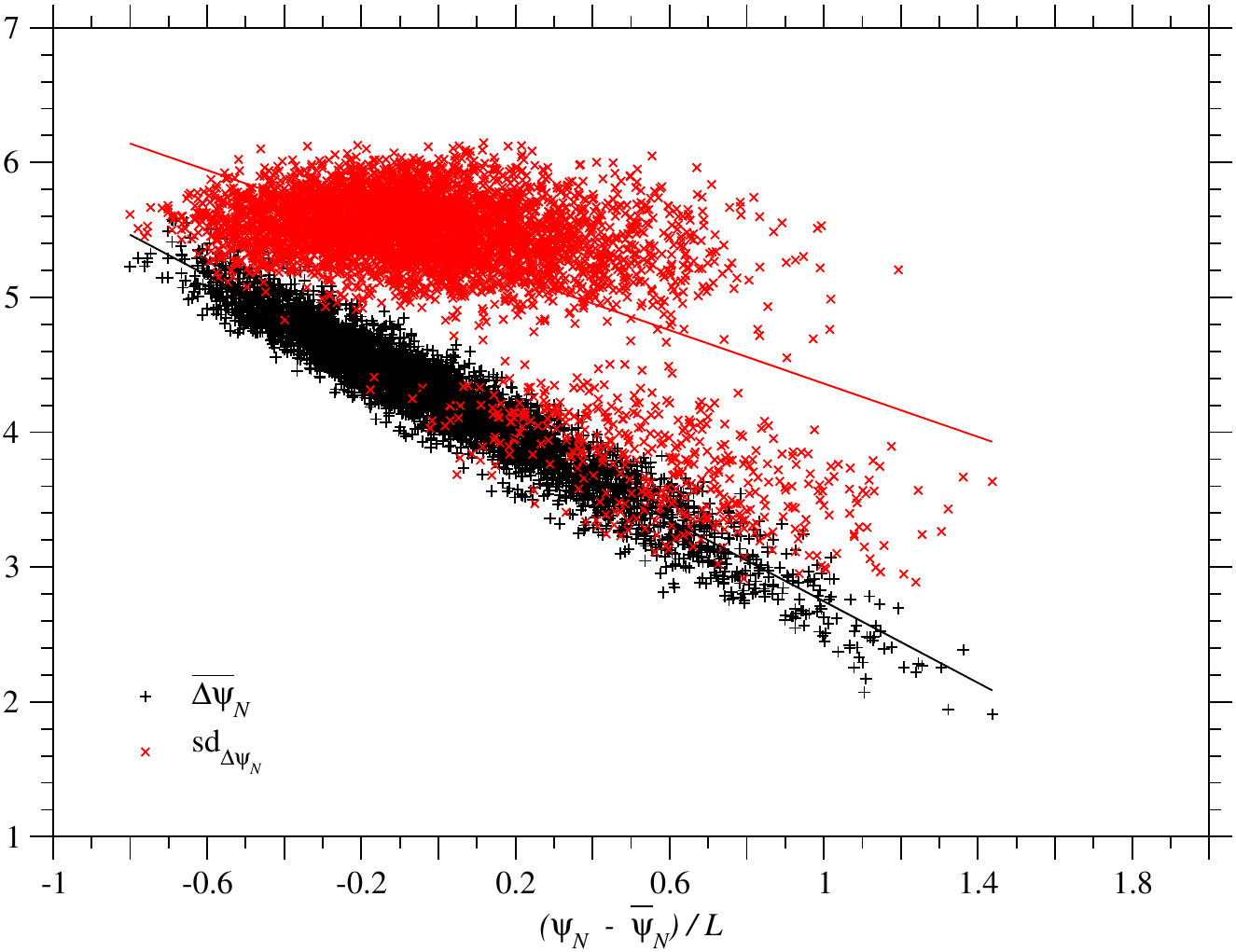}
\includegraphics*[width=82mm,angle=0]{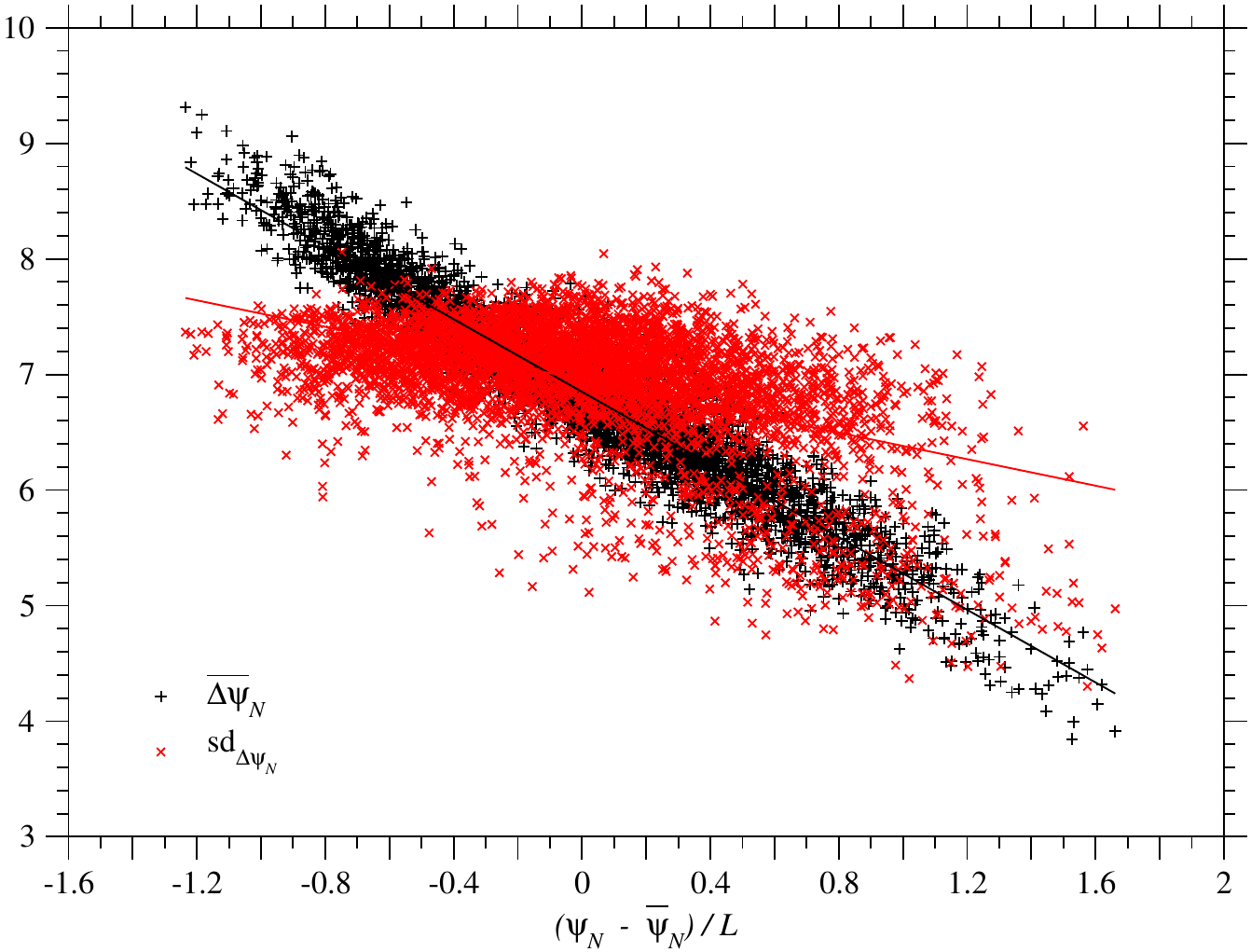}
}
}%  FigureInLegends
\vspace*{1em}
\caption{
\FigureLegends{
\label{sfig: 1r69-a:6-58.full_non_del.dca0_18.0_20.simple-gauge.ddPhi_at_opt}
\label{fig: 1r69-a:6-58.full_non_del.dca0_18.0_20.simple-gauge.ddPhi_at_opt}
\BF{
Correlation between $\Delta \psi_N$ due to single nucleotide nonsynonymous substitutions and
$\psi_N$ of homologous sequences in the HTH\_3 family of the domain, 1R69-A:6-58.
}
\protect
The left and right figures correspond to the cutoff distance $r_{\script{cutoff}} \sim 8$ and $15.5$ \AA,
respectively.
Each of the black plus or red cross marks corresponds to the mean or the standard deviation 
of $\Delta \psi_N$ due to
all types of single nucleotide nonsynonymous substitutions
over all sites in each of the homologous sequences.
Representatives of unique sequences,
which are at least 20\% different from each other, are employed;
the number of the representatives is almost equal to $M_{\script{eff}}$ in \Table{\ref{\TBL: Proteins_studied}}.
The solid lines show the regression lines for the mean and the standard deviation of $\Delta \psi_N$.
% End of figures_mean_and_sd_legends.tex
}%  FigureLegends
}
\end{figure*}

\CLEARPAGE
\begin{figure*}[h!]
\FigureInLegends{
\noindent
\hspace*{1em} (a) $r_{\script{cutoff}} \sim 8$ \AA\  \hspace*{14em} (b) $r_{\script{cutoff}} \sim 15.5$ \AA\ 

\centerline{
\includegraphics*[width=82mm,angle=0]{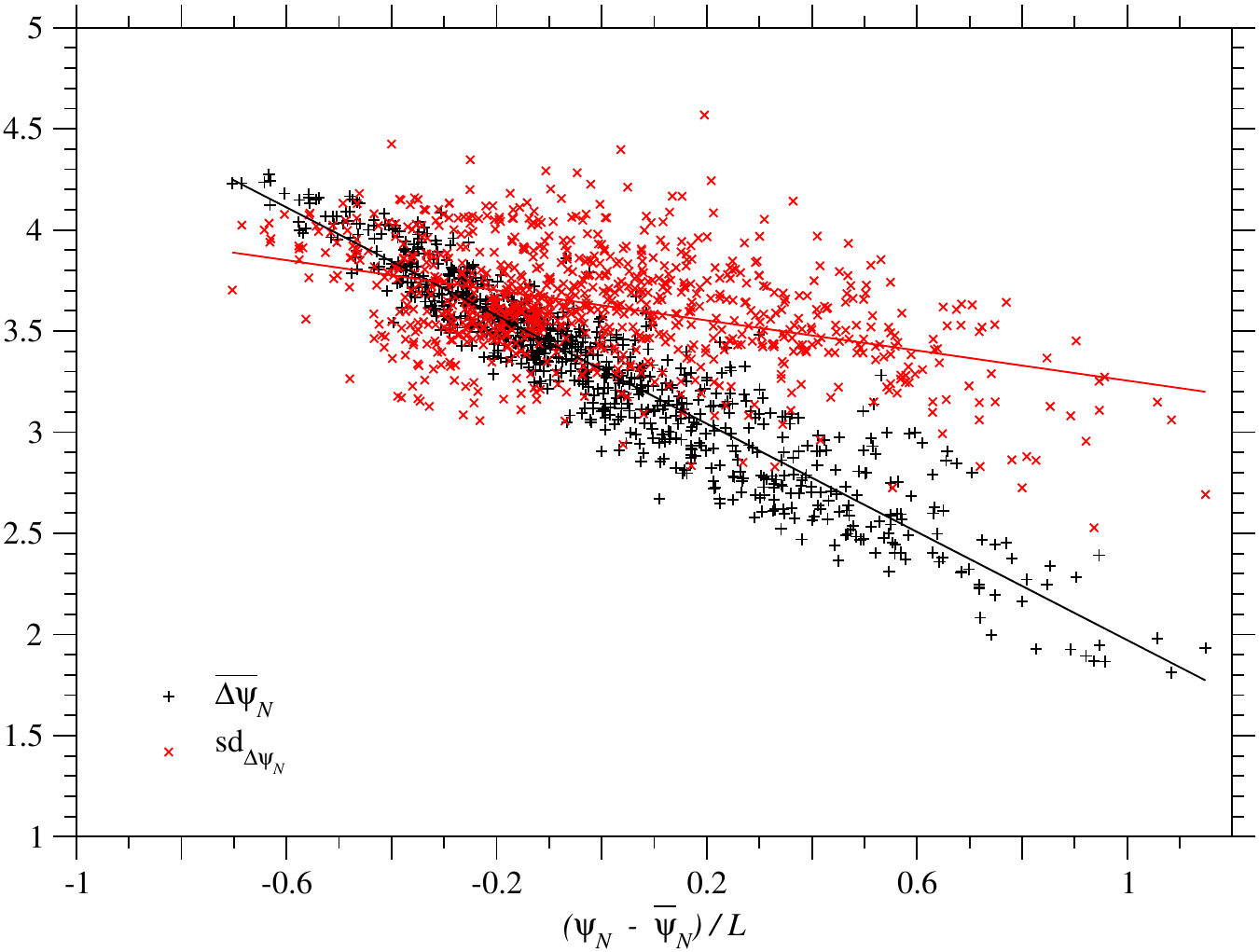}
\includegraphics*[width=82mm,angle=0]{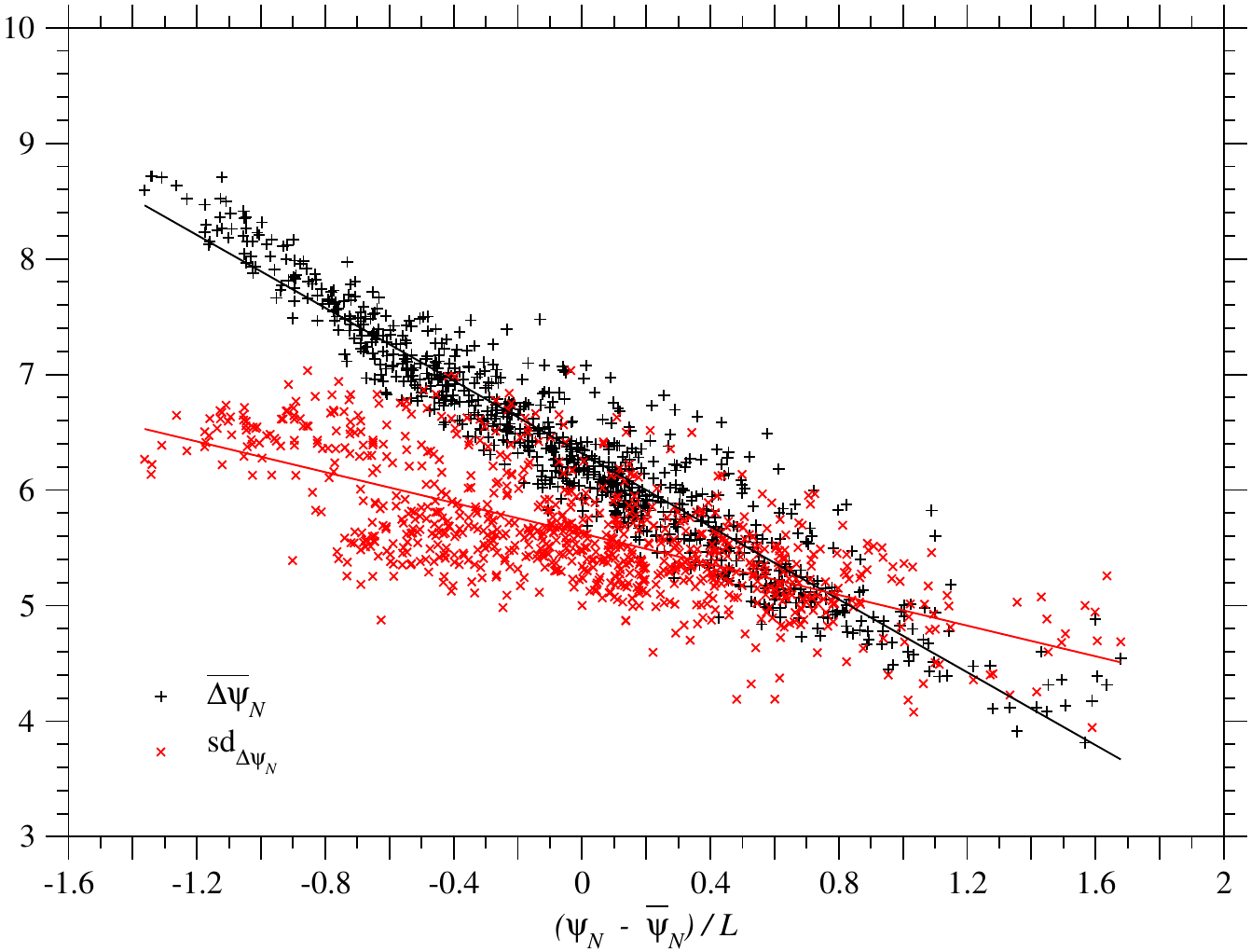}
}
}%  FigureInLegends
\vspace*{1em}
\caption{
\FigureLegends{
\label{sfig: 3e10:4-76.full_non_del.dca0_23.0_20.simple-gauge.AB.ddPhi_at_opt}
\label{fig: 3e10:4-76.full_non_del.dca0_23.0_20.simple-gauge.AB.ddPhi_at_opt}
\BF{
Correlation between $\Delta \psi_N$ due to single nucleotide nonsynonymous substitutions and
$\psi_N$ of homologous sequences in the Nitroreductase family of the domain, 3E10-A/B:4-76.
}
\protect
The left and right figures correspond to the cutoff distance $r_{\script{cutoff}} \sim 8$ and $15.5$ \AA,
respectively.
Each of the black plus or red cross marks corresponds to the mean or the standard deviation 
of $\Delta \psi_N$ due to
all types of single nucleotide nonsynonymous substitutions
over all sites in each of the homologous sequences.
Representatives of unique sequences,
which are at least 20\% different from each other, are employed;
the number of the representatives is almost equal to $M_{\script{eff}}$ in \Table{\ref{\TBL: Proteins_studied}}.
The solid lines show the regression lines for the mean and the standard deviation of $\Delta \psi_N$.
% End of figures_mean_and_sd_legends.tex
}%  FigureLegends
}
\end{figure*}

\CLEARPAGE
\begin{figure*}[h!]
\FigureInLegends{
\noindent
\hspace*{1em} (a) $r_{\script{cutoff}} \sim 8$ \AA\  \hspace*{14em} (b) $r_{\script{cutoff}} \sim 15.5$ \AA\ 

\centerline{
\includegraphics*[width=82mm,angle=0]{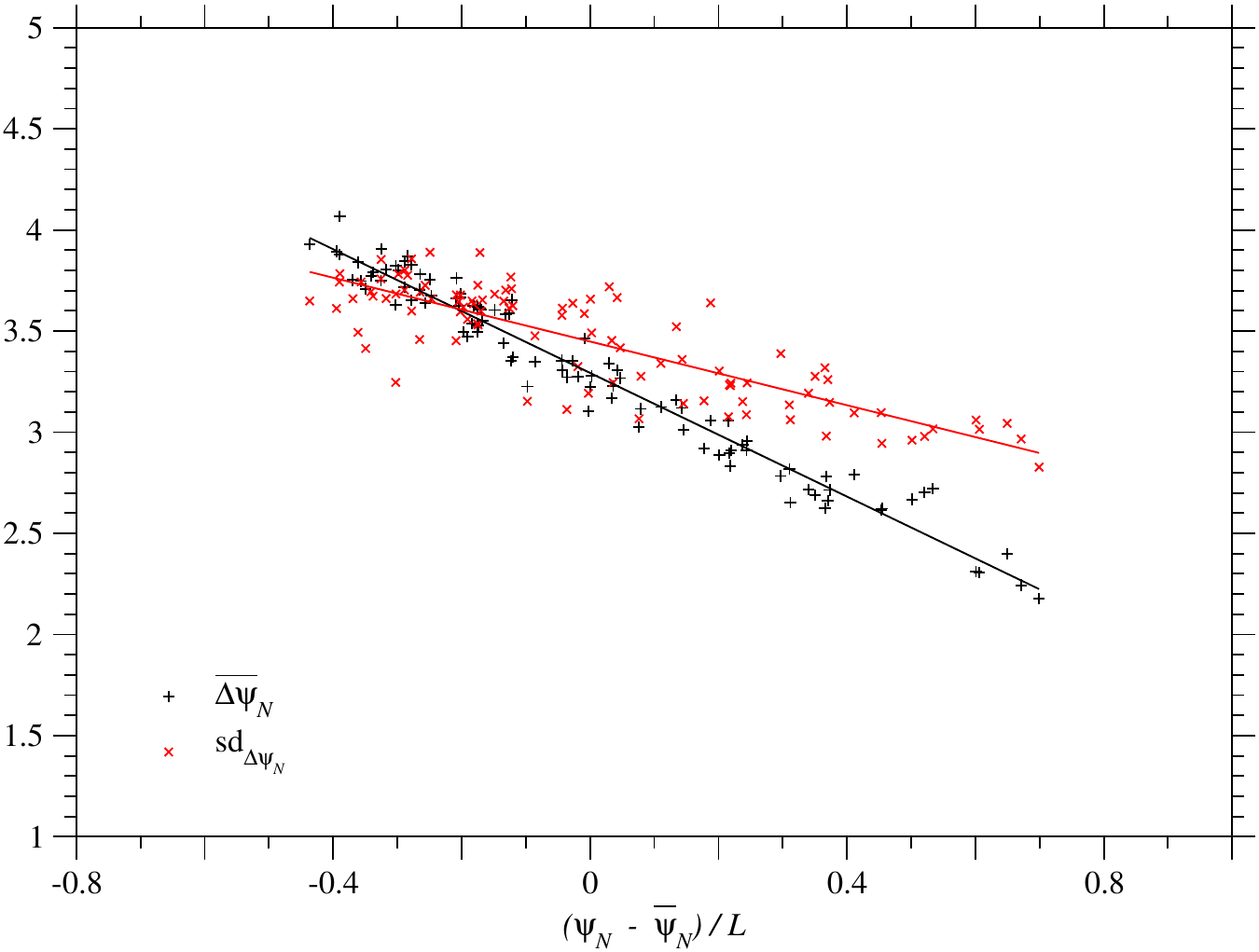}
\includegraphics*[width=82mm,angle=0]{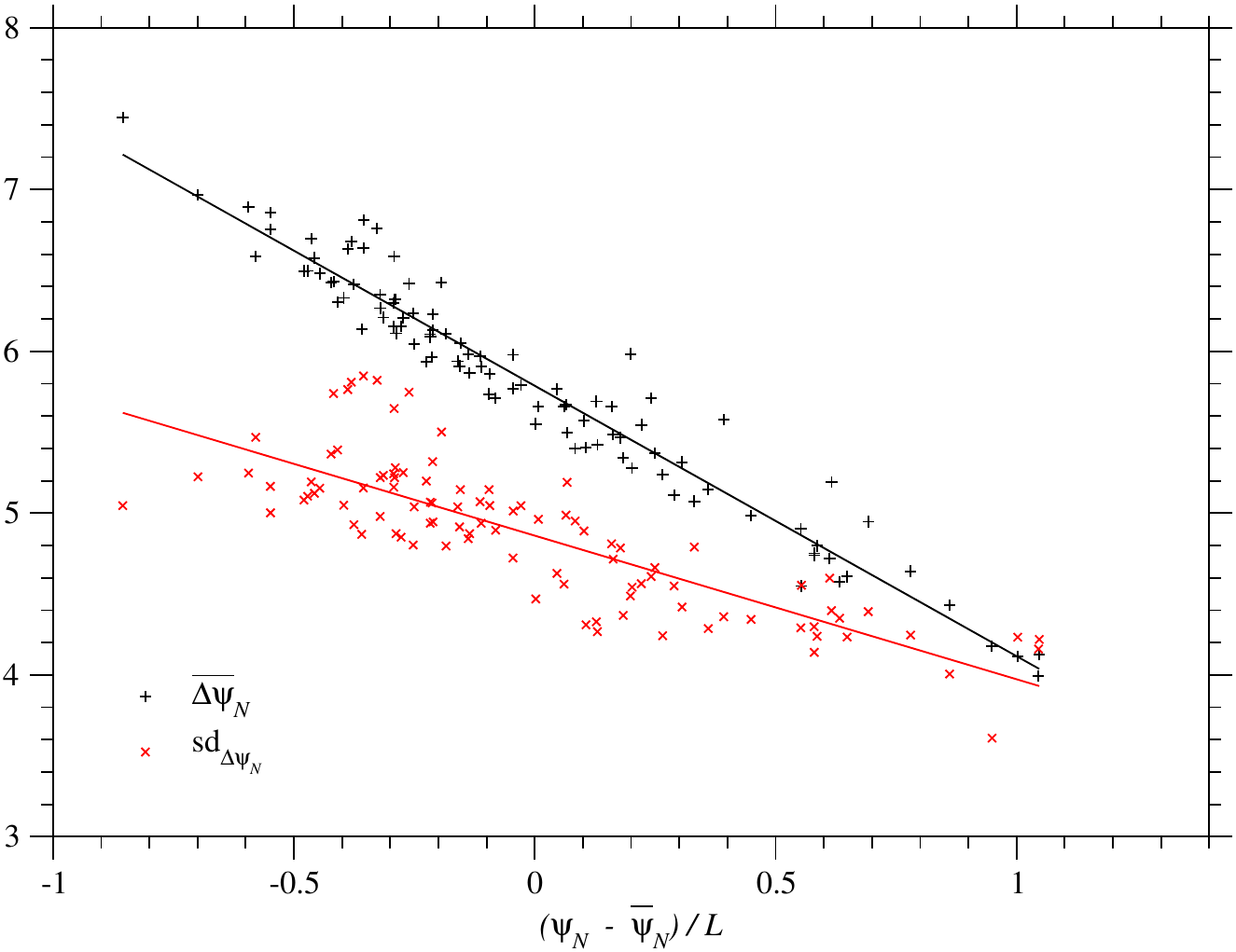}
}
\centerline{
\includegraphics*[width=82mm,angle=0]{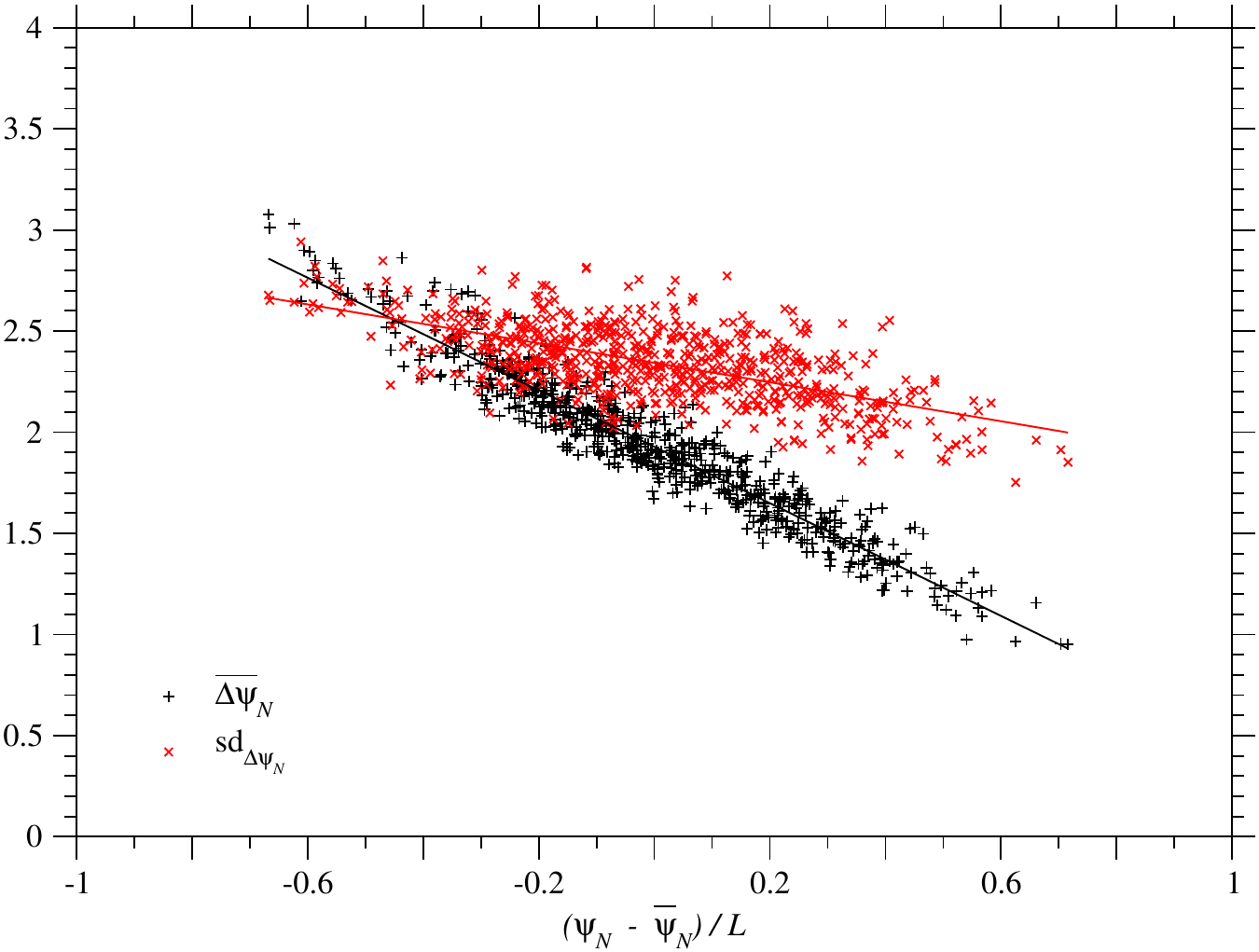}
\includegraphics*[width=82mm,angle=0]{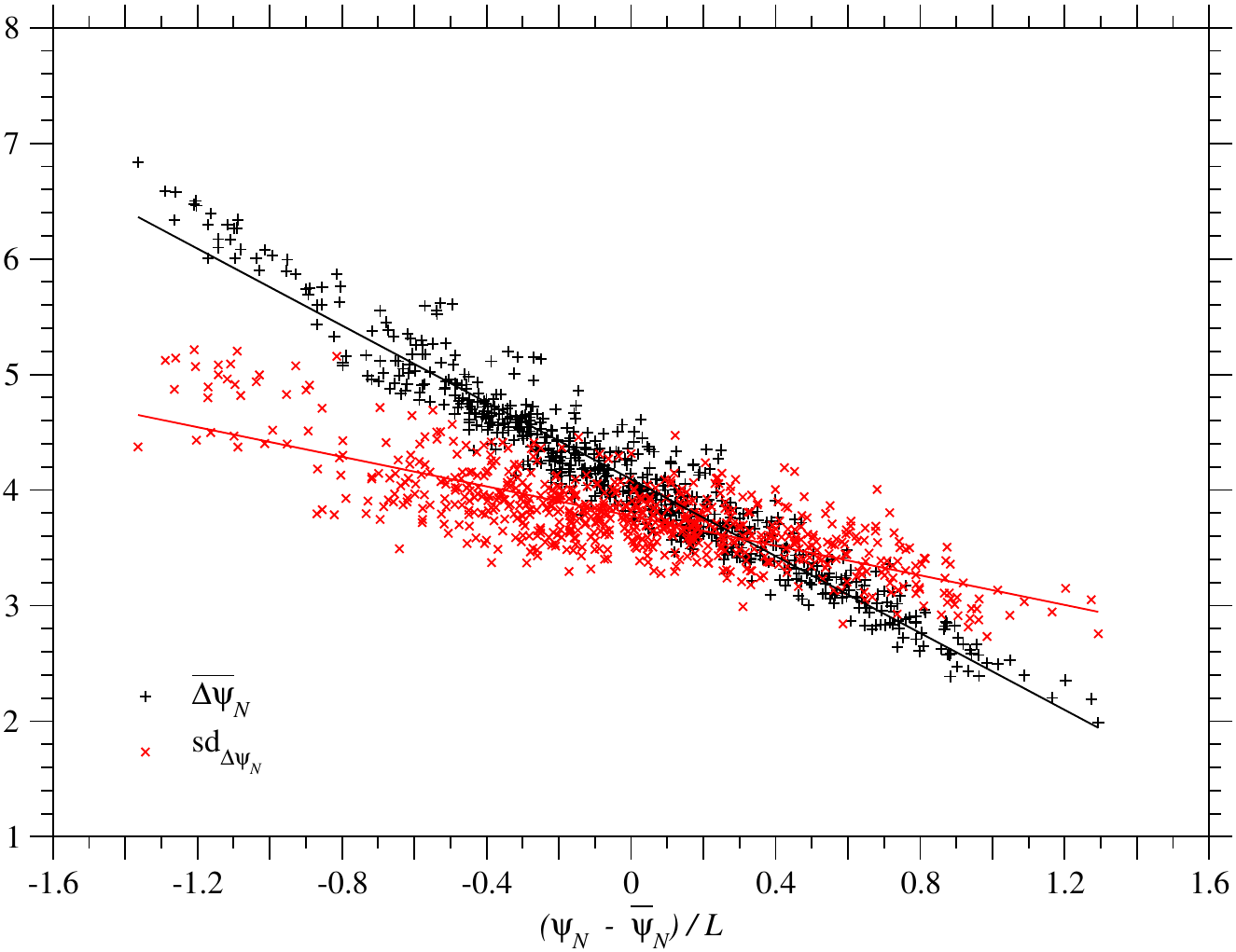}
}
}%  FigureInLegends
\vspace*{1em}
\caption{
\FigureLegends{
\label{sfig: 1wdn-a:5-222.full_non_del.dca0_25.0_20.simple-gauge.ddPhi_at_opt}
\label{fig: 1wdn-a:5-222.full_non_del.dca0_25.0_20.simple-gauge.ddPhi_at_opt}
\BF{
Correlation between $\Delta \psi_N$ due to single nucleotide nonsynonymous substitutions and
$\psi_N$ of homologous sequences in the SBP\_bac\_3 family of the domains, 1WDN-A:5-222 (upper) and 1WDN-A:89-182 (lower).
}
\protect
The left and right figures correspond to the cutoff distance $r_{\script{cutoff}} \sim 8$ and $15.5$ \AA,
respectively.
Each of the black plus or red cross marks corresponds to the mean or the standard deviation 
of $\Delta \psi_N$ due to
all types of single nucleotide nonsynonymous substitutions
over all sites in each of the homologous sequences.
Representatives of unique sequences,
which are at least 20\% different from each other, are employed;
the number of the representatives is almost equal to $M_{\script{eff}}$ in \Table{\ref{\TBL: Proteins_studied}}.
The solid lines show the regression lines for the mean and the standard deviation of $\Delta \psi_N$.
% End of figures_mean_and_sd_legends.tex
}%  FigureLegends
}
\end{figure*}

\CLEARPAGE
\begin{figure*}[h!]
\FigureInLegends{
\noindent
\hspace*{1em} (a) $r_{\script{cutoff}} \sim 8$ \AA\  \hspace*{14em} (b) $r_{\script{cutoff}} \sim 15.5$ \AA\ 

\centerline{
\includegraphics*[width=82mm,angle=0]{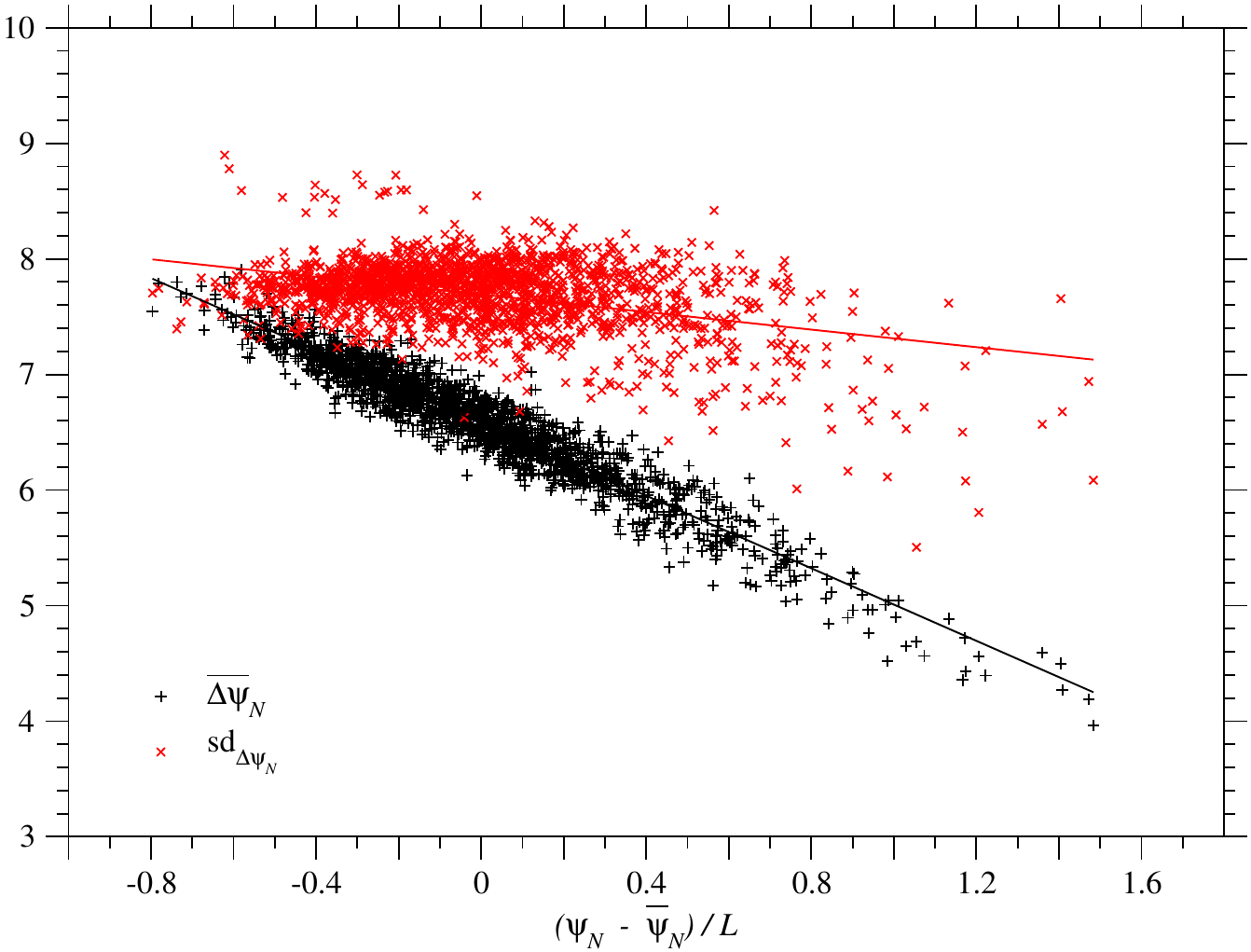}
\includegraphics*[width=82mm,angle=0]{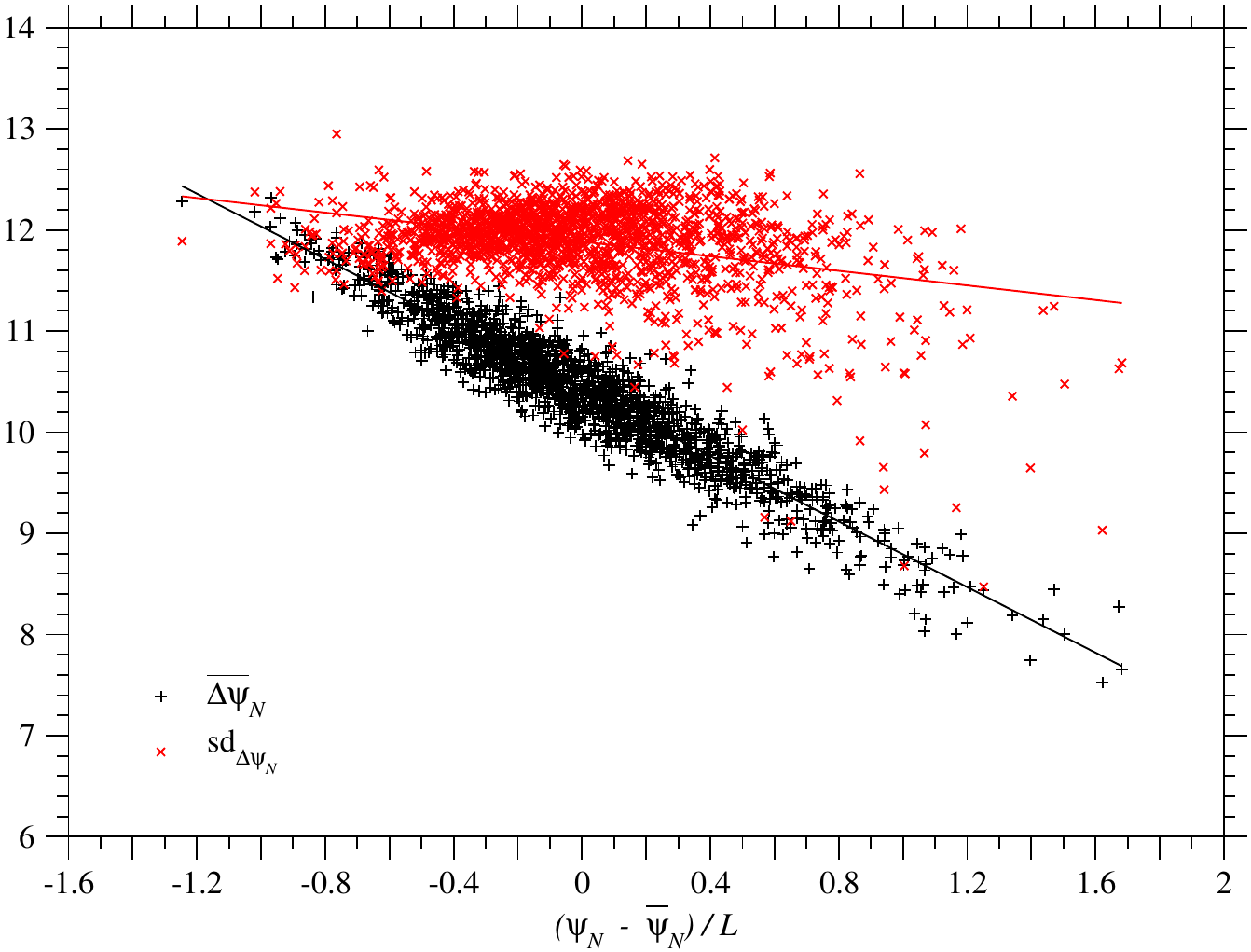}
}
}%  FigureInLegends
\vspace*{1em}
\caption{
\FigureLegends{
\label{sfig: 1oap-a:52-146.full_non_del.dca0_169.0_20.simple-gauge.ddPhi_at_opt}
\label{fig: 1oap-a:52-146.full_non_del.dca0_169.0_20.simple-gauge.ddPhi_at_opt}
\BF{
Correlation between $\Delta \psi_N$ due to single nucleotide nonsynonymous substitutions and
$\psi_N$ of homologous sequences in the OmpA family of the domains, 1OAP-A:52-146.
}
\protect
The left and right figures correspond to the cutoff distance $r_{\script{cutoff}} \sim 8$ and $15.5$ \AA,
respectively.
Each of the black plus or red cross marks corresponds to the mean or the standard deviation 
of $\Delta \psi_N$ due to
all types of single nucleotide nonsynonymous substitutions
over all sites in each of the homologous sequences.
Representatives of unique sequences,
which are at least 20\% different from each other, are employed;
the number of the representatives is almost equal to $M_{\script{eff}}$ in \Table{\ref{\TBL: Proteins_studied}}.
The solid lines show the regression lines for the mean and the standard deviation of $\Delta \psi_N$.
% End of figures_mean_and_sd_legends.tex
}%  FigureLegends
}
\end{figure*}

\CLEARPAGE
\begin{figure*}[h!]
\FigureInLegends{
\noindent
\hspace*{1em} (a) $r_{\script{cutoff}} \sim 8$ \AA\  \hspace*{14em} (b) $r_{\script{cutoff}} \sim 15.5$ \AA\ 

\centerline{
\includegraphics*[width=82mm,angle=0]{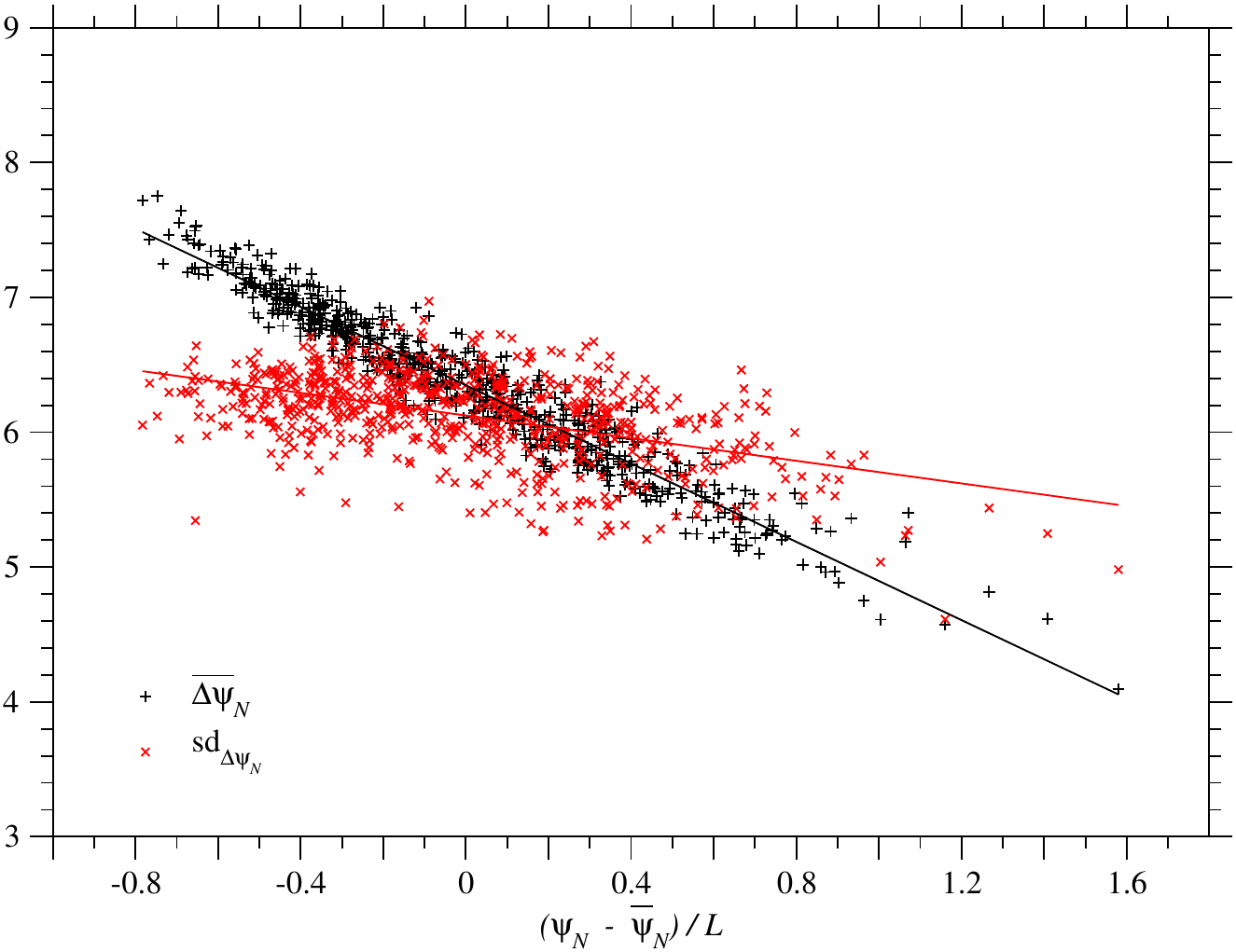}
\includegraphics*[width=82mm,angle=0]{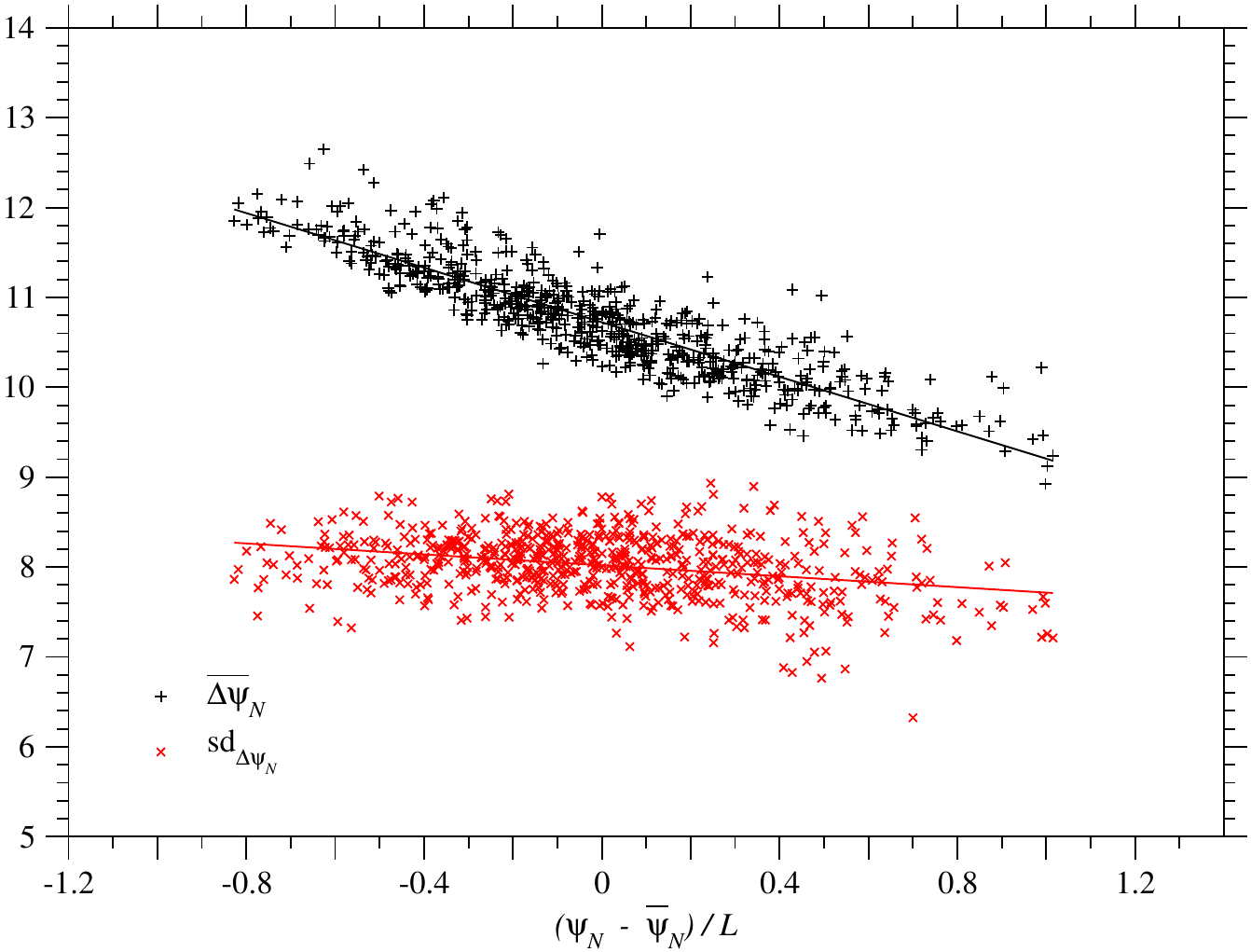}
}
}%  FigureInLegends
\vspace*{1em}
\caption{
\FigureLegends{
\label{sfig: 1jwe-a:30-127.full_non_del.dca0_235.0_20.simple-gauge.ddPhi_at_opt}
\label{fig: 1jwe-a:30-127.full_non_del.dca0_235.0_20.simple-gauge.ddPhi_at_opt}
\BF{
Correlation between $\Delta \psi_N$ due to single nucleotide nonsynonymous substitutions and
$\psi_N$ of homologous sequences in the DnaB family of the domains, 1JWE-A:30-127.
}
\protect
The left and right figures correspond to the cutoff distance $r_{\script{cutoff}} \sim 8$ and $15.5$ \AA,
respectively.
Each of the black plus or red cross marks corresponds to the mean or the standard deviation 
of $\Delta \psi_N$ due to
all types of single nucleotide nonsynonymous substitutions
over all sites in each of the homologous sequences.
Representatives of unique sequences,
which are at least 20\% different from each other, are employed;
the number of the representatives is almost equal to $M_{\script{eff}}$ in \Table{\ref{\TBL: Proteins_studied}}.
The solid lines show the regression lines for the mean and the standard deviation of $\Delta \psi_N$.
% End of figures_mean_and_sd_legends.tex
}%  FigureLegends
}
\end{figure*}

\CLEARPAGE
\begin{figure*}[h!]
\FigureInLegends{
\noindent
\hspace*{1em} (a) $r_{\script{cutoff}} \sim 8$ \AA\  \hspace*{14em} (b) $r_{\script{cutoff}} \sim 15.5$ \AA\ 

\centerline{
\includegraphics*[width=82mm,angle=0]{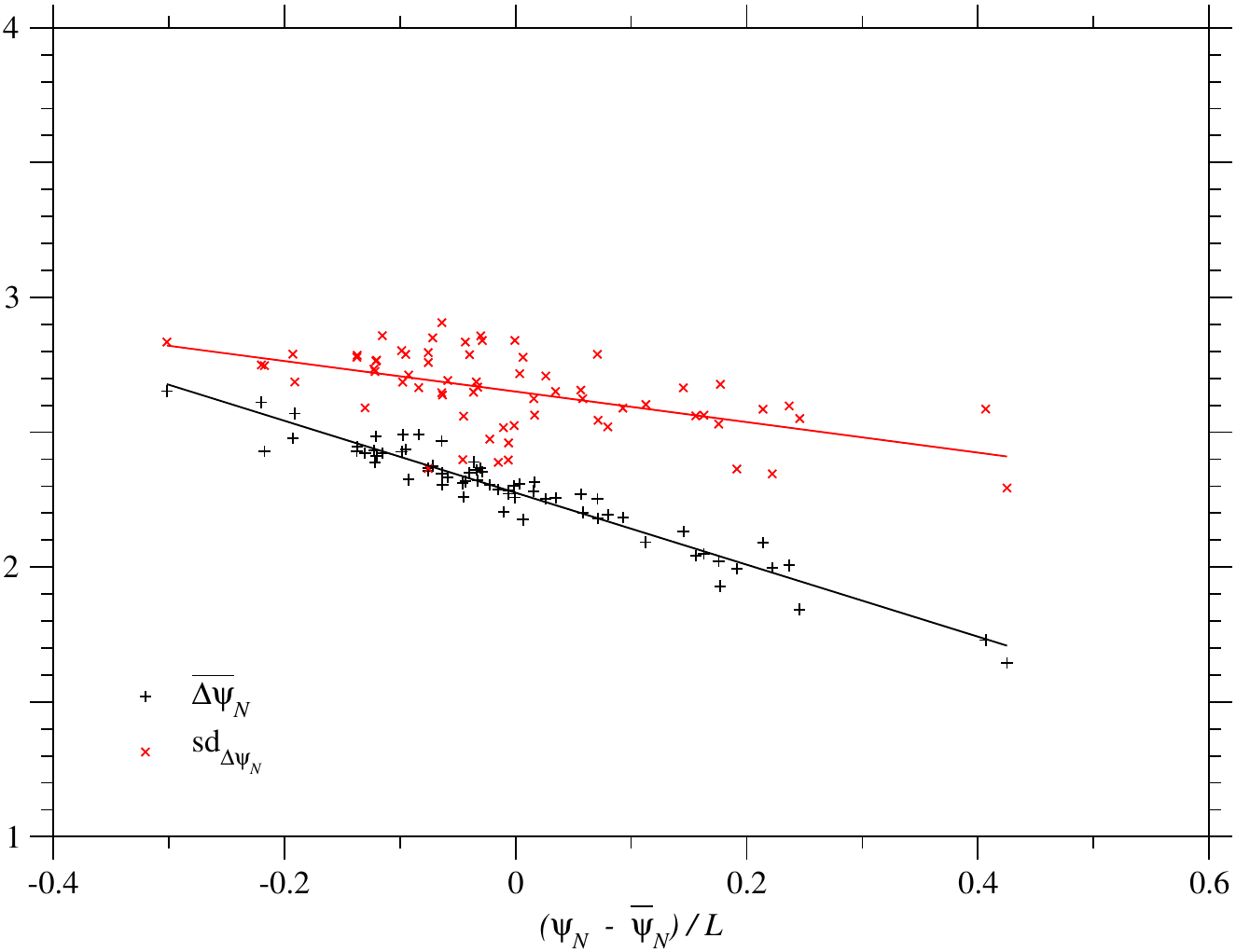}
\includegraphics*[width=82mm,angle=0]{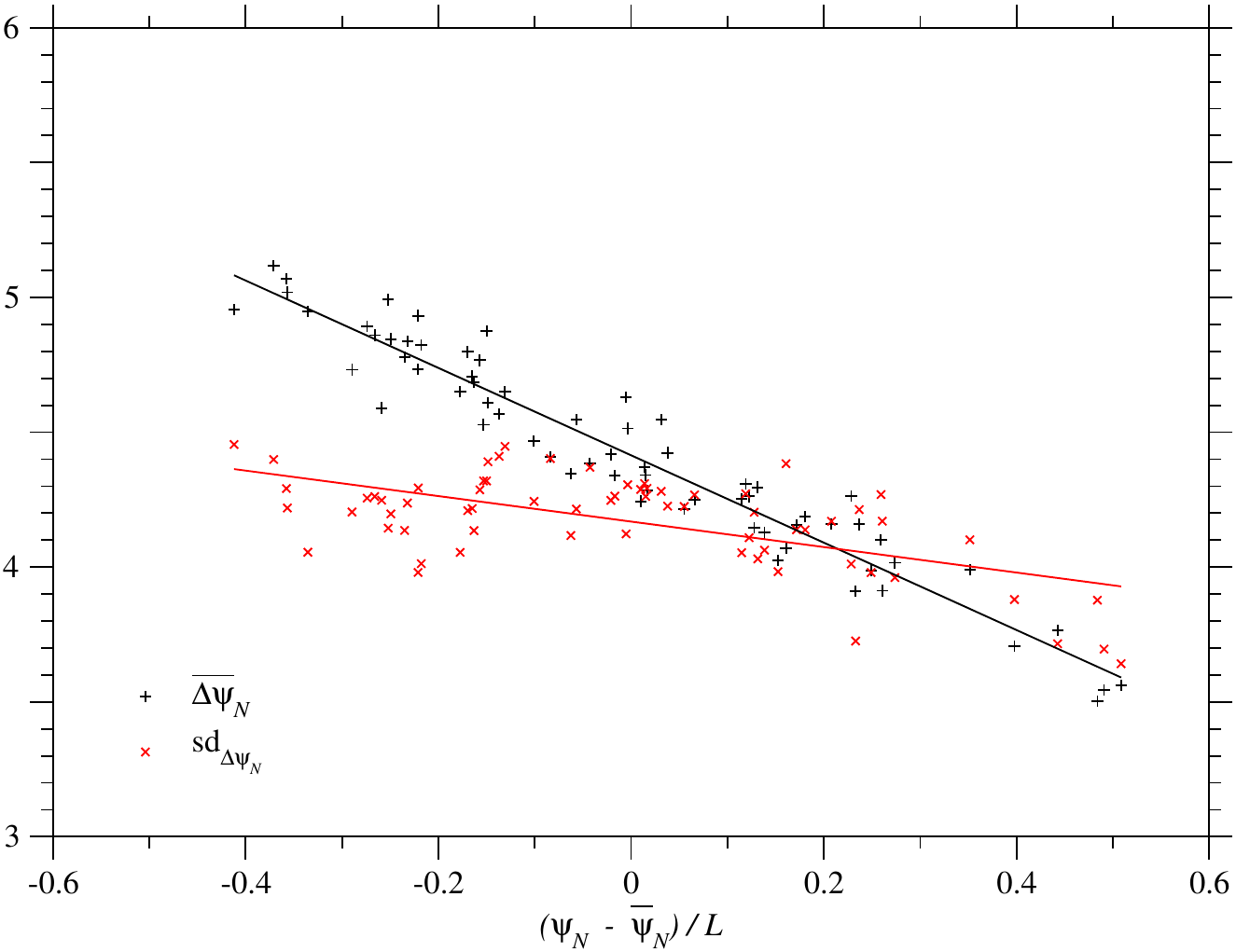}
}
\centerline{
\includegraphics*[width=82mm,angle=0]{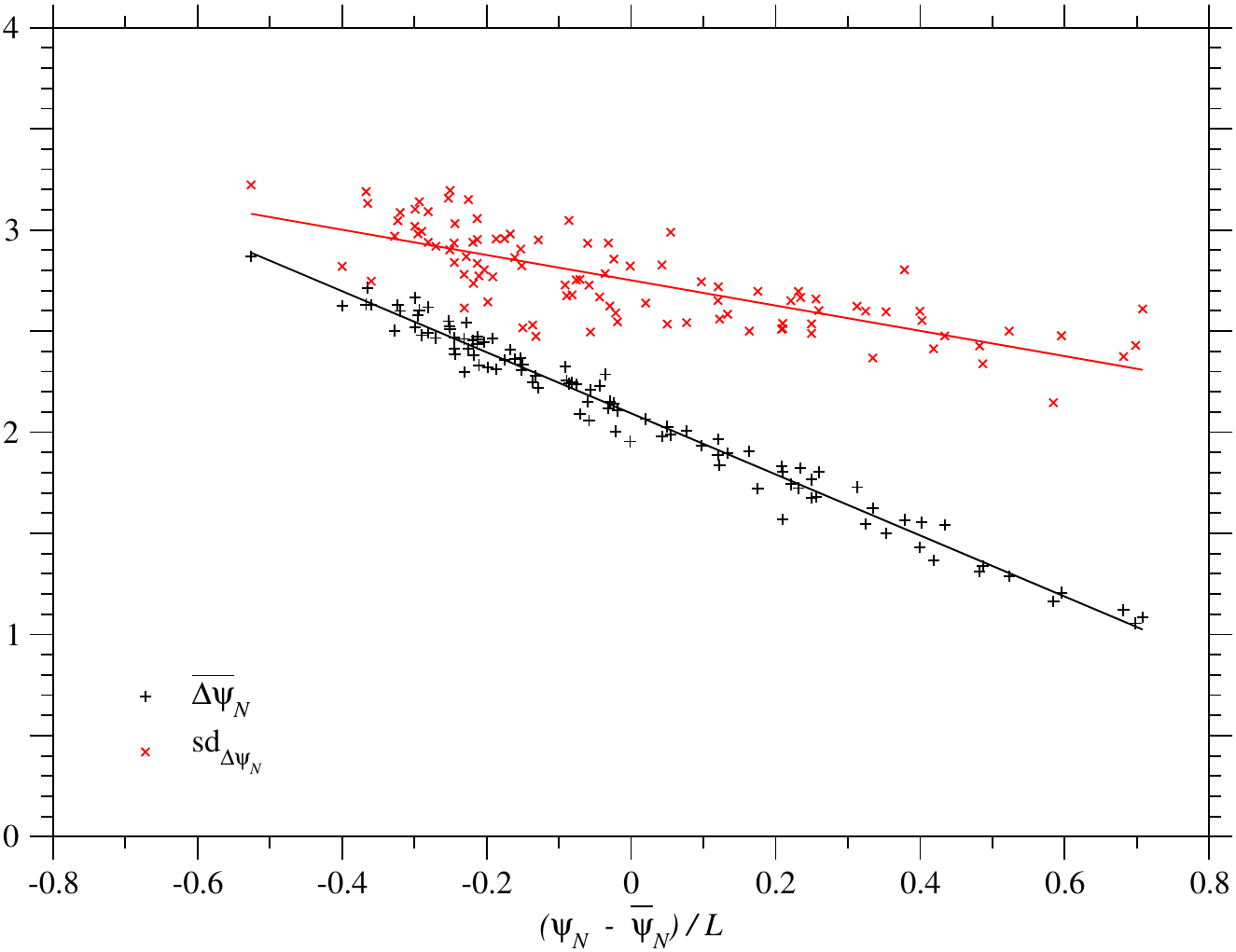}
\includegraphics*[width=82mm,angle=0]{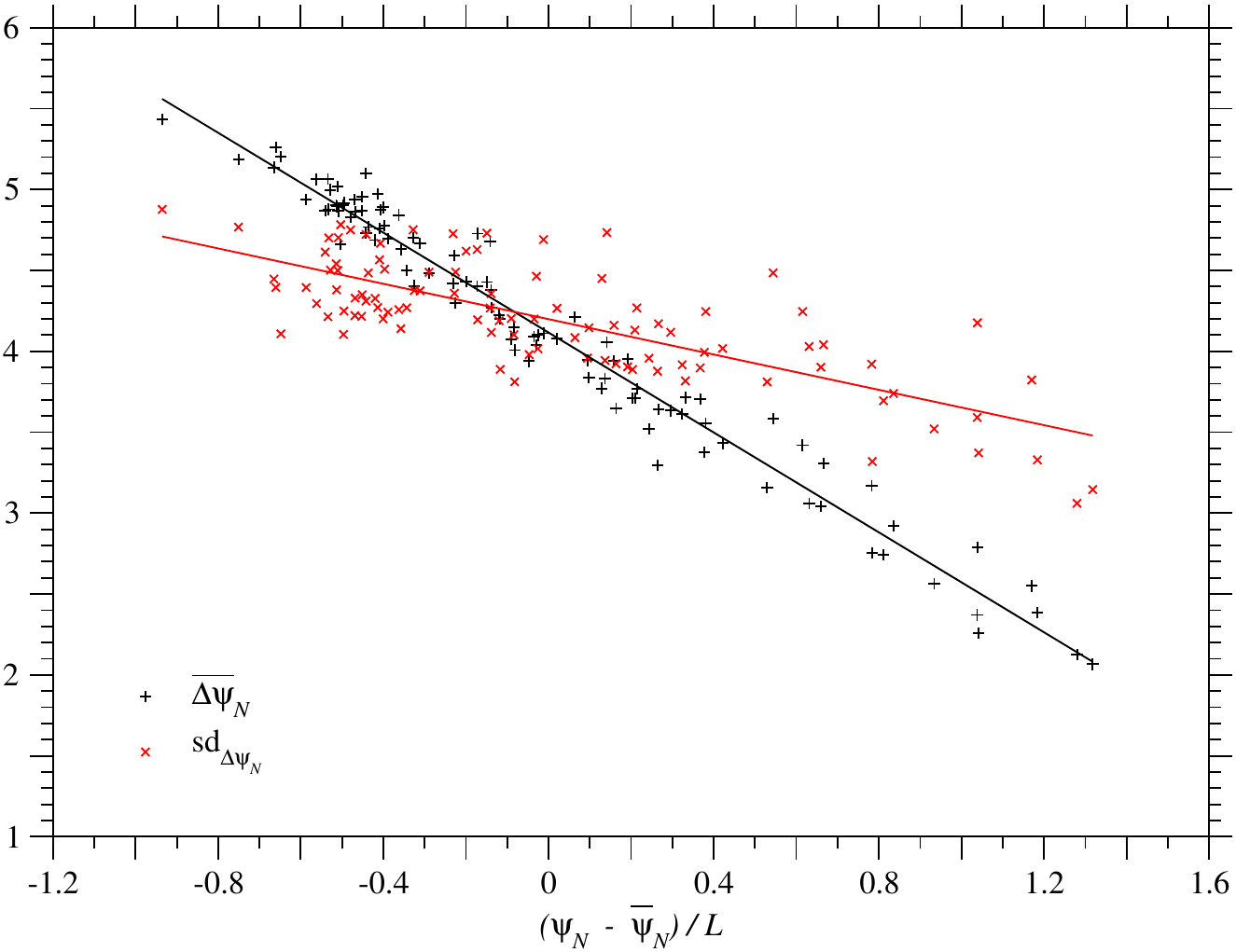}
}
}%  FigureInLegends
\vspace*{1em}
\caption{
\FigureLegends{
\label{sfig: 2f6g:90-280.full_non_del.dca0_235.0_20.simple-gauge.AB.ddPhi_at_opt}
\label{fig: 2f6g:90-280.full_non_del.dca0_235.0_20.simple-gauge.AB.ddPhi_at_opt}
\BF{
Correlation between $\Delta \psi_N$ due to single nucleotide nonsynonymous substitutions and
$\psi_N$ of homologous sequences in the LysR\_substrate family of the domains, 2F6G-A:90-280 (above) and 2F6G-A:163-265 (below).
}
\protect
The left and right figures correspond to the cutoff distance $r_{\script{cutoff}} \sim 8$ and $15.5$ \AA,
respectively.
Each of the black plus or red cross marks corresponds to the mean or the standard deviation 
of $\Delta \psi_N$ due to
all types of single nucleotide nonsynonymous substitutions
over all sites in each of the homologous sequences.
Representatives of unique sequences,
which are at least 20\% different from each other, are employed;
the number of the representatives is almost equal to $M_{\script{eff}}$ in \Table{\ref{\TBL: Proteins_studied}}.
The solid lines show the regression lines for the mean and the standard deviation of $\Delta \psi_N$.
% End of figures_mean_and_sd_legends.tex
}%  FigureLegends
}
\end{figure*}

\CLEARPAGE
\begin{figure*}[h!]
\FigureInLegends{
\noindent
\hspace*{1em} (a) $r_{\script{cutoff}} \sim 8$ \AA\  \hspace*{14em} (b) $r_{\script{cutoff}} \sim 15.5$ \AA\ 

\centerline{
\includegraphics*[width=82mm,angle=0]{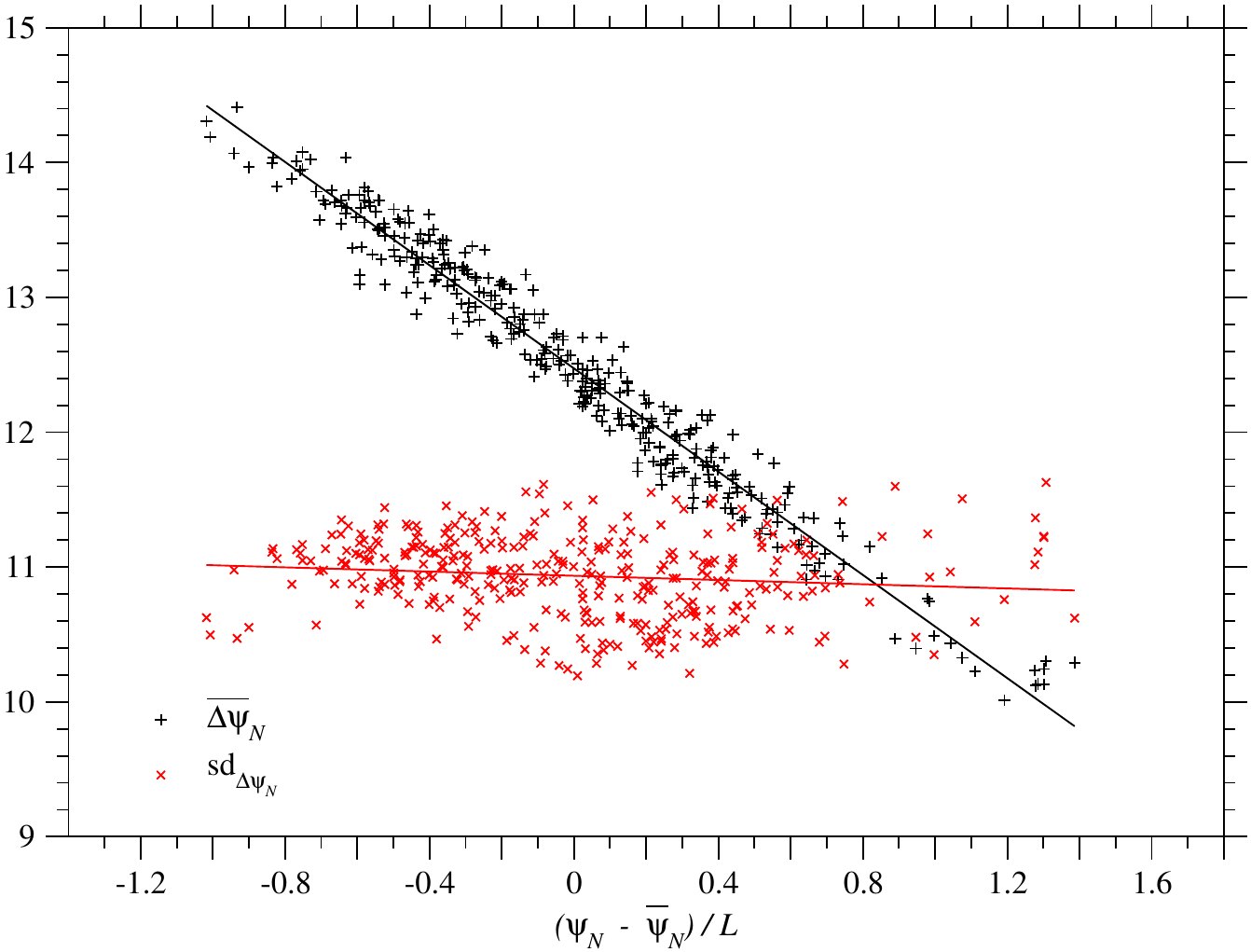}
\includegraphics*[width=82mm,angle=0]{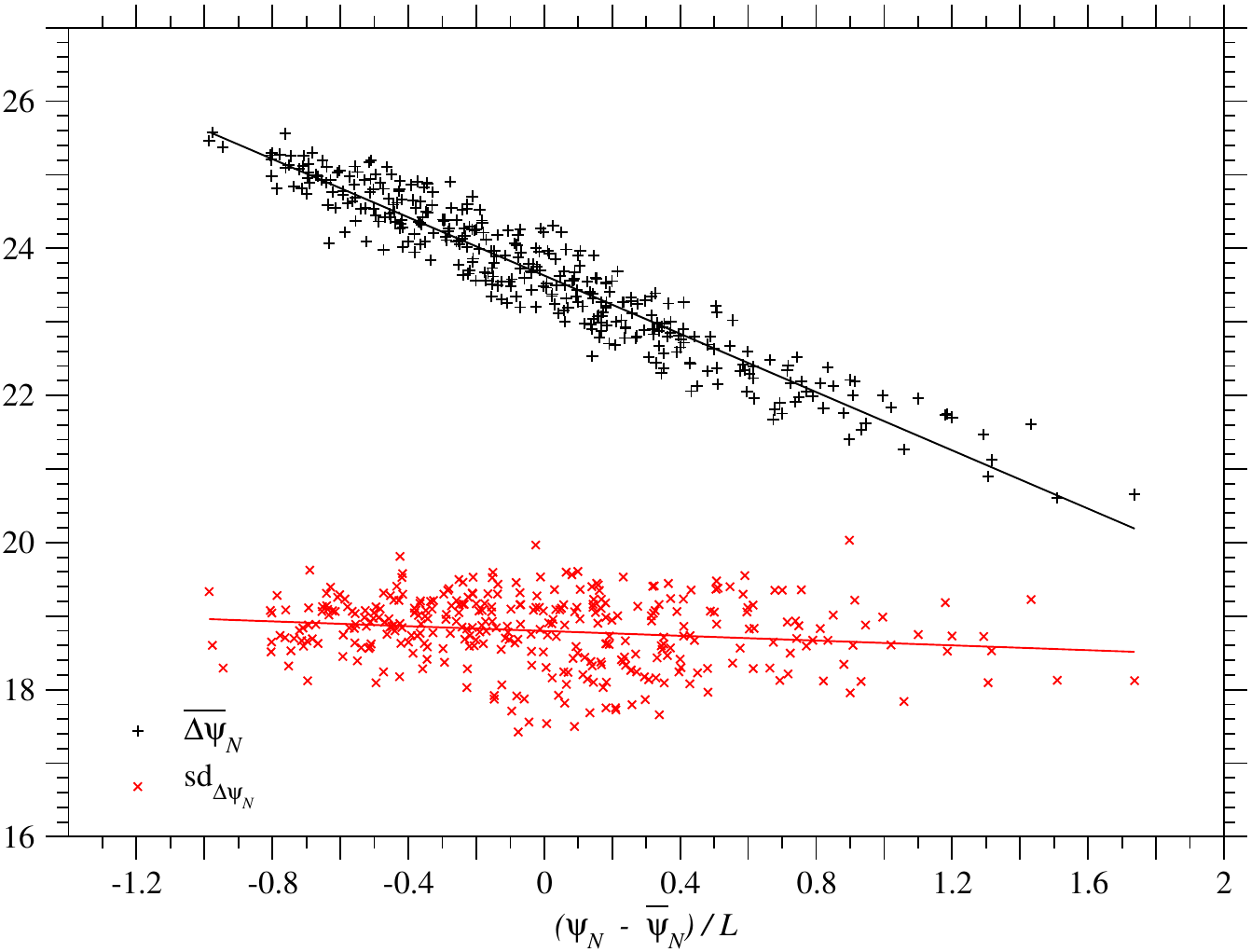}
}
\centerline{
\includegraphics*[width=82mm,angle=0]{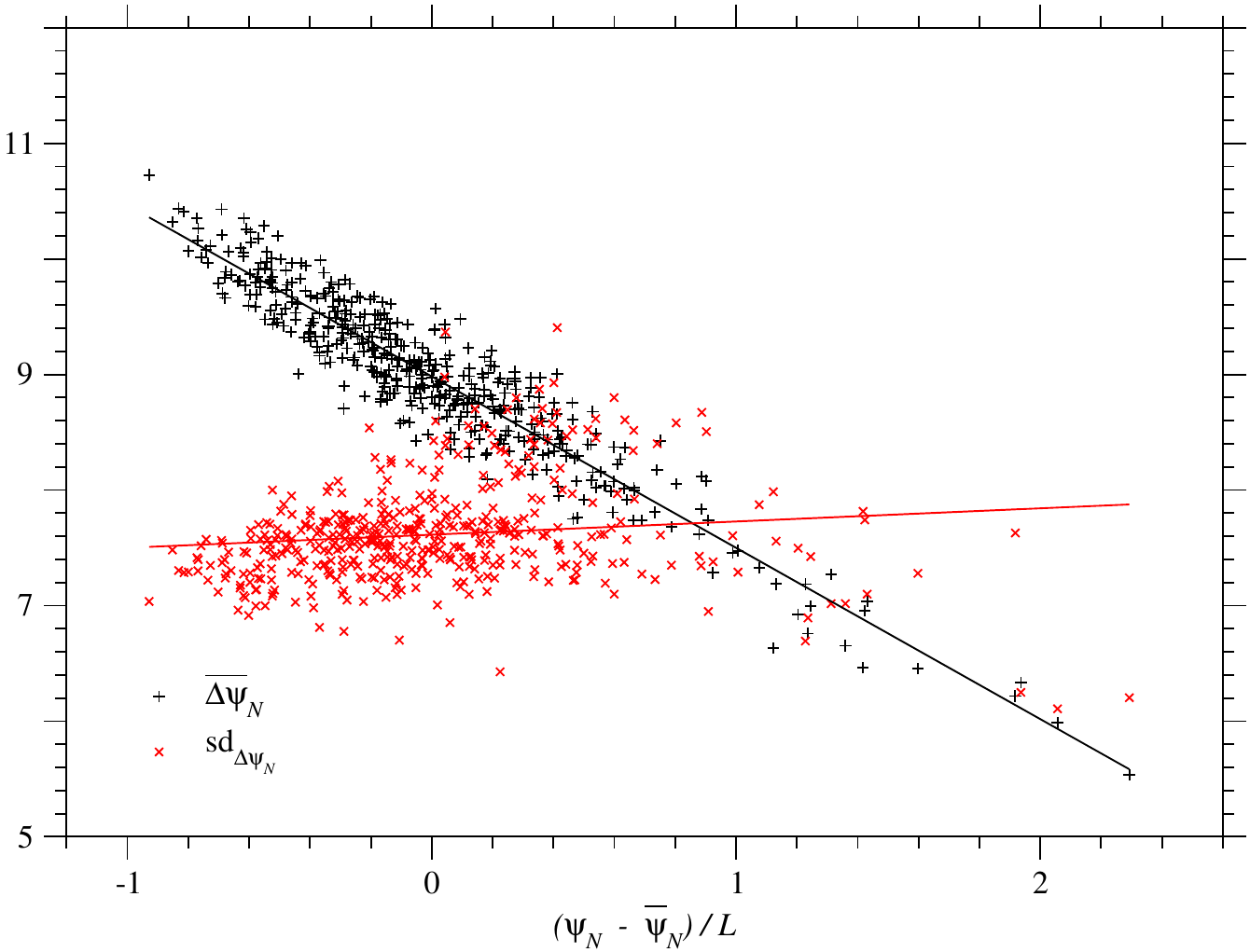}
\includegraphics*[width=82mm,angle=0]{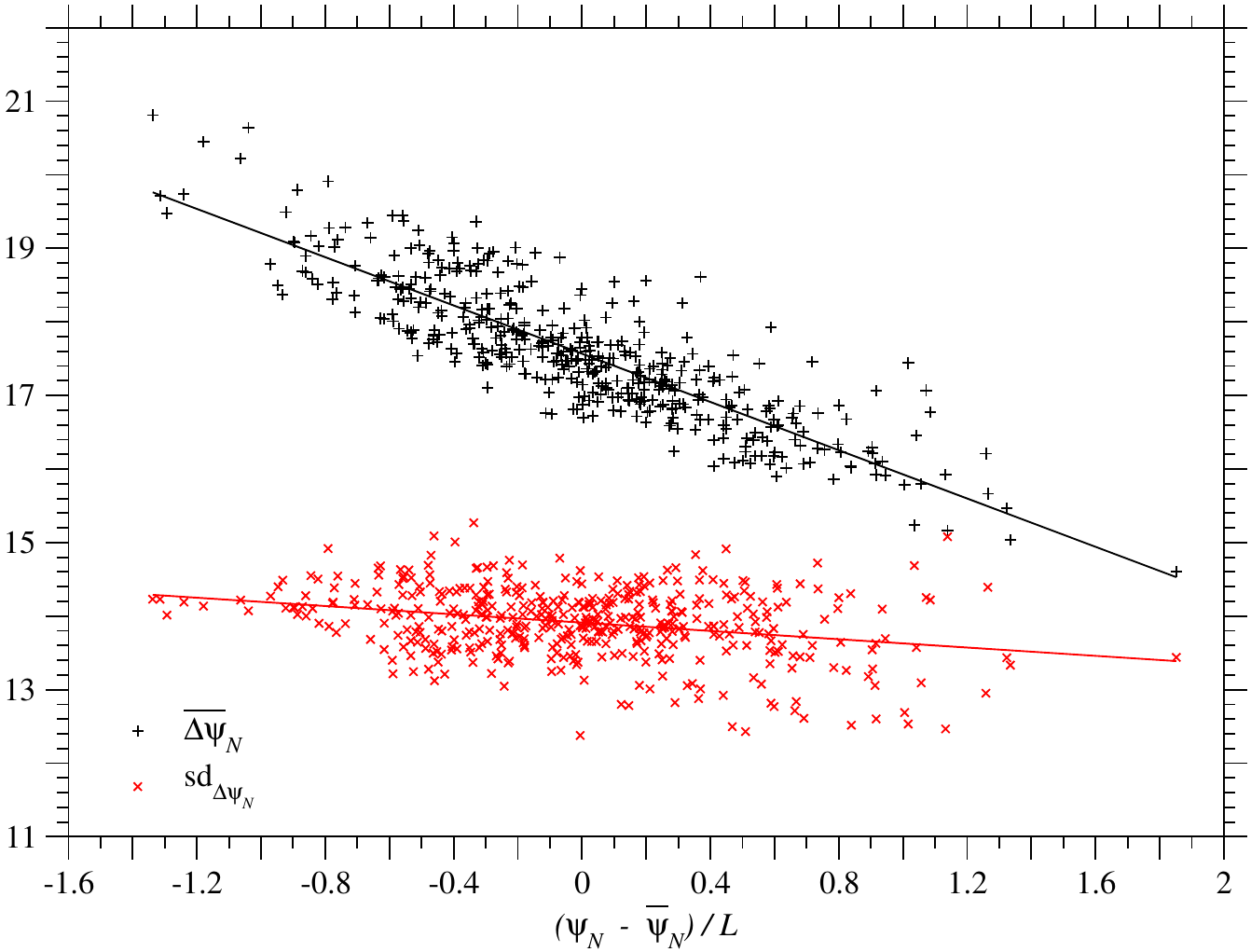}
}
}%  FigureInLegends
\vspace*{1em}
\caption{
\FigureLegends{
\label{sfig: 1n2x-a:8-292.full_non_del.dca0_13.0_20.simple-gauge.ddPhi_at_opt}
\label{fig: 1n2x-a:8-292.full_non_del.dca0_13.0_20.simple-gauge.ddPhi_at_opt}
\BF{
Correlation between $\Delta \psi_N$ due to single nucleotide nonsynonymous substitutions and
$\psi_N$ of homologous sequences in the Methyltransf\_5 family of the domains, 1N2X-A:8-292 (above) and 1N2X-A:137-216 (below).
}
\protect
The left and right figures correspond to the cutoff distance $r_{\script{cutoff}} \sim 8$ and $15.5$ \AA,
respectively.
Each of the black plus or red cross marks corresponds to the mean or the standard deviation 
of $\Delta \psi_N$ due to
all types of single nucleotide nonsynonymous substitutions
over all sites in each of the homologous sequences.
Representatives of unique sequences,
which are at least 20\% different from each other, are employed;
the number of the representatives is almost equal to $M_{\script{eff}}$ in \Table{\ref{\TBL: Proteins_studied}}.
The solid lines show the regression lines for the mean and the standard deviation of $\Delta \psi_N$.
% End of figures_mean_and_sd_legends.tex
}%  FigureLegends
}
\end{figure*}

\CLEARPAGE
\begin{figure*}[h!]
\FigureInLegends{
\noindent
\hspace*{1em} (a) $r_{\script{cutoff}} \sim 8$ \AA\  \hspace*{14em} (b) $r_{\script{cutoff}} \sim 15.5$ \AA\ 

\centerline{
\includegraphics*[width=82mm,angle=0]{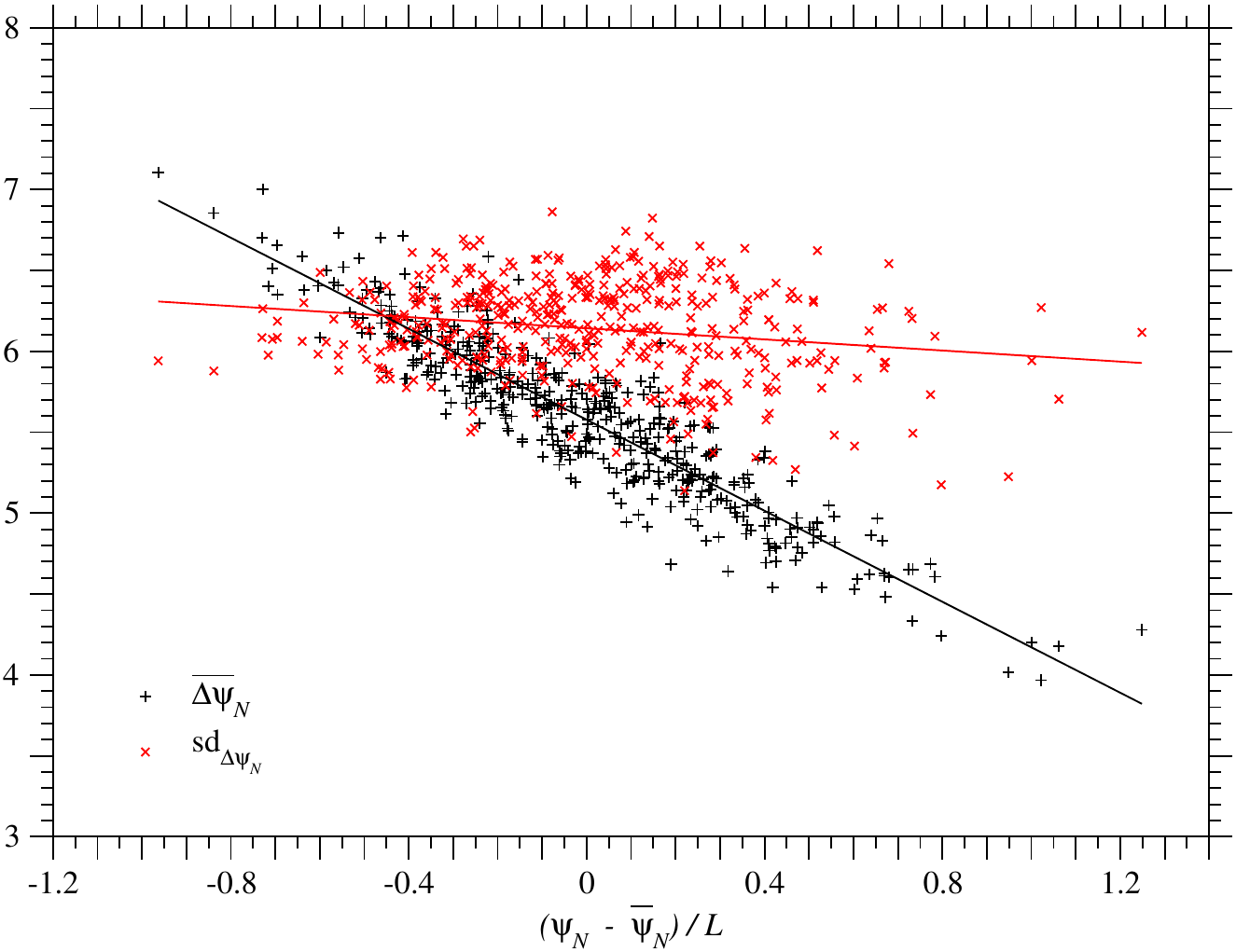}
\includegraphics*[width=82mm,angle=0]{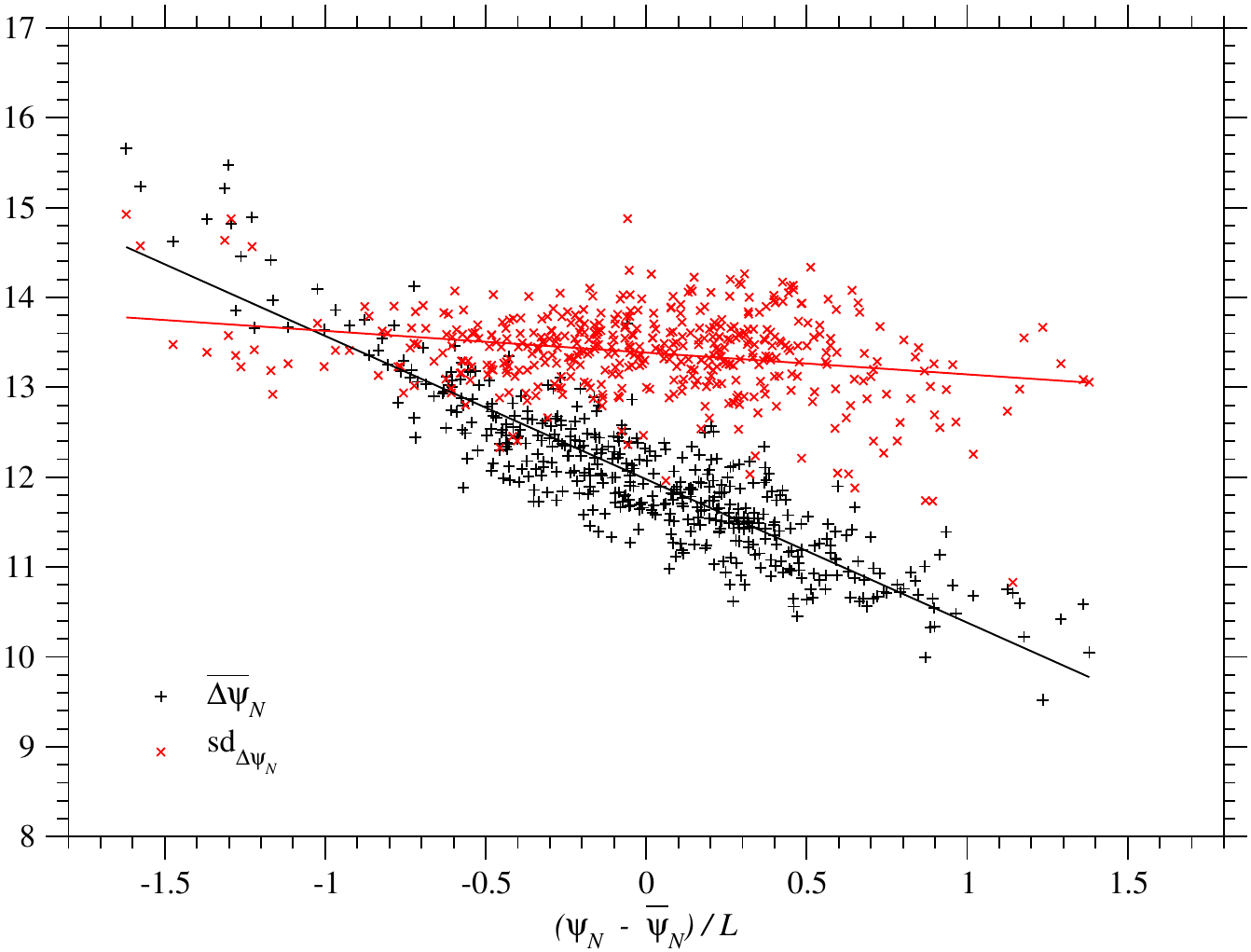}
}
}%  FigureInLegends
\vspace*{1em}
\caption{
\FigureLegends{
\label{sfig: 1fmk-a:87-134.full_non_del.dca0_14.0_20.simple-gauge.ddPhi_at_opt}
\label{fig: 1fmk-a:87-134.full_non_del.dca0_14.0_20.simple-gauge.ddPhi_at_opt}
\BF{
Correlation between $\Delta \psi_N$ due to single nucleotide nonsynonymous substitutions and
$\psi_N$ of homologous sequences in the SH3\_1 family of the domain, 1FMK-A:87-134.
}
\protect
The left and right figures correspond to the cutoff distance $r_{\script{cutoff}} \sim 8$ and $15.5$ \AA,
respectively.
Each of the black plus or red cross marks corresponds to the mean or the standard deviation 
of $\Delta \psi_N$ due to
all types of single nucleotide nonsynonymous substitutions
over all sites in each of the homologous sequences.
Representatives of unique sequences,
which are at least 20\% different from each other, are employed;
the number of the representatives is almost equal to $M_{\script{eff}}$ in \Table{\ref{\TBL: Proteins_studied}}.
The solid lines show the regression lines for the mean and the standard deviation of $\Delta \psi_N$.
% End of figures_mean_and_sd_legends.tex
}%  FigureLegends
}
\end{figure*}

\CLEARPAGE
\begin{figure*}[h!]
\FigureInLegends{
\noindent
\hspace*{1em} (a) $r_{\script{cutoff}} \sim 8$ \AA\  \hspace*{14em} (b) $r_{\script{cutoff}} \sim 15.5$ \AA\ 

\centerline{
\includegraphics*[width=82mm,angle=0]{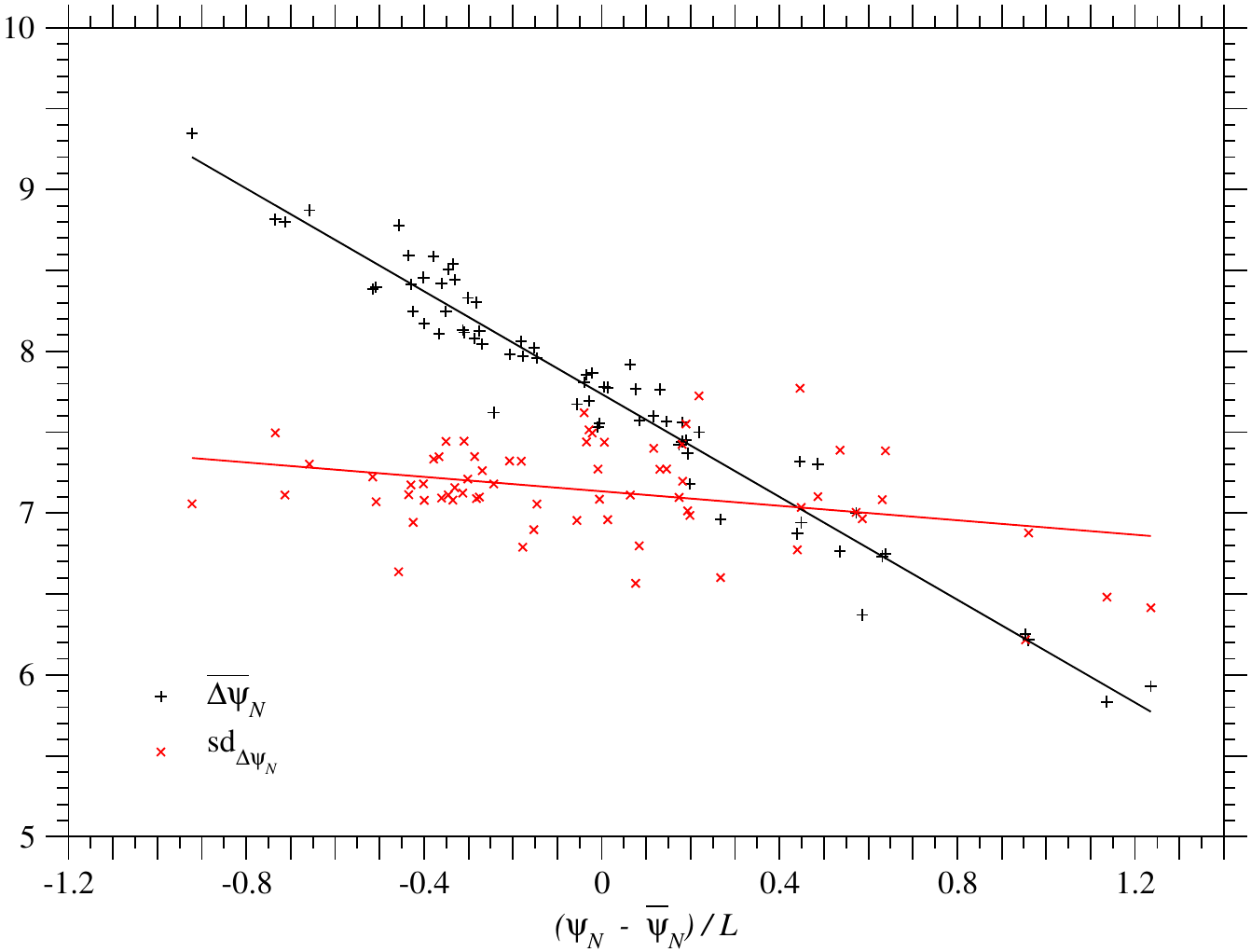}
\includegraphics*[width=82mm,angle=0]{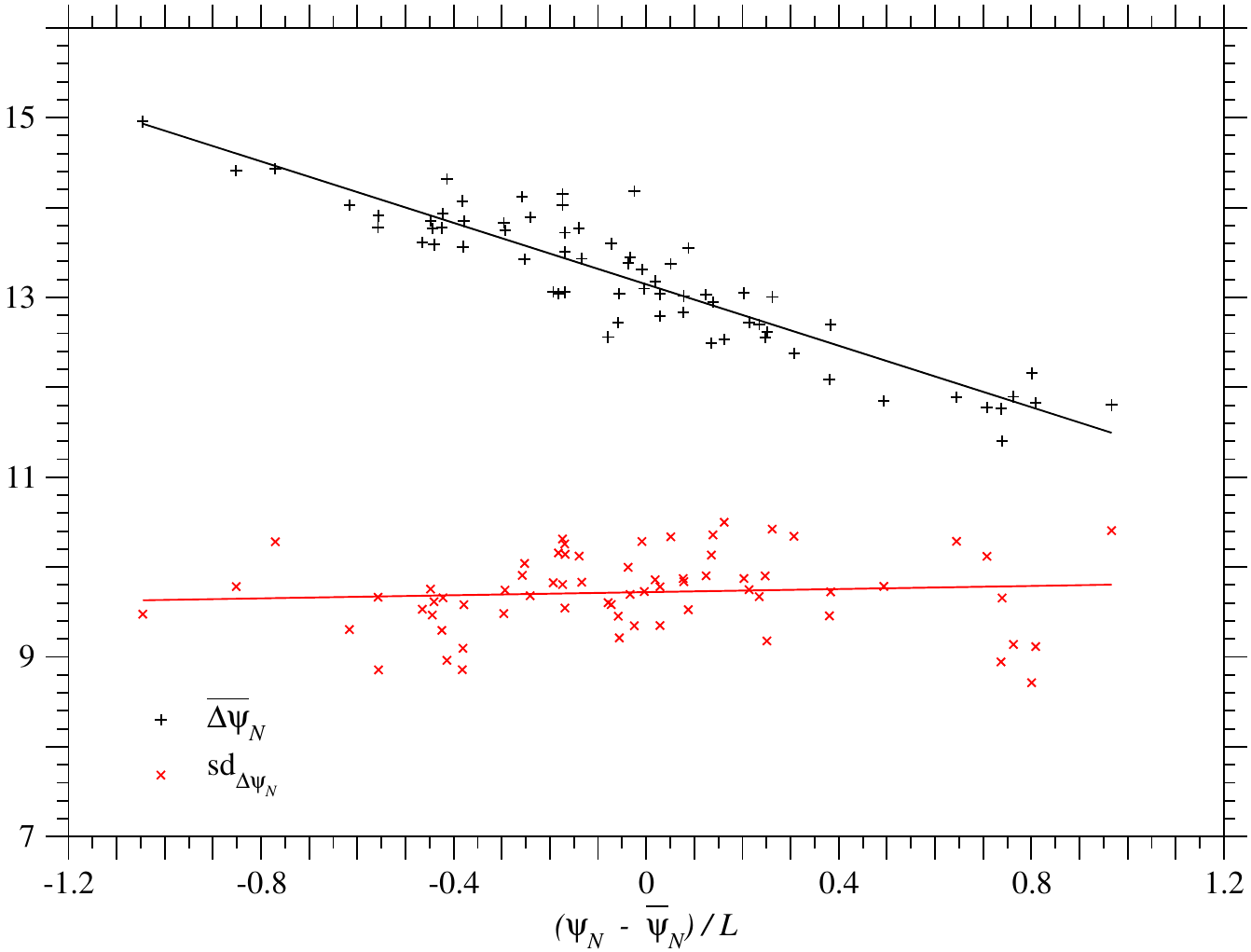}
}
}%  FigureInLegends
\vspace*{1em}
\caption{
\FigureLegends{
\label{sfig: 2abd-a:2-81.full_non_del.dca0_22.0_20.simple-gauge.ddPhi_at_opt}
\label{fig: 2abd-a:2-81.full_non_del.dca0_22.0_20.simple-gauge.ddPhi_at_opt}
\BF{
Correlation between $\Delta \psi_N$ due to single nucleotide nonsynonymous substitutions and
$\psi_N$ of homologous sequences in the ACBP family of the domain, 2ABD-A:2-81.
}
\protect
The left and right figures correspond to the cutoff distance $r_{\script{cutoff}} \sim 8$ and $15.5$ \AA,
respectively.
Each of the black plus or red cross marks corresponds to the mean or the standard deviation 
of $\Delta \psi_N$ due to
all types of single nucleotide nonsynonymous substitutions
over all sites in each of the homologous sequences.
Representatives of unique sequences,
which are at least 20\% different from each other, are employed;
the number of the representatives is almost equal to $M_{\script{eff}}$ in \Table{\ref{\TBL: Proteins_studied}}.
The solid lines show the regression lines for the mean and the standard deviation of $\Delta \psi_N$.
% End of figures_mean_and_sd_legends.tex
}%  FigureLegends
}
\end{figure*}

}%  SUPPLEMENT

\FigA{

\CLEARPAGE
 
\begin{figure*}[h!]
\FigureInLegends{
\noindent
\SUPPLEMENT{
\hspace*{1em} (a) $r_{\script{cutoff}} \sim 8$ \AA\  \hspace*{14em} (b) $r_{\script{cutoff}} \sim 15.5$ \AA\ 

}%  SUPPLEMENT
\TEXT{
\centerline{
{\small{$r_{\script{cutoff}} \sim 8$ \AA\ }}
}
}%  TEXT
\SUPPLEMENT{
\centerline{
\includegraphics*[width=82mm,angle=0]{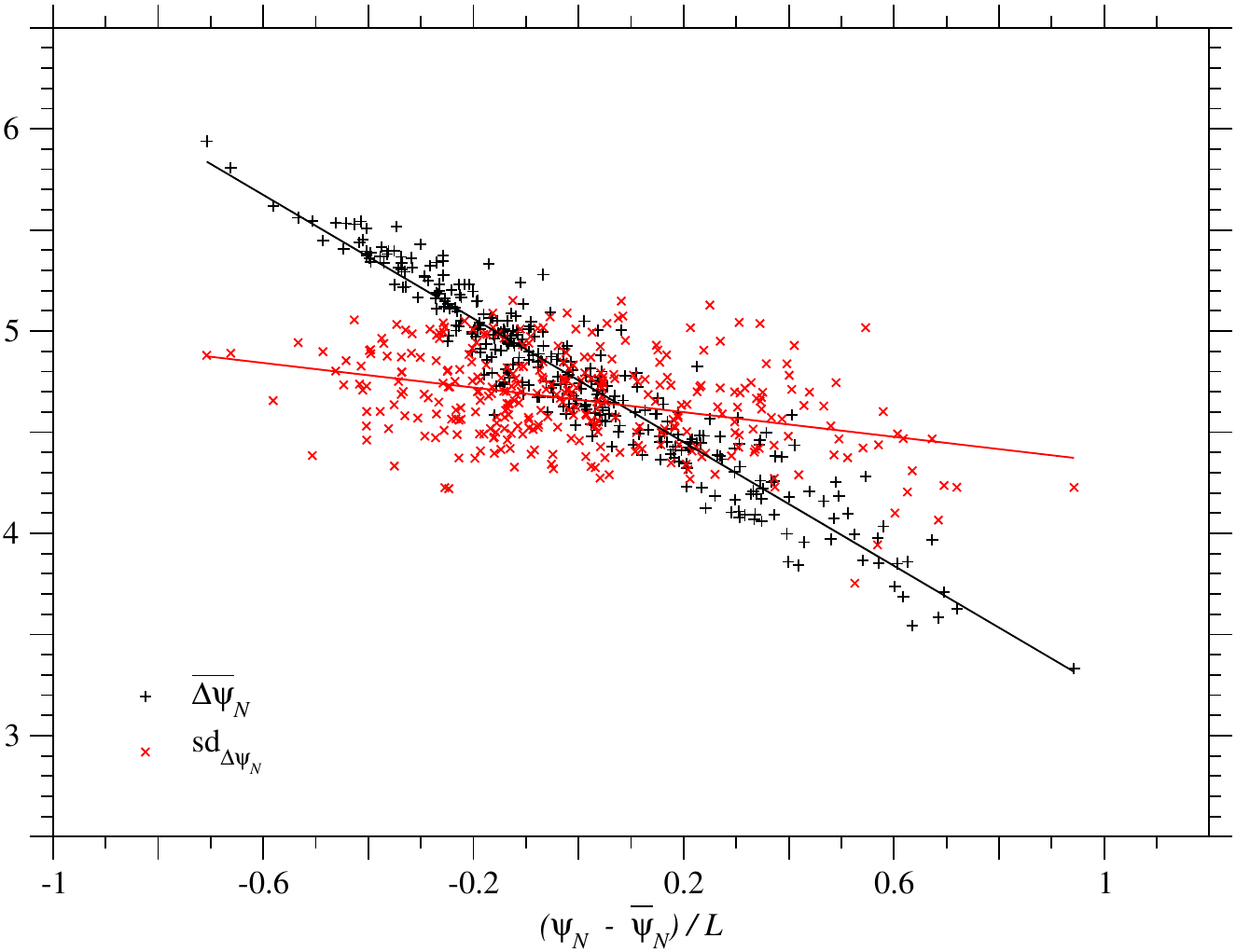}
\includegraphics*[width=82mm,angle=0]{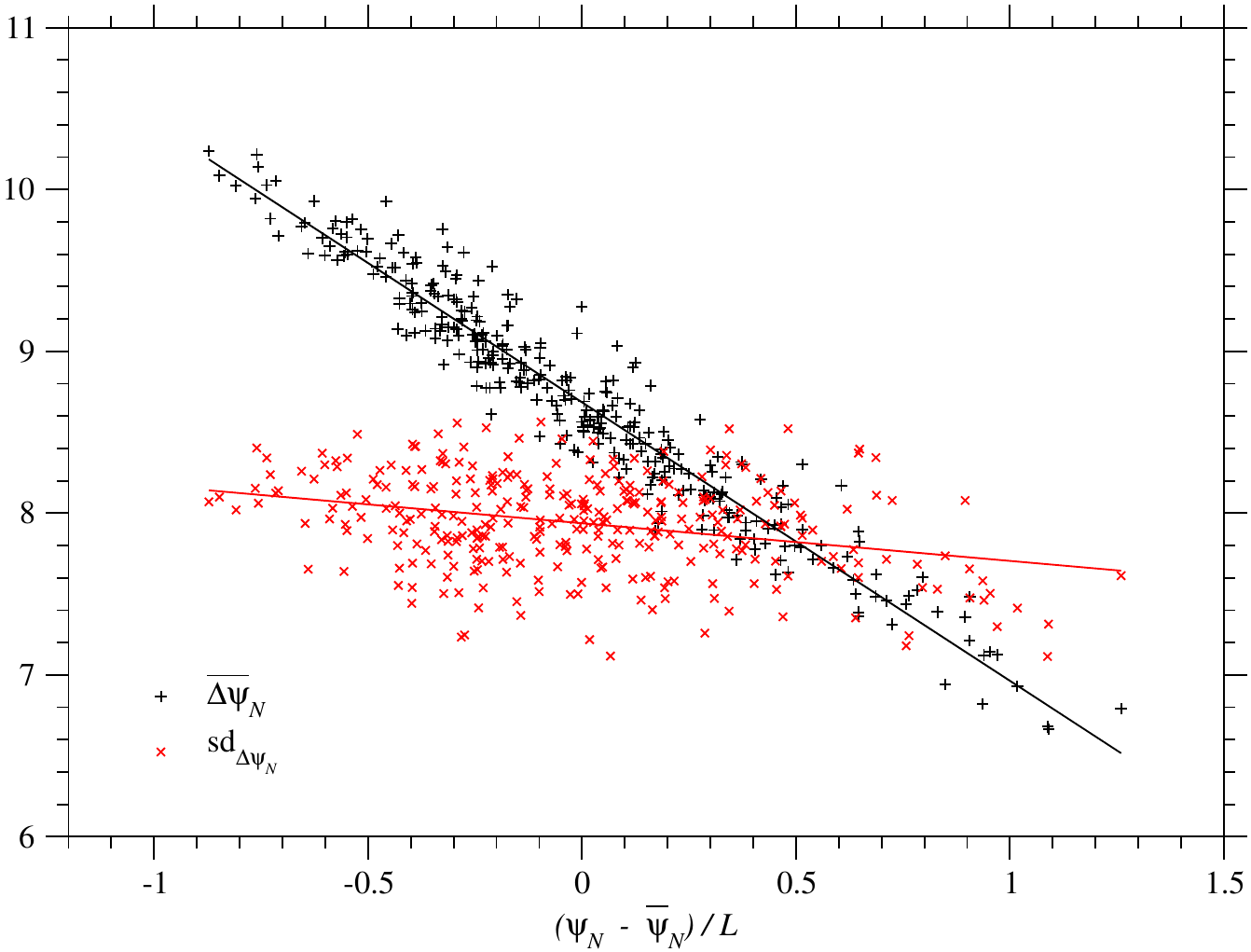}
}
}%  SUPPLEMENT
\TEXT{
\centerline{
\includegraphics*[width=82mm,angle=0]{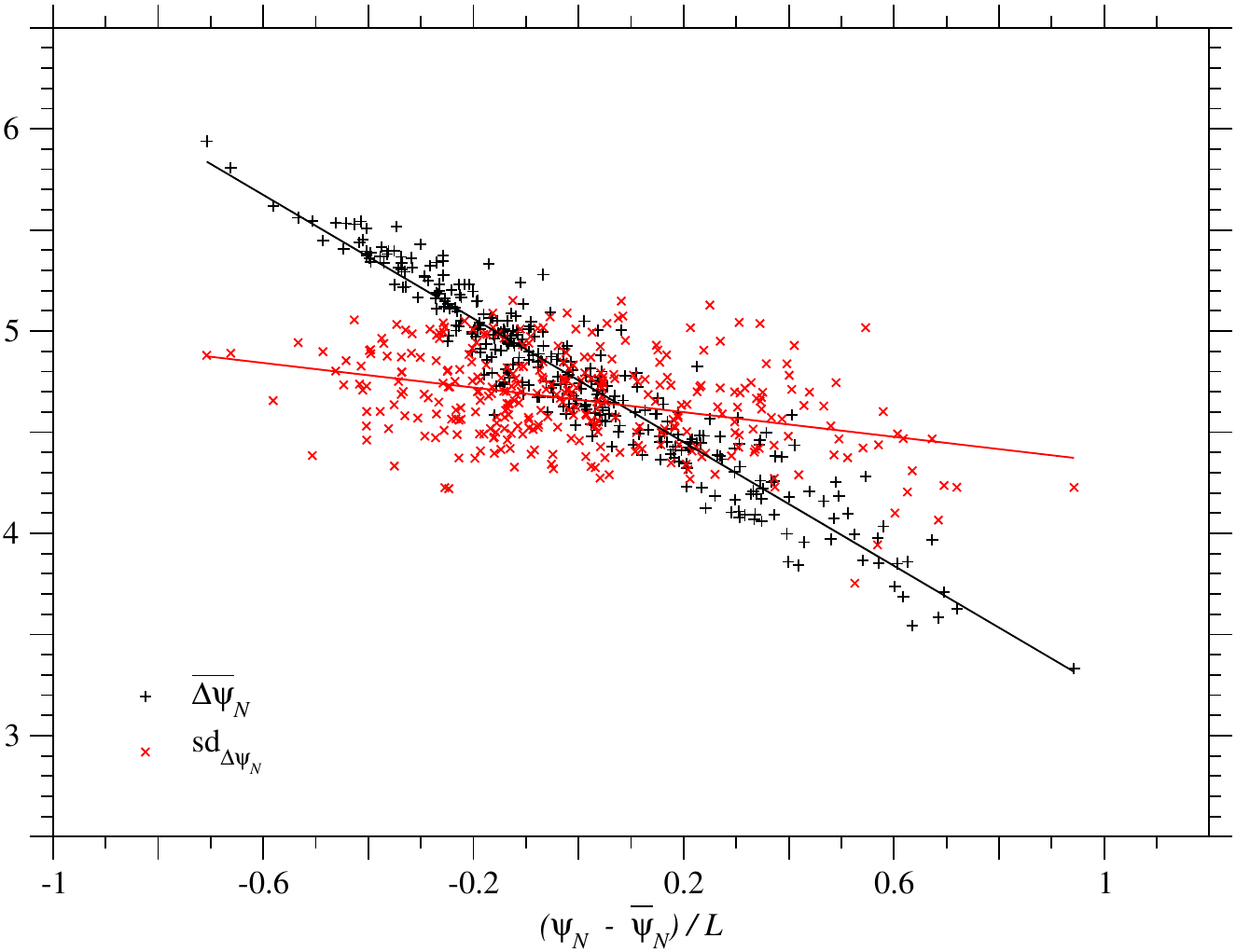}
}
}%  TEXT
}%  FigureInLegends
\vspace*{1em}
\caption{
\FigureLegends{
\SUPPLEMENT{
\label{sfig: 1gm1-a:16-96_dca0_205_simple-gauge_ddPhi_at_opt}
\label{sfig: 1gm1-a:16-96_dca0_33_simple-gauge_ddPhi_at_opt}
}%  SUPPLEMENT
\TEXT{
\label{fig: 1gm1-a:16-96_dca0_205_simple-gauge_ddPhi_at_opt}
}%  TEXT
\BF{
Correlation between $\Delta \psi_N$ due to single nucleotide nonsynonymous substitutions and
$\psi_N$ of homologous sequences in the PDZ domain family.
}
\SUPPLEMENT{
The left and right figures correspond to the cutoff distance $r_{\script{cutoff}} \sim 8$ and $15.5$ \AA,
respectively.
}%  SUPPLEMENT
\TEXT{
This figure corresponds to the cutoff distance $r_{\script{cutoff}} \sim 8$ \AA;
see \Fig{\ref{sfig: 1gm1-a:16-96_dca0_33_simple-gauge_ddPhi_at_opt}} for $r_{\script{cutoff}} \sim 15.5$ \AA.
}%  TEXT
Each of the black plus or red cross marks corresponds to the mean or the standard deviation 
of $\Delta \psi_N$ due to
all types of single nucleotide nonsynonymous substitutions
over all sites in each of the homologous sequences of the PDZ domain family.
Only 335 representatives of unique sequences with no deletions, which are at least 20\% different from each other, are employed;
the number of the representatives is almost equal to $M_{\script{eff}}$ in \Table{\ref{\TBL: Proteins_studied}}.
The solid lines show the regression lines for the mean and the standard deviation of $\Delta \psi_N$.
}%  FigureLegends
}
\end{figure*}
}%  FigA

\SUPPLEMENT{

\CLEARPAGE
\begin{figure*}[h!]
\FigureInLegends{
\noindent
\hspace*{1em} (a) $r_{\script{cutoff}} \sim 8$ \AA\  \hspace*{14em} (b) $r_{\script{cutoff}} \sim 15.5$ \AA\ 

\centerline{
\includegraphics*[width=82mm,angle=0]{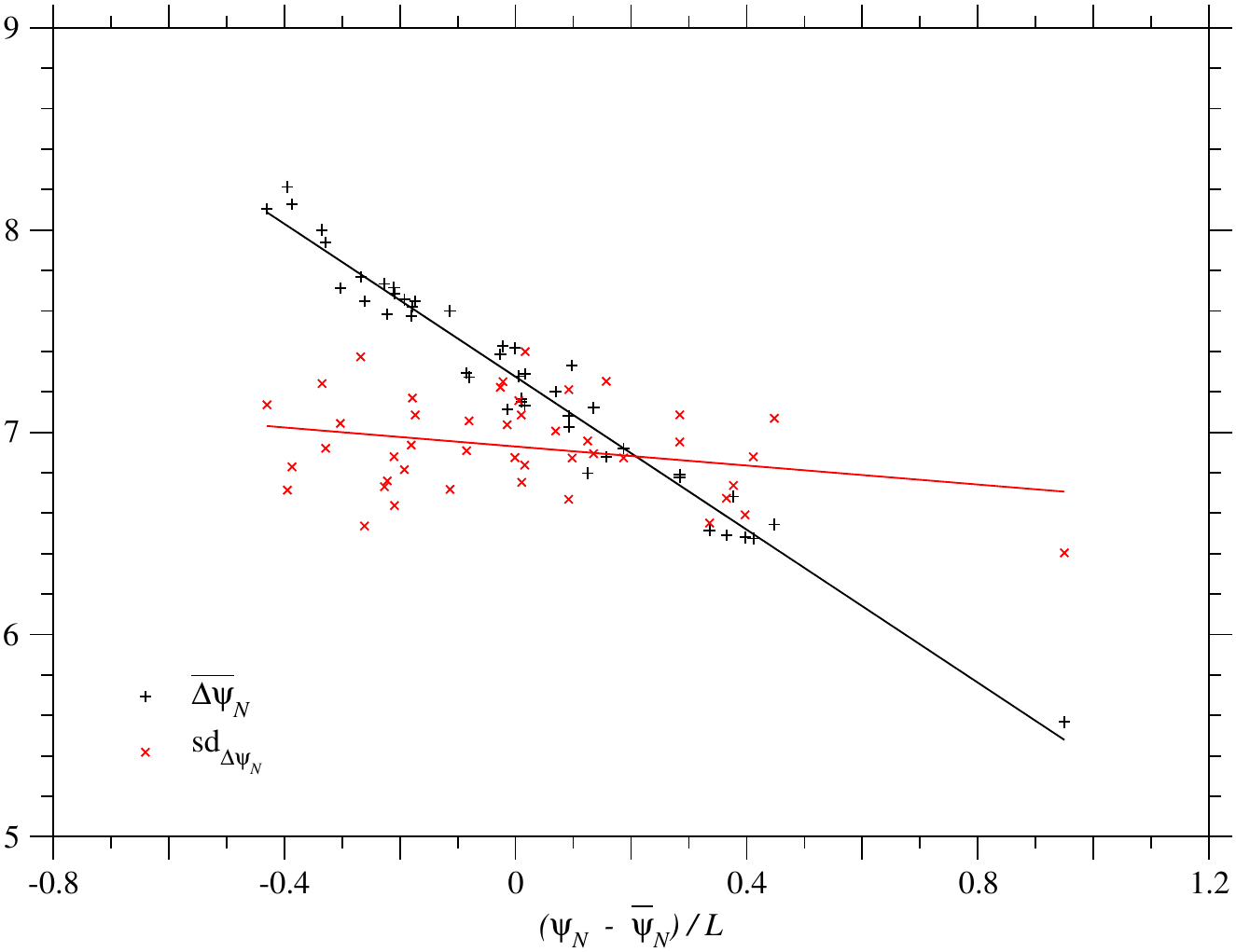}
\includegraphics*[width=82mm,angle=0]{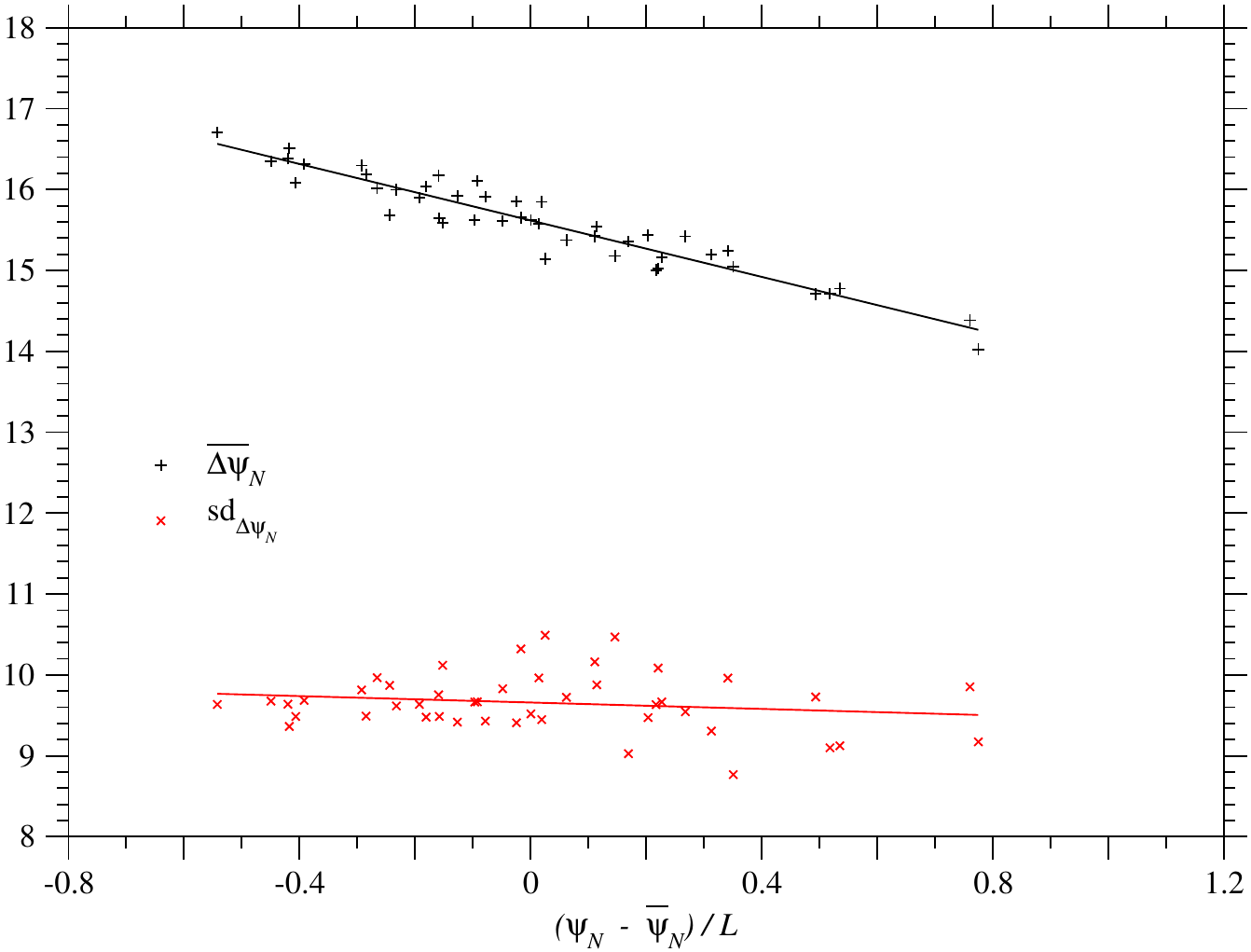}
}
}%  FigureInLegends
\vspace*{1em}
\caption{
\FigureLegends{
\label{sfig: 5azu-a:4-128.full_non_del.dca0_23.0_20.simple-gauge.BD.ddPhi_at_opt}
\label{fig: 5azu-a:4-128.full_non_del.dca0_23.0_20.simple-gauge.BD.ddPhi_at_opt}
\BF{
Correlation between $\Delta \psi_N$ due to single nucleotide nonsynonymous substitutions and
$\psi_N$ of homologous sequences in the Copper-bind family of the domain, 5AZU-B/D:4-128.
}
\protect
The left and right figures correspond to the cutoff distance $r_{\script{cutoff}} \sim 8$ and $15.5$ \AA,
respectively.
Each of the black plus or red cross marks corresponds to the mean or the standard deviation 
of $\Delta \psi_N$ due to
all types of single nucleotide nonsynonymous substitutions
over all sites in each of the homologous sequences.
Representatives of unique sequences,
which are at least 20\% different from each other, are employed;
the number of the representatives is almost equal to $M_{\script{eff}}$ in \Table{\ref{\TBL: Proteins_studied}}.
The solid lines show the regression lines for the mean and the standard deviation of $\Delta \psi_N$.
% End of figures_mean_and_sd_legends.tex
}%  FigureLegends
}
\end{figure*}

}%  SUPPLEMENT

% End of figures_mean_and_sd.tex

\FigB{
\CLEARPAGE

\begin{figure*}[h!]
\FigureInLegends{
\noindent
\SUPPLEMENT{
(a) $r_{\script{cutoff}} \sim 8$ \AA\  \hspace*{13em} (b) $r_{\script{cutoff}} \sim 15.5$ \AA\ 

}%  SUPPLEMENT
\TEXT{
\centerline{
{\small{$r_{\script{cutoff}} \sim 8$ \AA\ }}
}
}%  TEXT
\SUPPLEMENT{
\centerline{
\includegraphics*[height=75mm,angle=0]{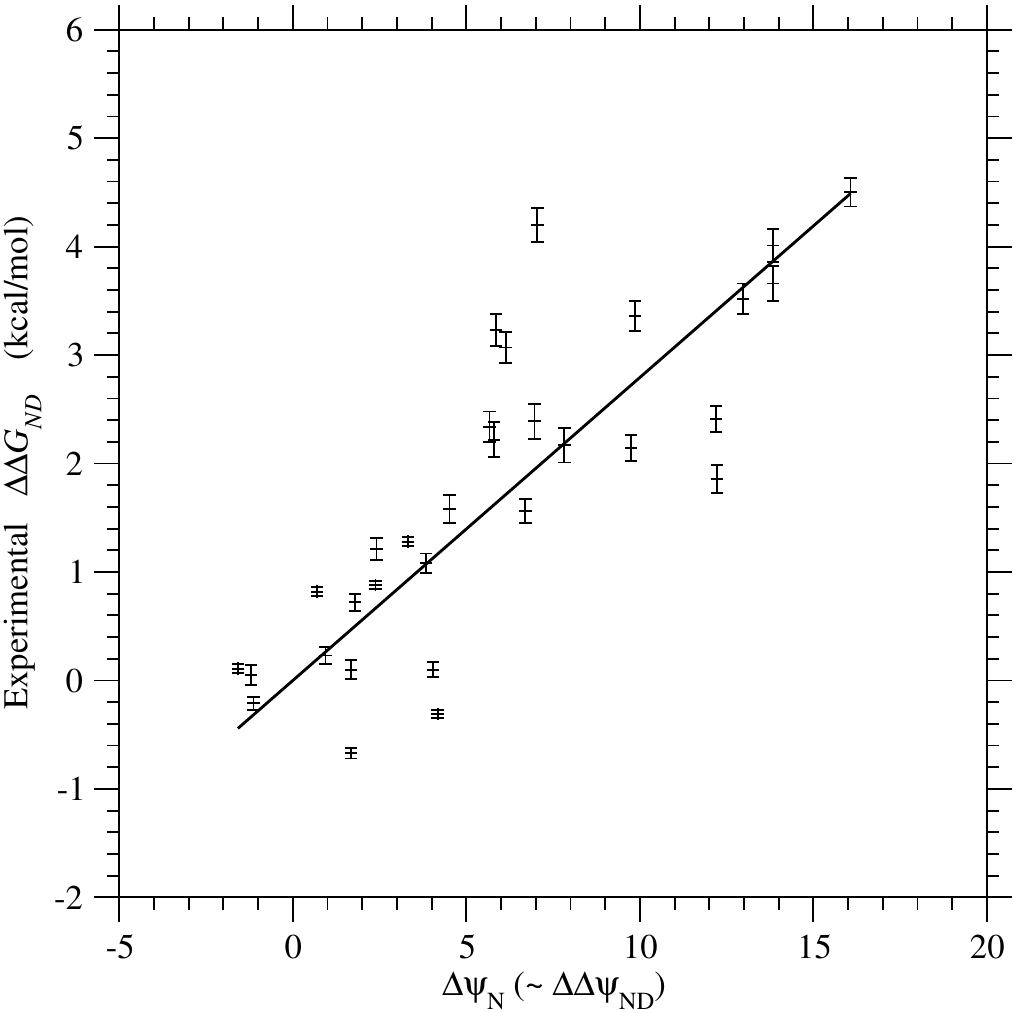}
\includegraphics*[height=75mm,angle=0]{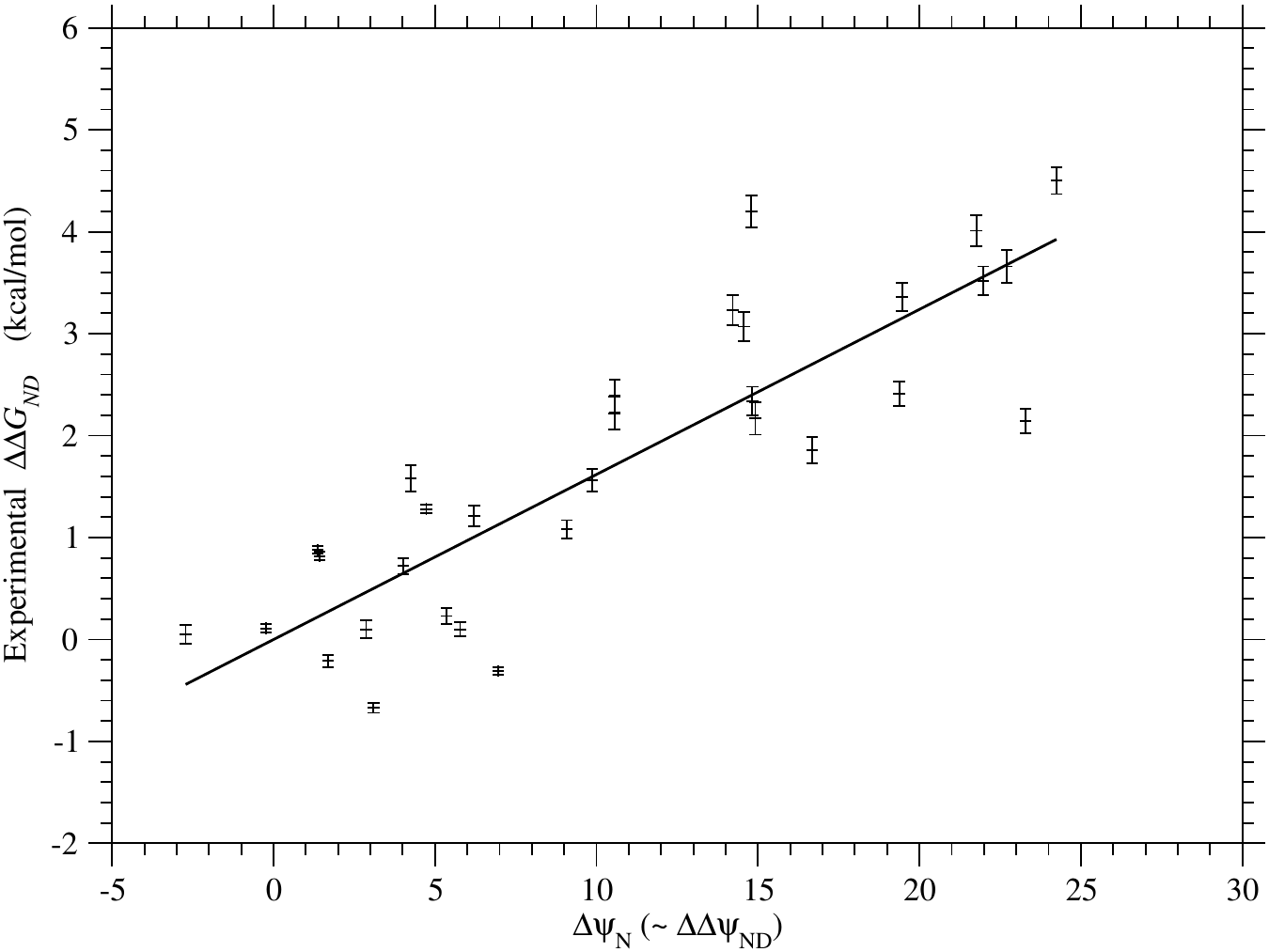}
}
}%  SUPPLEMENT
\TEXT{
\centerline{
\includegraphics*[height=75mm,angle=0]{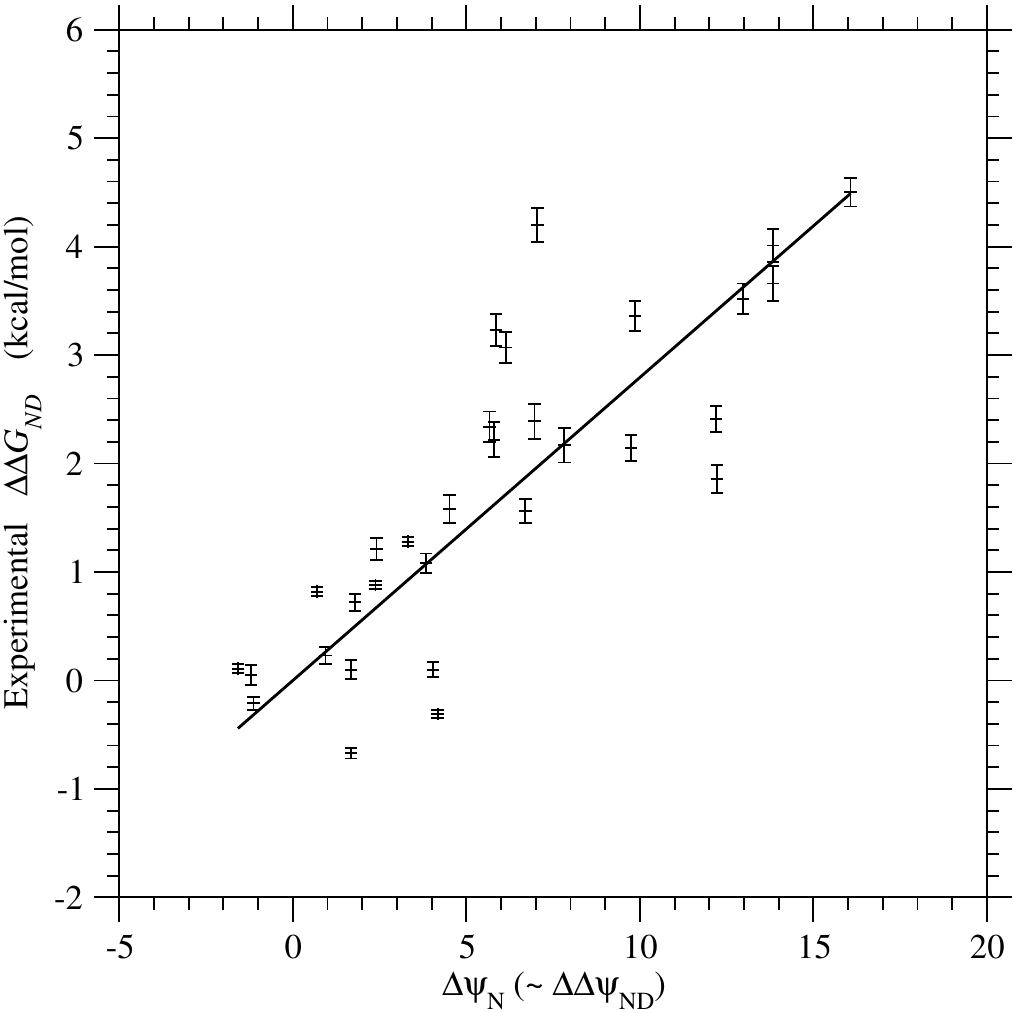}
}
}%  TEXT
}%  FigureInLegends
\vspace*{1em}
\caption{
\FigureLegends{
\SUPPLEMENT{
\label{sfig: 1gm1-a:16-96_dca0_205_simple-gauge_ddG-dPhi_at_opt}
\label{sfig: 1gm1-a:16-96_dca0_33_simple-gauge_ddG-dPhi_at_opt}
}%  SUPPLEMENT
\TEXT{
\label{fig: 1gm1-a:16-96_dca0_205_simple-gauge_ddG-dPhi_at_opt}
}%  TEXT
\BF{
Regression of the experimental values\CITE{GGCJVTVB:07} of folding free energy changes ($\Delta\Delta G_{ND}$) 
due to single amino acid substitutions on
$\Delta \psi_N (\simeq \Delta\Delta \psi_{ND})$ for the same types of substitutions in the PDZ domain. 
}
\SUPPLEMENT{
The left and right figures correspond to the cutoff distance $r_{\script{cutoff}} \sim 8$ and $15.5$ \AA, respectively.
The solid lines show the least-squares regression lines through the origin with the slopes, $0.279$ kcal/mol for $r_{\script{cutoff}} \sim 8$ \AA\ 
and $0.162$ kcal/mol for $r_{\script{cutoff}} \sim 15.5$ \AA, which are the estimates of $k_B T_s$.
The reflective correlation coefficients for them are equal to $0.93$ and $0.94$, respectively.
}%  SUPPLEMENT
\TEXT{
This figure corresponds to the cutoff distance $r_{\script{cutoff}} \sim 8$ \AA;
see \Fig{\ref{sfig: 1gm1-a:16-96_dca0_33_simple-gauge_ddG-dPhi_at_opt}} for $r_{\script{cutoff}} \sim 15.5$ \AA.
The solid line shows the least-squares regression line through the origin with the slope, $0.279$ kcal/mol, which is the estimates of $k_B T_s$.
The reflective correlation coefficient is equal to $0.93$.
}%  TEXT
The free energies are in kcal/mol units.
}%  FigureLegends
}
\end{figure*}
}%  FigB

\SUPPLEMENT{
\FigC{

\CLEARPAGE
 
\begin{figure*}[h!]
\FigureInLegends{
\centerline{
\includegraphics*[width=82mm,angle=0]{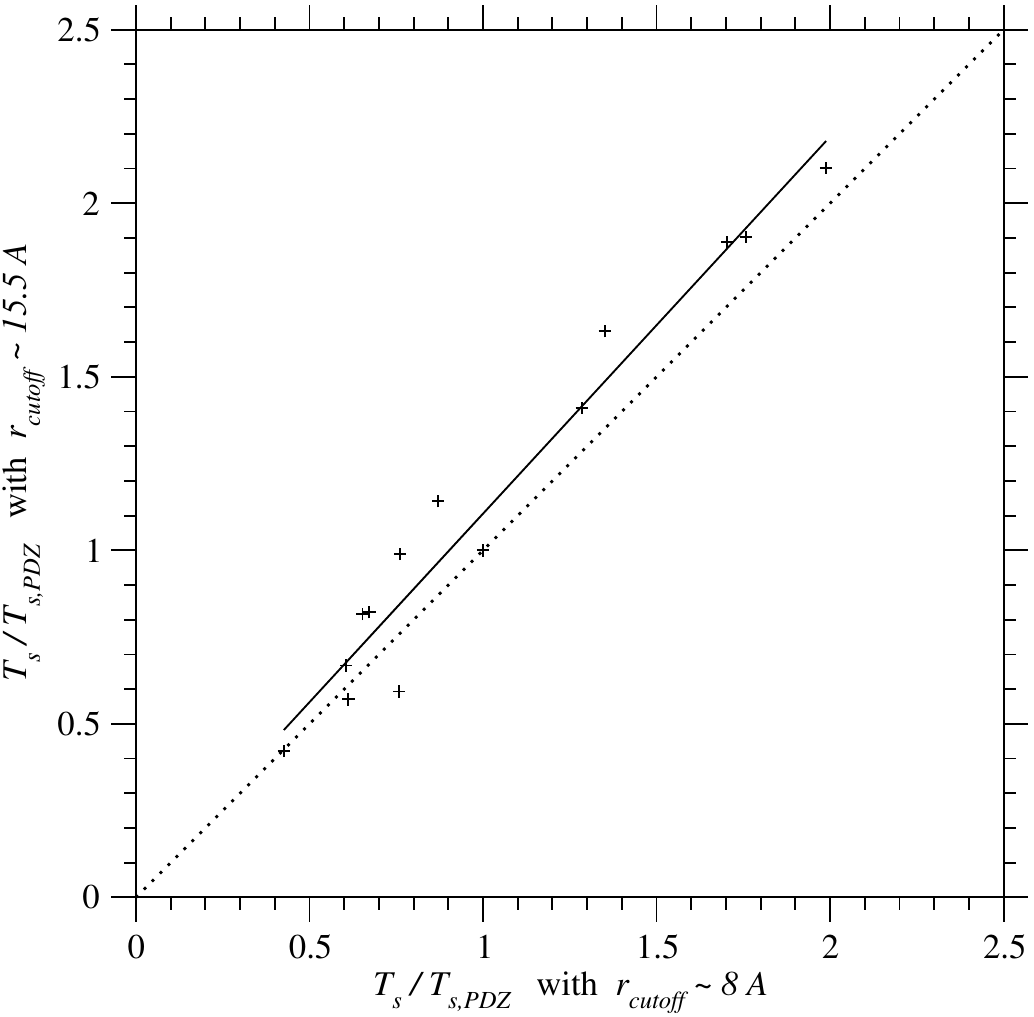}
\includegraphics*[width=82mm,angle=0]{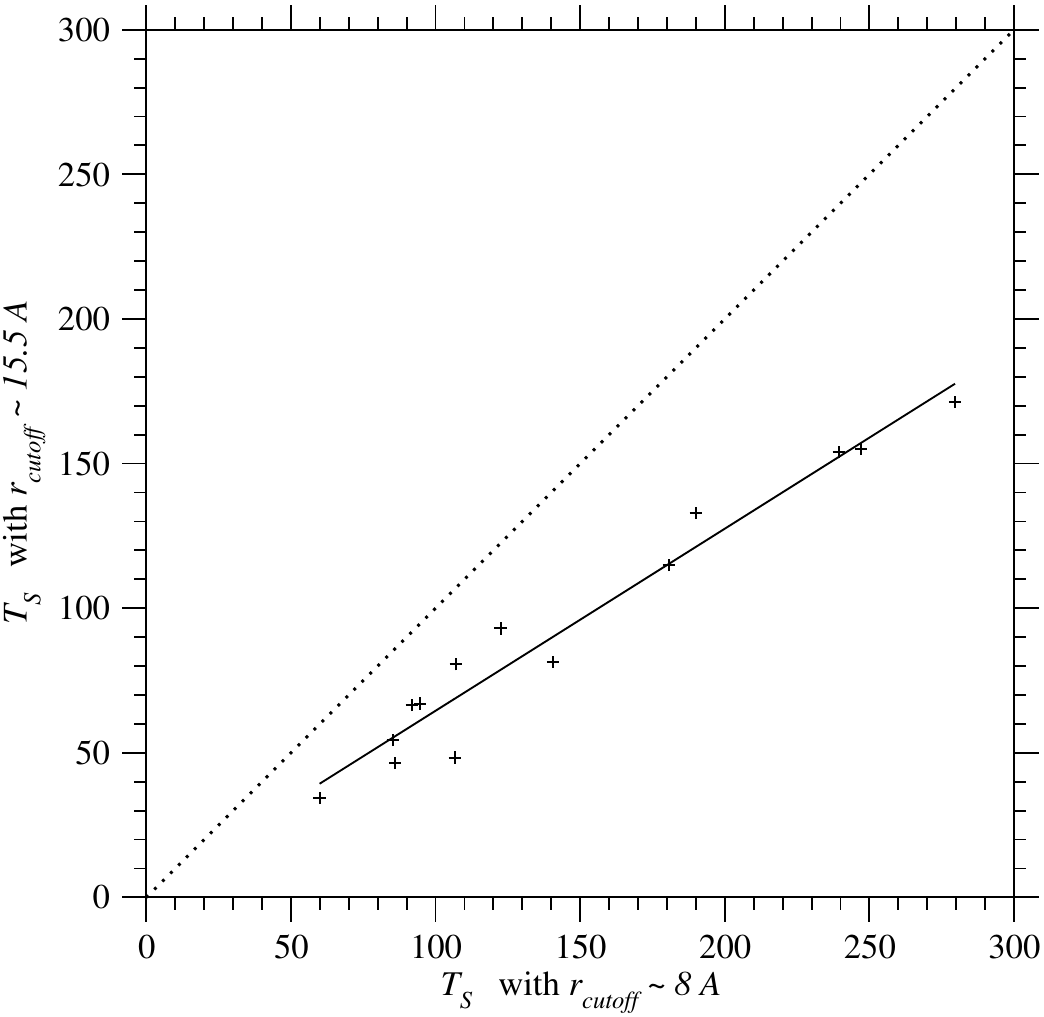}
}
}%  FigureInLegends
\vspace*{1em}
\caption{
\SUPPLEMENT{
\label{sfig: Ts_relative_to_Tpdz_8_vs_16A}
\label{sfig: Ts_8_vs_16A}
\label{fig: Ts_relative_to_Tpdz_8_vs_16A}
\label{fig: Ts_8_vs_16A}
}%  SUPPLEMENT
\TEXT{
\label{fig: Ts_relative_to_Tpdz_8_vs_16A}
\label{fig: Ts_8_vs_16A}
}%  TEXT
\FigureLegends{
\BF{Comparison of selective temperatures ($T_s$)
estimated with different cutoff distances by the present method. 
}
The abscissa and ordinate correspond to the cases of $r_{\script{cutoff}} \sim 8$ and $15.5$ \AA,
respectively.
The $T_s$ is in $^\circ$K units.
The solid lines show the regression lines,
$(T_s/T_{s,PDZ})_{15.5A} = 1.09 (T_s/T_{s,PDZ})_{8A}  + 0.02$ and
$(T_s)_{15.5A} = 0.630 (T_s)_{8A}  + 1.57$.
The correlation coefficients are equal to 0.98 for both.
}%  FigureLegends
}
\end{figure*}

}%  FigC
}%  SUPPLEMENT

\SUPPLEMENT{

\CLEARPAGE
 
\begin{figure*}[h!]
\FigureInLegends{
\centerline{
\includegraphics*[width=82mm,angle=0]{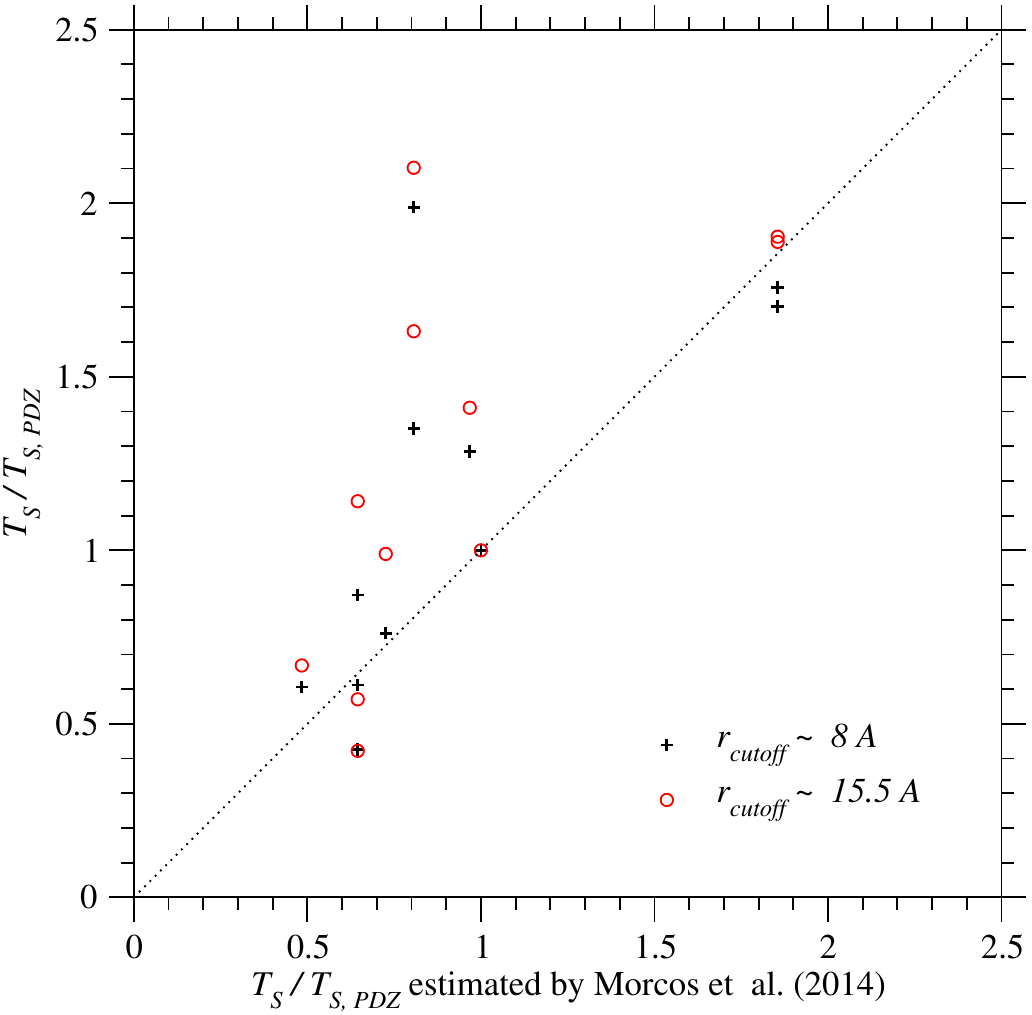}
\includegraphics*[width=82mm,angle=0]{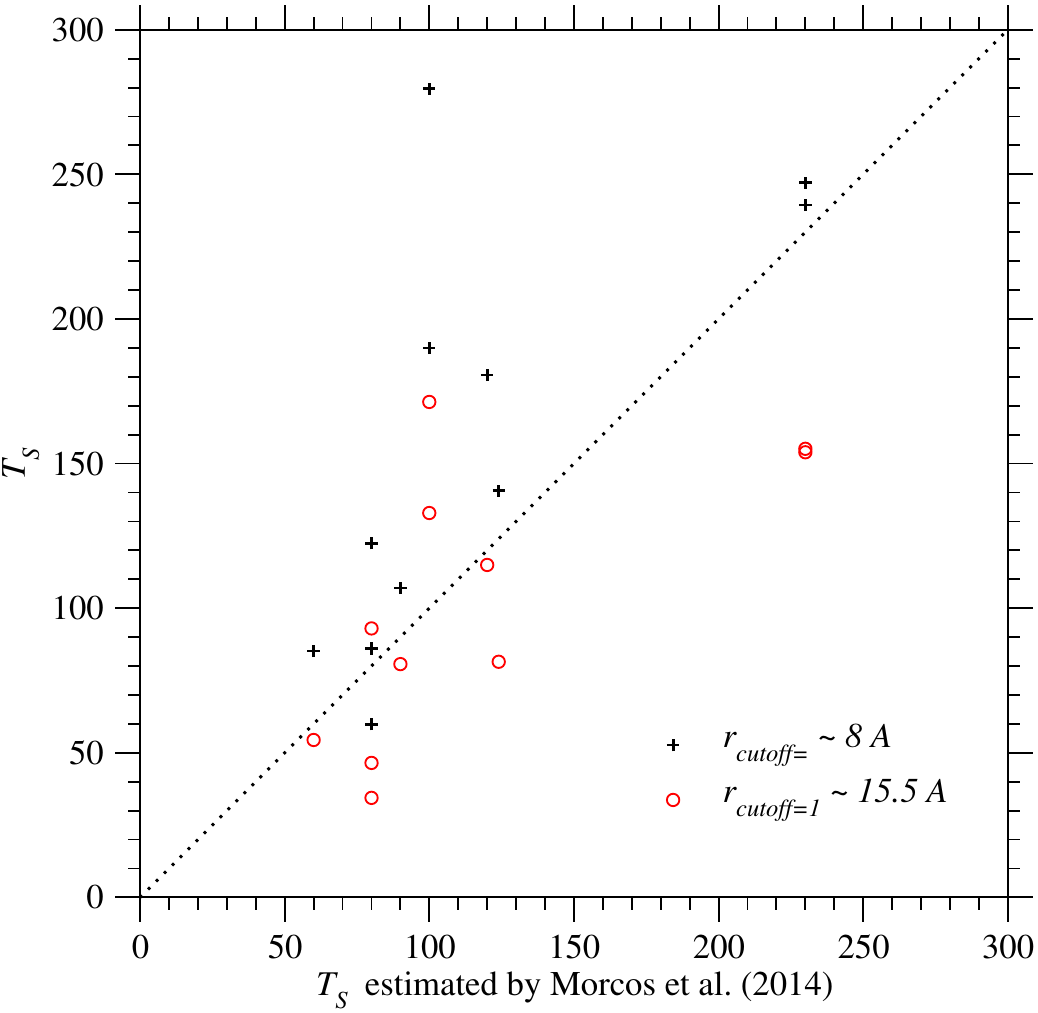}
}
}%  FigureInLegends
\vspace*{1em}
\caption{
\label{sfig: Ts_relative_to_Tpdz_Wolynes_vs_8_and_16A}
\label{fig: Ts_relative_to_Tpdz_Wolynes_vs_8_and_16A}
\label{sfig: Ts_Wolynes_vs_8_and_16A}
\label{fig: Ts_Wolynes_vs_8_and_16A}
\FigureLegends{
\BF{Selective temperatures ($T_s$) 
estimated by the present method are plotted against those estimated by Morcos et al.\CITE{MSCOW:14};
their estimated values of $T_s$ tend to fall between the upper ($r_{\script{cutoff}} \sim 8$) 
and lower ($r_{\script{cutoff}} \sim 15.5$ \AA) estimates of $T_s$. 
Plus and open circle marks correspond to
the cases of $r_{\script{cutoff}} \sim 8$ and $15.5$ \AA,
respectively.
}
}%  FigureLegends
}
\end{figure*}

}%  SUPPLEMENT

\SUPPLEMENT{

\CLEARPAGE
 
\begin{figure*}[h!]
\FigureInLegends{
\centerline{
\includegraphics*[width=82mm,angle=0]{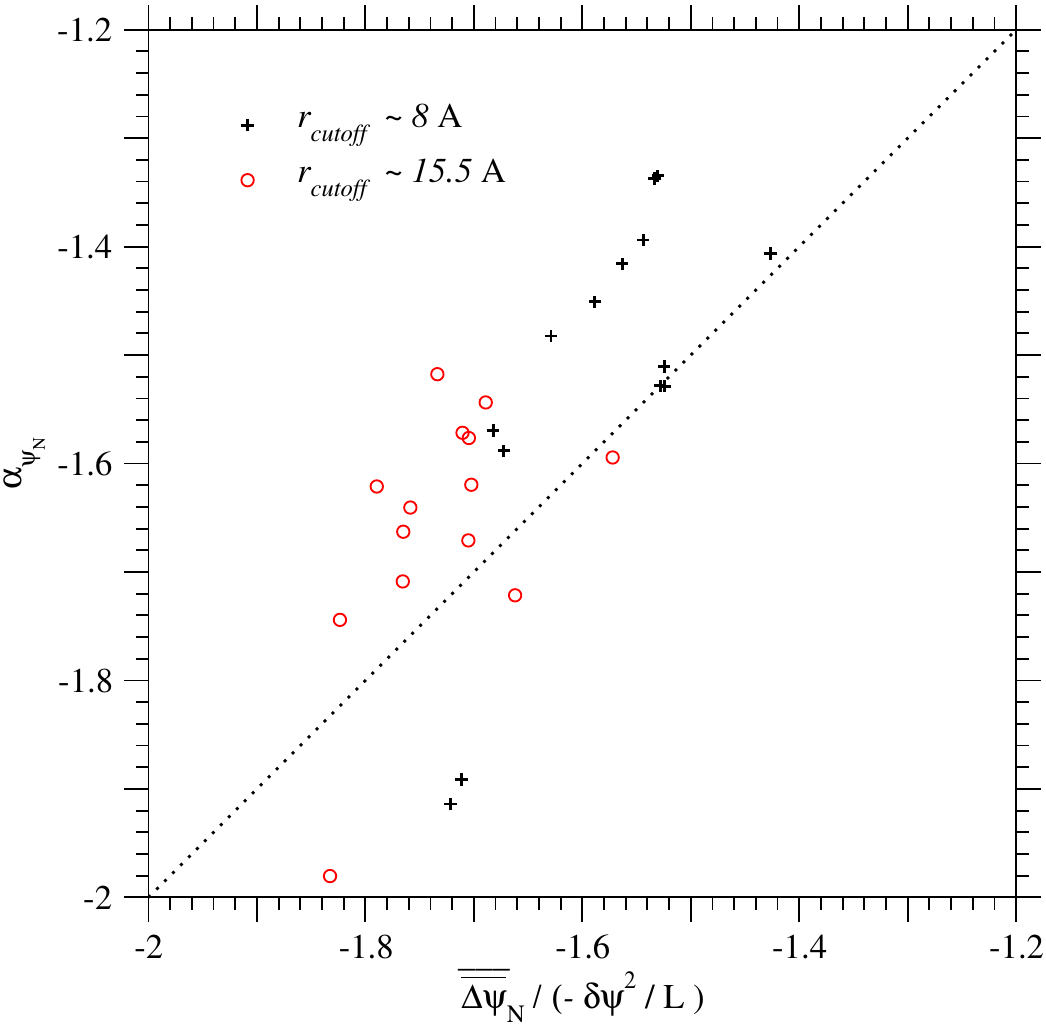}
}
}%  FigureInLegends
\vspace*{1em}
\caption{
\label{sfig: mean_over_dpsi2_vs_slope}
\label{fig: mean_over_dpsi2_vs_slope}
\FigureLegends{
\BF{
Comparison of $\alpha_{\psi_N}$,
which is the regression coefficient of $\overline{\Delta \psi_N}$ on $\psi_N / L$,
with $\overline{\overline{\Delta \psi_N}} / ( -{\delta \psi}^2 / L) $
for each protein family.
}
Plus and open circle marks correspond to
the cases of $r_{\script{cutoff}} \sim 8$ and $15.5$ \AA,
respectively.
}%  FigureLegends
}
\end{figure*}

}%  SUPPLEMENT

\FigD{

\CLEARPAGE
 
\begin{figure*}[h!]
\FigureInLegends{
\SUPPLEMENT{
\centerline{
\includegraphics*[width=82mm,angle=0]{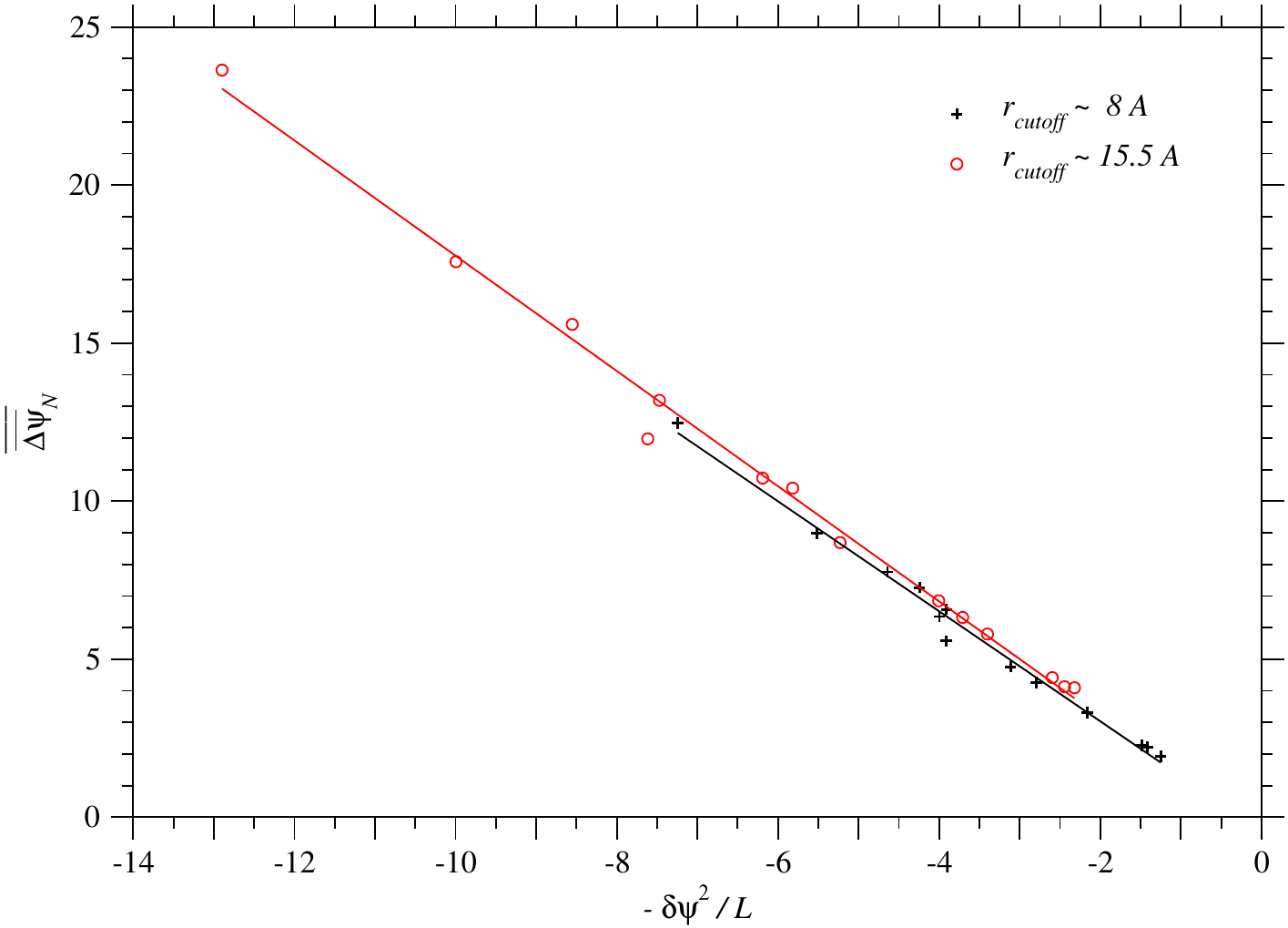}
}
}%  SUPPLEMENT
\TEXT{
\centerline{
\includegraphics*[width=82mm,angle=0]{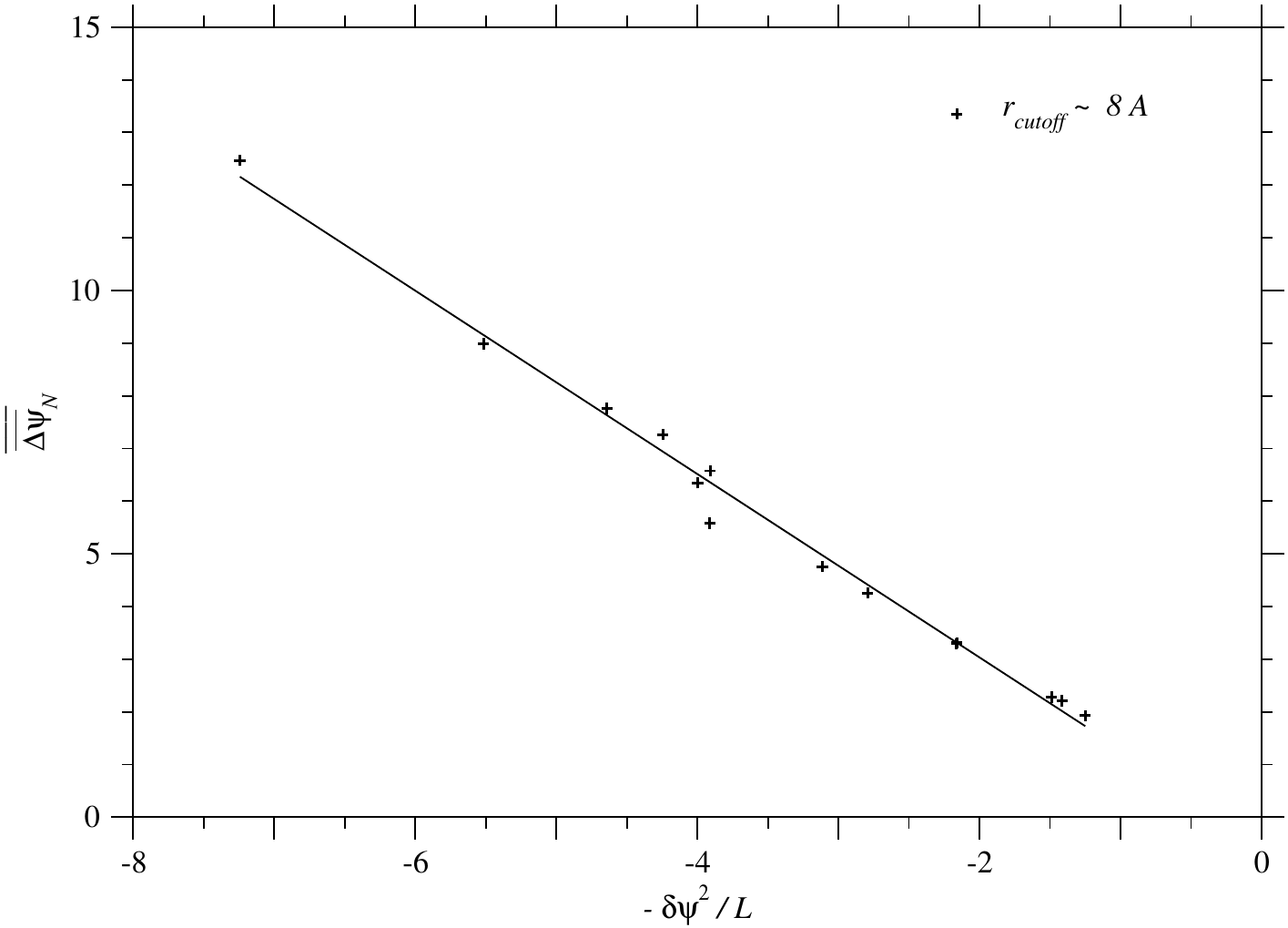}
}
}%  TEXT
}%  FigureInLegends
\vspace*{1em}
\caption{
\SUPPLEMENT{
\label{sfig: dpsi2_over_L_vs_ddPsi}
\label{sfig: dpsi2_over_L_vs_dPsi}
}%  SUPPLEMENT
\TEXT{
\label{fig: dpsi2_over_L_vs_ddPsi}
\label{fig: dpsi2_over_L_vs_dPsi}
}%  TEXT
\FigureLegends{
\BF{
Dependence of the average of
$\overline{\Delta \psi_N}$ 
due to single nucleotide nonsynonymous substitutions 
over homologous sequences
on $-{\delta \psi}^2 / L$ across protein families.
}
\SUPPLEMENT{
Plus and open circle marks indicate the values for
each protein family 
in the cases of $r_{\script{cutoff}} \sim 8$ and $15.5$ \AA,
respectively.
}%  SUPPLEMENT
\TEXT{
Plus marks indicate the value for
each protein family 
in the case of $r_{\script{cutoff}} \sim 8$ \AA.
The correlation coefficient is
equal to 0.995,
and the regression line is
$\overline{\overline{ \Delta \psi_N(\sigma^N_{j \neq i}, \sigma^N_i\rightarrow \sigma_i) } }
= - 1.74 (- {\delta \psi}^2 / L) - 0.445$.
See \Fig{\ref{sfig: dpsi2_over_L_vs_ddPsi}} for $r_{\script{cutoff}} \sim 15.5$ \AA.
}%  TEXT
\SUPPLEMENT{
In the case of the cutoff distance 8 \AA, 
the correlation coefficient is
equal to 0.995,
and the regression line is
$\overline{\overline{ \Delta \psi_N(\sigma^N_{j \neq i}, \sigma^N_i\rightarrow \sigma_i) } }
= - 1.74 (- {\delta \psi}^2 / L) - 0.445$.
In the case of $r_{\script{cutoff}} \sim 15.5$ \AA, 
the correlation coefficient is
equal to 0.996,
and the regression line is
$\overline{\overline{ \Delta \psi_N(\sigma^N_{j \neq i}, \sigma^N_i\rightarrow \sigma_i) } }
= - 1.82 (- {\delta \psi}^2 / L) - 0.466$.
}%  SUPPLEMENT
}%  FigureLegends
}
\end{figure*}

}%  FigD

\FigE{

\CLEARPAGE
 
\begin{figure*}[h!]
\FigureInLegends{
\SUPPLEMENT{
\centerline{
\includegraphics*[width=82mm,angle=0]{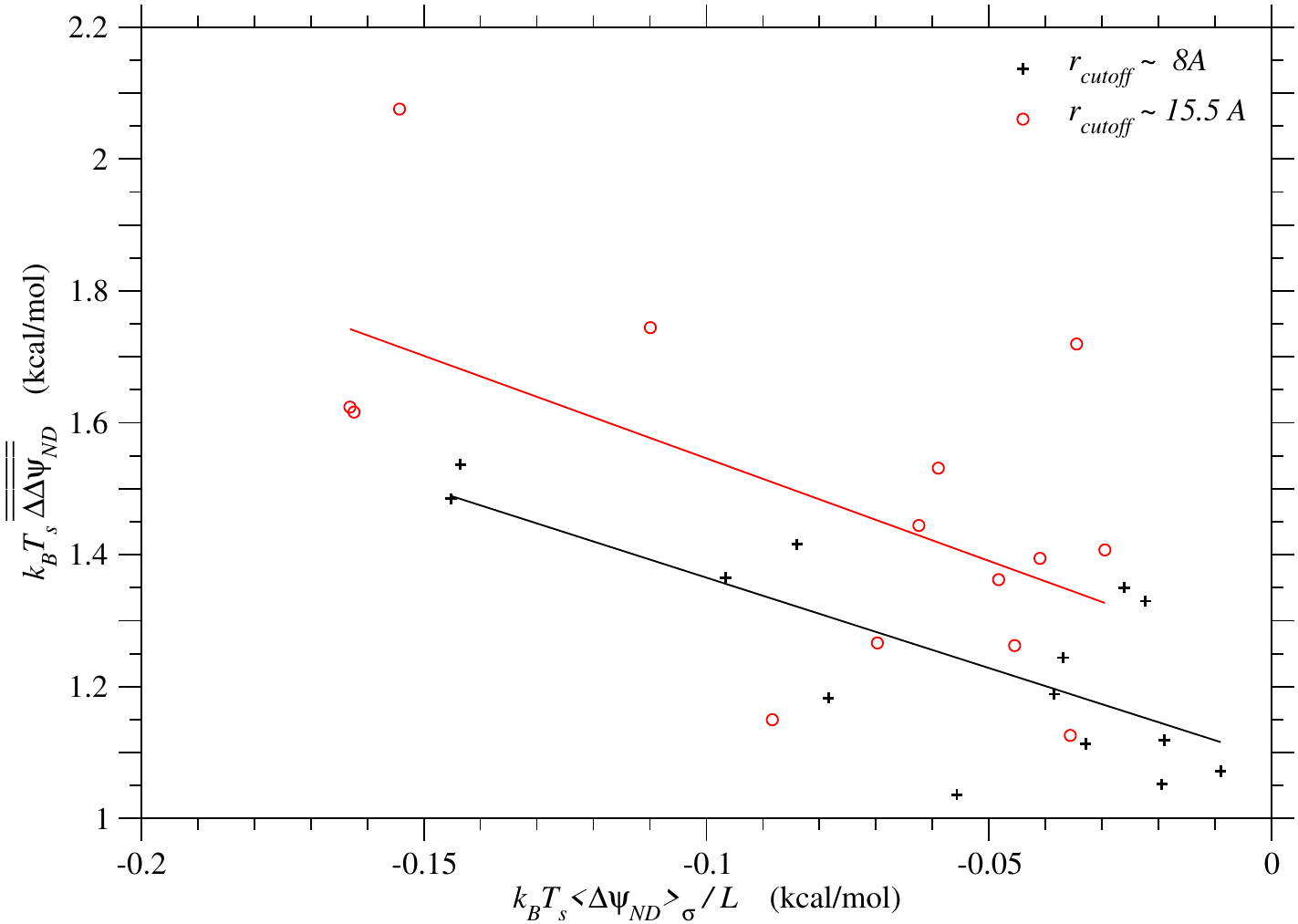}
}
}%  SUPPLEMENT
\TEXT{
\centerline{
\includegraphics*[width=82mm,angle=0]{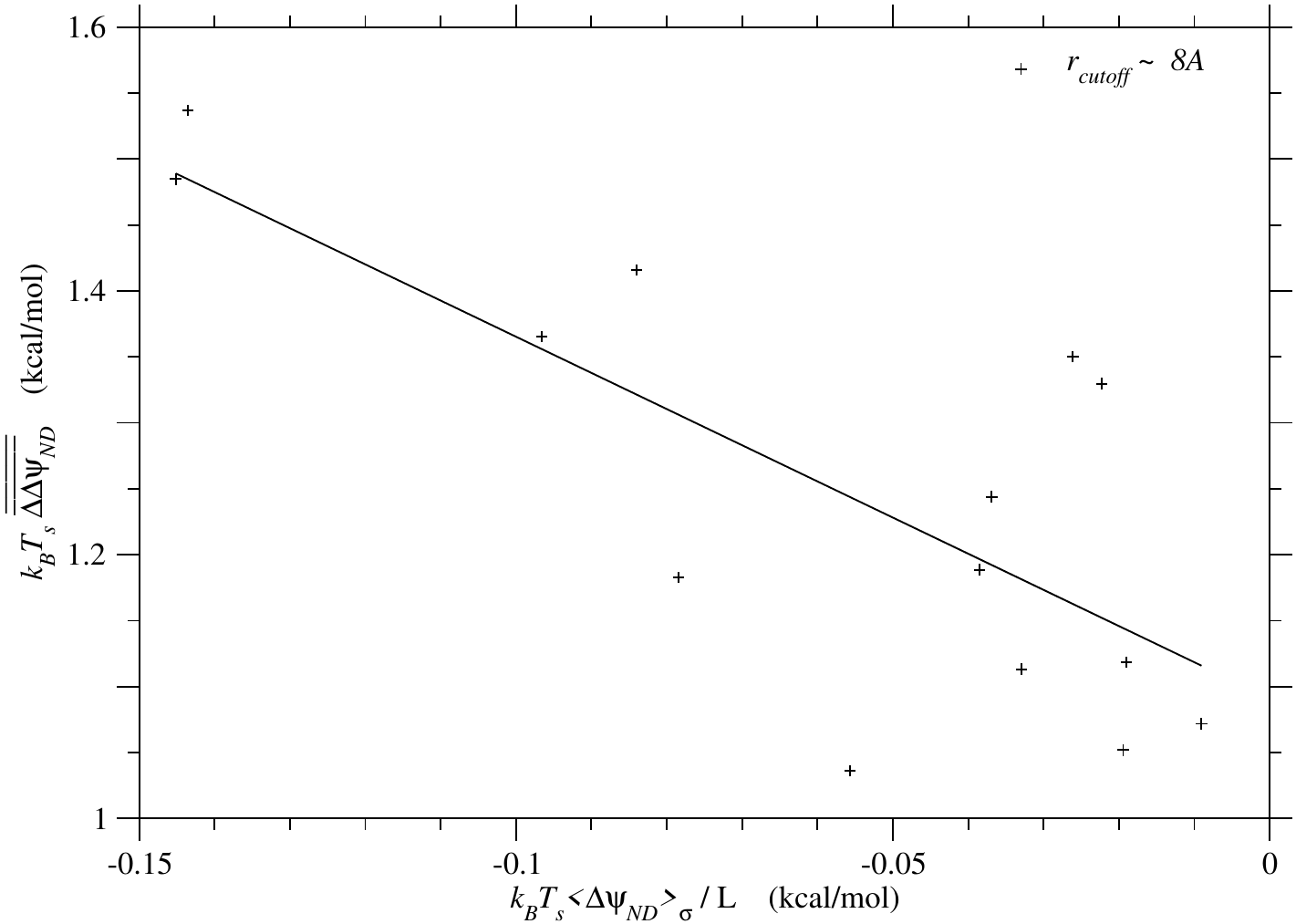}
}
}%  TEXT
}%  FigureInLegends
\vspace*{1em}
\caption{
\SUPPLEMENT{
\label{sfig: dG_over_L_vs_ddG}
}%  SUPPLEMENT
\TEXT{
\label{fig: dG_over_L_vs_ddG}
}%  TEXT
\FigureLegends{
\BF{
The sample average of folding free energy change, 
$\overline{\overline{\Delta\Delta G_{ND}}} \simeq k_B T_s \overline{\overline{\Delta\Delta \psi_{ND}}}$,
is plotted against the ensemble average of folding free energy per residue, 
$\langle \Delta G_{ND} \rangle_{\VEC{\sigma}} / L \simeq k_B T_s \langle \Delta \psi_{ND} \rangle_{\VEC{\sigma}} / L$,
for each protein family.
}
\SUPPLEMENT{
In the case of the cutoff distance 8 \AA, 
the correlation coefficient is $r = -0.75$, and the regression line is 
$\overline{\overline{ \Delta \Delta G_{ND}(\sigma^N_{j \neq i}, \sigma^N_i\rightarrow \sigma_i) } }
= - 2.74 \langle \Delta G_{ND} \rangle_{\VEC{\sigma}} / L + 1.09$.
In the case of $r_{\script{cutoff}} \sim 15.5$ \AA, 
the correlation coefficient is $r = -0.59$, and the regression line is 
$\overline{\overline{ \Delta \Delta G_{ND}(\sigma^N_{j \neq i}, \sigma^N_i\rightarrow \sigma_i) } }
= - 3.11 \langle \Delta G_{ND} \rangle_{\VEC{\sigma}} / L + 1.24$.
}%  SUPPLEMENT
\TEXT{
The correlation coefficient is $r = -0.75$, and the regression line is 
$\overline{\overline{ \Delta \Delta G_{ND}(\sigma^N_{j \neq i}, \sigma^N_i\rightarrow \sigma_i) } }
= - 2.74 \langle \Delta G_{ND} \rangle_{\VEC{\sigma}} / L + 1.09$.
See \Fig{\ref{sfig: dG_over_L_vs_ddG}} for $r_{\script{cutoff}} \sim 15.5$ \AA.
}%  TEXT
The free energies are in kcal/mol units.
}%  FigureLegends
}
\end{figure*}

}%  FigE

\FigF{

\CLEARPAGE
 
\begin{figure*}[h!]
\FigureInLegends{
\SUPPLEMENT{
\centerline{
\includegraphics*[width=82mm,angle=0]{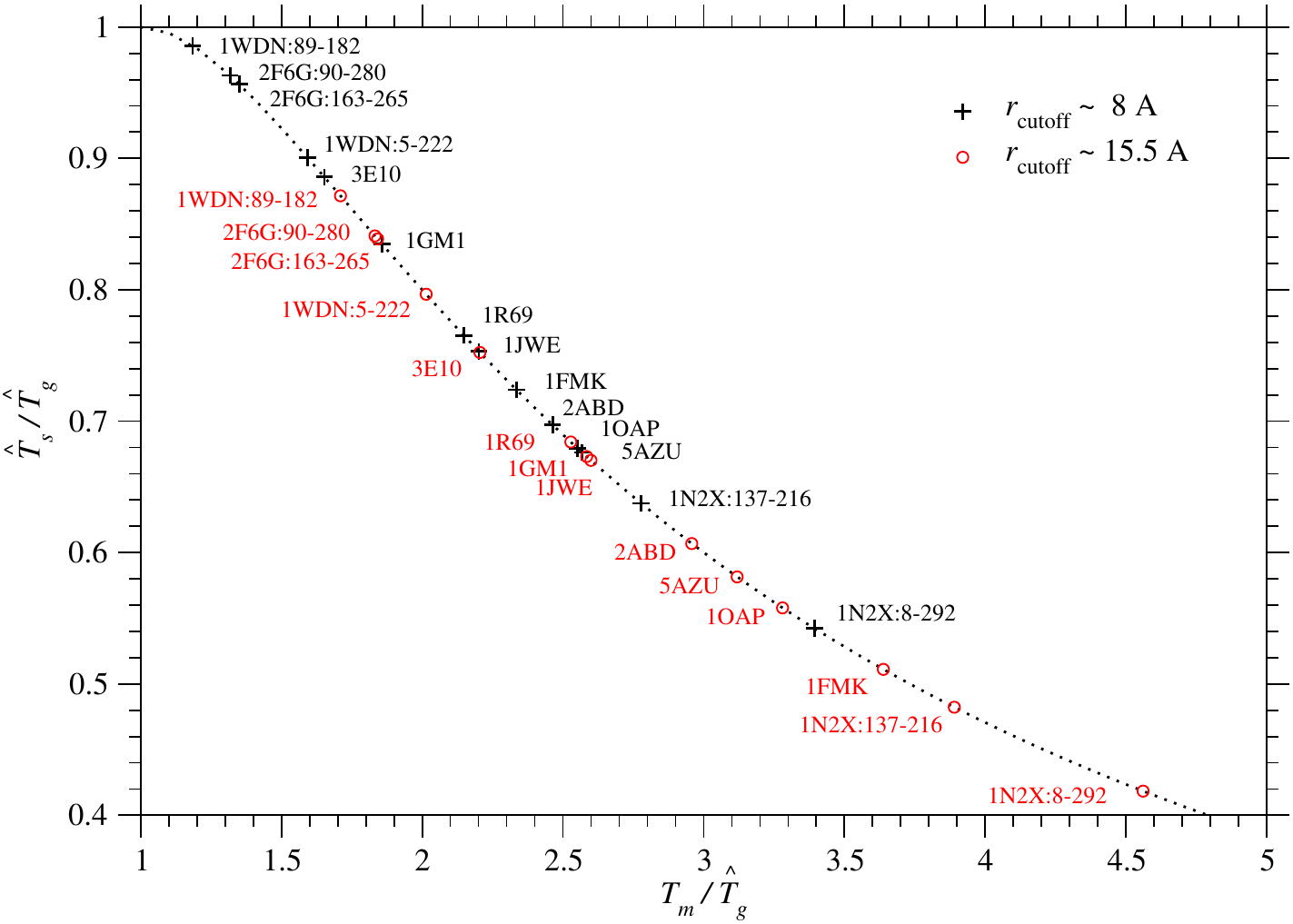}
}
}%  SUPPLEMENT
\TEXT{
\centerline{
\includegraphics*[width=82mm,angle=0]{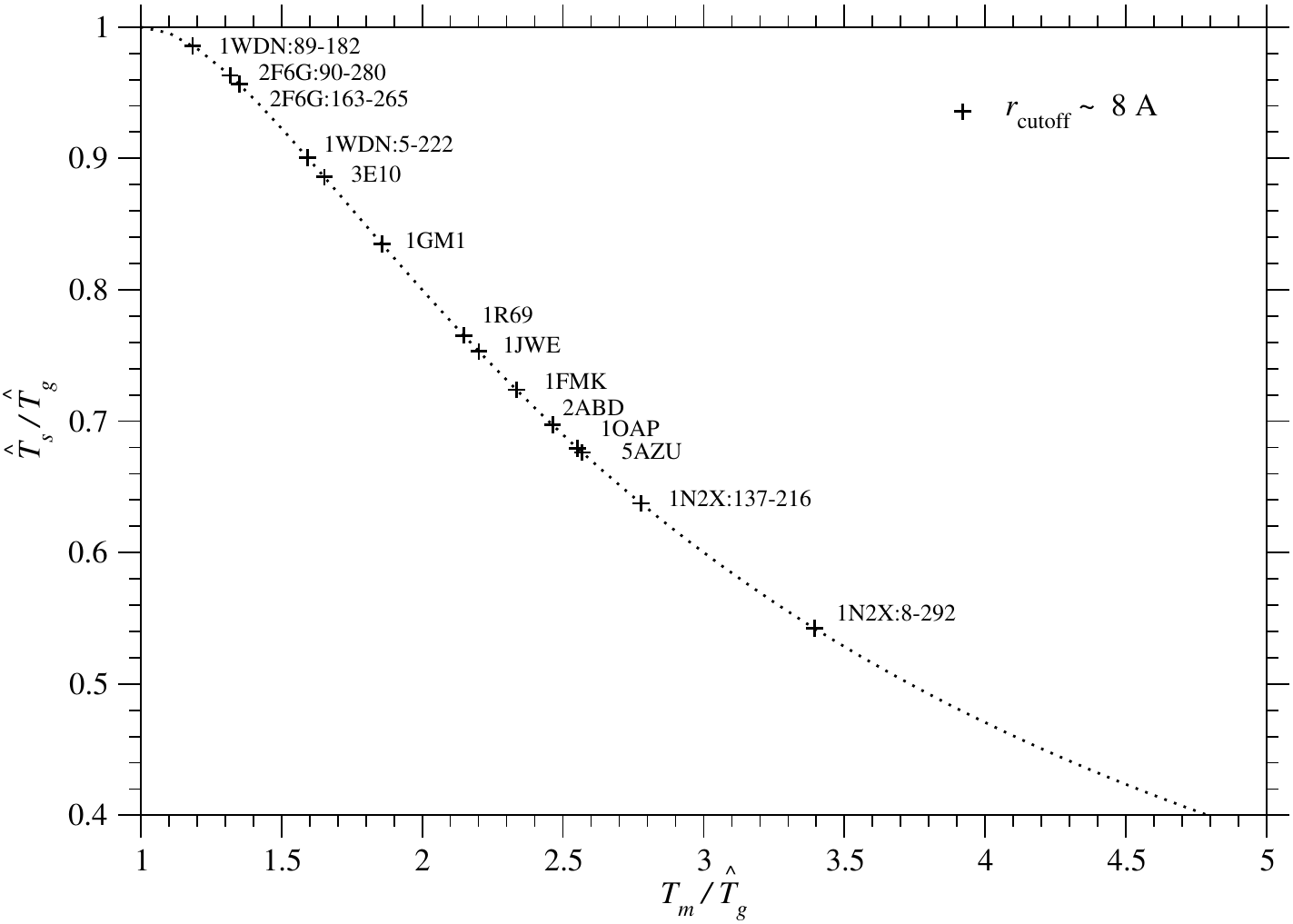}
}
}%  TEXT
}%  FigureInLegends
\vspace*{1em}
\caption{
\SUPPLEMENT{
\label{sfig: Tm_over_Tg_vs_Ts_over_Tg}
}%  SUPPLEMENT
\TEXT{
\label{fig: Tm_over_Tg_vs_Ts_over_Tg}
}%  TEXT
\FigureLegends{
\BF{$\hat{T}_s/\hat{T}_g$ is plotted against $T_m / \hat{T}_g$ for each protein domain.
}
A dotted curve corresponds to 
\Eq{\ref{\EQ: relationship_among_characteristic_T}},
$\hat{T}_s/\hat{T}_g = 2 (T_m / \hat{T}_g) / ((T_m/\hat{T}_g)^2 + 1)$.
\SUPPLEMENT{
Plus and open circle marks indicate the values estimated with $r_{\script{cutoff}} \sim 8$ and $15.5$ \AA,
respectively.
}%  SUPPLEMENT
\TEXT{
Plus marks indicate the values estimated with $r_{\script{cutoff}} \sim 8$ \AA.
See \Fig{\ref{sfig: Tm_over_Tg_vs_Ts_over_Tg}} for $r_{\script{cutoff}} \sim 15.5$ \AA.
}%  TEXT
The effective temperature $T_s$ for selection and glass transition temperature $T_g$ must satisfy $T_s < T_g < T_m$ 
for proteins to be able to fold into unique native structures.
}%  FigureLegends
}
\end{figure*}

}%  FigF

\FigG{

\CLEARPAGE
 
\begin{figure*}[h!]
\FigureInLegends{
\SUPPLEMENT{
\centerline{
\includegraphics*[width=82mm,angle=0]{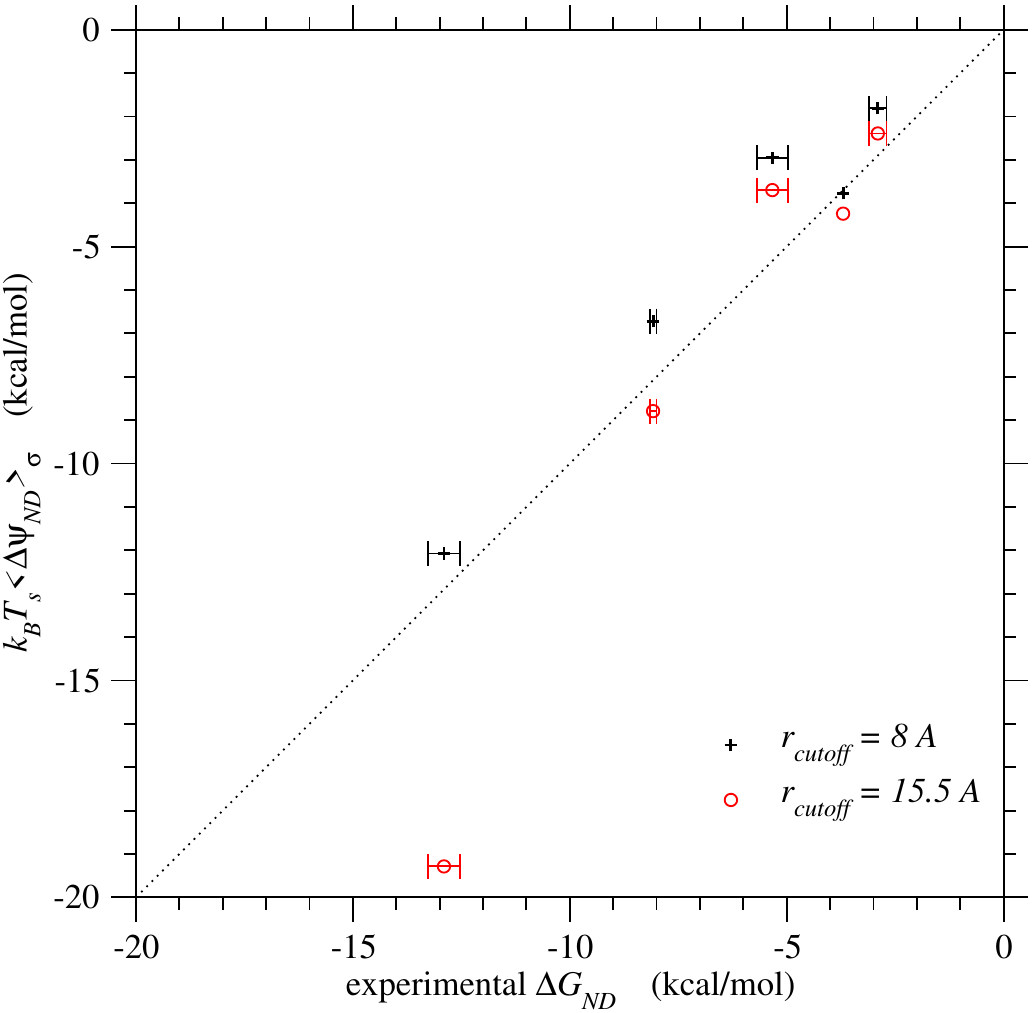}
}
}%  SUPPLEMENT
\TEXT{
\centerline{
\includegraphics*[width=82mm,angle=0]{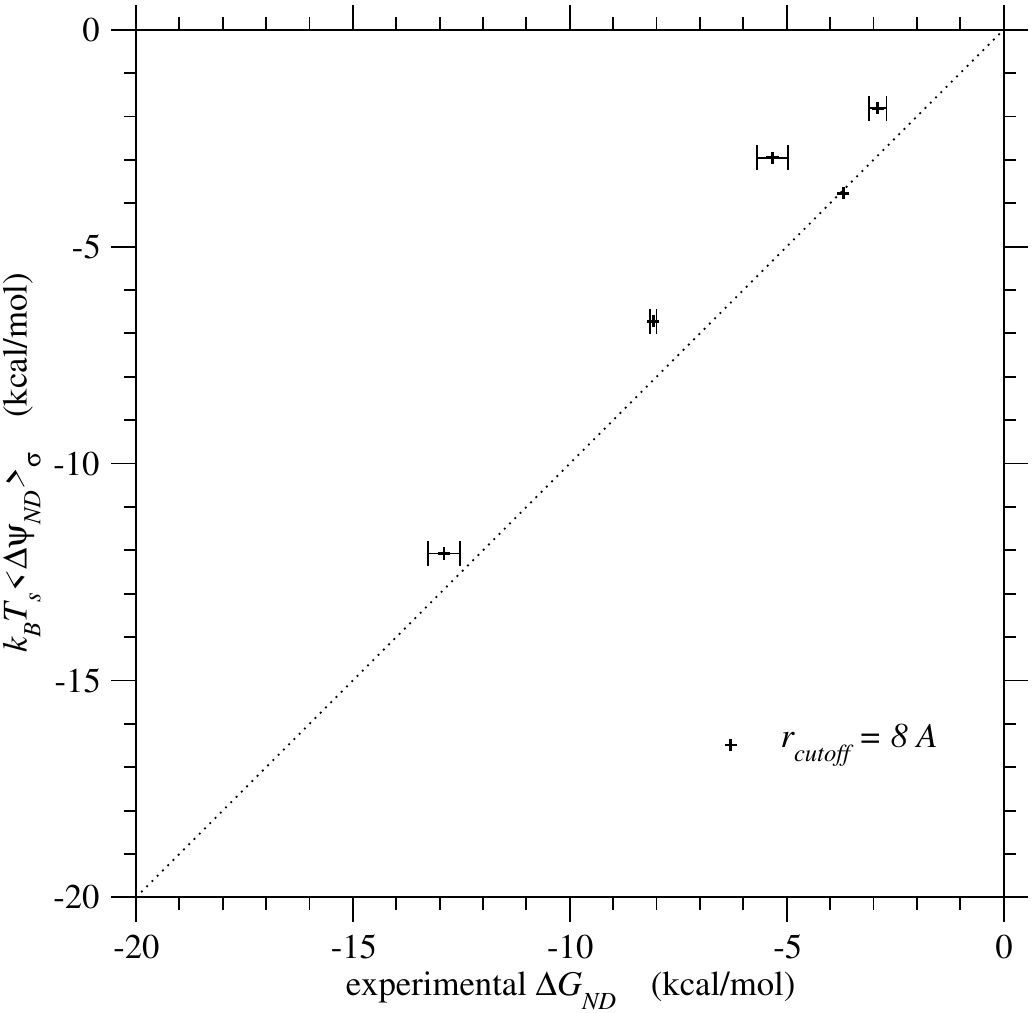}
}
}%  TEXT
}%  FigureInLegends
\vspace*{1em}
\caption{
\SUPPLEMENT{
\label{sfig: dG_exp_vs_8_and_16A}
}%  SUPPLEMENT
\TEXT{
\label{fig: dG_exp_vs_8_and_16A}
}%  TEXT
\FigureLegends{
\BF{Folding free energies, 
$\langle \Delta G_{ND} \rangle_{\VEC{\sigma}} \simeq k_B T_s \langle \Delta \psi_{ND} \rangle_{\VEC{\sigma}}$,
predicted by the present method are plotted against their experimental values, $\Delta G_{ND}(\VEC{\sigma_N})$.
}
\SUPPLEMENT{
Plus and open circle marks 
indicate the values estimated with $r_{\script{cutoff}} \sim 8$ and $15.5$ \AA,
respectively.
}%  SUPPLEMENT
\TEXT{
Plus marks 
indicate the values estimated with $r_{\script{cutoff}} \sim 8$ \AA.
See \Fig{\ref{sfig: dG_exp_vs_8_and_16A}} for $r_{\script{cutoff}} \sim 15.5$ \AA.
}%  TEXT
The free energies are in kcal/mol units.
}%  FigureLegends
}
\end{figure*}

}%  FigG

% End of figures_JTB_1.tex
% \input{figures_JTB_2.tex}

\TextFig{

\ifdefined\FigH
\else
\NoFigureInText{
\newcommand{\FigH}[1]{#1}
}%  NoFigureInText
\FigureInText{
\newcommand{\FigH}[1]{}
}%  FigureInText
\fi

\ifdefined\FigI
\else
\NoFigureInText{
\newcommand{\FigI}[1]{#1}
}%  NoFigureInText
\FigureInText{
\newcommand{\FigI}[1]{}
}%  FigureInText
\fi

\ifdefined\FigJ
\else
\NoFigureInText{
\newcommand{\FigJ}[1]{#1}
}%  NoFigureInText
\FigureInText{
\newcommand{\FigJ}[1]{}
}%  FigureInText
\fi

\ifdefined\FigK
\else
\NoFigureInText{
\newcommand{\FigK}[1]{#1}
}%  NoFigureInText
\FigureInText{
\newcommand{\FigK}[1]{}
}%  FigureInText
\fi

\ifdefined\FigL
\else
\NoFigureInText{
\newcommand{\FigL}[1]{#1}
}%  NoFigureInText
\FigureInText{
\newcommand{\FigL}[1]{}
}%  FigureInText
\fi

\ifdefined\FigM
\else
\NoFigureInText{
\newcommand{\FigM}[1]{#1}
}%  NoFigureInText
\FigureInText{
\newcommand{\FigM}[1]{}
}%  FigureInText
\fi

\ifdefined\FigN
\else
\NoFigureInText{
\newcommand{\FigN}[1]{#1}
}%  NoFigureInText
\FigureInText{
\newcommand{\FigN}[1]{}
}%  FigureInText
\fi

\ifdefined\FigO
\else
\NoFigureInText{
\newcommand{\FigO}[1]{#1}
}%  NoFigureInText
\FigureInText{
\newcommand{\FigO}[1]{}
}%  FigureInText
\fi

\ifdefined\FigP
\else
\NoFigureInText{
\newcommand{\FigP}[1]{#1}
}%  NoFigureInText
\FigureInText{
\newcommand{\FigP}[1]{}
}%  FigureInText
\fi

\renewcommand{\SUPPLEMENT}[1]{}

\ifdefined\CLEARPAGE

\NoFigureInText{
\renewcommand{\CLEARPAGE}{\FigureLegends{\clearpage\newpage}}
}%  NoFigureInText
\FigureInText{
\renewcommand{\CLEARPAGE}{}
}%  FigureInText

\else

\NoFigureInText{
\newcommand{\CLEARPAGE}{\FigureLegends{\clearpage\newpage}}
}%  NoFigureInText
\FigureInText{
\newcommand{\CLEARPAGE}{}
}%  FigureInText

\fi

}%  TextFig

\SupFig{

\ifdefined\FigH
\renewcommand{\FigH}[1]{#1}
\else
\newcommand{\FigH}[1]{#1}
\fi

\ifdefined\FigI
\renewcommand{\FigI}[1]{#1}
\else
\newcommand{\FigI}[1]{#1}
\fi

\ifdefined\FigJ
\renewcommand{\FigJ}[1]{#1}
\else
\newcommand{\FigJ}[1]{#1}
\fi

\ifdefined\FigK
\renewcommand{\FigK}[1]{#1}
\else
\newcommand{\FigK}[1]{#1}
\fi

\ifdefined\FigL
\renewcommand{\FigL}[1]{#1}
\else
\newcommand{\FigL}[1]{#1}
\fi

\ifdefined\FigM
\renewcommand{\FigM}[1]{#1}
\else
\newcommand{\FigM}[1]{#1}
\fi

\ifdefined\FigN
\renewcommand{\FigN}[1]{#1}
\else
\newcommand{\FigN}[1]{#1}
\fi

\ifdefined\FigO
\renewcommand{\FigO}[1]{#1}
\else
\newcommand{\FigO}[1]{#1}
\fi

\ifdefined\FigP
\renewcommand{\FigP}[1]{#1}
\else
\newcommand{\FigP}[1]{#1}
\fi

\renewcommand{\SUPPLEMENT}[1]{#1}
\ifdefined\CLEARPAGE
\renewcommand{\CLEARPAGE}{\FigureLegends{\clearpage\newpage}}
\else
\newcommand{\CLEARPAGE}{\FigureLegends{\clearpage\newpage}}
\fi
}%  SupFig

\renewcommand{\SkipFigure}[1]{}

\FigH{

\CLEARPAGE
 
\begin{figure*}[h!]
\FigureInLegends{
\noindent
\SUPPLEMENT{
\hspace*{1em} (a) $r_{\script{cutoff}} \sim 8$\AA\ 
\hspace*{14em} (b) $r_{\script{cutoff}} \sim 15.5$\AA\ 

}%  SUPPLEMENT
\TEXT{
\centerline{
{\small{$r_{\script{cutoff}} \sim 8$\AA\ }}
}
}%  TEXT
\SUPPLEMENT{
\centerline{
\includegraphics*[width=82mm,angle=0]{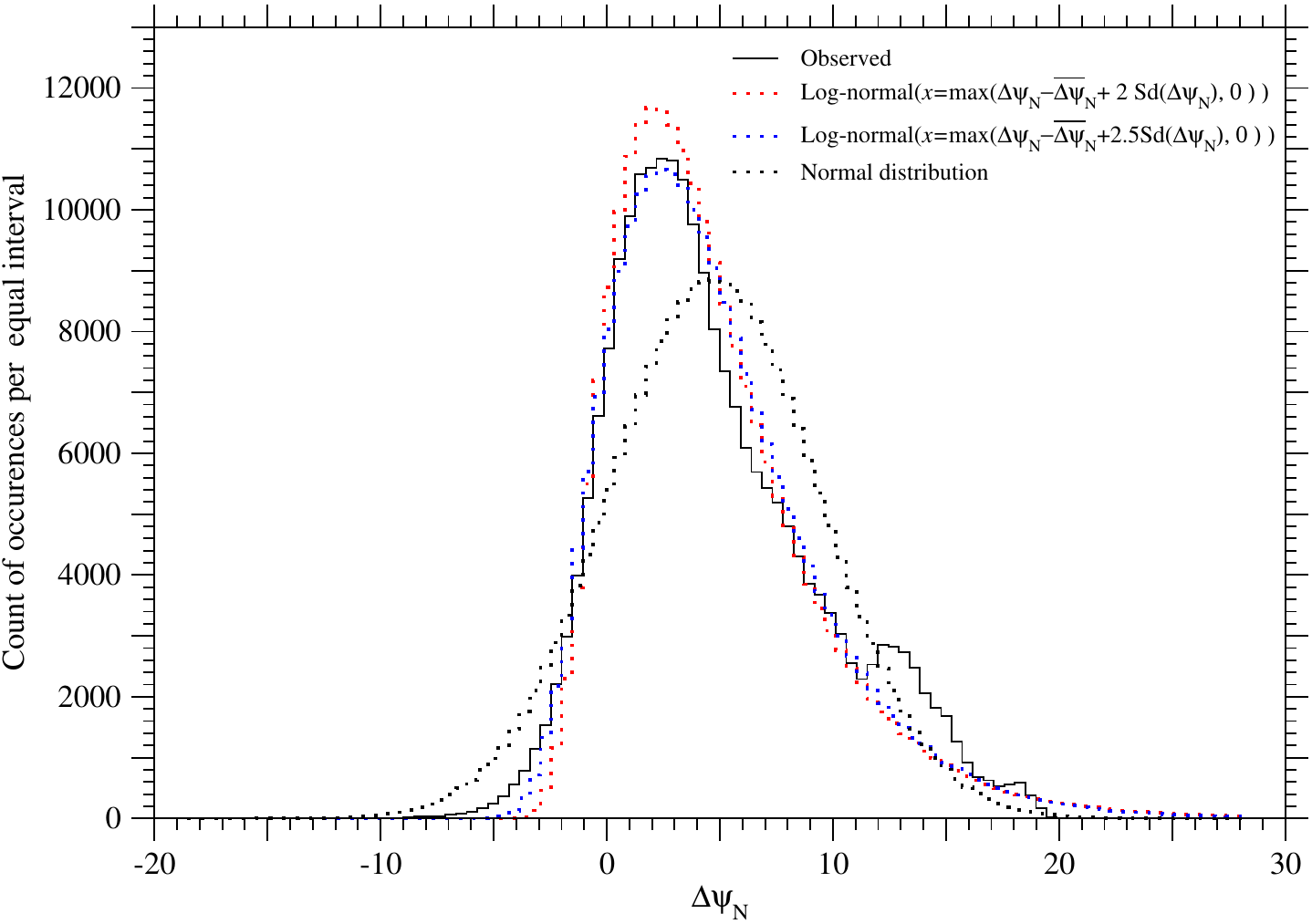}
\includegraphics*[width=82mm,angle=0]{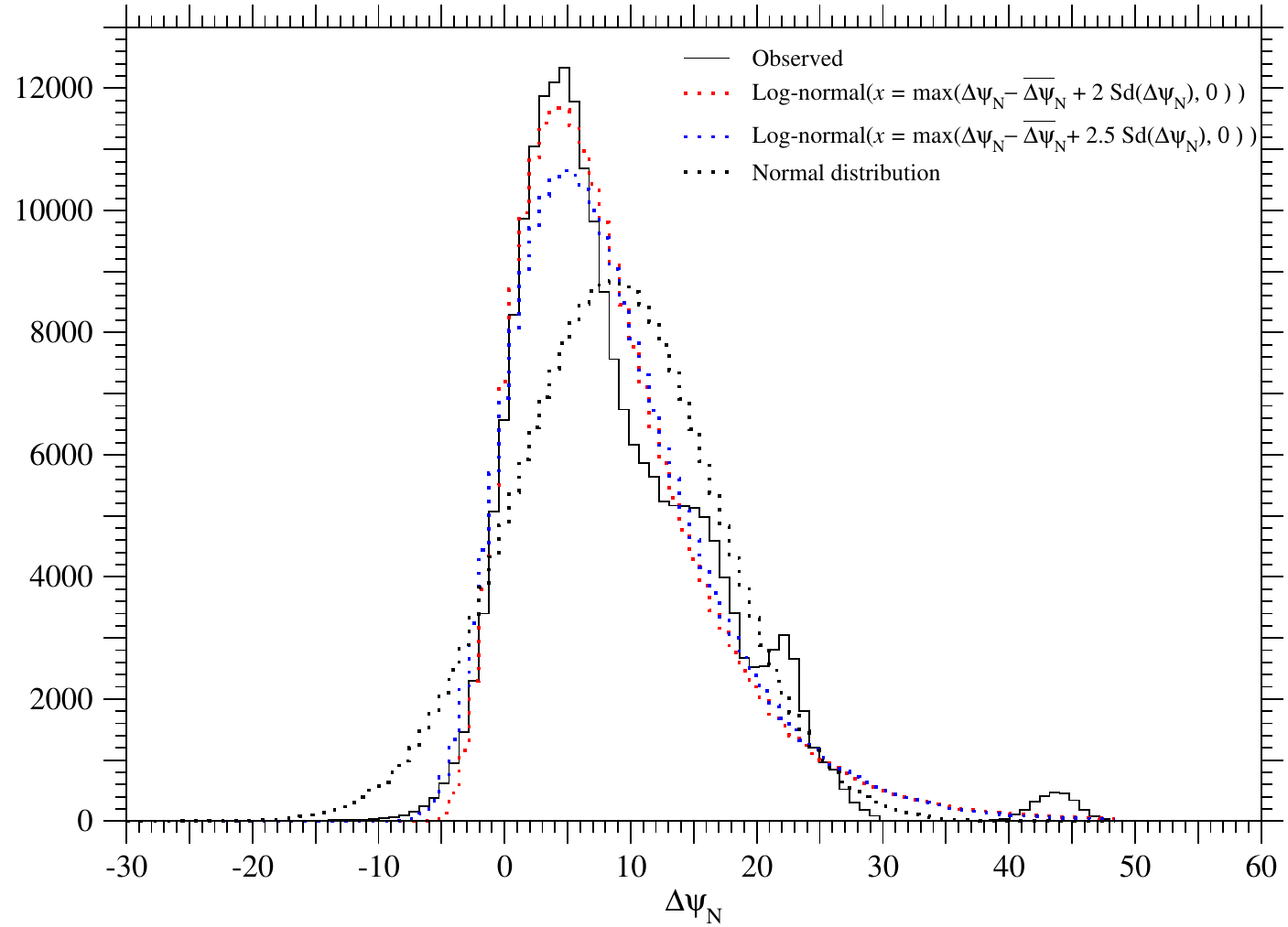}
}
}%  SUPPLEMENT
\TEXT{
\centerline{
\includegraphics*[width=82mm,angle=0]{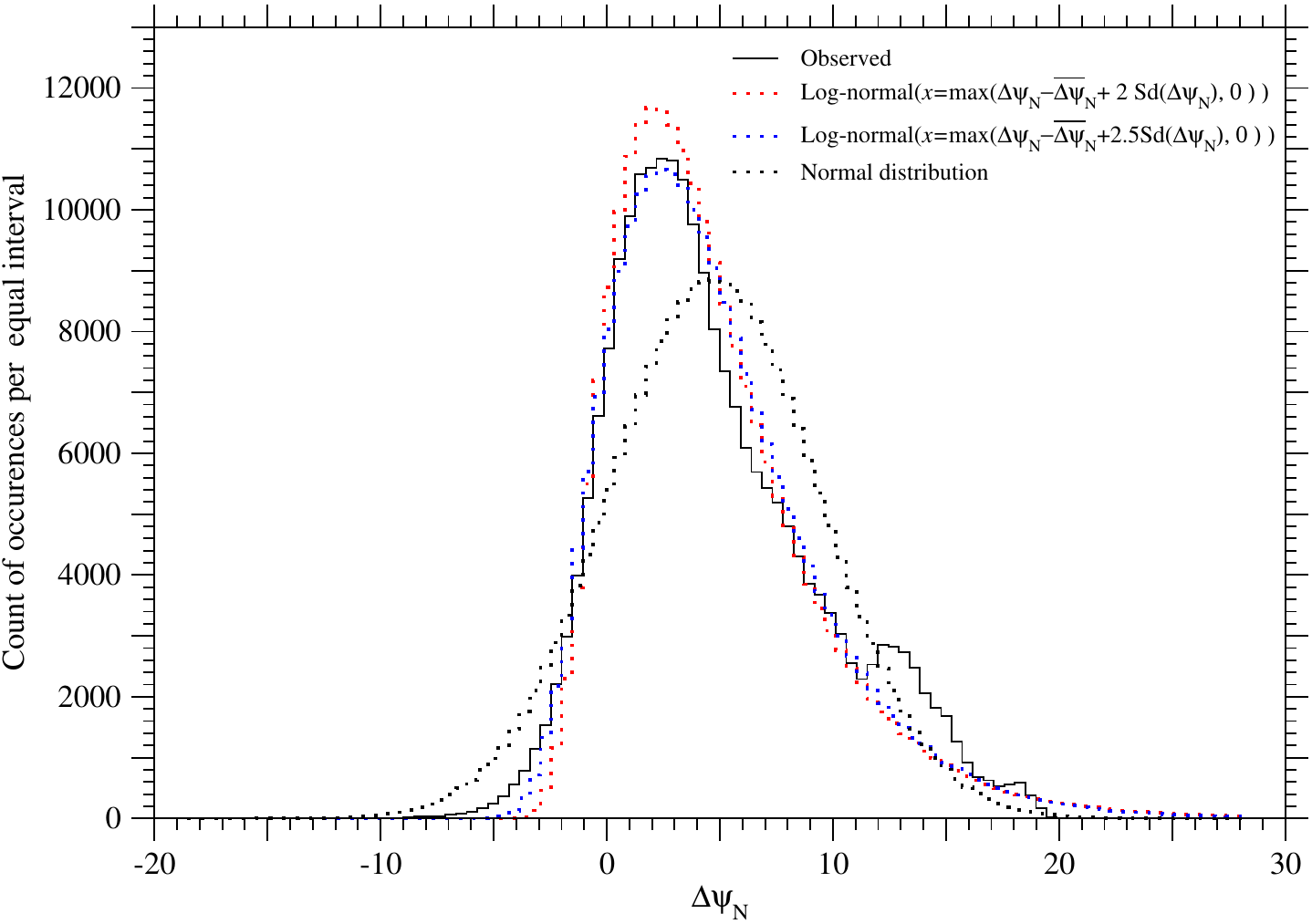}
}
}%  TEXT
}%  FigureInLegends
\vspace*{1em}
\caption{
\SUPPLEMENT{
\label{sfig: 1gm1-a:16-96_full_non_del_dca0_205_0_20_simple-gauge_dPhiN_distr}
\label{sfig: 1gm1-a:16-96_full_non_del_dca0_33_0_20_simple-gauge_dPhiN_distr}
}%  SUPPLEMENT
\TEXT{
\label{fig: 1gm1-a:16-96_full_non_del_dca0_205_0_20_simple-gauge_dPhiN_distr}
}%  TEXT
\FigureLegends{
\BF{
The observed frequency distribution and the fitted distributions of $\Delta \psi_N$ in the PDZ protein family.
}
A black solid line indicates the observed frequency distribution of $\Delta \psi_N$ per equal interval in homologous sequences of the PDZ protein family,
and red dotted and blue dotted lines indicate the total frequencies of log-normal distributions with $n_{\script{shift}} = 2$ or $2.5$
and parameters estimated with the mean and variance of the observed distribution for each protein;
see \Eqs{\REF{\EQ: log-normal} to \REF{\EQ: statistics_for_log-normal}}.
A black dotted line indicates the total frequencies of normal distributions the mean and variance of which are equal to those of 
the observed distribution for each protein.
Only representatives of unique sequences with no deletions, which are at least 20\% different from each other, are employed;
the total count is equal to 
222,466 over 335 homologous sequences,
which is almost equal to $M_{\script{eff}}$ in \Table{\ref{\TBL: Proteins_studied}}.
\TEXT{
See \Fig{\ref{sfig: 1gm1-a:16-96_full_non_del_dca0_33_0_20_simple-gauge_dPhiN_distr}} for $r_{\script{cutoff}} \sim 15.5$ \AA.
}%  TEXT
}%  FigureLegends
}
\end{figure*}

}%  FigH

\SUPPLEMENT{

\CLEARPAGE
 
\begin{figure*}[h!]
\FigureInLegends{
\noindent
\hspace*{1em}
(a) $r_{\script{cutoff}} \sim 8$\AA\ 
\hspace*{14em}
(b) $r_{\script{cutoff}} \sim 15.5$\AA\ 

\centerline{
\includegraphics*[width=82mm,angle=0]{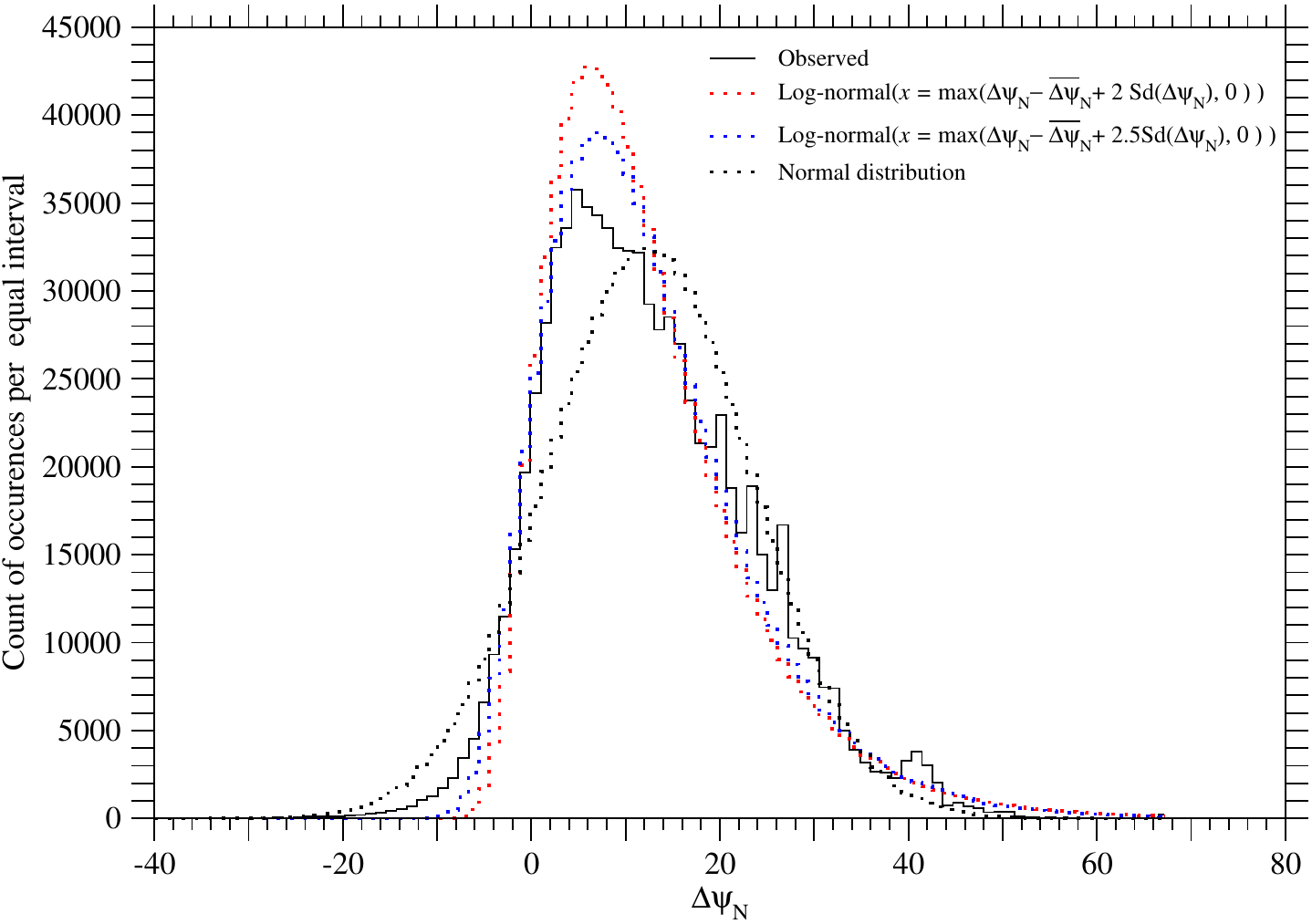}
\includegraphics*[width=82mm,angle=0]{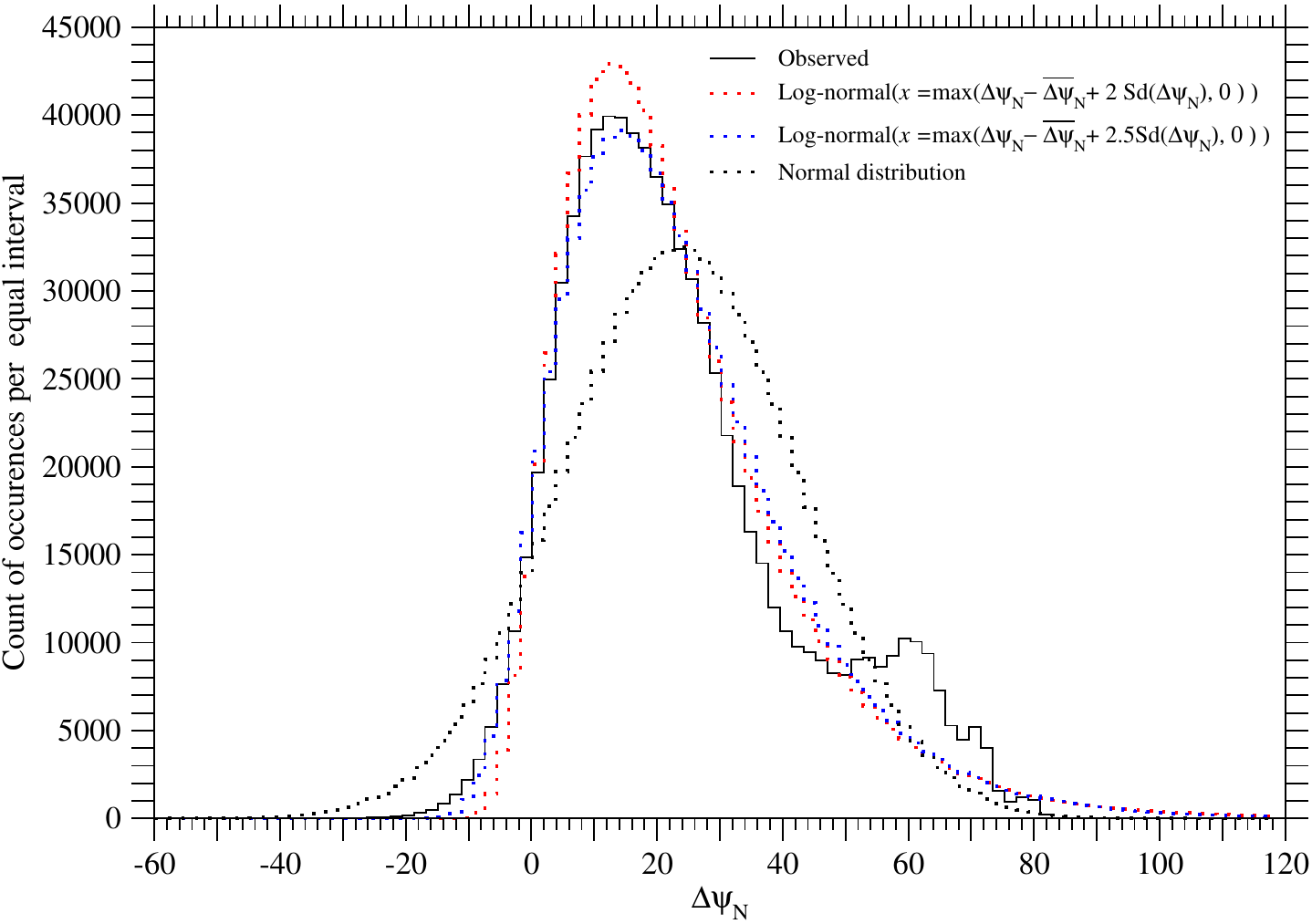}
}
}%  FigureInLegends
\vspace*{1em}
\caption{
\label{sfig: 1n2x-a:8-292_full_non_del_dca0_13_0_20_simple-gauge_dPhiN_distr}
\label{fig: 1n2x-a:8-292_full_non_del_dca0_13_0_20_simple-gauge_dPhiN_distr}
\label{sfig: 1n2x-a:8-292_full_non_del_dca0_175_0_20_simple-gauge_dPhiN_distr}
\label{fig: 1n2x-a:8-292_full_non_del_dca0_175_0_20_simple-gauge_dPhiN_distr}
\FigureLegends{
\BF{
The observed frequency distribution and the fitted distribution of $\Delta \psi_N$ in the Methyltransf\_5 family of the domain, 1N2X-A:8-292.
}
A black solid line indicates the observed frequency distribution of $\Delta \psi_N$ per equal interval 
in homologous sequences of the Methyltransf\_5 protein family,
and red dotted and blue dotted lines indicate the total frequencies of log-normal distributions 
with $n_{\script{shift}} = 2$ or $2.5$
and parameters estimated with the mean and variance of the observed distribution for each protein;
see \Eqs{\REF{\EQ: log-normal} to \REF{\EQ: statistics_for_log-normal}}.
A black dotted line indicates the total frequencies of normal distributions the mean and variance of which are equal to those of 
the observed distribution for each protein.
Only representatives of unique sequences, which are at least 20\% different from each other, are employed;
the total count is equal to 
814549 over 354 homologous sequences,
which is almost equal to $M_{\script{eff}}$ in \Table{\ref{\TBL: Proteins_studied}}.
}%  FigureLegends
}
\end{figure*}

}%  SUPPLEMENT

\FigI{

\CLEARPAGE
 
\begin{figure*}[h!]
\FigureInLegends{
\centerline{
\includegraphics*[width=82mm,angle=0]{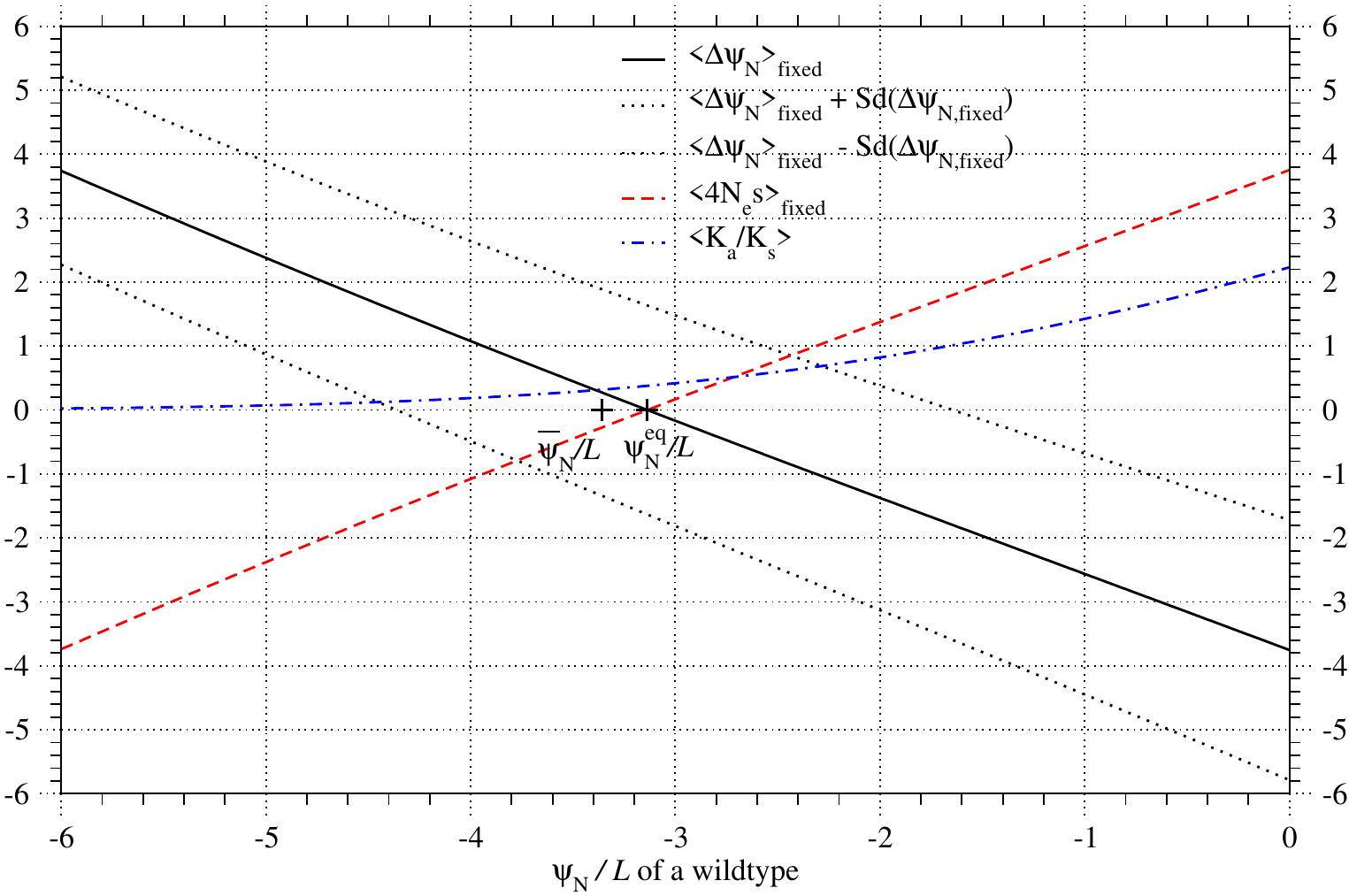}
}
}%  FigureInLegends
\vspace*{1em}
\caption{
\SUPPLEMENT{
\label{sfig: ave_ddPhi_vs_dPhiN_of_fixed_mutants_logG-2sd_PDZ_8A}
}%  SUPPLEMENT
\TEXT{
\label{fig: ave_ddPhi_vs_dPhiN_of_fixed_mutants_logG-2sd_PDZ_8A}
}%  TEXT
\FigureLegends{
\BF{
The average 
of $\Delta \psi_{N} (\simeq \Delta\Delta \psi_{ND})$ 
over fixed single nucleotide nonsynonymous mutations versus $\psi_{N} / L$ of 
a wildtype for the PDZ protein family.
}
The averages of $\Delta \psi_{N} (\simeq \Delta\Delta \psi_{ND})$ 
and $4N_e s$ over the fixed mutants,
and the average of $K_a/K_s (\equiv u(s)/u(0))$ over all the mutants are
plotted against $\psi_{N} / L$ of a wildtype by solid, broken, and dash-dot lines,
respectively; $q_m = 1 / (2 \times 10^6)$ is assumed.
Dotted lines show the values of $\langle \Delta \psi_{N} \rangle_{\script{fixed}} \pm \textrm{sd}$,
where the $\textrm{sd}$ is the standard deviation of $\Delta \psi_{N}$ over fixed mutants.
Fixation probability has been calculated with $\Delta\Delta \psi_{ND} \simeq \Delta \psi_{N}$;
see \Eqs{\REF{\EQ: fixation_probability} and \REF{\EQ: selective_advantage_and_dPsi_N}}.
Here the empirical relationships of \Eqs{\REF{\EQ: regression_of_dPsi_on_Psi} and \REF{\EQ: var_of_dPsi}}
are assumed; that is,
the mean of $\Delta \psi_N$ linearly decreases as $\psi_N$ increases,
but the standard deviation of $\Delta \psi_N$ is constant irrespective of $\psi_N$.
The slope ($\alpha_{\psi_N}$) and intercept 
($- \alpha_{\psi_N}\overline{\psi_N}/L + \overline{\overline{\Delta \psi_N}}$)
and the average of $\text{Sd}(\Delta \psi_N)$ over homologous sequences  
that are estimated with $r_{\script{cutoff}} \sim 8$\AA\ for the PDZ
and listed in \Table{\ref{\TBL: ddPsi_with_8A}} are employed here. 
The distribution of $\Delta \psi_N$ due to single nucleotide nonsynonymous mutations is approximated by
a log-normal distribution with $n_{\script{shift}} = 2.0$;
see \Eqs{\REF{\EQ: log-normal} to \REF{\EQ: statistics_for_log-normal}}.
The $\psi_{N}^{\script{eq}}$, where $\langle \Delta\Delta \psi_{ND} \rangle_{\script{fixed}} \simeq \langle \Delta \psi_{N} \rangle_{\script{fixed}} = 0$, 
is the stable equilibrium value
of $\psi_{N}$ in the protein evolution of the PDZ protein family.
The $\psi_{N}^{\script{eq}}$ is close to the average of $\psi_N$ over homologous sequences ($\overline{\psi_N}$),
indicating that the present approximations for $\psi_{N}^{\script{eq}}$ and
for $\overline{\psi_N} = \langle \psi_N \rangle_{\VEC{\sigma}}$ are consistent to each other.
}%  FigureLegends
}
\end{figure*}

}%  FigI

\SUPPLEMENT{

\CLEARPAGE
 
\begin{figure*}[h!]
\FigureInLegends{
\centerline{
\includegraphics*[width=82mm,angle=0]{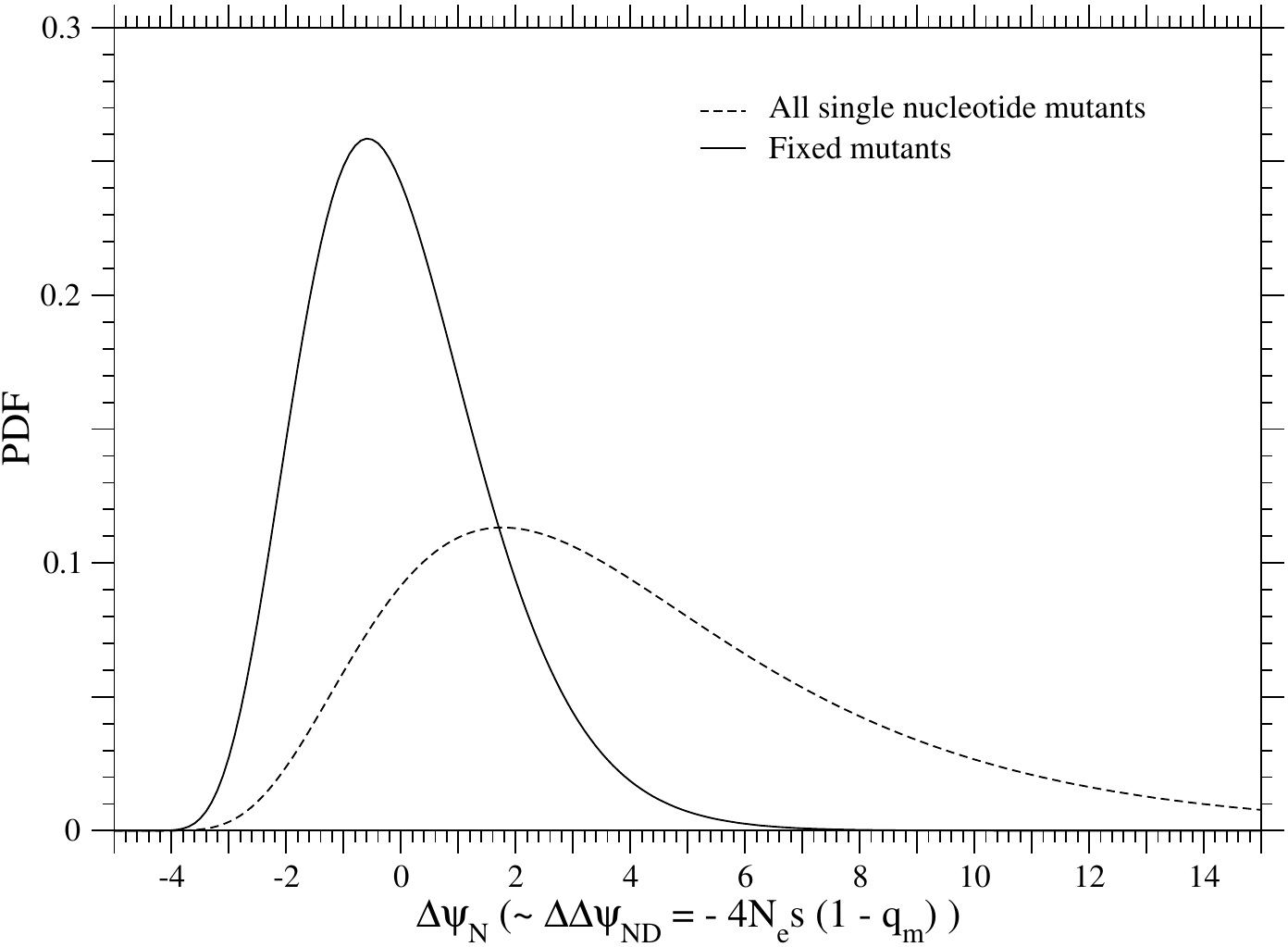}
\includegraphics*[width=82mm,angle=0]{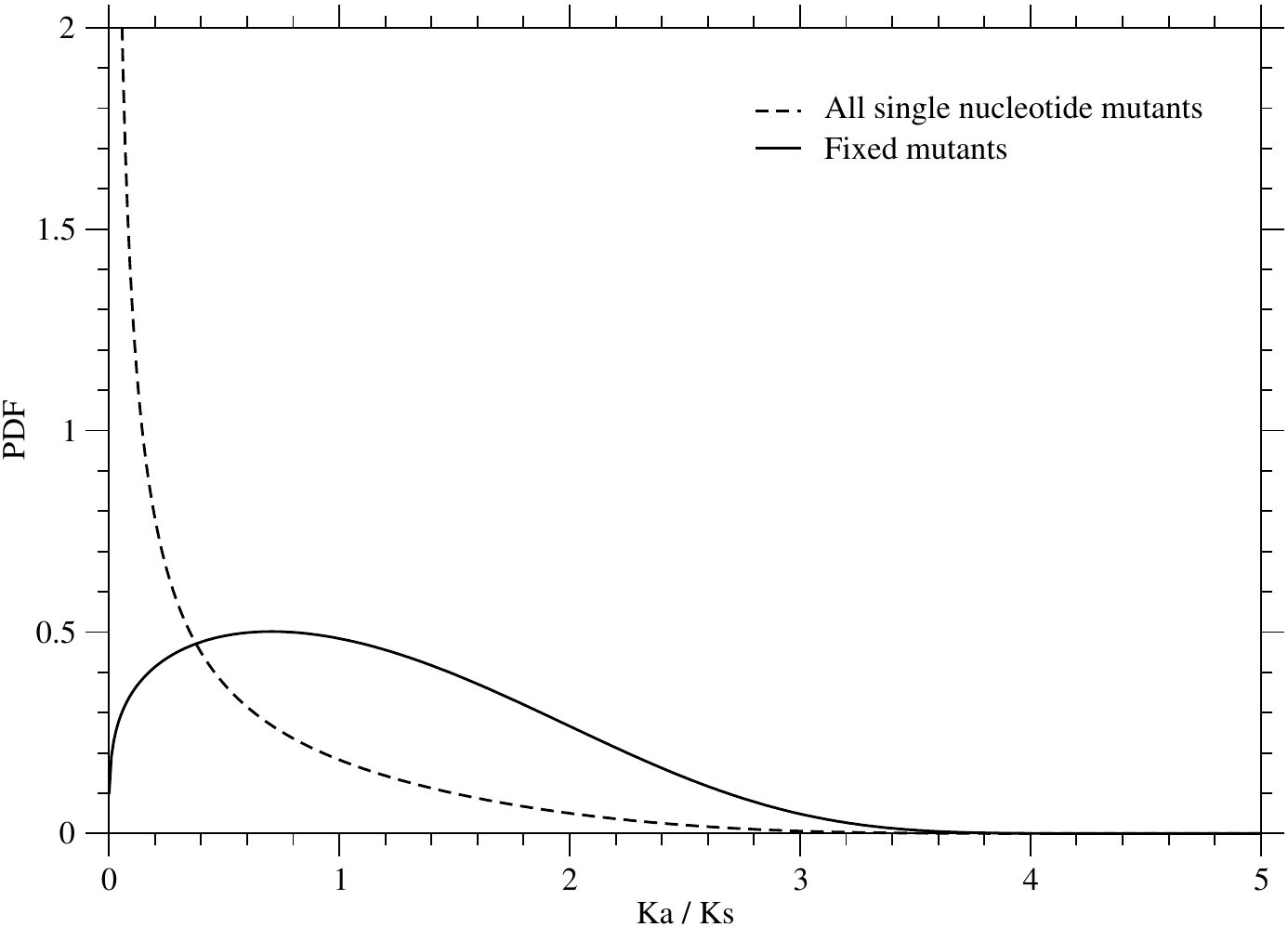}
}
}%  FigureInLegends
\vspace*{1em}
\caption{
\label{sfig: pdf_of_ddPhi_at_PhiNe_logG-2sd_PDZ_8A}
\label{fig: pdf_of_ddPhi_at_PhiNe_logG-2sd_PDZ_8A}
\label{sfig: pdf_of_ka_over_ks_at_PhiNe_logG-2sd_PDZ_8A}
\label{fig: pdf_of_ka_over_ks_at_PhiNe_logG-2sd_PDZ_8A}
\FigureLegends{
\BF{
PDFs of $\Delta \psi_N (\simeq \Delta\Delta \psi_{ND} = - 4N_e s (1 - q_m) )$ 
and of $K_a/K_s$ 
for all single nucleotide nonsynonymous mutants and for their fixed mutants 
at equilibrium 
($\langle \Delta \psi_N \rangle_{\script{fixed}} = 0$)
for the PDZ protein family.
}
$K_a/K_s$ is defined as the ratio of nonsynonymous to synonymous substitution rate per site, $u(s)/u(0)$;
see \Eq{\ref{seq: def_Ka_over_Ks}}.
Fixation probability has been calculated with $\Delta\Delta \psi_{ND} \simeq \Delta \psi_{N}$;
see \Eqs{\REF{\EQ: fixation_probability} and \REF{\EQ: selective_advantage_and_dPsi_N}}.
The equilibrium value $\psi_N^{\script{eq}}$, 
where $\langle \Delta\Delta \psi_{ND} \rangle_{\script{fixed}} \simeq \langle \Delta \psi_{N} \rangle_{\script{fixed}} = 0$, 
is calculated
by using 
the linear dependency 
of $\overline{\Delta \psi_N}$ on $\psi_N$ (\Eq{\ref{\EQ: regression_of_dPsi_on_Psi}})
and 
estimated values
with $r_{\script{cutoff}} \sim 8$\AA\  for the PDZ 
in \Tables{\ref{\TBL: ddPsi_with_8A}}.
The standard deviation of $\Delta \psi_N$ is approximated to be constant
and equal to $\overline{\mbox{Sd}(\Delta \psi_N)}$; see \Eq{\ref{\EQ: var_of_dPsi}}.
The distribution of $\Delta \psi_N$ due to single nucleotide nonsynonymous mutations is approximated by
a log-normal distribution with $n_{\script{shift}} = 2.0$;
see \Eqs{\REF{\EQ: log-normal} to \REF{\EQ: statistics_for_log-normal}}.
}%  FigureLegends
}
\end{figure*}

}%  SUPPLEMENT

\FigJ{

\CLEARPAGE
 
\begin{figure*}[h!]
\FigureInLegends{
\noindent
\SUPPLEMENT{
\hspace*{1em}
(a) $r_{\script{cutoff}} \sim 8$\AA\ 
\hspace*{14em}
(b) $r_{\script{cutoff}} \sim 15.5$\AA\ 

}%  SUPPLEMENT
\TEXT{
\centerline{
{\small{$r_{\script{cutoff}} \sim 8$\AA\ }}
}
}%  TEXT
\SUPPLEMENT{
\centerline{
\includegraphics*[width=82mm,angle=0]{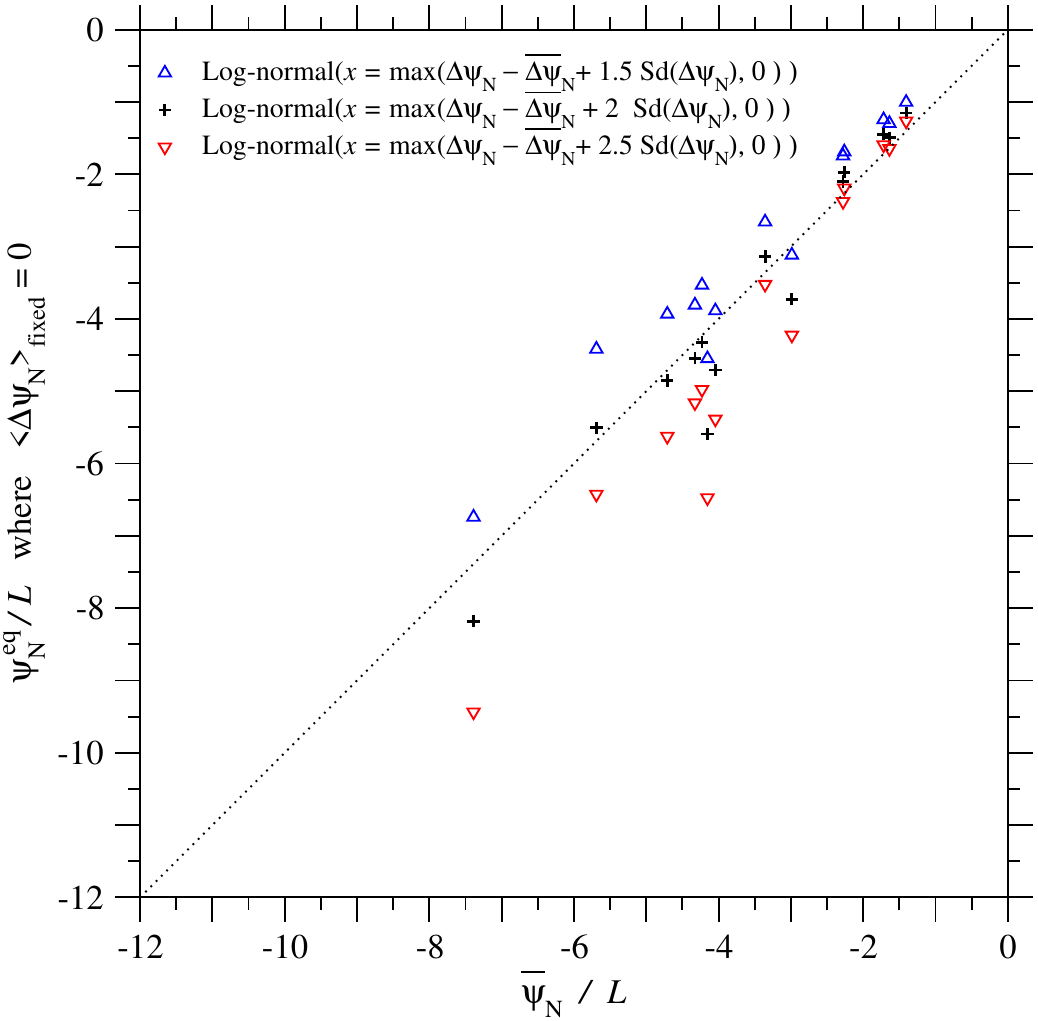}
\includegraphics*[width=82mm,angle=0]{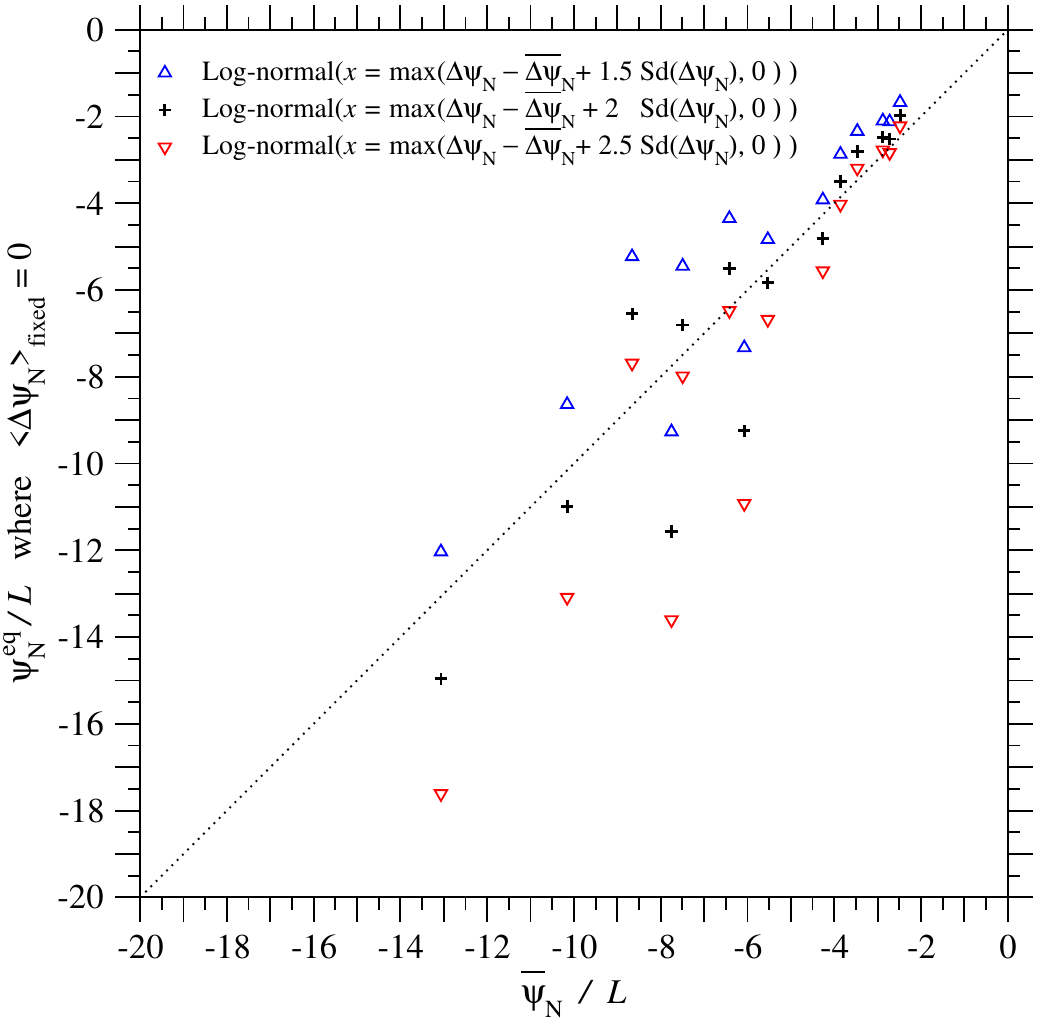}
}
}%  SUPPLEMENT
\TEXT{
\centerline{
\includegraphics*[width=82mm,angle=0]{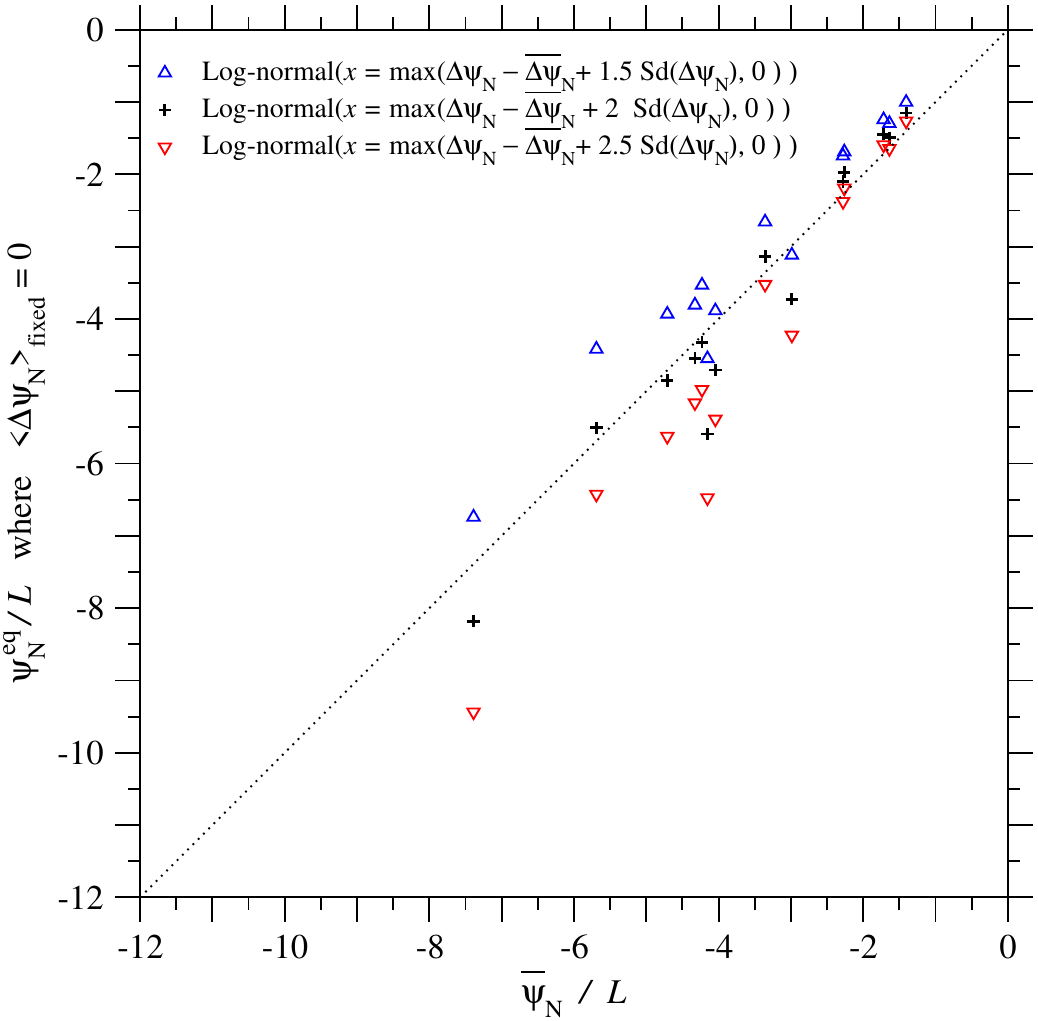}
}
}%  TEXT
}%  FigureInLegends
\vspace*{1em}
\caption{
\SUPPLEMENT{
\label{sfig: PhiNe_obs_vs_exp_logG_8A}
\label{sfig: PhiNe_obs_vs_exp_logG_16A}
}%  SUPPLEMENT
\TEXT{
\label{fig: PhiNe_obs_vs_exp_logG_8A}
}%  TEXT
\FigureLegends{
\BF{
The equilibrium value of $\psi_N / L$, where $\langle \Delta \psi_N \rangle_{\script{fixed}} = 0$,
is plotted against the average of $\psi_N / L$ over homologous sequences for each protein family.
}
\SUPPLEMENT{
The cutoff distances, 
(a) $r_{\script{cutoff}} = 8$\AA\ and (b) $r_{\script{cutoff}} = 15.5$\AA, are employed to estimate $\psi_N$ of
each protein family.
}%  SUPPLEMENT
\TEXT{
The cutoff distance
$r_{\script{cutoff}} = 8$\AA\ is employed to estimate $\psi_N$ of each protein family;
see \Fig{\ref{sfig: PhiNe_obs_vs_exp_logG_16A}} for $r_{\script{cutoff}} = 15.5$\AA.
}%  TEXT
The equilibrium values $\psi_N^{\script{eq}}$, 
where $\langle \Delta \psi_N \rangle_{\script{fixed}} = 0$,
are calculated
by using 
the linear dependency 
of $\overline{\Delta \psi_N}$ on $\psi_N$ (\Eq{\ref{\EQ: regression_of_dPsi_on_Psi}})
and
estimated values 
\SUPPLEMENT{
with $r_{\script{cutoff}} \sim 8$ or $15.5$\AA\  
in \Tables{\ref{\TBL: ddPsi_with_8A} or \ref{\TBL: ddPsi_with_16A}}.
}%  SUPPLEMENT
\TEXT{
with $r_{\script{cutoff}} \sim 8$
in \Tables{\ref{\TBL: ddPsi_with_8A}}.
}%  TEXT
The standard deviation of $\Delta \psi_N$ is approximated to be constant
and equal to $\overline{\mbox{Sd}(\Delta \psi_N)}$; see \Eq{\ref{\EQ: var_of_dPsi}}.
Plus, upper triangle, and lower triangle marks indicate the cases of
log-normal distributions with $n_{\script{shift}} = 1.5, 2.0$, and $2.5$ 
employed to approximate
the distribution of $\Delta \psi_N$, respectively; 
see \Eqs{\REF{\EQ: log-normal} to \REF{\EQ: statistics_for_log-normal}}. 
}%  FigureLegends
}
\end{figure*}

}%  FigJ

\FigK{

\CLEARPAGE
 
\begin{figure*}[h!]
\FigureInLegends{
\SUPPLEMENT{
\centerline{
\includegraphics*[width=82mm,angle=0]{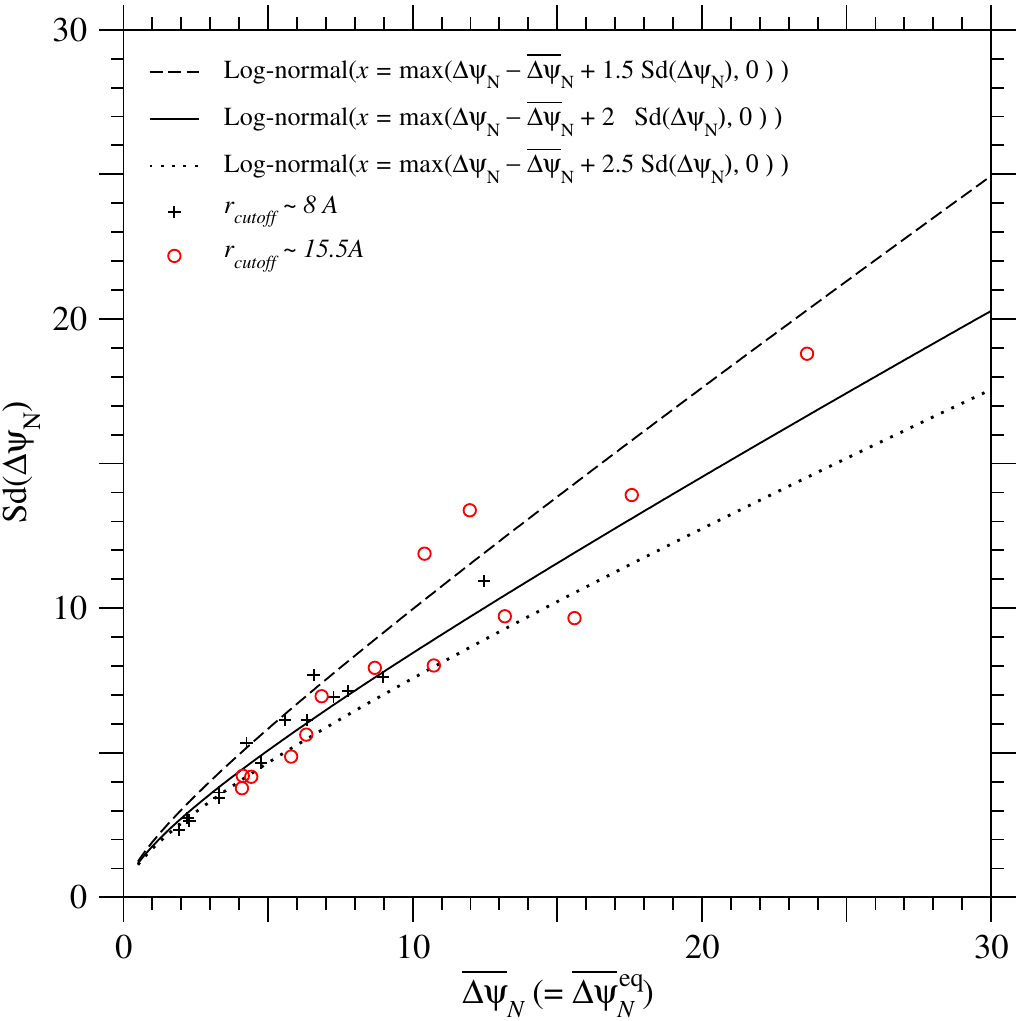}
}
}%  SUPPLEMENT
\TEXT{
\centerline{
\includegraphics*[width=82mm,angle=0]{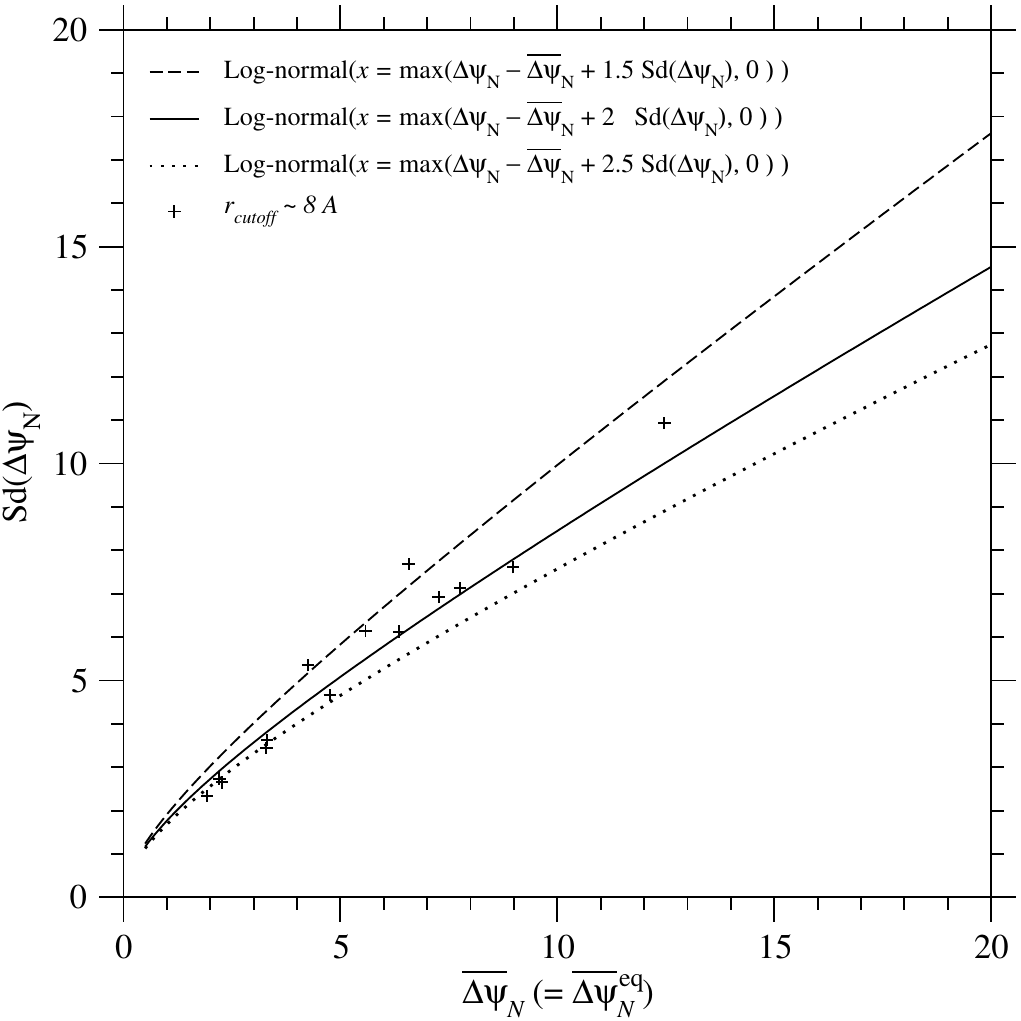}
}
}%  TEXT
}%  FigureInLegends
\vspace*{1em}
\caption{
\SUPPLEMENT{
\label{sfig: ddPhi_mean_vs_sd_at_equil}
}%  SUPPLEMENT
\TEXT{
\label{fig: ddPhi_mean_vs_sd_at_equil}
}%  TEXT
\FigureLegends{
\BF{
Relationship between the mean and the standard deviation of $\Delta \psi_N$ due to single nucleotide nonsynonymous mutations 
at equilibrium,
$\langle \Delta \psi_{N} \rangle_{\script{fixed}} = 0$
}
The standard deviation of $\Delta \psi_N$ 
that satisfies $\langle \Delta \psi_{N} \rangle_{\script{fixed}} = 0$
is plotted against
its mean, $\overline{\Delta \psi_N}$.
Broken, solid, and dotted lines indicate the cases of
log-normal distributions with $n_{\script{shift}} = 1.5, 2.0$ and $2.5$ 
employed to approximate the distribution of $\Delta \psi_N$, respectively; 
see \Eqs{\REF{\EQ: log-normal} to \REF{\EQ: statistics_for_log-normal}}. 
\SUPPLEMENT{
Plus and open circle marks indicate the averages, 
$\overline{\overline{\Delta \psi_N}}$ and $\overline{\text{Sd}(\Delta \psi_N)}$,
over homologous sequences in each protein family
for $r_{\script{cutoff}} \sim 8$ and $15.5$\AA , respectively; 
see \Tables{\ref{\TBL: ddPsi_with_8A} and \ref{\TBL: ddPsi_with_16A}}.
}%  SUPPLEMENT
\TEXT{
Plus marks indicate the averages, 
$\overline{\overline{\Delta \psi_N}}$ and $\overline{\text{Sd}(\Delta \psi_N)}$,
over homologous sequences in each protein family
for $r_{\script{cutoff}} \sim 8$ \AA, 
which are listed in \Tables{\ref{\TBL: ddPsi_with_8A}}.
See \Fig{\ref{sfig: ddPhi_mean_vs_sd_at_equil}} for $r_{\script{cutoff}} \sim 15.5$ \AA.
}%  TEXT
}%  FigureLegends
}
\end{figure*}

}%  FigK

\FigL{

\CLEARPAGE
 
\begin{figure*}[h!]
\FigureInLegends{
\SUPPLEMENT{
\centerline{
\includegraphics*[width=82mm,angle=0]{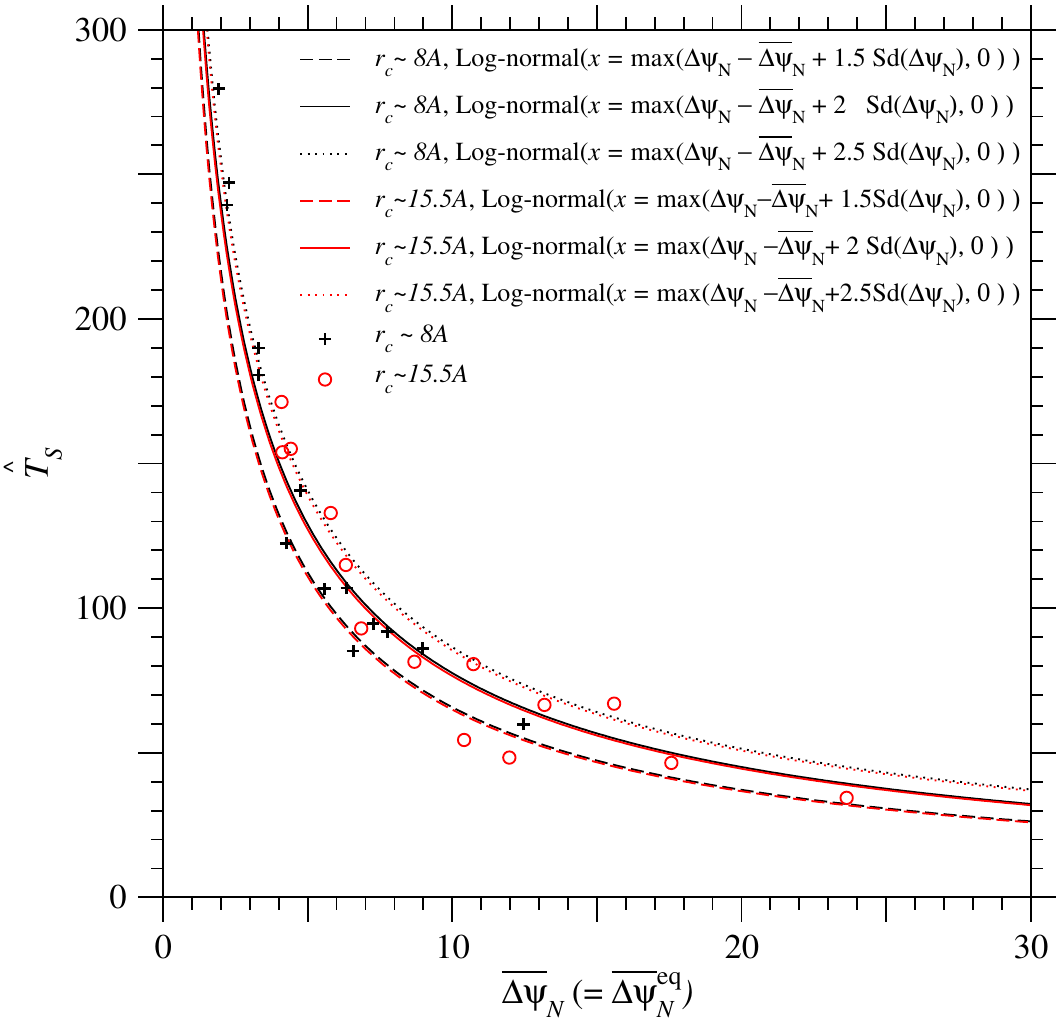}
\includegraphics*[width=82mm,angle=0]{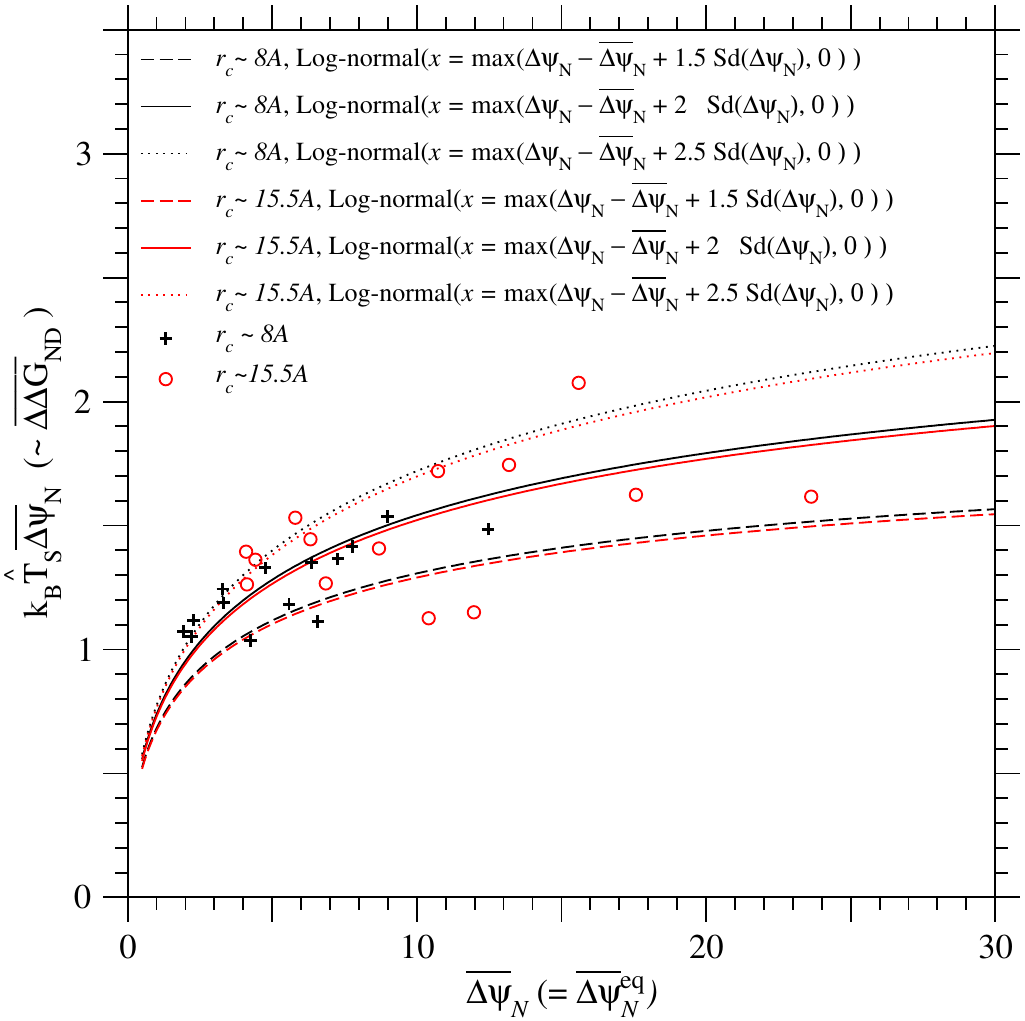}
}
}%  SUPPLEMENT
\TEXT{
\centerline{
\includegraphics*[width=82mm,angle=0]{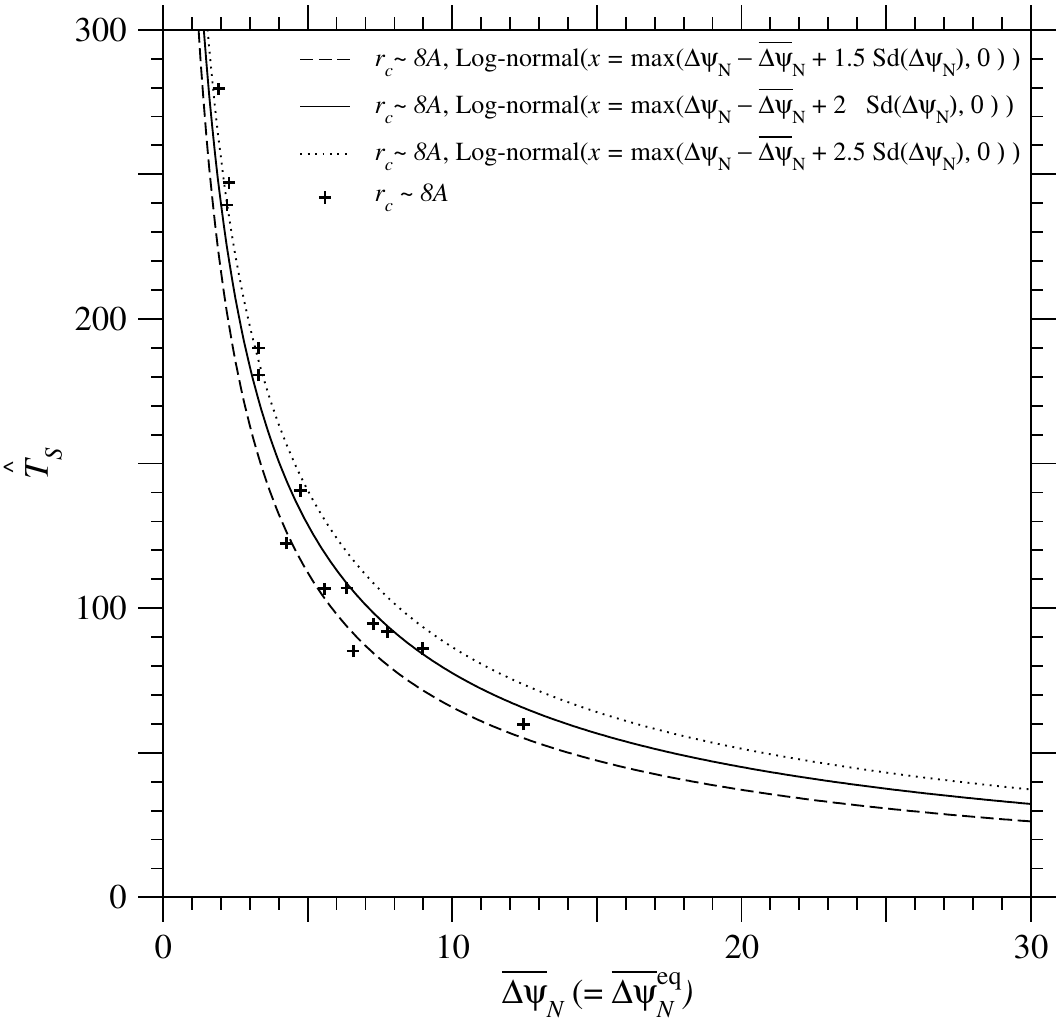}
\includegraphics*[width=82mm,angle=0]{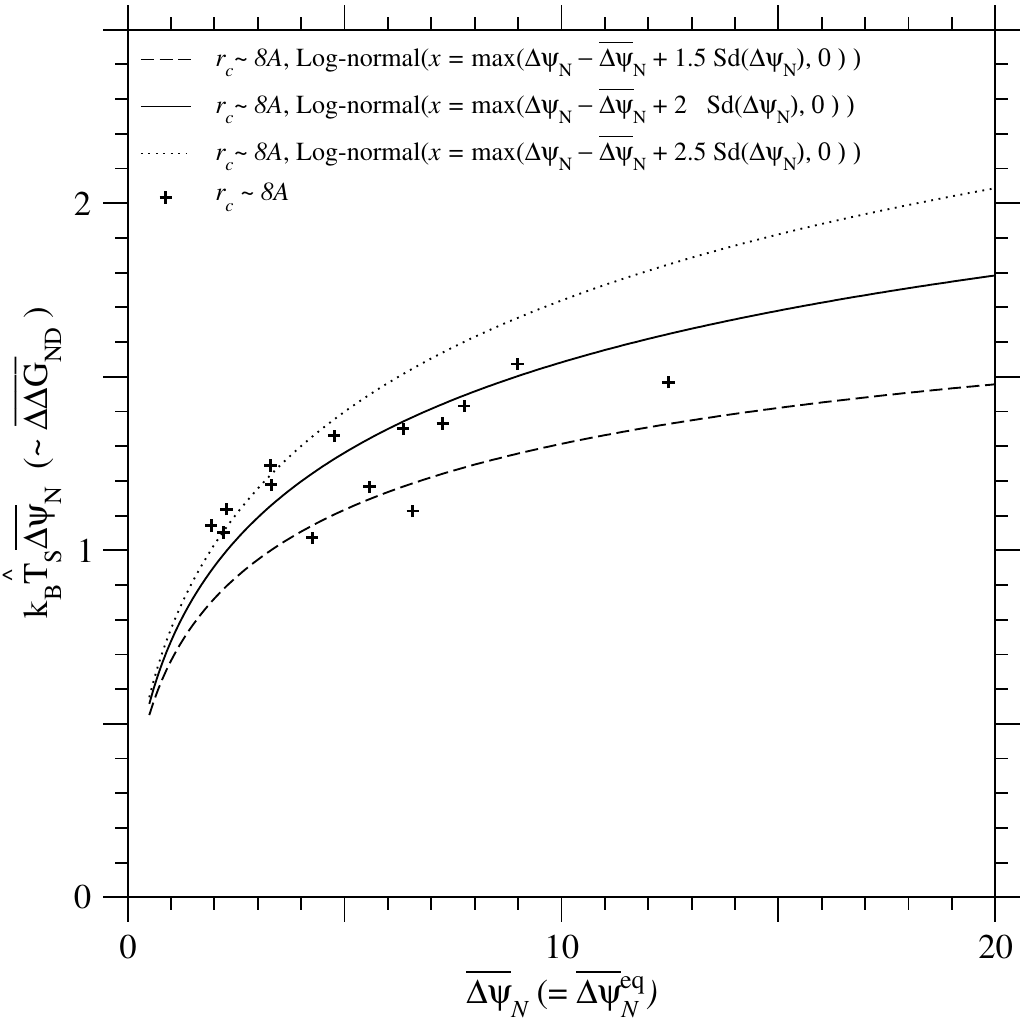}
}
}%  TEXT
}%  FigureInLegends
\vspace*{1em}
\caption{
\SUPPLEMENT{
\label{sfig: ddPhi_mean_vs_Ts_at_equil}
\label{sfig: ddPhi_mean_vs_ddG_at_equil}
}%  SUPPLEMENT
\TEXT{
\label{fig: ddPhi_mean_vs_Ts_at_equil}
\label{fig: ddPhi_mean_vs_ddG_at_equil}
}%  TEXT
\FigureLegends{
\BF{
Relationships between $\hat{T}_s$ and $\overline{\Delta \psi_N}$ and between 
$k_B \hat{T}_s \overline{\Delta \psi_N} (\simeq \overline{\Delta\Delta G_{ND}} )$ 
and $\overline{\Delta \psi_N}$
at equilibrium, 
$\langle \Delta \psi_{N} \rangle_{\script{fixed}} = 0$.
}
The estimate $\hat{T}_s (= (\hat{T}_s \overline{Sd}(\Delta \psi_N))_{PDZ}/Sd(\Delta \psi_N))$ of effective temperature for selection 
and the estimate of mean folding free energy change,
\SUPPLEMENT{
$k_B \hat{T}_s \overline{\Delta \psi_N} (= k_B (\hat{T}_s \overline{Sd}(\Delta \psi_N))_{PDZ}/Sd(\Delta \psi_N) \cdot \overline{\Delta \psi_N} 
\simeq \overline{\Delta\Delta G_{ND}} )$,
}%  SUPPLEMENT
\TEXT{
$k_B \hat{T}_s \overline{\Delta \psi_N} (\simeq \overline{\Delta\Delta G_{ND}} )$,
}%  TEXT
are plotted against $\overline{\Delta \psi_N}$
under the condition of 
$\langle \Delta \psi_{N} \rangle_{\script{fixed}} = 0$.
The $T_s$ is estimated in relative to the $T_s$ of the PDZ family   
in the approximation
that the standard deviation of $\Delta G_{N}$ due to single nucleotide
nonsynonymous mutations is constant irrespective of protein families;
see \Eq{\ref{\EQ: var_of_ddG}}.
Broken, solid, and dotted lines indicate the cases of
log-normal distributions with $n_{\script{shift}} = 1.5, 2.0$ and $2.5$ 
employed to approximate the distribution of $\Delta \psi_N$, respectively; 
see \Eqs{\REF{\EQ: log-normal} to \REF{\EQ: statistics_for_log-normal}}. 
\SUPPLEMENT{
Plus and open circle marks indicate
those estimates against 
the average of $\overline{\Delta \psi_N}$ over homologous sequences
for each protein family
with $r_{\script{cutoff}} \sim 8$ and $15.5$\AA , respectively; 
see \Tables{\ref{\TBL: ddPsi_with_8A} and \ref{\TBL: ddPsi_with_16A}}.
The curves for $r_{\script{cutoff}} \sim 8$ and $15.5$\AA\ almost overlap with each other,
because the estimates of $(\hat{T}_s \overline{Sd}(\Delta \psi_N))_{PDZ}$ for the PDZ 
with $r_{\script{cutoff}} \sim 8$ and $15.5$\AA\  are almost equal to each other.
}%  SUPPLEMENT
\TEXT{
Plus marks indicate
those estimates against 
the average of $\overline{\Delta \psi_N}$ over homologous sequences
for each protein family
with $r_{\script{cutoff}} \sim 8$\AA, 
which are listed in \Tables{\ref{\TBL: ddPsi_with_8A} and \ref{\TBL: Ts_with_8A}}.
See \Fig{\ref{sfig: ddPhi_mean_vs_Ts_at_equil}} for $r_{\script{cutoff}} \sim 15.5$\AA.
}%  TEXT
}%  FigureLegends
}
\end{figure*}

}%  FigL

\FigM{

\CLEARPAGE
 
\begin{figure*}[h!]
\FigureInLegends{
\centerline{
\includegraphics*[width=82mm,angle=0]{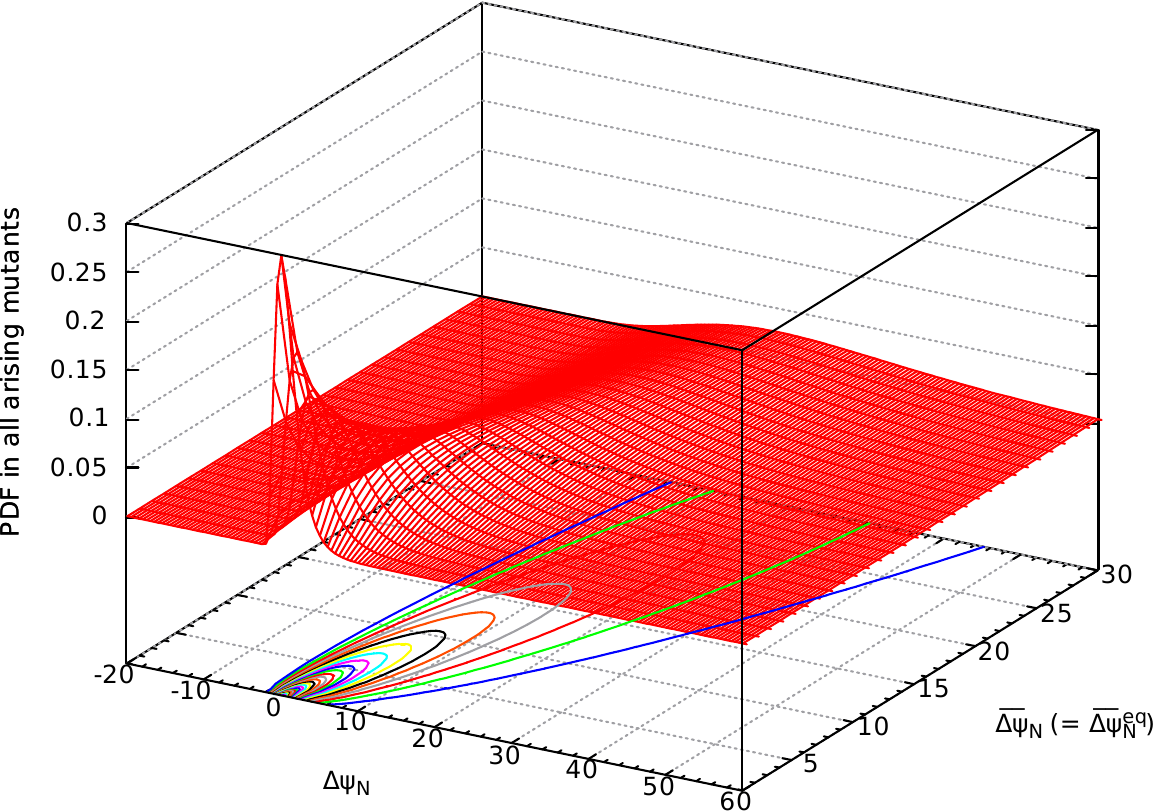}
\includegraphics*[width=82mm,angle=0]{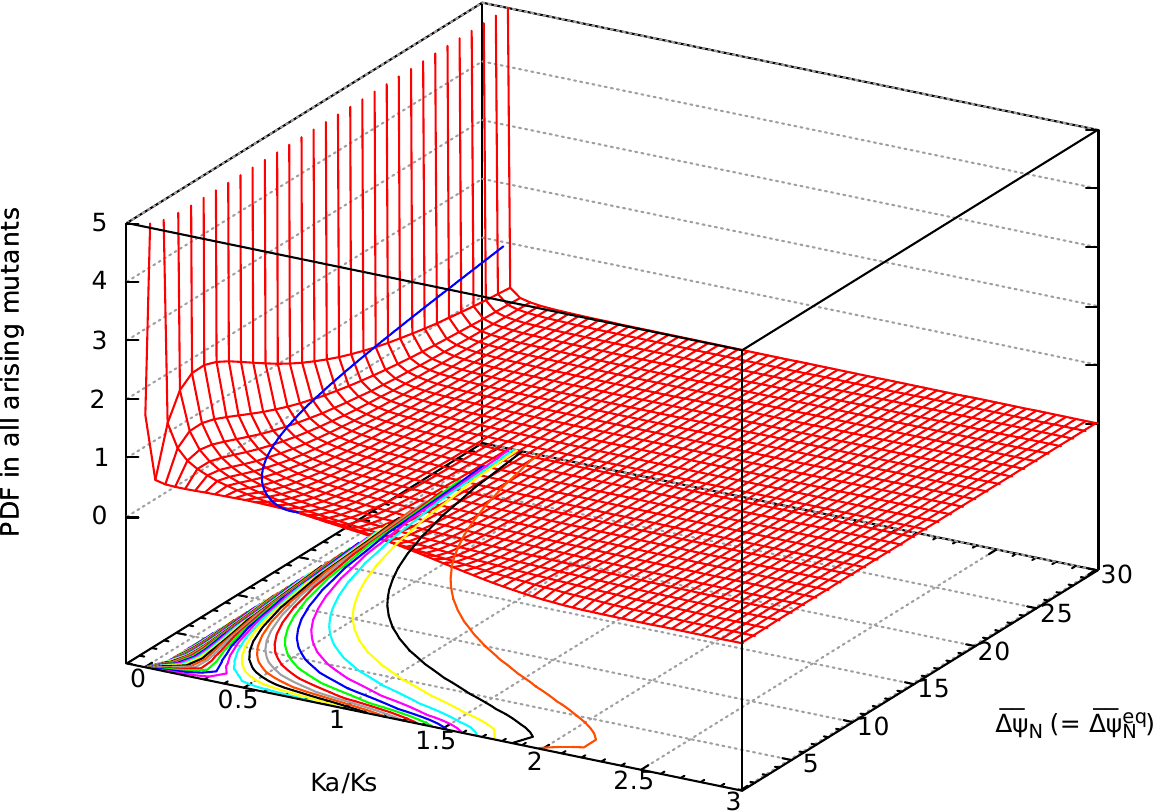}
}
\vspace*{1em}
\centerline{
\includegraphics*[width=82mm,angle=0]{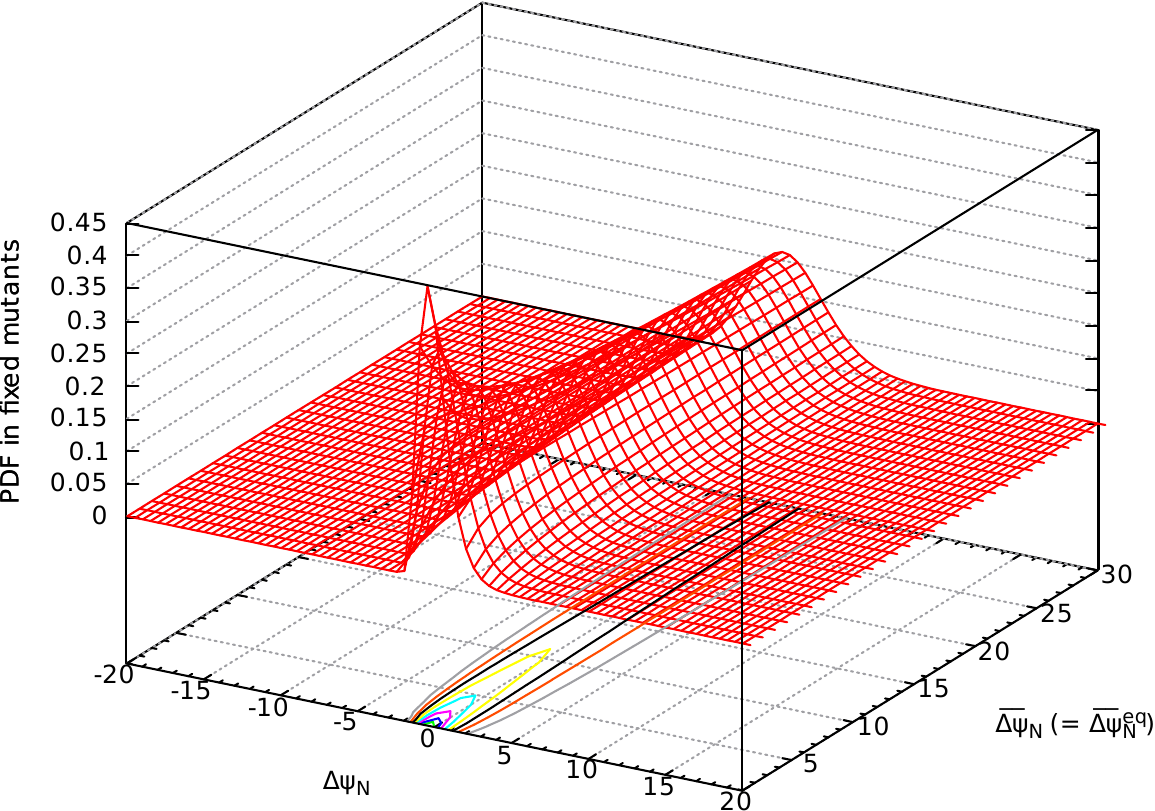}
\includegraphics*[width=82mm,angle=0]{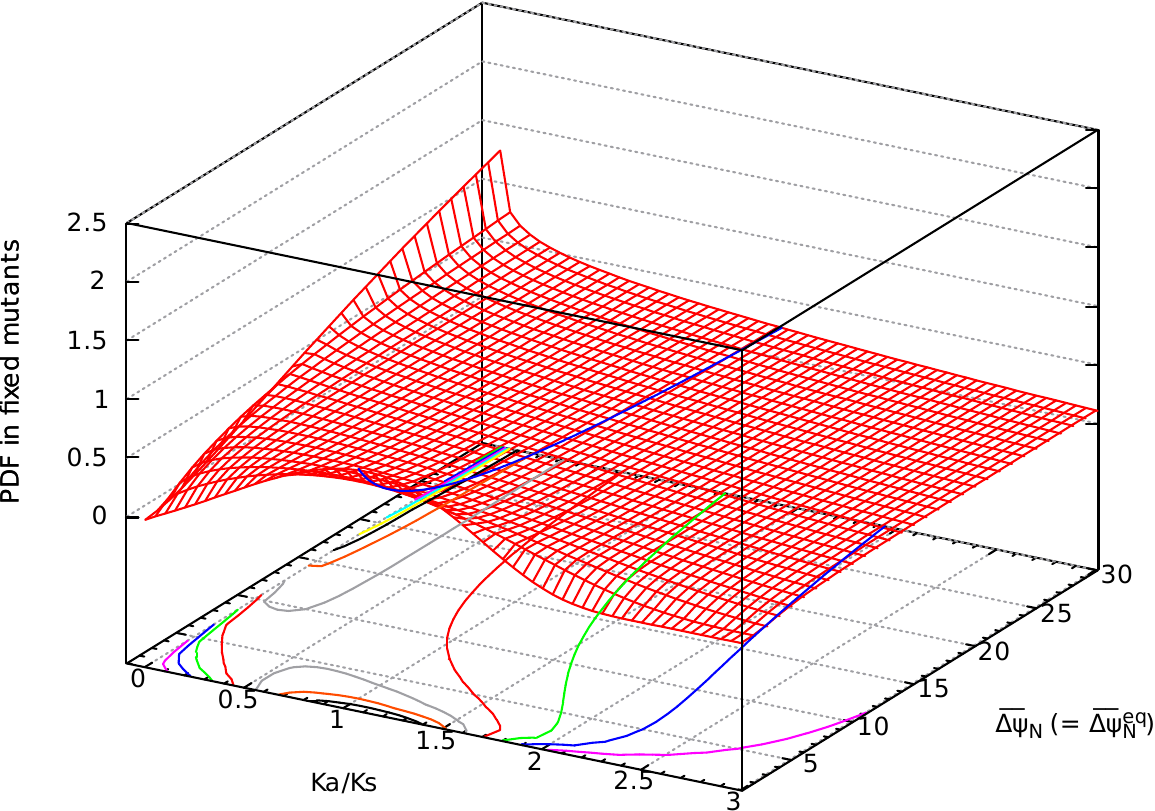}
}
}%  FigureInLegends
\vspace*{1em}
\caption{
\SUPPLEMENT{
\label{sfig: pdf_of_ddPhi_at_equil_for_mean_ddPhi}
\label{sfig: pdf_of_ddPhi_fixed_at_equil_for_mean_ddPhi}
\label{sfig: pdf_of_ka_over_ks_at_equil_for_mean_ddPhi}
\label{sfig: pdf_of_ka_over_ks_fixed_at_equil_for_mean_ddPhi}
}%  SUPPLEMENT
\TEXT{
\label{fig: pdf_of_ddPhi_at_equil_for_mean_ddPhi}
\label{fig: pdf_of_ddPhi_fixed_at_equil_for_mean_ddPhi}
\label{fig: pdf_of_ka_over_ks_at_equil_for_mean_ddPhi}
\label{fig: pdf_of_ka_over_ks_fixed_at_equil_for_mean_ddPhi}
}%  TEXT
\FigureLegends{
\BF{
PDFs of $\Delta \psi_N$ (left) and $K_a/K_s$ (right)
in all singe nucleotide nonsynonymous mutants (upper) and in their fixed mutants (lower)
as a function of 
$\overline{\Delta \psi_N}$
at equilibrium, $\langle \Delta \psi_{N} \rangle_{\script{fixed}} = 0$.
}
Fixation probability has been calculated with $\Delta\Delta \psi_{ND} \simeq \Delta \psi_{N}$;
see \Eqs{\REF{\EQ: fixation_probability} and \REF{\EQ: selective_advantage_and_dPsi_N}}.
The distribution of $\Delta \psi_N$ due to single nucleotide nonsynonymous mutations is approximated by
a log-normal distribution with $n_{\script{shift}} = 2.0$;
see \Eqs{\REF{\EQ: log-normal} to \REF{\EQ: statistics_for_log-normal}}.
The standard deviation of $\Delta \psi_N$ is determined to
satisfy $\langle \Delta \psi_{N} \rangle_{\script{fixed}} = 0$ at
$\overline{\Delta \psi_N} = \overline{\Delta \psi}_N^{\script{eq}}$. 
}%  FigureLegends
}
\end{figure*}

}%  FigM

\FigN{

\CLEARPAGE
 
\begin{figure*}[h!]
\FigureInLegends{
\centerline{
\includegraphics*[width=82mm,angle=0]{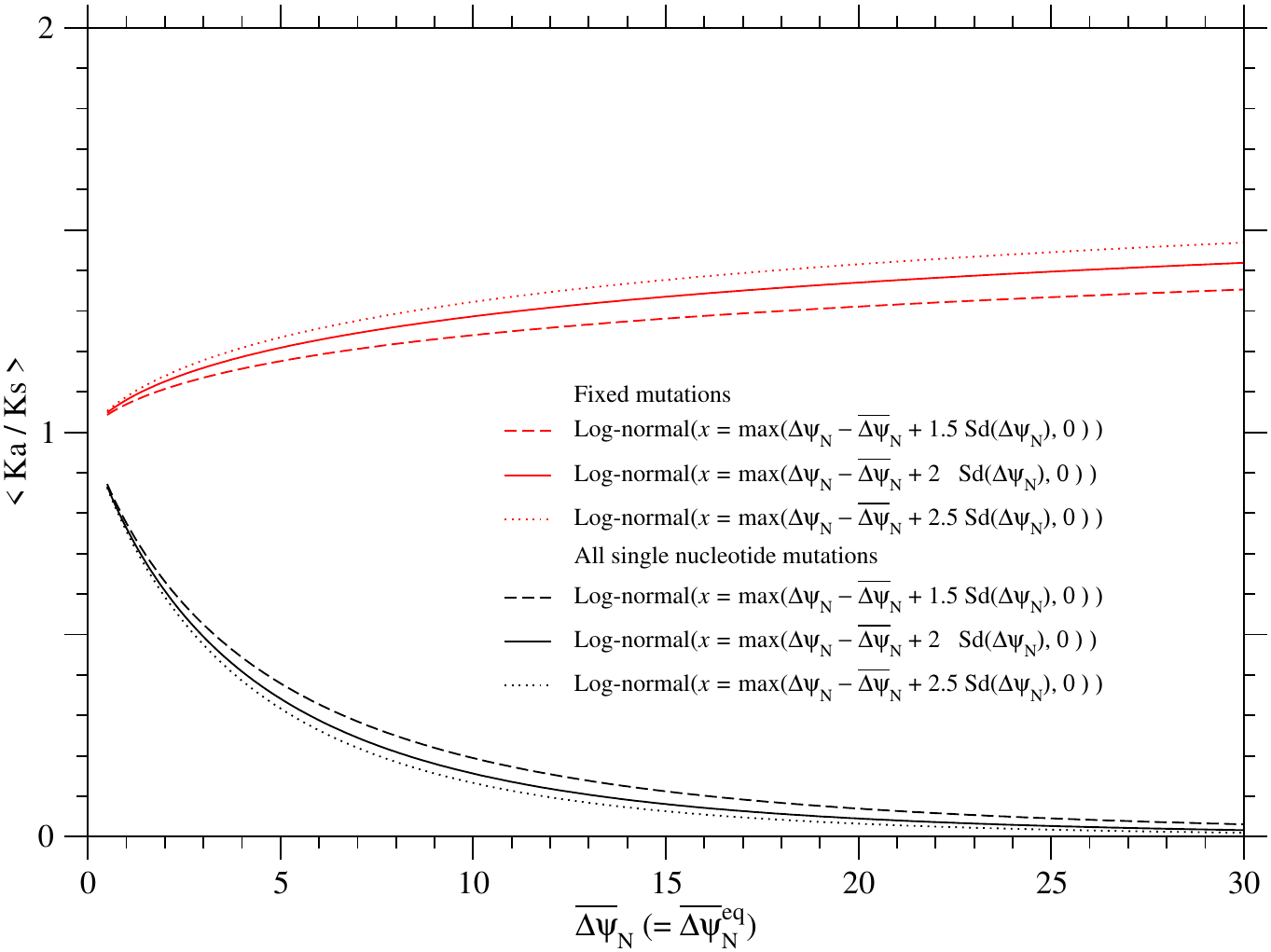}
}
}%  FigureInLegends
\vspace*{1em}
\caption{
\SUPPLEMENT{
\label{sfig: ave_ka_over_ks_at_equil_for_mean_ddPhi}
\label{sfig: ave_ka_over_ks_fixed_at_equil_for_mean_ddPhi}
}%  SUPPLEMENT
\TEXT{
\label{fig: ave_ka_over_ks_at_equil_for_mean_ddPhi}
\label{fig: ave_ka_over_ks_fixed_at_equil_for_mean_ddPhi}
}%  TEXT
\FigureLegends{
\BF{
The averages of $K_a/Ks$ over all single nucleotide nonsynonymous mutations and over
their fixed mutations as a function of $\overline{\Delta \psi_N}$
at equilibrium, $\langle \Delta \psi_{N} \rangle_{\script{fixed}} = 0$.
}
Black and red lines indicate $\langle K_a/K_s \rangle$ and $\langle K_a/K_s \rangle_{\script{fixed}}$,
respectively.
Fixation probability has been calculated with $\Delta\Delta \psi_{ND} \simeq \Delta \psi_{N}$;
see \Eqs{\REF{\EQ: fixation_probability} and \REF{\EQ: selective_advantage_and_dPsi_N}}.
Broken, solid, and dotted lines indicate the cases of
log-normal distributions with $n_{\script{shift}} = 1.5, 2.0$ and $2.5$ 
employed to approximate the distribution of $\Delta \psi_N$, respectively; 
see \Eqs{\REF{\EQ: log-normal} to \REF{\EQ: statistics_for_log-normal}}. 
The standard deviation of $\Delta \psi_N$ is determined to
satisfy 
$\langle \Delta \psi_{N} \rangle_{\script{fixed}} = 0$ at 
$\overline{\Delta \psi_N} = \overline{\Delta \psi}_N^{\script{eq}}$. 
}%  FigureLegends
}
\end{figure*}

}%  FigN

\FigO{

\CLEARPAGE
 
\begin{figure*}[h!]
\FigureInLegends{
\centerline{
\includegraphics*[width=82mm,angle=0]{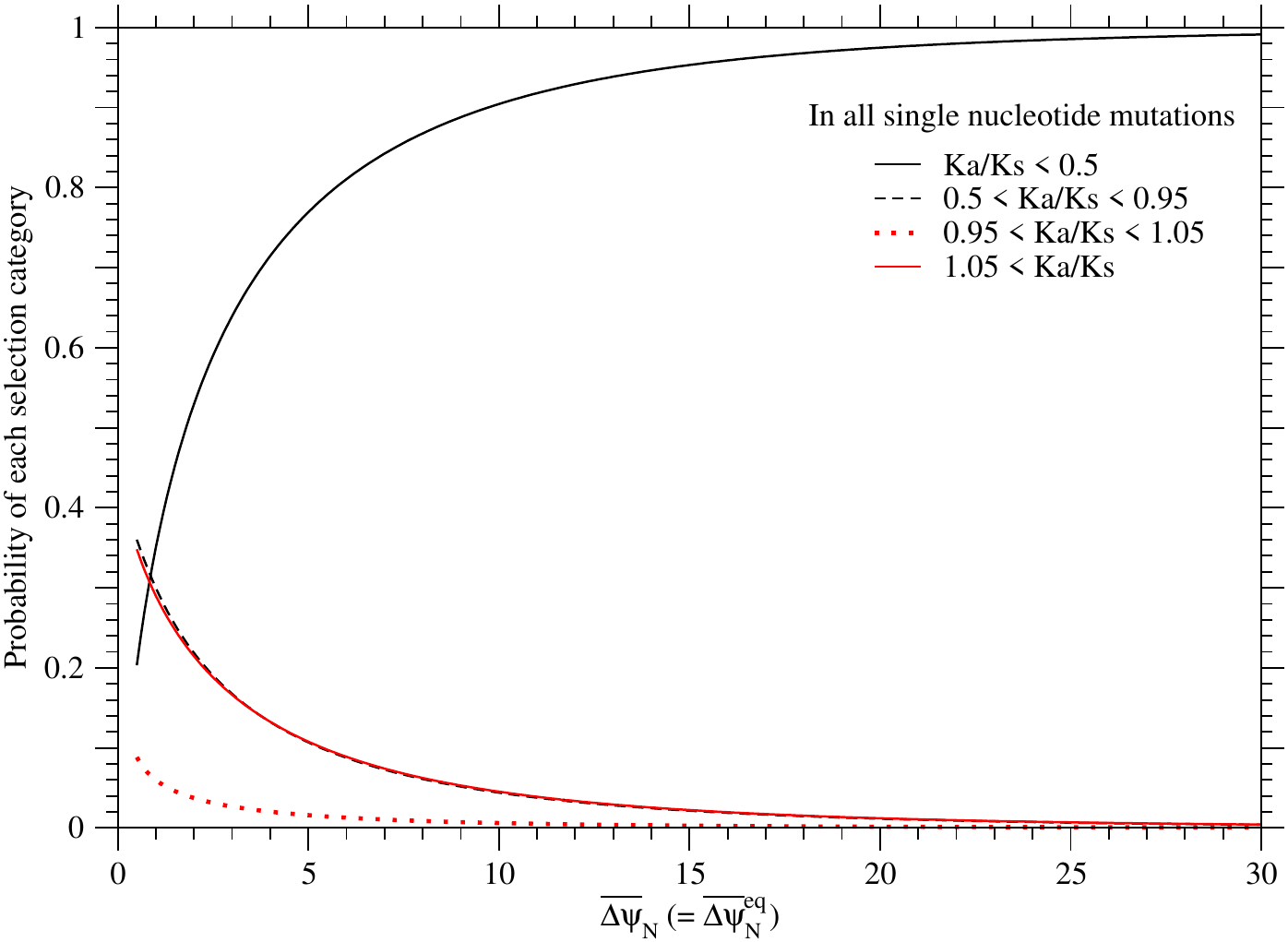}
\includegraphics*[width=82mm,angle=0]{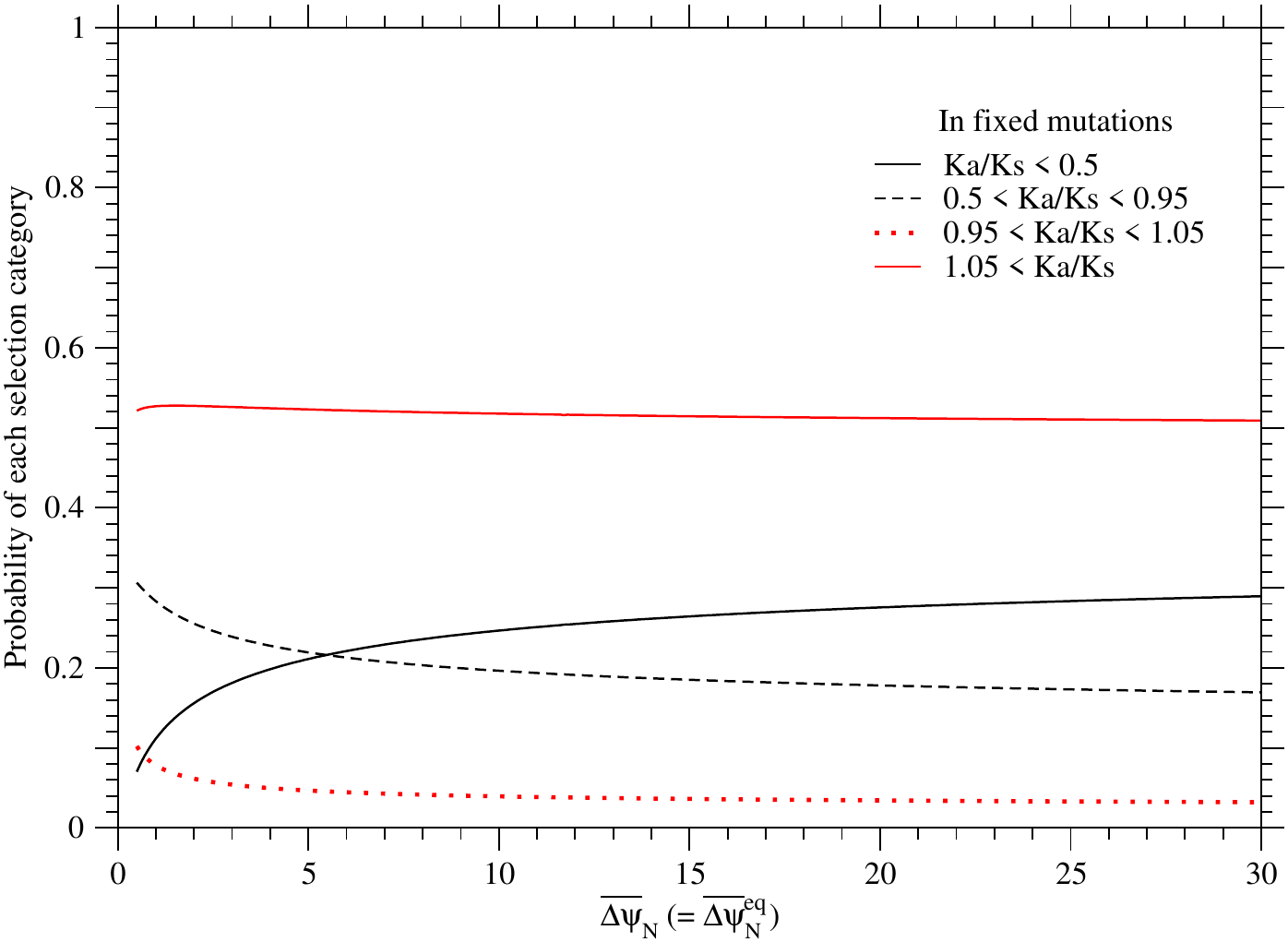}
}
}%  FigureInLegends
\vspace*{1em}
\caption{
\SUPPLEMENT{
\label{sfig: prob_of_each_selection_category_at_equil_for_mean_ddPhi}
\label{sfig: prob_of_each_selection_category_fixed_at_equil_for_mean_ddPhi}
}%  SUPPLEMENT
\TEXT{
\label{fig: prob_of_each_selection_category_at_equil_for_mean_ddPhi}
\label{fig: prob_of_each_selection_category_fixed_at_equil_for_mean_ddPhi}
}%  TEXT
\FigureLegends{
\BF{
The probabilities of each selection category in all single nucleotide nonsynonymous mutations and in
their fixed mutations
as a function of $\overline{\Delta \psi_N}$ 
at equilibrium, $\langle \Delta \psi_{N} \rangle_{\script{fixed}} = 0$.
}
The left and right figures are for single nucleotide nonsynonymous mutations and
for their fixed mutations, respectively.
Red solid, red dotted, black broken, and black solid lines indicate 
positive, neutral, slightly negative and negative selection categories, respectively;
the values of $K_a/K_s$ are divided arbitrarily into four categories,
$K_a/K_s > 1.05$, $1.05 > K_a/K_s > 0.95$, $0.95 > K_a/K_s > 0.5$, and $0.5 > K_a/K_s$,
which correspond to their selection categories, respectively.
Fixation probability has been calculated with $\Delta\Delta \psi_{ND} \simeq \Delta \psi_{N}$;
see \Eqs{\REF{\EQ: fixation_probability} and \REF{\EQ: selective_advantage_and_dPsi_N}}.
The distribution of $\Delta \psi_N$ due to single nucleotide nonsynonymous mutations is approximated by
a log-normal distribution with $n_{\script{shift}} = 2.0$;
see \Eqs{\REF{\EQ: log-normal} to \REF{\EQ: statistics_for_log-normal}}.
The standard deviation of $\Delta \psi_N$ is determined to
satisfy
$\langle \Delta \psi_{N} \rangle_{\script{fixed}} = 0$ at 
$\overline{\Delta \psi_N} (= \overline{\Delta \psi}_N^{\script{eq}} )$.
}%  FigureLegends
}
\end{figure*}

}%  FigO

\FigP{

\CLEARPAGE
 
\begin{figure*}[h!]
\FigureInLegends{
\centerline{
\includegraphics*[width=82mm,angle=0]{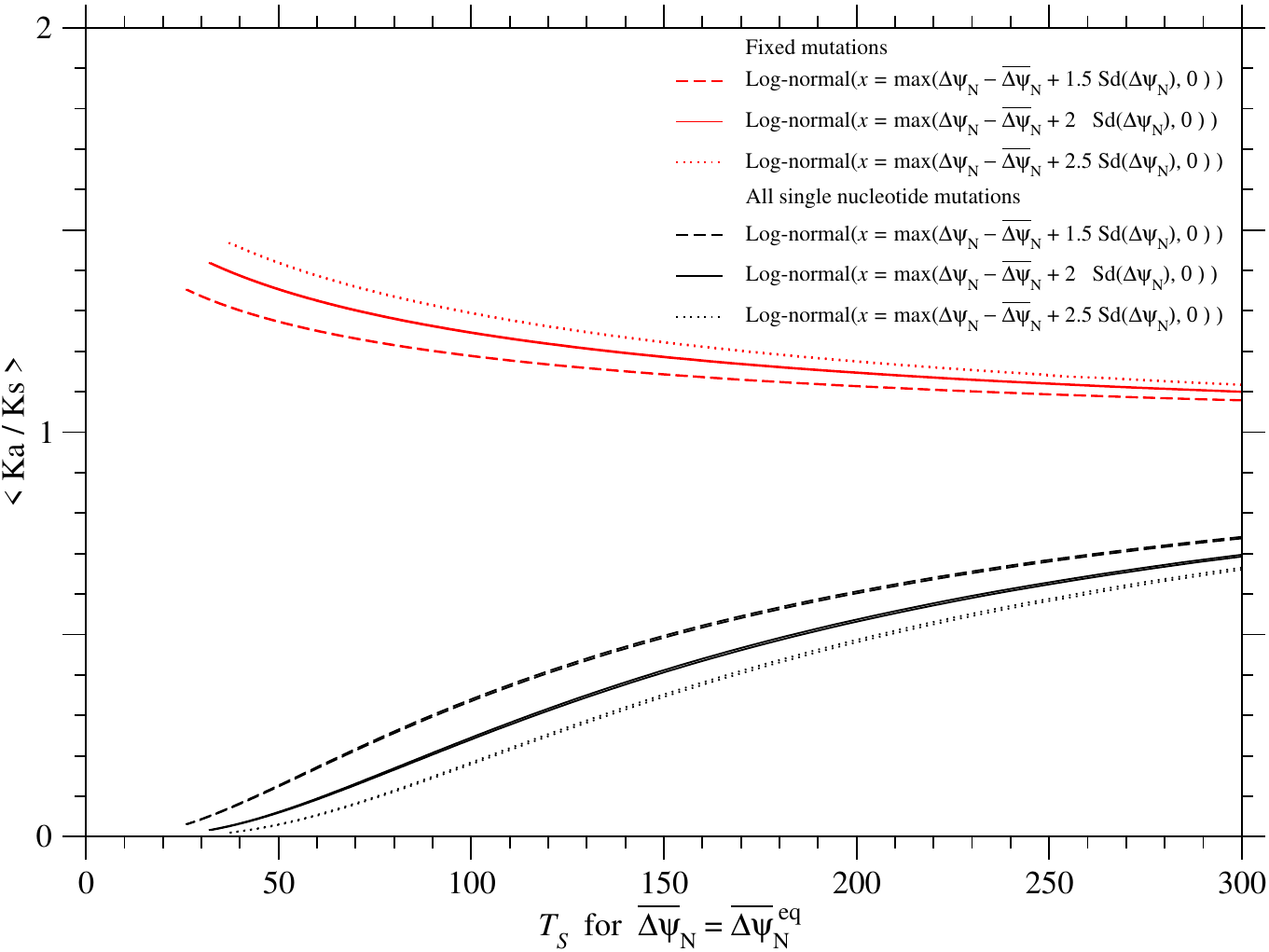}
}
}%  FigureInLegends
\vspace*{1em}
\caption{
\SUPPLEMENT{
\label{sfig: ave_ka_over_ks_at_equil_for_Ts}
\label{sfig: ave_ka_over_ks_fixed_at_equil_for_Ts}
}%  SUPPLEMENT
\TEXT{
\label{fig: ave_ka_over_ks_at_equil_for_Ts}
\label{fig: ave_ka_over_ks_fixed_at_equil_for_Ts}
}%  TEXT
\FigureLegends{
\BF{
The averages of $K_a/Ks$ over all single nucleotide nonsynonymous mutations and over
their fixed mutations as a function of the effective temperature of selection, 
$T_s (= (T_s \overline{Sd}(\Delta\psi_N))_{PDZ} /Sd(\Delta\psi_N) )$, 
at equilibrium, 
$\langle \Delta \psi_{N} \rangle_{\script{fixed}} = 0$.
}
Black and red lines indicate $\langle K_a/K_s \rangle$ and $\langle K_a/K_s \rangle_{\script{fixed}}$,
respectively.
Fixation probability has been calculated with $\Delta\Delta \psi_{ND} \simeq \Delta \psi_{N}$;
see \Eqs{\REF{\EQ: fixation_probability} and \REF{\EQ: selective_advantage_and_dPsi_N}}.
The distribution of $\Delta \psi_N$ due to single nucleotide nonsynonymous mutations is approximated by
a log-normal distribution with $n_{\script{shift}} = 2.0$;
see \Eqs{\REF{\EQ: log-normal} to \REF{\EQ: statistics_for_log-normal}}.
The standard deviation of $\Delta \psi_N$ is determined to
satisfy 
$\langle \Delta \psi_{N} \rangle_{\script{fixed}} = 0$ at 
$\overline{\Delta \psi_N} (= \overline{\Delta \psi}_N^{\script{eq}} )$.
The $T_s$ is estimated in the scale relative to the $T_s$ of the PDZ family   
in the approximation
that the standard deviation of $\Delta G_{N}$ due to single nucleotide
nonsynonymous mutations is constant irrespective of protein families;
see \Eq{\ref{\EQ: var_of_ddG}}.
Broken, solid, and dotted lines indicate the cases of
log-normal distributions with $n_{\script{shift}} = 1.5, 2.0$ and $2.5$ 
employed to approximate the distribution of $\Delta \psi_N$, respectively; 
see \Eqs{\REF{\EQ: log-normal} to \REF{\EQ: statistics_for_log-normal}}. 
The curves for $r_{\script{cutoff}} \sim 8$ and $15.5$\AA\ almost overlap with each other,
because the estimates of $(\hat{T}_s \overline{Sd}(\Delta \psi_N))_{PDZ}$ for the PDZ
with $r_{\script{cutoff}} \sim 8$ and $15.5$\AA\  are almost equal to each other.
}%  FigureLegends
}
\end{figure*}

}%  FigP

% End of figures_JTB_2.tex
% End of sfigures_1+2.tex

}%  SupplementaryMaterial

% End of supplement_1+2.tex

\end{document}